\documentclass[prb,aps,nofootinbib,amssymb,twocolumn,superscriptaddress,10pt]{revtex4-2}
\usepackage{times}
\usepackage{graphicx}
\usepackage{float}
\usepackage{latexsym,amsmath,amssymb,bm,euscript}
\usepackage{color}
\usepackage{subfigure}
\usepackage{epstopdf}
\usepackage[colorlinks=true,linkcolor=blue,citecolor=blue,urlcolor=blue]{hyperref}
\usepackage{hyperref}
\usepackage{comment}
\usepackage{cleveref}
\usepackage{siunitx}
\usepackage{multirow}

\usepackage{physics}

\usepackage{multirow}
\usepackage{bbm}
\usepackage{tabularx}
\usepackage{ltablex}
\usepackage{placeins}
\usepackage{tikz}
\usetikzlibrary{decorations.pathreplacing}
\usepackage{ytableau}
\usepackage{dcolumn}
\usepackage{bm}
\usepackage{array,booktabs,tabularx}
\usepackage[USenglish]{babel}
\usepackage{xcolor}
\usepackage{amsfonts}

\usepackage{placeins}

\let\vec\mathbf

\makeatletter
\renewcommand\onecolumngrid{
	\do@columngrid{one}{\@ne}%
	\def\set@footnotewidth{\onecolumngrid}
	\def\footnoterule{\kern-6pt\hrule width 1.5in\kern6pt}%
}

\def\ba#1\ea{\begin{align}#1\end{align}}  
\let\vec\mathbf

{\catcode`\|=\active
	\xdef\set{\protect\expandafter\noexpand\csname set \endcsname}
	\expandafter\gdef\csname set \endcsname#1{\mathinner
		{\lbrace\,{\mathcode`\|32768\let|\midvert #1}\,\rbrace}}
	\xdef\Set{\protect\expandafter\noexpand\csname Set \endcsname}
	\expandafter\gdef\csname Set \endcsname#1{\left\{%
		\ifx\SavedDoubleVert\relax \let\SavedDoubleVert\|\fi
		\:{\let\|\SetDoubleVert
			\mathcode`\|32768\let|\SetVert
			#1}\:\right\}}
}
\def\midvert{\egroup\mid\bgroup}
\def\SetVert{\@ifnextchar|{\|\@gobble}
	{\egroup\;\mid@vertical\;\bgroup}}
\def\SetDoubleVert{\egroup\;\mid@dblvertical\;\bgroup}

\begingroup
\edef\@tempa{\meaning\middle}
\edef\@tempb{\string\middle}
\expandafter \endgroup \ifx\@tempa\@tempb
\def\mid@vertical{\middle|}
\def\mid@dblvertical{\middle\SavedDoubleVert}
\else
\def\mid@vertical{\mskip1mu\vrule\mskip1mu}
\def\mid@dblvertical{\mskip1mu\vrule\mskip2.5mu\vrule\mskip1mu}
\fi

\definecolor{colorhhy}{rgb}{0.9, 0.17, 0.31}
\definecolor{colorhhy}{rgb}{0.9, 0.17, 0.31}

\allowdisplaybreaks

\crefname{appendix}{Appendix}{Appendices}
\crefname{equation}{Eq.}{Eqs.}
\crefname{figure}{Fig.}{Figs.}
\crefname{table}{Table}{Tables}
\crefname{section}{Section}{Sections}
\crefname{enumi}{Case}{Cases}
\creflabelformat{appendix}{[#2#1#3]}
\crefrangelabelformat{figure}{#3#1#4--(#5\crefstripprefix{#1}{#2}#6}
\crefmultiformat{figure}{Figs.~#2#1\xdef\mycreffirstarg{#1}#3}%
 { and~#2#1#3}{, #2#1#3}{ and~#2#1#3}

\makeatletter     
\renewcommand\onecolumngrid{
\do@columngrid{one}{\@ne}%
\def\set@footnotewidth{\onecolumngrid}
\def\footnoterule{\kern-6pt\hrule width 1.5in\kern6pt}%
}

\makeatletter
\AtBeginDocument{\let\LS@rot\@undefined}
\makeatother

\crefname{appendix}{Appendix}{Appendices}
\crefname{equation}{Eq.}{Eqs.}
\crefname{figure}{Fig.}{Figs.}
\crefname{table}{Table}{Tables}
\crefname{section}{Section}{Sections}
\creflabelformat{appendix}{[#2#1#3]}

\makeatletter
\AtBeginDocument{\let\LS@rot\@undefined}
\makeatother

\makeatletter
\renewcommand\onecolumngrid{\do@columngrid{one}{\@ne}\def\set@footnotewidth{\onecolumngrid}\def\footnoterule{\kern-6pt\hrule width 1.5in\kern6pt}}

\newcommand{\siSection}{appendix}
\newcommand{\citeSI}[1]{(see \cref{#1})}
\newcommand{\refCiteSI}[1]{\cref{#1}}
\allowdisplaybreaks

\begin{document}
\title{Kagome Materials I: SG 191, ScV$_6$Sn$_6$. Flat Phonon Soft Modes and Unconventional CDW Formation: Microscopic and Effective Theory}
\author{Haoyu Hu}
	\thanks{These authors contributed equally to this work.}
	\affiliation{Donostia International Physics Center (DIPC), Paseo Manuel de Lardizábal. 20018, San Sebastián, Spain}

	\author{Yi Jiang}
	\thanks{These authors contributed equally to this work.}
	\affiliation{Beijing National Laboratory for Condensed Matter Physics, and Institute of Physics, Chinese Academy of Sciences, Beijing 100190, China}
	\affiliation{University of Chinese Academy of Sciences, Beijing 100049, China}
	
	\author{Dumitru C\u{a}lug\u{a}ru}
	\thanks{These authors contributed equally to this work.}
	\affiliation{Department of Physics, Princeton University, Princeton, NJ 08544, USA}
	
	\author{Xiaolong Feng}
	\thanks{These authors contributed equally to this work.}
	\affiliation{Donostia International Physics Center (DIPC), Paseo Manuel de Lardizábal. 20018, San Sebastián, Spain}
	
	\affiliation{Max Planck Institute for Chemical Physics of Solids, 01187 Dresden, Germany}

	\author{David Subires}
	\affiliation{Donostia International Physics Center (DIPC), Paseo Manuel de Lardizábal. 20018, San Sebastián, Spain}

	\author{Maia G.~Vergniory}
	\affiliation{Donostia International Physics Center (DIPC), Paseo Manuel de Lardizábal. 20018, San Sebastián, Spain}
	\affiliation{Max Planck Institute for Chemical Physics of Solids, 01187 Dresden, Germany}
		
	\author{Claudia Felser}
	\affiliation{Max Planck Institute for Chemical Physics of Solids, 01187 Dresden, Germany}

	\author{Santiago Blanco-Canosa}
	\affiliation{Donostia International Physics Center (DIPC), Paseo Manuel de Lardizábal. 20018, San Sebastián, Spain}
	\author{B.~Andrei Bernevig}
	\email{bernevig@princeton.edu}
	\affiliation{Donostia International Physics Center (DIPC), Paseo Manuel de Lardizábal. 20018, San Sebastián, Spain}
	\affiliation{Department of Physics, Princeton University, Princeton, NJ 08544, USA}
	\affiliation{IKERBASQUE, Basque Foundation for Science, 48013 Bilbao, Spain}

\begin{abstract}
Kagome Materials with flat bands exhibit wildly different physical properties depending on symmetry group, and electron number. Their complicated physics and even the one-particle "spaghetti" of electron/phonon bands are so far amenable only to phenomenological interpretation. In this first paper of a series, we analyze the case of the kagome 166 material ScV$_6$Sn$_6$ in SG 191, with Fermi level away from the spaghetti flat bands (with a hidden structure~\cite{JIA23}) endemic of these materials. Experimentally, a $\sim$95K charge density wave (CDW) at vector $\bar{K}=(\frac{1}{3},\frac{1}{3},\frac{1}{3})$ exists, with no nesting/peaks in the electron susceptibility at $\bar{K}$. We show that ScV$_6$Sn$_6$ has a collapsed phonon mode at $H=(\frac{1}{3},\frac{1}{3},\frac{1}{2})$ and an imaginary flat phonon band in the $H$-vicinity. The soft phonon is supported on the triangular Sn (Sn$^T$) $z$-directed mirror-even vibrations. A faithful, three-degree-of-freedom simple force constant model describes the entire soft phonon dispersion. At high temperatures, our arguments and \textit{ab-initio} calculations show a very flat in-plane phonon band at most $k_z$’s, with increasing $z$-dispersion as a function of $k_z$ away from $H$. Sn$^T$ $p_z$ orbitals contribute to the Fermi level bands and a strong $p_z$ Sn$^T$ electron-even Sn$^T$ phonon coupling softens the $H$ mode. We model it by a new (Gaussian) approximation of the hopping parameter~\cite{YU23}, and show that the resulting field-theoretical renormalization of the phonon frequency reproduces the collapse of the $H$ phonon and induces small in-plane dispersion away from $H$. To explain the appearance of the $\bar{K}$ charge density wave (CDW) we build an effective model of two order parameters (OPs) -- one at the collapsed phonon $H$ and one at the CDW $\bar{K}$. Comparing with experimental data~\cite{KOR23}, we show that the $H$ OP undergoes a second-order phase transition; however its flatness around $H$ induces large fluctuations. The $\bar{K}$ OP is first order, competes with $H$ and its transition is induced by the large fluctuations of $H$ OP. We construct CDW OPs in the electron and phonon fields that match the \textit{ab-initio} calculations. We furthermore develop the theory for the very similar compound YV$_6$Sn$_6$ which however does not have a CDW phase; our approach explains this difference. In YV$_6$Sn$_6$ the out-of-plane phonon acquires weight on the heavier Y atom; Y does not participate in the susceptibility, hence lowering the electron renormalization of the phonon band. Our results not only explain the CDW in ScV$_6$Sn$_6$, but show an unprecedented level of modeling of complex electronic systems that open new collaborations of \textit{ab-initio} and analytics.
\end{abstract}
\maketitle

\begin{figure*}[!t]
    \centering
    \includegraphics[width=2.0\columnwidth,draft=false]{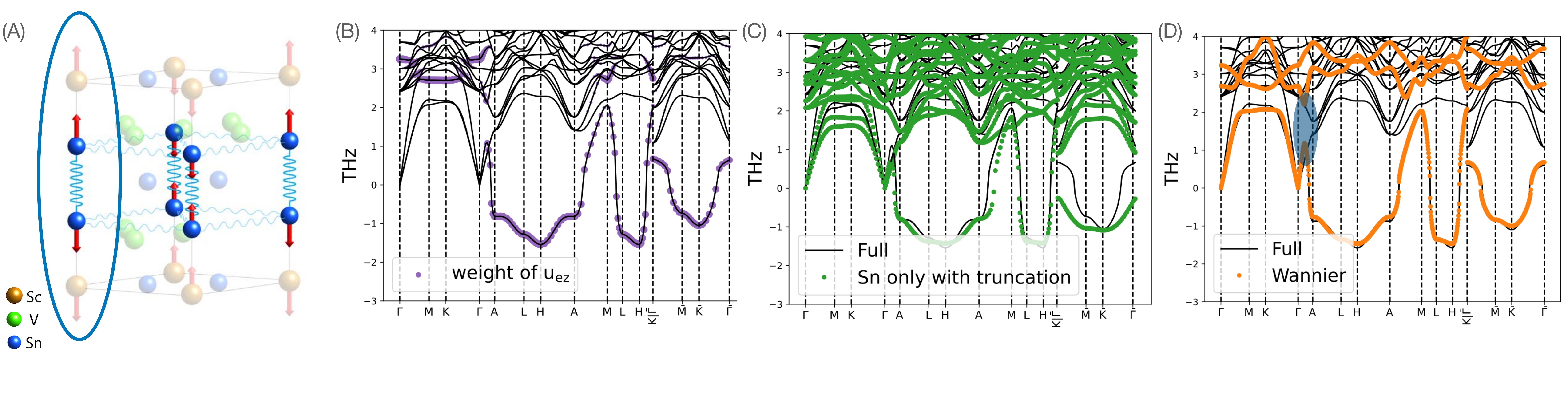}
    \caption{{\bf Imaginary flat phonon modes.} (A) Vibration mode of the lowest-energy phonon band at $H$, with the blue circle marking the effective 1D phonon model. (B) DFT-calculated phonon spectrum. 
(C) Comparison between the full phonon spectrum from DFT calculation (black solid line) and spectrum from the Sn-only model (green dots). (D) Comparison between the full phonon spectrum from DFT calculations (black solid line) and the simple three-orbital phonon model (orange dots) derived via Wannier construction. The simple three-orbital phonon model has captured the imaginary flat phonon bands and also the avoided level crossing along $\Gamma$-$A$ line (blue shaded area).
    $\bar{\Gamma} = (0,0,\frac{1}{3}), \bar{M} = (\frac{1}{2},0,\frac{1}{3})$
    }
    \label{fig:main:pho_spec}
\end{figure*}

\emph{Introduction.}  Due to the frustrated geometry and non-trivial electronic properties~\cite{ORT19,CHO21b,DEN21,HER20a,ISH21,JIA23,ORT20,KAN21a,LI21b,LI21c,LIN21,LIU23,NEU22,ORT21,PAL22,ZHA21b,ZHA22b}, kagome materials have been found to exhibit rich physics, including charge density waves~\cite{CHE22a,DIE21,FER22,JIA21,KEN21a,LI21a,LI23,LIA21a,LIU21b,LUO22,MIE22,RAT21,SHU21,SON21a,SET21,TAN21,TSI22,TSV23,UYK21,UYK22,WAN21b,WAN21c,WAN21d,WAN21g,YU21,ZHU22,ZHA21b} and superconductivity~\cite{CHE21a,CHE21b,DU21,DUA21,FEN21,KAN23a,LI22a,LIU21d,MU21,NAK21,NI21,ORT21a,SHR22,SON21b,WAN21e,WAN21f,WAN23,WU21b,XIA21,XU21b,YIN21,YIN21a,YU21a,ZHA22a}. However, due to the complicated lattice structures and numerous electronic degrees of freedom, it has been a challenge to provide a faithful theoretical understanding beyond the phenomenological theory and rough scenarios; most works treat the electrons in these materials as $s$-wave kagome flat bands, an inaccurate description of the kagome $d$-orbitals at the Fermi level mixed with the hexagonal and triangular $p_{x,y,z}$ orbitals which, in general, would \emph{not} result in flat bands \cite{JIA23}. In this work, the first part of a series, we perform a systematic theoretical study of the ScV$_6$Sn$_6$ -- one of the kagome materials in the 166 family~\cite{POK21,ROM11,YIN20a,CHE21c} that has recently attracted intense attention~\cite{KOR23, ARA22,TAN23,ZHA22a,HU23d,LEE23,TUN23,CAO23,HU23e,KAN23,GU23,GUG23, CHE23,MOZ23,YI23}. In this material, a soft flat phonon band has been experimentally observed~\cite{KOR23,CAO23}, whose second-order collapse happens at the same time as a first-order phase transition to a \emph{different momentum} CDW phase~\cite{KOR23}.
By combining \textit{ab-initio} and analytical methods, we overcome the complexity of the material, and provide a comprehensive understanding of the experimental observations, \emph{from microscopics to effective models}. Moreover, the range of techniques developed here can be applied to other kagome materials~\cite{JIA23}, and opens a new route to explain their complicated properties.

To be specific, we study the imaginary flat phonon, CDW transitions, and the properties of CDW phase of ScV$_6$Sn$_6$. We first identify an imaginary (soft) phonon mode with a flat dispersion near the $H$ point. Its softening is attributed to the electron-phonon coupling and the charge fluctuations of the mirror-even electron orbitals formed by the $p_z$ orbitals of two Sn atoms.
As the $H$ phonon collapses, a strong fluctuation, induced by the flatness of phonon bands, stabilizes a CDW phase at $\bar{K}$ via a first-order phase transition. 
We then discuss the properties of the CDW phase from its atomic displacements and electronic order parameters.

\emph{Imaginary (soft) almost-flat phonon bands.} We first perform an \textit{ab-initio} DFT calculation of the phonon spectrum at zero temperature. We observe an imaginary (soft) phonon mode with leading order instability at $H$ (Fig.~\ref{fig:main:pho_spec} (B)). The vibration mode of the lowest-energy phonon at $H$ point is shown in Fig.~\ref{fig:main:pho_spec} (A). The soft phonon band is relatively flat near the $H$ point, which leads to a soft phonon in a large region of the Brillouin zone (BZ) (Fig.~\ref{fig:main:pho_spec} (B))~\cite{KOR23}.
The phonon spectrum can be quantitatively captured by a force-constant model with only Sn atoms (the heaviest in the material) and truncated coupling, as shown in Fig~\ref{fig:main:pho_spec} (C)~\citeSI{app:sec:phonon_spec}. An even simpler three-``orbital'' phonon model could also be derived via a Wannier construction, where the three orbitals refer to the $z$-directed vibrations of two triangular Sn atoms and a third orbital corresponding to the collective $z$-directed displacement of all the other atoms~\citeSI{app:sec:three_band_model}.

The three-orbital phonon model can be further simplified by dropping the third orbital.
The resulting model describes a series of one-dimensional (1D) phonon chains formed by two triangular Sn$^T$ atoms, as illustrated in Fig.~\ref{fig:main:pho_spec} (A). 
The effective 1D phonon chain is characterized by a dynamical matrix $D_{1D}$ taking the form of a Su–Schrieffer–Heeger chain~\cite{SU79}: 
\begin{eqnarray}
&&D_{1D}(\mathbf{k}) = 
\begin{bmatrix}
    d_1 +d_2 & - d_1e^{ik_z(1-2z_0)}  -d_2 e^{ik_z(-2z_0)} \\ 
\text{h.c.}
     & d_1 +d_2 
\end{bmatrix} \nonumber\\ 
\label{eq:main_1d_dynmat}
\end{eqnarray}
where $z_0$ and $1-2z_0$ characterize the $z$ coordinates of the two triangular Sn atoms. $d_1(<0)$ and $d_2(>0)$ denote the intra- and inter-unit-cell coupling between $z$-directed vibration of two triangular Sn atom~\citeSI{app:sec:phonon_spec}. 
The lowest eigenvalue $(2d_1)$ of the dynamical matrix is negative and produces an imaginary flat phonon mode at $k_z=\pi$ plane. 
The eigenvector of the imaginary phonon is characterized by the mirror-even vibration mode $u_{ez}(\mathbf{q}=\pi)$, which is also confirmed by the DFT calculations (Fig.~\ref{fig:main:pho_spec} (B)). In real space, this mirror-even mode takes the form of
\begin{eqnarray} 
u_{ez}(\mathbf{R}) =\frac{1}{\sqrt{2}}(u_{Sn^T_1,z}(\mathbf{R}) -u_{Sn^T_2,z}(\mathbf{R}))
\label{eq:main_ph_even}
\end{eqnarray} 
where ${u}_{Sn^T_{i},z}(\mathbf{R}), i=1,2$ denote
the $z$-direction movements of two triangular Sn atoms at unit cell $\mathbf{R}$.

\begin{figure*}[!t]
	\centering
	\includegraphics[width=2.0\columnwidth,draft=false]{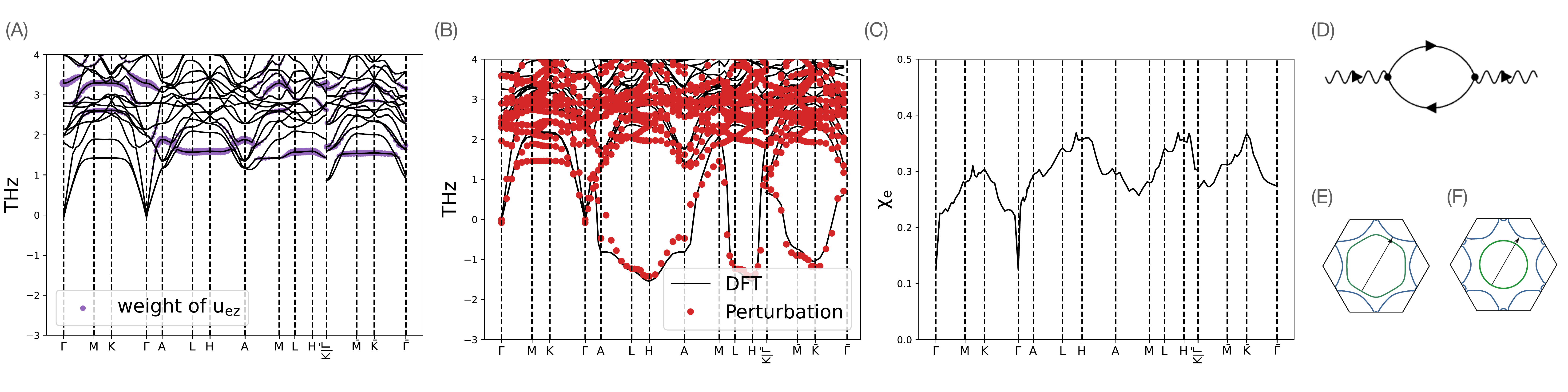}
	\caption{{\bf Origin of imaginary phonon mode.} (A) High-temperature (non-interacting) phonon spectrum from DFT calculations at $T=1.2$eV. (B) Low-temperature (zero-temperature) phonon spectra from DFT and perturbation calculations. (C) Charge susceptibility of the mirror-even electron $c_{\mathbf{R},e,\sigma}$. (D) One-loop corrections to the phonon self-energy. The solid and curly lines denote the non-interacting electron propagator and non-interaction phonon propagator respectively. (E) Weak Fermi surface nesting between the Fermi surface at $k_z/(2\pi)=-0.32$ (green line) and $k_z/(2\pi)=0.18$ (blue line) planes with nesting vector near $H$. (F) Weak Fermi surface nesting between the Fermi surface at $k_z/(2\pi)=0.20$ (blue line) and $k_z/(2\pi)=0.53$ (green line) planes with nesting vector $\bar{K}$.}
	\label{fig:main:ele_ph_corr}
\end{figure*}

\emph{Origin of the imaginary phonon mode.} To understand the origin of the imaginary phonon mode, we calculate the electron corrections to the phonon propagator via perturbation theory with the corresponding one-loop Feynman diagram shown in Fig.~\ref{fig:main:ele_ph_corr} (D)~\citeSI{app:sec:ele_phon_model}. Our calculation is based on a recently developed Gaussian approximation of electron-phonon coupling~\cite{YU23}, the DFT-calculated electron bands, and the DFT-calculated high-temperature phonon spectrum. The latter can be understood as a ``non-interacting'' phonon spectrum with vanishing electron corrections.
We find a great match between the DFT and the perturbation calculations as shown in Fig.~\ref{fig:main:ele_ph_corr} (B).
Crucially, our perturbation calculation identifies the main driving force behind the imaginary phonon model as we now discuss.

In the high-temperature phonon spectrum, the lowest-energy phonon band is mostly formed by $u_{ez}$ and is extremely flat (Fig.~\ref{fig:main:ele_ph_corr} (A)). 
At low temperature, we find the imaginary phonon mode is mostly driven by the electron-phonon coupling between the mirror-even phonon fields $u_{ez}(\mathbf{R})$ and the mirror-even Sn$^T$ electron fields $c_{\mathbf{R},e,\sigma}$ 
, defined as:
\begin{equation}
c_{\mathbf{R}, e,\sigma} = \frac{1}{\sqrt{2}} \bigg( c_{\mathbf{R}, (Sn_1^T,p_z), \sigma} - c_{\mathbf{R}, (Sn_2^T,p_z), \sigma}
\bigg), 
\label{eq:main_ele_even}
\end{equation} 
where $c_{\mathbf{R}, (Sn_{i=1,2}^T,p_z), \sigma} $ are the annihilation operators of $p_z$-orbital with spin $\sigma$ of two Sn$^T$ atoms respectively. 
The corresponding electron-phonon coupling 
is~\citeSI{app:sec:ele_phon_model}
\begin{equation}
\tilde{g}\sum_{\mathbf{R},\sigma } u_{ez}(\mathbf{R}) 
c_{\mathbf{R},e,\sigma}^\dag  c_{\mathbf{R},e,\sigma}
\label{eq:main_ele_ph}
\end{equation} 
Even though the electron-phonon coupling takes the form of an ``on-site'' coupling, both the phonon fields (Eq.~\ref{eq:main_ph_even}) and the electron fields (Eq.~\ref{eq:main_ele_even}) are ``molecular orbitals'' formed by the superposition of phonon fields and electron fields of two triangular Sn atoms, respectively. Via electron-phonon coupling, the charge fluctuations of $c_{\mathbf{R},e,\sigma}$, which is characterized by the charge susceptibility $\chi_e$, induce a normalization to the $u_{ez}(\mathbf{R})$ phonon fields. In Fig.~\ref{fig:main:ele_ph_corr} (C), we show the behavior of $\chi_e$. We find a weak momentum-dependency in $\chi_{e}$ with two weak peaks: one near $H$ point and the other one at $\bar{K}$. Both peaks in $\chi_e$ come from the weak Fermi-surface nestings, as illustrated in Fig.~\ref{fig:main:ele_ph_corr} (E), (F).  Since the momentum dependency of $\chi_e$ is weak, $\chi_e$ can be approximately described by the following ansatz in the real space
\begin{equation}
    \chi_e(\mathbf{R}) \approx \chi^{on\text{-}site} \delta_{\mathbf{R},\bm{0}}+ \sum_{i=1,...,6}\chi^{xy}\delta_{\mathbf{R},\mathbf{R}_i^{xy}} +\sum_{i=1,2}\chi^{z}\delta_{\mathbf{R},\mathbf{R}_i^{z}}
\end{equation}
where $\chi^{on\text{-}site}$ denotes the dominant on-site contribution, and $\chi^{xy}$ and $\chi^z$ denote relatively weak in-plane and out-of-plane nearest-neighbor contributions, respectively, with $\mathbf{R}_{i}^{xy},\mathbf{R}_i^z$ labeling the in-plane and out-of-plane nearest-neighbors~\citeSI{app:sec:ele_corr_svs}. 
The strong on-site term $\chi^{on\text{-}site}$ lowers the energy of $u_{ez}$ 
and produce an imaginary phonon mode.
The weak in-plane charge correlations characterized by $\chi^{xy}$ lead to a weak-in-plane dispersion of the phonon mode with the leading-order instability at $H$ point. The weak $\chi^{z}$ does not
change the behaviors of the phonon spectrum qualitatively.

\emph{Comparison to the non-CDW compound YV$_6$Sn$_6$}. Based on the same methods, we perform calculations on YV$_6$Sn$_6$, which has the same structure as ScV$_6$Sn$_6$ but with Sc replaced by Y~\citeSI{app:sec:yvs}. Since Y is heavier than Sc, Y contributes to the lowest-energy phonon band and reduces the weight of $u_{ez}$. Moreover, YV$_6$Sn$_6$ also has a weaker charge susceptibility $\chi_e$. Consequently, the electron corrections to the phonon bands are reduced and we do not observe any imaginary phonon~\citeSI{app:sec:yvs}.

\begin{figure*}
    \centering
    \includegraphics[width=2.0\columnwidth,draft=false]{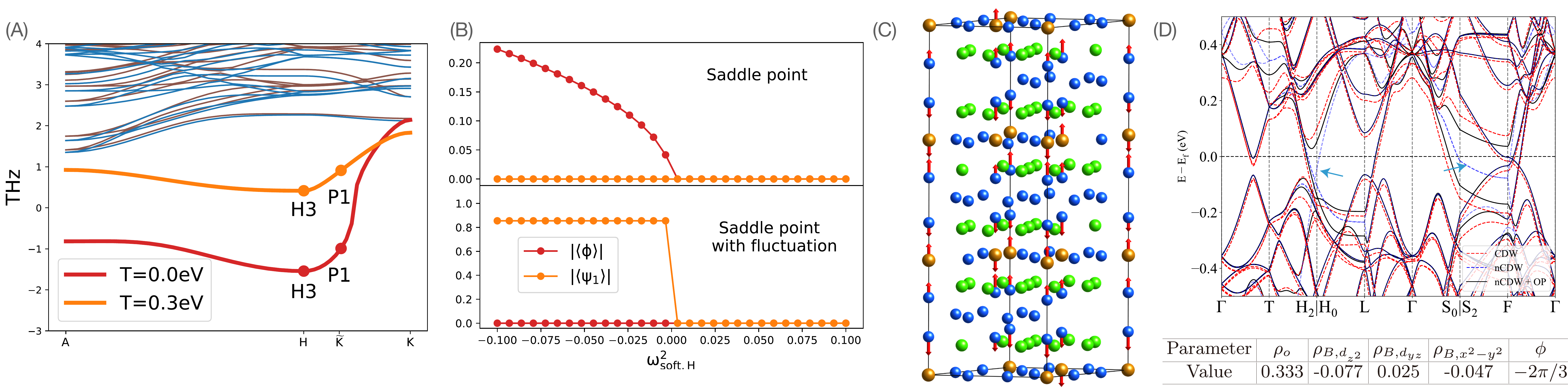}
    \caption{{\bf CDW phase transition and CDW phase} (A) Flat phonon bands at zero temperature and intermediate temperature (near transition point) $T=0.3$eV, where the $T$ is the electron temperature. The lowest-energy phonon bands have been marked by red and orange color respectively, and the irreducible representations of the phonon fields at $H,\bar{K}$ have been labeled. 
(B) Phase transition in saddle-point approximation (top panel) and saddle-point approximation with Gaussian fluctuations (bottom panel). In the saddle-point approximation, we observe a second-order phase transition to the $H$-order phase. However, after including Gaussian fluctuations, we observe a first-order transition to the $\bar{K}$ phase. 
Parameters of the mode are given in \refCiteSI{app:sec:LG_theory}.
    (C) The displacement of atoms between the CDW phase and the pristine phase. The arrow labels the movement of atoms from the pristine phase to CDW phase. 
    (D) Band structure of CDW phase (red line), non-CDW phase (blue line), where we can observe the gap opening as marked by blue arrows. The CDW band can be reproduced by adding the mean-field order parameter (Eq.~\ref{eq:main_op}) to the non-CDW Hamiltonian, $H_{nCDW}+H_{OP}$ (black line).
}
    \label{fig:main:cdw}
\end{figure*}
\emph{CDW phase transition.}  We now discuss the CDW phase transition driven by the collapsing of the phonon mode~\cite{KOR23}. 
We build an effective model which contains two types of fields/order parameters: $\phi$ and $\psi_{i=1,2}$. The $\phi$ field represents the lowest-energy phonon near $H$ point. The $\phi$ phonons collapse as we lower the temperature and form an $H_3$ irreducible representation~\cite{zotero-4159,BRA17,ELC17,VER17}  of the little group at $H$ (Fig.~\ref{fig:main:cdw} (A)). 
$\psi_{i=1,2}$ denote the lowest-energy phonon with wave vector $\bar{K}$ (CDW wavevector) and form a $P_1$ irreducible representation of the little group at $\bar{K}$~\cite{zotero-4159,BRA17,ELC17,VER17}{}. 
We denote the phase with $\langle \phi \rangle \ne 0 $ as the $H$-order phase, and the phase with $\langle \psi_i\rangle \ne 0$ as the $\bar{K}$-order phase. In addition, we focus on the experimentally-observed single-$\bar{K}$ phase, where only one of $\psi_{i=1,2}$ develops a non-zero expectation value (here, we take $\langle \psi_1\rangle \ne 0$)~\citeSI{app:sec:LG_theory}.

Based on symmetry, we construct an effective theory for the
$\phi$ and $\psi_i$ fields~\citeSI{app:sec:LG_theory}. 
The Lagrangian of $\phi$ fields, $L_\phi$, takes the form of the $\phi^4$ theory with bilinear and quadratic terms. 
The Lagrangian of $\psi$ fields, $L_\psi$, contains bilinear, cubic and quadratic terms.
The two fields are coupled with each other via a quadratic coupling $L_{\phi\psi}$~\citeSI{app:sec:LG_theory}. The final Lagrangian is $L= L_\phi +L_\psi +L_{\psi\phi} $.

We let $\omega_{soft,H}^2$ be the ``mass'' of $\phi$ field, which is also the square of the phonon energy at $H$ point. We treat $\omega_{soft,H}^2$ as the tuning parameter where the phonon collapses at $\omega_{soft,H}^2=0$.
At the saddle-point approximation, we can derive the following Ginzburg-Landau free energy
\begin{eqnarray} 
&&F = \omega_{soft,H}^2 |{\phi^c}|^2 + u_\phi |{\phi^c}|^4 +
+\omega^2_{CDW,\bar{K}}|\psi_1^c|^2\nonumber\\ 
&&+\gamma ((\psi_1^c)^3 + (\psi^{c,\dag}_1)^{3}) 
+u_{\phi\psi} |\phi^c|^2 |\psi^c_1|^2 
\end{eqnarray} 
where $u_{\phi}, \omega_{CDW,\bar{K}},\gamma,u_{\phi\psi}$ are parameters of the model~\citeSI{app:sec:LG_theory}.
${\phi^c},{\psi^c}$ are the expectation values (saddle-point solutions) of the $\phi$, $\psi$ fields respectively. 
At the saddle-point level, we observe a second-order phase transition between $H$-order phase 
and disorder phase
(Fig.~\ref{fig:main:cdw} (B)).
However, since the phonon band is flat near the $H$ point (Fig.~\ref{fig:main:cdw} (A)), there is a strong fluctuation of the $\phi$ field. 
Hence, we incorporate the Gaussian fluctuations of the $\phi$ field into the saddle-point solution~\cite{AUE94}. 
We then identify another first-order transition from the $H$-order phase to the $\bar{K}$-order phase~\citeSI{app:sec:LG_theory}.
The transition happens at 
\begin{equation}
\omega_{soft,H}^2=\omega_{trans}^2  \approx  - D^2 \bigg( 
 C - \frac{1}{3}\log(D/M)
 \bigg) , 
 \label{eq:main_cdw_trans}
\end{equation} 
where $D $ characterizes the bandwidth of the soft (almost) flat phonon near $H$ point, and $C$ and $M$ are constants~\citeSI{app:sec:LG_theory}. 
As the phonon band becomes extremely flat $D\approx 0$, the transition to the $\bar{K}$-order happens almost immediately after the phonon collapsing with $\omega_{trans}^2 \approx 0$ (Fig.~\ref{fig:main:cdw} (B)), which is consistent with the experimental observation.

We comment that, due to the flatness of the phonon bands, all phonons near the $H$ point (including $\bar{K}$) are likely to generate the CDW instabilities. 
However, among all the momentum points near $H$, $\bar{K}$ is the only momentum at which symmetry allows 
a cubic interaction term. 
The cubic term produces a first-order transition that
suppresses the fluctuations and 
stabilize 
$\bar{K}$-order.
Hence $\bar{K}$ is a natural choice of the CDW wave vector. 
Finally, we point out that high-order fluctuations beyond Gaussian fluctuations could also make contributions, and a numerical exact simulation is required to find the exact phase diagrams near the transition point.

\emph{Properties of the CDW phase}. We discuss the properties of the CDW phase from two aspects: (1) displacement of atoms; (2) electronic order parameters. From X-ray experiments in Ref.~\cite{KOR23}, the CDW phase has $R\bar{3}m$ structure with a symmetry group SG 166. However, a slightly different structure (SG 155 $R32$), with an additional weak inversion symmetry breaking has also been reported\cite{ARA22}. Here, we focus on the $R\bar{3}m$ structure since the inversion breaking is weak.

We extract the displacement of atoms ($\Delta u$) between the CDW phase and pristine phase from experimental data~\citeSI{app:sec:CDW_ops}.
In Fig.~\ref{fig:main:cdw} (C), we illustrate the displacements of each atom. 
We find $\Delta u$ corresponds to a single-$\bar{K}$ phase, exhibiting a $98\%$ overlap with our theoretically calculated lowest-energy phonon field at $\bar{K}$.

The displacement of atoms leads to a change of hopping amplitudes. We extract the mean-field electronic order parameter ($H_{OP}$) by comparing the tight-binding models of pristine ($H_{nCDW}$) and CDW ($H_{CDW}$) phases and requiring $H_{CDW} \approx H_{nCDW} +H_{OP}$. We find that $H_{OP}$~\citeSI{app:sec:CDW_ops} can be described by an on-site term of $c_{\mathbf{R},e,\sigma}$ and a nearest-neighbor bond term between $c_{\mathbf{R},e,\sigma}$ and V $d$ orbitals ($d_{\mathbf{R}',i,\alpha,e,\sigma}$ where $\alpha=z^2,yz,x^2-y^2$ denote orbital, $i=1,2,3$ denote sublattice, and $e$ denotes the mirror-even state). 
\begin{eqnarray}
&H_{OP}= \sum_{\mathbf{R},\sigma }\rho_o \cos(\bar{K}\cdot \mathbf{R}+\phi)   c_{\mathbf{R},e,\sigma}^\dag c_{\mathbf{R},e,\sigma} 
+ 
\nonumber\\ 
&
\sum_{(\mathbf{R},\mathbf{R}',i)\in NN,\alpha,\sigma}\rho_{B,\alpha} \cos(\bar{K}\cdot\mathbf{R} +\phi) c_{\mathbf{R},e,\sigma}^\dag d_{\mathbf{R},i,\alpha,e,\sigma}   
\label{eq:main_op}
\end{eqnarray}
where $NN$ denotes the set of nearest-neighbor bonds. The values of the coefficient are given in Fig.~\ref{fig:main:cdw} (D). The CDW band structure can be approximately reproduced by $H_{nCDW} +H_{OP}$ (Fig.~\ref{fig:main:cdw} (D)).
Finally, we provide the selection rules which have been used~\cite{KOR23} to match the DFT-calculated band structures and the ARPES photoemission spectra~\citeSI{app:sec:arpes}.

\emph{Summary and discussion.}
We have performed a comprehensive study on the phonons, electrons, electron-phonon coupling and CDW phase of the Kaomge material ScV$_6$Sn$_6$. We demonstrate that electron-phonon coupling produces an imaginary, almost-flat phonon band with a minimum at $H$. We have shown that the imaginary phonon and the strong fluctuations of its flat band induce a CDW phase at wavevector $\bar{K}$ via a first-order phase transition. We have also discussed the nature of the CDW phase, from the displacement of atoms and the electronic CDW order parameters. 

Our work, for the first time, provides a detailed theoretical understanding of the experimental observations in ScV$_6$Sn$_6$~\cite{KOR23}. Moreover, the methods spearheaded in this work, including perturbation-theory calculations of the phonon spectrum, Wannier constructions of phonon models, identifying the CDW order parameters, and comparing theoretical-calculated and ARPES-observed band structures via selection rules, could directly be applied to other materials. These techniques allow us to uncover and understand the physics behind the usual ``messy'' band structures of electrons and phonons in materials by concentrating on the relevant degrees of freedom. This opens a new phase in the collaborations of analytical field-theory analysis, numerical \textit{ab-initio} calculations, and experimental observations.

\emph{Note added.} During the writing of this long manuscript, we learned about the results of Hengxin Tan and Binghai Yan in Ref.~\cite{TAN23}.  

\emph{Acknowledgments.}
H.H. was supported by the European Research Council (ERC) under the European Union’s Horizon 2020 research and innovation program (Grant Agreement No. 101020833). D.C. acknowledges the hospitality of the Donostia International Physics Center, at which this work was carried out. D.C. and B.A.B. were supported by the European Research Council (ERC) under the European Union’s Horizon 2020 research and innovation program (grant agreement no. 101020833) and by the Simons Investigator Grant No. 404513, the Gordon and Betty Moore Foundation through Grant No. GBMF8685 towards the Princeton theory program, the Gordon and Betty Moore Foundation’s EPiQS Initiative (Grant No. GBMF11070), Office of Naval Research (ONR Grant No. N00014-20-1-2303), Global Collaborative Network Grant at Princeton University, BSF Israel US foundation No. 2018226, NSF-MERSEC (Grant No. MERSEC DMR 2011750). B.A.B. and C.F. are also part of the SuperC collaboration. 
D.S. and S.B-C. acknowledge financial support from the MINECO of Spain through the project PID2021-122609NB-C21 and by MCIN and by the European Union Next Generation EU/PRTR-C17.I1, as well as by IKUR Strategy under the collaboration agreement between Ikerbasque Foundation and DIPC on behalf of the Department of Education of the Basque Government.

\let\oldaddcontentsline\addcontentsline
\renewcommand{\addcontentsline}[3]{}

\let\addcontentsline\oldaddcontentsline

\clearpage

\renewcommand{\thetable}{S\arabic{table}}
\renewcommand{\thefigure}{S\arabic{figure}}
\renewcommand{\theequation}{S\arabic{section}.\arabic{equation}}
\onecolumngrid
\pagebreak
\thispagestyle{empty}
\newpage
\begin{center}
	\textbf{\large Supplemental Material: Kagome Materials I: SG 191, ScV$_6$Sn$_6$. Flat Phonon Soft Modes and Unconventional CDW Formation: Microscopic and Effective Theory}\\[.2cm]
\end{center}

\appendix
\renewcommand{\thesection}{\Roman{section}}
\tableofcontents
\let\oldaddcontentsline\addcontentsline
\newpage

\section{Crystal structure}\label{app:sec:crystal_struct_nCDW}
\begin{figure}[t]
    \centering
    \includegraphics[width=1\textwidth]{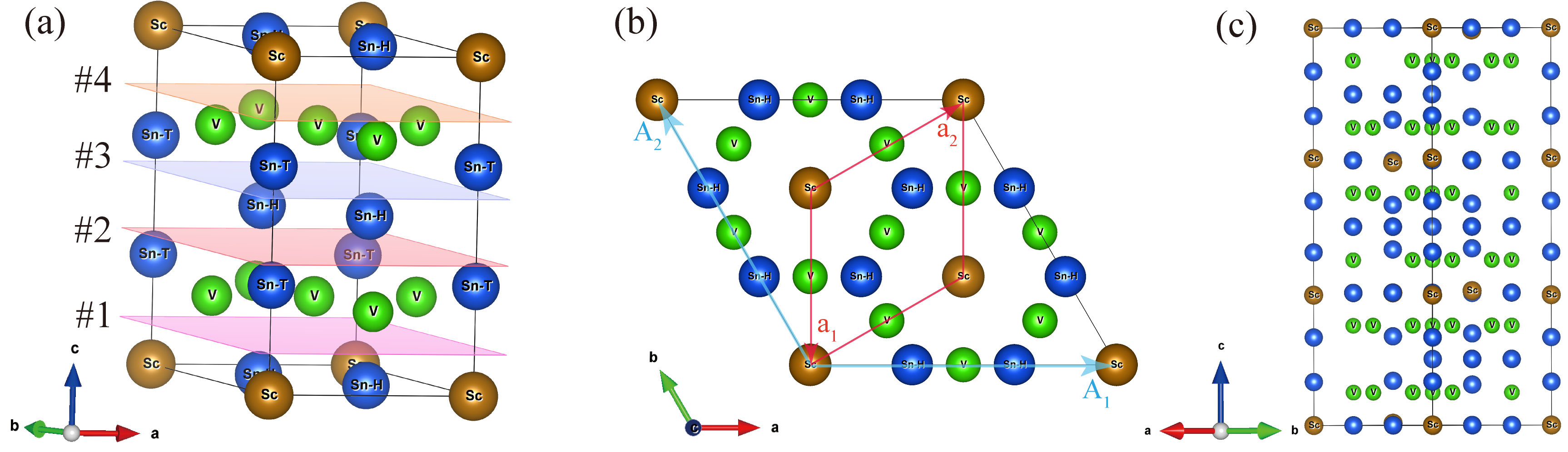}
    \caption{(a) The crystal structure ScV$_6$Sn$_6$ in the pristine phase. We mark four planes with \#1-\#4 in the unit cell, which give four possible surface terminations atoms below each plane form the surface when cleaving perpendicularly to the $\hat{\vec{z}}$ direction. (b) The top view of the conventional cell of the $(\frac{1}{3},\frac{1}{3},\frac{1}{3})$ CDW structure\cite{ARA22}. The blue $A_1, A_2$ denotes the conventional cell basis of the CDW phase, while the red $a_1, a_2$ denotes the pristine cell basis. (c) The side view of the conventional cell of the CDW structure. }
    \label{fig:lattie_config}
\end{figure}

ScV$_6$Sn$_6$ has a high-temperature structure of space group (SG) 191 $P6/mmm$, as shown in \cref{fig:lattie_config}(a). We consider two slightly different crystal structures, one being the experimental structure measured at 280 $K$ given in Ref.~\cite{ARA22}, and the other being the relaxed structure 
In \cref{Tab:latt_const_compare}, we compare their lattice constants and atomic positions. 
The lattice constants measured in this work are \textit{a}=\textit{b}=5.47 \r{A} and \textit{c}=9.17 \r{A}, which is close to the experimental structure in Ref.~\cite{ARA22}.
The primitive vectors of the lattice are defined as 
\ba 
\bm{a}_1 = a(1,0,0),\quad \bm{a}_2 = a(-\frac{1}{2},\frac{\sqrt{3}}{2},0),\quad \bm{a}_3 =c(0,0,1)
\ea 
Each unit cell contains one Sc atom, six V atoms, and six Sn atoms. The Sc atom forms a triangular lattice for each $xy$ plane. Six V atoms from two layers of Kagome lattice. Two of Sn atoms form two layers of triangular lattices and the remaining four Sn atoms form two layers of the honeycomb lattice. The position of the atom in the unit cell is shown in \cref{Tab:atom_pos_compare}.

\begin{table}[t]
	\global\long\def\arraystretch{1.12}
	\setlength{\tabcolsep}{5mm}{
		\begin{tabular}{c|c|c|c}
			\hline\hline
			Structure & $a$~(\AA) & $c$ (\AA)  & $d_{\text{Sn}^{\text{T}}-\text{Sn}^{\text{T}}}$~(\AA) \\\hline
			Experimental\cite{ARA22} & 5.4669 & 9.1594 & 3.218
			\\\hline
			Relaxed & 5.5001 & 9.2150  & 3.363
			\\\hline\hline
	\end{tabular}}
	\caption{\label{Tab:latt_const_compare}Comparsion of the experimental and relaxed structure of ScV$_6$Sn$_6$. In the last column, we give the shortest distance between two triangular Sn atoms. 
}
\end{table}

\begin{table}[t]
\begin{tabular}{c|c|c}
\hline\hline
Structure & Sc & V  \\\hline
Experimental\cite{ARA22}  & $(0,0,0)$ 
& $(\frac{1}{2},0,0.2475)$,$(0,\frac{1}{2},0.2475)$ ,$(\frac{1}{2},\frac{1}{2},0.2475)$,
$(\frac{1}{2},0,0.7525)$,$(0,\frac{1}{2},0.7525)$ ,$(\frac{1}{2},\frac{1}{2},0.7525)$ 
\\\hline
Relaxed & $(0,0,0)$ & 
$(\frac{1}{2},0,0.2473)$,$(0,\frac{1}{2},0.2473)$ ,$(\frac{1}{2},\frac{1}{2},0.2473)$,
$(\frac{1}{2},0,0.7527)$,$(0,\frac{1}{2},0.7527)$ ,$(\frac{1}{2},\frac{1}{2},0.7527)$ 
\\\hline\hline
Structure   & Sn$^T$ & Sn$^H$  \\\hline
Experimental\cite{ARA22} 
&   $(0,0, 0.3243)$,$(0,0, 0.6757)$ & 
$(\frac{1}{3},\frac{2}{3},0), (\frac{2}{3},\frac{1}{3},0)$
,$(\frac{1}{3},\frac{2}{3},\frac{1}{2}), (\frac{2}{3},\frac{1}{3},\frac{1}{2})$
\\\hline
Relaxed & $(0,0,0.3175)$, $(0,0,0.6825)$
&$(\frac{1}{3},\frac{2}{3},0), (\frac{2}{3},\frac{1}{3},0)$
,$(\frac{1}{3},\frac{2}{3},\frac{1}{2}), (\frac{2}{3},\frac{1}{3},\frac{1}{2})$
\\\hline\hline
\end{tabular}
\caption{\label{Tab:atom_pos_compare}Comparsion of the experimental and relaxed structure of ScV$_6$Sn$_6$, where we use Sn$^T$ and Sn$^H$ to denote triangular and honeycomb Sn, respectively. $(i,j,m)$ in the table denote the position vector $i \bm{a}_1 +j\bm{a_2} +m\bm{a}_3$.}
\end{table}

\subsection{Crystal structure of CDW phase}\label{sec:cdw_structure}
We discuss the crystal structure of the charge-density-wave (CDW) phase. 
The $(\frac{1}{3},\frac{1}{3},\frac{1}{3})$ CDW leads to a $\sqrt{3}\times\sqrt{3}\times3$ supercell~\cite{ARA22}. 
We compare two slightly different CDW structures. The first is from our experimental measurement, which has SG 166 $R\overline{3}m$ with inversion symmetry preserved. The other one is from Ref.~\cite{ARA22} measured at 50 $K$, which belongs to SG 155 $R32$ where inversion is broken. In \cref{Tab:atom_pos_compare_CDW} and \cref{Tab:atom_pos_compare_CDW_allatom}, the lattice constants, the nearest distance between Sn$^T$ atoms, and atom positions in these two CDW structures are compared, which are very close. The nearest distance between two Sn$^{T}$ atoms in three layers of the CDW conventional unit cell is different, where in layers 1 and 3 the distance is shortened while in layer 2 the distance is enlarged compared with the non-CDW structure. The inversion symmetry breaking in structure\cite{ARA22} is caused by the small displacements of V and honeycomb Sn atoms.  
These two CDW structures produce almost identical band structures near the Fermi level $E_f$, which are shown in \cref{fig:compare_cdw_bands} in \cref{sec:cdw_band_structure}.

\begin{table}[t]
\global\long\def\arraystretch{1.12}
\setlength{\tabcolsep}{10mm}{
\begin{tabular}{c|c|c}
\hline
\hline
CDW Structure & This work & Experimental~\cite{ARA22}  \\
\hline
$a$~ (\AA) & $9.4693$ & $9.4561$   \\
$c$~ (\AA) &  $27.5153$ & $27.4124$    \\
\hline
$d_{\text{Sn}^{\text{T}}-\text{Sn}^{\text{T}}}$, layer 1, 3  (\AA) & $3.0924$ & $3.073$ \\
\hline
$d_{\text{Sn}^{\text{T}}-\text{Sn}^{\text{T}}}$, layer 2  (\AA) & $3.5093$ & $3.530$ \\
\hline
\hline
\end{tabular}}
\caption{\label{Tab:atom_pos_compare_CDW} Comparison of the CDW structure of ScV$_6$Sn$_6$ in this work and Ref.~\cite{ARA22}. We list the nearest distance between two Sn$^{T}$ atoms in three different layers of the CDW conventional unit cell, where in layers 1 and 3 the distance is shortened while in layer 2 the distance is enlarged compared with the non-CDW structure.
}
\end{table}

\begin{table}[t]
\tiny
\begin{tabular}{c|c}
\hline\hline
Structure & Sc  \\\hline
Ref.~\cite{ARA22}  & $(0,0,0)$ 
\\
& $(0.3280, 0.3280, 0.3280)$ 
\\
& $(0.6720, 0.6720, 0.6720)$ 
\\\hline
This work & $(0,0,0)$ 
\\
&$(0.3285, 0.3285, 0.3285)$ 
\\
& $(0.6715, 0.6715, 0.6715)$ 
\\
\hline\hline
Structure & V  \\\hline
Ref.~\cite{ARA22}
& $(0.0846, 0.5834, 0.0843) (0.5834, 0.0843, 0.0846) (0.5824, 0.5829, 0.0819)(0.2491, 0.7491, 0.2491)(0.7491, 0.2491, 0.2491)(0.7509, 0.7509, 0.2509)$
\\
& $(0.9181, 0.4171, 0.4176)(0.4176, 0.9181, 0.4171)(0.9154, 0.9157, 0.4166)(0.0819, 0.5824, 0.5829)(0.5829, 0.0819, 0.5824)(0.0843, 0.0846, 0.5834)$
\\
& $(0.7509, 0.2509, 0.7509)(0.2509, 0.7509, 0.7509)(0.2491, 0.2491, 0.7491)(0.9157, 0.4166, 0.9154)(0.4166, 0.9154, 0.9157)(0.4171, 0.4176, 0.9181)$
\\\hline
This work & 
$(0.0843, 0.5834, 0.0843)(0.5834, 0.0843, 0.0843)(0.5827, 0.5827, 0.0821)(0.2491, 0.7493, 0.2491)(0.7493, 0.2491, 0.2491)(0.7509, 0.7509, 0.2507)$
\\
&$(0.9179, 0.4173, 0.4173)(0.4173, 0.9179, 0.4173)(0.9157, 0.9157, 0.4166)(0.0821, 0.5827, 0.5827)(0.5827, 0.0821, 0.5827)(0.0843, 0.0843, 0.5834)$
\\
&$(0.7509, 0.2507, 0.7509)(0.2507, 0.7509, 0.7509)(0.2491, 0.2491, 0.7493)(0.9157, 0.4166, 0.9157)(0.4166, 0.9157, 0.9157)(0.4173, 0.4173, 0.9179)$
\\
\hline\hline
Structure   & Sn$^T$ \\\hline
Ref.~\cite{ARA22} 
&   $(0.1093, 0.1093, 0.1093),(0.2214, 0.2214, 0.2214)$ 
\\
& $(0.4356, 0.4356, 0.4356)(0.5644, 0.5644, 0.5644)$ 
\\
& $(0.7786, 0.7786, 0.7786), (0.8907, 0.8907, 0.8907)$ 
\\
\hline
This work & $(0.1093, 0.1093, 0.1093),(0.2216, 0.2216, 0.2216)$ 
\\
& $(0.4362, 0.4362, 0.4362),(0.5638, 0.5638, 0.5638) $
\\
& $(0.7784, 0.7784, 0.7784),(0.8907, 0.8907, 0.8907)$ 
\\
\hline\hline
Structure   & Sn$^H$  \\\hline
Ref.~\cite{ARA22} 
&  
$(0.6670, 0.3330, 0.0000)(0.3333, 0.6667, 0.0000)(0.8323, 0.5000, 0.1677)(0.5000, 0.8322, 0.1678)$
\\
& 
$(0.6670, 0.3330, 0.0000)(0.3333, 0.6667, 0.0000)(0.8323, 0.5000, 0.1677)(0.5000, 0.8322, 0.1678)$
\\
& 
$(0.3330, 0.0000, 0.6670)(0.0000, 0.3333, 0.6667)(0.5000, 0.1677, 0.8323)(0.1678, 0.5000, 0.8322)$
\\
\hline
This work & 
$(0.3330, 0.6670, 0.0000)(0.6670, 0.3330, 0.0000)(0.8323, 0.5000, 0.1677)(0.5000, 0.8323, 0.1677)$
\\
& 
$(0.6670, 0.0000, 0.3330)(0.0000, 0.6670, 0.3330)(0.1677, 0.8323, 0.5000)(0.8323, 0.1677, 0.5000)$
\\
& 
$(0.3330, 0.0000, 0.6670)(0.0000, 0.3330, 0.6670)(0.5000, 0.1677, 0.8323)(0.1677, 0.5000, 0.8323)$
\\\hline\hline
\end{tabular}
\caption{\label{Tab:atom_pos_compare_CDW_allatom} Comparsion of the $(\frac{1}{3},\frac{1}{3},\frac{1}{3})$ CDW structure of ScV$_6$Sn$_6$ in this work and from Ref.~\cite{ARA22}. We use Sn$^T$ and Sn$^H$ to denote triangular and honeycomb Sn, respectively. $(i,j,m)$ in the table denote the position vector $i \bm{P}_1 +j\bm{P}_2 +m\bm{P}_3$, where $\bm{P}_{i=1,2,3}$ are the primitive cell basis defined in \cref{eq:basis_transf_a}, and lattice constants are given in \cref{Tab:atom_pos_compare_CDW}.}
\end{table}

A top view and a side view of the conventional cell of the CDW phase are shown in \cref{fig:lattie_config}(b)(c).
The lattice constants $a',c'$ in the CDW phase roughly satisfy $a'=\sqrt{3}a, c'=3c$ compared with the pristine (non-CDW) lattice constants $a$ and $c$, with errors less than 0.3 \AA. We first set $a'=\sqrt{3}a, c'=3c$ and construct the basis transformation.
We take the conventional basis $A_i$ of the CDW phase and conventional $a_i$ of the non-CDW phase as
\begin{equation}
\begin{aligned}
    \bm{A}_1&=a'(1,0,0),~ \bm{A}_2=a'(-\frac{1}{2},\frac{\sqrt{3}}{2},0),~ \bm{A}_3=c'(0,0,1)\\
    \bm{a}_1&=a'(0,-\frac{1}{\sqrt{3}},0),~ \bm{a}_2=a'(\frac{1}{2},\frac{1}{2\sqrt{3}},0),~ \bm{a}_3=c'(0,0,\frac{1}{3})
\end{aligned}
\end{equation}
Denote the primitive cell (rhombohedral lattice) basis of the CDW phase as $\bm{P}_i$. The three sets of basis are related by:
\begin{equation}
\begin{bmatrix}
    \bm{P}_1\\ 
     \bm{P}_2\\ 
     \bm{P}_3 
\end{bmatrix}= \frac{1}{3}
\begin{bmatrix}
    2 & 1 & 1\\ 
    -1 & 1 & 1 \\ 
    -1 & -2 & 1 
\end{bmatrix}
\begin{bmatrix}
     \bm{A}_1\\ 
    \bm{A}_2\\ 
    \bm{A}_3 
\end{bmatrix},\quad
\begin{bmatrix}
    \bm{A}_1\\ 
    \bm{A}_2\\ 
    \bm{A}_3 
\end{bmatrix}= 
\begin{bmatrix}
    1 & 2 & 0\\ 
    -2 & -1 & 0 \\ 
    0 & 0 & 3 
\end{bmatrix}
\begin{bmatrix}
    \bm{a}_1\\ 
    \bm{a}_2\\ 
    \bm{a}_3 
\end{bmatrix}\Rightarrow
\begin{bmatrix}
     \bm{P}_1\\ 
     \bm{P}_2\\ 
     \bm{P}_3 
\end{bmatrix}=
\begin{bmatrix}
    0 & 1 & 1\\ 
    -1 & -1 & 1 \\ 
    1 & 0 & 1 
\end{bmatrix}
\begin{bmatrix}
    \bm{a}_1\\ 
    \bm{a}_2\\ 
    \bm{a}_3 
\end{bmatrix}.
\label{eq:basis_transf_a}
\end{equation}

\begin{figure}[t]
    \centering
    \includegraphics[width=0.6\textwidth]{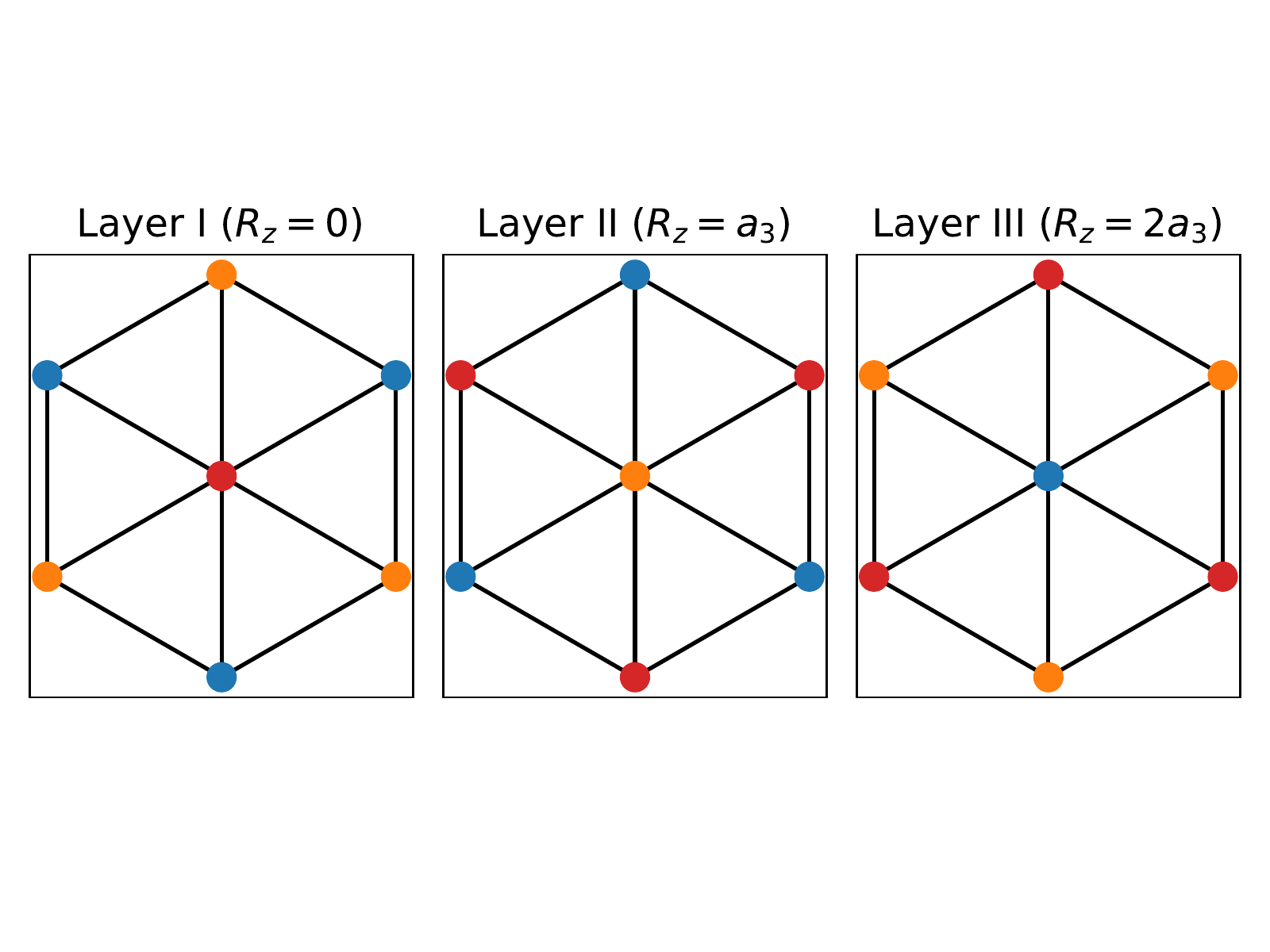}
    \caption{Pattern of CDW phase. Red, blue, and orange dots denote three types of non-CDW unit cells. For each layer, the unit cell is tripled (formed by the combination of the red, blue, and orange unit cells). Along $z$ direction, there will also be a $k_z=2\pi/3$ modulation. As shown in the Figure,  moving along $z$ direction, the unit cell changes as red$\rightarrow$ orange $\rightarrow$ blue and then repeats. }
    \label{fig:cdw_patt}
\end{figure}
We also discuss the patterns of the CDW phase here. In the non-CDW phase, the unit-cell forms a triangular lattice. In the CDW phase, the unit cell will be tripled with modulation in both $xy$ plane and $z$ direction. In the CDW phase, for each $xy$ plane, we label three different types of the (non-CDW) unit cells by different colors (red, orange, blue) as shown in \cref{fig:cdw_patt}. Moving along $z$ direction, the materials will change as "Layer I" $\rightarrow$ "Layer II" $\rightarrow$ "Layer III" and then repeat. However, we note that one can realize the "Layer II" pattern by performing a translational transformation (shifting by $\bm{a}_1$ or $\bm{a}_2$ or $-\bm{a}_1-\bm{a}_1$) on the "Layer I" pattern. Equivalently, one can perform the same translational transformation and change the "Layer II" pattern to the "Layer III" pattern or change the "Layer III" pattern to the "Layer I" pattern. 
This indicates the translational symmetry of the CDW phase, which also gives three primitive cell vectors of CDW phase as we defined in \cref{eq:basis_transf_a} and also shown here
\ba 
\bm{P}_1 = \bm{a}_2 +\bm{a}_3 ,\quad 
\bm{P}_2 = -\bm{a}_1 -\bm{a}_2+\bm{a}_3 
,\quad 
\bm{P}_3 = \bm{a}_1 +\bm{a}_3 
\ea 
where the additional $\bm{a}_3$ corresponds to the transformation between different layers.

\section{Phonon spectrum} 
\label{app:sec:phonon_spec}
\subsection{Dynamical matrix}
We let $u_{i \mu}(\mathbf{R},t)$ denote the $\mu$-direction displacement of atom $i$ at time $t$ and position $\mathbf{R}+\mathbf{r}_i$, where $\mathbf{R}$ denote the position vector of unit-cell and $r_i$ denote the location of $i$-th atom within the unit cell. We label the atoms with $i=1,2,...,13$. $i=1$ denote the Sc, $i = 2,3,...,7$ denote the V, and $i = 8,9$ denote triangular Sn,
and $i=10,11,12,13$ denote Honeycomb Sn. The corresponding $\mathbf{r}_i$ are (for the relaxed structure)
\ba 
&\text{Sc}: \quad  \mathbf{r}_1 = (0,0,0) \nonumber \\ 
&\text{V}:  \quad \mathbf{r}_2 =(\frac{1}{2},0,0.2475),\quad \mathbf{r}_3=(0,\frac{1}{2},0.2473),\quad \mathbf{r}_4=(\frac{1}{2},\frac{1}{2},0.2473)\nonumber \\
&\text{V}:  \quad \mathbf{r}_5=(\frac{1}{2},0,0.7523),\quad \mathbf{r}_6=(0,\frac{1}{2},0.7527),\quad \mathbf{r}_7=(\frac{1}{2},\frac{1}{2},0.7527)\nonumber\\ 
&\text{Sn}^T: \quad  \mathbf{r}_8=(0,0,0.3175), \quad \mathbf{r}_9=(0,0,0.6825)\nonumber \\ 
&\text{Sn}^H: \quad 
\mathbf{r}_{10}=(\frac{1}{3},\frac{2}{3},0),\quad \mathbf{r}_{11}=(\frac{2}{3},\frac{1}{3},0),\quad 
\mathbf{r}_{12}=(\frac{1}{3},\frac{2}{3},\frac{1}{2}),\quad \mathbf{r}_{13}=(\frac{2}{3},\frac{1}{3},\frac{1}{2}) \label{app:eqn:atom_position_vecs}
\ea

Since $u$ describes the displacement of atoms, $u$ is a real field. We use $M_{i}$ to denote the mass of atom $i$. Then the equations of motion of displacement fields are 
\ba 
M_{i}\partial_t^2 u_{i \mu}(\mathbf{R},t) = \sum_{\mathbf{R}', \nu,j}\Phi_{i\mu,j \nu}(\mathbf{R}-\mathbf{R}')u_{j \nu}(\mathbf{R}',t)
\label{eq:eom_u}
\ea 
where $\Phi_{i\mu, j\nu}(\mathbf{R}-\mathbf{R}')$ are the matrices of force constants. $\Phi_{i \mu,j \nu}(\mathbf{R}-\mathbf{R}')u_{j\nu}(\mathbf{R}',t)$ denotes the force felt by $u_{i\mu}(\mathbf{R},t)$ induced by the displacement $u_{j\nu}(\mathbf{R}',t)$. 

The solutions of \cref{eq:eom_u} take the following generic formula
\ba 
u_{i \mu}(\mathbf{R},t) = \frac{1}{\sqrt{N}}\sum_{\mathbf{k}} v_{n}^{i \mu}(\mathbf{k})e^{i\mathbf{k}(\mathbf{R}+\mathbf{r}_i)} e^{-i\omega_{n,\mathbf{k}}t }
\label{eq:mod_exp}
\ea
where $N$ is the total number of unit cells, $v_n^{i\mu}(\mathbf{k})$ is the vector that characterizes the $n$-th vibration mode, and $\mathbf{k}$ and $\omega_{n,\mathbf{k}}$ are the momentum and frequency of the $n$-th vibration mode. Combining \cref{eq:eom_u} and \cref{eq:mod_exp}, we obtain the following equations that determine the values of $v_{n}^{i \mu}(\mathbf{k})$ and $\omega_{n,\mathbf{k}}$ 
\ba 
M_\alpha \omega_{n,\mathbf{k}}^2 v_n^{i \mu}(\mathbf{k}) = \sum_{j\nu} \Phi_{i \mu,j\nu}(\mathbf{k}) v_n^{j \nu}(\mathbf{k})
\label{eq:eom_u_2}
\ea 
with $\Phi_{i \mu,j \nu}(\mathbf{k}) =\sum_{\mathbf{R}}\Phi_{i\mu,j\nu}(\mathbf{R})e^{-i \mathbf{k} (\mathbf{R} +\mathbf{r}_j -\mathbf{r}_i) } $. It is more convenient to rescale the $v_n^{i \mu}(\mathbf{k})$ by $\frac{1}{\sqrt{M_i}}$ and let 
\ba 
\eta_n^{i \mu}(\mathbf{k}) = \frac{1}{\sqrt{M_i}} v_n^{i \mu}(\mathbf{k}) 
\label{eq:rescale_dis}
\ea 
Then using \cref{eq:eom_u_2} and \cref{eq:rescale_dis}, we have 
\ba 
(\omega_{n,\mathbf{k}})^2\eta_n^{i  \mu}(\mathbf{k}) = \sum_{\gamma \nu } D_{i \mu,j \nu}(\mathbf{k}) \eta_n^{j \nu}(\mathbf{k}) 
\label{eq:eom_eta}
\ea 
where we introduce the dynamical matrix 
\ba 
D_{i\mu,j\nu}(\mathbf{k}) = \Phi_{i  \mu,j \nu} (\mathbf{k}) \frac{1}{\sqrt{M_{i }M_{j}}},\quad 
D_{i\mu,j \nu}(\mathbf{R}) = \Phi_{i \mu,j \nu} (\mathbf{R}) \frac{1}{\sqrt{M_{i }M_{j }}}
\ea 
Therefore, the phonon spectrum $\omega_{n,\mathbf{k}}$ is obtained by solving the eigen equations of the dynamical matrix (\cref{eq:eom_eta}).

We next discuss the symmetry properties. We remark that, from a symmetry standpoint, the displacements of the $i$-th atom along the three Cartesian directions can be thought as $p_x$, $p_y$, and $p_z$ orbitals. Restricting ourselves to symmorphic symmetry operations, we let $g$ denote a certain (unitary or antiunitary) symmetry of the phonon Hamiltonian. The action of $g$ on the displacement  is given by
\begin{align}
	g u_{i \mu} \left( \mathbf{R},t \right) g^{-1} &= \sum_{j \nu} \mathcal{D}^g_{j\nu,i\mu} u_{j \nu} \left( g \left(\mathbf{R}+ \mathbf{r}_i \right) - \mathbf{r}_j , gt\right)
\end{align}
where the orthogonal representation matrix $\mathcal{D}^g_{j\nu,i\mu}$ is given by
\begin{equation}
	\label{app:eqn:repres_mat_sym}
	\mathcal{D}^g_{j\nu,i\mu} = \begin{cases}
		\left[\mathcal{R}(g)\right]_{\nu\mu}, & \quad \text{if } g \left(\mathbf{R}+ \mathbf{r}_i \right) - \mathbf{r}_j \text{ is a lattice vector}\\
		0, & \quad \text{otherwise}
	\end{cases}.
\end{equation}
In \cref{app:eqn:repres_mat_sym}, $\mathcal{R}(g)$ denotes the matrix representation of the symmetry $g$ on the Cartesian coordinates (in particular, for $g$ being the time reversal symemtry $\mathcal{T}$, $\mathcal{R}(g)$ is identity). As a result of $g$ being a symmetry, we find that the force constant matrix and dynamical matrix obey
\begin{equation}
	\mathcal{D}^g \Phi^{(*)}\left(\mathbf{k} \right) \mathcal{D}^{Tg} = \Phi\left( g \mathbf{k} \right), \quad
	\mathcal{D}^g D^{(*)}\left(\mathbf{k} \right) \mathcal{D}^{Tg} = D\left( g \mathbf{k} \right),
 \label{eq:symmetry_const}
\end{equation}
where ${}^{(*)}$ denotes a complex conjugation operation in the cases where $g$ is antiunitary, while $g\mathbf{k}$ denotes the action of the symmetry $g$ on the momentum $\mathbf{k}$. In addition, we mention that the symmetry operation can only map the atom to another atom of the same type. Therefore, the force constant matrix and dynamical matrix have similar symmetry properties.

Since the symmetry transformation could only map Sc to Sc, V to V, triangular Sn to triangular Sn, and honeycomb Sn to honeycomb Sn, the representation matrix $D^g$ is block diagonalized. We then write the representation matrix as
\ba 
\mathcal{D}^g_{\mu\nu,ij} = 
\begin{bmatrix}
    \mathcal{D}^{Sc,g} \\ 
    & \mathcal{D}^{V,g} \\ 
    & & \mathcal{D}^{Sn^T,g} \\ 
    & & &\mathcal{D}^{Sn^H,g} 
\end{bmatrix}_{\alpha\gamma} \oplus \mathcal{D}^{u,g}_{ij}
\ea 
where $\mathcal{D}^{Sc,g}\mathcal{D}^{V,g},\mathcal{D}^{Sn^T,g},\mathcal{D}^{Sn^H,g}$ describes the transformation of different groups of atoms, and $\mathcal{D}^{u,g}$ describes the transformation of $x,y,z$ direction vibrations. 

The generators of symmetry group in the non-CDW phase are $C_{3z},C_{2z},C_{2,110}$ rotational symmetries and $i$ inversion symmetry. $C_{2,110}$ denotes the $C_2$ rotation along $\bm{a_1}+\bm{a_2}$ axis. We find the following representation matrix
\ba 
&\mathcal{D}^{Sc,C_{3z}} = \begin{bmatrix}
    1
\end{bmatrix},\quad 
\mathcal{D}^{V,C_{3z}} = 
\begin{bmatrix}
    0 & 1 & 0 \\ 
    0 & 0 & 1 \\ 
    1 & 0 & 0 \\ 
 &&&   0 & 1 & 0 \\ 
 &&&   0 & 0 & 1 \\ 
 &&&   1 & 0 & 0 \\ 
\end{bmatrix} ,\quad \mathcal{D}^{Sn^T,C_{3z}} =\mathbb{I}_{2\times 2},\quad 
\mathcal{D}^{Sn^H,C_{3z}}
=\mathbb{I}_{4\times 4},\quad 
\mathcal{D}^{u,C_{3z}} = 
\begin{bmatrix}
    -\frac{1}{2} & \frac{\sqrt{3}}{2} & 0 \\ 
    -\frac{\sqrt{3}}{2} & -\frac{1}{2} & 0 \\ 
    0 & 0 & 1 
\end{bmatrix} \nonumber \\ 
&\mathcal{D}^{Sc,C_{2z}} = \begin{bmatrix}
    1
\end{bmatrix},\quad 
\mathcal{D}^{V,C_{2z}} = \mathbb{I}_{6\times 6 } ,\quad 
\mathcal{D}^{Sn^T,C_{2z}} =\mathbb{I}_{2\times 2},\quad 
\mathcal{D}^{Sn^H,C_{2z}}
=\begin{bmatrix}
    0 &1 \\ 
    1& 0 \\ 
    & & 0 & 1 \\ 
    & & 1 & 0
\end{bmatrix},\quad 
\mathcal{D}^{u,C_{2z}} = 
\begin{bmatrix}
    -1& 0& 0 \\ 
   0 & -1  & 0 \\ 
    0 & 0 & 1 
\end{bmatrix} \nonumber \\
&\mathcal{D}^{Sc,C_{2,110}} = \begin{bmatrix}
    1
\end{bmatrix},\quad 
\mathcal{D}^{V,C_{2,110}} =
\begin{bmatrix}
    & & & 0 & 1 & 0 \\ 
    & & & 1 & 0 & 0 \\ 
    & & & 0 & 0 & 1 \\ 
   0 & 1 & 0 \\ 
    1 & 0 & 0 \\ 
    0 & 0 & 1 
\end{bmatrix}
,\quad 
\mathcal{D}^{Sn^T,C_{2,110}} =\begin{bmatrix}
    0 & 1 \\ 
    1 & 0 
\end{bmatrix},\quad 
\mathcal{D}^{Sn^H,C_{2,110}}
=\begin{bmatrix}
   & &  0 & 1 \\ 
   & &   1& 0 \\ 
     0 & 1 \\ 
    1 & 0
\end{bmatrix},\nonumber \\ 
&
\mathcal{D}^{u,C_{2,110}} = 
\begin{bmatrix}
    0& 1& 0 \\ 
   1 & 0  & 0 \\ 
    0 & 0 & -1 
\end{bmatrix} \nonumber \\ 
&\mathcal{D}^{Sc,i} = \begin{bmatrix}
    1
\end{bmatrix},\quad 
\mathcal{D}^{V,i} =
\begin{bmatrix}
    & & & 1 & 0 & 0 \\ 
    & & & 0 & 1 & 0 \\ 
    & & & 0 & 0 & 0 \\ 
   1 & 0 & 0 \\ 
    0 & 1 & 0 \\ 
    0 & 0 & 1 
\end{bmatrix}
,\quad 
\mathcal{D}^{Sn^T,i} =\begin{bmatrix}
    0 & 1 \\ 
    1 & 0 
\end{bmatrix},\quad 
\mathcal{D}^{Sn^H,i}
=\begin{bmatrix}
   & &  0 & 1 \\ 
   & &   1& 0 \\ 
     0 & 1 \\ 
    1 & 0
\end{bmatrix},\quad 
\mathcal{D}^{u,i} = 
\begin{bmatrix}
    -1 & 0& 0 \\ 
   0 & -1  & 0 \\ 
    0 & 0 & -1 
\end{bmatrix} \nonumber \\ 
&\mathcal{D}^{Sc,i} = \begin{bmatrix}
    1
\end{bmatrix},\quad 
\mathcal{D}^{V,i} =
\begin{bmatrix}
    & & & 1 & 0 & 0 \\ 
    & & & 0 & 1 & 0 \\ 
    & & & 0 & 0 & 0 \\ 
   1 & 0 & 0 \\ 
    0 & 1 & 0 \\ 
    0 & 0 & 1 
\end{bmatrix}
,\quad 
\mathcal{D}^{Sn^T,i} =\begin{bmatrix}
    0 & 1 \\ 
    1 & 0 
\end{bmatrix},\quad 
\mathcal{D}^{Sn^H,i}
=\begin{bmatrix}
   & &  0 & 1 \\ 
   & &   1& 0 \\ 
     0 & 1 \\ 
    1 & 0
\end{bmatrix},\quad 
\mathcal{D}^{u,i} = 
\begin{bmatrix}
    -1 & 0& 0 \\ 
   0 & -1  & 0 \\ 
    0 & 0 & -1 
\end{bmatrix}
\label{eq:symmetry_transf}
\ea

Besides the symmetry constraints, the force constant matrix always satisfies the following equation
\ba 
&\Phi_{i \mu,j \nu}(\mathbf{R}-\mathbf{R}') = \Phi_{j\nu,i\mu}(\mathbf{R}'-\mathbf{R}),\quad 
\sum_{\mathbf{R}',j} \Phi_{i\mu,j\nu}(\mathbf{R}-\mathbf{R}') = 0
\label{eq::phono_sum_rule_1}
\ea 
Equivalently, in the momentum space, we have 
\ba 
&\Phi_{j \mu,j \nu}(\mathbf{k}) = \Phi_{j\nu,i\mu}(-\mathbf{k}),\quad 
\sum_{j} \Phi_{i\mu,j \nu}(\mathbf{k}=0) = 0
\label{eq::phono_sum_rule_2}
\ea 
In terms of the dynamic matrix, we find 
\ba 
D_{i\mu,j\nu}(\mathbf{k}) = D_{j\nu,i\mu}(-\mathbf{k}) 
,\quad  \sum_{j} D_{i\mu,j \nu}(\mathbf{k}=0)\sqrt{M_j} = 0 
\label{eq:sumrule_dynmat}
\ea 
The second equation of \cref{eq:sumrule_dynmat} also indicates three zero modes at $\mathbf{k}=0$. We use $\eta_{x,y,z}(\mathbf{k}=0)$ to denote the eigenvectors of three zero modes. They take the form of
\ba 
\eta_{x}^{i\mu}(\mathbf{k}=0) = \sqrt{M_\alpha} \delta_{\mu,x}\nonumber\\
\eta_{y}^{\alpha\mu}(\mathbf{k}=0) = \sqrt{M_\alpha} \delta_{\mu,y}\nonumber\\
\eta_{z}^{\alpha\mu}(\mathbf{k}=0) = \sqrt{M_\alpha} \delta_{\mu,z}
\label{eq:acoustic_zero_q}
\ea 
We note that the three eigenvectors describe three acoustic modes at $\mathbf{k}=0$ and satisfy \cref{eq:eom_eta} with eigenvalues $\omega_{x,\mathbf{k}=0}=\omega_{y,\mathbf{k}=0}=\omega_{z,\mathbf{k}=0}=0$.

\subsection{Low and high temperature phonon spectrum from ab initio calculation} \label{app:sec:phonon_spectrum:ab_initio}
We numerically obtain the phonon spectrum by calculating the force constant matrix via density functional theory (DFT) as implemented in Vienna \textit{ab-initio} Simulation Package (VASP)~\cite{KRE94,KRE96c,KRE96b} and Quantum Espresso (QE)~\cite{GIA09}. The generalized gradient approximation (GGA) with Perbew-Burke-Ernzerhof (PBE) scheme~\cite{PER96a} is adopted for the exchange-correlation functional. The structure was fully relaxed before the phonon calculation with a force convergence criteria of $10^{-3}$ eV/\AA. The phonon spectra and force constants by VASP are extracted via PHONOPY code with density functional perturbation theory (DFPT)~\cite{TOG15}, using a $3\times 3\times 2$ supercell with a $\Gamma$-centered $3\times3\times2$ $k$-mesh. The energy cutoff is set to be 320 eV. For calculations in QE, we adopted the standard solid-state pseudopotentials (SSSP)~\cite{PRA18} and the kinetic-energy cutoff for wavefunctions is set to be 90 Ry with a Fermi-Dirac smearing of 0.005 Ry. For the self-consistent calculations, a $12\times12\times6$ Monkhorst-Pack k-mesh was implemented with an energy convergence criteria of $10^{-8}$ Ry. For electron-phonon coupling (EPC) calculations, a denser $24\times24\times12$ $k$-mesh was employed with an energy convergence criteria of $10^{-13}$ Ry. Spin-orbit coupling is not included in the structural relaxation and phonon calculations.

\begin{figure}[h]
	\centering
    \includegraphics[angle=0, width=0.95\textwidth]{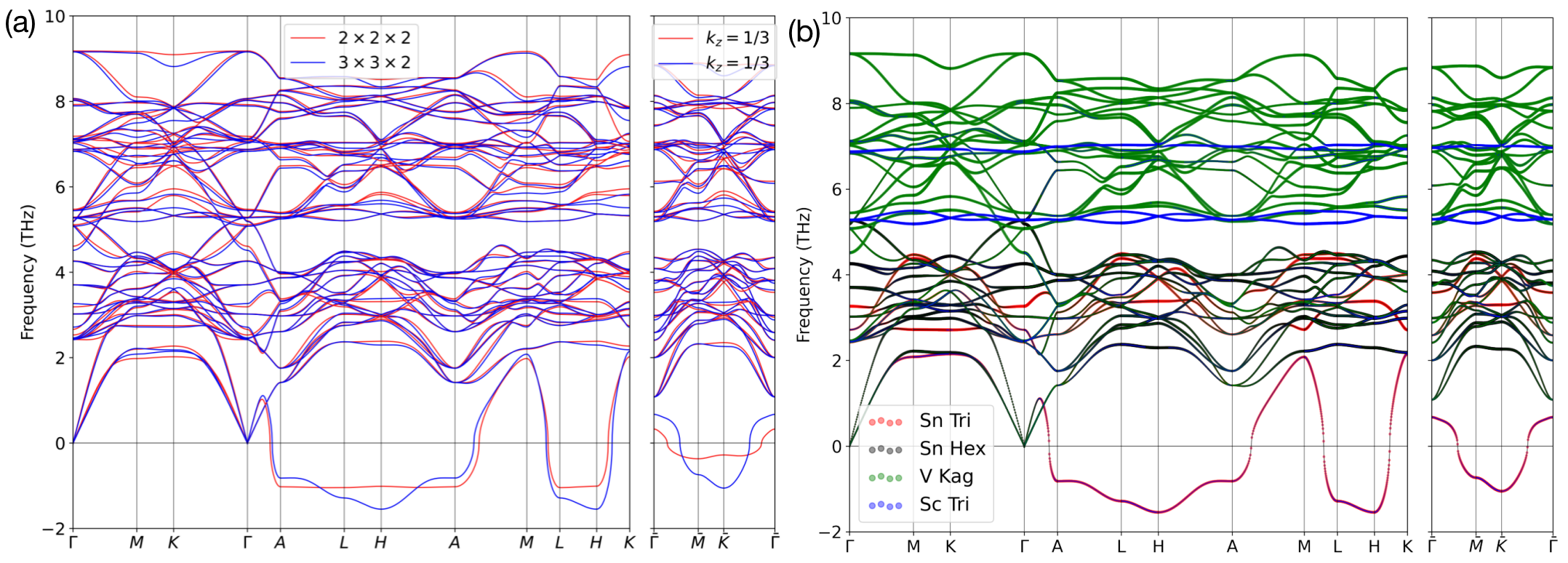}
	\caption{Phonon spectrum of ScV$_6$Sn$_6$ by VASP. (a) The blue and red lines denote the results calculated with $3\times3\times2$ and $2\times2\times2$ supercells.  The path at $k_z = 1/3$ plane is plotted on the right side. (b) The phonon spectrum is projected according to the site projection, including the triangular Sn (red), hexagonal Sn (black), kagome V (green), and triangular Sc (blue). The imaginary phonon band in the $k_z=\pi$ plane is mostly contributed by Sn$^{\text{T}}$ atoms. If we further resolve the modes to vibrations in $x$, $y$, and $z$ directions, we can see the instability is caused by the out-of-plane vibration of Sn$^{\text{T}}$ atoms.}
	\label{Fig-Pho-Sc-highT}
\end{figure}

\begin{figure}[h]
	\centering
    \includegraphics[angle=0, width=0.5\textwidth]{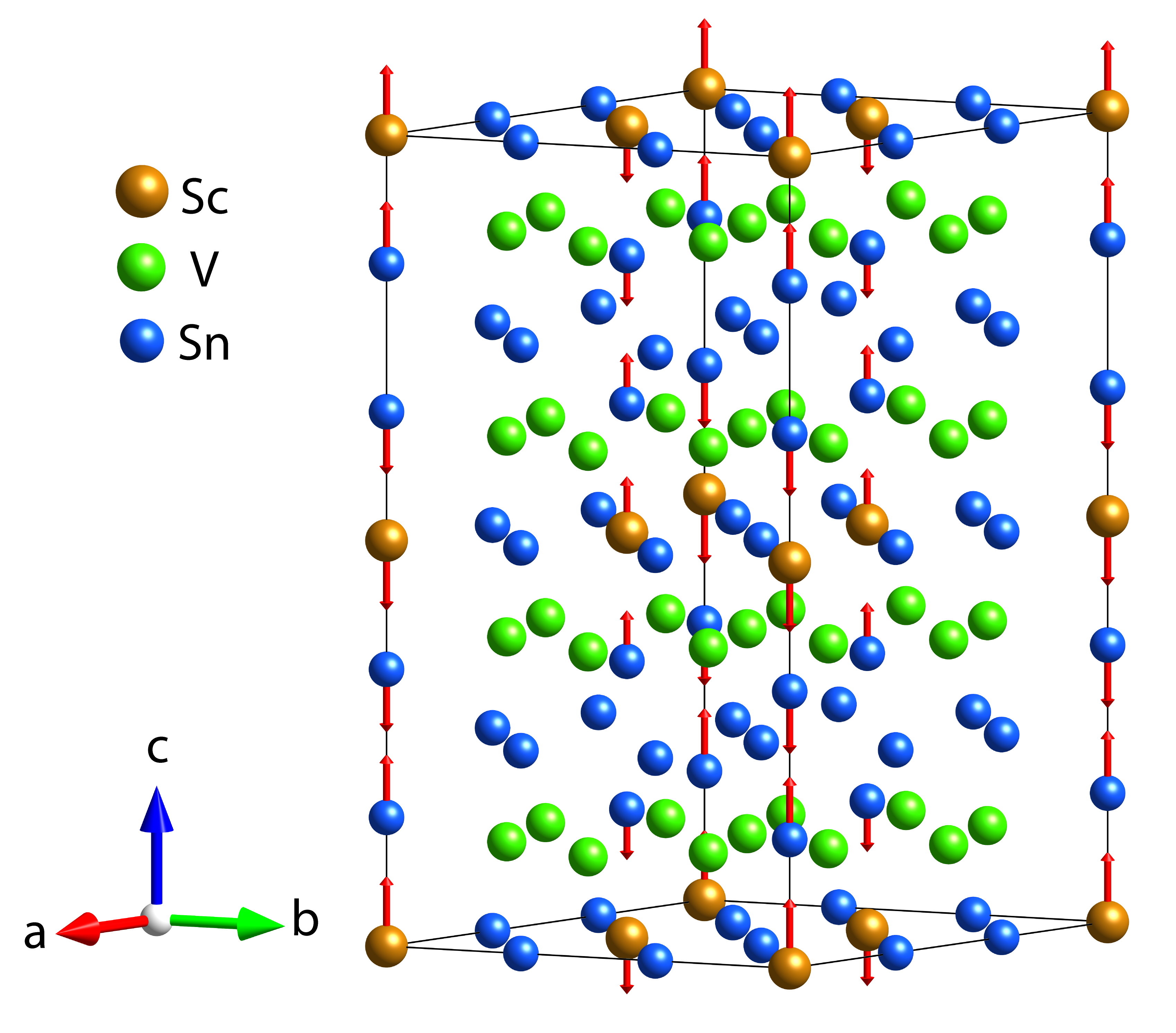}
	\caption{Vibration mode corresponding to $q=(\frac{1}{3},\frac{1}{3},\frac{1}{2})$, which is mainly contributed by the out-of-plane vibration of Sn$^{\text{T}}$ atoms and Sc atoms.}
	\label{Fig-Sc-movement}
\end{figure}
From the phonon spectrum of ScV$_6$Sn$_6$ in \cref{Fig-Pho-Sc-highT} (a), one can identify the instability in the $k_z=\pi$ plane, which contains the most negative squared frequency (the reason for this will be explained analytically later in \cref{sec:ele_corr_Sc}).
The atom projection reveals that the dominant contribution is the out-of-plane vibration of triangular Sn atoms, as plotted in \cref{Fig-Pho-Sc-highT} (b).

\begin{figure}[ht]
	\centering
	\includegraphics[angle=0, width=0.95\textwidth]{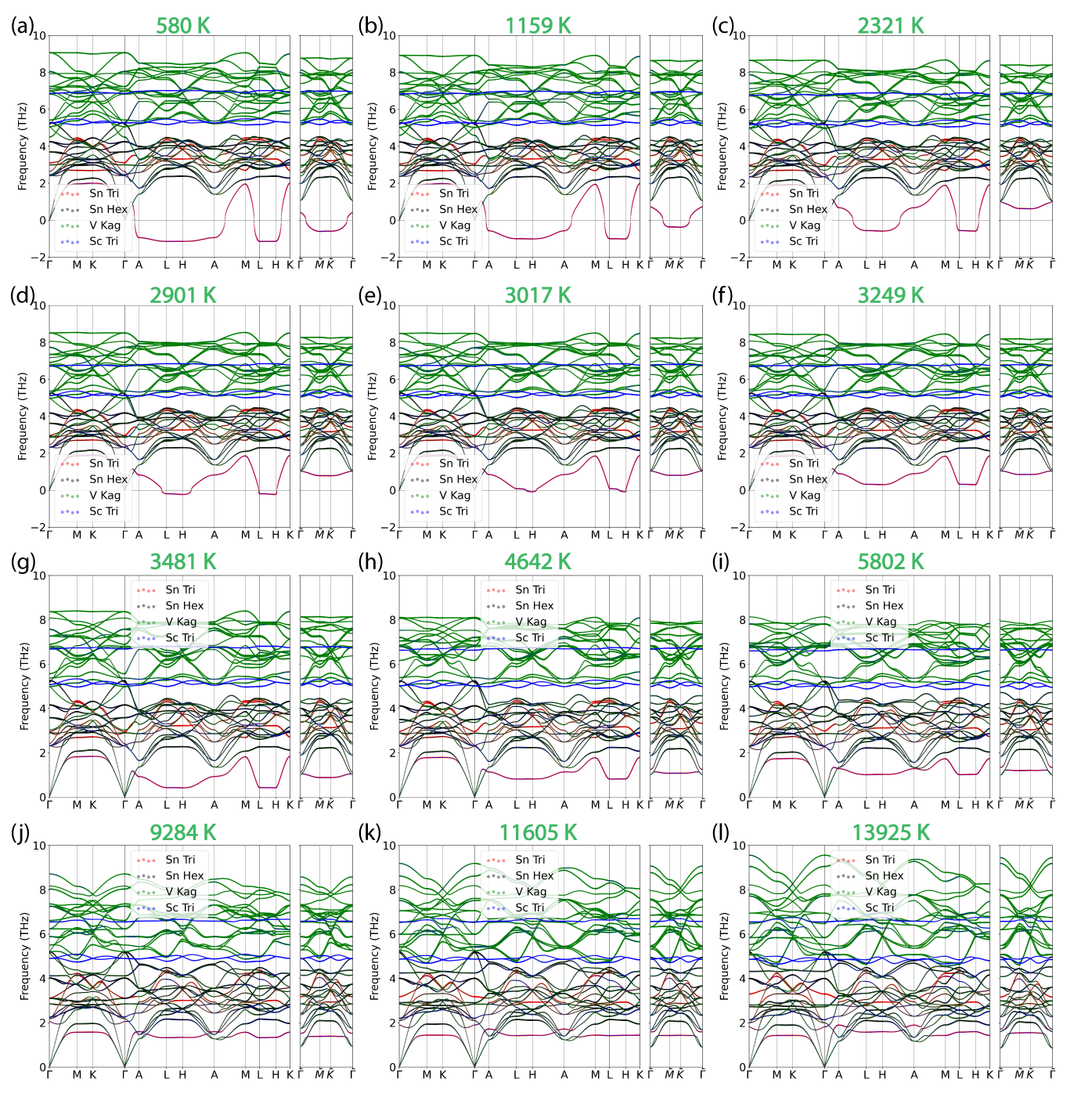}
	\caption{Temperature-dependent phonon spectrum with atomic projections, calculated via smearing method within harmonic approximation. Here, we adopted the Fermi-Dirac smearing, which corresponds to the occupations of electrons, with a set of smearing values: (a) $0.05$ eV, (b) $0.1$ eV, (c) $0.2$ eV, (d) $0.25$ eV, (e) $0.26$ eV, (f) $0.28$ eV, (g) $0.3$ eV, (h) $0.4$ eV, (i) $0.5$ eV, (j) $0.8$ eV, (k) $1.0$ eV, (l) $1.2$ eV. }
	\label{Fig-Pho-Sc-smear}
\end{figure}

To understand better the evolution of the phonon spectrum we perform calculations at different temperatures. The temperature-dependent phonon spectra are calculated via a smearing method within the harmonic approximation. The Fermi-Dirac smearing is adopted, which indicates the physical occupation of electrons, reflecting the electronic temperature of the system. While the smearing method can simulate the finite-temperature phonon within harmonic approximation, the electronic temperature here could be far away from the experimental transition temperature since the contribution of ions is not included. As plotted in \cref{Fig-Pho-Sc-smear}, the estimated electronic transition temperature would be ~$3000$ K. We however employ the stable, positive, phonon spectrum at high temperatures to later analytically compute the field theory correction to the phonon frequency, and show that it leads to its renormalization to imaginary frequency. This gives a strong analytic matching of the low temperature phonon spectrum.

\begin{figure}[h]
	\centering
	\includegraphics[angle=0, width=0.95\textwidth]{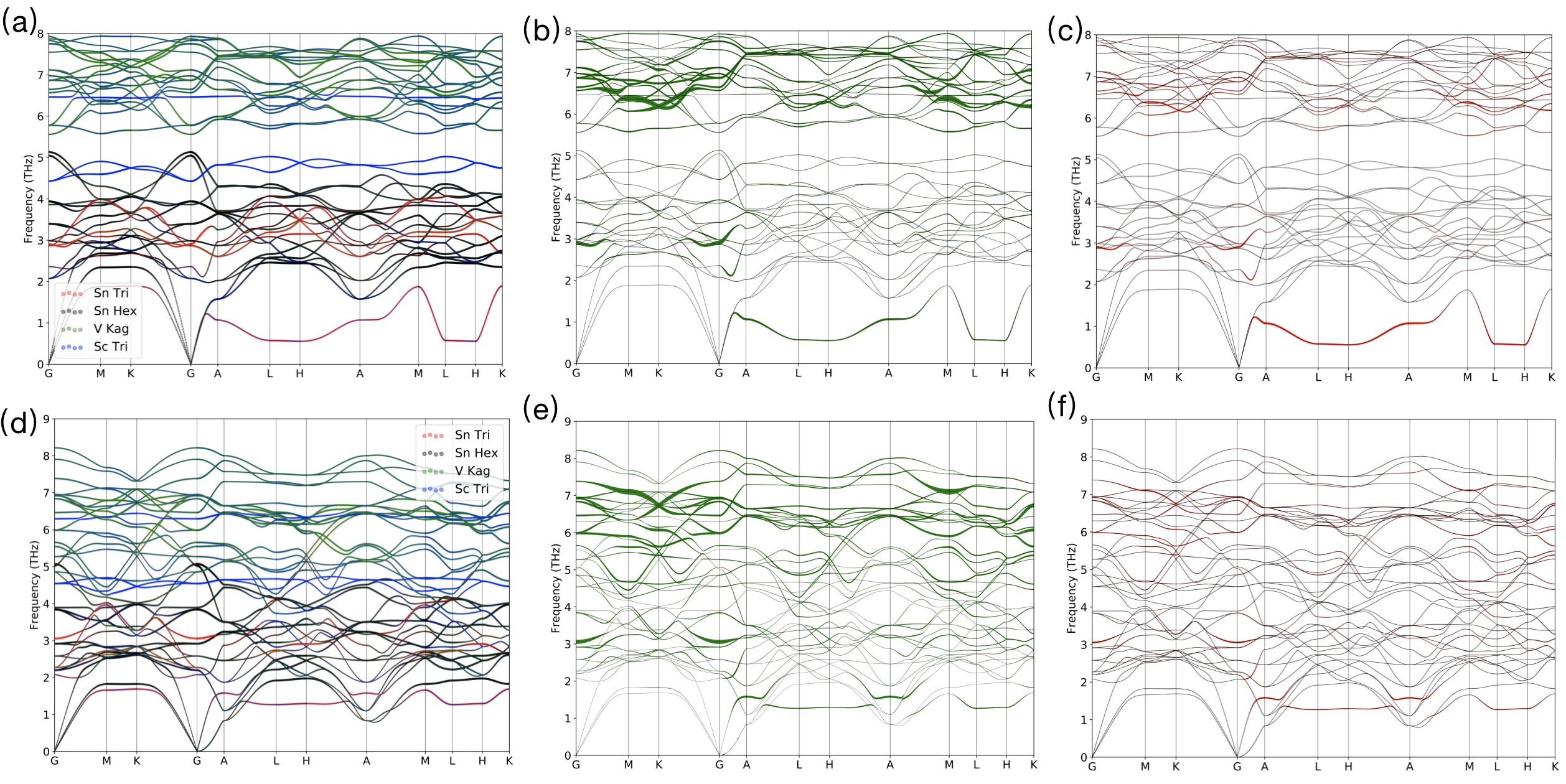}
	\caption{Phonon spectrum calculated by QE, weighted by atomic projection (left), phonon linewidth (middle) and electron-phonon coupling (right) with smearing values of $0.4$ eV (top) and $0.8$ eV (bottom). Evidently, the previously unstable mode has a relatively large contribution to EPC, which overlaps with the projection of triangular Sn.}
	\label{Fig-Pho-Sc-epc}
\end{figure}

In \cref{Fig-Pho-Sc-epc}, we also plot the phonon spectrum weighted by the magnitude of $\lambda_{\mathbf{q}\nu}$ which characterize the strength of electron-phonon coupling by integrating it  over the whole BZ.  $\lambda_{\mathbf{q}\nu}$ is given by
\begin{align}
\lambda_{\mathbf{q}\nu} = \frac{\gamma_{\mathbf{q}\nu}}{\pi \hbar N(\varepsilon_{\mathrm{F}}) \omega_{\mathbf{q}\nu}},
\label{lam_gam}
\end{align}
where $N(\varepsilon_{\mathrm{F}})$ is the density of states at the Fermi surface, $\omega_{\mathbf{q}\nu}$ is the phonon frequency. $\gamma_{\mathbf{q}\nu}$ is the linewidth, given by
\begin{align}
\gamma_{\mathbf{q}\nu}=2\pi\omega_{\mathbf{q}\nu}\sum_{i,j}\int \frac{d^3k}{\Omega}|g_{\mathbf{k},\mathbf{q}}^{ij,\nu}|^2 \delta(\varepsilon_{\mathbf{q},i}-\varepsilon_{\mathrm{F}}) \delta(\varepsilon_{\mathbf{k+q},j}-\varepsilon_{\mathrm{F}}),
\label{gamma}
\end{align}
where $g_{\mathbf{q}\nu}^{\mathbf{k},i,j}$ is the EPC matrix element, and can be calculated by
\begin{align}
g_{\mathbf{k},\mathbf{q}}^{ij,\nu}= (\frac{\hbar}{2M\omega_{\mathbf{q}\nu}})^{\frac{1}{2}} \langle \psi_{\mathbf{k},i}|\frac{dV_{\mathrm{SCF}}}{du_{\mathbf{q}\nu}} \eta_{\mathbf{q}\nu}|\psi_{\mathbf{k+q},j} \rangle.
\label{epcmatele}
\end{align}
Here, $\psi_{k,i}$ is the electronic wave function at momentum k, of the band $i$, $V_{\mathrm{SCF}}$ is the Kohn-Sham potential, $u$ is the atomic displacement, and $\eta_{\mathbf{q}\nu}$ is the phonon eigenvector. From \cref{Fig-Pho-Sc-epc}, we observe a strong normalization of the low-branch phonon mode at $k_z=\pi$ plane. In \cref{sec:ele_corr_Sc}, we will show analytically how the electron correction drives the instability in the phonon modes. 

\begin{figure}[h]
	\centering
	\includegraphics[angle=0, width=0.95\textwidth]{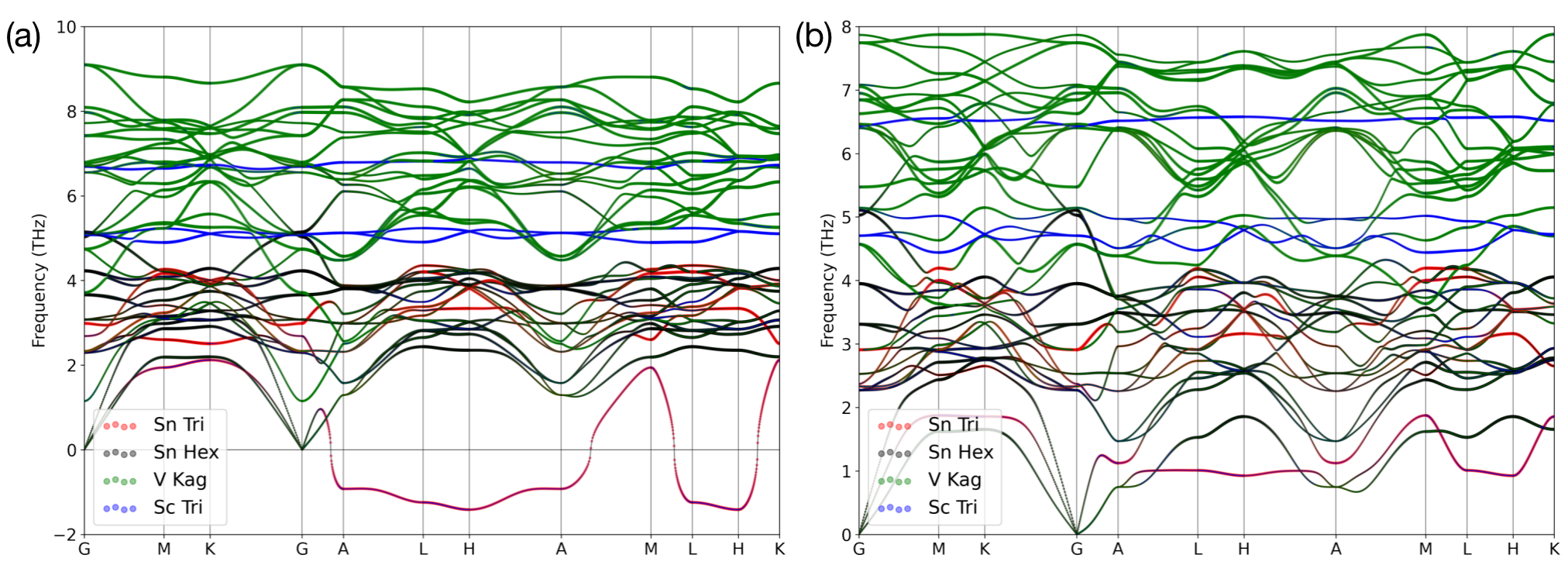}
	\caption{Phonon spectrum calculated by QE with a $3\times3\times2$ $q$-mesh, weighted by atomic projection with smearing values of $0.014$ eV and $0.4$ eV to simulate the low-T and high-T phonon.}
	\label{Fig-Pho-Sc-qe332}
\end{figure}

\begin{figure}[h]
	\centering
	\includegraphics[angle=0, width=0.95\textwidth]{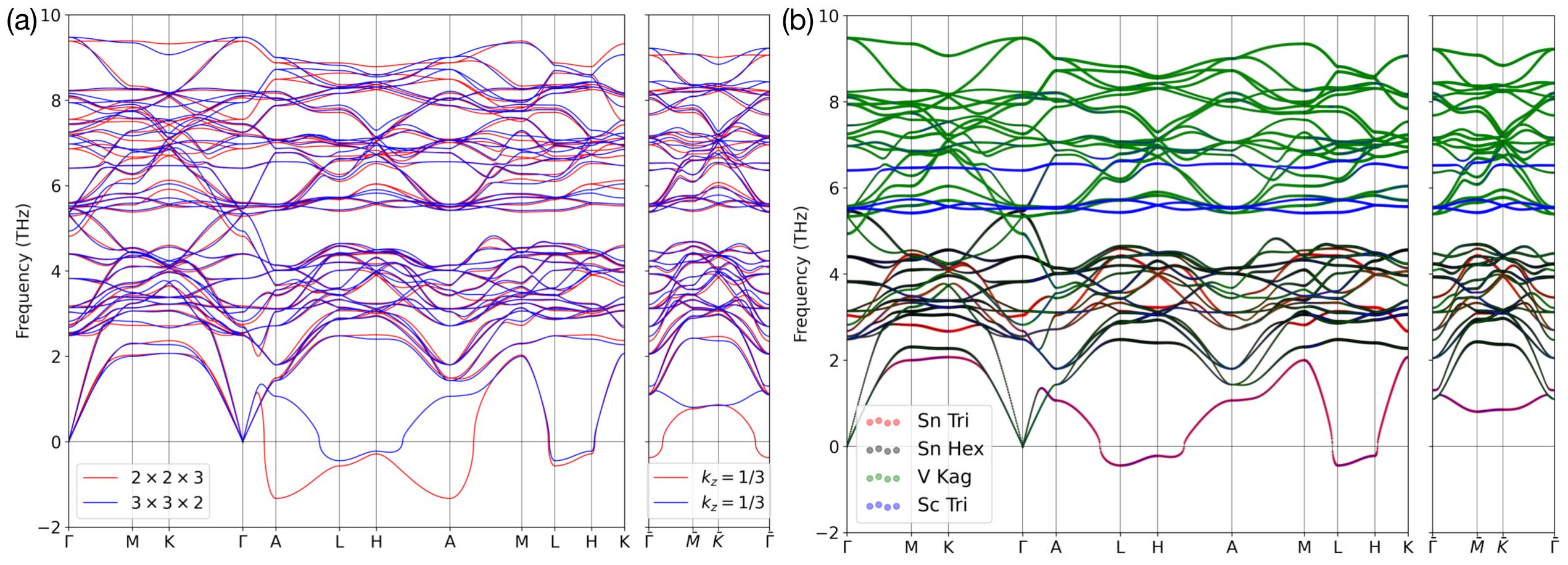}
	\caption{Phonon spectrum of the experimental ScV$_6$Sn$_6$ structure by VASP. (a) The blue and red lines denote the spectrum calculated with $3\times3\times2$ and $2\times2\times3$ supercells. (b) The phonon spectrum is projected according to the site projection, including the triangular Sn (red), hexagonal Sn (black), kagome V (green), and triangular Sc (blue).}
	\label{Fig-Pho-Sc-Exp}
\end{figure}

Here, we also calculated the phonon spectrum with EPC by QE with a $3\times3\times2$ $q$-mesh, as shown in \cref{Fig-Pho-Sc-qe332}, with Fermi-Dirac smearing of $0.014$ eV and $0.4$ eV, respectively. Furthermore, the phonon spectrum of the experimental structures were also calculated as shown in \cref{Fig-Pho-Sc-Exp}.

\subsection{Perturbation theory and effective phonon analytic model with only Sn atoms} 
In this \siSection{}, based on the force-constant matrix calculated from DFT calculation, we derive an effective analytic phonon model that only involves the Sn atoms. This will further be used later to analyze the renormalized phonon frequency.

There are three types of atoms Sn, V, Sc in these materials with mass 
\ba 
M_{Sn} = 118.71 u,\quad M_{Sc} = 44.96 u,\quad M_V = 50.94 u . 
\ea 
Sn is the heaviest atom with a mass at least one time larger than V and Sc. Therefore, there is a separation of scales, the low-energy phonon spectrum is mainly contributed by Sn, and we can treat the effect of Sc and V via perturbation (by contrast, the electronic spectrum is dominated by V, with some Sn mixture). To perform perturbation calculations, we rewrite force constant matrix $\Phi_{i\mu,j \nu}(\mathbf{k})$ as 
\ba 
\Phi_{i\mu,j \nu}(\mathbf{k}) = 
\begin{bmatrix}
    [\Phi^{Sn-Sn}] &  [\Phi^{Sn-V}] &  [\Phi^{Sn-Sc}] \\ 
     [\Phi^{V-Sn}] &  [\Phi^{V-V}] &  [\Phi^{V-Sc}] \\ 
     [\Phi^{Sc-Sn}] &  [\Phi^{Sc-V}] &  [\Phi^{Sc-Sc}] 
\end{bmatrix}_{\alpha\mu,\gamma\nu} 
\ea 
where $\Phi^{a-b}$ denote the force-constant matrix between atom $a$ and atom $b$, with $a,b \in \{$Sn,V,Sc $\}$. The corresponding dynamical matrix is 
\ba 
D_{i\mu,j \nu}(\mathbf{k}) = 
\begin{bmatrix}
    \frac{1}{M_{Sn}}[\Phi^{Sn-Sn}] &  \frac{1}{\sqrt{M_{Sn}M_{V}}}[\Phi^{Sn-V}] &  \frac{1}{\sqrt{M_{Sn}M_{Sc}} }[\Phi^{Sn-Sc}] \\ 
    \frac{1}{\sqrt{M_{V}M_{Sn}} } [\Phi^{V-Sn}] &   \frac{1}{M_V}[\Phi^{V-V}] & \frac{1}{\sqrt{M_VM_{Sn}}} [\Phi^{V-Sc}] \\ 
     \frac{1}{\sqrt{M_{Sc}M_{Sn}}}[\Phi^{Sc-Sn}] &  \frac{1}{\sqrt{M_{Sc}M_{V}}} [\Phi^{Sc-V}] &  \frac{1}{M_{Sc}} [\Phi^{Sc-Sc}] 
\end{bmatrix}_{i\mu,j\nu} 
\ea 
Since Sn is the heaviest atom, we can treat $\alpha =\frac{1}{\sqrt{M_{Sn}}}$ as a small parameter (note that this is not a dimensionless parameter; it corresponds to the dimensionless ratios of $M_V/M_{Sn}$ and $M_{Sc}/M_{Sn}$ being small). Then we write the dynamical matrix as 
\ba 
D(\mathbf{k}) = 
\begin{bmatrix}
    \alpha^2 D_2(\mathbf{k}) &\alpha  D_1(\mathbf{k})  \\ 
    \alpha D_1'(\mathbf{k}) &  D_0(\mathbf{k}) 
\end{bmatrix}
\ea 
where  
\ba 
&D_0(\mathbf{k})  =\begin{bmatrix}
  \frac{1}{M_V}[\Phi^{V-V}] & \frac{1}{\sqrt{M_VM_{Sn}}} [\Phi^{V-Sc}] \\ 
  \frac{1}{\sqrt{M_{Sc}M_{V}}} [\Phi^{Sc-V}] &  \frac{1}{M_{Sc}} [\Phi^{Sc-Sc}] 
\end{bmatrix},\quad
D_2(\mathbf{k}) = 
  [\Phi^{Sn-Sn}]
\nonumber \\ 
&D_1(\mathbf{k})= \begin{bmatrix}
   \frac{1}{\sqrt{M_{V}}}[\Phi^{Sn-V}] &  \frac{1}{\sqrt{M_{Sc}} }[\Phi^{Sn-Sc}] 
\end{bmatrix},\quad 
D_1'(\mathbf{k})= \begin{bmatrix}
    \frac{1}{\sqrt{M_{Sn}} } [\Phi^{V-Sn}]  \\ 
     \frac{1}{\sqrt{M_{Sn}}}[\Phi^{Sc-Sn}] 
\end{bmatrix}
\ea 
The eigenvalues $\lambda$ of $D(\mathbf{k})$ satisfies 
\ba 
0=\text{det}[\lambda \mathbb{I} -D(\mathbf{k})] = \det\bigg(\lambda \mathbb{I}-D_0(\mathbf{k}) \bigg) \det\bigg( \lambda\mathbb{I} - \alpha^2 D_2(\mathbf{k}) - \alpha^2 D_1(\mathbf{k}) [\lambda \mathbb{I}-D_0(\mathbf{k})]^{-1} D_1'(\mathbf{k}) \bigg)
\ea 
Since we are interested in the low-energy mode formed by Sn with the corresponding eigenvalues $\lambda \sim \alpha^2$, we only need to consider the second determinant 
\ba 
\det\bigg( \lambda\mathbb{I} - \alpha^2 D_2(\mathbf{k}) - \alpha’^2 D_1(\mathbf{k}) [\lambda \mathbb{I}-D_0(\mathbf{k})]^{-1} D_1'(\mathbf{k}) \bigg) = 0
\ea 
The eigenvalue problems can be solved by performing expansion of $\lambda = \alpha^2 \lambda_0 +  O(\alpha^3)$, where the leading-order contributions $\lambda_0$ satisfies 
\ba 
&\det\bigg( \alpha^2\lambda_0 \mathbb{I} -D_{eff}(\mathbf{k}) \bigg)  =0\nonumber\\
&D_{eff}(\mathbf{k}) =  \alpha^2 D_2(\mathbf{k}) -  \alpha^2 D_1(\mathbf{k})D_0(\mathbf{k})^{-1} D_1'(\mathbf{k}) 
\label{eq:egval_eff}
\ea 
Therefore, we can find the eigenvalues by diagonalizing the effective dynamical matrix $D_{eff}$. 

We next discuss the corresponding eigenvector of the dynamical matrix. 
The eigenvectors $\eta$ of the dynamical matrix $D(\mathbf{k})$ with eigenvalue $\alpha^2 \lambda_0$ 
should satisfy 
\ba 
\begin{bmatrix}
    \alpha^2[\lambda_0\mathbb{I}- D_2(\mathbf{k})] &-\alpha  D_1(\mathbf{k})  \\ 
    -\alpha D_1'(\mathbf{k}) & \alpha^2\lambda_0\mathbb{I}- D_0(\mathbf{k}) 
\end{bmatrix}\ 
\begin{bmatrix}
    &\eta_0 + \alpha \eta_1 +\alpha^2\eta_2\\
    &\eta'_0 +\alpha \eta'_1+\alpha^2\eta_2'
\end{bmatrix}= O(\alpha^3)
\ea 
were we expand the eigenvector $\eta$ in powers of $\alpha$. More explicitly, we find
\ba 
&\alpha^2\bigg[ \lambda_0\mathbb{I}-D_2(\mathbf{k}))\eta_0 
-D_1(\mathbf{k})\eta_1'\bigg]- \alpha D_1(\mathbf{k}) \eta_0' =O(\alpha^3)\nonumber \\
&-D_0(\mathbf{k})\eta_0' + \alpha [-D_1'(\mathbf{k})\eta_0 -D_0\eta_1' ] + \alpha^2[-D_1'(\mathbf{k})\eta_1' +\lambda_0\eta_0' -D_0\eta_2'] = O(\alpha^3)
\label{eq:pert_exp_egv}
\ea 
From the second line, canceling the coefficients of powers in $\alpha$ separately, we have 
\ba 
\eta_0' = 0, \quad \eta_1' = -D_0^{-1}D_1'(\mathbf{k})\eta_0 
\ea 
Combining the above equations with the first line of \cref{eq:pert_exp_egv}, we find 
\ba 
\bigg[ D_2(\mathbf{k}) \eta_0 -D_1(\mathbf{k})D_0^{-1}D_1'(\mathbf{k})\bigg] \eta_0 = \alpha^2 \lambda_{0}\eta_0 \Rightarrow D_{eff}(\mathbf{k}) \eta_0 = \alpha^2\lambda_0 \eta_0 
\label{eq:egvec_eff}
\ea 
Clearly, finding the eigenvector perturbatively is equivalent to finding the eigenvector of $D_{eff}(\mathbf{k})$.

From \cref{eq:egval_eff} and \cref{eq:egvec_eff}, we find that solving the system in the small $\alpha$ limit is equivalent to solving the eigenvectors and eigenvalues of the following effective dynamical matrix of Sn
\ba 
D^{eff}(\mathbf{k}) =& \alpha^2D_2(\mathbf{k}) -\alpha^2 D_1'(\mathbf{k})D_0^{-1}D_1(\mathbf{k}) =  \nonumber\\
=&\frac{1}{M_{Sn}}\Phi^{Sn-Sn} - \frac{1}{M_{Sn}}
\begin{bmatrix}
    \frac{1}{\sqrt{M_V}}\Phi^{Sn-V} & \frac{1}{\sqrt{M_Sc}}\Phi^{Sn-Sc} 
\end{bmatrix}
\begin{bmatrix}
  \frac{1}{M_V}[\Phi^{V-V}] & \frac{1}{\sqrt{M_VM_{Sn}}} [\Phi^{V-Sc}] \\ 
  \frac{1}{\sqrt{M_{Sc}M_{V}}} [\Phi^{Sc-V}] &  \frac{1}{M_{Sc}} [\Phi^{Sc-Sc}] 
\end{bmatrix}
\begin{bmatrix}
    \frac{1}{\sqrt{M_V}}\Phi^{V-Sn} \\ \frac{1}{\sqrt{M_{Sc}}}\Phi^{Sc-Sn} 
\end{bmatrix}
\label{eq:Deff}
\ea 
Of course,  $D_{eff}$ is nothing but the perturbed Sn-Sn matrix by the Sn-V and Sn-Sc force constants. 

\begin{figure}
    \centering
    \includegraphics[width=1.0\textwidth]{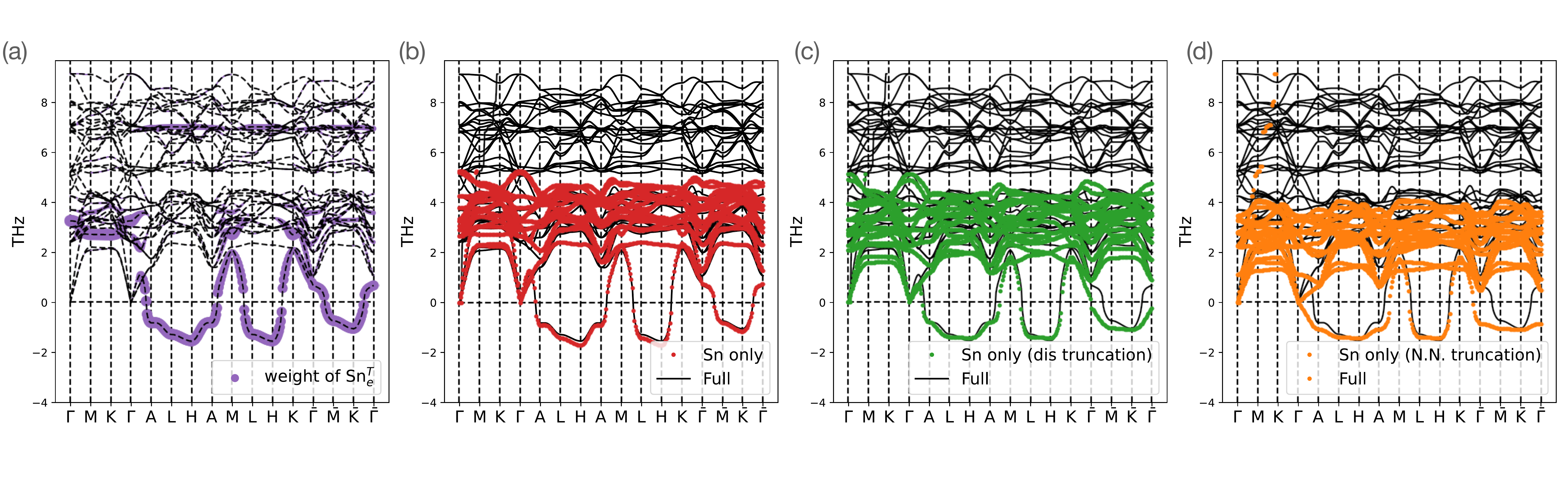}
    \caption{(a) Phonon spectrum of the full model, where the purple dots mark the weights of the mirror-even modes of triangular Sn modes along $z$-driection ($\frac{1}{\sqrt{2}}(u_{Sn_1^T,z}(\mathbf{R}) - u_{Sn_2^T,z}(\mathbf{R}) )$). (b) Comparison between the spectrum of the full model and the spectrum of the effective model with only Sn atoms. (c) Comparison between the spectrum of the full model and the spectrum of the effective model with only Sn atoms and truncated coupling (where the couplings between atoms with a distance smaller than $5.6\AA$ has been included). 
    (d) Comparison between the spectrum of the full model and the spectrum of the effective model with only Sn atoms and truncated coupling (where only the nearest-neighbor couplings between different types of atoms have been included). 
    Here $\bar{\Gamma}=(0,0,\frac{1}{3}),\bar{M} = (\frac{1}{2},0,\frac{1}{3}), \bar{K} = (\frac{1}{3},\frac{1}{3},\frac{1}{3})$. 
}
    \label{fig:Sn_only_phonon_model}
\end{figure}
We now derive the effective phonon model with only Sn atoms. We take the full model derived from our DFT calculation and use \cref{eq:Deff} to derive the effective dynamic matrix $D^{eff}$ that only involves Sn atoms. We make a further approximation and truncate the coupling. \textcolor{black}{Here, two types of truncations have been considered: (1) distance truncation where all the couplings between atoms with a distance smaller than $5.6\AA$ are included; (2) nearest-neighbor truncation where only the nearest-neighbor couplings between different types of atoms are included. 
This leads to an even simpler model $D^{eff,trunc}$. In \cref{fig:Sn_only_phonon_model} (b), (c), (d), we compare the spectrum obtained by the full model, the effective model with only Sn atoms, and the effective model with truncated coupling.} For the low-energy phonon mode, we observe a good agreement between the full model and the effective Sn-only model. The truncated effective Sn-only model produces a worse match than the effective Sn-only model (the latter of which is and excellent approximation) but captures very well the qualitative (and quantitative) behavior of imaginary phonon mode. \textcolor{black}{Moreover, the distance truncation produces a better match than the nearest-neighbor truncation, since distance truncation includes one more next-nearest-neighbor coupling between atoms than the nearest-neighbor truncation.}

We now explicitly write down the dynamic matrix of the truncated effective model(distance truncation) \textcolor{black}{where we have included all the couplings between the atoms with a distance smaller than $5.6\AA$.  The couplings within this distance contain the nearest-neighbor coupling between different types of atoms and also a next-nearest-neighbor coupling between the two honeycomb Sn atoms from different layers (this next-nearest-neighbor coupling is absent in the nearest-neighbor truncation). If we only include the nearest-neighbor coupling, the avoided level-crossing behavior along $\Gamma$-$A$ line (see \cref{sec:level_crossing}) will not be very well reproduced as shown in \cref{fig:Sn_only_phonon_model} (d) vs (c).}
We label the six Sn atoms as $1,2,3,4,5,6$ with their position vector in the unit cells are 
\ba 
&\mathbf{r}^{Sn}_1 = z_0\bm{a}_3 ,\quad \mathbf{r}^{Sn}_2 = (1-z_0)\bm{a}_3 \nonumber \\ 
&\mathbf{r}^{Sn}_{3} = \frac{2}{3}\bm{a}_1 + \frac{1}{3}\bm{a}_2 ,\quad 
\mathbf{r}^{Sn}_{4} = \frac{1}{3}\bm{a}_1 + \frac{2}{3}\bm{a}_2, \quad 
\mathbf{r}^{Sn}_{5} = \frac{2}{3}\bm{a}_1 + \frac{1}{3}\bm{a}_2 +\frac{1}{2}\bm{a_3},\quad 
\mathbf{r}^{Sn}_{6} = \frac{1}{3}\bm{a}_1 + \frac{2}{3}\bm{a}_2+\frac{1}{2}\bm{a_3}
\label{eq:atom_pos}
\ea 
where $z_0=0.3243$ for experimental structure and $z_0=0.3175$ for relaxed structure. $1,2$ label two triangular Sn and $3,4,5,6 $ label four honeycomb Sn. 
The truncated effective dynamic matrices of Sn are shown below (with unit THz$^2$ that has been omitted)
\begin{itemize}
    \item In-plane nearest-neighbor interaction between triangular Sn$^{T}$. 
    \ba 
    D^{eff}_{11, \mu\nu}(\bm{a}_1) =D^{eff}_{22,\mu\nu}(\bm{a}_1) = \begin{bmatrix}
    -1.81 & 0 & 0.23 \\
    0&-0.22&0 \\
    -0.23&0&0.17 
    \end{bmatrix}_{\mu\nu}
    \label{eq:int_tig_sn_tirg_n_xy}
    \ea 
    \item Out-of-plane nearest-neighbor interaction between two triangular Sn$^T$ along $z$ directions.
    \ba 
    D^{eff}_{12,\mu \nu}(\bm{0}) = 
    \begin{bmatrix}
        -0.14 & 0& 0 \\
        0& -0.14 & 0 \\
        0 & 0 &2.61 
    \end{bmatrix},\quad 
    D^{eff}_{12}(-\bm{a}_3) = 
    \begin{bmatrix}
        0.04& 0& 0 \\
        0& 0.04 & 0 \\
        0 & 0 &-4.06
    \end{bmatrix} 
    \label{eq:int_trig_sn}
    \ea 
    \item In-plane nearest-neighbor interaction between honeycomb Sn$^H$. 
    \ba 
    D^{eff}_{34}(\bm{0})
    =\begin{bmatrix}
        -2.27& 1.12 & 0 \\
        1.12 &-0.98 & 0 \\
        0 & 0 & -0.97
    \end{bmatrix},\quad 
     D^{eff}_{56}(\bm{0})
    =\begin{bmatrix}
        -2.36 & 1.54 & 0 \\
        1.54 & -0.58 & 0 \\
        0 & 0 & -0.88
    \end{bmatrix}
    \ea 
    \item Out-of-plane nearest-neighbor interaction between two honeycomb Sn$^H$ layers. 
    \ba 
    D^{eff}_{35}(\bm{0} )  = 
    \begin{bmatrix}
    0.18 & 0 & 0 \\
        0&0.18 & 0 \\
        0& 0 & -0.75
    \end{bmatrix}
    \ea 
    \textcolor{black}{ 
    \item Out-of-plane next-nearest-neighbor interaction between two honeycomb Sn$^H$ layers. 
    \ba 
    D^{eff}_{43}(-\bm{a}_2 )  = 
    \begin{bmatrix}
    -0.03 & -0.0 & 0.0 \\
        -0.0 & -0.57 & -0.78 \\
        0.0&-0.86 & -1.53
    \end{bmatrix}
    \ea 
    }
    \item Out-of-plane nearest-neighbor interaction between honeycomb Sn$^H$ and triangular Sn$^T$.
    \ba 
    D^{eff}_{31}(\bm{0})  = 
    \begin{bmatrix}
        -0.33 & -0.14& 0.23 \\
        -0.14 &  -0.16 & 0.13 \\
        0.36 & 0.21 & -0.32 
    \end{bmatrix},\quad 
     D^{eff}_{51}(\bm{0})  = 
    \begin{bmatrix}
        -0.8 & -0.38& -0.27 \\
        -0.38 &  -0.36 & -0.16 \\
       -0.04& -0.02 & -0.49
    \end{bmatrix}
    \label{eq:int_trig_sn_h_sn}
    \ea 
\end{itemize}
The dynamical matrix along other bonds can be generated by symmetry transformations as introduced in \cref{eq:symmetry_transf}. 

However, we cannot eliminate the honeycomb Sn atoms with the same procedure. This is because both triangular and honeycomb Sn atoms contribute to the low-energy phonon spectrum and a direct perturbation theory fails. However, an even simpler phonon model with only three Wannier orbitals could indeed be obtained via the Wannier construction procedure, as we will discuss in \cref{app:sec:three_band_model}. Before going to the three-band phonon model, we will first discuss the properties of the phonon spectrum based on our Sn-only model in the next \siSection{} \cref{sec:imag_flat}. We will first show the imaginary flat band can be understood from an effective 1D model (\cref{sec:imag_flat}), which only contains $z$-direction vibration of the two triangular Sn modes. \textcolor{black}{The effective 1D model cannot quantitatively reproduce the phonon spectrum but could qualitatively describe the imaginary flat phonon mode.}
Then we analyze the behaviors of the acoustic phonon (\cref{sec:acoustic}). Finally, we show the hybridization between the 1D phonon mode and the acoustic mode produces an avoided level crossing along $\Gamma$-$A$ line (\cref{sec:level_crossing}).

\subsection{Imaginary flat mode}
\label{sec:imag_flat}
From the zero temperature DFT calculated phonon spectrum (\cref{Fig-Pho-Sc-highT}), we observe that there is a relatively flat mode with imaginary frequency at $k_z=\pi$ plane. Furthermore, the imaginary phonon mode describes the out-of-plane vibration of triangular Sn$^T$ with the opposite direction 
($\frac{1}{\sqrt{2}}(u_{Sn^T_1 z}(\mathbf{R}) - u_{Sn^T_1 z}(\mathbf{R}))$) 
as shown in \cref{fig:Sn_only_phonon_model} (a). From our truncated effective model, the $z$ direction displacement of Sn$^T$ only weakly couples to the nearby Sn$^H$ (\cref{eq:int_trig_sn_h_sn}) and the Sn$^T$ on the same plane (\cref{eq:int_tig_sn_tirg_n_xy}), but will strongly couple to the Sn$^T$ that are located at the same $x,y$ coordinates but with different $z$ coordinates (\cref{eq:int_trig_sn}). The origin of the weak in-plane dispersion has been discussed later in \cref{sec:ele_corr_Sc}.
This allows us to treat the $z$ direction displacement field of Sn $^T$ via an effective one-dimensional(1D) model along $z$ direction. 
\textcolor{black}{The effective 1D model cannot quantitively reproduce the phonon spectrum but could provide a qualitative understanding of the imaginary phonon mode.}
The corresponding dynamical matrix that only contains the $z$ directional movement of two triangular Sn atoms are 
\ba 
D_{1D}(\mathbf{k}) = &
\begin{bmatrix}
    d_1 +d_2 & - d_1e^{ik_z(1-2z_0)}  -d_2 e^{ik_z(-2z_0)} \\ 
     - d_1e^{-ik_z(1-2z_0)}  -d_2 e^{ik_z(2z_0)} & d_1 +d_2 
\end{bmatrix}\nonumber\\ 
= &
\begin{bmatrix}
    d_1 +d_2 & - d_1e^{ik_z\Delta r_z}  -d_2 e^{ik_z (-1+\Delta r_z)} \\ 
     - d_1e^{-ik_z\Delta r_z}  -d_2 e^{ik_z(1-\Delta r_z)} & d_1 +d_2 
\end{bmatrix} 
\label{eq:1d_dyn_mat}
\ea 
where $d_1$ describes the 
coupling \textcolor{black}{in the dynamical matrix} between two triangular Sn within the same unit cell and $d_2$ describes the coupling \textcolor{black}{in the dynamical matrix} between two triangular Sn atoms with one located at the unit cell at $\mathbf{R}$ and the other one located at the unit cell at $\mathbf{R}+\bm{a}_3$. Here $k_z = \bm{\mathbf{k}} \cdot \bm{a}_3 $. 
$\Delta r_z =1-2z_0$ describes the distance between two Sn with $z_0$ defined in \cref{eq:atom_pos}. 
From 
\cref{eq:int_trig_sn}, $d_1 =- 2.61 <0$ and $d_2=4.06>0$, values which will later (\cref{sec:ele_corr_Sc}) 
justified and obtained analytically through a (Gaussian) approximation for the electron-phonon coupling and renormalization of the phonon frequency. We now show that the negative $d_1$ leads to an imaginary mode. 
From \cref{eq:eom_eta}, the equation of motion of the 1D model is 
\ba 
\omega(\mathbf{k})^2  \begin{bmatrix}
    \eta^{1z}(\mathbf{k}) \\ \eta^{2z}(\mathbf{k}) 
\end{bmatrix}_i 
= \sum_j D_{1D}(\mathbf{k})_{ij}\begin{bmatrix}
    \eta^{1z}(\mathbf{k}) \\ \eta^{2z}(\mathbf{k}) 
\end{bmatrix}_j
\label{eq:eom_1d}
\ea 
where $\eta^{1z}(\mathbf{k})$ and $\eta^{2z}(\mathbf{k})$ describe the $z$ directional movement of two triangular Sn atoms.

We next introduce the following two new basis vectors which are even $\eta^{e,z}$ and odd $\eta^{o,z}$ under mirror $z$ transformation respectively
\ba 
\eta^{e,z}(\mathbf{R}) = \frac{1}{\sqrt{2}}[\eta^{1 z}(\mathbf{R}) - \eta^{2 z}(\mathbf{R})] \nonumber \\
\eta^{o,z}(\mathbf{R}) = \frac{1}{\sqrt{2}}[\eta^{1 z}(\mathbf{R}) + \eta^{2 z}(\mathbf{R})]
\ea 
The Wannier center of the two modes locates at $\mathbf{r}_e = \mathbf{r}_o = \frac{1}{2}\bm{a}_z$. We then consider the corresponding Fourier transformation
\ba 
&\eta^{e,z}(\mathbf{k}) = \frac{1}{\sqrt{N}}\sum_\mathbf{R} e^{-i\mathbf{k}\cdot (\mathbf{R}+\mathbf{r}_e)}\eta^{e,z}(\mathbf{R}) = \frac{1}{\sqrt{2}}[\eta^{1 z}(\mathbf{k})e^{i\mathbf{k}_z \Delta r } - \eta^{2 z}(\mathbf{k})e^{-i\mathbf{k}_z \Delta r}] \nonumber\\ 
&
\eta^{o,z}(\mathbf{k}) =\frac{1}{\sqrt{N}}\sum_\mathbf{R} e^{-i\mathbf{k}\cdot (\mathbf{R}+\mathbf{r}_o)}\eta^{o,z}(\mathbf{R})= \frac{1}{\sqrt{2}}[\eta^{1 z}(\mathbf{k})e^{i\mathbf{k}_z \Delta r} + \eta^{2 z}(\mathbf{k})e^{-i\mathbf{k}_z \Delta r}] 
\label{eq:flat_mode_basis}
\ea 
where $\Delta r = z_0-\frac{1}{2}$ denote the distance between the atoms and the Wannier centers. 

In the new basis, the equation of motion (\cref{eq:eom_1d}) becomes 
\ba 
&\omega(\mathbf{k})^2  \begin{bmatrix}
    \eta^{e,z}(\mathbf{k}) \\ \eta^{o,z}(\mathbf{k}) 
\end{bmatrix}_i 
= \sum_j \tilde{D}_{1D}(\mathbf{k})_{ij}\begin{bmatrix}
     \eta^{e,z}(\mathbf{k}) \\ \eta^{o,z}(\mathbf{k})
\end{bmatrix}_j  \nonumber \\ 
&\tilde{D}_{1D}(\mathbf{k}) = 
\begin{bmatrix}
    2d_1+d_2+d_2\cos(k_z) & -i d_2\sin(k_z) \\ 
    id_2\sin(k_z)  & d_2 -d_2 \cos(k_z) 
\end{bmatrix}
\label{eq:eom_1d_new}
\ea 

The two eigenvalues are 
\ba 
\omega_1(\mathbf{k})^2  = d_1+d_2-\sqrt{d_1^2+d_2^2+2d_1d_2\cos(k_z)} ,\quad 
\omega_2(\mathbf{k})^2  = d_1+d_2+\sqrt{d_1^2+d_2^2+2d_1d_2\cos(k_z)} 
\label{eq:eig_1d_mode}
\ea 

Since $d_2$ is positive and $d_1$ is negative, this naturally leads to an imaginary phonon mode. $\omega_1(\mathbf{k})^2$ is most negative at $k_z=\pi$ with
\ba 
\omega_1^2(k_z=\pi) = 2d_1 <0 
\ea 
which indicates the leading order instability at $k_z=\pi$. The reason for this is the nearest neighbor nature of the dynamical matrix: all nearest neighbor $1D$ dynamical matrices with $d_1<0$ (instability) would give a minimum at $k_z=\pi$ 
The corresponding eigenvector is (in the new basis) 
\ba 
\begin{bmatrix}
   1 & 0 
\end{bmatrix}^T 
\ea 
or equivalently $\eta^{e,z}(k_z=\pi)$. 
In real space, this leads to the following displacement
\ba 
\eta^{1z}(\mathbf{R}) = -\eta_0 (-1)^{i_z},\quad \eta^{2z}(\mathbf{R}) = \eta_0(-1)^{i_z} 
\ea 
where we use $i_z$ to denote $i_z$-th unit cell along $z$ directions and $\eta_0$ is a real number. This pattern of movement is shown in \cref{Fig-Sc-movement}.

\subsection{Acoustic modes} 
\label{sec:acoustic}
We next analyze the acoustic modes near the $\Gamma$ points. We focus on the small momentum near $\Gamma$ point and project the dynamical matrix of the Sn model (\cref{eq:Deff}) to a subspace spanned by the following three acoustic modes (\cref{eq:acoustic_zero_q})
\ba 
\eta^{\alpha \mu}_{x}(\mathbf{k}) = \frac{1}{\sqrt{6}} \delta_{\mu,x}  \nonumber \\ 
\eta^{\alpha \mu}_{y}(\mathbf{k}) = \frac{1}{\sqrt{6}} \delta_{\mu,y}  \nonumber \\ 
\eta^{\alpha \mu}_{z}(\mathbf{k}) = \frac{1}{\sqrt{6}} \delta_{\mu,z} 
\label{eq:acoustic_basis}
\ea 
with $\mu \in \{x,y,z\}$ and $\alpha=1,...,6$ labels six Sn atoms. At $\mathbf{k}=0$, $\eta_{x,y,z}(\mathbf{k}=0)$ describes three acoustic modes with zero energy and satisfy 
\ba 
\sum_{\gamma} D^{eff}_{\alpha\gamma,\mu\nu} \eta_x^{\gamma\mu} =\sum_{\gamma} D^{eff}_{\alpha\gamma,\mu\nu} \eta_y^{\gamma\mu}
=\sum_{\gamma} D^{eff}_{\alpha\gamma,\mu\nu} \eta_z^{\gamma\mu}=0
\ea 
where $D^{eff}$ is the effective dynamical matrix with only Sn atoms (\cref{eq:Deff}). 

We next project the effective dynamical matrix to the subspace spanned by three phonon modes. We let 
\ba 
D_{mn}^{accous}(\mathbf{k}) = \sum_{\alpha \mu \gamma\nu} \eta^{\alpha\mu}_{m} (\mathbf{k})D^{eff}_{\alpha\mu,\gamma\nu }(\mathbf{k})\eta^{\gamma \nu,*}_n(\mathbf{k}) ,\quad m,n \in \{x,y,z\}
\ea 

The dispersion of the acoustic modes can then be derived by finding the eigenvectors of $D^{accous}(\mathbf{k})$. We can also perform a small $|\mathbf{k}|$ expansion of $D^{\mu\nu}_{acous}(\mathbf{k})$
\ba 
D_{\mu\nu}^{acous}(\mathbf{k}) \approx D^{accous,0}_{\mu\nu} + \sum_{\delta } D^{accous,1,\delta}_{\mu\nu}k^\delta  + \sum_{\delta \sigma  } D^{accous,2,\delta \sigma}_{\mu\nu} k^\delta k^\sigma 
\ea 
Since at $\mathbf{k}=0$, three acoustic modes are eigenstates with zero energy, $D^{accous,0}=0$. 
We note that the $D_{\mu\nu}^{acous}(\mathbf{k})$ should follow the similarly symmetry constraints as \cref{eq:symmetry_const}. Here we give the transformation matrix of three acoustic modes under symmetry. 
\ba 
&D^{accous,C_{3z}} = 
\begin{bmatrix}
    -\frac{1}{2} & \frac{\sqrt{3}}{2} \\ 
    -\frac{\sqrt{3}}{2} & -\frac{1}{2} \\ 
    & & 1 
\end{bmatrix},\quad 
D^{accous,C_{2z}} = 
\begin{bmatrix}
   -1  \\ 
   &-1\\ 
    & & 1 
\end{bmatrix},
\quad 
D^{accous,C_{2,120}} = 
\begin{bmatrix}
    -1 &0  &  \\ 
  0 &1 \\ 
    & & -1 
\end{bmatrix},\nonumber \\ 
& 
D^{accous,i} = 
\begin{bmatrix}
   -1& &  \\ 
  &-1 \\ 
    & & -1 
\end{bmatrix}
\ea 
The symmetry-allowed $D^{accous}(\mathbf{k})$ at small $|\mathbf{k}|$ takes the form of
\textcolor{black}{
\ba 
&D^{accous}_{\mu\nu}(\mathbf{k}) \approx 
\begin{bmatrix}
    \alpha_{xy,1} k_x^2 + \alpha_{xy,2}k_y^2 +\alpha_{xy,3}k_z^2& (\alpha_{xy,1} - \alpha_{xy,2})k_xk_y  & \alpha_{xz} k_xk_z\\ 
    (\alpha_{xy,1} - \alpha_{xy,2})k_xk_y  & \alpha_{xy,1}k_y^2 + \alpha_{xy,2}k_x^2 +\alpha_{xy,3}k_z^2& \alpha_{xz} k_yk_z \\ 
    \alpha_{xz} k_xk_z & \alpha_{xz} k_yk_z& \alpha_{zz,1}(k_x^2+k_y^2) +\alpha_{zz,2}k_z^2
\end{bmatrix} +O(|\mathbf{k}|^3)
\label{eq:accous_kp}
\ea 
}
where $\alpha_{xy,1},\alpha_{xy,2},\alpha_{xy,3},\alpha_{zz,1},\alpha_{zz,2},\alpha_{xz}$ are real numbers that characterize the acoustic phonon modes. The corresponding phonon modes take the dispersion of
\textcolor{black}{ 
\ba 
(\omega_{1}(\mathbf{k}))^2 = &(k_x^2+k_y^2)\alpha_{xy,2} + \alpha_{xy,3}k_z^2 \nonumber \\ 
(\omega_{2}(\mathbf{k}))^2 = &
\frac{1}{2}\bigg\{ 
(\alpha_{xy,1}+\alpha_{zz,1})(k_x^2+k_y^2) +(\alpha_{xy,3} +\alpha_{zz,2})k_z^2 \nonumber\\ 
&
+
\sqrt{
\bigg[ 
(\alpha_{xy,1}-\alpha_{zz,1})(k_x^2+k_y^2) +(\alpha_{xy,3} -\alpha_{zz,2})k_z^2 
\bigg]^2 + 4\alpha_{xz}^2(k_x^2+k_y^2)k_z^2
} 
\bigg]
\bigg\}
\nonumber \\ 
(\omega_{3}(\mathbf{k}))^2 = &
\frac{1}{2}\bigg\{ 
(\alpha_{xy,1}+\alpha_{zz,1})(k_x^2+k_y^2) +(\alpha_{xy,3} +\alpha_{zz,2})k_z^2 \nonumber\\ 
&
-
\sqrt{
\bigg[ 
(\alpha_{xy,1}-\alpha_{zz,1})(k_x^2+k_y^2) +(\alpha_{xy,3} -\alpha_{zz,2})k_z^2 
\bigg]^2 + 4\alpha_{xz}^2(k_x^2+k_y^2)k_z^2
} 
\bigg]
\bigg\}
\label{eq:accous_disp}
\ea 
By fitting the dynamical matrix with \cref{eq:accous_kp} near $\mathbf{k} =\bm{0}$, we find $\alpha_{xy,1} = 48.0\text{THz}^2\AA^2$, $\alpha_{xy,2} = 19.8\text{THz}^2\AA^2$,
$\alpha_{xy,3} =17.3\text{THz}^2\AA^2$,
$\alpha_{zz,1} =25.5\text{THz}^2\AA^2$,
$\alpha_{zz,2} =100.0\text{THz}^2\AA^2$,
$\alpha_{xz} =30.8\text{THz}^2\AA^2$,
}
We observe that along $\Gamma$-$A$ line with $k_x=k_y=0$, we have 
\ba 
(\omega_1((0,0,k_z))^2 = \alpha_{xy,3}k_z^2,\quad 
(\omega_2((0,0,k_z))^2 = \alpha_{xy,3}k_z^2,\quad (\omega_3((0,0,k_z))^2 = \alpha_{zz,2}k_z^2
\ea 
and $\omega_1,\omega_2$ modes are degenerate along this line.

\subsection{Coupling between acoustic modes and imaginary modes} 
\label{sec:level_crossing}
We next study the coupling between acoustic modes and imaginary modes along $\Gamma-A$ line. \textcolor{black}{We will show there is an avoided level-crossing along this line.}
We consider the Hilbert space spanned by three acoustic modes $\eta^{x}(\mathbf{k}),\eta^y(\mathbf{k}),\eta^z(\mathbf{k})$ (\cref{eq:acoustic_basis}), and the flat mode $\eta^{e,z}(\mathbf{k})$ (\cref{eq:flat_mode_basis}). We have dropped the $\eta^{o,z}(\mathbf{k})$ mode here for the following reason: $\eta^{o,z}(\mathbf{k})$ mode has non-zero overlap with acoustic modes (since it describes the same direction movements of two triangular Sn); then the coupling between $\eta^{o,z}(\mathbf{k})$ and $\eta^{e,z}(\mathbf{k})$ mode will be partially captured by the coupling between acoustic mode and $\eta^{e,z}(\mathbf{k})$ modes.
\textcolor{black}{Moreover, since $|d_2|$ is much larger than $|d_1|$, the low-energy phonon mode is dominant by the $\eta^{e,z}(\mathbf{k})$ orbital for most $\mathbf{k}$. Even though at $k_z=0$, the low-energy phonon mode is completely formed by $\eta^{a,z}(\mathbf{k})$, as we move away from $k_z$, the low-energy phonon mode will be quickly dominant by $\eta^{e,z}(\mathbf{k})$. }
Thus we have omitted $\eta_{o,z}$ in the current consideration.
We introduce the following basis 
\ba 
U_1(\mathbf{k}) =
    \eta^x (\mathbf{k}) ,\quad U_2(\mathbf{k})=  \eta^y (\mathbf{k}), \quad U_3(\mathbf{k})  = \eta^z (\mathbf{k}) ,\quad   U_4(\mathbf{k}) = \eta^{e,z} (\mathbf{k}) 
\ea 
We consider the dynamical matrix defined in the $U_n(\mathbf{k})$ basis, which takes the form of 
\ba 
&D_U(\mathbf{k}) = 
\begin{bmatrix}
    [D_{acous}(\mathbf{k})]_{3\times 3} & v_\mathbf{k} \\ 
    v_\mathbf{k}^\dag & D_{e,z}(\mathbf{k})
\end{bmatrix},\quad 
D_{e,z}(\mathbf{k}) = \begin{bmatrix}
    \epsilon_{e,z}(\mathbf{k})
\end{bmatrix}
\label{eq:dyn_acous_flat}
\ea 
where $D_{acous}(\mathbf{k})$ describes the acoustic modes (\cref{eq:accous_kp}) and $D_{e,z}(\mathbf{k}) $ (the top left element of the matrix in \cref{eq:eom_1d_new}) describes the dispersions of mirror-even phonon fields $\eta^{e,z}$ with $\epsilon_{e,z}(\mathbf{k}) = 2d_1 + d_2(1+\cos(k_z))$. However, due to the sum rule (\cref{eq::phono_sum_rule_1} or \cref{eq::phono_sum_rule_2}), the coupling between the mirror-even phonon mode and the acoustic mode could introduce a $\mathbf{k}$-\textcolor{black}{dependent} term to the bottom left element of $D_U$ (\cref{eq:dyn_acous_flat}) 
. 
To capture this effect, we normalize $d_1$ such that $\epsilon_{e,z}(\mathbf{k}=0)$ matches its numerical value from DFT calculations, which gives $d_1 \approx -0.6\text{THZ}^2$. We also point out that, at $k_z=\pi$ plane, $\eta^{e,z}$ is just the imaginary phonon mode we identified from the simple 1D model (\cref{eq:eom_1d}). 
$v_\mathbf{k}$ describes the coupling between acoustic modes and flat modes. We focus on the high symmetry line $\Gamma-A$, where the symmetry allowed $v_{\mathbf{k}}$ are 
\ba 
v_{\mathbf{k} =(0,0,k_z)} = \begin{bmatrix}
    0 & 0 & i\lambda_{k_z} 
\end{bmatrix}^T
\ea 
with $\lambda_{k_z} = -\lambda_{-k_z}$. We then assume, to first order, $\lambda_{k_z} \approx \lambda k_z$ and we numerically find $\lambda=18.2\text{THZ}^2\AA$ The coupling between $\eta^{a,z}$ and $\eta^z$ leads to two dispersive modes with 
\ba 
\omega_{1,2}(\mathbf{k})^2 = \frac{ \epsilon_z(\mathbf{k})  + \epsilon_{e,z}(\mathbf{k}) }{2} \pm \sqrt{\lambda^2 k_z^2 + \frac{ ( \epsilon_z(\mathbf{k})  - \epsilon_{e,z}(\mathbf{k}) )^2 }{4}} 
 \ea 
 where from $\epsilon_{z}(\mathbf{k}) = \alpha_{zz,1}(k_x^2+k_y^2) + \alpha_{zz,2}k_z^2$ is the dispersion (\cref{eq:accous_disp}) of the third acoustic modes(\cref{eq:accous_kp}).
 
\begin{figure}
    \centering   
    \includegraphics[width=0.5\textwidth]{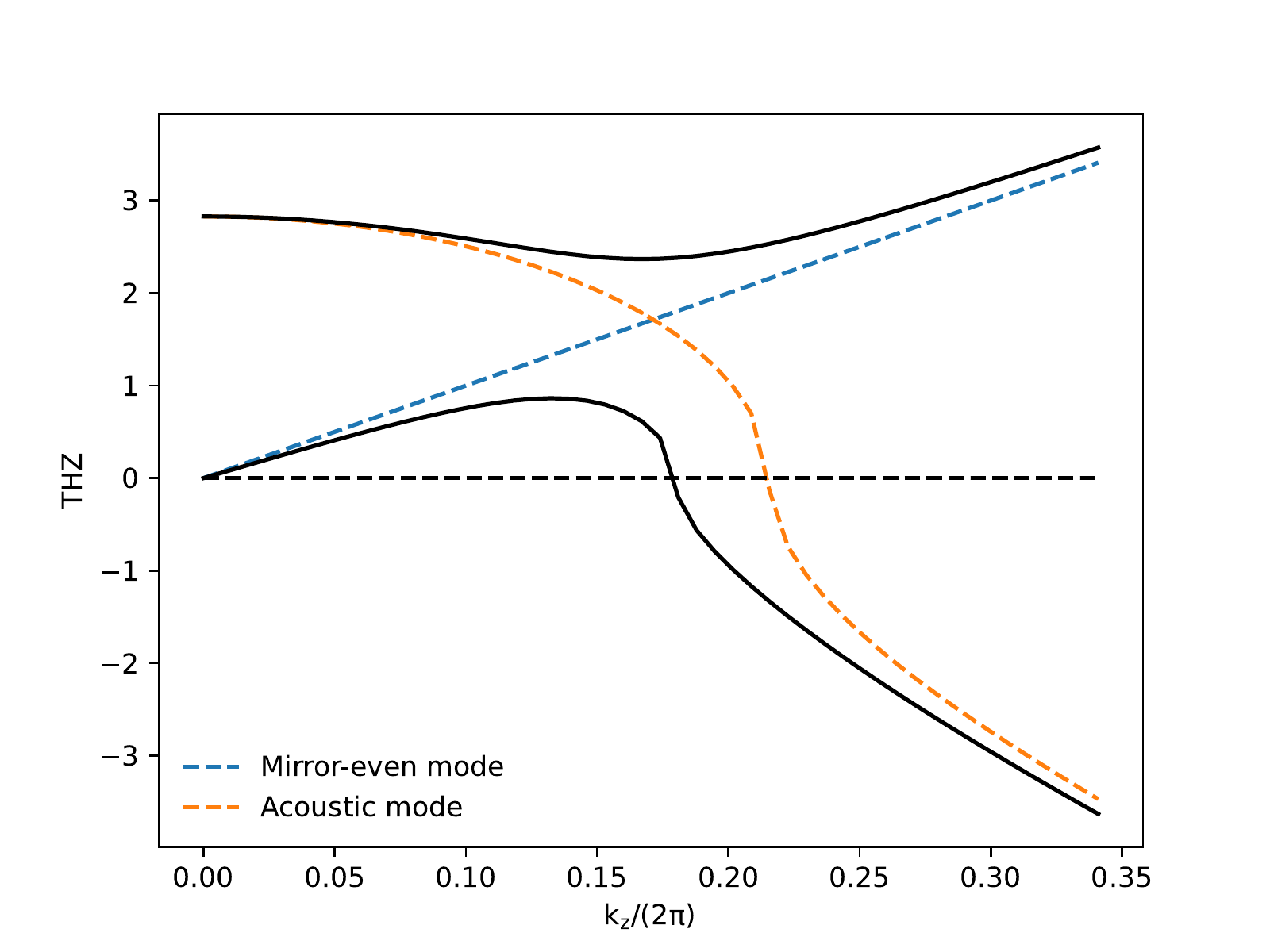}
    \caption{\textcolor{black}{Avoided level-crossing along $\Gamma$-$A$ line. The orange and blue dashed line marks the phonon frequency of acoustic mode $(\epsilon_z(\mathbf{k}))$ and mirror-even $\eta^{e,z}$ mode $(\epsilon_{e,z}(\mathbf{k}))$ in the absence of the coupling between two modes $(\lambda=0)$. The solid black lines show the phonon frequency (\cref{eq:level_cross}) at the nonzero coupling $\lambda\ne 0$ between two modes (acoustic, and mirror-eve $\eta^{e,z}$). We observe an avoided level crossing after turning on $\lambda$. }}
    \label{fig:avoid_level_crossing}
\end{figure}

We consider $\mathbf{k}=(0,0,k_z)$ and perform a small $|k_z|$ expansion. We find 
\ba 
&\omega_1^2(k_z)\approx (\alpha_{zz,2} - \frac{\lambda^2}{2(d_1+d_2)})k_z^2
- \frac{\lambda^2}{8(d_1+d_2)^2}\bigg(2\alpha_{zz,2}+d_2  - \frac{\lambda^2}{d_1+d_2} \bigg)k_z^4 = v_z^2k_z^2 - \alpha_{zz,4} k_z^4  \nonumber \\
&v_z^2 = (\alpha_{zz,2} - \frac{\lambda^2}{2(d_1+d_2)}),\quad 
\alpha_{zz,4} = \frac{\lambda^2}{8(d_1+d_2)^2}\bigg(2\alpha_{zz,2}+d_2  - \frac{\lambda^2}{d_1+d_2} \bigg) \nonumber \\ 
&\omega_2^2 (k_z)\approx 2(d_1+d_2) -\frac{1}{2}(d_2 - \frac{\lambda^2}{2(d_1+d_2)})k_z^2
+ \frac{\lambda^2}{24(d_1+d_2)^2}\bigg(6\alpha_{zz,2}+4d_2  -3 \frac{\lambda^2}{d_1+d_2} \bigg)k_z^4
\label{eq:level_cross}
\ea 
Since $d_1+d_2>0$, $\omega_2$ describes a gapped mode near $\mathbf{k}=\bm{0}$. $\omega_1^2$ describes an acoustic mode with velocity $v_z$. \textcolor{black}{In practice, we have $\alpha_{zz,4}\approx 0.05\text{THz}^2\AA^4>0$}
. Therefore the quadratic contribution to the $\omega_1(k_z)^2$ is negative. However, the bilinear contribution to $\omega_1(k_z)^2$ is positive $(v_z^2k_z^2)$. Then the different signs of quadratic term and bilinear term produce a small peak structure near $\mathbf{k}=0$ along $\Gamma$-$A$ line that has been observed in the phonon spectrum in \cref{fig:Sn_only_phonon_model} (a) and also illustrated in \cref{fig:avoid_level_crossing}.

\begin{figure}
    \centering
    \includegraphics[width=0.8\textwidth]{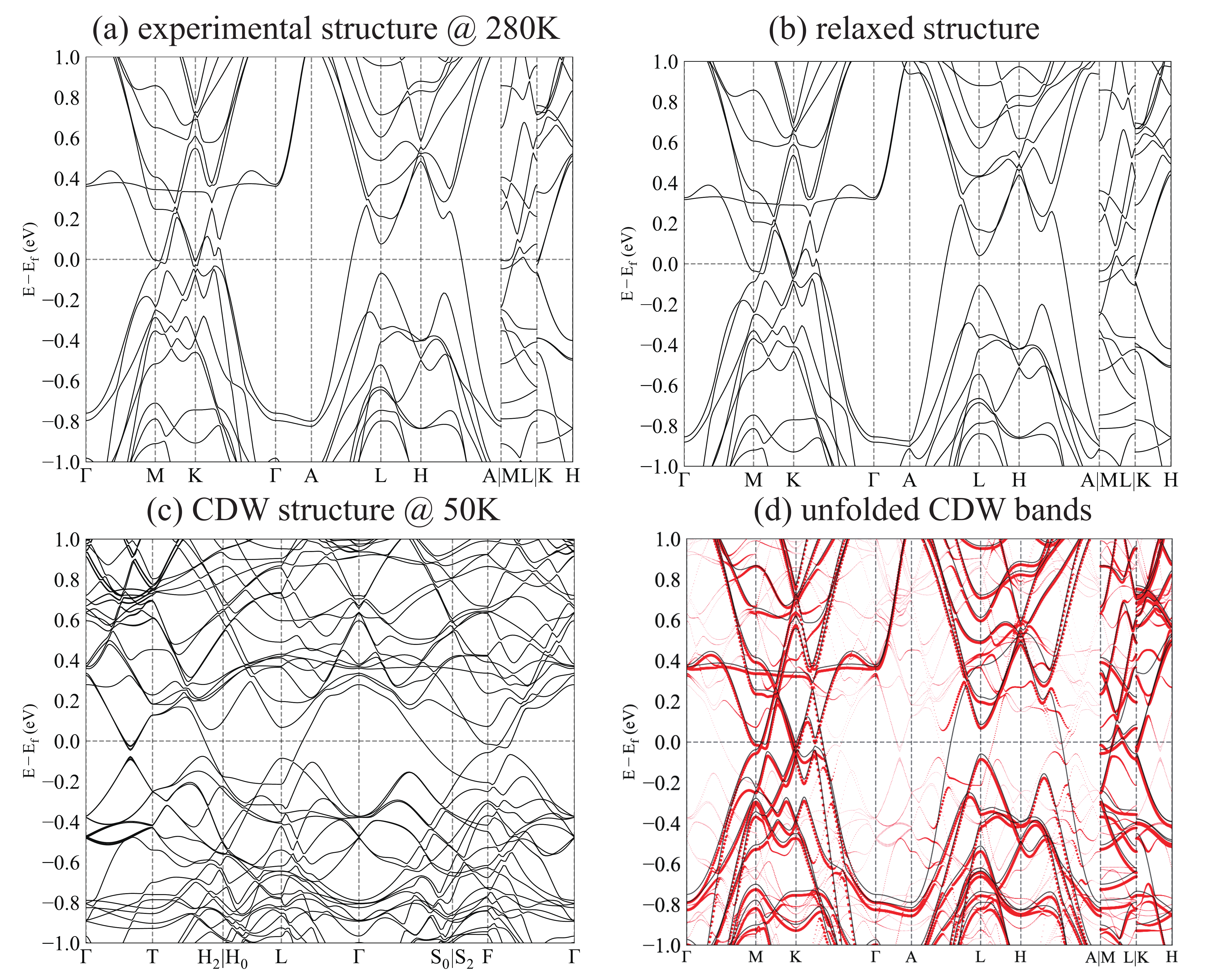}
    \caption{The band structure of ScV$_6$Sn$_6$ in the pristine and  $(\frac{1}{3},\frac{1}{3},\frac{1}{3})$ CDW phase. 
    (a)(b) The band structure with SOC of the experimental\cite{ARA22} and relaxed pristine structure, respectively. The Fermi energy of the relaxed structure is about 30 meV higher on the $k_z=0$ plane.
    (c) The band structure of the CDW phase in the BZ of the primitive cell in SG 155. (d) Unfolded band structure of the CDW phase to the pristine BZ of SG 191, where the red dots are unfolded bands while the grey lines are bands of the experimental pristine structure. 
    }
\label{fig:ScV6Sn6_struct_band}
\end{figure}

\begin{figure}
    \centering
    \includegraphics[width=1\textwidth]{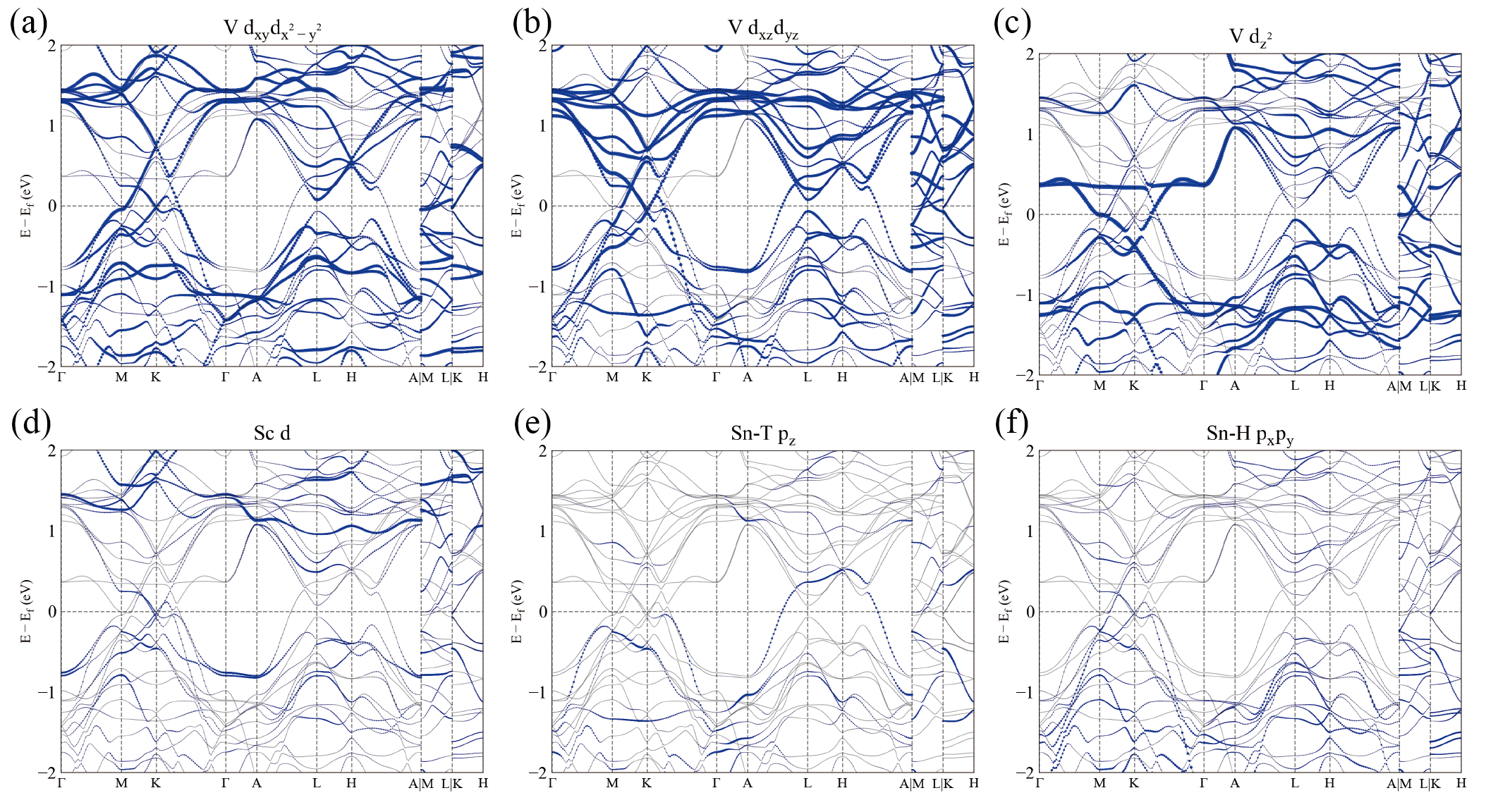}
    \caption{The orbital projections of bands of ScV$_6$Sn$_6$ of the experimental pristine phase.
    (a) $d_{xy},d_{x^2-y^2}$ orbitals of V. (b) $d_{xz}, d_{yz}$ of V. (c) $d_{z^2}$ of V. (d) $d$ of Sc. (e) $p_z$ of triangular Sn. (f) $p_x,p_y$ of honeycomb Sn.
    }
    \label{fig:fatband}
\end{figure}

\begin{figure}
    \centering
    \includegraphics[width=1\textwidth]{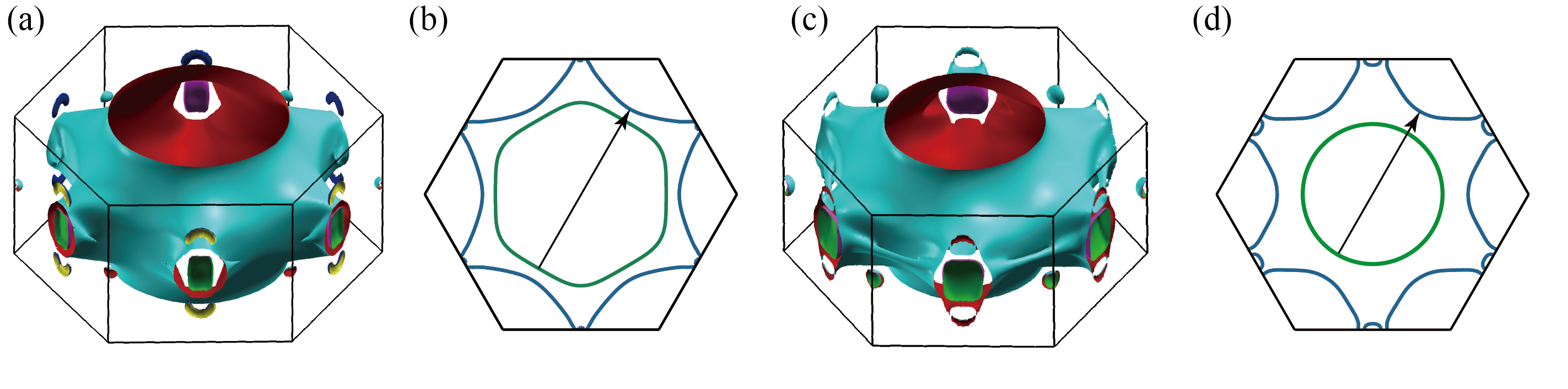}
    \caption{The Fermi surfaces (FS) of pristine ScV$_6$Sn$_6$.  (a) FS of the experimental structure\cite{ARA22}. (b) Two slices of FS, with the green line being $k_z=-0.32$ and the blue line being $k_z=0.18$. The arrow denotes a weak FS nesting of $\bm{q}/(2\pi)=(0.4, 0.4, 0.5)$. (c) FS of the relaxed structure. (d) Two slices of FS, with the green line being $k_z/(2\pi)=0.20$ and the blue line being $k_z/(2\pi)=0.53$. The arrow denotes a weak FS nesting of $\bm{q}/(2\pi)=(\frac{1}{3}, \frac{1}{3}, \frac{1}{3})$.}
    \label{fig:fermi_surface}
\end{figure}

\section{Electronic band structure}
We also perform a DFT calculation and obtain the electronic band structure of the system. 
In \cref{fig:ScV6Sn6_struct_band}(a)(b), we compute the band structure with spin-orbital coupling (SOC). The Fermi energy $E_f$ of the relaxed structure on $k_z=0$ plane is about 30 meV higher than the experimental one. In \cref{fig:fatband}, we show the orbital projections of the experimental structure. We only show the orbitals with non-negligible distributions near $E_f$. The band that crosses $E_f$ along $A$-$L$ and $H$-$A$ and forms part of the large Fermi surface is mainly $d_{xz}, d_{yz}$ of V and $p_z$ of triangular Sn. 

The density of the states of different orbitals at Fermi energy are shown in \cref{tab:dos_Sc}. We can clearly see only $d$ orbitals of V and $p_z$ orbitals of triangular Sn are the relevant low-energy degrees of freedom. 
\begin{table}
    \centering
    \begin{tabular}{c|c|c |c |c |c}
    \hline\hline
        Orbitals & Sc, $d$ orbital & V, $d$ orbital & Triangular Sn, $p_z$ orbital & Triangular Sn, $p_{x,y}$ orbital & Honeycomb Sn, $p$ orbital\\
        \hline 
        DOS@E$_\text{f}$ (relaxed structure) & 0.19 & 2.94 & 0.39 & 0.05 & 0.14\\
          \hline 
        DOS@E$_\text{f}$ (experimental structure) & 0.13 & 2.87 & 0.35 & 0.05 & 0.10\\
    \hline\hline
    \end{tabular}
    \caption{Density of states (DOS) of different orbitals at the Fermi energy E$_\text{F}$.}
    \label{tab:dos_Sc}
\end{table}

In \cref{fig:fermi_surface}(a)(c), we compute the Fermi surfaces of two structures, which also share similar features, i.e., a large connected Fermi surface plus some small disconnected ones. In \cref{fig:fermi_surface}(b)(d), we plot several $k_z$-slices of Fermi surfaces and mark the possible nesting vector, which will be discussed in detail in the \cref{sec:bare_sus}.

\subsection{Band structure of CDW phase}\label{sec:cdw_band_structure}
In \cref{fig:ScV6Sn6_struct_band}(c), we compute by ab-initio the band structure of the CDW phase in the primitive BZ, while in \cref{fig:ScV6Sn6_struct_band}(d), the bands are unfolded to the pristine BZ of SG 191, with a comparison with the bands of experimental pristine structure. It can be seen that the major difference comes from the bands along $A$-$L$ and $H$-$A$, where the pristine bands cross the $E_f$ and form the large Fermi surface, while the unfolded bands have small distributions at $E_f$.

We remark that \cref{fig:ScV6Sn6_struct_band}(c) is computed in the BZ of the primitive CDW cell which is 3 times larger than the non-CDW unit cell. One can also compute the band structure in the conventional CDW cell which is 9 times larger than the non-CDW cell and thus has a smaller BZ compared with the primitive CDW cell, leading to further folded bands. However, the unfolded bands in \cref{fig:ScV6Sn6_struct_band}(d) are in the pristine BZ of the non-CDW structure.

\begin{figure}
    \centering
    \includegraphics[width=0.5\textwidth]{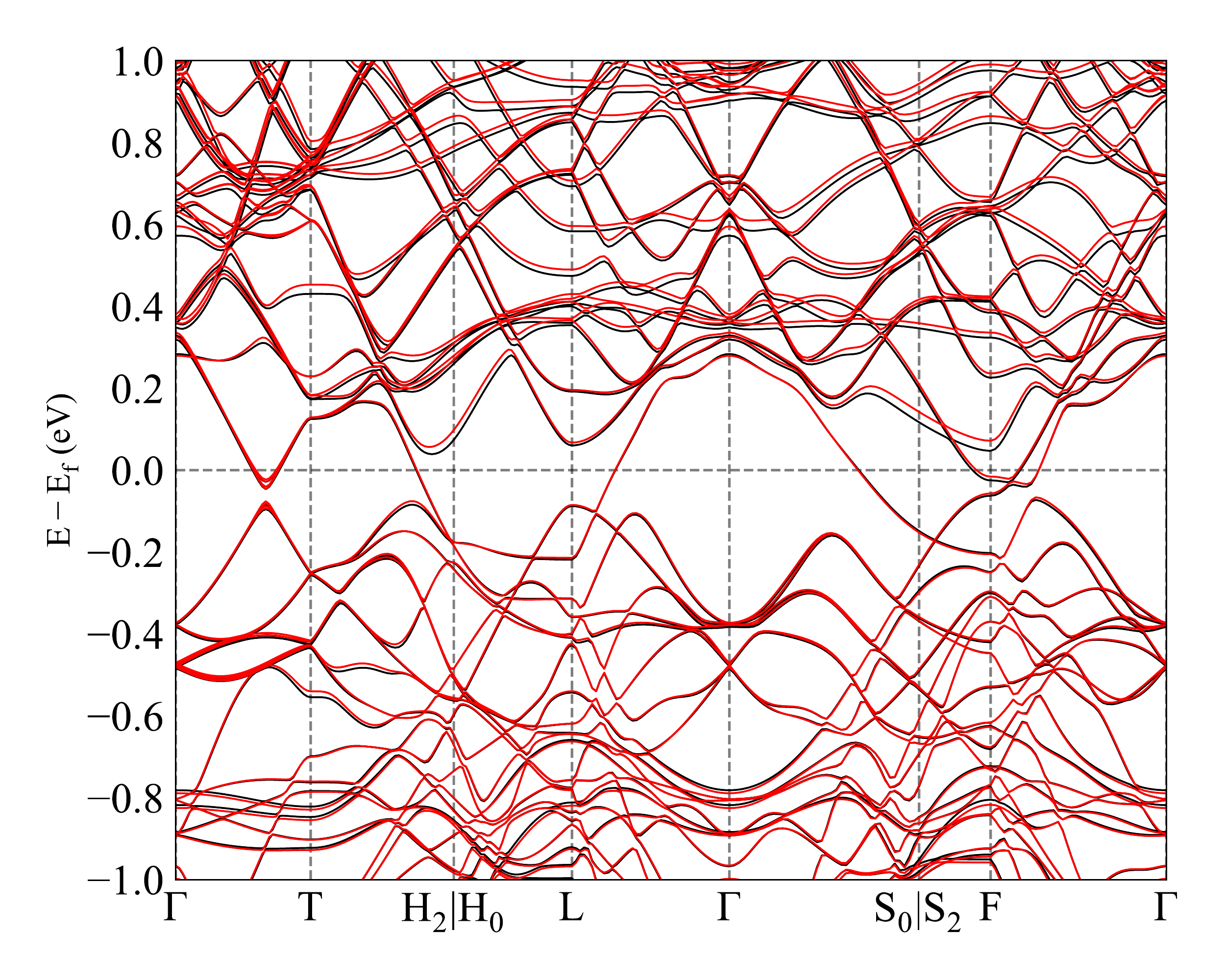}
    \caption{Comparison of the band structures in the CDW phase, with red bands computed from the CDW structure given by Ref.~\cite{ARA22}, and the black bands from our experiments. The differences between these two slightly different CDW structures are described in \cref{sec:cdw_structure}. The two band structures are very close near $E_f$.}
    \label{fig:compare_cdw_bands}
\end{figure}

We also compare the band structures of two slightly different CDW structures from Ref.~\cite{ARA22} and our experiments in \cref{fig:compare_cdw_bands}, which are very close near $E_f$. The inversion symmetry is weakly broken in the CDW structure from Ref.~\cite{ARA22}, as described in \cref{sec:cdw_structure}.

\subsection{First-principle computational details}\label{app:sec:electronic_band_structure:comp_details}

The \textit{ab-initio} band structures are computed using the Vienna Ab-initio Simulation Package (VASP)\cite{KRE93, KRE93a, KRE94, KRE96c,KRE96b}, where the generalized gradient approximation (GGA) of the Perdew–Burke–Ernzerhof (PBE)-type~\cite{PER96a} exchange-correlation potential is used. A $8\times 8\times 6$ $\Gamma$-centered Monkhorst–Pack grid is adopted with a plane-wave energy cutoff of $400$ eV. 
We compute maximally localized Wannier functions (MLWFs) using Wannier90\cite{MAR97c, SOU01b, MAR12a, PIZ20a} to obtain onsite energies and hoppings parameters, by considering the Sc $d$, V $d$, and Sn $p$ orbitals. \textit{WannierTools}\cite{WU18b} is used to compute the eigenvalues and eigenfunctions. The Fermi surface is computed using \textit{WannierTools} and $k_z$ slices are plotted using \textit{iFermi}\cite{GAN21}.

\section{Electron-phonon model}\label{app:sec:ele_phon_model}

In this \siSection{}, we study the effect of electron-phonon coupling. We consider a generic system with electrons and phonons, which is described by the following Hamiltonian 
\ba 
&H = H_c +H_{ph} +H_{g} \nonumber\\
&H_c = \sum_{\mathbf{k},\alpha \gamma,\sigma }t_{\alpha \gamma }(\mathbf{k})c_{\mathbf{k},\alpha \sigma}^\dag c_{\mathbf{k},\gamma\sigma } \nonumber \\
&H_{ph} = \sum_{\mathbf{R},i,\mu } \frac{ (P_{i\mu}(\mathbf{R}))^2}{2M_\alpha }
+ \sum_{\mathbf{R}',\mathbf{R}'i \mu,j \nu} \frac{1}{2}\Phi_{i \mu,j\nu}(\mathbf{R}-\mathbf{R}')u_{i \mu}(\mathbf{R}) u_{j \nu}(\mathbf{R}')
\nonumber \\
&H_{g} = \sum_{\mathbf{q},\mathbf{k},\alpha,\gamma,\delta,\sigma} 
\frac{g_{\mathbf{k},\mathbf{q}}^{\alpha \gamma i \mu}}{\sqrt{N} }
u_{i\mu}(\mathbf{q})c_{\mathbf{k}+\mathbf{q},\alpha\sigma}^\dag c_{\mathbf{k} ,\gamma \sigma } 
\label{eq:model_ele_ph}
\ea 
$c_{\mathbf{k},\alpha \sigma}^\dag$ creates an electron with momentum $\mathbf{k}$, orbital $\alpha$ and spin $\sigma$.
$t_{\alpha\gamma}(\mathbf{k})$ is the hopping matrix of electrons in the momentum space. 
$u_{i \mu}(\mathbf{R})$ denotes the \textcolor{black}{$\mu\in\{x,y,z\}$} direction displacement  field of $i$-th atom and $P_{i \mu}(\mathbf{R})$ denotes the corresponding momentum of displacement field.
$g_{\mathbf{k},\mathbf{q}}^{\alpha \gamma,i\mu}$ is the tensor that characterize the electron-phonon couplings. $N$ is the total number of unit cells.

\subsection{Phonon Hamiltonian}\label{app:sec:ele_phon_model:hamiltonian} 
We first consider the phonon Hamiltonian ${ \hat{H} }_{ph}$. We note that $P$ and $u$ follows the canonical commutation relation
\ba 
[u_{i\mu}(\mathbf{R}) , P_{i'\mu'}(\mathbf{R}')] = i \delta_{i,i'}\delta_{\mu,\mu'}\delta_{\mathbf{R},\mathbf{R}'}
\ea 
It is more convenient to introduce the new set of operators
\ba 
&\tilde{u}_{i\mu}(\mathbf{R}) ={\sqrt{M_i}}u_{\alpha\mu}(\mathbf{R}),\quad \tilde{P}_{i\mu}(\mathbf{R}) =\frac{1}{\sqrt{M_i}}P_{i\mu}(\mathbf{R})\nonumber \\
&[\tilde{u}_{i\mu}(\mathbf{R}) , \tilde{P}_{i'\mu'}(\mathbf{R}')] = i \delta_{i,i'}\delta_{\mu,\mu'}\delta_{\mathbf{R},\mathbf{R}'}
\label{eq:def_tilde_up}
\ea 
The Hamiltonian in terms of the new set of operators become
\ba 
&H_{ph} = \sum_{\mathbf{R},i,\mu } \frac{ (\tilde{P}_{i\mu}(\mathbf{R}))^2}{2 }
+ \sum_{\mathbf{R},\mathbf{R}',i\mu,j\nu} \frac{1}{2}D_{i\mu,j\nu}(\mathbf{R}-\mathbf{R}')\tilde{u}_{i\mu}(\mathbf{R}) \tilde{u}_{j\nu}(\mathbf{R}')
\ea 
where 
\begin{equation}
	\label{app:eqn:dyn_matrix}
	D_{i\mu,j\nu}(\mathbf{R}-\mathbf{R}')=\Phi_{i\mu,j\nu}(\mathbf{R}-\mathbf{R}')/\sqrt{M_iM_j}
\end{equation} 
is the dynamic matrix. We transform to the momentum space using the following Fourier transformation 
\textcolor{black}{ \ba 
&\tilde{u}_{i\mu}(\mathbf{q}) = \frac{1}{\sqrt{N}} 
\sum_{\mathbf{R}} \tilde{u}_{i\mu}(\mathbf{R} ) e^{-i\mathbf{q} \cdot (\mathbf{R} +\mathbf{r}_i )} \nonumber\\ 
&\tilde{P}_{i \mu}(\mathbf{q}) = \frac{1}{\sqrt{N}}\sum_{\mathbf{R}}\tilde{P}_{i\mu}(\mathbf{R})e^{i\mathbf{q}\cdot (\mathbf{R} +\mathbf{r}_i)} \nonumber\\ 
&D_{i \mu,j \nu}(\mathbf{q}) =\sum_{\mathbf{R}}D_{i\mu,j\nu}(\mathbf{R}) e^{-i\mathbf{q}\cdot (\mathbf{R}+\mathbf{r}_i -\mathbf{r}_j) }  \label{app:eqn:ft_convention_phonon}
\ea 
}
with $\mathbf{r}_i$ denote the position of $i$-th atom in the unit cell. 
Then we have 
\ba 
H_{ph} = \sum_{\mathbf{q},i,\mu } \frac{ \tilde{P}_{i\mu}(\mathbf{q})\tilde{P}_{i\mu}(-\mathbf{q})}{2 }
+ \sum_{\mathbf{q},i\mu,j  \nu} \frac{1}{2}D_{i\mu,j\nu}(\mathbf{q})\tilde{u}_{i \mu}(-\mathbf{q}) \tilde{u}_{j \nu}(\mathbf{q})
\ea 
We introduce the eigenvector and eigenvalue of dynamical matrix
\ba 
\sum_{j,\nu} D_{i \mu,j \nu}(\mathbf{q}) U_{j\nu,n}(\mathbf{q}) = \omega^2_n(\mathbf{q}) U_{i \mu,n}(\mathbf{q}), 
\quad \text{with} \quad 
\sum_{i,\mu} U^{*}_{i \mu,m} (\mathbf{q}) U_{i \mu,n} (\mathbf{q}) = \delta_{mn},
\label{app:eqn:eigs_of_phon_dyn_mat}
\ea 
and the corresponding operator in the band basis
\ba 
&x_n(\mathbf{q}) = \sum_{i \mu} U^*_{i \mu,n }(\mathbf{q}) \tilde{u}_{i \mu}(\mathbf{q}),\quad 
p_n(\mathbf{q}) = \sum_{i\mu} U_{i\mu,n }(\mathbf{q}) \tilde{P}_{i\mu}(\mathbf{q}) \nonumber \\
&[x_n(\mathbf{q}),p_m(\mathbf{q}')] = i\sum_{i\mu}U^*_{i\mu,n}(\mathbf{q})U_{i\mu,m}(\mathbf{q})\delta_{\mathbf{q},\mathbf{q}'} = i\delta_{n,m}\delta_{\mathbf{q},\mathbf{q}'}
\label{eq:def_xp}
\ea 
We also assume the eigenvalues $\omega^2_n(\mathbf{q})$ of the bare dynamical matrix (without corrections from electron-phonon coupling - or, in ab-initio, the high temperature dynamical matrix) is real and positive, otherwise, the bare phonon excitation will already introduce an instability of the system. Without loss of generality, we pick $\omega_n(\mathbf{q})\ge 0$.

The phonon Hamiltonian becomes
\ba 
H_{ph}& = \sum_{\mathbf{q},n} \frac{1}{2}\bigg( p_n(\mathbf{q})p_n(-\mathbf{q}) + \omega^2_n(\mathbf{q})x_n(\mathbf{q}) x_n(-\mathbf{q})\bigg) \nonumber \\ 
&
=  \sum_{\mathbf{q},n} \omega_n(\mathbf{q})\bigg(\frac{\sqrt{\omega_n(\mathbf{q})}}{\sqrt{2}}x_n(\mathbf{q})+\frac{i}{\sqrt{2\omega_n(\mathbf{q})}}p_n(-\mathbf{q})\bigg)
\bigg(\frac{\sqrt{\omega_n(\mathbf{q})}}{\sqrt{2}}x_n(-\mathbf{q})-\frac{i}{\sqrt{2\omega_n(\mathbf{q})}}p_n(\mathbf{q})\bigg)
\ea 
where we assume the frequency of the phonon is positive and real (not imaginary) otherwise the bare phonon excitation will lead to an instability of the system. We then define the bosonic operator
\ba 
&b_{\mathbf{q},n} =\frac{\sqrt{\omega_n(\mathbf{q})}}{\sqrt{2}}x_n(\mathbf{q})+\frac{i}{\sqrt{2\omega_n(\mathbf{q})}}p_n(-\mathbf{q})
,\quad 
b^\dag_{\mathbf{q},n} =\frac{\sqrt{\omega_n(\mathbf{q})}}{\sqrt{2}}x_n(-\mathbf{q})-\frac{i}{\sqrt{2\omega_n(\mathbf{q})}}p_n(\mathbf{q})\nonumber\\ 
&[b_{\mathbf{q},n},b_{\mathbf{q}',n'}^\dag] =\delta_{\mathbf{q},\mathbf{q}'}\delta_{n,n'}
\label{eq:def_b}
\ea 
Then the phonon Hamiltonian describes non-interacting bosons
\ba 
H_{ph} = \sum_{\mathbf{q},n} \omega_{n}(\mathbf{q}) b_{\mathbf{q},n}^\dag b_{\mathbf{q},n}
\ea 
Combining \cref{eq:def_tilde_up}, \cref{eq:def_xp}, \cref{eq:def_b}, the original displacement fields of atom can be written as
\ba 
&u_{\alpha\mu}(\mathbf{q}) = \sum_n\sqrt{\frac{1}{2\omega_n(\mathbf{q})M_{i}}}U_{i\mu,n}(\mathbf{q}) 
\bigg(b_{\mathbf{q},n} + b_{-\mathbf{q},n}^\dag\bigg) \nonumber\\ 
&
P_{\alpha\mu}(\mathbf{q}) = -\sum_n i\sqrt{\frac{\omega_n(\mathbf{q})M_i}{2}}U_{i\mu,n}^*(\mathbf{q})\bigg( b_{-\mathbf{q},n}-b_{\mathbf{q},n}^\dag \bigg) 
\label{eq:u_to_b}
\ea 
then the electron-phonon coupling can be written as 
\ba 
H_{g} = \sum_{\mathbf{q},\mathbf{k},\alpha,\gamma,i,\mu, n,\sigma} 
\frac{g_{\mathbf{k},\mathbf{q}}^{\alpha \gamma i \mu}U_{i\mu,n}(\mathbf{q})}{\sqrt{N 2\omega_n(\mathbf{q})M_i} }
(b_{\mathbf{q},n}+b_{-\mathbf{q},n}^\dag)c_{\mathbf{k}+\mathbf{q},\alpha\sigma}^\dag c_{\mathbf{k} ,\gamma \sigma } 
\label{eq:ele_ph_b}
\ea 

In practice, one can either work with the $b$, $b^\dag$ basis or with the $u$, $P$ basis.

\subsection{Path integral formula}

We now introduce the path integral of the system. We first perform a Trotter expansion of the Hamiltonian
\ba 
Z = \lim_{N\rightarrow \infty}\text{Tr}[\prod_{i=1}^N e^{-\Delta \tau H}]
\ea
where $\Delta \tau = \beta/N$ with $\beta$ the inverse temperature. 
We insert the identity operator at each time slice \textcolor{black}{$\tau_i=i\beta/N$}
\ba 
I(\tau_i) = \int D[c(\tau_i),c^\dag(\tau_i), b(\tau_i),b^\dag(\tau_i)] |c(\tau_i)\rangle  |b(\tau_i)\rangle \langle b(\tau_i) | \langle c(\tau_i) | 
\ea 
\textcolor{black}{where $c(\tau_i),b(\tau_i)$ denote the electron and phonon fields at time slice $\tau_i$.}
Then we have
\ba 
Z =&\lim_{N\rightarrow \infty}\prod_{i=1}^N \text{Tr}[\prod_{i=1}^N e^{-\Delta \tau H}I(\tau_i)]\nonumber \\ 
=&\lim_{N\rightarrow \infty}\prod_{i=1}^N 
\int D[c(\tau_i),c^\dag(\tau_i),b(\tau_i),b^\dag(\tau_i)] 
\langle c(\tau_{i+1})| \langle b(\tau_{i+1})| e^{-\Delta \tau H} | b(\tau)\rangle |c(\tau)\rangle
\label{eq:trotter}
\ea
where $\tau_i=i\beta/N$. 
For each time slice, we have 
\ba 
&\langle c(\tau+\Delta\tau))| \langle b(\tau+\Delta \tau)|e^{-\Delta \tau H} | b(\tau)\rangle |c(\tau)\rangle  \nonumber \\ 
\approx &\langle c(\tau+\Delta\tau))| \langle b(\tau+\Delta \tau)| b(\tau)\rangle |c(\tau)\rangle -\Delta \tau 
\langle c(\tau+\Delta\tau))| \langle u(\tau+\Delta \tau)|
H
| u(\tau)\rangle |c(\tau)\rangle  \nonumber\\
\approx &1 + \Delta \tau \sum_{\mathbf{k},\alpha \sigma} \partial_\tau c_{\mathbf{k},\alpha\sigma}^\dag(\tau) c_{\mathbf{k},\alpha\sigma}(\tau) + \Delta \tau \sum_{\mathbf{q},n} \partial_\tau b_{\mathbf{q},n}^\dag(\tau) b_{\mathbf{q},n}(\tau) - H(\tau)\Delta \tau \nonumber \\ 
\approx &\exp\bigg[ \Delta \tau \bigg(  \sum_{\mathbf{k},\alpha \sigma} \partial_\tau c_{\mathbf{k},\alpha\sigma}^\dag(\tau) c_{\mathbf{k},\alpha\sigma}(\tau) +  \sum_{\mathbf{q},n}\partial_\tau b_{\mathbf{q},n}^\dag(\tau) b_{\mathbf{q},n}(\tau) - H(\tau) \bigg) \bigg] 
\label{eq:each_time}
\ea 
where $H(\tau)$ can be obtained by replacing $c,c^\dag, b,b^\dag$ with $c(\tau),c^\dag(\tau),b(\tau),b^\dag(\tau)$
\ba 
&H(\tau) = H_c(\tau) + H_{ph}(\tau)+H_g(\tau) \nonumber \\ 
&H_c(\tau) =  \sum_{\mathbf{k},\alpha \gamma \sigma}c_{\mathbf{k},\alpha\sigma}^\dag(\tau) t_{\alpha\gamma}(\mathbf{k})  c_{\mathbf{k},\gamma\sigma}(\tau) \nonumber \\ 
&H_{ph}(\tau) =  \sum_{\mathbf{q},n} \omega_{n}(\mathbf{q}) b_{\mathbf{q},n}^\dag(\tau) b_{\mathbf{q},n}(\tau) \nonumber \\ 
&H_g(\tau) = \sum_{\mathbf{q},\mathbf{k},\alpha,\gamma,i,\mu, n,\sigma} 
\frac{g_{\mathbf{k},\mathbf{q}}^{\alpha \gamma i \mu}U_{i\mu,n}(\mathbf{q})}{\sqrt{N 2\omega_n(\mathbf{q})M_i} }
(b_{\mathbf{q},n}(\tau)+b_{-\mathbf{q},n}^\dag(\tau))c_{\mathbf{k}+\mathbf{q},\alpha\sigma}^\dag(\tau) c_{\mathbf{k} ,\gamma \sigma } (\tau) 
\label{eq:Htau}
\ea 
Combining \cref{eq:trotter} and \cref{eq:each_time}, we have
\ba 
Z = &\int D[b,b^\dag ,c,c^\dag] \exp\bigg[ 
-\int_0^\beta   
\bigg( -\sum_{\mathbf{k},\alpha \sigma} \partial_\tau c_{\mathbf{k},\alpha\sigma}^\dag(\tau) c_{\mathbf{k},\alpha\sigma}(\tau) -  \sum_{\mathbf{q},n}\partial_\tau b_{\mathbf{q},n}^\dag(\tau) b_{\mathbf{q},n}(\tau)+H(\tau)
\bigg) 
d\tau 
\bigg] \nonumber \\ 
=&\int D[b,b^\dag ,c,c^\dag] \exp\bigg[ 
-\int_0^\beta   
\bigg( \sum_{\mathbf{k},\alpha \sigma} c_{\mathbf{k},\alpha\sigma}^\dag(\tau) \partial_\tau c_{\mathbf{k},\alpha\sigma}(\tau) +  \sum_{\mathbf{q},n}b_{\mathbf{q},n}^\dag(\tau) \partial_\tau b_{\mathbf{q},n}(\tau)+H(\tau)
\bigg) 
d\tau 
\bigg]
\ea 
For future convenience, we transform $b,b^\dag$ to the original $u,P$ basis. For each time slice, we utilize \cref{eq:u_to_b} and find
\ba 
&\int_0^\beta \sum_{\mathbf{q},n}b_{\mathbf{q},n}^\dag(\tau) \partial_\tau b_{\mathbf{q},n}(\tau) d\tau \nonumber \\ 
=&\int_0^\beta \sum_{\mathbf{q},n,i \mu,j\nu} \bigg( 
\sqrt{\frac{\omega_n(\mathbf{q})M_i}{2} } U_{i\mu,n}(\mathbf{q}) u_{i\mu}(\mathbf{q},\tau) 
- i\sqrt{ \frac{1}{2\omega_n(\mathbf{q})M_i} } U^*_{i\mu,n}(-\mathbf{q}) P_{i\mu}(\mathbf{q},\tau) 
\bigg) \nonumber \\ 
&\partial_\tau  
\bigg( 
\sqrt{\frac{\omega_n(\mathbf{q})M_j}{2} } U_{j\nu,n}^*(\mathbf{q}) u_{j\nu}(\mathbf{q},\tau) 
+ i\sqrt{ \frac{1}{2\omega_n(\mathbf{q})M_j} } U_{j\nu,n}(-\mathbf{q}) P_{j\nu}(-\mathbf{q},\tau) 
\bigg)  d\tau \nonumber \\ 
=&\int_0^\beta  \sum_{\mathbf{q}, i,\mu} \bigg[ \frac{\omega_n(\mathbf{q})M_i}{2}u_{i\mu}(-\mathbf{q},\tau)\partial_\tau u_{i\mu}(\mathbf{q},\tau) + \frac{1}{2\omega_n(\mathbf{q})M_i}P_{i\mu}(\mathbf{q},\tau) \partial_\tau P_{i\mu}(-\mathbf{q},\tau) -i P_{i\mu}(\mathbf{q},\tau) u_{i\mu}(\mathbf{q},\tau)\bigg] d\tau \nonumber \\
=&\int_0^\beta  \sum_{\mathbf{q}, i,\mu} \bigg[ \frac{\omega_n(\mathbf{q})M_i}{4}\partial_\tau(u_{i\mu}(-\mathbf{q},\tau) u_{i\mu}(\mathbf{q},\tau)) + \frac{1}{4\omega_n(\mathbf{q})M_i}\partial_\tau( P_{i\mu}(\mathbf{q},\tau)  P_{i\mu}(-\mathbf{q},\tau)) -i P_{i\mu}(\mathbf{q},\tau) u_{i\mu}(\mathbf{q},\tau)\bigg] d\tau \nonumber \\
=&\int_0^\beta  \sum_{\mathbf{q}, i,\mu} ( -i) P_{i\mu}(\mathbf{q},\tau) u_{i\mu}(\mathbf{q},\tau)d\tau 
\ea 
where we drop the total derivative in the final line. The Hamiltonian of the phonon becomes
\ba 
H_{ph}(\tau) =&\sum_{\mathbf{q},n}\omega_n(\mathbf{q})b_{\mathbf{q},n}^\dag b_{\mathbf{q},n} \nonumber \\ 
=&\sum_{\mathbf{q},n,i\mu,j\nu}\omega_n(\mathbf{q})
 \bigg( 
\sqrt{\frac{\omega_n(\mathbf{q})M_i}{2} } U_{i\mu,n}(\mathbf{q}) u_{i\mu}(\mathbf{q},\tau) 
- i\sqrt{ \frac{1}{2\omega_n(\mathbf{q})M_i} } U^*_{i\mu,n}(-\mathbf{q}) P_{i\mu}(\mathbf{q},\tau) 
\bigg) \nonumber \\ 
&  
\bigg( 
\sqrt{\frac{\omega_n(\mathbf{q})M_j}{2} } U_{j\nu,n}^*(\mathbf{q}) u_{j\nu}(\mathbf{q},\tau) 
+ i\sqrt{ \frac{1}{2\omega_n(\mathbf{q})M_j} } U_{j\nu,n}(-\mathbf{q}) P_{j\nu}(-\mathbf{q},\tau) 
\bigg) \nonumber \\ 
=&\sum_{\mathbf{q},n,i\mu,j\nu}\omega_n(\mathbf{q})\bigg[ \frac{\omega_n(\mathbf{q})M_iM_j}{2}U_{i\mu,n}(\mathbf{q})U_{j\nu,n}^*(\mathbf{q}) u_{i\mu}(\mathbf{q},\tau) u_{j\nu}(\mathbf{q},\tau) \nonumber\\ 
&
+ \frac{1}{2M_iM_j\omega_n(\mathbf{q})}U_{i\mu,n}^*(-\mathbf{q})U_{j\nu,n}(-\mathbf{q})P_{i\mu}(\mathbf{q},\tau)P_{j\nu}(-\mathbf{q},\tau) 
\bigg]  \nonumber \\ 
=&\sum_{\mathbf{q},i\mu,j\nu}\bigg[ \frac{\sqrt{M_iM_j}D_{i\mu,j\nu}(\mathbf{q})}{2}u_{i\mu}(\mathbf{q},\tau) u_{j\nu}(\mathbf{q},\tau) 
+\delta_{i,j}\delta_{\mu,\nu} \frac{1}{2M_iM_i}P_{i\mu}(\mathbf{q},\tau)P_{i\mu}(-\mathbf{q},\tau) 
\bigg]  \nonumber \\  
=&\sum_{\mathbf{q}}\bigg[\sum_{i\mu,j\nu} \frac{\Phi_{i\mu,h\nu}(\mathbf{q})}{2}u_{i\mu}(\mathbf{q},\tau) u_{j\nu}(\mathbf{q},\tau) 
+ \sum_{i\mu}\frac{1}{2M_iM_i}P_{i\mu}(\mathbf{q},\tau)P_{i\mu}(-\mathbf{q},\tau) 
\bigg] 
\ea 

Then the action and the partition function of the system can be written as 
\ba 
Z = &\int D[u,P,c,c^\dag] e^{-S}\nonumber\\ 
S = &S_c+S_{ph} +S_g \nonumber \\ 
S_c =& \int_0^\beta \sum_{\mathbf{k},\alpha \gamma \sigma}c_{\mathbf{k},\alpha\sigma}^\dag(\tau)\bigg( \partial_\tau \delta_{\alpha,\gamma} + t_{\alpha\gamma}(\mathbf{k}) \bigg) c_{\mathbf{k},\gamma\sigma}(\tau)d\tau  \nonumber \\ 
S_{ph}=&
\int_0^\beta \sum_{\mathbf{q}, i \mu} 
\bigg[ iP_{i\mu}(\mathbf{q},\tau)\partial_\tau u_{i\mu}(\mathbf{q},\tau)  + \sum_{\mathbf{q},i\mu} \frac{P_{i\mu}(\mathbf{q},\tau)P_{i\mu}(-\mathbf{q},\tau)}{2M_i} + \frac{1}{2}\sum_{\mathbf{q},i\mu,j\nu} \Phi_{i\mu,j\nu}(\mathbf{q})u_{i\mu}(-\mathbf{q},\tau)u_{j\nu}(\mathbf{q},\tau)\bigg]
d\tau 
\nonumber \\ 
S_g=&\int_0^\beta \sum_{\mathbf{q},\mathbf{k},i,\mu,\alpha\gamma, \sigma,}  \bigg(\frac{g_{\mathbf{k},\mathbf{q}}^{\alpha \gamma ,i \mu }}{\sqrt{N} }
u_{i \mu}(\mathbf{q},\tau)c_{\mathbf{k}+\mathbf{q},\alpha}^\dag(\tau) c_{\mathbf{k} ,\gamma \sigma }(\tau)\bigg) d\tau 
\label{eq:def_S}
\ea

It is more convenient to consider the band basis of electrons. We diagonalize the electron-hopping matrix 
\ba 
\sum_{\gamma} t_{\alpha\gamma}(\mathbf{k})U_{\gamma n}^c(\mathbf{k}) = \epsilon_{\mathbf{k},n}U_{\alpha n}^c(\mathbf{k})
\ea 
and let 
\ba 
\gamma_{\mathbf{k},n\sigma} =\sum_{\alpha}U_{\alpha n}^{c,*}(\mathbf{k}) c_{\mathbf{k},\alpha\sigma}
\label{eq:band_basis_gam}
\ea 
Then the action of the electron $S_c$ and the action of electron-phonon coupling now take the form of  
\ba 
&S_c = \int_0^\beta \sum_{\mathbf{k}, n, \sigma} \gamma_{\mathbf{k},n\sigma}^\dag(\tau)(\partial_\tau + \epsilon_{\mathbf{k},n})\gamma_{\mathbf{k},n\sigma}(\tau) d\tau \nonumber \\ 
&S_g = \int_0^\beta \sum_{\mathbf{k},\mathbf{q},i \mu, n,m,\sigma} \frac{h_{\mathbf{k},\mathbf{q}}^{nm, i \mu} }{\sqrt{N}}u_{i \mu}(\mathbf{q}, \tau) \gamma_{\mathbf{k}+\mathbf{q},n\sigma}^\dag(\tau) \gamma_{\mathbf{k},m\sigma}(\tau) d\tau 
 \ea 
 where the new coupling tensor is
 \ba
 h_{\mathbf{k},\mathbf{q}}^{nm,i \mu} = \sum_{\alpha\gamma} U_{\alpha n}^{c,*}(\mathbf{k}+\mathbf{q})U_{\gamma m}^c(\mathbf{k})g_{\mathbf{k},\mathbf{q}}^{\alpha\gamma,i \mu}
 \ea 
Here, we will work with displacement field $u$ and momentum $P$. 
The partition function now becomes
\ba 
Z = \int D[\gamma,\gamma^\dag,u,P] e^{-(S_c+S_{ph}+S_g)}
\ea 
\textcolor{black}{We will later (\cref{sec:ele_corr_to_phonon}) use the electron-phonon model to calculate the correction to the phonon propagator by integrating out the electron fields. Since it is more convenient to integrate out electron fields in the band basis of electrons, so we introduce electron-phonon coupling ($h_{\mathbf{k},\mathbf{q}}^{nm,i\mu}$) in the band basis of the electron. However, we still work with atomic basis instead of band basis for the phonon fields, which allows us to identify the electron correction to the vibration mode of each atom.
}

\subsection{Gaussian approximation of electron-phonon coupling}
In this \siSection{}, we describe how to estimate the electron-phonon coupling from Gaussian approximation. 
The tight-binding model of electrons can be written as 
\ba 
{ \hat{H} }_c = \sum_{\mathbf{R},\mathbf{R}',\alpha,\gamma } c_{\mathbf{R},\alpha \sigma}^\dag t_{\alpha\gamma}(\mathbf{R}-\mathbf{R}') c_{\mathbf{R}',\gamma \sigma}
\ea 
Following a recently developed model for electron-phonon coupling~\cite{YU23},
we assume the hopping can be described by a Gaussian function which decays exponentially
\ba 
t_{\alpha\gamma}(\mathbf{R}-\mathbf{R}') = t_{0,\alpha\gamma} e^{-\gamma_{\alpha\gamma} |\mathbf{R}-\mathbf{R}' +\mathbf{r}_\alpha-\mathbf{r}_\gamma|^2/2 }
\label{eq:gaus_approx}
\ea 
We next introduce atomic displacement $\bm{u}_{i_\alpha}$ to the hopping Hamiltonian which gives
\ba 
{ \hat{H} }_c' = \sum_{\mathbf{R},\mathbf{R}',\alpha,\gamma}  t_{\alpha\gamma}(\mathbf{R}-\mathbf{R}'+ \bm{u}_{i_\alpha}(\mathbf{R}) - \bm{u}_{i_\gamma}(\mathbf{R}') ) c_{\mathbf{R},\alpha \sigma}^\dag  c_{\mathbf{R}',\gamma \sigma}
\ea 
where the vector field is defined as $\bm{u}_{i_\alpha}(\mathbf{R}) =\begin{bmatrix}
    u_{i_\alpha x}(\mathbf{R}) & u_{i_\alpha y}(\mathbf{R}) & u_{i_\alpha z}(\mathbf{R})
\end{bmatrix} $ and $i_{\alpha}$ denote the atom index of orbital $\alpha$. We then expand in powers of $\bm{u}$ fields and keep the zeroth and linear order term. 
\ba 
{ \hat{H} }_c' \approx &  \sum_{\mathbf{R},\mathbf{R}',\alpha,\gamma}  t_{\alpha\gamma}(\mathbf{R}-\mathbf{R}'  ) c_{\mathbf{R},\alpha \sigma}^\dag t(\mathbf{R}-\mathbf{R}'+\mathbf{r}_\alpha -\mathbf{r}_\gamma) c_{\mathbf{R}',\gamma \sigma} \nonumber \\ 
& +  \sum_{\mathbf{R},\mathbf{R}',\alpha,\gamma}  \partial_{R^\mu}t_{\alpha\gamma}(\mathbf{R}-\mathbf{R}'  )(u_{i_\alpha \mu}(\mathbf{R})-u_{i_\gamma \mu}(\mathbf{R}')) c_{\mathbf{R},\alpha \sigma}^\dag  c_{\mathbf{R}',\gamma \sigma}
\label{eq:ele_ph_couple_v0}
\ea 
The zeroth order term is the original tight-binding Hamiltonian of electrons. The linear-order term produces the electron-phonon coupling ${ \hat{H} }_g$ which is ~\cite{YU23}
\ba 
{ \hat{H} }_g = \sum_{\mathbf{R},\mathbf{R}',\alpha,\gamma}  \partial_{R^\mu}t_{\alpha\gamma}(\mathbf{R}-\mathbf{R}'  )(u_{i_\alpha \mu}(\mathbf{R})-u_{i_\gamma \mu}(\mathbf{R}')) c_{\mathbf{R},\alpha \sigma}^\dag  c_{\mathbf{R}',\gamma \sigma}
\ea 
We use the ansatz in \cref{eq:gaus_approx} and find
\ba 
\partial_{R^\mu}t_{\alpha\gamma}(\mathbf{R}) = -t_{0,\alpha\gamma} \gamma_{\alpha\gamma} (\mathbf{R}+\mathbf{r}_\alpha -\mathbf{r}_\gamma)^\mu e^{-\gamma_{\alpha\gamma}|\mathbf{R}+\mathbf{r}_\alpha -\mathbf{r}_\gamma|^2 /2} = 
-\gamma_{\alpha\gamma}(\mathbf{R}+\mathbf{r}_\alpha -\mathbf{r}_\gamma)^\mu t_{\alpha\gamma}(\mathbf{R})
\label{eq:der_hop}
\ea 
\textcolor{black}{This gives the electron-phonon coupling defined in the real space
\ba 
&H_g = \sum_{\mathbf{R},\mathbf{R}',\mathbf{R}'',i,\alpha,\gamma} G^{\alpha\gamma}(\mathbf{R}-\mathbf{R}') (\delta_{i,i_\alpha}\delta_{\mathbf{R}'',\mathbf{R}} -\delta_{i,i_\gamma}\delta_{\mathbf{R}'',\mathbf{R}'})u_{i\mu}(\mathbf{R}'')c_{\mathbf{R},\alpha\sigma}^\dag c_{\mathbf{R}',\gamma\sigma} \nonumber\\ 
&G^{\alpha\gamma,\mu}(\mathbf{R}) = -\gamma_{\alpha\gamma}(\mathbf{R}+\mathbf{r}_\alpha -\mathbf{r}_\gamma)^\mu t_{\alpha\gamma}(\mathbf{R})
\label{eq:ele_ph_real_space}
\ea 
where $i_\alpha$ denotes the atom index of the electron orbital $\alpha$, $G^{\alpha\gamma,\mu}(\mathbf{R})$ characterize the strength of electron-phonon coupling in the real space.}

Equivalently, we can work with momentum space by performing Fourier transformation 
\ba 
t_{\alpha\gamma}(\mathbf{k}) = \sum_{\mathbf{R}} t_{\alpha\gamma}(\mathbf{R})e^{-i\mathbf{k}\cdot(\mathbf{R}+\mathbf{r}_\alpha -\mathbf{r}_\gamma)}
\ea 
(It is worth mentioning the periodic condition of $t_{\alpha\gamma}(\mathbf{k})$ is $t_{\alpha\gamma}(\mathbf{k}+\mathbf{G}) =t_{\alpha\gamma}(\mathbf{k}) e^{-i\mathbf{G}\cdot (\mathbf{r}_\alpha-\mathbf{r}_\gamma)}$ where $\mathbf{G} \in \mathbb{Z}\bm{b}_1 +\mathbb{Z}\bm{b}_2 +\mathbb{Z}\bm{b}_3 $).
By performing Fourier transformation to \cref{eq:der_hop}, we obtain
\ba
&-\sum_\mathbf{R} \gamma_{\alpha\gamma}(\mathbf{R}+\mathbf{r}_\alpha -\mathbf{r}_\gamma)^\mu t_{\alpha\gamma}(\mathbf{R})e^{-i\mathbf{k}\cdot(\mathbf{R}+\mathbf{r}_\alpha-\mathbf{r}_\gamma)}
=-\gamma_{\alpha\gamma}i\partial_{k^\mu} t_{\alpha\gamma}(\mathbf{k})
\ea 
The Fourier transformation of electron and phonon fields reads
\ba 
c_{\mathbf{k},\alpha\sigma} = \frac{1}{\sqrt{N}}\sum_{\mathbf{R}} c_{\mathbf{R},\alpha}e^{-i\mathbf{k} \cdot (\mathbf{R}+\mathbf{r}_\alpha)},\quad 
u_{i\mu}(\mathbf{q}) = \frac{1}{\sqrt{N}}\sum_{\mathbf{R}}u_{i\mu}(\mathbf{R}) e^{-i\mathbf{q} \cdot (\mathbf{R}+\mathbf{r}_i)} 
\ea

Then the electron-phonon coupling in the momentum space becomes
\ba 
{ \hat{H} }_g=&\sum_{\mathbf{R},\mathbf{R}',\alpha,\gamma,\sigma}  \partial_{R^\mu}t_{\alpha\gamma}(\mathbf{R}-\mathbf{R}'  )(u_{i_\alpha \mu}(\mathbf{R})-u_{i_\gamma \mu}(\mathbf{R}')) c_{\mathbf{R},\alpha \sigma}^\dag  c_{\mathbf{R}',\gamma \sigma} \nonumber\\ 
 =&\frac{1}{N^2\sqrt{N}}\sum_{\mathbf{R},\mathbf{R}',\alpha\gamma,\sigma} 
 \sum_{\mathbf{q},\mathbf{k},\mathbf{k}_1,\mathbf{k}_2} \bigg(-\gamma_{\alpha\gamma} i\partial_{k^\mu} t_{\alpha\gamma}(\mathbf{k})\bigg)\bigg(u_{i_\alpha \mu}(\mathbf{q})e^{i\mathbf{q} \cdot (\mathbf{R}+\mathbf{r}_\alpha)} -u_{i_\gamma\mu}(\mathbf{q})e^{i\mathbf{q}\cdot (\mathbf{R}'+\mathbf{r}_\gamma)} \bigg) c_{\mathbf{k}_1,\alpha\sigma}^\dag c_{\mathbf{k}_2,\gamma\sigma} \nonumber \\ 
 &e^{i\mathbf{k}_2\cdot (\mathbf{R}'+\mathbf{r}_\gamma) -i\mathbf{k}_1\cdot(\mathbf{R}+\mathbf{r}_\alpha) + i\mathbf{k} \cdot(\mathbf{R}-\mathbf{R}'+\mathbf{r}_\alpha-\mathbf{r}_\gamma)} \nonumber\\ 
 =& \frac{1}{\sqrt{N}}\sum_{\mathbf{k},\mathbf{q}}\sum_{\alpha\gamma,\sigma}
 \bigg[-\gamma_{\alpha\gamma} i\partial_{k^\mu} t_{\alpha\gamma}(\mathbf{k})u_{i_\alpha\mu}(\mathbf{q})c_{\mathbf{k}+\mathbf{q},\alpha\sigma}^\dag c_{\mathbf{k},\gamma\sigma} +
 \gamma_{\alpha\gamma} i\partial_{k^\mu} t_{\alpha\gamma}(\mathbf{k})u_{i_\gamma\mu}(\mathbf{q})c_{\mathbf{k},\alpha\sigma}^\dag c_{\mathbf{k}-\mathbf{q},\gamma\sigma}\bigg] \nonumber \\ 
 =&\frac{1}{\sqrt{N}}
 \sum_{\mathbf{k},\mathbf{q},\alpha\gamma,\sigma} 
 i\gamma_{\alpha\gamma}\bigg(-\partial_{k^\mu} t_{\alpha\gamma} (\mathbf{k})
u_{i_\alpha\mu}(\mathbf{q})  
 +\partial_{k^\mu}t_{\alpha\gamma}(\mathbf{k}+\mathbf{q}) 
 u_{i_\gamma \mu}(\mathbf{q}) \bigg)c_{\mathbf{k}+\mathbf{q},\alpha\sigma}^\dag c_{\mathbf{k},\gamma\sigma}\nonumber\\
 =&\frac{1}{\sqrt{N}}
 \sum_{\mathbf{k},\mathbf{q},\alpha\gamma,\sigma} 
 i\gamma_{\alpha\gamma}\bigg(-\partial_{k^\mu} t_{\alpha\gamma} (\mathbf{k})
u_{i_\alpha \mu}(\mathbf{q})  
 +\partial_{k^\mu}t_{\alpha\gamma}(\mathbf{k}+\mathbf{q}) 
 u_{i_\gamma \mu}(\mathbf{q}) \bigg)c_{\mathbf{k}+\mathbf{q},\alpha\sigma}^\dag c_{\mathbf{k},\gamma\sigma} \nonumber \\ 
 =& \sum_{\mathbf{q},\mathbf{k},\alpha,\gamma,i,\sigma} 
\frac{g_{\mathbf{k},\mathbf{q}}^{\alpha \gamma ,i \mu}}{\sqrt{N} }
u_{i\mu}(\mathbf{q})c_{\mathbf{k}+\mathbf{q},\alpha\sigma}^\dag c_{\mathbf{k} ,\gamma \sigma } 
\ea 
where the electron-phonon coupling is defined as
\ba 
g_{\mathbf{k},\mathbf{q}}^{\alpha \gamma ,i \mu} = i\gamma_{\alpha\gamma}(-\partial_{k^\mu}t_{\alpha\gamma}(\mathbf{k}) \delta_{i_\alpha,i} + \partial_{k^\mu}t_{\alpha\gamma}(\mathbf{k}+\mathbf{q})\delta_{i_\gamma,i})
 \label{eq:ele_ph_from_gaussian} 
\ea 

Here we comment that, in the ansatz given in \cref{eq:gaus_approx}, we have ignored the angular dependency of the hopping amplitude. For example, for $p_z$ orbitals, the behaviors of hopping along $z$ directions and $xy$ directions are different. A more accurate treatment should take this angular dependency into consideration~\cite{YU23} 
. Here, as an approximation, we ignore the angular dependency and use current ansatz to estimate the electron-phonon coupling.  
In practice, we separate the orbitals into three groups: I: $p_z$ orbitals of triangular Sn atoms; II: $p_{x,y}$ orbitals of Honeycomb Sn atoms; III: $d_{z^2,xz,yz,x^2-y^2,xy}$ orbitals of V atoms. 
\textcolor{black}{For given two groups $A,B$, we build the following dataset
\ba 
T_{AB} = \bigg\{ \bigg(\mathbf{R}+\mathbf{r}_a-\mathbf{r}_b, t_{ab}(\mathbf{R})\bigg) \bigg| a\in A\text{ and }b\in B \bigg\} \cup  \bigg\{ \bigg(\mathbf{R}+\mathbf{r}_a-\mathbf{r}_b, t_{ab}(\mathbf{R})) \bigg| a\in B\text{ and }b\in A \bigg\} 
\ea 
where the hopping parameter $t_{ab}(\mathbf{R})$ are obtained from DFT calculations. We then fit the data in each data set with the ansatz given in \cref{eq:gaus_approx} and obtain the decaying factor $\gamma_{AB}$ for each dataset.}
We find $\gamma_{I,I}=0.31$eV$\text{\AA}^{-2}$, 
$\gamma_{II,II}=0.89$eV$\text{\AA}^{-2}$, $\gamma_{III,III}=0.36$eV$\text{\AA}^{-2}$, $\gamma_{I,II}=\gamma_{II,I}=0.12$eV$\text{\AA}^{-2}$,
$\gamma_{I,III}=\gamma_{III,I}=0.85$eV$\text{\AA}^{-2}$,
$\gamma_{II,III}=\gamma_{II,III}=0.24$eV$\text{\AA}^{-2}$. We then estimate the electron-phonon coupling via \cref{eq:ele_ph_from_gaussian}.

\begin{figure}
    \centering
    \includegraphics[width=0.5\textwidth]{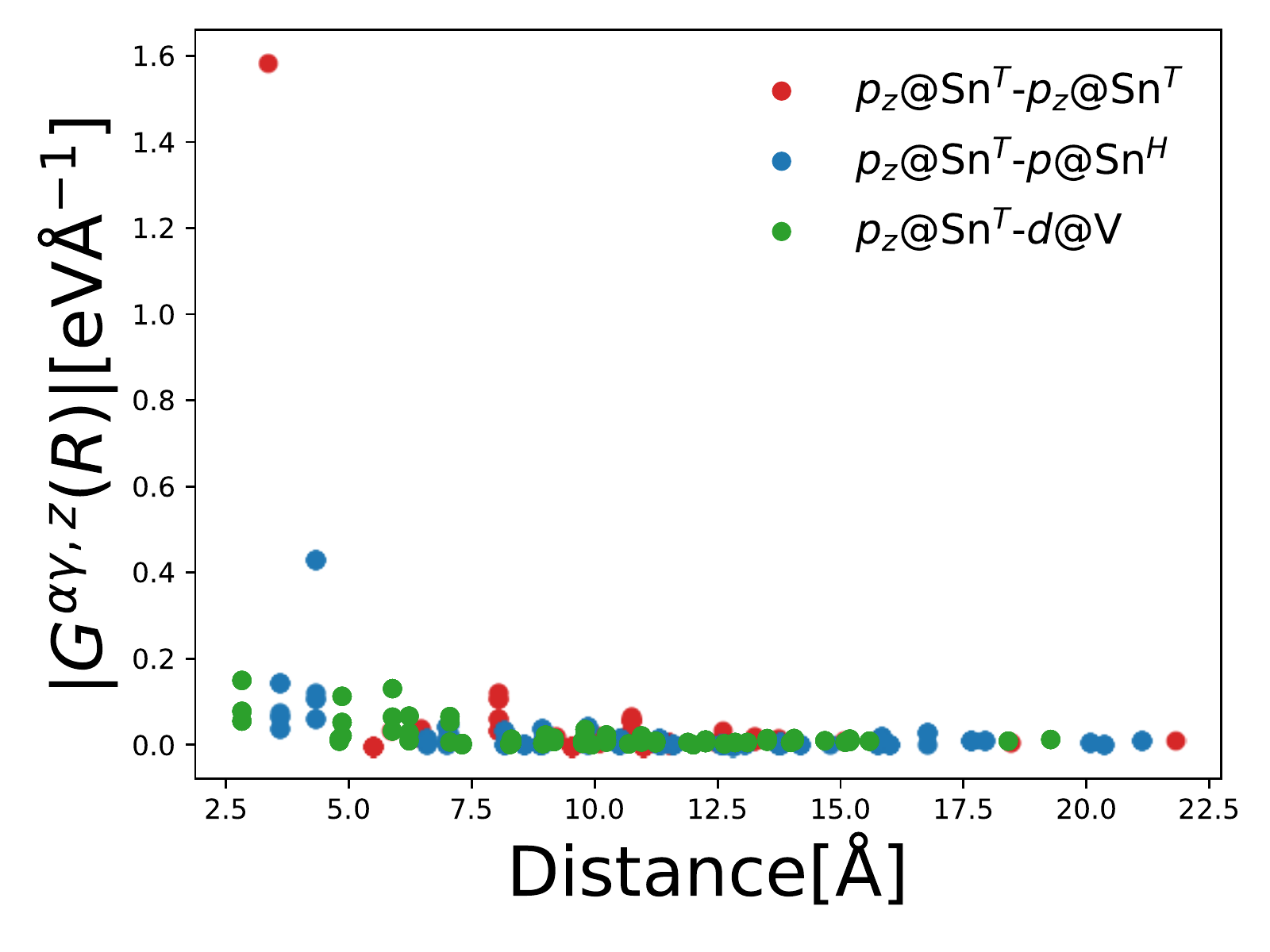}
    \caption{Strength of electron-phonon coupling \textcolor{black}{$|G^{\alpha\gamma,\mu=z}(\mathbf{R})|$ (\cref{eq:ele_ph_real_space}) } that involves $z$-direction movement of triangular Sn atoms. The horizontal axis labels the distance $|\mathbf{R} +\mathbf{r}_{\alpha}-\mathbf{r}_\gamma |$. Red points denote the couplings induced by the hopping between the $p_z$ orbital at the triangular Sn atom and $p_z$ orbital at the triangular Sn atom, \textcolor{black}{$|G^{\alpha\gamma,z}(\mathbf{R})|$ with $\alpha =\gamma= p_z\text{@Sn}^T$}.  Blue points denote the coupling induced by the hopping between $p_z$ orbital at the triangular Sn atom and $p_{x,y,z}$ orbitals at the Honeycomb Sn atom \textcolor{black}{$|G^{\alpha\gamma,z}(\mathbf{R})|$ with $\alpha = p_z\text{@Sn}^T,\gamma = p_{x,y,z}\text{@Sn}^H$}. 
     Green points denote the coupling induced by the hopping between $p_z$ orbital at the triangular Sn atom and $d_{z^2,xz,yz,x^2-y^2,xy}$ orbitals at the V atom,
     \textcolor{black}{$|G^{\alpha\gamma,z}(\mathbf{R})|$ with $\alpha = p_z\text{@Sn}^T,\gamma = d_{z^2,xz,yz,x^2-y^2,xy}\text{@V}$}. The other hoppings will generate much weaker interactions and are not shown here. 
We find that the dominant coupling is generated by the intra-unit-cell hopping between $p_z$-orbital electrons of two triangular Sn atoms. }
    \label{fig:gaus_approx}
\end{figure}

Since the imaginary phonon mode is mainly formed by the triangular Sn atoms, we focus on the electron-phonon coupling induced by them. We note that the electron coupling induced by triangular Sn atoms is induced from the hopping between electrons on the triangular Sn atoms and other orbitals. 
In \cref{fig:gaus_approx}, we plot the coupling strength that involves $z$-direction vibration of triangular Sn atoms (which is the main source of imaginary mode). We observe that the dominant coupling is induced by the intra-unit cell hopping between two $p_z$ orbitals of the triangular Sn
\ba 
t_{(Sn^T,p_z)}\sum_{\mathbf{R},\sigma} \bigg[c_{\mathbf{R}, (Sn_1^T,p_z), \sigma} ^\dag c_{\mathbf{R},(Sn_2^T,p_z),\sigma}+\text{h.c.}\bigg]
\label{eq:ee_hop}
\ea 
where $c_{\mathbf{R},(Sn_1^T,p_z),\sigma}, c_{\mathbf{R},(Sn^T_2,p_z),\sigma}$ denote the electron operators of $p_z$ orbitals at two triangular Sn atoms respectively. The corresponding electron-phonon coupling is
\ba 
 &-t_{(Sn^T,p_z)}\gamma_{ ( Sn^T,p_z)}\sum_{\mathbf{R},\sigma}\Delta r^z \bigg(u_{Sn^T_1,z}(\mathbf{R}) - u_{Sn^T_2,z}(\mathbf{R}) )\bigg)\bigg(c_{\mathbf{R},(Sn_1^T,p_z),\sigma}^\dag c_{\mathbf{R},(Sn_2^T,p_z),\sigma} + c_{\mathbf{R},(Sn_2^T,p_z),\sigma}^\dag c_{\mathbf{R},(Sn_1^T,p_z),\sigma} \bigg)
 \label{eq:ee_ele_ph}
\ea  
\textcolor{black}{where $u_{Sn_1^T,z}(\mathbf{R})-u_{Sn_2^T,z}(\mathbf{R})$ is equivalent to the mirror even mode $\eta^{e,z}$ we introduced near \cref{eq:flat_mode_basis}.}
where $\gamma_{( Sn^T,p_z)}$ is the decaying factor of the hopping between electrons at $p_z$ orbital of triangular Sn and $\Delta r^z$ is the distance between two triangular Sn atoms within the same unit cell (note that two triangular Sn atoms have the same $x,y$ coordinates). 
In addition, we also find the coupling to the $V$ $d$ orbitals and $Sn^H$ $p$ orbitals are relatively weak \textcolor{black}{as shown in \cref{fig:gaus_approx}}.

\subsection{Electron correction to phonon propagators} 
\label{sec:ele_corr_to_phonon}
Due to the electron-phonon coupling, the electron will introduce a correction to the phonon propagators. To capture this effect, we derive an effective theory of phonon fields by integrating out the electron fields $\gamma$ 
\ba 
Z =& \int D[u,P,\gamma,\gamma^\dag ] e^{-(S_{ph}+S_c +S_{g})} 
=\int D[u,P]e^{-S_{ph}} \int D[\gamma,\gamma^\dag] e^{-S_{g}} e^{-S_c} \nonumber \\ 
\propto &\int D[u,P]e^{-S_{ph}} \langle  e^{-S_{g}} \rangle_{S_c}
= \int D[u,P]e^{-\bigg( S_{ph} - \log  \langle  e^{-S_{g}} \rangle_{S_c}  \bigg)
} 
\label{eq:eff_part}
\ea 
The expectation value $\langle \cdot \rangle_{S_c}$ is taken with respect to the non-interacting system of $\gamma$ electrons ($S_c$ or $H_c$). We next perform the following expansion in powers of electron-phonon coupling and truncate to the second order
\ba 
- \log  \langle  e^{-S_{g}} \rangle_{S_c} \approx - \log  \langle  1-S_{g} +\frac{1}{2}S_{g}^2 \rangle_{S_c} \approx 
\langle S_{g}\rangle_{S_c} - \frac{1}{2} 
\bigg[ \langle S_g^2 \rangle_{S_c} - \bigg(\langle S_g\rangle_{S_c}\bigg)^2
\bigg]
\label{eq:log_exp}
\ea 
Combining \cref{eq:eff_part} and \cref{eq:log_exp}, we find 
\ba 
Z \approx  \int D[u,P] \exp\bigg[ 
-\bigg[ S_{ph} + \langle S_g \rangle_{S_c} -\frac{1}{2}\langle S_g^2 \rangle_{S_c} +\frac{1}{2}\bigg(\langle S_g\rangle_{S_c}\bigg)^2
\bigg] \bigg] 
\label{eq:partition_eff}
\ea 
We first calculate the first-order term
\ba 
\langle S_g \rangle_{S_c} =& \int_0^\beta \sum_{\mathbf{k},\mathbf{q},i \mu, n,m,\sigma} \frac{h_{\mathbf{k},\mathbf{q}}^{nm, i \mu} }{\sqrt{N}}u_{i \mu}(\mathbf{q}, \tau) \langle \gamma_{\mathbf{k}+\mathbf{q},n\sigma}^\dag(\tau) \gamma_{\mathbf{k},m\sigma}(\tau) \rangle_{S_c} d\tau 
\ea 
Due to the translational symmetry, only $\mathbf{q}=0$ components remain non-zero. \textcolor{black}{Moroever, only elements with $m=n$ are non-zero due to the fact that, in the band basis $\gamma_{\mathbf{k},n\sigma}$, only electrons with the same band index are coupled with each other.}
In addition, we let $n^\gamma_{\mathbf{k},n\sigma} = \langle \gamma_{\mathbf{k},n\sigma}^\dag \gamma_{\mathbf{k},n\sigma}\rangle_{S_c}$ denote the filling of $\gamma_{\mathbf{k},n\sigma}$ electrons. Then
\ba 
\langle S_g \rangle_{S_c}  = \beta \sum_{\mathbf{k},i\mu,\sigma}u_{i\mu}(\mathbf{q}=0,\tau)  \frac{h_{\mathbf{k},\mathbf{q}=0}^{nn, i \mu} }{\sqrt{N}} n^\gamma_{\mathbf{k},n\sigma} 
\ea 
We note that $\langle S_g \rangle_{S_c}\ne 0$ indicates the system is unstable. Because $\langle S_g\rangle $ is a linear term in the phonon fields. Thus the expectation value $\langle u_{i\mu}(\mathbf{q}=0)\rangle $ with respect to the full Hamiltonian cannot be zero due to the existence of linear in $u$ term. This means a condensation of $u_{i\mu}(\mathbf{q}=0)$ bosons, and an instability of the current atomic configuration. After the condensation, the atomic position is redistributed which makes the new effective theory has $\langle S_g\rangle_{S_c} = 0$ (otherwise, further redistribution of atom positions is required). 
However, since this condensation happens for the phonon fields with $\mathbf{q}=0$. This redistribution of atomic positions will not indicate a CDW transition. In practice, in the DFT calculation of the phonon spectrum, the atomic configurations are relaxed and satisfy $\langle u_{i\mu}(\mathbf{q}=0)\rangle =0 $ which also indicates the vanishing of $\langle S_g\rangle_{S_c}$. 
 Therefore, for what follows, we work with the assumption that $\langle S_g\rangle_{S_c=0}$.

We next consider the second-order terms
\ba 
&-\frac{1}{2}\bigg[ \langle S_g^2\rangle_{S_c} -\bigg(\langle S_g\rangle_{S_c}\bigg)^2\bigg] \nonumber \\ 
=&-\int_0^\beta \int_0^\beta \frac{1}{2}\sum_{i \mu,i'\mu', n,m,n',m',\sigma,\sigma',q,q',\mathbf{k},\mathbf{k}'} 
\frac{h_{\mathbf{k},\mathbf{q}}^{nm,i \mu}h_{\mathbf{k}',\mathbf{q}'}^{n'm',i'\mu'}}{N}
u_{\delta \mu}(\mathbf{q},\tau) u_{\delta'\mu'}(\mathbf{q}', \tau')  \nonumber \\ 
&\bigg[ \langle 
\gamma_{\mathbf{k}+\mathbf{q},n \sigma}^\dag(\tau) \gamma_{\mathbf{k},m\sigma}(\tau)  
\gamma_{\mathbf{k}'+\mathbf{q}',n' \sigma'}^\dag(\tau') \gamma_{\mathbf{k}',m'\sigma'}(\tau')  
\rangle_{S_c}
-\langle \gamma_{\mathbf{k}+\mathbf{q},n \sigma}^\dag(\tau) \gamma_{\mathbf{k},m\sigma}(\tau) \rangle_{S_c} \langle \gamma_{\mathbf{k}'+\mathbf{q}',n' \sigma'}^\dag(\tau') \gamma_{\mathbf{k}',m'\sigma'}(\tau')  
\rangle_{S_c}
\bigg] 
d\tau d \tau'\nonumber \\ 
=&-\int_0^\beta \int_0^\beta \frac{1}{2}\sum_{i \mu,i'\mu', n,m,n',m',\sigma,\sigma',q,q',\mathbf{k},\mathbf{k}'} 
\frac{h_{\mathbf{k},\mathbf{q}}^{nm,i \mu}h_{\mathbf{k}',\mathbf{q}'}^{n'm',i'\mu'}}{N}
u_{\delta \mu}(\mathbf{q},\tau) u_{\delta'\mu'}(\mathbf{q}', \tau')  \nonumber \\ 
&\bigg[ \langle 
:\gamma_{\mathbf{k}+\mathbf{q},n \sigma}^\dag(\tau) \gamma_{\mathbf{k},m\sigma}(\tau)  : 
:\gamma_{\mathbf{k}'+\mathbf{q}',n' \sigma'}^\dag(\tau') \gamma_{\mathbf{k}',m'\sigma'}(\tau'):    
\rangle_{S_c}
\bigg] 
d\tau d \tau'\nonumber \\ 
=&-\int_0^\beta \int_0^\beta \frac{1}{2}\sum_{i\mu,i'\mu', n,m,n',m',\mathbf{q},\mathbf{k},\mathbf{k}'} 
\frac{h_{\mathbf{k},\mathbf{q}}^{nm,i  \mu}h_{\mathbf{k}'+\mathbf{q},-\mathbf{q}}^{n'm',i'\mu'}}{N}
u_{i \mu}(\mathbf{q}, \tau) u_{i'\mu'}(-\mathbf{q}, \tau')
\chi^{nm,n'm'}_0(\mathbf{k},\mathbf{k}',\mathbf{q}, \tau-\tau'))d\tau d\tau' 
\label{eq:2nd_order}
\ea 
where we have introduced the susceptibility
\ba 
\chi_{0,\gamma}^{nm,n'm'}(\mathbf{k},\mathbf{k}',\mathbf{q},\tau-\tau') = \sum_{\sigma,\sigma'} \langle :\gamma^\dag_{\mathbf{k}+\mathbf{q},n\sigma}(\tau) \gamma_{\mathbf{k},m\sigma}(\tau): :\gamma_{\mathbf{k}',n'\sigma}^\dag(\tau') \gamma_{\mathbf{k}'+\mathbf{q},m'\sigma}(\tau'):\rangle_{S_c}
\label{eq:sus_band}
\ea 
\textcolor{black}{and we have used the fact that $\langle S_g\rangle_{S_c}=0$.} \textcolor{black}{$\sigma,\sigma'$ are spin indices that have been summed over.}
The normal ordering is defined as
\ba 
:O: = O- \langle O\rangle_{S_c}
\label{eq:normal_order}
\ea. 

For the fermion bilinear system $S_c$, we have 
\ba 
 &\chi_{0,\gamma}^{nm,n'm'}(\mathbf{k},\mathbf{k}',\mathbf{q},\tau-\tau')\nonumber \\ = &\sum_{\sigma,\sigma'} \langle :\gamma^\dag_{\mathbf{k}+\mathbf{q},n\sigma}(\tau) \gamma_{\mathbf{k},m\sigma}(\tau): :\gamma_{\mathbf{k}',n'\sigma}^\dag(\tau') \gamma_{\mathbf{k}'+\mathbf{q},m'\sigma}(\tau'):\rangle_{S_c} \nonumber 
 \\ 
 =&\sum_{\sigma,\sigma'} \langle :\gamma^\dag_{\mathbf{k}+\mathbf{q},n\sigma}(\tau) \gamma_{\mathbf{k},m\sigma}(\tau):\rangle_{S_c} \langle  :\gamma_{\mathbf{k}',n'\sigma}^\dag(\tau') \gamma_{\mathbf{k}'+\mathbf{q},m'\sigma}(\tau'):\rangle_{S_c} \nonumber 
 \\ 
 &+\langle  \gamma_{\mathbf{k}+\mathbf{q},n\sigma}^\dag(\tau) \gamma_{\mathbf{k}'+\mathbf{q},m'\sigma'}(\tau') \rangle_{S_c} 
 \langle  \gamma_{\mathbf{k},m\sigma}(\tau) \gamma^\dag_{\mathbf{k}',n'\sigma'}(\tau') \rangle_{S_c}  \nonumber  \\ 
 =& \sum_{\sigma} \delta_{\mathbf{k},\mathbf{k}'} \delta_{n,m'}\delta_{m,n'} \langle  \gamma_{\mathbf{k}+\mathbf{q},n\sigma}^\dag(\tau) \gamma_{\mathbf{k}+\mathbf{q},n\sigma}(\tau') \rangle_{S_c} 
 \langle \gamma_{\mathbf{k},m\sigma}(\tau) \gamma^\dag_{\mathbf{k},m\sigma}(\tau') \rangle_{S_c} 
 \label{eq:wick_thm}
 \ea
For a non-interacting system $S_c$, we note that~\cite{HU23a} 
\ba 
&\langle \gamma_{\mathbf{k},n\sigma}^\dag(\tau) \gamma_{\mathbf{k},n\sigma} (0)\rangle_{S_c} = \frac{e^{-(\beta-\tau)\epsilon_{\mathbf{k},n}}}{1+e^{-\beta \epsilon_{\mathbf{k},n}} },\quad \tau>0 \nonumber\\ 
&\langle  \gamma_{\mathbf{k},n\sigma}(\tau) \gamma_{\mathbf{k},n\sigma}^\dag (0)\rangle_{S_c} = \frac{e^{-\tau\epsilon_{\mathbf{k},n}}}{1+e^{-\beta \epsilon_{\mathbf{k},n}} },\quad \tau>0
\label{eq:ele_non_int_prop}
\ea 
Then, for $\tau-\tau'\ge 0$, combining \cref{eq:wick_thm} and \cref{eq:ele_non_int_prop}, we find
\ba 
&\chi_{0,\gamma}^{nm,n'm'}(\mathbf{k},\mathbf{k}',\mathbf{q},\tau-\tau') =\sum_{\sigma} \delta_{\mathbf{k},\mathbf{k}'} \delta_{n,m'}\delta_{m,n'} \frac{ e^{-(\beta -\tau) \epsilon_{\mathbf{k}+\mathbf{q},n}  -\tau \epsilon_{\mathbf{k},m} }
}{
(1+e^{-\beta \epsilon_{\mathbf{k},m}(\mathbf{k})})(1+e^{-\beta \epsilon_{\mathbf{k}+\mathbf{q},n} })
}
\ea 
\textcolor{black}{where $\sigma$ is the spin index that has been summed over}.
The Fourier transformation gives 
\ba 
 \chi_{0,\gamma}^{nm,n'm'}(\mathbf{k},\mathbf{k}',\mathbf{q},i\Omega_n) = &\int_0^\beta \chi_0^{nm,n'm'}(\mathbf{k},\mathbf{k}',\mathbf{q},\tau) e^{i\Omega_n\tau}d\tau \nonumber\\ 
 =&\sum_{\sigma} \delta_{\mathbf{k},\mathbf{k}'} \delta_{n,m'}\delta_{m,n'} \frac{-n_F(\epsilon_{\mathbf{k}+\mathbf{q},n}) +n_F(\epsilon_{\mathbf{k},m})}{i\Omega_n + \epsilon_{\mathbf{k}+\mathbf{q},n} - \epsilon_{\mathbf{k},m} }
\label{eq:sus_non_int}
\ea 
where $n_F(x) = 1/(1+e^{\beta x})$ is the Fermi-Dirac distribution function. 
\textcolor{black}{ The inverse Fourier transformation gives 
\ba 
\chi_{0,\gamma}^{nm,n'm'}(\mathbf{k},\mathbf{k}',\mathbf{q},\tau) = \frac{1}{\beta}\sum_{i\Omega_n}  \chi_{0,\gamma}^{nm,n'm'}(\mathbf{k},\mathbf{k}',\mathbf{q},i\Omega_n)e^{-i\Omega_n \tau} 
\label{eq:sus_inv_ft}
\ea 
Combining \cref{eq:2nd_order}, \cref{eq:sus_non_int} and \cref{eq:sus_inv_ft}, we find 
\ba 
&-\frac{1}{2}\bigg[ \langle S_g^2\rangle_{S_c} -\bigg(\langle S_g\rangle_{S_c}\bigg)^2\bigg] \nonumber\\ 
=&-\int_0^\beta \int_0^\beta \frac{1}{2\beta }\sum_{i\mu,i'\mu', n,m,n',m',\mathbf{q},\mathbf{k},\mathbf{k}',i\Omega_n} 
\frac{h_{\mathbf{k},\mathbf{q}}^{nm,i  \mu}h_{\mathbf{k}'+\mathbf{q},-\mathbf{q}}^{n'm',i'\mu'}}{N}
u_{i \mu}(\mathbf{q}, \tau) u_{i'\mu'}(-\mathbf{q}, \tau')
\chi^{nm,n'm'}_0(\mathbf{k},\mathbf{k}',\mathbf{q}, i\Omega_n))d\tau d\tau' \nonumber\\ 
=& -\int_0^\beta \int_0^\beta \frac{1}{2\beta }\sum_{i\mu,i'\mu', n,m,\mathbf{q},\mathbf{k},i\Omega_n} 
\frac{h_{\mathbf{k},\mathbf{q}}^{nm,i  \mu}h_{\mathbf{k}+\mathbf{q},-\mathbf{q}}^{mn,i'\mu'}}{N}
u_{i \mu}(\mathbf{q}, \tau) u_{i'\mu'}(-\mathbf{q}, \tau')
\frac{-n_F(\epsilon_{\mathbf{k}+\mathbf{q},n}) +n_F(\epsilon_{\mathbf{k},m})}{i\Omega_n + \epsilon_{\mathbf{k}+\mathbf{q},n} - \epsilon_{\mathbf{k},m} }e^{-i\Omega_n(\tau-\tau')}d\tau d\tau' 
\ea 
We ignore the $\Omega_n$-dependency in the denominators and find }
\ba 
&-\frac{1}{2}\bigg[ \langle S_g^2\rangle_{S_c} -\bigg(\langle S_g\rangle_{S_c}\bigg)^2\bigg] \nonumber \\ 
\approx &-\int_0^\beta \int_0^\beta \frac{1}{2}\sum_{i\mu,i'\mu', n,m,\mathbf{q},\mathbf{k},\sigma,i\Omega_n} 
\frac{h_{\mathbf{k},\mathbf{q}}^{nm,i \mu}h_{\mathbf{k}+\mathbf{q},-\mathbf{q}}^{mn,i'\mu'}}{N}
u_{i \mu}(\mathbf{q}, \tau) u_{i\mu'}(-\mathbf{q}, \tau')
\frac{1}{\beta}
\frac{-n_F(\epsilon_{\mathbf{k}+\mathbf{q},n}) +n_F(\epsilon_{\mathbf{k},m})}{\epsilon_{\mathbf{k}+\mathbf{q},n} - \epsilon_{\mathbf{k},m} }
e^{-i\Omega_n(\tau-\tau')}d\tau d\tau' \nonumber \\ 
=& -\int_0^\beta \sum_{i \mu,i'\mu', n,m,\mathbf{q},\mathbf{k},\sigma} 
\frac{h_{\mathbf{k},\mathbf{q}}^{nm,i \mu}h_{\mathbf{k}+\mathbf{q},-\mathbf{q}}^{mn,i'\mu'}}{2N}
u_{i \mu}(\mathbf{q}, \tau) u_{i'\mu'}(-\mathbf{q}, \tau)
\frac{-n_F(\epsilon_{\mathbf{k}+\mathbf{q},n}) +n_F(\epsilon_{\mathbf{k},m})}{\epsilon_{\mathbf{k}+\mathbf{q},n} - \epsilon_{\mathbf{k},m} }
d\tau 
\label{eq:2nd_order_Seff}
\ea 
\textcolor{black}{where the summation over $\Omega_n$ gives a $\delta$-function of $\tau-\tau'$.}
Via \cref{eq:partition_eff} and \cref{eq:partition_eff}, we introduce the effective action of the system 
\ba 
Z = &\int D[u,P]e^{-S_{eff}} \nonumber \\ 
S_{eff} = &S_{ph} -\frac{1}{2}\langle S_g^2\rangle_{S_c} \nonumber \\ 
\approx & \int_0^\beta 
\bigg[ - \sum_{\mathbf{q}, i \mu} iP_{i\mu}(-\mathbf{q},\tau)\partial_\tau u_{ia\mu}(\mathbf{q},\tau)  + \sum_{\mathbf{q},i\mu} \frac{P_{i\mu}(\mathbf{q},\tau)P_{i\mu}(-\mathbf{q},\tau)}{2M_\alpha} + \frac{1}{2}\sum_{\mathbf{q},\alpha\mu,\gamma\nu} \Phi^{eff}_{i\mu,i'\nu}(\mathbf{q})u_{i\mu}(-\mathbf{q},\tau)u_{i'\nu}(\mathbf{q},\tau)\bigg]
d\tau 
\ea 
where the effective force-constant matrix that has been re-normalized by electrons is (from \cref{eq:2nd_order_Seff})
\ba 
\Phi^{eff}_{i\mu,i'\nu}(\mathbf{q}) = \Phi_{i\mu,i'\nu}(\mathbf{q}) -\sum_{n,m, \sigma, \mathbf{k}} 
\frac{h_{\mathbf{k},\mathbf{q}}^{nm, i'\nu} h_{\mathbf{k}+\mathbf{q},-\mathbf{q}}^{mn,i \mu}}{N}\frac{-n_F(\epsilon_{\mathbf{k}+\mathbf{q},n}) +n_F(\epsilon_{\mathbf{k},m})}{\epsilon_{\mathbf{k}+\mathbf{q},n} - \epsilon_{\mathbf{k},m} }
\label{eq:eff_force_const}
\ea 
where the electron-phonon coupling introduces an additional contribution to the force-constant matrix
\ba 
\Phi^{corr}_{i\mu,i'\nu}(\mathbf{q}) = -\sum_{n,m, \sigma, \mathbf{k}} 
\frac{h_{\mathbf{k},\mathbf{q}}^{nm, i'\nu} h_{\mathbf{k}+\mathbf{q},-\mathbf{q}}^{mn,i \mu}}{N}\frac{-n_F(\epsilon_{\mathbf{k}+\mathbf{q},n}) +n_F(\epsilon_{\mathbf{k},m})}{\epsilon_{\mathbf{k}+\mathbf{q},n} - \epsilon_{\mathbf{k},m} }
\label{eq:corr_force_const}
\ea 

\textcolor{black}{ 
In practice, one could do one step more to integrate the $P$ fields and find an effective theory with only $u$ fields
\ba 
Z = &\int D[u,P]e^{-S_{eff}} \nonumber \\ 
=& \int D[u,P] 
\exp\bigg\{ -
\int_0^\beta  
\bigg[ - \sum_{\mathbf{q},i \mu}iP_{i\mu}(-\mathbf{q},\tau)\partial_\tau u_{i\mu}(\mathbf{q},\tau)  + \sum_{\mathbf{q},i\mu} \frac{P_{i\mu}(\mathbf{q},\tau)P_{i\mu}(-\mathbf{q},\tau)}{2M_i}
\bigg]d\tau \bigg\} \nonumber\\ 
& \exp\bigg\{ 
-
 \int_0^\beta \frac{1}{2}\sum_{\mathbf{q},i\mu,i'\nu} \Phi^{eff}_{i\mu,i'\nu}(\mathbf{q})u_{i\mu}(-\mathbf{q},\tau)u_{i'\nu}(\mathbf{q},\tau)\bigg]
d\tau 
\bigg\} 
\ea 
We could integrate over $P$ fields by using the Gaussian integral, which gives
\ba 
Z =  &\int D[u,P]e^{-S_{eff}}\propto \int D[u] e^{-S_{eff}^u }
\nonumber \\ 
S_{eff}^u = &\int_0^\beta \bigg[ \sum_{\mathbf{q}, i\mu } \frac{M_{i}}{2 } \partial_\tau u_{i \mu}(\mathbf{q},\tau) \partial_\tau u_{i\mu}(-\mathbf{q},\tau) +  \frac{1}{2}\sum_{\mathbf{q},i\mu,i'\nu} \Phi^{eff}_{i\mu,i'\nu}(\mathbf{q})u_{i\mu}(-\mathbf{q},\tau)u_{i'\nu}(\mathbf{q},\tau)\bigg]d\tau
\ea 
We also give the expression in the Matsubara frequency domain. We let 
\ba 
u_{i\mu}(\mathbf{q}, i\Omega_n) = \int_0^\beta u_{i\mu}(\mathbf{q},\tau)e^{i\Omega_n\tau }d\tau 
\ea 
Then 
\ba 
S_{eff}^u = &\frac{1}{\beta} \sum_{\mathbf{q},i\Omega_n,i\mu,i'\nu} \frac{1}{2}\bigg(M_i \Omega_n^2\delta_{i,i'}\delta_{\mu,\mu'} 
+ \Phi^{eff}_{i\mu,i'\nu}(\mathbf{q})\bigg) u_{i\mu}(\mathbf{q},i\Omega_n)u_{i'\mu'}(-\mathbf{q},-i\Omega_n)
\ea 
One could also work with the rescaled fields as introduced in \cref{eq:def_tilde_up}, and find
\ba 
\tilde{u}_{i\mu}(\mathbf{q},i\Omega_n) = &\sqrt{M_i}u_{i \mu}(\mathbf{R},i\Omega_n) \\ 
Z =  &\int D[\tilde{u}] e^{-S_{eff}^{\tilde{u} }}
\nonumber \\ 
S_{eff}^{\tilde{u}} = &\frac{1}{\beta} \sum_{\mathbf{q},i\Omega_n,i\mu,i'\nu} \frac{1}{2}\bigg( \Omega_n^2\delta_{i,i'}\delta_{\mu,\mu'} 
+ D^{eff}_{i\mu,i'\nu}(\mathbf{q})\bigg) u_{i\mu}(\mathbf{q},i\Omega_n)u_{i'\mu'}(-\mathbf{q},-i\Omega_n)
\ea 
where the effective dynamical matrix is 
\ba 
 D^{eff}_{i\mu,i'\nu}(\mathbf{q}) = \frac{1}{\sqrt{M_iM_i'}}\Phi^{eff}_{i\mu,i'\nu}(\mathbf{q})
 \label{eq:seff_tilde_u}
\ea 
}

\textcolor{black}{ 
Finally, we describe the high-order contribution, which will lead to the interaction between phonon fields. Our effective theory of phonon is derived by the cumulant expansion in \cref{eq:log_exp}. In practice, we could perform expansion to higher orders, which gives 
\ba 
&- \log  \langle  e^{-S_{g}} \rangle_{S_c} \approx - \log  \langle  1-S_{g} +\frac{1}{2}S_{g}^2 -\frac{1}{6} S_g^3 + \frac{1}{24}S_g^4\rangle_{S_c} \nonumber
 \\ 
 \approx & 
\langle S_{g}\rangle_{S_c} - \frac{1}{2} 
\bigg[ \langle S_g^2 \rangle_{S_c} - \bigg(\langle S_g\rangle_{S_c}\bigg)^2
\bigg] \nonumber\\ 
& + \frac{1}{6}
\bigg[ 
\langle S_g^3\rangle_{S_c} - 3\langle S_g\rangle_{S_c}\langle S_g^2\rangle_{S_c}
+2 \langle S_g\rangle_{S_c}^3 
\bigg] +
\frac{1}{24}
\bigg[ -\langle S_g^4 \rangle_{S_c} +4 \langle S_g\rangle_{S_c}\langle S_g^3\rangle_{S_c}
+3\langle S_g^2 \rangle_{S_c}^2 
-12 \langle S_c\rangle_{S_c}^2\langle S_c^2\rangle_{S_c} + 6\langle S_g\rangle_{S_c}^4
\bigg]
\label{eq:log_exp_high_order}
\ea 
Since $S_g$ is linear in phonon fields $u$, the term in the final line of \cref{eq:log_exp_high_order} produces terms that are cubic in $u$ and quadratic in $u$ which describes the interaction between phonon fields induced by the electrons. One could also include even higher-order terms to capture even higher-order (in $u$) interactions. We note that this interaction effect will be considered when we study the CDW phase transition and build the Landau Ginzburg free energy in \cref{sec:cdw_transition}. A detailed evaluation of high-order terms based on \cref{eq:log_exp_high_order} is in general complicated and has to be done numerically. But the interaction terms given by \cref{eq:log_exp_high_order} are required to be invariant under symmetry transformation (since $S_c$ and $S_g$ are all invariant under symmetry transformation). This allows us to write down the symmetry-allowed terms without giving the explicit value of each term (\cref{sec:cdw_transition}) 
}

\subsection{Electron correction to phonon propagators (Feynmann diagrams)}
In this \siSection{}, we provide an alternative derivation of \cref{eq:corr_force_const} based on Feynman diagrams. 
The Hamiltonian of the system using $b,b^\dag$ fields (\cref{eq:def_b}) is
\ba 
&H = H_c +H_{ph} +H_{g} \nonumber\\
&H_c = \sum_{\mathbf{k},\alpha \gamma,\sigma }t_{\alpha \gamma }(\mathbf{k})c_{\mathbf{k},\alpha \sigma}^\dag c_{\mathbf{k},\gamma\sigma } \nonumber \\
&H_{ph} = \sum_{\mathbf{q},n }\omega_n(\mathbf{q}) b^\dag_{\mathbf{q},\delta} b_{\mathbf{q},\kappa }\nonumber \\
&H_{g} = \sum_{\mathbf{q},n,\mathbf{k},\alpha,\gamma,\delta,\sigma} 
\bigg( \frac{\tilde{g}_{\mathbf{k},\mathbf{q}}^{\alpha \gamma ,n }}{\sqrt{N} }
b_{\mathbf{q},n}c_{\mathbf{k}+\mathbf{q},\alpha\sigma}^\dag c_{\mathbf{k} ,\gamma \sigma }+\text{h.c.}\bigg) 
\ea 
where (\cref{eq:ele_ph_b})
\ba 
\tilde{g}_{\mathbf{k},\mathbf{q}}^{\alpha \gamma ,n } = \sum_{i\mu} \frac{g_{\mathbf{k},\mathbf{q}}^{\alpha \gamma i \mu}U_{i\mu,n}(\mathbf{q})}{\sqrt{2\omega_n(\mathbf{q})M_i} }
\ea 
and the corresponding action is 
\ba 
S = \int_0^\beta \sum_{\mathbf{k},\alpha \sigma} c_{\mathbf{k},\alpha \sigma}^\dag (\tau) \partial_\tau c_{\mathbf{k},\alpha \sigma}(\tau)+ \int_0^\beta \sum_{\mathbf{q},n} b_{\mathbf{q},n}^\dag(\tau)\partial_\tau b_{\mathbf{q},n}(\tau)d\tau +\int_0^\beta H(\tau) d\tau 
\ea 
where $\tau$ is the imaginary time, $c_{\mathbf{k},\alpha \sigma}(\tau),b_{\mathbf{q},n}(\tau)$ are the electron and phonon fields at imaginary time $\tau$ respectively, and $H(\tau)$ is the corresponding Hamiltonian at imaginary time $\tau$.

We next introduce the Fourier transformation and work with Matsubara frequency
\ba 
c&_{\mathbf{k},\alpha\sigma}(i\omega_n) = \int_0^\beta c_{\mathbf{k},\alpha \sigma}(\tau)e^{i\omega_n\tau} d\tau,\quad \omega_n = (2n+1)\pi/\beta, n\in\mathbb{Z}\nonumber\\ 
&b_{\mathbf{q},n}(i\Omega_n) = \int_0^\beta b_{\mathbf{q},n}(\tau)e^{i\Omega_n\tau} d\tau,\quad \Omega_n = 2n\pi/\beta, n\in\mathbb{Z}\nonumber\\ 
\ea 
We can write the action as
\ba 
S = &S_0 +S_g \nonumber \\ 
S_0 =& \frac{1}{\beta}\sum_{\mathbf{k},\alpha \gamma \sigma,i\omega_n}c_{\mathbf{k},\alpha\sigma}^\dag(i\omega_n)\bigg(-i\omega_n \delta_{\alpha,\gamma} + t_{\alpha\gamma}(\mathbf{k}) \bigg) c_{\mathbf{k},\gamma\sigma}(i\omega_n)+\frac{1}{\beta}\sum_{\mathbf{q},n,i\Omega_n}
b^\dag_{\mathbf{q},n}(i\Omega_n)\bigg(-i\Omega_n+\omega_{n}(\mathbf{q})
\bigg) b_{\mathbf{q},n}(i\Omega) \nonumber \\ 
S_g=&\frac{1}{\beta^2}\sum_{\mathbf{q},\mathbf{k},\alpha \gamma,n,\sigma,i\omega_n,i\Omega_n}  \bigg(\frac{\tilde{g}_{\mathbf{k},\mathbf{q}}^{\alpha \gamma ,n}}{\sqrt{N} }
b_{\mathbf{q},n}(i\Omega_n)c_{\mathbf{k}+\mathbf{q},\alpha}^\dag(i\omega_n+i\Omega_n) c_{\mathbf{k} ,\gamma \sigma }(i\omega_n)+
\text{h.c.}\bigg) 
\ea

It is more convenient to consider the band basis of electrons. We use band-basis operators  (\cref{eq:band_basis_gam}) of electrons
Then 
\ba 
S = &S_0 +S_g \nonumber \\ 
S_0 =& \frac{1}{\beta}\sum_{\mathbf{k},n\sigma,i\omega_n}\gamma_{\mathbf{k},n\sigma}^\dag(i\omega_n) \bigg(-i\omega_n+ \epsilon_{\mathbf{k},n} \bigg) \gamma_{\mathbf{k},n\sigma}(i\omega_n)+\frac{1}{\beta}\sum_{\mathbf{q},n,i\Omega_n}
b^\dag_{\mathbf{q},n}(i\Omega_n)\bigg(-i\Omega_n+\omega_{n}(\mathbf{q})
\bigg) b_{\mathbf{q},n}(i\Omega)  \nonumber \\ 
S_g=&\frac{1}{\beta^2}\sum_{\mathbf{q},i,\mathbf{k},n,m,\sigma,i\omega_n,i\Omega_n}  \bigg[ \frac{h_{\mathbf{k},\mathbf{q}}^{nm,i}}{\sqrt{N} }
b_{\mathbf{q},i}(i\Omega_n)\gamma_{\mathbf{k}+\mathbf{q},n}^\dag(i\omega_n+i\Omega_n) \gamma_{\mathbf{k} ,m\sigma }(i\omega_n)+\text{h.c.}\bigg] 
\label{eq:e_ph_action}
\ea 
where 
\ba 
\tilde{h}_{\mathbf{k},\mathbf{q}}^{nm,i} = \sum_{\alpha \gamma,}\tilde{g}_{\mathbf{k},\mathbf{q}}^{\alpha \gamma,i} U_{\alpha n}^{c,*}(\mathbf{k})U_{\gamma m}^c(\mathbf{k})
\label{eq:tilde_h}
\ea 

\begin{figure}
    \centering
    \includegraphics[width=0.6\textwidth]{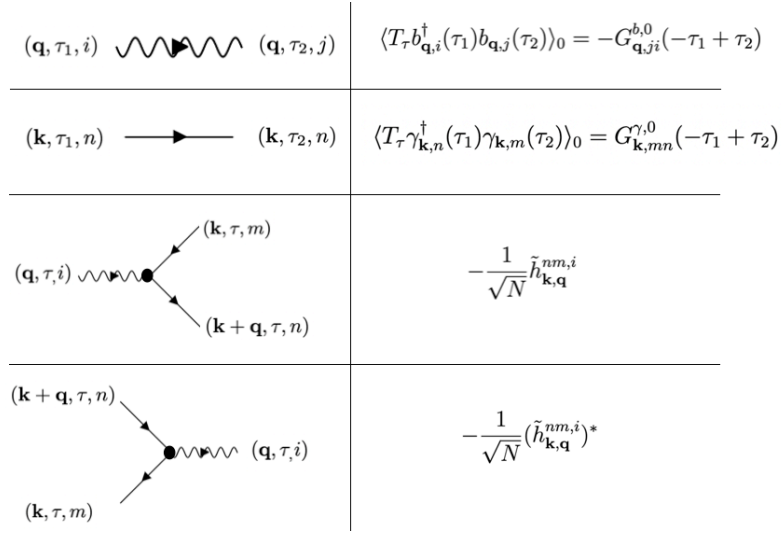}
    \caption{Feynman diagrams of the electron-phonon model. 
    \textcolor{black}{The minus sign before the interaction vertex in the Feynman rule comes from the minus sign in the exponential $e^{-S_g}$ before $S_g$.} 
}
    \label{fig:feynman_rule}
\end{figure}

\begin{figure}
    \centering
    \includegraphics[width=0.8\textwidth]{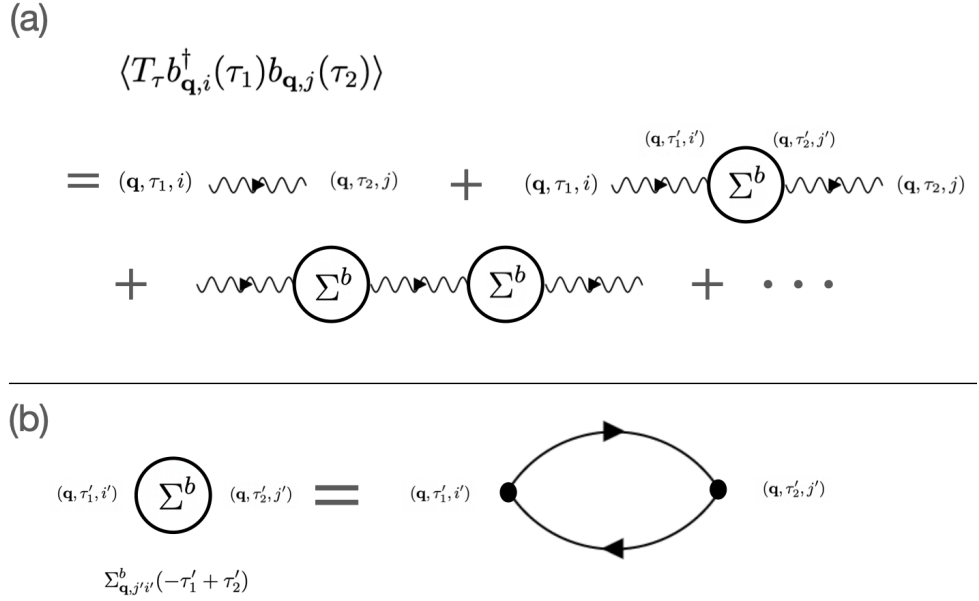}
    \caption{(a)Relation between phonon propagators $G^b$ and phonon self-energy $\Sigma^b$. (b) One-loop contributions to the phonon self-energy.}
    \label{fig:phonon_prop}
\end{figure}

We now calculate the correction to the phonon propagators. 
The bare phonon propagator and electron propagator with respect to the non-interacting system $S_0$ (\cref{eq:e_ph_action}) are 
\ba 
G_{\mathbf{q},ij}^{b,0}(\tau-\tau') =- \langle T_\tau b_{\mathbf{q},i}(\tau)b_{\mathbf{q},j}^\dag(\tau')\rangle_0 ,
\quad 
G_{\mathbf{k},nm}^{\gamma,0}(\tau-\tau') = - \langle T_\tau \gamma_{\mathbf{k},i}(\tau)\gamma_{\mathbf{k},j}^\dag(\tau')\rangle_0 
\ea 
We can perform Fourier transformation and work with Matsubara frequency 
\ba 
&G^{b,0}_{\mathbf{q},ij}(i\Omega_n) = \int_0^\beta G_{\mathbf{q},ij}^{b,0}(\tau) e^{i\Omega_n\tau} d\tau ,\quad 
G^{\gamma,0}_{\mathbf{k},nm}(i\omega_n) =\int_0^\beta G_{\mathbf{k},nm}^{\gamma,0}(\tau) e^{i\omega_n\tau} d\tau \,.
\ea 
 Using the non-interaction action $S_0$ (\cref{eq:e_ph_action}), we have
\ba 
&G^{b,0}_{\mathbf{q},ij}(i\Omega_n) =\delta_{i,j}\frac{1}{i\Omega_n-\omega_{\mathbf{q},i}} ,\quad \quad 
G^{\gamma,0}_{\mathbf{k},nm}(i\omega_n) = \delta_{n,m}\frac{1}{i\omega_n-\epsilon_{\mathbf{k},n}} 
\ea 
The corresponding rules of the Feynman diagram are illustrated in \cref{fig:feynman_rule}.

The phonon propagator can be calculated via the diagram shown in \cref{fig:phonon_prop} (a) and the Feynman rule shown in \cref{fig:feynman_rule}:
\ba 
&-G^b_{\mathbf{q},ji}(-\tau_1+\tau_2) = \langle T_\tau b_{\mathbf{q},i}^\dag (\tau_1) b_{\mathbf{q},j}(\tau_2)\rangle \nonumber \\ 
=&-G^{b,0}_{\mathbf{q},ji}(-\tau_1+\tau_2) 
+\int_0^{\beta} \int_0^\beta \sum_{i'j'}\bigg( -G^{b,0}_{\mathbf{q},i'i}(-\tau_1+\tau_1') \bigg) \Sigma^b_{\mathbf{q},j'i'}(-\tau_1'+\tau_2') 
\bigg( -G^{b,0}_{\mathbf{q},jj'}(-\tau_2'+\tau_2) \bigg) d\tau_1'd\tau_2'  + ...  \nonumber \\ 
=& -G^{b,0}_{\mathbf{q},ji}(-\tau_1+\tau_2) 
+\int_0^{\beta} \int_0^\beta  \sum_{i'j'} \bigg( -G^{b,0}_{\mathbf{q},i'i}(-\tau_1+\tau_1') \bigg) \Sigma^b_{\mathbf{q},j'i'}(-\tau_1'+\tau_2') 
\bigg[ -G^{b,0}_{\mathbf{q},jj'}(-\tau_2'+\tau_2) + ... \bigg] d\tau_1'd\tau_2' \nonumber \\ 
=& -G^{b,0}_{\mathbf{q},ji}(-\tau_1+\tau_2) 
+\int_0^{\beta} \int_0^\beta  \sum_{i'j'} \bigg( -G^{b,0}_{\mathbf{q},i'i}(-\tau_1+\tau_1') \bigg) \Sigma^b_{\mathbf{q},j'i'}(-\tau_1'+\tau_2') 
\bigg( -G^{b}_{\mathbf{q},jj'}(-\tau_2'+\tau_2)\bigg) d\tau_1'd\tau_2' 
\label{eq:self_enegy_exp}
\ea 
\textcolor{black}{where $...$ denotes the high-order term and the final line gives the Dyson equation of the phonon propagator.} 
$\Sigma^b$ is the self-energy of the phonon that denotes the summation of all the diagrams that can not be split into two parts by removing one phonon propagator line. 
We next perform Fourier transformation to \cref{eq:self_enegy_exp} and find
\ba 
&\frac{1}{\beta}\int_0^\beta \int_0^\beta \bigg[-G^b_{\mathbf{q},ji}(-\tau_1+\tau_2)\bigg] e^{i\Omega_n(-\tau_1+\tau_2)} d\tau_1 d\tau_2  \nonumber \\ 
=&\frac{1}{\beta}\int_0^\beta \int_0^\beta \bigg[-G^{b,0}_{\mathbf{q},ji}(-\tau_1+\tau_2)\bigg] e^{i\Omega_n(-\tau_1+\tau_2)} d\tau_1 d\tau_2 \nonumber \\ 
&
+ \frac{1}{\beta} 
\int_0^{\beta} \int_0^\beta  \int_0^{\beta} \int_0^\beta  \sum_{i'j'} \bigg( -G^{b,0}_{\mathbf{q},i'i}(-\tau_1+\tau_1') \bigg) \Sigma^b_{\mathbf{q},j'i'}(-\tau_1'+\tau_2') 
\bigg( -G^{b}_{\mathbf{q},jj'}(-\tau_2'+\tau_2)\bigg) d\tau_1'd\tau_2' e^{i\Omega_n(-\tau_1+\tau_2)} d\tau_1 d\tau_2   
\label{eq:self_ft_0}
\ea 

We then introduce the following identities
\ba 
&\frac{1}{\beta}\int_0^\beta \int_0^\beta \bigg[-G^b_{\mathbf{q},ji}(-\tau_1+\tau_2)\bigg] e^{i\Omega_n(-\tau_1+\tau_2)} d\tau_1 d\tau_2 \nonumber \\ 
=& \frac{1}{\beta^2}\int_0^\beta \int_0^\beta \bigg[-\sum_{i\Omega_n'}G^b_{\mathbf{q},ji}(i\Omega_n')e^{-i\Omega_n'(-\tau_1+\tau_2)}\bigg] e^{i\Omega_n(-\tau_1+\tau_2)} d\tau_1 d\tau_2 \nonumber \\ 
=& -\sum_{i\Omega_n'} G^{b}_{\mathbf{q},ji}(i\Omega_n') \delta_{i\Omega_n',i\Omega_n} =- G^b_{\mathbf{q},ji}(i\Omega_n) 
\label{eq:self_ft_1}
\ea 
and 
\ba 
&\frac{1}{\beta}\int_0^\beta \int_0^\beta \bigg[-G^{b,0}_{\mathbf{q},ji}(-\tau_1+\tau_2)\bigg] e^{i\Omega_n(-\tau_1+\tau_2)} d\tau_1 d\tau_2=- G^{b,0}_{\mathbf{q},ji}(i\Omega_n) 
\label{eq:self_ft_2}
\ea 
and 
\ba 
&\frac{1}{\beta} \int_0^{\beta} \int_0^\beta  \int_0^{\beta} \int_0^\beta  \sum_{i'j'} \bigg( -G^{b,0}_{\mathbf{q},i'i}(-\tau_1+\tau_1') \bigg) \Sigma^b_{\mathbf{q},j'i'}(-\tau_1'+\tau_2') 
\bigg( -G^{b}_{\mathbf{q},jj'}(-\tau_2'+\tau_2)\bigg) d\tau_1'd\tau_2' e^{i\Omega_n(-\tau_1+\tau_2)} d\tau_1 d\tau_2   \nonumber \\ 
=&\frac{1}{\beta^4} \int_0^{\beta} \int_0^\beta  \int_0^{\beta} \int_0^\beta \sum_{i\Omega_n',i\Omega_n'',i\Omega_n'''}  \sum_{i'j'}  
\bigg( -G^{b,0}_{\mathbf{q},i'i}(i\Omega_n') \bigg) \Sigma^b_{\mathbf{q},j'i'}(i\Omega_n'') 
\bigg( -G^{b}_{\mathbf{q},jj'}(i\Omega_n'')\bigg)\nonumber \\  
&e^{-i\Omega_n'(-\tau_1+\tau_1') -i\Omega_n''(-\tau_1'+\tau_2') -i\Omega_n'''(-\tau_1'+\tau_2)}
e^{i\Omega_n(-\tau_1+\tau_2)} 
d\tau_1'd\tau_2'  d\tau_1 d\tau_2  \nonumber \\ 
=&\sum_{i'j'}G_{\mathbf{q},i'i}^{b,0}(i\Omega_n)\Sigma_{\mathbf{q},j'i'}^b(i\Omega_n)G_{\mathbf{q},jj'}^{b}(i\Omega_n) 
\label{eq:self_ft_3} 
\ea 
Combining \cref{eq:self_ft_0}, \cref{eq:self_ft_1}, \cref{eq:self_ft_2} and \cref{eq:self_ft_3}, we have 
\ba 
-G_{\mathbf{q},ji}^b(i\Omega_n) = -G_{\mathbf{q},ji}^{b,0}(i\Omega_n)  +\sum_{i'j'}G_{\mathbf{q},i'i}^{b,0}\Sigma_{\mathbf{q},j'i'}^b(i\Omega_n) G_{\mathbf{q},jj'}^b(i\Omega_n) 
\label{eq:green_self_energy_phonon_summation}
\ea 
We then treat $G_{\mathbf{q},ji}^{b}(i\Omega_n),G_{\mathbf{q},ji}^{b,0}(i\Omega_n),.\Sigma_{\mathbf{q},ji}^{b}(i\Omega_n)$ as matrix with row and column indices $j,i$ respectively. Then \cref{eq:green_self_energy_phonon_summation} can be written as
\ba 
&G_\mathbf{q}^b(i\Omega_n) = G_{\mathbf{q}}^{b,0}(i\Omega_n) -  G_{\mathbf{q}}^{b}(i\Omega_n) \cdot \Sigma_\mathbf{q}^b(i\Omega_n) \cdot G_\mathbf{q}^{b,0}(i\Omega_n) \nonumber \nonumber \\ 
\Rightarrow& G_\mathbf{q}^b(i\Omega_n)  = \bigg[\bigg(G_\mathbf{q}^{b,0}(i\Omega_n)\bigg)^{-1} + \Sigma^b_\mathbf{q}(i\Omega_n) \bigg]^{-1}
\label{eq:green_self_energy_phonon}
\ea 
where $\cdot$ denotes matrix product and $ ^{-1}$ denotes matrix inverse.  

As we mentioned before, $\Sigma^b$ includes the summation of all the diagrams that can not be split into two parts by removing one phonon propagator line. The leading-order contribution (one-loop contribution) to the self-energy $\Sigma^b$ comes from the Fermi bubble diagram shown in \cref{fig:phonon_prop}(b). The one-loop contribution is
\ba 
[\Sigma^b_{\mathbf{q},j'i'}(-\tau_1'+\tau_2')]_{\text{1-loop}} = -\sum_{\mathbf{k},nn'mm'} \frac{\tilde{h}_{\mathbf{k},\mathbf{q}}^{nm,i'}}{\sqrt{N}}
\frac{(\tilde{h}_{\mathbf{k},\mathbf{q}}^{n'm',j'})^*}{\sqrt{N}}
\bigg(G^{\gamma,0}_{\mathbf{k},mm'}(\tau_1'-\tau_2')\bigg)\bigg(G^{\gamma,0}_{\mathbf{k}+\mathbf{q},n'n}(\tau_2'-\tau_1')\bigg)
\ea 
The additional minus sign comes from the loop formed by the fermion propagators \textcolor{black}{(note that, for each loop formed by the fermion propagators, one needs to commute odd number times of fermion operators to transform the expectation value into the form of $\langle \gamma \gamma^\dag \rangle_{S_c} \langle \gamma \gamma^\dag \rangle_{S_c}...$, and then the odd number of commutation gives an overall minus sign for each fermion loop).} 
Performing Fourier transformation, we find the one-loop correction to the self-energy is 
\ba 
&[\Sigma^b_{\mathbf{q},j'i'}(i\Omega_n)]_{\text{1-loop}} \nonumber \\ 
=&\frac{1}{\beta} \int_0^\beta 
[\Sigma^b_{\mathbf{q},j'i'}(-\tau_1'+\tau_2')]_{\text{1-loop}}e^{i\Omega_n(-\tau_1'+\tau_2')}
 d\tau_1'd\tau_2' \nonumber \\ 
 =&\frac{-1}{\beta N} \frac{1}{\beta^2}\int_0^\beta \int_0^\beta 
\sum_{\mathbf{k},nn'mm'}\sum_{i\omega_n',i\omega_n''} \tilde{h}_{\mathbf{k},\mathbf{q}}^{nm,i'}
(\tilde{h}_{\mathbf{k},\mathbf{q}}^{n'm',j'})^*G^{\gamma,0}_{\mathbf{k},mm'}(i\omega_n')G^{\gamma,0}_{\mathbf{k}+\mathbf{q},n'n}(i\omega_n'')e^{i\Omega_n(-\tau_1'+\tau_2') -i\omega_n'(\tau_1'-\tau_2') -i\omega_n''(\tau_2'-\tau_2')}
 d\tau_1'd\tau_2' \nonumber \\ 
 =&\frac{-1}{\beta N}  
\sum_{\mathbf{k},i\omega_n',nm} \tilde{h}_{\mathbf{k},\mathbf{q}}^{nm,i'}
(\tilde{h}_{\mathbf{k},\mathbf{q}}^{nm,j'})^*G^{\gamma,0}_{\mathbf{k},mm}(i\omega_n')G^{\gamma,0}_{\mathbf{k}+\mathbf{q},nn}(i\omega_n'+i\Omega_n)
\label{eq:one_loop_self_energy}
\ea 
We could also directly evaluate the summation in $i\omega_n$ which gives
\ba
&\frac{1}{\beta} \sum_{i\omega_n} G^{\gamma,0}_{\mathbf{k},mm}(i\omega_n')G^{\gamma,0}_{\mathbf{k}+\mathbf{q},nn}(i\omega_n'+i\Omega_n) \nonumber\\ 
=&\frac{1}{\beta} \sum_{i\omega_n'}\frac{1}{i\omega_n'-\epsilon_{\mathbf{k},m} }\frac{1}{i\omega_n' +i\Omega_n-\epsilon_{\mathbf{k}+\mathbf{q},n}} = \frac{n_F(\epsilon_{\mathbf{k},m}) -n_f(\epsilon_{\mathbf{k},n})}{i\Omega_n+\epsilon_{\mathbf{k},m} - \epsilon_{\mathbf{k}+\mathbf{q},n}}
\ea
Then 
\ba 
[\Sigma^b_{\mathbf{q},j'i'}(i\Omega_n)]_{\text{1-loop}}
=&\frac{1}{ N}  
\sum_{\mathbf{k},i\omega_n',nm} \tilde{h}_{\mathbf{k},\mathbf{q}}^{nm,i'}
(\tilde{h}_{\mathbf{k},\mathbf{q}}^{nm,j'})^*\frac{-n_F(\epsilon_{\mathbf{k},m}) +n_f(\epsilon_{\mathbf{k},n})}{i\Omega_n+\epsilon_{\mathbf{k},m} - \epsilon_{\mathbf{k}+\mathbf{q},n}}
\label{eq:one_loop_sigb}
\ea 

Combining \cref{eq:green_self_energy_phonon} and \cref{eq:one_loop_self_energy}, we are able to calculate the one-loop correction to the phonon propagator. This also leads to an effective phonon action
\ba 
S^b_{eff} = \frac{1}{\beta}\sum_{\mathbf{k},i\Omega_n, i,j}b_{i,\mathbf{q}}^\dag(i\Omega_n) (-1)G^{b,-1}_{\mathbf{q},ij}(i\Omega_n) b_{j,\mathbf{q}} = \frac{1}{\beta}\sum_{\mathbf{k}, i\Omega_n, i,j}b_{i,\mathbf{q}}^\dag(i\Omega_n) \bigg(-i\Omega_n \delta_{i,j} +\omega_{i}(\mathbf{q})\delta_{ij} -\Sigma_{\mathbf{q},ij}^b(i\Omega_n=0)\bigg) b_{j,\mathbf{q}}(i\Omega_n)  
\ea 
where we only keep the $i\Omega_n=0$ contribution of the self-energy correction. 
We transform to the $u,P$ basis via \cref{eq:def_b}, and also using \cref{eq:tilde_h}, \cref{eq:one_loop_sigb}, we realize the same effective action $S_{eff}^u$ given in \cref{eq:seff_tilde_u}, in other words $S_{eff}^u =S_{eff}^b$.

\section{Electron corrections to the phonon spectrum in $\text{ScV$_6$Sn$_6$}$}
\label{app:sec:ele_corr_svs}
In this \siSection{}, we describe the correction to the phonon propagator from the electron-phonon coupling. We comment that in the DFT calculations and DFT codes of the dynamical matrix, the correction of electrons has been automatically included - giving the renormalization of the phonon frequency from high to low temperatures. However, as it is a black-box, it is hard to analyze the connection between the imaginary phonon mode and the electron-phonon coupling from the final results of DFT. To get a physical and analytical understanding of the role of electron-phonon coupling, we directly calculate the electron corrections to the phonon propagator using our analytic \cref{eq:eff_force_const}. \textcolor{black}{ 
Before moving to the calculations and results, we first summarize our key observations and results here. 
\begin{itemize}
    \item At high temperatures where the electron corrections to the phonon spectrum are small, the lowest-energy phonon mode is extremely flat. This indicates the weak dispersions of lowest-energy phonon modes at low temperature comes from the electron correction. 
    \item The electron correction to the phonon modes mainly comes from the orbital-resolved charge fluctuations (charge susceptibility) of mirror-even electron orbital (as we introduced later in \cref{eq:ele_even_odd}). However, the charge susceptibility of this mirror-even electron orbital shows  some (weak) momentum dependency due to the weak Fermi surface nesting. At low temperatures, this weak momentum dependency leads to a flat imaginary phonon mode with leading instability at $H$ point, which has been observed experimentally.
    \item The phonon dispersions of the flat imaginary phonon modes at low temperature can also be understood analytically from a triangular lattice phonon model.
\end{itemize}
}

\label{sec:ele_corr_Sc}
\begin{figure}
    \centering
    \includegraphics[width=1.0\textwidth]{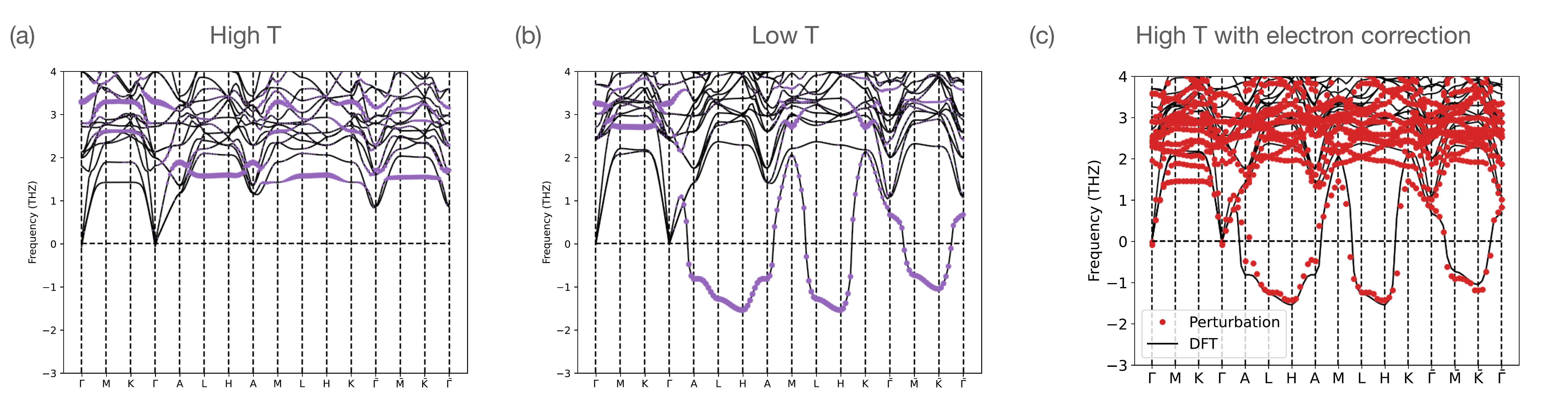}
    \caption{(a) Phonon spectrum from DFT calculation at high temperature $T=$1.2eV. (B) Phonon spectrum from DFT calculation at low-temperature (zero temperature). (C) Comparison between phonon spectrum derived by adding electron correction to the high-temperature phonon spectrum (red dots) and low-temperature phonon spectrum derived from DFT. In (a), (b), the purple dots mark the weight of the $z$-direction mirror-even vibration mode of the triangular Sn atoms ($\frac{1}{\sqrt{2}}(u_{Sn^T_1 z}(\mathbf{R}) - u_{Sn^T_2 z}(\mathbf{R}))$). }
    \label{fig:ele_corr_phonon}
\end{figure}
\textcolor{black}{We now prove the above statements via our calculations.} In practice, we take the force-constant matrix of phonons derived from DFT calculations at high temperatures  $\Phi_{i\mu,i'\nu}(\mathbf{q})$ as our no-interacting force-constant matrix. 
We point out that, at high temperatures, the correction from electron-phonon coupling is reduced and vanishes at infinite high temperatures. This can be seen from \cref{eq:corr_force_const} and the fact that $\lim_{T\leftarrow\infty} n_F(\epsilon)=1/2$. Clearly, the correction vanishes at infinite high-temperature $\lim_{T\leftarrow \infty} \Phi^{corr}_{\alpha\mu,\gamma\nu}(\mathbf{q}) = 0 $. Thus, we take the force-constant matrix derived from DFT at high temperatures as our ``non-interacting" force-constant matrix (the force-constant matrix without correction from electrons). 
 Next, we calculate the $\Phi^{corr}_{\alpha\mu,\gamma\nu}(\mathbf{q})$ via \cref{eq:corr_force_const} and analyze the structure of electron phonon corrections. The final spectrum of phonons with electron correction can be derived from the effective force-constant matrix (\cref{eq:eff_force_const}). It is remarkable that the simple analytic model we use can determine very accurately the phonon frequency renormalization and its soft modes.  

We first describe the setup of our electron-phonon model (\cref{eq:model_ele_ph}). The hopping matrix $t_{\alpha\gamma}(\mathbf{k})$ of electrons are derived from DFT calculations without spin-orbit coupling and with relaxed structure; the force constant matrix $\Phi_{\alpha\mu,\gamma\nu}(\mathbf{q})$ are calculated from DFT at high electron temperature $T=1.2$eV in a $2\times 2 \times 3$ supercell and with relaxed structure (we note that the electron temperature is not equivalent to the lattice temperature that is being  measured experimentally); the electron-phonon coupling $g_{\mathbf{k},\mathbf{q}}^{\alpha\gamma,i \mu}$ is derived from the Gaussian approximation introduced in \cref{eq:ele_ph_from_gaussian}. 
We also notice that Sn is the heaviest atom of the system and the low-energy modes including the imaginary mode are formed by the Sn atoms. 
We then only consider the corrections to the Sn atoms to simplify the calculations. 
In addition, as we showed in \cref{fig:gaus_approx}, the coupling between triangular Sn phonons (which is the origin of the imaginary mode) only weakly couples to the V $d$ electrons, so we also drop the corresponding electron-phonon coupling. 
In addition, we perform calculations by only considering the bands within the energy window $[E_f-20\text{meV}, E_f+20\text{meV}]$ ($E_f$ is the Fermi energy) at temperature $T=1$meV (note that, this is the electron temperature) with a $20\times 20 \times 20$ $k$ mesh.

In \cref{fig:ele_corr_phonon}, we show the high-temperature phonon spectrum (derived from DFT), low-temperature spectrum (derived from DFT) and the comparison between the high-temperature spectrum with electron correction and the low-temperature spectrum. We observe that our electron-correction calculation correctly and remarkably produces the imaginary mode with leading order instability at $H$ point. This indicates the imaginary phonon is indeed driven by the electron-phonon coupling and the corresponding electron correction to the phonon propagators.

\subsection{Origin of imaginary phonon}
\begin{figure}
    \centering
    \includegraphics[width=0.8\textwidth]{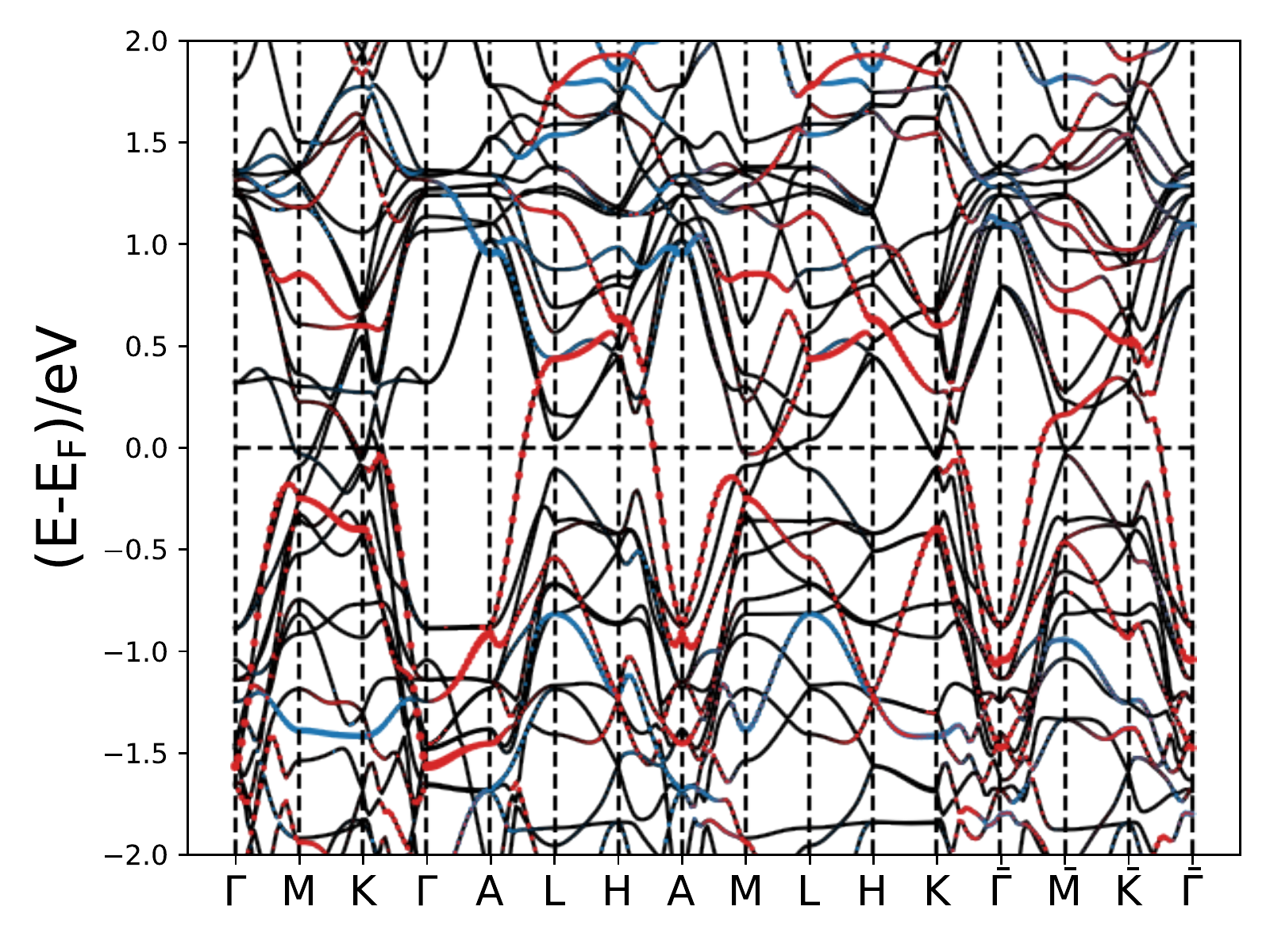}
    \caption{Electron band structure along the high-symmetry line. Red and blue dots label the orbital weights of mirror-even ($c_{\mathbf{R},e,\sigma}$) and mirror-odd ($c_{\mathbf{R},o,\sigma}$) orbitals respectively. We observe that only mirror-even orbital produce non-negligible contributions near Fermi energy. }
    \label{fig:disp_mirror_weight}
\end{figure}

We next perform a detailed analysis of the imaginary mode. As shown in \cref{fig:ele_corr_phonon} and also discussed in \cref{sec:imag_flat}, the imaginary mode is formed by the $z$-direction movement of triangular Sn atoms. Furthermore, as shown in \cref{eq:gaus_approx} and discussed in \cref{eq:ee_ele_ph}, the $z$-direction movement of triangular Sn atoms are strongly coupled to the electrons at the $p_z$ orbitals of triangular Sn. After a careful investigation of the electron band structures, we find the electrons at the $p_z$ orbitals of two triangular Sn form mirror-even/mirror-odd (bonding/anti-bonding) states due to the strong intra-unit-cell hopping ($t_{(Sn^T,p_z)}$ in \cref{eq:ee_hop}) between $p_z$ orbitals of two triangular Sn.
\textcolor{black}{This also relates to the fact that the inter-unit-cell hopping between triangular Sn $p_z$ orbitals along $z$-direction is weak because there is an additional Sc atoms between two triangular Sn atoms from two different cells. The inter-unit-cell hopping is an assisted hopping and thus weaker than the intra-unit-cell one which is a direct hopping.} 
The mirror-even/mirror-odd (with mirror plane $r_z = \frac{a_3}{2}$) states of two triangular Sn $p_z$ orbitals are defined as
\ba 
c_{\mathbf{R}, e,\sigma} = \frac{1}{\sqrt{2}} \bigg( c_{\mathbf{R}, (Sn_1^T,p_z), \sigma} - c_{\mathbf{R}, (Sn_2^T,p_z), \sigma}
\bigg) ,\quad c_{\mathbf{R}, o,\sigma} = \frac{1}{\sqrt{2}} \bigg( c_{\mathbf{R}, (Sn_1^T,p_z), \sigma} - c_{\mathbf{R}, (Sn_2^T,p_z), \sigma}
\bigg) 
\label{eq:ele_even_odd}
\ea 
where $c_{\mathbf{R},e,\sigma}$ and $c_{\mathbf{R},o,\sigma}$ are even and odd under mirror $z$ transformation respectively. By investigating the orbital weights of the two orbitals (\cref{fig:disp_mirror_weight}), we find only $c_{\mathbf{R},e,\sigma}$ has non-negligible weight near Fermi energy and we can drop the contribution of $ c_{\mathbf{R}, o,\sigma}$. 

Similarly, we could also introduce the mirror even $u_{ez}$ and mirror odd $u_{oz}$ basis for phonon fields
\ba 
&u_{e z}(\mathbf{R}) = \frac{1}{\sqrt{2}}(u_{Sn^T_1,z}(\mathbf{R})
-u_{Sn^T_2,z}(\mathbf{R})),\quad u_{o z}(\mathbf{R}) = \frac{1}{\sqrt{2}}(u_{Sn^T_1,z}(\mathbf{R})
+u_{Sn^T_2,z}(\mathbf{R}))
\label{eq:ph_even_odd}
\ea 
We comment that this even-odd basis is equivalent to the even-odd basis we introduced in \cref{eq:eom_1d}. The difference is, here, we are working with the field operators. In \cref{eq:eom_1d}, we are dealing with the eigenvectors of the dynamic matrix. \textcolor{black}{If we take the product between the eigenvectors ($\eta^{e,z}(\mathbf{q}),\eta^{o,z}(\mathbf{q})$) and the phonon operators (in the atomic basis), the resulting operators are just the $u_{ez},u_{oz}$ operator we considered here.}

We next investigate the corresponding electron-phonon couplings. The dominant electron-phonon couplings given in \cref{eq:ee_ele_ph} can be written as
\ba
&-t_{(Sn^T,p_z)}\gamma_{ ( Sn^T,p_z)}\sum_{\mathbf{R},\sigma}\Delta r^z \bigg(u_{Sn^T_1,z}(\mathbf{R}) - u_{Sn^T_2,z}(\mathbf{R}) )\bigg)\bigg(c_{\mathbf{R},e,\sigma}^\dag c_{\mathbf{R},e,\sigma} + c_{\mathbf{R},o,\sigma}^\dag c_{\mathbf{R},o,\sigma} \bigg)\nonumber\\ 
\approx &\tilde{g} \sum_{\mathbf{R},\sigma}u_{e,z}(\mathbf{R}) 
c_{\mathbf{R},e,\sigma}^\dag c_{\mathbf{R},e,\sigma}
\label{eq:even_odd_g_v0}
\ea 
where we drop the contribution from mirror-odd orbital ($c_{\mathbf{R},o}$) in the last line since $c_{\mathbf{R},o}$ is irrelevant to the low-energy physics \textcolor{black}{because its band is not close to the Fermi energy and does not contribute 
 the Fermi surface as shown in \cref{fig:disp_mirror_weight}}.
The corresponding electron-phonon coupling (in \cref{eq:even_odd_g_v0}) is $\tilde{g} = -\sqrt{2}t_{(Sn^T,p_z)} \Delta r^z$.

We next derive the corresponding correction to the phonon propagator. 
We let the electron-phonon coupling term be
\ba 
S_g = \int_0^\beta \tilde{g} \sum_{\mathbf{R},\sigma } u_{ez}(\mathbf{R},\tau) 
c_{\mathbf{R},e,\sigma}^\dag (\tau) c_{\mathbf{R},e,\sigma}(\tau) d\tau  
\ea 
The corresponding correction has been derived in \cref{eq:2nd_order} and is
\ba 
&-\frac{1}{2}\bigg[ \langle S_g ^2\rangle_{S_c} -\bigg(\langle S_g\rangle_{S_c}\bigg) ^2\bigg] \nonumber\\ 
=&-\int_0^\beta \int_0^\beta \frac{1}{2} \tilde{g}^2 \sum_{\mathbf{R},\mathbf{R}',\sigma,\sigma'} u_{ez}(\mathbf{R},\tau) u_{ez}(\mathbf{R}',\tau')  
\langle :c_{\mathbf{R},e,\sigma}^\dag (\tau) c_{\mathbf{R},e,\sigma}(\tau): :c_{\mathbf{R}',e,\sigma'}^\dag (\tau) c_{\mathbf{R}',e,\sigma'}(\tau):\rangle_{S_c} d\tau d\tau' 
\label{eq:correc_ee}
\ea 
We introduce the charge susceptibility of mirror even-orbital 
\ba 
\chi_e(\mathbf{R}-\mathbf{R}',\tau-\tau') = \frac{1}{N}\sum_{\sigma,\sigma'} \langle :c_{\mathbf{R},e,\sigma}^\dag (\tau) c_{\mathbf{R},e,\sigma}(\tau): :c_{\mathbf{R}',e,\sigma}^\dag (\tau) c_{\mathbf{R}',e,\sigma}(\tau):\rangle_{S_c}
\label{eq:chi_e_def}
\ea 
From \cref{eq:ele_even_odd} and \cref{eq:ph_even_odd} 
,  
the Wannier centers of both the mirror-even electrons $c_{\mathbf{R},e,\sigma}$ and the mirror-even phonons $u_{ez}(\mathbf{R})$ locate at $\mathbf{r}_e = \frac{1}{2}\bm{a}_3$. 
Correspondingly, we introduce the following Fourier transformation
\ba 
&u_{ez}(\mathbf{q},i\Omega_n) = \frac{1}{\sqrt{N}}\int_0^\beta \sum_{\mathbf{k}} e^{-i\mathbf{q}\cdot (\mathbf{R}+\mathbf{r}_e)+i\Omega_n\tau } u_{ez}(\mathbf{R},\tau)d\tau 
,\quad \nonumber\\ 
&c_{\mathbf{k},e,\sigma}(i\omega_n)\frac{1}{\sqrt{N}}\int_0^\beta \sum_{\mathbf{k}} e^{-i\mathbf{k}\cdot (\mathbf{R}+\mathbf{r}_e)+i\Omega_n\tau }  c_{\mathbf{R},e,\sigma}(\tau)d\tau 
\label{eq:ee_ft}
\ea 
Similarly, we also introduce Fourier transformation of the susceptibility
\ba 
\chi_e(\mathbf{q},i\Omega_n) = \sum_{\mathbf{R}}\int_0^\beta  \chi_{e}(\mathbf{R},\tau)e^{i\Omega_n\tau-i\mathbf{q}\cdot\mathbf{R}} d\tau 
\label{eq:chi_e_ft}
\ea
Combining \cref{eq:correc_ee}, \cref{eq:chi_e_ft}, \cref{eq:ee_ft} and dropping the frequency($i\Omega_n$)-dependency of $\chi_e$, we find the following correction in the action of phonon fields 
\ba 
-\frac{1}{2} \int_0^\beta \frac{1}{\beta}\sum_{\mathbf{q},i\Omega_n}u_{ez}(-\mathbf{q},-i\Omega_n)u_{ez}(\mathbf{q},i\Omega_n) \tilde{g}^2 \chi_e(\mathbf{q},i\Omega_n=0)
\label{eq:corr_even_odd}
\ea 
Correspondingly, the correction to the force constant matrix is 
\ba 
\Phi^{corr}_{ez,ez}(\mathbf{q}) = -\tilde{g}^2 \chi_e(\mathbf{q},i\Omega_n=0)
\ea 
We now transform back to the original Sn basis. Using \cref{eq:ee_ele_ph}, we find 
\ba
u_{ez}(\mathbf{q}) =& \frac{1}{\sqrt{N}}\sum_\mathbf{R} e^{-i\mathbf{q}\cdot(\mathbf{R}+\mathbf{r}_e)}u_{ez}(\mathbf{R},\tau) \nonumber\\ 
= &\frac{1}{\sqrt{2N}}\sum_\mathbf{R} e^{-i\mathbf{q}\cdot(\mathbf{R}+\mathbf{r}_e)}\sum_{\mathbf{q}'} \bigg( u_{Sn_1^T,z}(\mathbf{q}') e^{i\mathbf{q}'\cdot(\mathbf{R}+\mathbf{r}_{Sn_1^T})}-
u_{Sn_2^T,z}(\mathbf{q}') e^{i\mathbf{q}'\cdot(\mathbf{R}+\mathbf{r}_{Sn_2^T})}-\bigg) \nonumber\\ 
=&\frac{1}{\sqrt{2}}\bigg( u_{Sn_1^T,z}(\mathbf{q})e^{i\mathbf{q}\cdot(\Delta \mathbf{r}_1)} -
u_{Sn_2^T,z}(\mathbf{q})e^{i\mathbf{q}\cdot(\Delta \mathbf{r}_2)}\bigg) 
\ea
where $\Delta {\mathbf{r}_1} = \mathbf{r}_{Sn_1^T} -\mathbf{r}_e,\Delta {\mathbf{r}_2} = \mathbf{r}_{Sn_2^T} -\mathbf{r}_e $ denote the deviation between two Sn atoms and the Wannier center of the mirror even \textcolor{black}{phonon} mode
Then the corresponding correction to the force constant matrix in the original basis \textcolor{black}{(triangular Sn basis)}
becomes 
\ba 
&\Phi^{corr}_{( Sn^T_1,z), (Sn^T_1,z)}(\mathbf{q})
=\Phi^{corr}_{( Sn^T_2,z), (Sn^T_2,z)}(\mathbf{q}) \nonumber\\ 
=&-\Phi^{corr}_{( Sn^T_1,z), (Sn^T_2,z)}(\mathbf{q}) e^{-i\mathbf{q}\cdot(-\Delta \mathbf{r}_1 +\Delta \mathbf{r}_2 )} 
=-\Phi^{corr}_{( Sn^T_2,z), (Sn^T_1,z)}(\mathbf{q})^{i\mathbf{q}\cdot(-\Delta \mathbf{r}_1 +\Delta \mathbf{r}_2 )}   = -\frac{1}{2}\tilde{g}^2 \chi_e(\mathbf{q},i\Omega_n=0)
\label{eq:corr_chie}
\ea 
\textcolor{black}{ 
where the corresponding term in the action is 
\ba 
\int_0^\beta \frac{1}{2} \sum_{\mathbf{q}}\sum_{i,j\in \{1,2\}}\Phi^{corr}_{ (Sn_i^T,z), (Sn_j^T,z)}(\mathbf{q}) u_{Sn_i^T,z}(-\mathbf{q}) u_{Sn_j^T,z}(-\mathbf{q}) d\tau 
\ea 
}

We analyze the effect of the correction term $\Phi^{corr}_{( Sn^T_i,z), (Sn^T_j,z)}(\mathbf{q})$ in \cref{eq:corr_chie}. In 
\cref{fig:high_T_chi_e_corr} (c), we plot the phonon spectrum derived by combining the high-temperature phonon spectrum ($T=1.2$eV) with the electron correction given in \cref{eq:corr_chie} (c). \textcolor{black}{This is different from \cref{fig:ele_corr_phonon} (c), in the sense that, only the correction induced by the charge fluctuations $(\chi_e)$ of $c_{\mathbf{R},e,\sigma}$ orbitals are included in \cref{eq:corr_chie} (c).}
We observe that the correction induced by $\chi_e$ is already strong enough to generate the imaginary phonon, which indicates the charge fluctuation of $c_{\mathbf{R},e,\sigma}$ is the main driven force of the imaginary phonon.

\subsection{Triangular Sn electron susceptibility}\label{sec:triangular_Sn_suscept}

\begin{figure}
    \centering
    \includegraphics[width=1.0\textwidth]{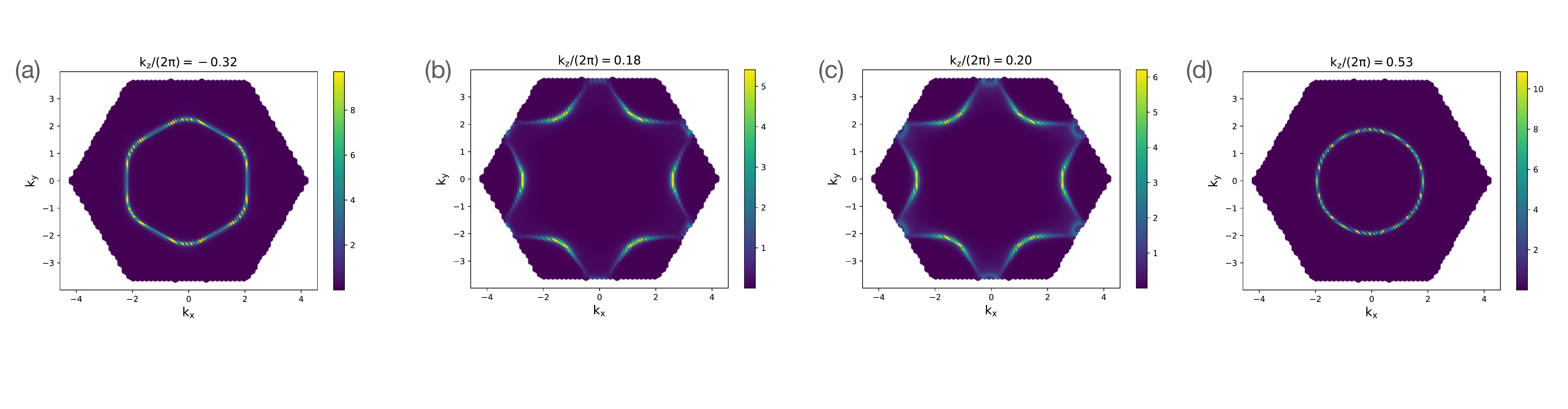}
    \caption{Density of states (at Fermi energy) of mirror-even orbitals $c_{\mathbf{k},e,\sigma}$ at different $k_z$ plane. We observe the non-negligible contributions of mirror-even orbitals to the Fermi surface. }
    \label{fig:fs_mirror_even}
\end{figure}

In this \siSection{}, we provide a detailed analysis of the behaviors of susceptibility $\chi_e(\mathbf{q},i\Omega_n=0)$. \textcolor{black}{Note that $\chi_e(\mathbf{q},i\Omega_n=0)$ is the orbital-resolved susceptibility that describes the charge susceptibility of mirror-even electron operators $c_{\mathbf{R},ez}$(\cref{eq:ele_even_odd}).} 
In \cref{fig:high_T_chi_e_corr} (a), we show the momentum dependency of $\chi_e(\mathbf{q},i\Omega_n=0)$. 
We observe there are peaks developed near both $H$ (the exact location of the peak is actually $(0.4,0.4,0.5)$, but the difference of intensities between $H$ point and the actual position of the peak is less than $1\%$) and $\bar{K}$ points. The origin of the peak could be understood via the weak nesting feature of the Fermi surfaces \textcolor{black}{that are formed by the $c_{\mathbf{R},e,\sigma}$ electrons.} 
\textcolor{black}{In \cref{fig:fs_mirror_even}, we plot the density of states of mirror-even electron orbitals $c_{\mathbf{R},e,\sigma}$ at four different $k_z$ planes, which produce the weak nesting of Fermi surface. Panel $a$ and panel $b$ of \cref{fig:fs_mirror_even} show two Fermi surfaces from $c_{\mathbf{R},e,\sigma}$ at $k_z/(2\pi)=-0.28$ and $k_z/(2\pi)=0.53$ respectively, which gives a nesting vector near $H$ point (as also shown in \cref{fig:fermi_surface} (a) where we plot two Fermi surfaces in the same figure and shows the nesting vector). Panel $c$ and panel $d$ of \cref{fig:fs_mirror_even} show two Fermi surfaces from $c_{\mathbf{R},e,\sigma}$ at $k_z/(2\pi)=0.20$ and $k_z/(2\pi)=0.53$ respectively, which gives a nesting vector near $\tilde{{{K}}}$ point (as also shown in \cref{fig:fermi_surface} (c) where we plot two Fermi surfaces in the same figure and shows the nesting vector). However, since the nesting is weak, we only observe a weak momentum dependency in the charge susceptibility ($\chi_{e}$) of $c_{\mathbf{R},e,\sigma}$ electrons with two peaks near $H$ and $\tilde{{{K}}}$ point as shown in \cref{fig:high_T_chi_e_corr}(a)}

Here, we also mention that $\chi_e$ has peaks near both $H$ and $\bar{K}$ points with similar strength. However, the lowest phonon-bands near $\bar{K}$ have less $u_{ez}$ weights as shown in \cref{fig:high_T_chi_e_corr}. Since the electron corrections are acting on the $u_{ez}$ phonons, the reduced weight leads to a reduced correction to the lowest phonon modes near $\bar{K}$. Consequently, the leading order instability (of the phonons) still happens near $H$. In addition, the peak of susceptibility near $H$ locates slightly away from $H$ points. However, since the deviation is very small and also the intensity at $H$ points is very close to the intensity at the peak, in reality, we do not expect this small deviation will lead to any instability at an incommensurate wave vector near $H$.

\textcolor{black}{Finally, we provide an approximate expression of $\chi_e(\mathbf{q},i\Omega_n=0)$. Since $\chi_e(\mathbf{q},i\Omega_n=0)$ only has a weak momentum dependency due to the weak Fermi surface nesting and its evolution as a function of momentum is similar to the dispersion of the tight-binding model on a triangular lattice. Moreover, the Wannier center of $c_{\mathbf{R},e,\sigma}$ electrons is at $\mathbf{r}= \frac{1}{2}\bm{a}_3$ for each unit cell and forms a triangular lattice. This motivates us to perform a Fourier transformation of $\chi_e$ and analyze its real space behavior
\ba 
\chi_e(\mathbf{R},i\Omega_n=0) =\frac{1}{N} \sum_{\mathbf{q}} \chi_e(\mathbf{q},i\Omega_n=0)e^{i\mathbf{q}\cdot\mathbf{R}}
\ea 
}
Indeed we find its real space behavior can be described by a strong on-site term and two relatively weak nearest-neighbor terms with one in-pane and the other one out-of-plane. More explicitly, we approximately have
\ba 
\chi_e(\mathbf{R},i\Omega_n=0) \approx \chi^{on\text{-}site} \delta_{\mathbf{R},\bm{0}}+ \sum_{i=1,...,6}\chi^{xy}\delta_{\mathbf{R},\mathbf{R}_i^{xy}} +\sum_{i=1,2}\chi^{z}\delta_{\mathbf{R},\mathbf{R}_i^{z}}
\label{eq:chi_e_real}
\ea 
where $\mathbf{R}_{i}^{xy}$ and $\mathbf{R}_i^z$ denotes in-plane and out-of-plane nearest neighbors respectively.  
The strong on-site term gives $\chi^{on\text{-}site}=0.289$. Two relatively weak nearest-neighbor terms are $\chi^{xy}=-0.009$ (in-plane) and $\chi^z=-0.016$ (out of plane). We mention that due to the weak momentum dependency, the susceptibility $\chi_e$ in the real space only has a sizeable value within a short distance (nearest-neighbor). \textcolor{black}{In the next \siSection{}, we will use the approximate behaviors of $\chi_e$ (\cref{eq:chi_e_real}) to build a simple model of the imaginary phonon.}

\subsection{An effective phonon model}
In order to obtain a better, analytic,  understanding of the phonon modes and electron corrections, we combine the effective 1D mode that we derived near \cref{eq:eom_1d_new} with the current electron-phonon corrections induced by the mirror-even electrons $c_{\mathbf{R},e,\sigma}$ (\cref{eq:corr_even_odd}).

We assume the high-temperature phonon spectrum (or non-interacting phonon spectrum) can be described by an effective 1D model. 
It is also worth mentioning that, as shown in \cref{fig:ele_corr_phonon} (a), the high-temperature phonon spectrum of the mirror even-modes is very flat, which further justifies the validity of the 1D model. The high-temperature phonon spectrum can be modeled by \cref{eq:eom_1d_new} with $d_1=0.67\text{THz}^2, d_2= 3.22\text{THz}^2$. We can see both $d_1,d_2$ are positive which indicates the absence of imaginary phonon (from \cref{eq:eig_1d_mode}).

\begin{figure}
    \centering
    \includegraphics[width=1.0\textwidth]{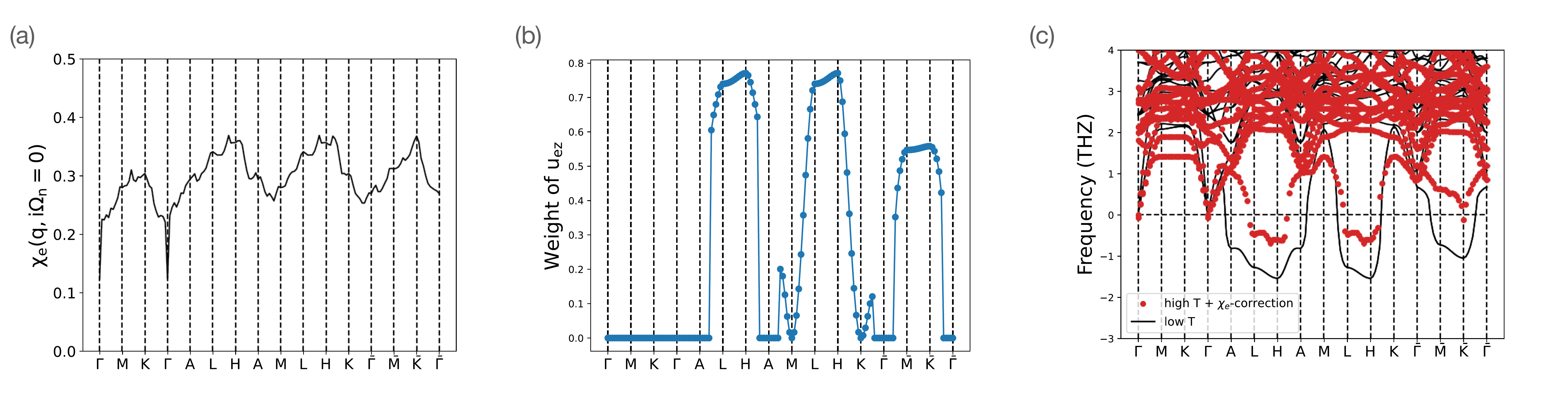}
    \caption{(a) \textcolor{black}{Orbital-resolved static susceptibility $\chi_e(\mathbf{q},i\Omega_n=0)$ (\cref{eq:chi_e_def})
    of $c_{\mathbf{R},e,\sigma}$ electrons. }
(b) Weight of the mirror-even phonon mode $u_{ez}$ of the lowest phonon modes in the high-temperature phonon spectrum. (c) Comparison between high-temperature phonon spectrum with $\chi_e$ corrections and low-temperature phonon spectrum (derived from DFT). We can observe significant renormalization introduced by $\chi_e$ which already introduces an imaginary phonon near $H$ and at $\bar{K}$. }
    \label{fig:high_T_chi_e_corr}
\end{figure}

We next include the correction generated by the charge fluctuations of mirror-even orbitals $c_{\mathbf{R},e,\sigma}$. 
Combining \cref{eq:eom_1d_new}, \cref{eq:corr_even_odd}, and \cref{eq:chi_e_real}, we derive the following effective dynamic matrix that describes the behaviors of the imaginary mode 
\ba 
D^{eff,eo}(\mathbf{q}) = \tilde{D}_{1D}(\mathbf{q}) +\begin{bmatrix}
    d_3(\mathbf{q})  &0 \\ 
    0 & 0 
\end{bmatrix}
= +\begin{bmatrix}
     2d_1+d_2+d_2\cos(q_z) + d_3(\mathbf{q})  & -i d_2\sin(q_z) \\ 
    id_2\sin(q_z)  &  d_2 -d_2 \cos(q_z) 
\end{bmatrix}
\ea 
where $\tilde{D}_{1D}(\mathbf{q})$ denotes the high-temperature 1D phonon model without electron corrections (\cref{eq:eom_1d_new}). $d_3(\mathbf{q})=-\tilde{g}^2 \chi_e(\mathbf{q},i\Omega_n=0)/M_{Sn}$ (from \cref{fig:high_T_chi_e_corr}) describes the correction induced by $c_{\mathbf{R},e,\sigma}$ electrons with \ba 
d_3(\mathbf{q}) = -\frac{ \tilde{g}^2 }{M_{Sn}}\chi_e(\mathbf{q},i\Omega_n=0) = -\frac{ \tilde{g}^2 }{M_{Sn}} \bigg( \chi^{on\text{-}site} +2 \chi^{xy}\bigg(\cos(q_1) + \cos(q_2) + \cos(q_1+q_2) \bigg)+ 2\chi^{z}\cos(q_z)\bigg) 
\label{eq:1d_correct_model_d3}
\ea 
with $q_1 = \mathbf{q}\cdot \bm{a}_1,q_2 = \mathbf{q}\cdot \bm{a}_2$ and $\tilde{g}= 2.23\AA^{-1}u^{-1}eV$. 
In addition, $D^{eff,eo}(\mathbf{q})$ is defined in the even and odd basis (\cref{eq:ele_even_odd}) with the first row/column corresponding to $u_{ez}$ and the second row/column corresponding to $u_{oz}$. It 

A direct diagonalization of $D^{eff,eo}$ leads to the following phonon spectrum 
\ba 
\omega_{1/2,\mathbf{q}}^2 =d_1+ d_2+\frac{d_3(\mathbf{q})}{2}  \pm 
\sqrt{d_2^2\sin(q_z)^2 +\bigg( d_1 +d_2\cos(q_z) + \frac{d_3(\mathbf{q})}{2}\bigg)^2 }
\label{eq:1d_correct_model_w12}
\ea 
Combining \cref{eq:1d_correct_model_d3} and \cref{eq:1d_correct_model_w12}, we can define 
\ba 
&\tilde{d}_1 = d_1 +\frac{-\tilde{g}^2}{2M_{Sn}} \chi^{on\text{-}site} , \nonumber\\ 
& d_{xyz}(\mathbf{q}) =
d_{xy} 
\bigg(\cos(q_1) + \cos(q_2) + \cos(q_1+q_2) \bigg) +d_z \cos(q_z),\quad d_{xy} = -\frac{\tilde{g}^2}{M_{Sn}}\chi_{xy},\quad d_z =- \frac{\tilde{g}^2}{M_{Sn}} \chi_z 
\label{eq:d1_dxyz}
\ea 
where $\tilde{d}_1 =-0.807\text{THz}^2$,
$d_{xy}=-0.081\text{THz}^2, d_z=-0.046\text{THz}^2$. 
Then the phonon spectrum in \cref{eq:1d_correct_model_w12} can be written as 
\ba 
\omega_{1/2,\mathbf{q}}^2 = \tilde{d}_1 +d_2 + \frac{ d_{xyz}(\mathbf{q})}{2} \pm \sqrt{ d_2\sin(q_z)^2 + (\tilde{d}_1 + d_2\cos(q_z) +\frac{d_{xyz}(\mathbf{q})}{2})^2}
\ea 
Since $d_{xyz}$ (\cref{eq:d1_dxyz}) refers to the weak momentum dependency term generated by the susceptibility and is small $\chi_e$, we can the spectrum expand in powers of $d_{xyz}$
\ba 
\omega_{1/2,\mathbf{q}}^2 \approx \bigg( \tilde{d}_1+ {d}_2 \pm 
\sqrt{{d}_2\sin(q_z)^2 +\bigg( \tilde{d}_1 +{d}_2\cos(q_z) \bigg)^2 }\bigg) +\frac{1}{2}\bigg(1 \pm \text{sgn}(\tilde{d}_1+d_2 \cos(q_z) )\bigg)d_{xyz}(\mathbf{q}) 
\label{eq:1d_correct_model_spec} 
\ea 
where $\text{sgn}(x)$ is the sign function. We observe that the first term in \cref{eq:1d_correct_model_spec} is just the spectrum of an effective 1D model (\cref{eq:eom_1d_new}) with a negative intra-unit-cell coupling $\tilde{d}_1 =-0.807\text{THz}^2$ and a positive inter-unit-cell coupling $d_2=3.22\text{THz}^2$. The negative $\tilde{d}_1$ comes from the electron corrections \cref{eq:d1_dxyz}. As we discussed near \cref{eq:eig_1d_mode}, the spectrum of the effective 1D phonon model has a minimum at $q_z=\pi$ when $\tilde{d}_1$ is negative. We now evaluate the phonon spectrum at $q_z=\pi$ plane
\ba 
\omega_{1, \mathbf{q}}^2\bigg|_{q_z=\pi} &= 2\tilde{d}_1 +d_{xyz}(\mathbf{q}) \nonumber\\ 
\omega_{2 \mathbf{q}}^2\bigg|_{q_z=\pi} &= 2d_2 
\ea 
where we find $\omega_{1,\mathbf{q}}^2|_{q_z=\pi}$ \textcolor{black}{gives the lowest eigenvalue} and is negative due to the negative $\tilde{d}_1$.
\textcolor{black}{$d_{xyz}$ (\cref{eq:d1_dxyz}) describes the weak momentum dependency generated by $\chi_{xy},\chi_z$ which is much weaker than $\tilde{d}_1$.} $\chi_{xy}$ (or $d_{xy}$) produces a weak in-plane dispersion of $\omega_{1,\mathbf{q}}^2$ mode which gives a minimum of $\omega_{1,\mathbf{q}}^2$ at $H$. \textcolor{black}{$\chi_z$ (or $d_z$) normalizes the $z$-direction dispersion of $\omega_{1,\mathbf{q}}^2$. However, since $|d_z| \sim 0.06|\tilde{d}_z|$, $d_z$ will not change the dispersion qualitatively.}

\section{Electron susceptibility} 
In this \siSection{}, we analyze the electron susceptibility of ScV$_6$Sn$_6$. In summary, we do not observe any singularity in both the total susceptibility and orbital resolved susceptibility. This indicates the CDW transition and also the collapsing of the phonon is not driven by a strong Fermi-surface nesting or the scattering between Van-Hove singularities where in both cases a strong enhancement of charge susceptibility is expected. 

\subsection{Calculation of susceptibility}
We first introduce the calculation susceptibility (without interaction/phonon normalization) of the system. The susceptibility of the non-interacting system is defined as
\ba 
\chi_0^{\alpha\gamma,\mu\nu }(\mathbf{R}-\mathbf{R}',\tau) = \sum_{\sigma\sigma'}\langle :c_{\mathbf{R},\alpha\sigma }^\dag(\tau) c_{\mathbf{R},\gamma \sigma}(\tau): :c_{\mathbf{R}',\mu \sigma'}^\dag(0) c_{\mathbf{R}',\nu \sigma'}(0): \rangle_{S_c} 
\ea 
where $S_c$ and the normal ordering have been defined in \cref{eq:def_S} and \cref{eq:normal_order} respectively. We have summed over spin indices since we are not interested in the spin order. 
We then perform Fourier transformation
\ba 
&\chi_0^{\alpha \gamma,\mu\nu}(\mathbf{q},i\Omega_n)\nonumber\\ 
= &\frac{1}{N}\sum_{\mathbf{R},\mathbf{R}'}\int_0^\beta \chi_0^{\alpha \gamma ,\mu\nu}(\mathbf{R}-\mathbf{R}',\tau)e^{-i\mathbf{q} \cdot (\mathbf{R}-\mathbf{R}') + i\Omega_n\tau }d\tau \nonumber \\ 
=& \frac{1}{N^3}\int_0^\beta \sum_{\mathbf{R},\mathbf{R}'}\sum_{\mathbf{k},\mathbf{k}',\mathbf{k}_1,\mathbf{k}_2, \sigma\sigma'} \langle :c_{\mathbf{k},\alpha\sigma}^\dag(\tau) c_{\mathbf{k}_1,\gamma\sigma }(\tau): :c_{\mathbf{k}_2,\mu \sigma'}^\dag(0)c_{\mathbf{k}',\nu\sigma'}(0):\rangle_{S_c} \nonumber \\ 
&
e^{-i\mathbf{q}\cdot(\mathbf{R}-\mathbf{R}') - i\mathbf{k} \cdot (\mathbf{R}+\mathbf{r}_\alpha) +i \mathbf{k}_1\cdot(\mathbf{R}+\mathbf{r}_\gamma) -i\mathbf{k}_2\cdot(\mathbf{R}'+\mathbf{r}_\mu) +i\mathbf{k}'\cdot (\mathbf{R}'+\mathbf{r}_\nu) }e^{ i\Omega_n\tau }
d\tau  \nonumber \\ 
=& \frac{1}{N^3}\int_0^\beta \sum_{\mathbf{k},\mathbf{k}',\mathbf{k}_1,\mathbf{k}_2, \sigma\sigma'} \langle  :c_{\mathbf{k},\alpha\sigma}^\dag(\tau) c_{\mathbf{k}_1,\gamma\sigma }(\tau): :c_{\mathbf{k}_2,\mu \sigma'}^\dag(0)c_{\mathbf{k}',\nu\sigma'}(0):\rangle_{S_c} \nonumber \\ 
&e^{-i\mathbf{k}\cdot r_\alpha + i(\mathbf{k}+\mathbf{q})\cdot r_\gamma -i(\mathbf{k}'+\mathbf{q})\cdot \mathbf{r}_\mu +i\mathbf{k}'\cdot r_\nu}\delta_{\mathbf{q}+\mathbf{k},\mathbf{k}_1}\delta_{\mathbf{q}+\mathbf{k}',\mathbf{k}_2}e^{ i\Omega_n\tau }
d\tau  \nonumber \\ 
=&\frac{1}{N}\int_0^\beta \sum_{\mathbf{k},\mathbf{k}', \sigma\sigma'} \langle:c_{\mathbf{k},\alpha\sigma}^\dag(\tau) c_{\mathbf{k}+\mathbf{q},\gamma\sigma }(\tau): :c_{\mathbf{k}'+\mathbf{q},\mu \sigma'}^\dag(0)c_{\mathbf{k}',\nu\sigma'}(0):\rangle_{S_c} 
e^{-i\mathbf{k}\cdot r_\alpha + i(\mathbf{k}+\mathbf{q})\cdot r_\gamma -i(\mathbf{k}'+\mathbf{q})\cdot \mathbf{r}_\mu +i\mathbf{k}'\cdot r_\nu}e^{ i\Omega_n\tau }d\tau 
\ea 
In practice, $\mathbf{k},\mathbf{k}',\mathbf{q}$ are taken from the first Brillouin zone. However, $\mathbf{k}+\mathbf{q}, \mathbf{k}'+\mathbf{q}$ do not necessarily belong to the first Brillouin zone. We then introducing a shifting vector $\mathbf{G}_{\mathbf{k}+\mathbf{q}} \in \mathbb{Z}\bm{b}_1+\mathbb{Z}\bm{b}_2+\mathbb{Z}\bm{b}_3$ such that $\mathbf{k}+\mathbf{q} + \mathbf{G}_{\mathbf{k}+\mathbf{q}}$ belongs to the first Brillouin zone. However, we note that the shifting of the momentum introduces an additional phase factor to the electron operators
\ba 
c_{\mathbf{k}+\mathbf{q},\alpha \sigma} = c_{\mathbf{k}+\mathbf{q}+\mathbf{G}_{\mathbf{k}+\mathbf{q}}, \alpha \sigma}e^{i\mathbf{G}_{\mathbf{k}+\mathbf{q}} \cdot \mathbf{r}_\alpha }
\ea 
Then we find 
\ba 
&\chi_0^{\alpha \gamma,\mu\nu}(\mathbf{q},i\Omega_n)\nonumber\\ 
=&\frac{1}{N}\int_0^\beta \sum_{\mathbf{k},\mathbf{k}', \sigma\sigma'} \langle :c_{\mathbf{k},\alpha\sigma}^\dag(\tau) c_{\mathbf{k}+\mathbf{q}+\mathbf{G}_{\mathbf{k}+\mathbf{q}},\gamma\sigma }(\tau): :c_{\mathbf{k}'+\mathbf{q}+\mathbf{G}_{\mathbf{k}'+\mathbf{q}},\mu \sigma'}^\dag(0)c_{\mathbf{k}',\nu\sigma'}(0):\rangle_{S_c} \nonumber\\ 
&
e^{-i\mathbf{k}\cdot \mathbf{r}_\alpha + i(\mathbf{k}+\mathbf{q})\cdot \mathbf{r}_\gamma -i(\mathbf{k}'+\mathbf{q})\cdot \mathbf{r}_\mu +i\mathbf{k}'\cdot \mathbf{r}_\nu +i\mathbf{G}_{\mathbf{k}+\mathbf{q}}\cdot \mathbf{r}_\gamma - i\mathbf{G}_{\mathbf{k}'+\mathbf{q}}\cdot \mathbf{r}_\mu }e^{ i\Omega_n\tau }d\tau 
\ea 

We then transform to the band basis using \cref{eq:band_basis_gam} and also shown below
\ba 
\gamma_{\mathbf{k},n\sigma} =\sum_{\alpha}U_{\alpha n}^{c,*}(\mathbf{k}) c_{\mathbf{k},\alpha\sigma}
\ea
The susceptibility can then be written as 
\ba 
&\chi_0^{\alpha \gamma,\mu\nu}(\mathbf{q},i\Omega_n)\nonumber \\ 
=&\frac{1}{N}\int_0^\beta \sum_{\mathbf{k},\mathbf{k}', nmn'm',\sigma\sigma'}U_{\mathbf{k},\alpha  n}^{c,*}U^{c}_{\mathbf{k}+\mathbf{q}+\mathbf{G}_{\mathbf{k}+\mathbf{q}},\gamma m}U^{c,*}_{\mathbf{k}'+\mathbf{q}+\mathbf{G}_{\mathbf{k}'+\mathbf{q}},\mu m'}U_{\mathbf{k}',\nu n'}^{c} \nonumber\\ 
&\langle :\gamma_{\mathbf{k},n\sigma}^\dag(\tau) \gamma_{\mathbf{k}+\mathbf{q}+\mathbf{G}_{\mathbf{k}+\mathbf{q}},m\sigma }(\tau): :\gamma_{\mathbf{k}'+\mathbf{q}+\mathbf{G}_{\mathbf{k}'+\mathbf{q}},m' \sigma'}^\dag(0)\gamma_{\mathbf{k}',n'\sigma'}(0):\rangle_{S_c} \nonumber\\
&e^{-i\mathbf{k}\cdot \mathbf{r}_\alpha + i(\mathbf{k}+\mathbf{q})\cdot \mathbf{r}_\gamma -i(\mathbf{k}'+\mathbf{q})\cdot \mathbf{r}_\mu +i\mathbf{k}'\cdot \mathbf{r}_\nu
+i\mathbf{G}_{\mathbf{k}+\mathbf{q}}\cdot \mathbf{r}_\gamma - i\mathbf{G}_{\mathbf{k}'+\mathbf{q}}\cdot \mathbf{r}_\mu
}e^{ i\Omega_n\tau }
d\tau 
\ea 
Using \cref{eq:sus_band} and \cref{eq:sus_non_int} where we have derived the susceptibility of the $\gamma$ electron, we find 
\ba 
&\chi_0^{\alpha \gamma,\mu\nu}(\mathbf{q},i\Omega_n)\nonumber \\
=&\frac{1}{N} \sum_{\mathbf{k},nm,\sigma}U_{\mathbf{k},\alpha  n}^{c,*}U^{c}_{\mathbf{k}+\mathbf{q}+\mathbf{G}_{\mathbf{k}+\mathbf{q}},\gamma m}U^{c,*}_{\mathbf{k}+\mathbf{q}+\mathbf{G}_{\mathbf{k}+\mathbf{q}},\mu m}U_{\mathbf{k},\nu n}^{c} 
\frac{n_F(\epsilon_{\mathbf{k}+\mathbf{q}+\mathbf{G}_{\mathbf{k}+\mathbf{q}},m\sigma}) - n_F(\epsilon_{\mathbf{k},n\sigma} ) }{i\Omega_n + \epsilon_{\mathbf{k},n\sigma} - \epsilon_{\mathbf{k}+\mathbf{q}+\mathbf{G}_{\mathbf{k}+\mathbf{q}},m\sigma}}\nonumber \\ 
&e^{-i\mathbf{k}\cdot \mathbf{r}_\alpha + i(\mathbf{k}+\mathbf{q})\cdot \mathbf{r}_\gamma -i(\mathbf{k}+\mathbf{q})\cdot \mathbf{r}_\mu +i\mathbf{k}\cdot \mathbf{r}_\nu  } \nonumber\\ 
=&\frac{1}{N} \sum_{\mathbf{k},nm,\sigma}U_{\mathbf{k},\alpha  n}^{c,*}U^{c}_{\mathbf{k}+\mathbf{q},\gamma m}U^{c,*}_{\mathbf{k}+\mathbf{q},\mu m}U_{\mathbf{k},\nu n}^{c} 
\frac{n_F(\epsilon_{\mathbf{k}+\mathbf{q},m\sigma}) - n_F(\epsilon_{\mathbf{k},n\sigma} ) }{i\Omega_n + \epsilon_{\mathbf{k},n\sigma} - \epsilon_{\mathbf{k}+\mathbf{q},m\sigma}}\nonumber \\ 
&e^{-i\mathbf{k}\cdot \mathbf{r}_\alpha + i(\mathbf{k}+\mathbf{q})\cdot \mathbf{r}_\gamma -i(\mathbf{k}+\mathbf{q})\cdot \mathbf{r}_\mu +i\mathbf{k}\cdot \mathbf{r}_\nu  } 
\ea

\subsection{Results of the susceptibility calculation}
\label{sec:bare_sus}
We now illustrate the result of the susceptibility calculation of ScV$_6$Sn$_6$ for both the experimental and relaxed pristine structures. 
Besides the susceptibility, we will also provide the results of ''nesting function" which characterizes the nesting of the Fermi surface
\begin{equation}
    \xi(\bm{q})=
\frac{1}{N} \sum_{m,n,\bm{k}}\delta(\epsilon_{\mathbf{k},n\sigma})\delta(\epsilon_{\mathbf{k}+\mathbf{q}, m\sigma}),
\end{equation}
In addition, we define the total susceptibility of the system
\ba 
\chi_0(\mathbf{q}) = \frac{1}{N} \sum_{n,m, \mathbf{k}} \frac{n_F(\epsilon_{\mathbf{k}+\mathbf{q},m\sigma}) - n_F(\epsilon_{\mathbf{k},n\sigma} ) }{\epsilon_{\mathbf{k},n\sigma} - \epsilon_{\mathbf{k}+\mathbf{q}, m\sigma}},
\ea 

However, we comment that the total susceptibility calculation loses the information on the orbital character of the bands, which turns out to be important. Thus, we also consider the charge susceptibility between a pair of orbitals
\begin{equation}
    \chi_0^{\alpha,\mu}(\bm{q}) = \chi_0^{\alpha \alpha,\mu\mu}(\bm{q})
    = \sum_{\sigma\sigma'}\langle T_\tau :c_{\mathbf{R},\alpha\sigma }^\dag(\tau) c_{\mathbf{R},\alpha \sigma}(\tau): :c_{\mathbf{R}',\mu \sigma'}^\dag(0) c_{\mathbf{R}',\mu \sigma'}(0): \rangle_{S_c} 
\end{equation}

\begin{figure}
    \centering
    \includegraphics[width=1\textwidth]{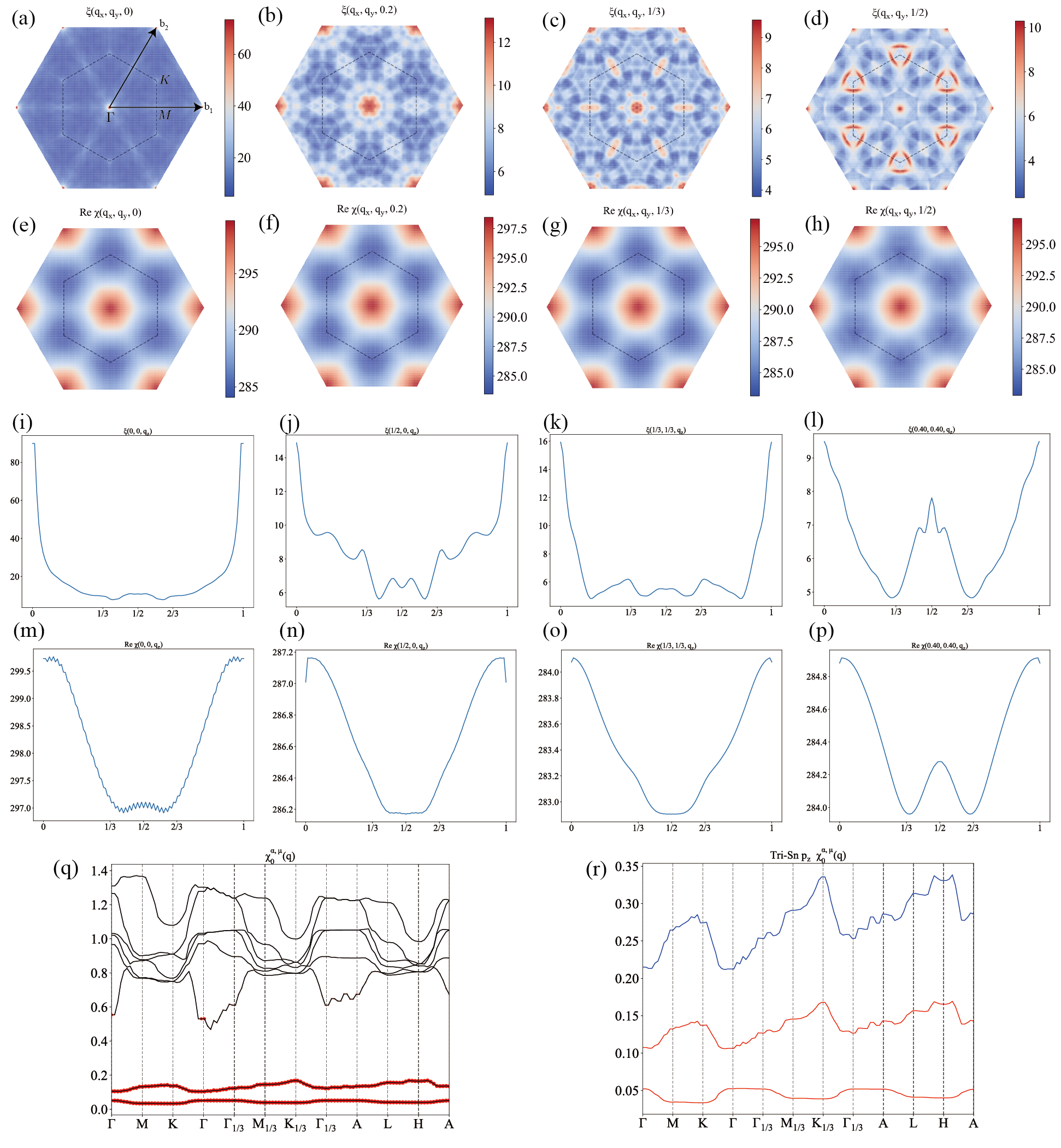}
    \caption{The susceptibility for the relaxed structure of ScV$_6$Sn$_6$. (a)-(d) and (e)-(h) are the nesting function $\xi(\mathbf{q})$ and the real part of the total susceptibility $\text{Re}~\chi_0(\mathbf{q})$ for fixed $q_z/(2\pi)=0,0.2,\frac{1}{3},\frac{1}{2}$ planes, respectively. (i)-(l) and (m)-(p) are the same but for fixed $(q_x,q_y)/(2\pi)=(0,0),(\frac{1}{2},0),(\frac{1}{3},\frac{1}{3}),(0.4,0.4)$ axis along $q_z$. (q) The eigenvalues of the orbital-resolved susceptibility matrix $\chi_0^{\alpha,\mu}(\mathbf{q})$, where the red dots denote the weight of two triangular Sn $p_z$ orbitals. (r) The two red lines are the eigenvalues of the $2\times 2$ orbital-resolved susceptibility matrix of two triangular Sn $p_z$ orbitals alone, and the blue line is the sum of the $2\times 2$ matrix. In (q) and (r), we use labels with subscript $\frac{1}{3}$ to denote high-symmetry-points (HSPs) on $q_z=\frac{1}{3}$ plane. 
    }
    \label{fig:suscept_relax}
\end{figure}

In \cref{fig:suscept_relax}(a)-(p), we first show the total susceptibility for the relaxed pristine structure of ScV$_6$Sn$_6$. \cref{fig:suscept_relax}(a)-(d) are the nesting function $\xi(\mathbf{q})$ and (e)-(h) are the total susceptibility $\chi_0(\mathbf{q})$ for fixed $q_z/(2\pi) =0,0.2,\frac{1}{3},\frac{1}{2}$ planes. \cref{fig:suscept_relax}(i)-(l) are the nesting function and (m)-(p) are the total susceptibility 
for fixed $\frac{1}{2\pi}(q_x,q_y)=(0,0),(\frac{1}{2},0),(\frac{1}{3},\frac{1}{3}),(0.4,0.4)$ axes along $q_z$. In \cref{fig:suscept_relax}(d) of $q_z=\pi$ plane, it can be seen that there exist $C_3$-symmetrical triangular-shaped peaks near $\mathbf{q}/(2\pi)=(\frac{1}{3},\frac{1}{3},\frac{1}{2})$ in the nesting function, with a representative peak at $\mathbf{q}/(2\pi)=(0.4,0.4,\frac{1}{2})$. These nesting peaks lead to a weak enhancement of the susceptibility near $\mathbf{q}/(2\pi)=(\frac{1}{3},\frac{1}{3},\frac{1}{2})$ as we also discussed near \cref{sec:triangular_Sn_suscept}. 
In \cref{fig:suscept_relax}(c) of $q_z=\frac{1}{3}$ plane, two main peaks of the nesting function appear at $(0,0,\frac{1}{3})$ and $(\frac{1}{2},0,\frac{1}{3})$, but no prominent inplane peak appears at $(\frac{1}{3},\frac{1}{3},\frac{1}{3})$, while in \cref{fig:suscept_relax}(k), a weak peak can be found along $q_z$ near $(\frac{1}{3},\frac{1}{3},\frac{1}{3})$. In \cref{fig:fermi_surface}(d), we show two slices of Fermi surfaces that have some parts connected by this $\bm{q}=(\frac{1}{3},\frac{1}{3},\frac{1}{3})$ nesting vector. 
For $\chi_0(\mathbf{q})$ of the total susceptibility on fixed $q_z$ plane in \cref{fig:suscept_relax}(e)-(h), the dominant in-plane peak appear at $(0,0)$, while only a small bump can be found near $(\frac{1}{3},\frac{1}{3},\frac{1}{3})$ along $q_z$ in \cref{fig:suscept_relax}(o).

We further calculate the orbital-resolved susceptibility. In \cref{fig:suscept_relax}(q), we show the eigenvalues of the orbital-resolved susceptibility matrix $\chi_0^{\alpha,\mu}(\mathbf{q})$, where we consider 8 sets of orbitals, i.e., the $d$ orbitals of 6 V atoms and $p_z$ orbitals of 2 triangular Sn atoms. We use red dots to denote the weight of two triangular Sn $p_z$ orbitals, which have weak hybridizations with $d$ orbitals of V, i.e., eigenvalues without red dots. 
In \cref{fig:suscept_relax}(r), the two red lines are the eigenvalues of the $2\times 2$ orbital-resolved susceptibility matrix of two triangular Sn $p_z$ orbitals only, and the blue line is the sum of the $2\times 2$ matrix. In \cref{fig:suscept_relax}(r), it can be seen 
that the susceptibility of two $p_z$ of Sn$^T$ has peaks near $K_{1/3}=(\frac{1}{3},\frac{1}{3},\frac{1}{3})$ and $H=(\frac{1}{3},\frac{1}{3},\frac{1}{2})$. We comment that this peak structure is consistent with the peak structure identified in the calculation of charge susceptibility of mirror-even orbital ($\chi_e(\mathbf{q})$) in \cref{sec:triangular_Sn_suscept} and \cref{fig:high_T_chi_e_corr}(a). 
Both susceptibilities (charge susceptibility of original triangular Sn $p_z$ orbitals and charge susceptibility of original triangular Sn mirror-even orbitals) arise from the fluctuations of mirror-even orbitals, and thus have a similar structure.

\begin{figure}
    \centering
    \includegraphics[width=1\textwidth]{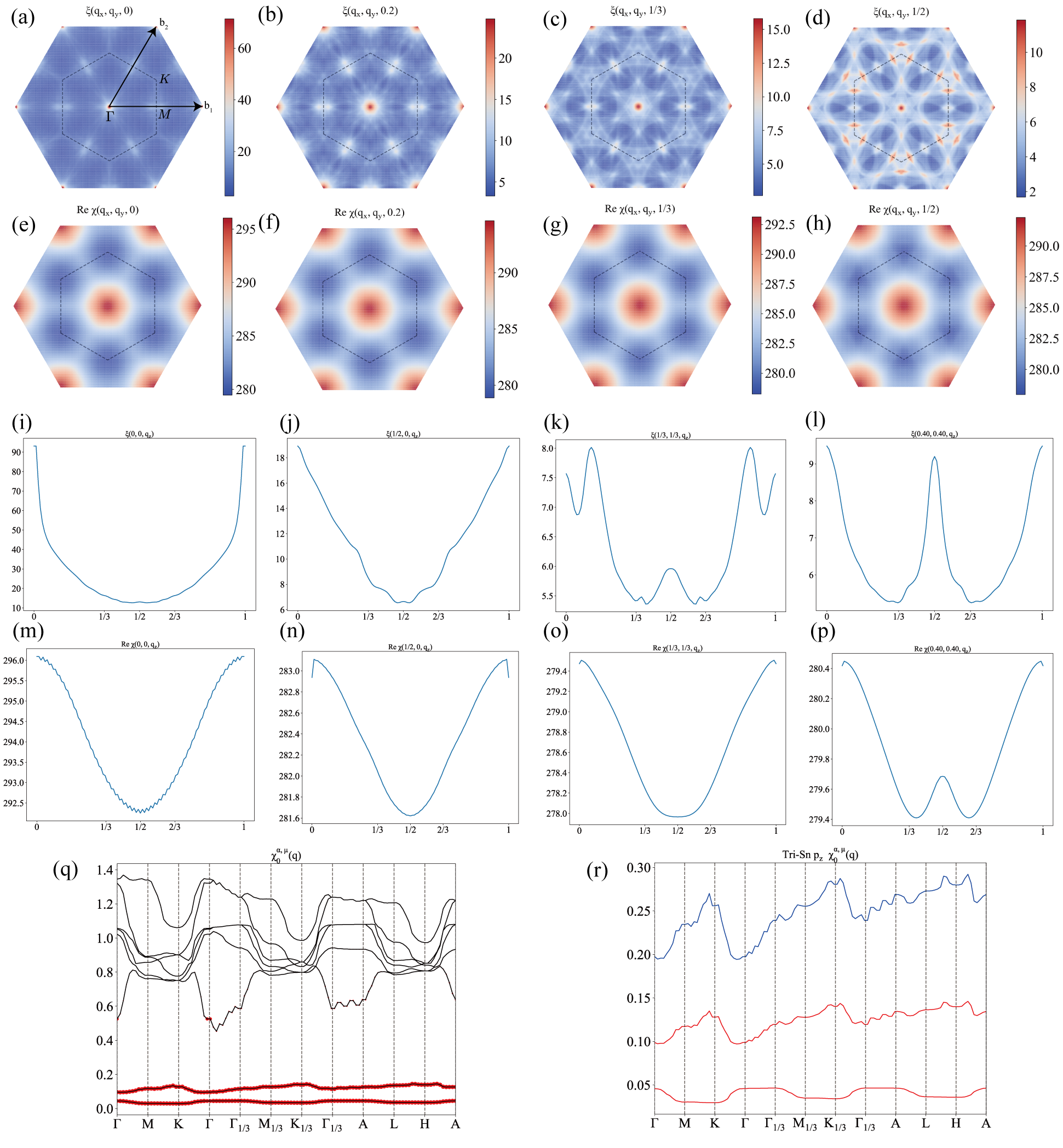}
    \caption{The susceptibility for the experimental structure of ScV$_6$Sn$_6$\cite{ARA22}. (a)-(d) and (e)-(h) are the nesting function $\xi(\mathbf{q})$ and the real part of the total susceptibility $\text{Re}~\chi_0(\mathbf{q})$ for fixed $q_z/(2\pi)=0,0.2,\frac{1}{3},\frac{1}{2}$ planes, respectively. (i)-(l) and (m)-(p) are the same but for fixed $(q_x,q_y)/(2\pi)=(0,0),(\frac{1}{2},0),(\frac{1}{3},\frac{1}{3}),(0.4,0.4)$ axis along $q_z$. (q) The eigenvalues of the orbital-resolved susceptibility matrix $\chi_0^{\alpha,\mu}(\mathbf{q})$, where the red dots denote the weight of two triangular Sn $p_z$ orbitals. (r) The two red lines are the eigenvalues of the $2\times 2$ orbital-resolved susceptibility matrix of two triangular Sn $p_z$ orbitals alone, and the blue line is the sum of the $2\times 2$ matrix. In (q) and (r), we use labels with subscript $\frac{1}{3}$ to denote HSPs on $q_z=\frac{1}{3}$ plane.  
    }
    \label{fig:suscept_exp}
\end{figure}

In addition, in \cref{fig:suscept_exp}, we show the computed susceptibility for the experimental structure. Compared with the relaxed structure, similar triangular-shaped in-plane peaks can be found on $q_z/(2\pi)=\frac{1}{2}$ plane near $q/(2\pi)=(\frac{1}{3},\frac{1}{3},\frac{1}{2})$ in the nesting function in \cref{fig:suscept_exp}(d), while the weak peak at $\mathbf{q}/(2\pi)=(\frac{1}{3},\frac{1}{3},\frac{1}{3})$ along $q_z$ almost disappears in \cref{fig:suscept_exp}(k). Also, no pronounced peak can be found in the total susceptibility except the dominant peak at $(0,0)$ on each fixed $q_z$ plane. 
In \cref{fig:fermi_surface}(b), we show two slices of Fermi surfaces that are connected by $\bm{q}/(2\pi)=(0.4,0.4,\frac{1}{2})$ (near $H$) nesting vector. In \cref{fig:suscept_exp}(r), the orbital-resolved susceptibility matrix for two triangular Sn $p_z$ orbitals has weaker peaks at $\mathbf{q}/(2\pi) =(\frac{1}{3},\frac{1}{3},\frac{1}{3})$ and near $\mathbf{q}/(2\pi) =(\frac{1}{3},\frac{1}{3},\frac{1}{2})$ compared with the relaxed structure, indicating the suppression of susceptibility when moving two triangular Sn closer. 
This can be verified from the phonon spectrum of the experimental structure shown in \cref{Fig-Pho-Sc-Exp}, which still has negative squared frequency but the value is higher compared with the relaxed structure in \cref{Fig-Pho-Sc-highT}.

\section{$\text{YV$_6$Sn$_6$}$} 
\label{app:sec:yvs}

In order to make sure our analysis of the phonon frequency renormalization in ScV$_6$Sn$_6$ is correct, we contrast it to the case of the stable, non-CDW compound YV$_6$Sn$_6$. Our aim is to 1. Show that our analytic calculation of the phonon frequency renormalization gives a stable phonon for YV$_6$Sn$_6$; 2. Understand, at a deep physical level,  the reasons for this stability here versus the instability in ScV$_6$Sn$_6$.

YV$_6$Sn$_6$ has a similar structure as ScV$_6$Sn$_6$ with Sc atoms replaced by Y. However, the mass of Y is 88.9u which is much larger than the mass of Sc (45.0u). More importantly, no CDW phase has been reported in this compound.  To understand the the absence of CDW phase, we perform a detailed theoretical study of the YV$_6$Sn$_6$.

\subsection{Crystal structure}
YV$_6$Sn$_6$ has the same structure with ScV$_6$Sn$_6$ of space group (SG) 191 $P6/mmm$, with slightly different lattice constants $a=5.523$ \AA, $c=9.186$ \AA, and the distance between two triangular Sn is $3.054$ \AA.

\subsection{Electronic structure}

\begin{figure}
	\centering
	\includegraphics[width=0.7\textwidth]{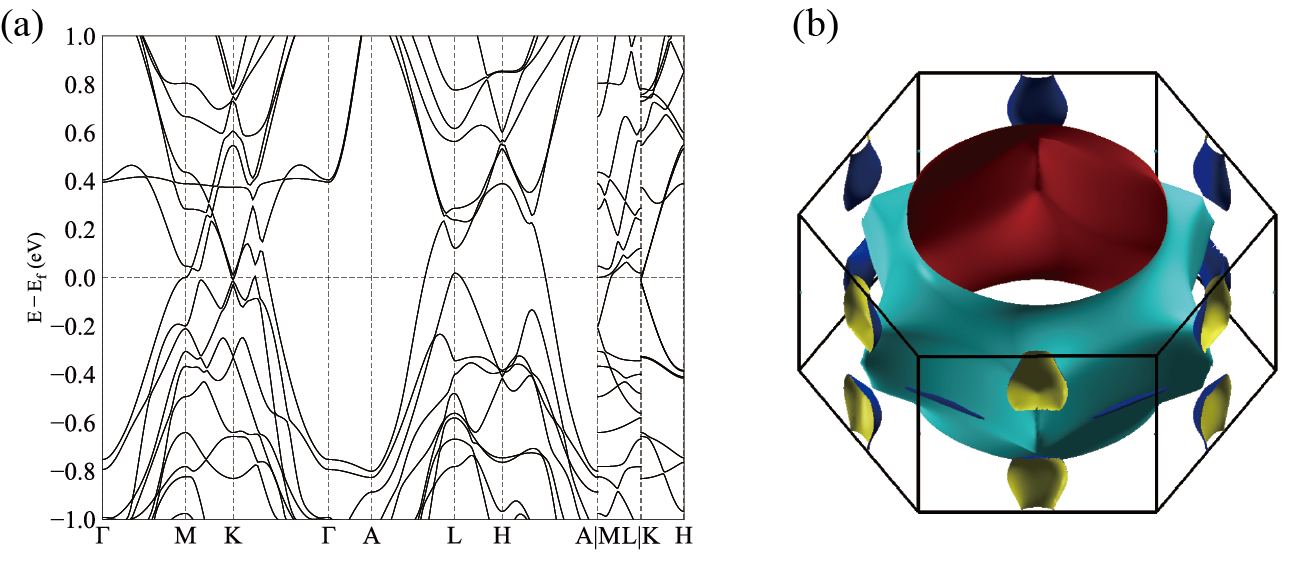}
	\caption{The (a) band structure and (b) Fermi surface of YV$_6$Sn$_6$.}
	\label{fig:YV6Sn6-bands}
\end{figure}

\begin{figure}
    \centering
    \includegraphics[width=0.4\textwidth]{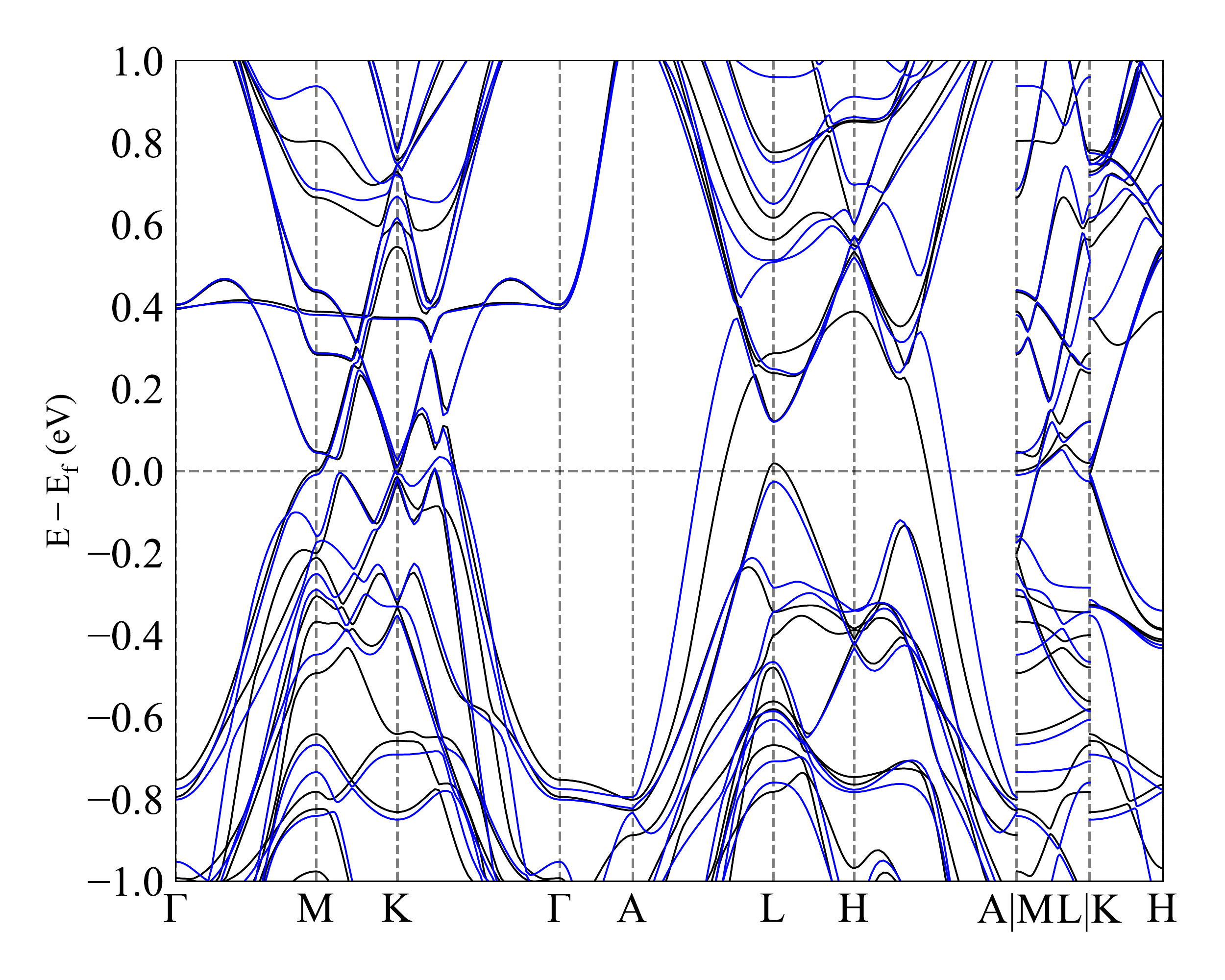}
    \caption{Comparison between band structure of YV$_6$Sn$_6$ (black line) and the band structure ScV$_6$Sn$_6$ with relaxed structure (blue line). The Fermi energy ScV$_6$Sn$_6$ has been increased by 80meV. We can observe a good match between two band structures at $k_z=0$ plane and small deviations at $k_z=\pi$ plane. }
    \label{fig:compare_bands}
\end{figure}

 Since both electrons of Y and electrons of Sc are irrelevant to the low-energy band structure, the two materials  YV$_6$Sn$_6$ and ScV$_6$Sn$_6$ have a very similar band structure, but with a Fermi energy shift. 
In \cref{fig:YV6Sn6-bands}(a), we show the band structure of YV$_6$Sn$_6$ with SOC, where the $E_f$ is about 80 meV lower (on $k_z=0$ plane) than the relaxed pristine structure of ScV$_6$Sn$_6$. In \cref{fig:YV6Sn6-bands}(b), we show the Fermi surface of YV$_6$Sn$_6$. We can observe the similarities between YV$_6$Sn$_6$ and ScV$_6$Sn$_6$ by comparing \cref{fig:YV6Sn6-bands} with \cref{fig:fatband}. In \cref{fig:compare_bands}, we compare the band structure of YV$_6$Sn$_6$ and the band structure of ScV$_6$Sn$_6$ with $80$meV shifting of the bands. We can observe a good match between two band structures at $k_z=0$ plane and \textcolor{black}{small deviations (but qualitatively the same) at $k_z=\pi$ plane. At $k_z=\pi$ plane, a smaller Fermi surface is observed for YV$_6$Sn$_6$ compared to ScV$_6$Sn$_6$.}

\subsection{Phonon spectrum} 
\begin{figure}[h]
	\centering
	\includegraphics[angle=0, width=0.55\textwidth]{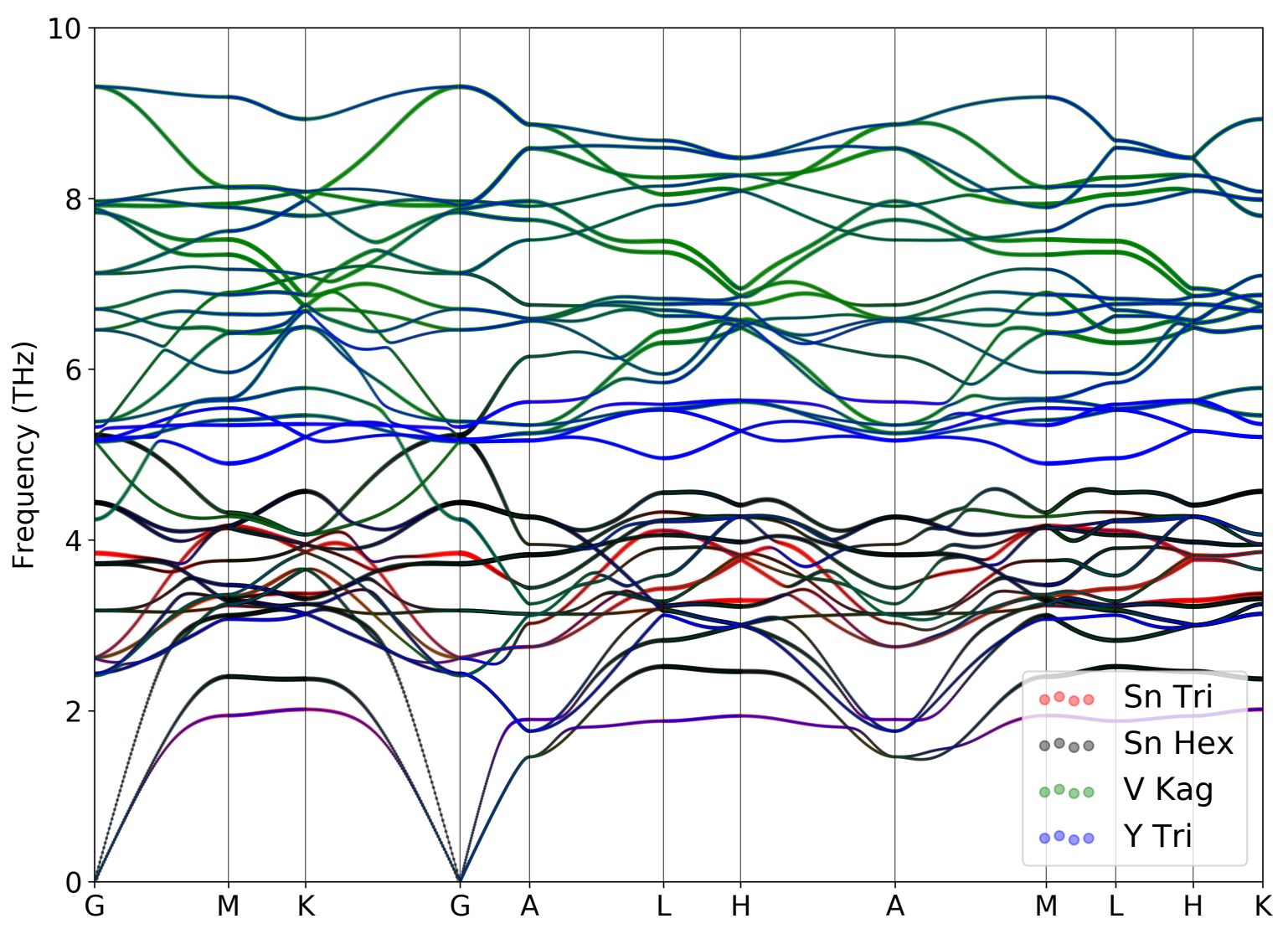}
	\caption{Phonon spectrum of YV$_6$Sn$_6$ calculated with $3\times3\times2$ $q$-grid by QE, weighted by atom sites, including the triangular Sn (red), hexagonal Sn (black), kagome V (green), and triangular Y (blue). The phonon spectrum shows no instability and the atom projection of lower branch shows the combination of triangular Y and triangular Sn atom, instead of the dominant triangular Sn atom in ScV$_6$Sn$_6$.}
	\label{Fig-Pho-yvs}
\end{figure}

Via DFT calculation, we first obtain the phonon spectrum of YV$_6$Sn$_6$ at zero temperature, which shows no instability as plotted in \cref{Fig-Pho-yvs}. Unlike the ScV$_6$Sn$_6$ where the lower branch of the phonon mode is contributed by triangular Sn atoms, the atom projection shows the lower branch of phonon modes of YV$_6$Sn$_6$ is contributed by both the Y atoms and triangular Sn atoms. This is expected and physically clear to us, because Y is much heavier than Sn: it will hence contribute more to the low-energy phonon mode. Moreover, similarly to ScV$_6$Sn$_6$, we find the low-energy phonon mode in YV$_6$Sn$_6$ is also relatively flat, thus the effective 1D-model description we built for ScV$_6$Sn$_6$ still holds for YV$_6$Sn$_6$.

\subsection{Electron-phonon coupling}
We next analyze the effect of electron-phonon coupling and discuss the absence of imaginary phonon mode. We perform similar calculations as those described in \cref{sec:ele_corr_Sc} for the ScV$_6$Sn$_6$. 

\begin{figure}
    \centering
    \includegraphics[width=1.0\textwidth]{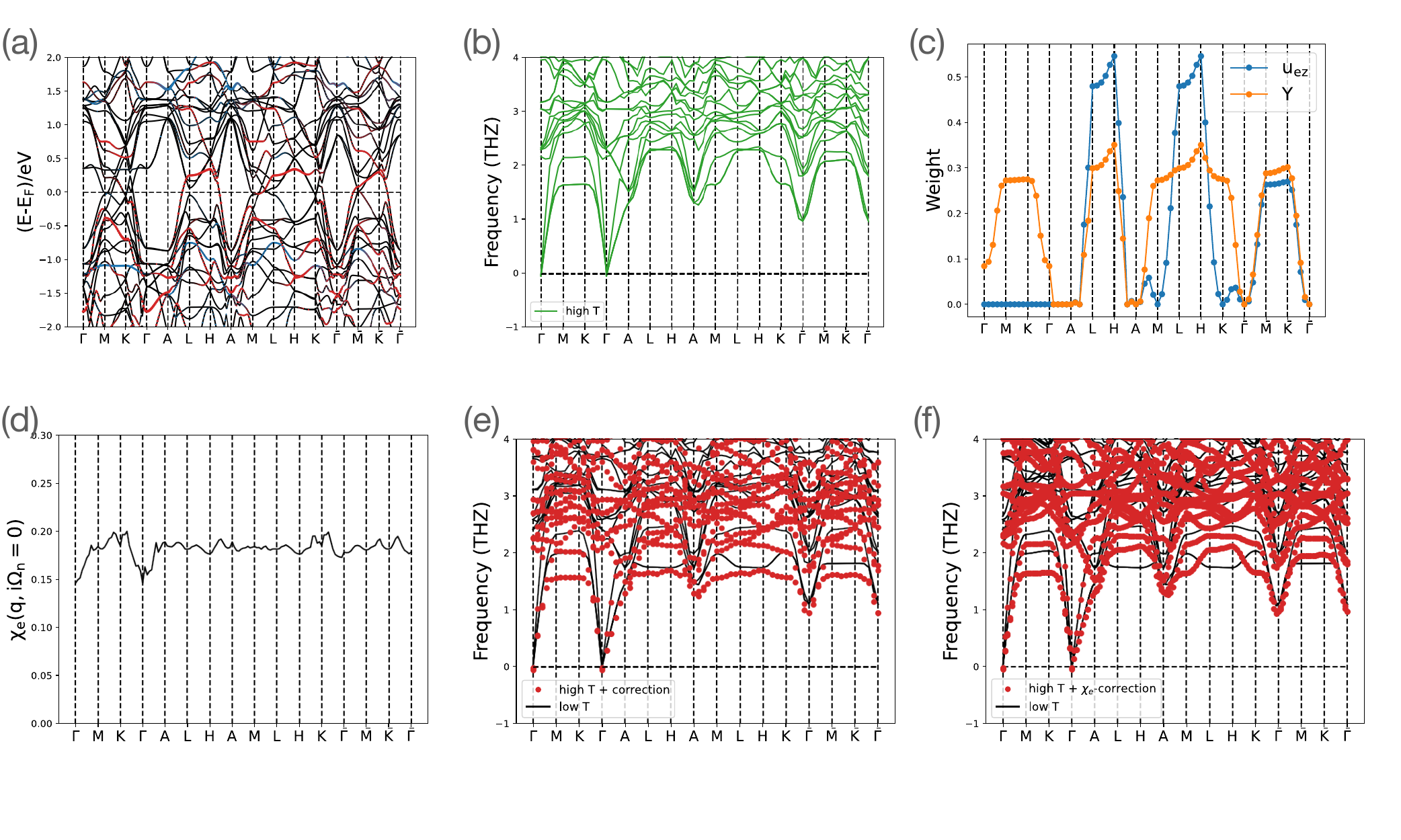}
    \caption{(a) Band structure of YV$_6$Sn$_6$ where red and blue denotes the weight of mirror-even and mirror-odd orbitals formed by triangular Sn $p_z$ orbitals. (b) High-temperature phonon spectrum derived from DFT ($T=0.8$eV). (c) Weights of the mirror-even vibrations of Sn atoms and the  vibration of Y atom of the lowest two phonon bands at high-temperature phase. (d) Charge susceptibility of mirror-even electron $c_{\mathbf{R},e,\sigma}$. (e) Comparison between high-temperature phonon spectrum with electron corrections and low-temperature phonon spectrum (derived from DFT at $T=0$eV). (F) Comparison between high-temperature phonon spectrum with $\chi_e$ corrections and low-temperature phonon spectrum (derived from DFT at $T=0$eV).}
    \label{fig:ele_ph_Y}
\end{figure}

In \cref{fig:ele_ph_Y} (a), we plot the band structure and orbital weights of the mirror-even and mirror-odd states ($c_{\mathbf{R},e/o,\sigma}$, \cref{eq:ele_even_odd}). 
We find the bands near Fermi energy are also mostly \textcolor{black}{formed by the mirror-even orbitals combing from the $p_z$ orbitals of two triangular Sn atoms (\cref{eq:ele_even_odd})}. \
In \cref{fig:ele_ph_Y} (b) and (c), we show the high-temperature phonon spectrum (which can be understood as the non-interacting phonon spectrum) and the corresponding weight of the low-energy modes.  Since the lowest \textcolor{black}{two} modes 
have very close energy \textcolor{black}{and are entangled (for example, along A-L-H line, \cref{fig:ele_ph_Y} (b))}, we plot the weights of mirror-even vibration modes \textcolor{black}{of Sn atoms}
$u_{ez}(\mathbf{R})$ (\cref{eq:ph_even_odd}) and $z$-direction vibrations of Y atoms of the lowest two bands. We observe that the lowest bands have non-negligible weights of Y, and the weight of the mirror-even phonon mode is reduced compared to ScV$_6$Sn$_6$ (\cref{fig:high_T_chi_e_corr} (b)). This is because Y has a much larger mass than Sc and thus makes more contribution to the low-energy phonon modes. 

\begin{figure}
    \centering
\includegraphics[width=1.0\textwidth]{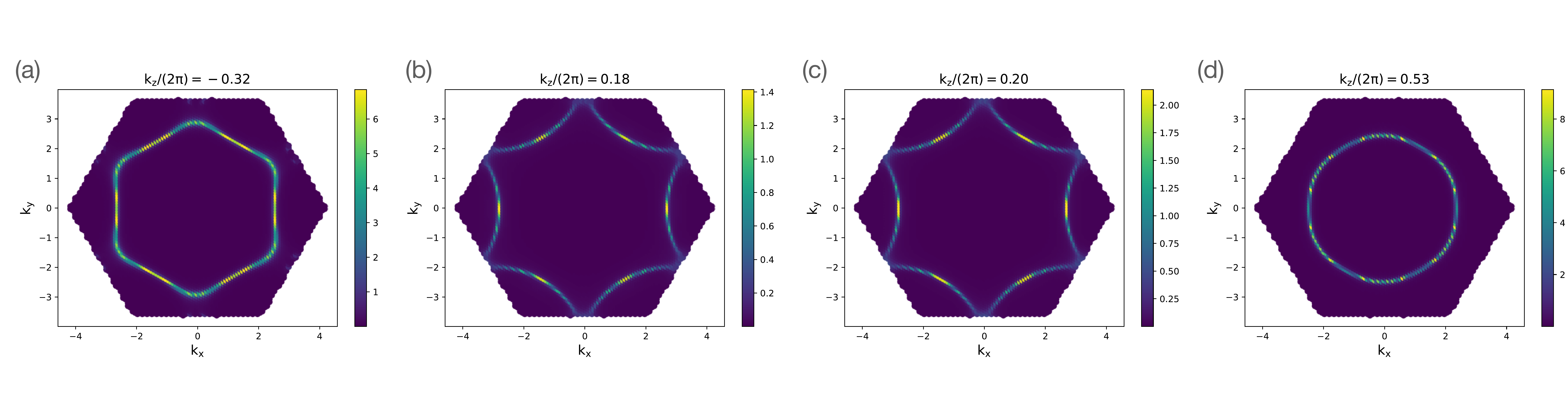}
    \caption{Density of states of mirror-even orbital at Fermi energy at different $k_z$ planes of YV$_6$Sn$_6$. We observe that the density of states at $k_z/(2\pi)=0.18,0.20$ are reduced compare with \cref{fig:fs_mirror_even}(ScV$_6$Sn$_6$).}
    \label{fig:fs_mirror_even_Y}
\end{figure}

To obtain the corrections \textcolor{black}{to the phonon spectrum} induced by the electrons 
we then investigate the charge susceptibility of the mirror even electrons $c_{\mathbf{R},e,\sigma}$. 
As shown in \cref{fig:ele_ph_Y} (d), the overall value of the susceptibility is smaller than the one in ScV$_6$Sn$_6$ (\cref{fig:high_T_chi_e_corr} (a)). In fact, we find the density of states of mirror-even orbital on the Fermi surface are reduced for $|k_z/(2\pi)|\lesssim 0.2$ due to the shifting of the Fermi energy and the change of band structures (\cref{fig:compare_bands}). In \cref{fig:fs_mirror_even_Y}, we show the density of states of the mirror even orbitals $(c_{\mathbf{R},e,\sigma}$) at four $k_z$ points (which are the $k_z$ points that give the nesting of Fermi surface in the ScV$_6$Sn$_6$ in \cref{fig:fs_mirror_even}). We can observe the reduction of the density of states of the mirror even orbitals $c_{\mathbf{R},e,\sigma}$ at the Fermi surfaces. \textcolor{black}{ 
In \cref{fig:fs_mirror_even_Y}, we find the density of states of mirror-even orbitals $c_{\mathbf{R},e,\sigma}$ at $k_z/(2\pi)=0.18,0.20$ planes (which are also planes that provide Fermi surface nesting) of YV$_6$Sn$_6$ is less than $1/3$ of the corresponding density of states of ScV$_6$Sn$_6$ (\cref{fig:fs_mirror_even}).
This reduced density of state produces an even weaker Fermi-surface nesting and a weaker orbital-resolved susceptibility ($\chi_e$) of $c_{\mathbf{R},e,\sigma}$ electrons.} \textcolor{black}{In \cref{fig:ele_ph_Y} (d), we can barely observe the momentum-dependency of the $\chi_e$, and the maximum value of  $\chi_e$ of YV$_6$Sn$_6$ is about $\sim 0.20 $, which is only $\sim 60\%$ of the maximum value ($\sim 0.36 $) of $\chi_e$ of ScV$_6$Sn$_6$ (\cref{fig:high_T_chi_e_corr}).}

Combining the above information, we mention the following two facts that lead to the absence of CDW in YV$_6$Sn$_6$, 
\begin{itemize}
    \item The low-branch phonon mode has less weight of mirror-even vibration modes $u_{ez}(\mathbf{R})$. Since the electron corrections are mostly acting on the $u_{ez}(\mathbf{R})$ mode, the correction to the low-brunch phonon will be reduced. 
    \item The density fluctuations of the mirror even electron ($c_{\mathbf{R},e,\sigma}$) is weak, which will also lead to weaker electron corrections to the phonon mode. 
\end{itemize}
Combining the above two effects, we conclude the electron corrections to the phonon mode are weaker in YV$_6$Sn$_6$. Consequently, we \textcolor{black}{do} not 
observe the CDW instability or imaginary phonons in YV$_6$Sn$_6$. 

To support this physical reasoning we  directly calculate 
the electron corrections \textcolor{black}{using both the method introduced in \cref{sec:ele_corr_to_phonon} and DFT}. In \cref{fig:high_T_chi_e_corr} (e) and (f), we show the comparison between the non-interacting (high-temperature) phonon spectrum with electron corrections and low-temperature phonon spectrum (derived from DFT). Again, we find a qualitatively very good agreement between our analytical calculation of the phonon frequency renormalization and ab-initio results and we do not observe any imaginary phonon mode in both calculations.

\subsection{Electron susceptibility}
\begin{figure}
    \centering
    \includegraphics[width=1\textwidth]{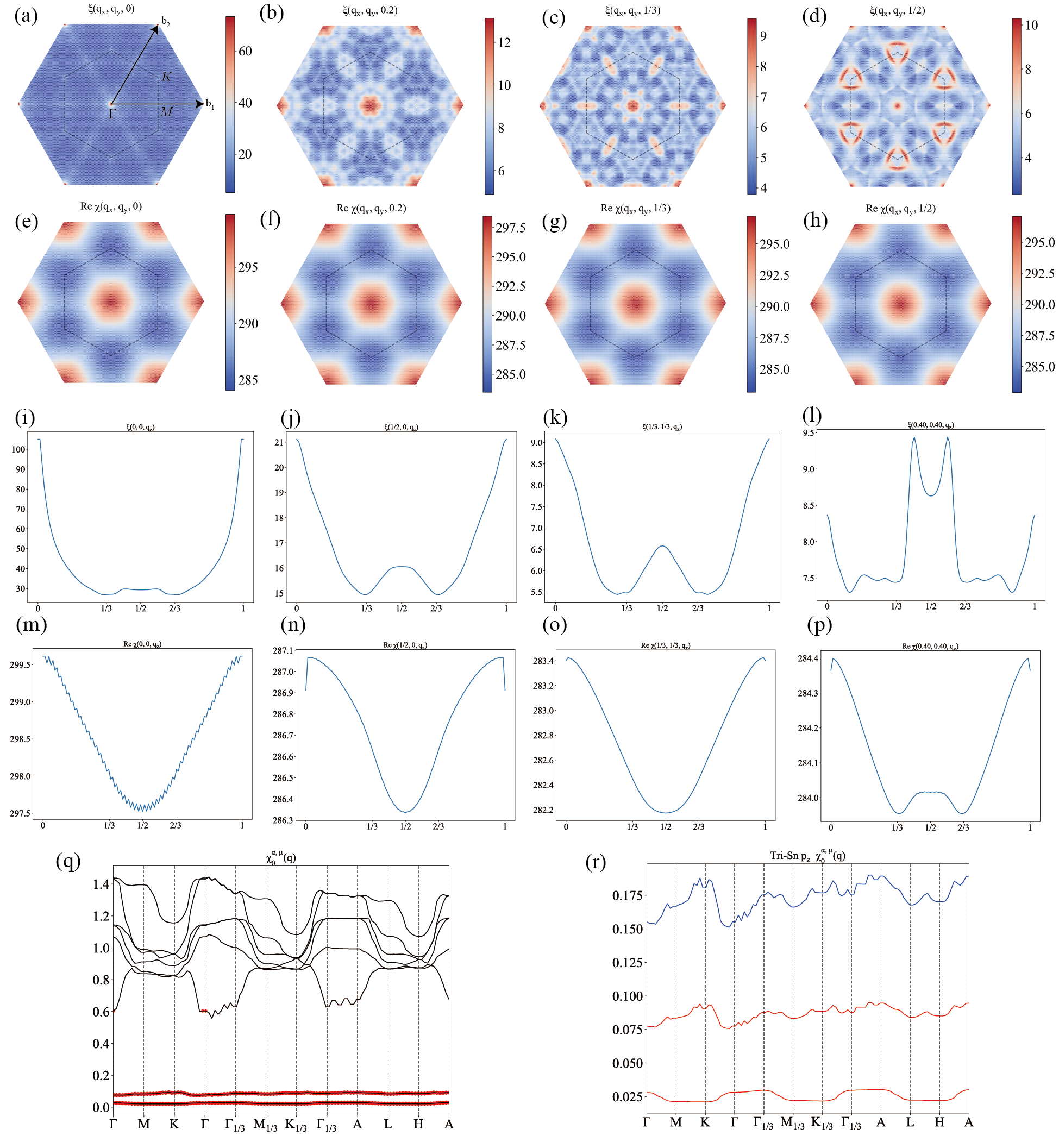}
    \caption{The susceptibility for YV$_6$Sn$_6$. (a)-(d) and (e)-(h) are the nesting function $\xi(\mathbf{q})$ and the real part of the total susceptibility $\text{Re}~\chi_0(\mathbf{q})$ for fixed $q_z/(2\pi)=0,0.2,\frac{1}{3},\frac{1}{2}$ planes, respectively. (i)-(l) and (m)-(p) are the same but for fixed $(q_x,q_y)/(2\pi)=(0,0),(\frac{1}{2},0),(\frac{1}{3},\frac{1}{3}),(0.4,0.4)$ axis along $q_z$. (q) The eigenvalues of the orbital-resolved susceptibility matrix $\chi_0^{\alpha,\mu}(\mathbf{q})$, where the red dots denote the weight of two triangular Sn $p_z$ orbitals. (r) The two red lines are the eigenvalues of the $2\times 2$ orbital-resolved susceptibility matrix of two triangular Sn $p_z$ orbitals alone, and the blue line is the sum of the $2\times 2$ matrix. In (q) and (r), we use labels with subscript $\frac{1}{3}$ to denote HSPs on $q_z=\frac{1}{3}$ plane. 
}
    \label{fig:suscept_YV6Sn6}
\end{figure}

We also compute the electron susceptibility of YV$_6$Sn$_6$ as shown in \cref{fig:suscept_YV6Sn6}, we show the computed susceptibility for YV$_6$Sn$_6$. Compared with the ScV$_6$Sn$_6$, the distribution of the nesting function and the total susceptibility is similar, except that the peak of the nesting function at $\bm{q}=(0.4,0.4,\frac{1}{2})$ split into two peaks along $q_z$ in \cref{fig:suscept_YV6Sn6}(l). 
Compared with ScV$_6$Sn$_6$, the orbital-resolved susceptibility matrix for two triangular Sn $p_z$ orbitals has no prominent peaks at $\mathbf{q}=(\frac{1}{3},\frac{1}{3},\frac{1}{3})$ or $\mathbf{q}=(\frac{1}{3},\frac{1}{3},\frac{1}{2})$ and is almost flat. \textcolor{black}{Since the susceptibility of $\chi_e$ also comes from the mirror-even orbital $c_{\mathbf{R},e,\sigma}$ of triangular Sn $p_z$ electrons, the behaviors of the charge susceptibility of Sn $p_z$ orbitals and the behaviors of $\chi_e$ are consistent where both shows a weak-momentum dependency and with a smaller amplitude comparing to the corresponding susceptibility in ScV$_6$Sn$_6$.}

\section{ CDW phase transition} \label{app:sec:LG_theory}
\label{sec:cdw_transition}
In this \siSection{}, we describe the corresponding effective theory of the CDW phase transition. We note that the phonon softening is observed at $H= \frac{1}{3} \frac{1}{3} \frac{1}{2}$. However, the CDW happens at $\bar{K} = \frac{1}{3}\frac{1}{3}\frac{1}{3}$. This indicates there are at least two types of order parameters that are important for low-energy physics, with one corresponding to the phonon softening with momentum $H$ and the other one corresponding to the CDW with momentum $\tilde{{{K}}}$. Here we explain the microscopic origin of the two order parameters, and why the softening phonon does not lead to a CDW at its wavevector; we also explain the origin of the CDW wavevector.  We use $\phi$ and $\psi$ to label the order parameters corresponding to the phonon softening and CDW respectively. We then build an effective theory of $\phi,\psi$ fields and demonstrate that a first-order phase transition to the CDW phase with wavevector $\tilde{{{K}}}$ can be induced by the strong fluctuation of $\phi$ fields. We show that these strong fluctuations, microscopically, are a consequence of flat phonon band near the $H$ point. In addition, the symmetry-allowed cubic term of $\psi$ fields ($\psi^3$) makes the transition first-order-like. 

This \siSection{} is organized as follows 
\begin{itemize}
    \item 1. We construct the effective action of $\phi$ fields and study the phase transition of $\phi$ fields at the saddle-point level. 
    \item 2. We construct the effective action of $\psi$ fields and study the phase transition of $\psi$ fields at the saddle-point level. 
    \item 3. We take both $\phi$ and $\psi$ fields into consideration and include the coupling between the $\phi$ and $\psi$ fields. We then study the corresponding phase transition at the saddle-point level.
    \item 4. Based on point 3, we include the Gaussian fluctuations on top of the saddle-point solution and study the resulting phase transition.
\end{itemize}

\subsection{$H$ order}
We first discuss the order parameter $\phi$ corresponding to the phonon softening. 
We use the phonon field with the lowest energy $\tilde{u}_{soft}(\mathbf{q})$ near $H$ as our order parameter. Since we have two non-equivalent $H$ points $H_1=2\pi(\frac{1}{3}\frac{1}{3}\frac{1}{2})$ and $H_2= 2\pi(\frac{-1}{3}\frac{-1}{3}\frac{-1}{2})$ in our first Brillouin zone, we need to introduce two order parameters which we denoted by $\phi_1,\phi_2$ where 
\ba 
&\phi_1(\mathbf{q}) = \sqrt{2}\tilde{u}_{soft}\bigg(2\pi(\frac{1}{3} \frac{1}{3} \frac{1}{2}) +\mathbf{q}  \bigg)
,\quad \quad \phi_2(\mathbf{q}) = \sqrt{2}\tilde{u}_{soft}\bigg( 2\pi(\frac{-1}{3} \frac{-1}{3} \frac{-1}{2}) +\mathbf{q} \bigg) 
\ea 
where $\mathbf{q}$ denotes the small momentum deviation from $H$,$-H$. In practice, we will only consider momentum $|\mathbf{q}| <\Lambda$ with $\Lambda$ a given small momentum cutoff. For convenience, we have also introduced a $\sqrt{2}$ factor which will cancel the $1/2$ factor in the effective action of the phonon (\cref{eq:seff_tilde_u}).

We now discuss the symmetry properties of $\phi$ fields. We first comment that, as shown in the \textit{ab-initio} phonon spectrum from \cref{fig:zero_temperature_phonon_model}, $\phi_1,\phi_2$ form the $\mathrm{H}_3$ irreducible representation of the little group at $H$ and $-H$ respectively. We next discuss the symmetry properties of the two fields.
Under a symmetry transformation $g$, we define the corresponding transformation matrix $D^{g,\phi}$ as
\ba 
g \phi_i(\mathbf{q}) g^{-1} = \sum_j D^{g,\phi}_{ij} \phi_j(g\mathbf{q}) 
\ea 
and we have
\ba 
D^{C_{3z},\phi} = \begin{bmatrix}
1 & 0 \\ 
0 & 1 
\end{bmatrix} ,\quad 
D^{C_{2z},\phi} = \begin{bmatrix}
0 & -1 \\ 
-1 & 0
\end{bmatrix} 
,\quad D^{C_{2,110},\phi} 
= \begin{bmatrix}
-1 & 0 \\ 
0 & -1 
\end{bmatrix} ,\quad 
 D^{i,\phi} 
= \begin{bmatrix}
 0 & 1 \\ 
1 & 0 
\end{bmatrix}
\ea 
 In addition, under translational transformation, we have 
\ba 
T_{\mathbf{R}} \phi_i(\mathbf{q}) T_\mathbf{R}^{-1} = e^{ i(H_i+\mathbf{q})\cdot\mathbf{R}} \phi_i(\mathbf{q}) 
\ea 
Under time reversal symmetry, we have 
\ba 
\mathcal{T} \phi_1(\mathbf{q}) \mathcal{T}^{-1} = \phi_2(-\mathbf{q}),\quad 
\mathcal{T} \phi_2(\mathbf{q}) \mathcal{T}^{-1} = \phi_1(-\mathbf{q})
\ea 
From construction, we have (this results from the fact that the phonon fields are real fields in the real space which indicates $u(\mathbf{q})^\dag = u(-\mathbf{q})$ for the phonon fields)
\ba 
\phi_1^\dag (\mathbf{q}) = \phi_2 (-\mathbf{q}) 
\label{eq:rel_phi}
\ea 
Therefore, it is sufficient to consider $\phi(\mathbf{q}) = \phi_1(\mathbf{q}) = \phi_2^\dag(-\mathbf{q})$ and then
\ba 
&C_{3z} \phi(\mathbf{q}) C_{3z}^{-1} = \phi(C_{3z}\mathbf{q}),\quad C_{2z}\phi(\mathbf{q}) C_{2z}^{-1} =-\phi^\dag(-C_{2z}\mathbf{q}) ,\quad C_{2,110} \phi(\mathbf{q}) C_{2,110}^{-1} = -\phi(C_{2,110}\mathbf{q}),\quad i \phi(\mathbf{q}) i = \phi^\dag(\mathbf{q}),\quad\nonumber\\ 
&\mathcal{T}\phi(\mathbf{q}) \mathcal{T}^{-1} = \phi^\dag(\mathbf{q}) ,\quad  T_\mathbf{R} \phi(\mathbf{q}) T_\mathbf{R}^{-1} =  e^{i(\text{H}_1+\mathbf{q})\cdot \mathbf{R}} \phi (\mathbf{q})
\ea 

The symmetry-allowed Langrangian of $\phi$ fields takes the form of 
\ba 
L_\phi[\phi(\mathbf{q},\tau)]  = \sum_{|\mathbf{q}|<\Lambda } \bigg[ \partial_\tau \phi(\mathbf{q},\tau) \partial_\tau \phi^\dag(\mathbf{q},\tau) + \omega_{soft, H+\mathbf{q}}^2 \phi(\mathbf{q},\tau) \phi^\dag(\mathbf{q},\tau) \bigg] + V_\phi[ \phi(\mathbf{q}, \tau)] 
\label{eq:Lphi_v0}
\ea 
where $\omega_{soft,H+\mathbf{q}}^2$ characterize the (rather flat) dispersion of the lowest-energy phonon fields near $H$ point. 
 $V_\phi[\phi(\mathbf{q},\tau)]$ is a function of $\phi$ fields denoting the interaction term that is generated by electron-phonon coupling, as described near \cref{eq:log_exp_high_order}.  $\phi(\mathbf{q},\tau)$ denote the $\phi(\mathbf{q})$ fields at time-slice $\tau$. We next expand $V_\phi$ in powers of $\phi$ fields and we truncate to quadratic order 
\ba 
V_\phi[\phi(\mathbf{q},\tau)] = \sum_{|\mathbf{q}_1|<\Lambda,|\mathbf{q}_2|<\Lambda,|\mathbf{q}_3|<\Lambda,|\mathbf{q}_4|<\Lambda}\frac{u_{\phi}}{N}  \phi(\mathbf{q}_1,\tau) \phi(\mathbf{q}_2,\tau) \phi^\dag(\mathbf{q}_3,\tau) \phi^\dag(\mathbf{q}_4,\tau) \delta_{\mathbf{q}_1+\mathbf{q}_2-\mathbf{q}_3 -\mathbf{q}_4, \bm{0} } 
\label{eq:Vphi_exp}
\ea 
Several remarks are in order
\begin{itemize}
    \item We have ignored the momentum dependency in the interaction coefficient for the interactions at all orders in $\phi$. 
    \item The second-order (bilinear) contributions have already been absorbed by the dispersion term $\omega_{soft,H}$.
    \item $V_\phi[\phi(\mathbf{q},\tau)]$ needs to follow the symmetry constraints (invariant under symmetry transformation). 
    \item The momentum-independent interaction coefficient of cubic term vanishes due to symmetry constraints (for ex, it would have finite $k_z$ momentum). 
\end{itemize}
Combining \cref{eq:Vphi_exp} and \cref{eq:Lphi_v0}, we reach the standard $\phi_4$ theory that describing the fluctuations of $H$-point phonon 
\ba 
L_\phi[\phi(\mathbf{q},\tau)]  = &\sum_{|\mathbf{q}|<\Lambda } \bigg[ \partial_\tau \phi(\mathbf{q},\tau) \partial_\tau \phi^\dag(\mathbf{q},\tau) + \omega_{soft, H+\mathbf{q}}^2 \phi(\mathbf{q},\tau) \phi^\dag(\mathbf{q},\tau) \bigg]\nonumber\\ 
&
+\sum_{|\mathbf{q}_1|<\Lambda,|\mathbf{q}_2|<\Lambda,|\mathbf{q}_3|<\Lambda,|\mathbf{q}_4|<\Lambda}\frac{u_{\phi}}{N} \phi(\mathbf{q}_1,\tau) \phi(\mathbf{q}_2,\tau) \phi^\dag(\mathbf{q}_3,\tau) \phi^\dag(\mathbf{q}_4,\tau) \delta_{\mathbf{q}_1+\mathbf{q}_2-\mathbf{q}_3 -\mathbf{q}_4, \bm{0} } 
\label{eq:Lphi}
\ea 
The corresponding partition function and free energy reads
\ba 
&Z_\phi = \int [D\phi] e^{-\int_0^\beta  L_\phi[\phi(\mathbf{q},\tau)]d\tau } \nonumber\\ 
&F_\phi =-\frac{1}{\beta}\log(Z_\phi)
\ea 

We first consider the phase transition by only considering the $\phi$ fields. 
To study the behavior of $\phi$ fields and the corresponding phase transition, we use the saddle-point approximation, where 
\ba 
&Z_\phi \approx e^{-\int_0^\beta L_\phi[ \phi^c(\mathbf{q},\tau)] d\tau }\nonumber\\ 
&F_\phi \approx \frac{1}{\beta}\int_0^\beta L_\phi[ \phi^c(\mathbf{q},\tau)] .
\ea 
$\phi^c$ denotes the solution of saddle-point equations
\ba 
\frac{\delta L_\phi[\phi(\mathbf{q},\tau)]}{\delta \phi(\mathbf{q},\tau)}\bigg|_{\phi = \phi^c } = 0
\label{eq:saddle_point_phi_tau}
\ea 
We take the following ansatz of the saddle-point solution which describes the condensation of $\phi$ fields
\ba 
\phi^c(\mathbf{q},\tau) = \sqrt{N}\rho_0 e^{i\theta_0 }\delta_{\mathbf{q},0}
\label{eq:phi_exp_val}
\ea 
Replacing the $\phi$ fields with the corresponding saddle-point solution, we construct the following free energy (Ginzburg-Landau free energy) 
\ba 
F_\phi[\rho_0,\theta_0] = \frac{1}{\beta} \int_0^\beta L_\phi[\phi^c(\mathbf{q},\tau)] = N(\omega_{soft,H}^2 \rho_0 +u_\phi \rho_0^4) 
\ea 
and the saddle-point equations in \cref{eq:saddle_point_phi_tau} 
are equivalent to the saddle-point equations of $F_\phi$ which are
\ba 
0=&\frac{\delta F_{\phi}[\rho_0,\theta_0]}{\delta \rho_0 } = N(2\omega_{soft,H}^2\rho_0 + 4u_\phi\rho_0^3) \nonumber\\ 
0=&\frac{\delta F_{\phi}[\rho_0,\theta_0]}{\delta \theta } =0
\ea 
We find the following two types of saddle point solutions (with the corresponding free energy)
\ba 
&\text{Disorder:}\quad \rho_0 = 0,\quad \quad F_{\phi} = 0 \nonumber\\ 
&H\text{-order:}\quad \rho_0 = \sqrt{ -\frac{\omega_{soft,H}^2}{2u_\phi } },\quad  \quad F_{\phi} =-N\frac{\omega_{soft,H}^4}{4 u_\phi } 
\label{eq:phi_sol}
\ea 
where the second solution with non-zero $\rho_0$ is only valid for when $\omega_{soft,H}^2 \le 0$ due to the square root. The phase of the system will be determined by the solution with lower free energy. By tuning $\omega_{soft,H}^2$, there will be a second-order phase transition between the ordered phase with $\rho_0 \ne 0$ at $\omega_{soft,H}^2<0$ and the disordered phase with $\rho_0 =0 $ at $\omega_{soft,H}^2>0$. The transition happens when the free energies of the two solutions are equal to each other, which is $\omega_{soft,H}^2 =0$. 
However, the saddle-point equations do not impose any condition of $\theta$. The $\theta$ dependency only appears in higher-order terms. To see this, we include the sixth-order term in the expansion of $V_\phi[\phi]$ (\cref{eq:Vphi_exp}). The symmetry-allowed $V[\phi]$ (up to sixth order) is 
\ba 
V_\phi[\phi(\mathbf{q},\tau)] =&\sum_{|\mathbf{q}_1|<\Lambda,|\mathbf{q}_2|<\Lambda,|\mathbf{q}_3|<\Lambda,|\mathbf{q}_4|<\Lambda}\frac{u_{\phi}}{N}  \phi(\mathbf{q}_1,\tau) \phi(\mathbf{q}_2,\tau) \phi^\dag(\mathbf{q}_3,\tau) \phi^\dag(\mathbf{q}_4,\tau) \nonumber\\ 
= &\sum_{|\mathbf{q}_1|<\Lambda, |\mathbf{q}_2|<\Lambda, |\mathbf{q}_3|<\Lambda, |\mathbf{q}_4|<\Lambda,|\mathbf{q}_5|<\Lambda,|\mathbf{q}_6|<\Lambda }\bigg\{ \nonumber\\ 
&v_{\phi,1} \phi^\dag(\mathbf{q}_1,\tau)\phi^\dag(\mathbf{q}_2,\tau)\phi^\dag(\mathbf{q}_3,\tau)
\phi(\mathbf{q}_4,\tau)\phi(\mathbf{q}_5,\tau)\phi(\mathbf{q}_6,\tau) \delta_{\mathbf{q}_1+\mathbf{q}_2+\mathbf{q}_3-\mathbf{q}_4-\mathbf{q}_5-\mathbf{q}_6,\bm{0}} \nonumber\\ 
&
+v_{\phi,2} \bigg[ \phi^\dag(\mathbf{q}_1,\tau)\phi^\dag(\mathbf{q}_2,\tau)\phi^\dag(\mathbf{q}_3,\tau)
\phi^\dag(\mathbf{q}_4,\tau)\phi^\dag(\mathbf{q}_5,\tau)\phi^\dag(\mathbf{q}_6,\tau) 
+\text{h.c.}\bigg] \delta_{\mathbf{q}_1+\mathbf{q}_2+\mathbf{q}_3+\mathbf{q}_4+\mathbf{q}_5+\mathbf{q}_6,\bm{0}}
\bigg\} 
\label{eq:Lphi_6}
\ea 
\textcolor{black}{We also comment that $\phi^\dag \phi^\dag \phi^\dag \phi^\dag$ type of terms are not allowed due to the momentum conservation.}
After including the sixth-order term, the free energy of the ansatz in \cref{eq:phi_exp_val} becomes
\ba 
F_{\phi}[\rho_0,\theta_0] = N(\omega_{soft,H}^2\rho_0^2 + u_\phi \rho_0^4 +v_{\phi,1} \rho_0^6) + Nv_{\phi,2}2\rho_0^6\cos(6\theta_0)
\ea 
This leads to the following saddle-point equations
\ba 
0=&\frac{\delta F_{\phi,0}}{\delta \rho_0 } = N(2\omega_{soft,H}^2\rho_0 + 4u_\phi\rho_0^3 + 6 v_{\phi,1}\rho_0^5 + 12 v_{\phi,2} \cos(6\theta_0)\rho_0^5 ) \nonumber\\ 
0=&\frac{\delta F_{\phi,0}}{\delta \theta } =- N12 v_{\phi,2} \rho_0^6\sin(6\theta_0)
\label{eq:saddle_point_phi}
\ea 
We first discuss the solution of the $\theta_0$. There are two types of solutions
\ba 
\cos(6\theta_0) = 1,\quad \text{or} \quad \cos(6\theta_0) = -1 
\ea 
The solution that makes $v_{\phi,2} \cos(6\theta_0) \le 0 $ will always lower free energy. We thus always pick such solutions, i.e.
\ba 
\cos(6\theta_0) =1,\quad \text{if}\quad v_{\phi,2} < 0\nonumber \\ 
\cos(6\theta_0) =-1,\quad \text{if}\quad v_{\phi,2} > 0
\label{eq:theta_saddle_point}
\ea 
Combining \cref{eq:saddle_point_phi} and \cref{eq:theta_saddle_point}, we obtain the new saddle-point equation of $\rho$
\ba 
0 = N\rho_0 ( 2\omega_{soft,H}^2 + 4 u_\phi \rho_0^2 + 6v_{\phi,1} \rho_0^4 - 12 v_{\phi,2}\rho_0^4) 
\ea 
There is a trivial (disordered) solution with $\rho_0=0$. For the $\rho_0\ne 0$ solution, we note $v_{\phi,1}, v_{\phi,2}$ are high-order and are expected to be small. We obtain non-zero $\rho_0$ (ordered) solution via perturbation in $v_\phi,v_{\phi,2}$. In practice, we expand the solutions at non-zero $v_{\phi,1},v_{\phi,2}$ near the order solution at $v_{\phi,1}=v_{\phi,2}=0$, which gives 
\ba 
\rho_0 \approx  \sqrt{-\frac{\omega_{soft,H}^2}{2u_\phi} } \bigg(1  + \frac{3v_{\phi,1} -2 v_{\phi,2}}{8u^2} \bigg) 
\ea 
The free energy of the trivial solution is $F_\phi=0$ and the free energy of the ordered solution is (expanding in powers of $v_{\phi,1},v_{\phi,2}$)
\ba 
F_\phi \approx  -\frac{\omega_{soft,H}^2}{4u\phi}\bigg(1 + \frac{3\omega_{soft,H}^2(v_{\phi,1}-2v_{\phi,2}) }{2u_\phi^2} \bigg)
\ea 
The transition point approximately locates at 
\ba 
-\frac{\omega_{soft,H}^2}{4u\phi}\bigg(1 + \frac{3\omega_{soft,H}^2(v_{\phi,1}-2v_{\phi,2}) }{2u_\phi^2} \bigg) = 0 \Rightarrow \omega_{soft,H}^2 =0 
\ea

\subsection{$\tilde{{{K}}}$ order} 
We next consider the order parameter $\psi$ that describes the CDW transition. We have 4 non-equivalent $\tilde{{{K}}}$ points $\tilde{{{K}}}_1=2\pi(\frac{1}{3}\frac{1}{3}\frac{1}{3}), \tilde{{{K}}}_2=2\pi(\frac{-1}{3}\frac{-1}{3}\frac{1}{3}),\tilde{{{K}}}_3=2\pi( \frac{-1}{3}\frac{-1}{3}\frac{-1}{3}),, \tilde{{{K}}}_4=2\pi( \frac{1}{3}\frac{1}{3}\frac{-1}{3})$ that are connected by symmetries. We use the corresponding phonon fields $\tilde{u}_{CDW}$ that generate CDW structure transition as our order parameters, \textcolor{black}{which are also the phonon fields of the lowest-energy phonon bands at $\tilde{{{K}}}$ (and symmetry related) point }
\ba 
&\psi_1(\mathbf{q}) =\sqrt{2} \tilde{u}_{CDW}(\tilde{{{K}}}_1+\mathbf{q} ),\quad \psi_2(\mathbf{q}) =\sqrt{2} \tilde{u}_{CDW}(\tilde{{{K}}}_2+\mathbf{q}),\quad \nonumber\\ 
&
\psi_3 (\mathbf{q}) =\sqrt{2}\tilde{u}_{CDW} (\tilde{{{K}}}_3+\mathbf{q}),\quad 
\psi_4(\mathbf{q}) =\sqrt{2} \tilde{u}_{CDW}(\tilde{{{K}}}_4+\mathbf{q} )
\ea 
where $\mathbf{q}$ with $|\mathbf{q}|<\Lambda$ denotes the small momentum deviated from $\tilde{{{K}}}_i$. Moreover, 
experimentally, $u_{CDW}$ (or equivalently the $\psi$ fields) form a $\mathrm{P}_1$ irreducible representation of the little group at the momentum $\tilde{{{K}}}_i$\footnote{Similarly, in the \textit{ab-initio} spectrum shown in \cref{fig:zero_temperature_phonon_model}, the irrep of the lowest-energy phonon band is $\mathrm{H}_3$ at the $\mathrm{H}$ point, which implies through the compatibility relations~\cite{zotero-4159,BRA17,ELC17,VER17}{} that its irrep at the $\tilde{{{K}}}$ point is $\mathrm{P}_1$.}. The properties under symmetry transformations are
\ba 
g \psi_i(\mathbf{q})  g^{-1} = \sum_j D^{g,\psi}_{ij} \psi_j (g\mathbf{q}) 
\ea 
where 
\ba 
&D^{C_{3z},\psi} = \begin{bmatrix}
1 \\
& 1 \\ 
&& 1 \\ 
&&& 1
\end{bmatrix} ,\quad 
D^{C_{2z},\psi} = \begin{bmatrix}
& 1 &  &  \\ 
1 & &  &  \\ 
 & & &1 \\ 
 &  & 1 
\end{bmatrix} 
,\quad D^{C_{2,110},\psi} = \begin{bmatrix}
& &  & 1 \\ 
&  &1   \\ 
&  1& \\ 
1 & 
\end{bmatrix}  
,\quad D^{{i},\psi} = \begin{bmatrix} 
 &  & 1  \\ 
 &  & & 1  \\ 
1 & &  &  \\ 
& 1 & & 
\end{bmatrix} .
\ea 
Moreover, we also have 
\ba 
&\mathcal{T} \psi_1(\mathbf{q}) \mathcal{T}^{-1} = \psi_3(-\mathbf{q}),\quad \mathcal{T} \psi_3(\mathbf{q}) \mathcal{T}^{-1} = \psi_1(-\mathbf{q}),\quad 
\mathcal{T} \psi_2(\mathbf{q}) \mathcal{T}^{-1} = \psi_4(-\mathbf{q}) ,\quad \mathcal{T} \psi_4 (\mathbf{q})\mathcal{T}^{-1} = \psi_2(-\mathbf{q}) \nonumber\\ 
&T_\mathbf{R} \psi_i(\mathbf{q}) T_\mathbf{R}^{-1} = e^{i(\tilde{{{K}}}_i+\mathbf{q})\cdot \mathbf{R}} \psi_i(\mathbf{q})  \nonumber\\ 
&
\psi_1^\dag(\mathbf{q}) = \psi_3(-\mathbf{q}),\quad \psi_2^\dag(\mathbf{q}) = \psi_4 (-\mathbf{q})
\label{eq:rel_psi}
\ea 
It is sufficient to only consider $\psi_1$ and $\psi_2$, since $\psi_3(-\mathbf{q})=\psi_1(\mathbf{q})^\dag, \psi_4(-\mathbf{q}) = \psi_2^\dag(\mathbf{q})$. We have 
\ba 
&C_{3z} \psi_1(\mathbf{q}) C_{3z}^{-1} = \psi_1(C_{3z}\mathbf{q})
,\quad 
C_{3z} \psi_2(\mathbf{q}) C_{3z}^{-1} = \psi_2(C_{3z}\mathbf{q})
\nonumber\\ 
 &C_{2z}\psi_1(\mathbf{q}) C_{2z}^{-1} =\psi_2(C_{2z}\mathbf{q}) ,\quad 
 C_{2z}\psi_2(\mathbf{q}) C_{2z}^{-1} =\psi_1(C_{2z}\mathbf{q})\nonumber\\ 
&C_{2,110} \psi_1(\mathbf{q}) C_{2,110}^{-1} =\psi_2^\dag(-C_{2,110}\mathbf{q}) ,\quad 
C_{2,110} \psi_2(\mathbf{q}) C_{2,110}^{-1} =\psi_1^\dag(-C_{2,110}\mathbf{q}) \nonumber\\ 
&i \psi_1(\mathbf{q})i^{-1}=\psi_1^\dag (\mathbf{q}) ,\quad 
i \psi_2(\mathbf{q}) i^{-1}  =\psi_2^\dag(\mathbf{q}) \nonumber\\
&\mathcal{T} \psi_1(\mathbf{q})\mathcal{T}^{-1}=\psi_1^\dag(\mathbf{q})  ,\quad 
\mathcal{T}\psi_2(\mathbf{q})\mathcal{T}^{-1}  =\psi_2^\dag(\mathbf{q}) \nonumber\\
&T_\mathbf{R} \psi_1(\mathbf{q}) T_\mathbf{R}^{-1}  = \psi_1(\mathbf{q})e^{i\tilde{{{K}}}_1\cdot \mathbf{R}} 
,\quad T_\mathbf{R} \psi_2(\mathbf{q}) T_\mathbf{R}^{-1}  = \psi_2(\mathbf{q}) e^{i\tilde{{{K}}}_2\cdot \mathbf{R}} 
\ea

The symmetry allowed the Langrangian of $\psi$ fields takes the form of 
\ba 
L_\psi[\psi_1(\mathbf{q},\tau),\psi_2(\mathbf{q},\tau)]  = \sum_{|\mathbf{q}|<\Lambda i=1,2} \bigg[ \partial_\tau \psi_i(\mathbf{q},\tau) \partial_\tau \psi_i^\dag(\mathbf{q},\tau) + \omega_{CDW, \tilde{{{K}}}_i+\mathbf{q}}^2 \psi_i(\mathbf{q},\tau) \psi_i^\dag(\mathbf{q},\tau) \bigg] + V_\psi[ \psi_1(\mathbf{q}, \tau), \psi_2(\mathbf{q},\tau)]
\label{eq:Lpsi_v0}
\ea 
where $\omega_{CDW,\mathbf{q}}$ denotes the dispersion of the $\tilde{u}_{CDW}(\mathbf{q})$ phonon. $V_\psi(\psi_1(\mathbf{q},\tau),\psi_2(\mathbf{q},\tau))$ is a function of $\psi_{1,2}$ fields denoting the interaction term that is generated by electron-phonon coupling (as discussed near \cref{eq:log_exp_high_order}). $\psi_i(\mathbf{q},\tau)$ denote the $\psi_i(\mathbf{q})$ fields at time-slice $\tau$. 
We next expand $V_\psi$ in powers of $\psi_i$ fields and we truncate to quadratic order 
\ba 
&V_\psi[\psi_1(\mathbf{q},\tau),\psi_2(\mathbf{q},\tau)] \nonumber\\
= & \sum_{|\mathbf{q}_1|<\Lambda,|\mathbf{q}_2|<\Lambda,|\mathbf{q}_3|<\Lambda}\frac{\gamma}{2N^{1/2}}\sum_{i=1,2}\bigg[ \psi_i(\mathbf{q}_1,\tau) \psi_i(\mathbf{q}_2,\tau) \psi_i(\mathbf{q}_3,\tau) +\psi^\dag_i(\mathbf{q}_1,\tau) \psi^\dag_i(\mathbf{q}_2,\tau) \psi^\dag_i(\mathbf{q}_3,\tau) \bigg]\delta_{\mathbf{q}_1+\mathbf{q}_2+\mathbf{q}_3,\bm{0}} \nonumber\\ 
& +
 \sum_{|\mathbf{q}_1|<\Lambda,|\mathbf{q}_2|<\Lambda,|\mathbf{q}_3|<\Lambda,|\mathbf{q}_4|<\Lambda}\frac{u_{\psi,1}}{N} \sum_i \psi_i(\mathbf{q}_1,\tau) \psi_i(\mathbf{q}_2,\tau) \psi_i^\dag(\mathbf{q}_3,\tau) \psi_i^\dag(\mathbf{q}_4,\tau) \delta_{\mathbf{q}_1+\mathbf{q}_2-\mathbf{q}_3 -\mathbf{q}_4, \bm{0}}  \nonumber\\ 
 &+
 \sum_{|\mathbf{q}_1|<\Lambda,|\mathbf{q}_2|<\Lambda,|\mathbf{q}_3|<\Lambda,|\mathbf{q}_4|<\Lambda}\frac{u_{\psi,2}}{N} \psi_1(\mathbf{q}_1,\tau) \psi_2(\mathbf{q}_2,\tau) \psi_1^\dag(\mathbf{q}_3,\tau) \psi_2^\dag(\mathbf{q}_4,\tau) \delta_{\mathbf{q}_1+\mathbf{q}_2-\mathbf{q}_3 -\mathbf{q}_4, \bm{0}}  \nonumber\\ 
\label{eq:Vpsi_exp}
\ea 
Several remarks are in order
\begin{itemize}
    \item We have ignored the momentum $\mathbf{q}$ dependency in the interaction coefficient for the interactions at all orders in $\psi$. 
    \item The second-order (bilinear) contributions have already been absorbed by the dispersion term $\omega_{CDW,\tilde{{{K}}}_i+\mathbf{q}}$
    \item The momentum-independent interaction coefficient of the cubic term $\gamma$ can now be non-zero. 
\end{itemize}

Combining \cref{eq:Lpsi_v0} and \cref{eq:Vpsi_exp}, we have 
\ba 
&L_\psi[\psi_1(\mathbf{q},\tau),\psi_2(\mathbf{q},\tau)] \nonumber\\ 
= &\sum_{|\mathbf{q}|<\Lambda i=1,2} \bigg[ \partial_\tau \psi_i(\mathbf{q},\tau) \partial_\tau \psi_i^\dag(\mathbf{q},\tau) + \omega_{CDW, \tilde{{{K}}}_i+\mathbf{q}}^2 \psi_i(\mathbf{q},\tau) \psi_i^\dag(\mathbf{q},\tau) \bigg] \nonumber\\ 
&+\sum_{|\mathbf{q}_1|<\Lambda,|\mathbf{q}_2|<\Lambda,|\mathbf{q}_3|<\Lambda}\frac{\gamma}{2N^{1/2}}\sum_{i=1,2}\bigg[ \psi_i(\mathbf{q}_1,\tau) \psi_i(\mathbf{q}_2,\tau) \psi_i(\mathbf{q}_3,\tau) +\psi^\dag_i(\mathbf{q}_1,\tau) \psi^\dag_i(\mathbf{q}_2,\tau) \psi^\dag_i(\mathbf{q}_3,\tau) \bigg]\delta_{\mathbf{q}_1+\mathbf{q}_2+\mathbf{q}_3,\bm{0}} \nonumber\\ 
& +
 \sum_{|\mathbf{q}_1|<\Lambda,|\mathbf{q}_2|<\Lambda,|\mathbf{q}_3|<\Lambda,|\mathbf{q}_4|<\Lambda}\frac{u_{\psi,1}}{N} \sum_i \psi_i(\mathbf{q}_1,\tau) \psi_i(\mathbf{q}_2,\tau) \psi_i^\dag(\mathbf{q}_3,\tau) \psi_i^\dag(\mathbf{q}_4,\tau) \delta_{\mathbf{q}_1+\mathbf{q}_2-\mathbf{q}_3 -\mathbf{q}_4, \bm{0}}  \nonumber\\ 
 &+
 \sum_{|\mathbf{q}_1|<\Lambda,|\mathbf{q}_2|<\Lambda,|\mathbf{q}_3|<\Lambda,|\mathbf{q}_4|<\Lambda}\frac{u_{\psi,2}}{N} \psi_1(\mathbf{q}_1,\tau) \psi_2(\mathbf{q}_2,\tau) \psi_1^\dag(\mathbf{q}_3,\tau) \psi_2^\dag(\mathbf{q}_4,\tau) \delta_{\mathbf{q}_1+\mathbf{q}_2-\mathbf{q}_3 -\mathbf{q}_4, \bm{0}} 
 \label{eq:Lpsi}
\ea

We first study the phase transitions of $\psi_1,\psi_2$ fields using \cref{eq:Lpsi_v0} only, where the effect of $\phi$ fields are dropped here. 
The partition function and free energy are
\ba 
&Z_\psi = \int D[\psi_1,\psi_2] e^{-\int_0^\beta  L_\psi[\psi_1(\mathbf{q},\tau),\psi_2(\mathbf{q},\tau)]d\tau } \nonumber\\ 
&F_\psi  =-\frac{1}{\beta}\log(Z_\psi ) 
\ea 

We use the saddle-point approximation, where 
\ba 
&Z_\psi  \approx e^{-\int_0^\beta L_\psi[ \psi_1^c(\mathbf{q},\tau),\psi_2^c(\mathbf{q},\tau)] d\tau }\nonumber\\ 
&F_\psi \approx \frac{1}{\beta}\int_0^\beta L_\psi[ \psi_1^c(\mathbf{q},\tau),\psi_2^c(\mathbf{q},\tau)]d\tau 
\ea 
and 
\ba 
\frac{\delta L_\psi[\psi_1(\mathbf{q},\tau),\psi_2(\mathbf{q},\tau)]}{\delta \phi(\mathbf{q},\tau)}\bigg|_{\psi_i = \psi_i^c } = 0
\label{eq:saddle_point_psi}
\ea 
We take the following ansatz of the saddle point solutions 
\ba 
\psi^c_{i} (\mathbf{q},\tau)= \sqrt{N}\rho_ie^{i\theta_i}\delta_{\mathbf{q},0}
\label{eq:psi_exp_val}
\ea 
where $\rho_i >0 $ and $\theta_i$ are the amplitude and phase that characterize $\psi_i$ fields.

The Ginzburg Landau free energy can be introduced as (using \cref{eq:Lpsi} and \cref{eq:psi_exp_val}) 
\ba 
F_\psi[\rho_1,\rho_2,\theta_1,\theta_2] =& L_\psi[ \psi_1= \psi_1^c, \psi_2 = \psi_2^c ] \nonumber\\ 
=&N \bigg( \sum_{i=1,2} (\omega_{CDW,\tilde{{{K}}}}^2\rho_i^2 + \gamma \rho_i ^3 \cos(3\theta_i) + u_{\psi,1} \rho_i^4 ) + u_{\psi,2} \rho_1^2 \rho_2^2\bigg) 
\label{eq:GL_thy_psi}
\ea 
Here we point out that the quadratic term can be written as 
\ba 
\begin{bmatrix}
    \rho_1^2 \\ \rho_2^2 
\end{bmatrix}
\begin{bmatrix}
    u_{\psi,1} & u_{\psi,2}/2 \\ 
    u_{\psi,2}/2 & u_{\psi,1}
\end{bmatrix} \begin{bmatrix}
    \rho_1^2 & \rho_2^2 
\end{bmatrix}
\ea 
and we require the matrix $\begin{bmatrix}
    u_{\psi,1} & u_{\psi,2}/2 \\ 
    u_{\psi,2}/2 & u_{\psi,1}
\end{bmatrix}$ is positive definite ($u_{\psi,1}^2 > (u_{\psi,2}/2)^2 )$, otherwise the free energy could go to negative infinity by developing an infinite large order parameter, which is unphysical. In practice, we will assume $2u_{\psi,1} > u_{\psi,2}$ to simplify the calculation. 

The saddle-point equations (\cref{eq:psi_exp_val}) which are also the saddle-point equations of $F_\psi$ can be written as
\ba 
&0 =\frac{1}{N} \frac{\delta F_\psi [\rho_1,\rho_2, \theta_1,\theta_2]}{\delta \rho_1} = 2 \omega_{CDW,\tilde{{{K}}}}^2\rho_1 +3 \gamma\cos(3\theta_1)\rho_1^2 + 4u_{\psi,1} \rho_1^3 + 2u _{\psi,2 }\rho_1 \rho_2^2 \nonumber \\ 
&0= \frac{1}{N}\frac{\delta F_\psi [\rho_1,\rho_2, \theta_1,\theta_2]}{\delta \rho_2} = 2 \omega_{CDW,\tilde{{{K}}}}^2\rho_2 +3 \gamma\cos(3\theta_2)\rho_2^2 + 4u_{\psi,1} \rho_2^3 + 2u _{\psi,2 }\rho_2 \rho_1^2 \nonumber\\ 
& 0= \frac{1}{N}\frac{\delta F_\psi [\rho_1,\rho_2, \theta_1,\theta_2]}{\delta \theta_i} = -3\gamma \rho_i^3 \sin(3\theta_i),\quad i=1,2
\ea 

We then aim to find the saddle-point solutions that minimize free energy $F_\psi$. We first consider the term containing $\theta_i$. We find the following solution will minimize the free energy 
\ba 
&\theta_i = 0,\quad \text{if} \quad  \gamma < 0\nonumber\\
&\theta_i = \frac{\pi}{3},\quad \text{if} \quad \gamma > 0 
\label{eq:theta_i_sol}
\ea  
\textcolor{black}{In addition, $\theta_i=2\pi/3$ is also a solution to the saddle-point equation but has higher energy compared to the choices in \cref{eq:theta_i_sol}. }
We use $\theta_c$ to label the optimal value in \cref{eq:theta_i_sol}, and then 
\ba 
\frac{1}{N}F_\psi[\rho_1,\rho_2, \theta_c,\theta_c] =& \sum_{i=1,2}( \omega_{CDW,\tilde{{{K}}}} \rho_i^2 + \gamma \rho_i^3\cos(3\theta_c) + u_{\psi,1}|\rho_i|^4) + u_{\psi,2}\rho_1^2 \rho_2^2 \nonumber\\ 
=&\sum_{i=1,2}( \omega_{CDW,\tilde{{{K}}}} \rho_i^2 -|\gamma|\rho_i^3 + u_{\psi,1}|\rho_i|^4) + u_{\psi,2}\rho_1^2 \rho_2^2 
\ea 
Then it is sufficient to solve the saddle-point equation of $\rho_1,\rho_2$ with $\gamma \cos(3\theta_i) = -|\gamma|$, i.e.
\ba 
&0  = 2 \omega_{CDW,\tilde{{{K}}}}^2\rho_1 -3 |\gamma|\rho_1^2 + 4u_{\psi,1} \rho_1^3 + 2u _{\psi,2 }\rho_1 \rho_2^2 \nonumber \\ 
&0=  2 \omega_{CDW,\tilde{{{K}}}}^2\rho_2 -3 |\gamma|\rho_2^2 + 4u_{\psi,1} \rho_2^3 + 2u _{\psi,2 }\rho_2 \rho_1^2 
\ea 
There are three types of solutions: trivial solution, single-$\tilde{{{K}}}$ solution and double $\tilde{{{K}}}$ solution.  

\textbf{ Trivial solution} that describes the non-ordered state 
    \ba 
    \rho_1 = \rho_2 =0 
    \ea 
    
\textbf{Single-$\tilde{{{K}}}$ order} with only one of $\rho_{1,2}$ is non-zero. There are four possible solutions
\ba 
(\rho_1,\rho_2) = (0, \rho_{c,1}) ,\quad (\rho_1,\rho_2) = (0, \rho_{c,2}), \quad (\rho_1,\rho_2) = (\rho_{c,1},0),\quad (\rho_1,\rho_2) = ( \rho_{c,2},0 ) 
\label{eq:single_k_sol}
\ea 
where 
\ba 
\rho_{c,1} = \frac{3|\gamma|-\sqrt{9\gamma^2-32u_{\psi,1}\omega_{CDW,\tilde{{{K}}}}^2}}{8u_{\psi,1}},\quad 
\rho_{c,2} = \frac{3|\gamma|+\sqrt{9\gamma^2+32u_{\psi,1}\omega_{CDW,\tilde{{{K}}}}^2}}{8u_{\psi,1}},\quad 
\ea 
The corresponding free energy is 
\ba 
&F_{\psi}[\rho_{c,1},0,\theta_{c},\theta_c] =F_{\psi}[0,\rho_{c,1},\theta_{c},\theta_c] = N\frac{(\Delta -3|\gamma|)^2[ (\Delta -3|\gamma|)|\gamma|+16u_{\psi,1} \omega_{soft, \tilde{{{K}}}}^2]}{2048u_{\psi,1}^3}\nonumber\\ 
&
F_{\psi}[\rho_{c,2},0,\theta_{c},\theta_c]=F_{\psi}[0,\rho_{c,2},\theta_{c},\theta_c] =N \frac{(\Delta -3|\gamma|)^2[- (\Delta +3|\gamma|)|\gamma|+16u_{\psi,1} \omega_{sof,t\tilde{{{K}}}}^2]}{2048u_{\psi,1}^3} \nonumber\\ 
&\Delta = \sqrt{9\gamma^2-32u_{\psi,1}\omega_{CDW,\tilde{{{K}}}}^2}
\ea 
We note that this solution in \cref{eq:single_k_sol} exists only when $\Delta^2>0$. We also note that  
\ba 
F_{\psi}[\rho_{c,1},0,\theta_{c},\theta_c] - F_{\psi}[\rho_{c,2},0,\theta_{c},\theta_c]  =N \frac{ |\gamma|\Delta^3}{256u_{\psi,1}^3} \ge 0 
\ea 
which indicates $\rho_{c,1}$ solution always has a higher energy than $\rho_{c,2}$ and it is sufficient to only consider the following $\rho_{c,2}$ solution
of 
\ba 
(\rho_1,\rho_2) = (\rho_{c,2},0),\quad (\rho_1,\rho_2) = ( 0,\rho_{c,2} ) 
\ea 
where has lower free energy than $\rho_{c,1}$

\textbf{Double-$\tilde{{{K}}}$ order} with both $\rho_1,\rho_2$ are non-zero. There are four solutions
    \ba 
    &(\rho_1,\rho_2) = (\rho_{c,3},\rho_{c,3}),\quad (\rho_1,\rho_2) =(\rho_{c,4},\rho_{c,4}) \nonumber\\ 
    &(\rho_1,\rho_2) = (\rho_{c,5}^a , \rho_{c,5}^b ),\quad  (\rho_1,\rho_2) = (\rho_{c,5}^b , \rho_{c,5}^a )
    \ea 
    We first analyze $(\rho_{c,3},\rho_{c,3}),(\rho_{c,4},\rho_{c,4})$, where we find
    \ba 
    \rho_{c,3} = \frac{3|\gamma| -\Delta_{c,3} }{4(2u_{\psi,1}+u_{\psi,2})},\quad \rho_{c,4} = \frac{3|\gamma| + \Delta_{c,3} }{4(2u_{\psi,1}+u_{\psi,2})},\quad  \Delta_{c,3}= \sqrt{ 9|\gamma|^2 - 16 \omega_{CDW, \tilde{{{K}}} }^2 (2u_{\psi,1} +u_{\psi,2})}
    \ea 
    where the solutions are only valid if
    \ba 
    9|\gamma|^2 \ge  16 \omega^2_{CDW,\tilde{{{K}}}}(2u_{\psi,1} + u_{\psi,2} )
    \ea 
    In addition, we also find 
    \ba 
    F[\rho_{c,3},\rho_{c,3} ,\theta_c, \theta_c] -   F[\rho_{c,4},\rho_{c,4} ,\theta_c, \theta_c]  = \frac{ N|\gamma |\Delta_{c,3} ^3}{16(2u_{\psi,1} +u_{\psi,2})^3 } > 0
    \ea 
    which means $(\rho_{c,4}, \rho_{c,4})$ is a better solution with lower free energy compared to $(\rho_{c,3},\rho_{c,3})$, and we do not need to consider $(\rho_{c,3},\rho_{c,3})$ solution. 
The condition that $\rho_{c,4}$ has a lower energy than the disorder-solution is
    \ba F[\rho_{c,4},\rho_{c,4},\theta_c,\theta_c] < 0 \Rightarrow  \omega^2_{CDW,\tilde{{{K}}}}(2u_{\psi,1} + u_{\psi,2} ) < \frac{\gamma^2}{2} 
    \ea

    The $\rho_{c,5}^a,\rho_{c,5}^b $ are 
    \ba 
  &  \rho_{c,5}^a  = \frac{3\gamma  + \Delta_{c,5} }{4(2u_{\psi,1} - u_{\psi,2})} ,\quad 
  \rho_{c,5}^b =   \frac{3\gamma  - \Delta_{c,5} }{4(2u_{\psi,1} - u_{\psi,2})} \nonumber\\ 
  &\Delta_{c,5} = \sqrt{\frac{9|\gamma|^2 (2u_{\psi,1}-3u_{\psi,2}) -16\omega_{CDW,\tilde{{{K}}}}^2(2u_{\psi,1}-u_{\psi,2})^2   }{ 2u_{\psi,1} + u_{\psi,2}}  }
    \ea 
    where the solutions are valid only when $\rho_{c,5}^{a},\rho_{c,5}^{b}$ are non-negative real numbers and
    \ba 
    9|\gamma|^2 (2u_{\psi,1}-3u_{\psi,2} ) \ge  16 \omega^2_{soft ,\tilde{{{K}}}}(2u_{\psi,1}-u_{\psi,2})^2 ,\quad 3\gamma \ge \Delta_{c,5}
    \ea 
We also find the condition where the ordered solution has a lower energy than the disordered solutions, which is
    \ba     &F[\rho_{c,5}^a,\rho_{c,5}^b,\theta_c,\theta_c]
    = F[\rho_{c,5}^b,\rho_{c,5}^a,\theta_c,\theta_c]
    \le 0 \nonumber \\ 
   \Rightarrow  & \omega_{CDW,\tilde{{{K}}}}^2 > \frac{9\gamma^2}{8(2u_{\psi_1}-u_{\psi,1})^{3/2}}\bigg[ 1 +\sqrt{ (2u_{\psi,1}+u_{\psi,2})/(6u_{\psi,1}-3u_{\psi,2})} \bigg]  \nonumber\\ 
  & \text{ or } \omega_{CDW,\tilde{{{K}}}}^2 < \frac{9\gamma^2}{8(2u_{\psi_1}-u_{\psi,1})^{3/2}}\bigg[ 1 -\sqrt{ (2u_{\psi,1}+u_{\psi,2})/(6u_{\psi,1}-3u_{\psi,2})} \bigg] 
    \ea

\textcolor{black}{To determine the phase of the system, one needs to compare the free energy of each solution and find the one with lowest free energy.}
We also comment that, due to the cubic term, the order-disorder phase transition we find here is, in general, a first-order phase transition with a jump in the order parameter.

\subsection{Effective field theory of both order parameters}
We next consider the coupling between two order parameters $L_{\phi\psi}$. Again, we perform an expansion in powers of $\phi, \psi_1,\psi_2$ and truncate to quadratic order, and we do not include the momentum-independency in the interaction coefficients. We find
\ba 
&L_{\phi \psi}[\phi(\mathbf{q},\tau),\psi_1(\mathbf{q},\tau),\psi_2(\mathbf{q},\tau)] = V_{\phi \psi}[\phi(\mathbf{q},\tau),\psi_1(\mathbf{q},\tau),\psi_2(\mathbf{q},\tau)] \nonumber\\ 
= & \sum_{|\mathbf{q}_1|<\Lambda,|\mathbf{q}_2|<\Lambda,|\mathbf{q}_3|<\Lambda,|\mathbf{q}_4|<\Lambda}\frac{u_{\phi\psi}}{N}\sum_{i=1,2} \phi(\mathbf{q}_1,\tau)\phi^\dag(\mathbf{q}_2,\tau) \psi_i(\mathbf{q}_3,\tau) \psi^\dag_i(\mathbf{q}_4,\tau) \delta_{\mathbf{q}_1-\mathbf{q}_2+\mathbf{q}_3-\mathbf{q}_4,\bm{0}} \nonumber\\ 
 & +\sum_{|\mathbf{q}_1|<\Lambda,|\mathbf{q}_2|<\Lambda,|\mathbf{q}_3|<\Lambda,|\mathbf{q}_4|<\Lambda}\frac{u_{\phi\psi,2}}{N} 
 \bigg[\phi^\dag(\mathbf{q}_1,\tau) \phi^\dag(\mathbf{q}_2,\tau)
 \psi_1(\mathbf{q}_3,\tau) \psi^\dag_{2}(\mathbf{q}_4,\tau) +\text{h.c.}\bigg]\delta_{\mathbf{q}_1+\mathbf{q}_2-\mathbf{q}_3+\mathbf{q}_4,\bm{0}}
\label{eq:Lphipsi}
\ea

Combing \cref{eq:Lphi}, \cref{eq:Lpsi} and \cref{eq:Lphipsi}, the final Lagrangian takes the form of 
\ba 
&L[\phi(\mathbf{q},\tau),\psi_1(\mathbf{q},\tau),\psi_2(\mathbf{q},\tau) ] 
= L_\phi[ \phi(\mathbf{q},\tau) ] + L_\psi[\psi_1(\mathbf{q},\tau),\psi_2(\mathbf{q},\tau)] + L_{\phi\psi}[\phi(\mathbf{q},\tau),\psi_1(\mathbf{q},\tau),\psi_2(\mathbf{q},\tau) ] 
\label{eq:Lall}
\ea 
The partition function and the free energy of the system can be written as 
\ba 
Z =& \int D[\phi, \psi_1,\psi_2] e^{-\int_0^\beta L[\phi(\mathbf{q},\tau),\psi_1(\mathbf{q},\tau),\psi_2(\mathbf{q},\tau)] d\tau }\nonumber \\
F =& -\frac{1}{\beta} \log(Z). 
\label{eq:part_Z}
\ea 

We again first study the model with saddle-point approximation 
\ba 
&Z \approx \exp\bigg( -\int_0^\beta L[\phi^c(\mathbf{q},\tau),\psi_1^c(\mathbf{q},\tau),\psi_2^c(\mathbf{q},\tau)]d\tau\bigg) \nonumber\\ 
&F \approx \frac{1}{\beta} \int_0^\beta L[\phi^c(\mathbf{q},\tau),\psi_1^c(\mathbf{q},\tau),\psi_2^c(\mathbf{q},\tau)]d\tau 
\ea 
We take the following ansatz (\cref{eq:phi_exp_val} and \cref{eq:psi_exp_val})
\ba 
\phi^c(\mathbf{q},\tau) =\sqrt{N}\rho_0 e^{i\theta_0}\delta_{\mathbf{q},0},\quad 
\psi_i^c(\mathbf{q},\tau) =\sqrt{N}\rho_i e^{i\theta_i}\delta_{\mathbf{q},0}
\label{eq:saddle_point_ansatz}
\ea 
The Landau free energy and the saddle-point equations are
 \ba 
 F[\rho_0,\theta_0,\rho_1\theta_1,\rho_2,\theta_2] = &N\bigg(\omega_{soft,H}^2 \rho_0 +u_\phi \rho_0^4 + \sum_{i=1,2} (\omega_{CDW,\tilde{{{K}}}}^2\rho_i^2 + \gamma \rho_i ^3 \cos(3\theta_i) + u_{\psi,1} \rho_i^4 ) + u_{\psi,2} \rho_1^2 \rho_2^2 \nonumber\\ 
 &+u_{\phi \psi} \rho_0^2(\rho_1^2+\rho_2^2)
 +2u_{\phi\psi,2}\rho_0^2 \rho_1\rho_2 \cos(2\theta_0-\theta_1+\theta_2)
 \bigg) 
 \label{eq:saddle_point_fe}
 \ea 
and the saddle-point equations are 
\ba 
0= \frac{\delta F[\rho_0,\theta_0,\rho_1\theta_1,\rho_2,\theta_2]}{\delta \rho_0} = & N\bigg[ 2\omega_{soft,H}^2\rho_0 + 4u_\phi\rho_0^3 +2 u_{\phi\psi}(\rho_1^2+\rho_2^2)\rho_0 
+4u_{\phi\psi,2}\rho_0\rho_1\rho_2 \cos(2\theta_0-\theta_1+\theta_2)
\bigg] \nonumber\\ 
0=\frac{\delta F[\rho_0,\theta_0,\rho_1\theta_1,\rho_2,\theta_2]}{\delta \theta_0} = &N\bigg[-4u_{\phi\psi,2} \rho_0^2\rho_1\rho_2\sin(2\theta_0-\theta_1+\theta_2)
\bigg] \nonumber\\ 
0=\frac{\delta F[\rho_0,\theta_0,\rho_1\theta_1,\rho_2,\theta_2]}{\delta \rho_1} =&N\bigg[2 \omega_{CDW,\tilde{{{K}}}}^2\rho_1 +3 \gamma\cos(3\theta_1)\rho_1^2 + 4u_{\psi,1} \rho_1^3 + 2u _{\psi,2 }\rho_1 \rho_2^2 
+2u_{\phi\psi}\rho_0^2 \rho_1 \nonumber\\ 
&
 +
2u_{\phi\psi,2}\rho_0^2\rho_2\cos(2\theta_0-\theta_1+\theta_2)
\bigg]  \nonumber\\ 
0=\frac{\delta F[\rho_0,\theta_0,\rho_1\theta_1,\rho_2,\theta_2]}{\delta \rho_2} =&N\bigg[ 2 \omega_{CDW,\tilde{{{K}}}}^2\rho_2 +3 \gamma\cos(3\theta_2)\rho_2^2 + 4u_{\psi,1} \rho_2^3 + 2u _{\psi,2 }\rho_2 \rho_1^2 
+2u_{\phi\psi}\rho_0^2 \rho_2 
\nonumber\\ 
&
 +
2u_{\phi\psi,2}\rho_0^2\rho_1\cos(2\theta_0-\theta_1+\theta_2)
\bigg]\nonumber\\ 
0= \frac{\delta F [[\rho_0,\theta_0,\rho_1\theta_1,\rho_2,\theta_2]}{\delta \theta_1} =& -3N\gamma \rho_i^3 \sin(3\theta_1) + 2Nu_{\phi\psi,2}\rho_0^2\rho_1\rho_2\sin(2\theta_0-\theta_1+\theta_2)
\nonumber\\
0= \frac{\delta F [[\rho_0,\theta_0,\rho_1\theta_1,\rho_2,\theta_2]}{\delta \theta_2} =& -3N\gamma \rho_i^3 \sin(3\theta_2) -2Nu_{\phi\psi,2}\rho_0^2\rho_1\rho_2\sin(2\theta_0-\theta_1+\theta_2)
\label{eq:saddel_point_eq}
\ea

Here we consider the cases with 
\ba 
u_{\phi\psi} \ge 0 
\label{eq:coexist}
\ea 
which will suppress the coexistent phase with both $\tilde{{{K}}}$ and $H$ orders. We note that if $u_{\phi\psi}<0$, one could reduce the energy via $u_{\phi\psi}$ term by making both $\rho_0 >0 $ and $\rho_1^2 +\rho_2^2>0$.
In other words, the coexistence phase will be favored. 
However, for $u_{\phi\psi} \ge 0 $, the energy is reduced by making either $\rho_0=0$ or $\rho_1^2 +\rho_2^2 =0 $. In practice, the coexistence phase could still exist (even though they are suppressed) with $u_{\phi \psi} \ge 0$, when the energy loss of the coexistence phase from $L_{\phi},L_\psi$ is larger than the energy gain of the coexistence phase from $L_{\phi\psi}$. {In addition, $u_{\phi \psi,2}$ term could also favor the coexistence phase with $\rho_0\ne 0, \rho_1\ne 0, \rho_2 \ne 0$ when the angle $\theta_{0,1,2}$ satisfies $u_{\phi\psi, 2}\cos(2\theta_0-\theta_1+\theta_2)<0$. But, whenever $u_{\phi\psi}$ is positive much larger than $|u_{\phi \psi,2}|$, the suppression of the coexistence phase from $u_{\phi\psi}$ is dominant.}  
Since the coexistence phase is not observed experimentally~\cite{KOR23}, we take $u_{\phi\psi}  \ge 0 $ and $u_{\phi\psi} >> |u_{\phi\psi,2}|$ to suppress the coexistence phase with both order parameters being non-zero.

In principle, there could be four types of phases
\begin{itemize}
    \item Disorder phase. $\rho_0=\rho_1=\rho_2=0$. 
    \item $H$-ordered phase: $\rho_0\ne 0, \rho_1=\rho_2=0$. 
    \item $\tilde{{{K}}}$-ordered phase: $\rho_0=0$, one of or both $\rho_1,\rho_2$ is non-zero. 
    \item Coexistence phase: $\rho_0 \ne 0$ and one of or both $\rho_1,\rho_2$ is non-zero. 
\end{itemize}
We note that, for the $H$ order solution with $\rho_1=\rho_2=0$, the saddle-point equations of $\rho_0$ in \cref{eq:saddel_point_eq} are the same as \cref{eq:saddle_point_phi}. For the $\tilde{{{K}}}$ order solution with $\rho_0=0$, the saddle-point equations of $\rho_1,\rho_2$ are the same as \cref{eq:saddle_point_psi}. 

\begin{figure}
    \centering
    \includegraphics[width=0.6\textwidth]{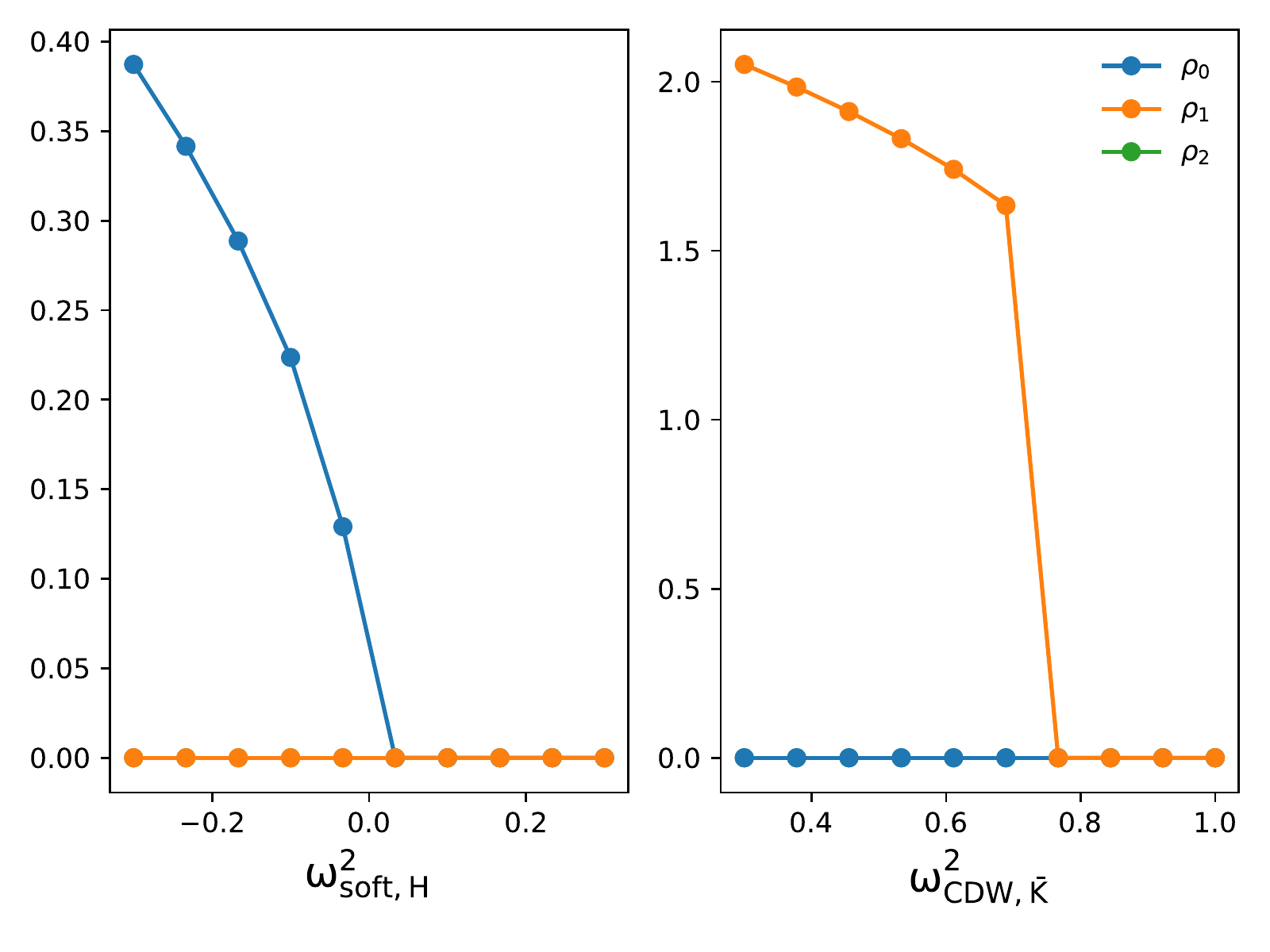}
    \caption{Evolution of order parameters at $u_\phi=1.0$, $u_{\psi,1}=0.33$, $u_{\psi,2}=0.5u_{\psi,1}$, $u_{\phi \psi}=1.0, u_{\phi\psi,2}=0.1u_{\phi\psi,2},\gamma=-1.0$. The order parameters at each point are obtained by minimizing the free energy at the saddle-point level. 
    (Left): Fix $\omega^2_{CDW,\tilde{{{K}}}}=0.5$ and tune $\omega^2_{soft,H}$. We observe a second-order phase transition from disorder phase $\rho_0=\rho_1=\rho_2=0$ to a $H$-order phase $\rho_0\ne 0, \rho_1=\rho_2=0$. (Right) Fix $\omega^2_{soft,H}=0.5 $ and tune $\omega^2_{CDW,\tilde{{{K}}}}$. We observe a first-order phase transition from a disordered phase $\rho_0=\rho_1=\rho_2=0$ to a $\tilde{{{K}}}$ order phase (single-$\tilde{{{K}}}$ order) with $\rho_0=\rho_2=0, \rho_1\ne 0$. }
    \label{fig:sad_pt_op}
\end{figure}

In \cref{fig:sad_pt_op}, we illustrate two phase transitions for given two sets of parameters as described in the caption: (1) second-order phase transition between disorder phase and $H$-order phase; (2) first-order phase transition between disorder phase and $\tilde{{{K}}}$-order (single-$\tilde{{{K}}}$ order) phase.

\subsection{Gaussian fluctuations} 
\begin{figure}
    \centering
    \includegraphics[width=0.6\textwidth]{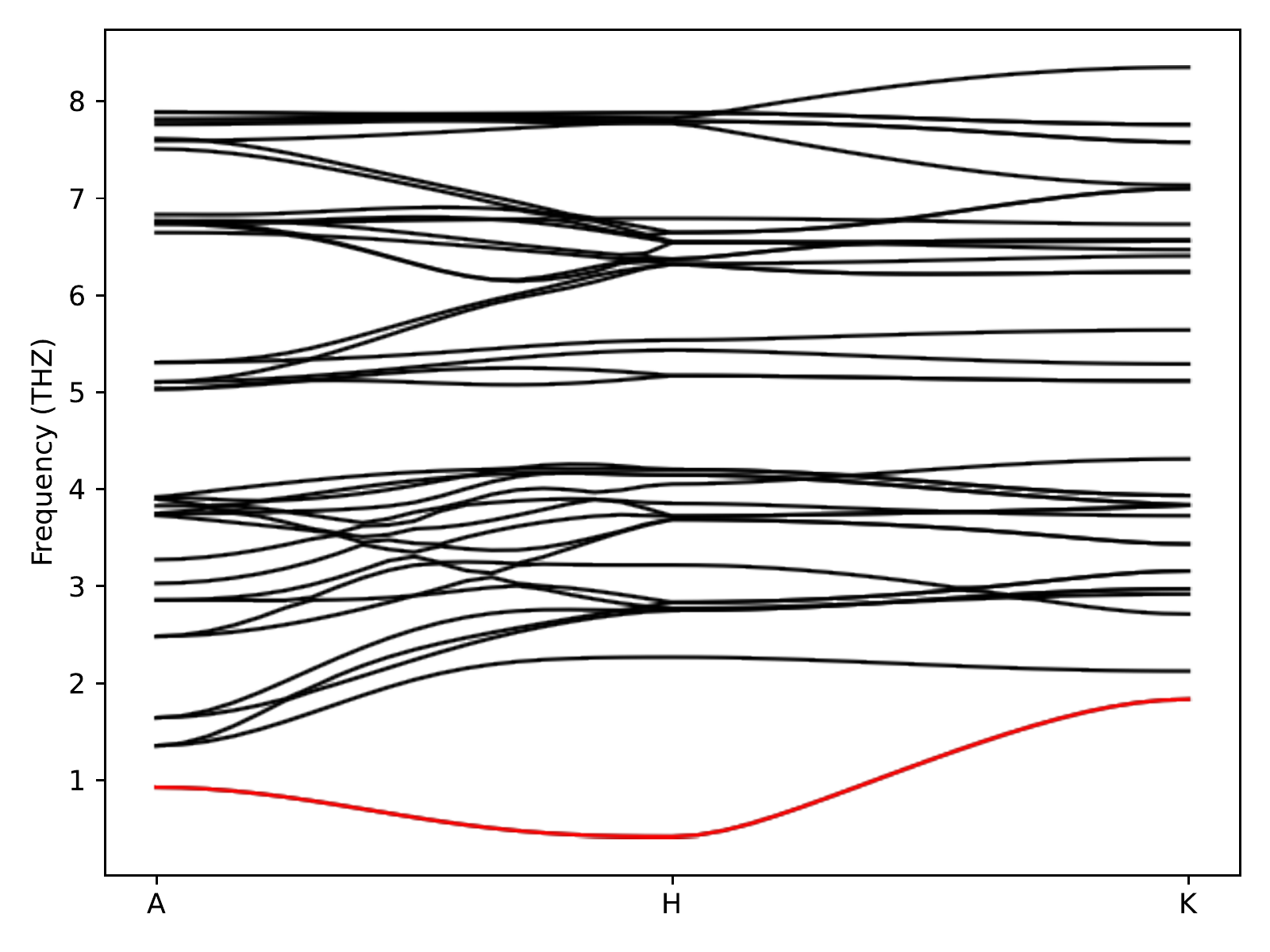}
    \caption{Phonon dispersion from DFT calculations near $H$ point at the temperature ($T=0.3$eV) close to the collapsing temperature of the $H$-point phonon. We can observe the lowest-energy phonon band (red) is extremely flat near $H$. }
    \label{fig:flat_phonon_H}
\end{figure}

We next study the phase transition when there is a strong fluctuation of $H$ order parameter. 
Experimentally, as the temperature is reduced, the energy gap of the phonon mode at $H$ point decreases (equivalently $\omega_{soft,H}$ decreases). Moreover, the calculated phonon bands are extremely flat near $H$ points (\cref{fig:flat_phonon_H}). Consequently, the fluctuations of $\phi$ fields are relatively strong.

To understand the effect of the fluctuations of $\phi$ fields, we include the Gaussian fluctuation on top of the saddle-point solutions. Here, we only consider the fluctuation of $\phi$ fields, which is much stronger than $\psi_1,\psi_2$ fields, as the phonon band acquires large dispersion around $\bar{K}$ but is relatively flat around $H$. This is because the phonon band reaches its minimum at $H$ point. Consequently, the phonon mode near $H$ ($\phi$ fields) has a smaller gap than the phonon mode near $\tilde{{{K}}}$ ($\psi_1,\psi_2$ fields).

We first separate $\phi$ into the saddle-point contribution and fluctuation 
\ba 
\phi(\mathbf{q},\tau) = \phi^{c}(\mathbf{q},\tau) + x(\mathbf{q},\tau) 
\ea 
where $\phi^{c}(\mathbf{q},\tau)$ denote saddle-point solutions of $\phi$ field and $x(\mathbf{q},\tau)$ denote fluctuations around saddle-point solutions. 
For the $\psi$ fields, we only include the saddle-point contribution
\ba 
\psi_i(\mathbf{q},\tau) =  \psi_i^{c}(\mathbf{q},\tau)
\ea 
where $\psi_c^c$ denote the saddle-point solutions of $\psi$ field. We take the following ansatz of the saddle-point solution which has already been used in \cref{eq:saddle_point_ansatz}
\ba 
\phi^c(\mathbf{q},\tau) =\sqrt{N}\rho_0 e^{i\theta_0}\delta_{\mathbf{q},0},\quad 
\psi_i^c(\mathbf{q},\tau) =\sqrt{N}\rho_i e^{i\theta_i}\delta_{\mathbf{q},0}
\ea

We then expand the Langrangian in powers of $x$
\ba 
L[\phi(\mathbf{q},\tau),\psi_1(\mathbf{q},\tau), \psi_2(\mathbf{q},\tau)] 
=& L[\rho_0,\theta_0,\rho_1,\theta_1,\rho_2,\theta_2] +L_x\nonumber\\ 
L_x=&  
\sum_{\mathbf{q}}M(\mathbf{q})  x^\dag(\mathbf{q},\tau)x(\mathbf{q},\tau) + \frac{1}{2}\sum_{\mathbf{q}} \bigg[N(\mathbf{q}) x(\mathbf{q},\tau) x(-\mathbf{q},\tau) +\text{h.c.}\bigg] 
\ea 
where $L_x$ denotes the effective action of the fluctuation fields
\ba 
M(\mathbf{q}) = \frac{\delta L[\phi(\mathbf{q},\tau),\psi_1(\mathbf{q},\tau),\psi_2(\mathbf{q},\tau)]}{\delta \phi(\mathbf{q},\tau) \delta \phi^\dag(\mathbf{q},\tau)}\bigg|_{\phi = \phi^c, \psi_1 = \psi_1^c, \psi_2 = \psi_2^c}
\nonumber\\ 
N(\mathbf{q}) = \frac{\delta L[\phi(\mathbf{q},\tau),\psi_1(\mathbf{q},\tau),\psi_2(\mathbf{q},\tau)]}{\delta \phi(\mathbf{q},\tau) \delta \phi(-\mathbf{q},\tau)}\bigg|_{\phi = \phi^c, \psi_1 = \psi_1^c, \psi_2 = \psi_2^c} \nonumber\\ 
\ea 
where $\phi^c, \psi_1^c, \psi_2^2$ denotes the saddle-point solutions. 
Written $M(\mathbf{q}),N(\mathbf{q})$ explicitly, we obtain the following effective action $L_x$
\ba 
L_x = & \sum_{|\mathbf{q}|<\Lambda }\bigg[ \partial_\tau x(\mathbf{q},\tau) \partial_\tau x^\dag(\mathbf{q},\tau) + \omega^2_{soft,H+\mathbf{q}} |x(\mathbf{q},\tau)|^2 \bigg] \nonumber\\ 
&
+ \sum_{|\mathbf{q}|<\Lambda} u_\phi \bigg[ \rho_0^2 e^{2i\theta_0} x^\dag(\mathbf{q},\tau)x^\dag(-\mathbf{q},\tau)  + \rho_0^2e^{-2i\theta_0} x(\mathbf{q},\tau)x(-\mathbf{q},\tau) + 4\rho_0^2 x^\dag(\mathbf{q},\tau) x(\mathbf{q},\tau) \bigg]  \nonumber\\ 
&+\sum_{|\mathbf{q}|<\Lambda}  u_{\psi \phi} \bigg(\rho_1^2+\rho_2^2\bigg) x^\dag(\mathbf{q},\tau)x(\mathbf{q},\tau)  \nonumber\\ 
&+ \sum_{|\mathbf{q}|<\Lambda} u_{\psi\phi,2} \bigg[ \rho_1\rho_2 e^{i(\theta_1-\theta_2)}x^\dag(\mathbf{q},\tau)x^\dag(-\mathbf{q},\tau)+\text{h.c.}\bigg]
\ea  
In practice, we observe that the experimental observed CDW phase corresponds to the single-$\tilde{{{K}}}$ phase with either $\rho_1 =0,\rho_2\ne 0 $ or $\rho_2 =0, \rho_1 \ne 0$. This leads to the fact that $\rho_1\rho_2 = 0$. Therefore, for the $\tilde{{{K}}}$-order, we will always focus on the saddle-point solution of single-$\tilde{{{K}}}$ order with $\rho_1\rho_2=0$. This allows us to drop the $u_{\psi\phi,2}$ term or, equivalently, set
\ba 
u_{\psi\phi,2}=0
\label{eq:upsiphi2}
\ea 
since $u_{\psi\phi,2}$ term is always proportional to $\rho_1\rho_2$ and vanishes. For what follows, we will always assume $u_{\psi\phi,2}=0$ in order to simplify the construction of the model. The effective Lagrangian becomes
\ba 
L_x = & \sum_{|\mathbf{q}|<\Lambda }\bigg[ \partial_\tau x(\mathbf{q},\tau) \partial_\tau x^\dag(\mathbf{q},\tau) + \omega^2_{soft,H+\mathbf{q}} |x(\mathbf{q},\tau)|^2 \bigg] \nonumber\\ 
&
+ \sum_{|\mathbf{q}|<\Lambda} u_\phi \bigg[ \rho_0^2 e^{2i\theta_0} x^\dag(\mathbf{q},\tau)x^\dag(-\mathbf{q},\tau)  + \rho_0^2e^{-2i\theta_0} x(\mathbf{q},\tau)x(-\mathbf{q},\tau) + 4\rho_0^2 x^\dag(\mathbf{q},\tau) x(\mathbf{q},\tau) \bigg]  \nonumber\\ 
&+\sum_{|\mathbf{q}|<\Lambda}  u_{\psi \phi} \bigg(\rho_1^2+\rho_2^2\bigg) x^\dag(\mathbf{q},\tau)x(\mathbf{q},\tau)  
\ea  
where we can always perform a global $U(1)$ transformation
\ba 
x(\mathbf{q},\tau) \rightarrow e^{ -i \theta_0} x(\mathbf{q},\tau) 
\label{eq:u1gauge}
\ea 
to remove the phase factor of $\theta_0$. Then we have 
\ba 
L_x = & \sum_{|\mathbf{q}|<\Lambda }\bigg[ \partial_\tau x(\mathbf{q},\tau) \partial_\tau x^\dag(\mathbf{q},\tau) + \omega^2_{soft,H+\mathbf{q}} |x(\mathbf{q},\tau)|^2 \bigg] \nonumber\\ 
&
+ \sum_{|\mathbf{q}|<\Lambda} u_\phi \rho_0^2
\bigg(x^\dag(\mathbf{q},\tau) x^\dag(-\mathbf{q},\tau) + x(\mathbf{q},\tau)x(-\mathbf{q},\tau) + 4 x^\dag(\mathbf{q},\tau) x(\mathbf{q},\tau) \bigg) 
\nonumber\\ 
&+\sum_{|\mathbf{q}|<\Lambda}  u_{\psi \phi} \bigg(\rho_1^2+\rho_2^2\bigg) x^\dag(\mathbf{q},\tau)x(\mathbf{q},\tau)  
\ea 

We now introduce two new fields
\ba 
y(\mathbf{q},\tau) = \frac{1}{2} (x^\dag(\mathbf{q},\tau) +x(-\mathbf{q},\tau) ) \nonumber\\ 
z(\mathbf{q},\tau) = \frac{i}{2} (x^\dag(\mathbf{q},\tau) -x(-\mathbf{q},\tau) ) 
\ea 
then
\ba 
L_x = & \sum_{|\mathbf{q}|<\Lambda }\bigg[ \partial_\tau y(\mathbf{q},\tau) \partial_\tau y(-\mathbf{q},\tau) + \bigg( \omega^2_{soft,H+\mathbf{q}} 
+6u_\phi \rho_0^2 + u_{\psi \phi} (\rho_1^2+\rho_2^2) 
\bigg) y(\mathbf{q},\tau) y(-\mathbf{q},\tau)\bigg] \nonumber\\ 
& \sum_{|\mathbf{q}|<\Lambda }\bigg[ \partial_\tau z(\mathbf{q},\tau) \partial_\tau z(-\mathbf{q},\tau) + \bigg( \omega^2_{soft,H+\mathbf{q}} +2u_{\phi}\rho_0^2 
+u_{\psi \phi} (\rho_1^2+\rho_2^2)
\bigg) z(\mathbf{q},\tau) z(-\mathbf{q},\tau)\bigg]  
\label{eq:gasu_thy}
\ea  
We next introduce Fourier transformation and work with bosonic Matsubara frequency
\ba 
&y(\mathbf{q},i\Omega_n) = \int_0^\beta y(\mathbf{q},\tau)e^{i\Omega_n\tau} d\tau ,\quad 
z(\mathbf{q},i\Omega_n) = \int_0^\beta z(\mathbf{q},\tau)e^{i\Omega_n\tau} d\tau 
\ea 
Then the action behaves as 
 \ba 
S_x =& \int_0^\beta L_xd\tau \nonumber\\ 
=&\frac{1}{\beta} \sum_{i\Omega_n, |\mathbf{q}|<\Lambda }
 \bigg[\bigg(
 \Omega_n^2 +\omega^2_{soft,H+\mathbf{q}}+6u_\phi \rho_0^2 + u_{\psi \phi} (\rho_1^2+\rho_2^2)\bigg) |y(\mathbf{q},i\Omega_n)|^2 \nonumber\\ 
&+\bigg(
 \Omega_n^2 +\omega^2_{soft,H+\mathbf{q}}+2u_\phi \rho_0^2 + u_{\psi \phi} (\rho_1^2+\rho_2^2)\bigg) |z(\mathbf{q},i\Omega_n)|^2\bigg] 
 \label{eq:Sx}
 \ea 
 
We are now in the position to calculate the free energy with Gaussian fluctuation included. The partition function now reads 
\ba 
Z = \int D[y,z] \exp\bigg( 
- \beta L[\rho_0e^{i\theta_0} , \rho_1e^{i\theta_1}, \rho_2 e^{i\theta_2} ] - S_x\bigg) 
\ea 
Since $S_x$ is just a bilinear Langraingian of $y,z$ fields, we could use Gaussian integral to integrate out $y,z$ fields which gives
\ba 
Z \propto&  e^{-\beta L[\rho_0e^{i\theta_0} , \rho_1e^{i\theta_1}, \rho_2 e^{i\theta_2} ] }\nonumber\\ 
&
\prod_{\Omega_n}\prod_\mathbf{q} \bigg( \Omega_n^2 + \omega_{soft,H+\mathbf{q}}^2 + 6u_\phi \rho_0^2 +u_{\psi \phi}(\rho_1^2 +\rho_2^2)\bigg)^{-1/2}
\bigg( \Omega_n^2 + \omega_{soft,H+\mathbf{q}}^2 + 2u_\phi \rho_0^2 +u_{\psi \phi}(\rho_1^2 +\rho_2^2)\bigg)^{-1/2}
\ea 
Then the free energy with Gaussian fluctuation reads
\ba 
&F[\rho_0,\theta_0,\rho_1,\theta_1,\rho_2,\theta_2] = -\frac{1}{\beta}\log(Z) \nonumber\\ 
=&L[\rho_0 e^{i\theta_0}, \rho_1e^{i\theta_1},\rho_2e^{i\theta_2}] 
\nonumber\\ 
&+ \frac{1}{2\beta} \sum_{\mathbf{q},\Omega_n}\bigg\{\log\bigg[ \Omega_n^2 +\omega^2_{soft,H+\mathbf{q}}+6u_\phi \rho_0^2 + u_{\psi \phi} (\rho_1^2+\rho_2^2)\bigg] +
\log\bigg[ \Omega_n^2 +\omega^2_{soft,H+\mathbf{q}}+2u_\phi \rho_0^2 + u_{\psi \phi} (\rho_1^2+\rho_2^2)\bigg] \bigg\} 
\label{eq:one_loop_free_energy}
\ea 
In practice, in order to simplify the calculation, we assumed the dispersion $\omega_{soft,H+\mathbf{q}}$ is isotropic with (small) velocity $c$
\ba 
\omega_{soft,H+\mathbf{q}}^2 = \omega_{soft,H}^2 + c^2|\mathbf{q}|^2 
\label{eq:pho_vel}
\ea 

Now, in order to determine the phase, we do the following
\begin{itemize}
    \item Find the saddle-point solutions by solving saddle-point equation \cref{eq:saddel_point_eq}.
    \item Calculate the free energy with Gaussian fluctuations \cref{eq:one_loop_free_energy}.
    \item The phase is determined by the saddle-point solution with the smallest free energy (\cref{eq:one_loop_free_energy}). 
\end{itemize}

We first describe three solutions we have at saddle-point level (here we assume we work with parameter region without coexistence phase): 1. disordered solution; 2. $H$-order solution; 3. $\tilde{{{K}}}$-order solution as introduced below \cref{eq:saddel_point_eq}. 
Three types of solutions are 
\begin{itemize}
    \item  Disorder: $\rho_0=0,\rho_1=0,\rho_2=0$. 
    \item  $H$-order: $\rho_0 = \sqrt{-\omega_{soft,H}^2 /(2u_\phi)}, \rho_1=\rho_2=0$ from \cref{eq:phi_sol}
    \item $\tilde{{{K}}}$-order: $\rho_0 = 0$, $\rho_1=\rho_{1,opt} ,\rho_2 =\rho_{2,opt}$. 
\end{itemize}
In practice, we could realize various $\tilde{{{K}}}$-order solutions. However, the experimentally observed CDW phase corresponds to the single-$\tilde{{{K}}}$ order. Therefore, we will assume we are in the parameter region, such that the single $\tilde{{{K}}}$-order has lower energy than the double-$\tilde{{{K}}}$ order to simplify our consideration.  
We use $\rho_{1,opt},\rho_{2,opt}$ to label the corresponding single-$\tilde{{{K}}}$ solution. \textcolor{black}{Since we are considering the single-$\tilde{{{K}}}$ phase, we always have $\rho_{1,opt}\rho_{opt,2}=0$.} 
The Ginzburg Landau free energies of three solutions (free energy at saddle-point level) are
\begin{itemize}
    \item Disorder: 
    \ba F_{dis,0} =0 
    \label{eq:f0dis}
    \ea 
    \item $H$-order: from \cref{eq:phi_sol}.
    \ba F_{H,0}=-N\frac{\omega_{soft,H}^4}{4 u_\phi }
    \label{eq:f0H}
    \ea 
    
    \item $\tilde{{{K}}}$-order: from \cref{eq:GL_thy_psi}
    \ba 
    F_{\tilde{{{K}}},0}=\omega_{CDW,\tilde{{{K}}}}^2 (|\rho_{1,opt}|^2 +|\rho_{2,opt}|^2) 
    -3|\gamma|(\rho_{1,opt}^2+\rho_{2,opt}^2)
    + u_{\psi,1}(|\rho_{1,opt}|^4+|\rho_{2,opt}|^4) 
    \label{eq:f0kb}
    \ea  
    \textcolor{black}{where we also use the fact that $\rho_{1,opt}\rho_{opt,2}=0$.} 
\end{itemize}

\begin{figure}
    \centering
    \includegraphics[width=1.0\textwidth]{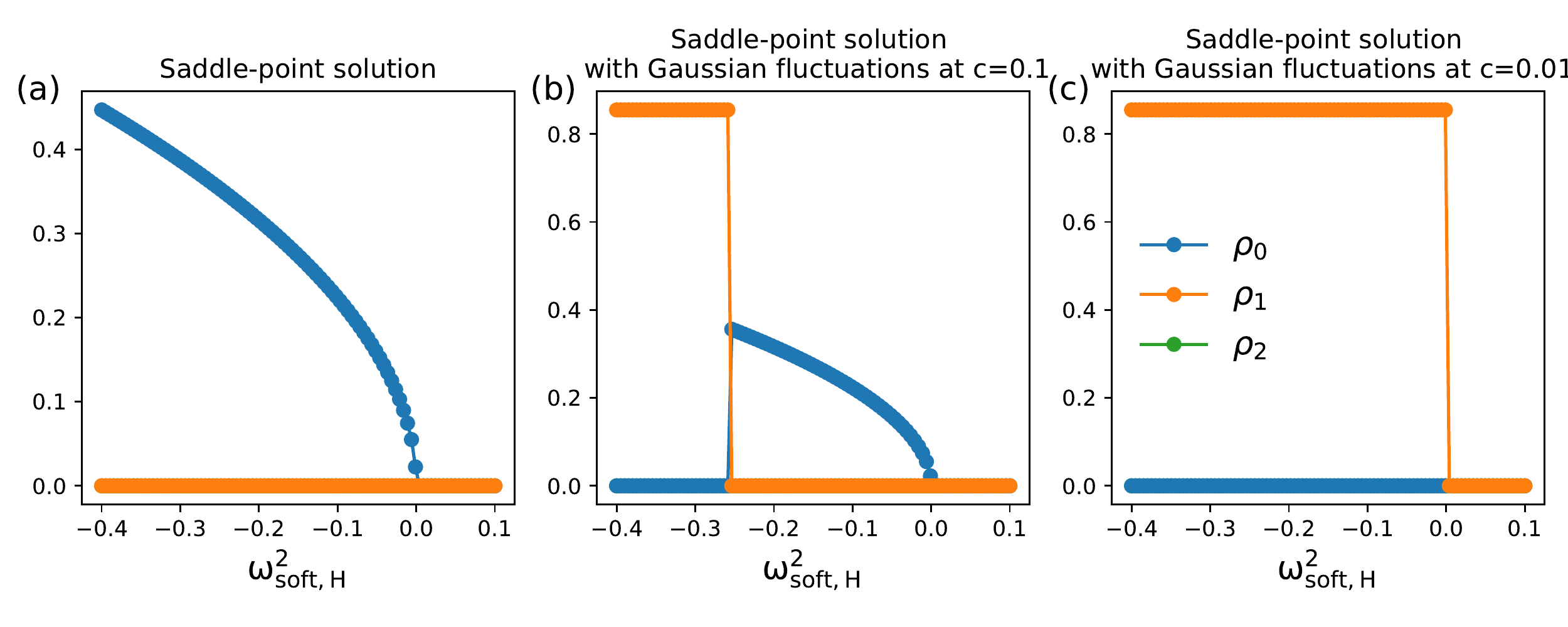}
    \caption{Solutions of the parameter setting given in \cref{eq:parameter}. (a) Saddle-point solutions without Gaussian fluctuations where a second-order phase transition from the disordered phase to the $H$ order phase has been observed. (b) (c) Saddle-point solution with Gaussian fluctuations included at different phonon velocities $c$. At a relatively large $c$, as we tune $\omega^2_{soft, H}$, we first observe a second-order transition to the $H$-order phase and then a first-order transition to the $\tilde{{{K}}}$-order (single-$\tilde{{{K}}}$) phase. At a relatively small $c$, the parameter region of $H$-order phase is suppressed (and is invisible in the right panel due to the finite number of parameter points). A first-order transition to the $\tilde{{{K}}}$-order (single-$\tilde{{{K}}}$) phase almost immediately happens after $\omega^2_{soft,H}$ reaching zero. This is likely the experimental situation. }
    \label{fig:op_fluct}
\end{figure}

We now discuss the parameters we considered. Our goal is to understand the experimentally observed phase transition driven by the collapsing of phonon at $H$. To mimic the experimental situations, we fix all the parameters except for $\omega_{soft,H}^2$. We treat $\omega_{soft,H}^2$, which is the lowest energy of phonon at $H$, as our tuning parameters. 
\textcolor{black}{We note that, for the $\bar{K}$-solution with $\rho_0=0$, the saddle-point equations (\cref{eq:saddel_point_eq}) and also the free energy of $\bar{K}$-solution (\cref{eq:saddle_point_fe}) do not depend on $\omega_{soft,H}^2$ due to the fact that $\rho_0=0$. Thus $\rho_{1,opt},\rho_{2,opt} $ and $F_{\tilde{{{K}}},0}$ do not depend on $\omega_{soft,H}^2$ and are also fixed.} 
We now assume that we are in the parameter region such that
\ba 
0=F_{dis,0} <F_{\tilde{{{K}}},0} 
\label{eq:pd_assumption}
\ea 
which means, at the current parameter setting and at saddle-point level, $\tilde{{{K}}}$ order is not energetically favored (otherwise we are already in the $\tilde{{{K}}}$-order phase without tuning any parameters). In practice,
pick the following set of parameters as an example to realize \cref{eq:pd_assumption}
\ba 
u_{\phi} = 1.0, \quad u_{\psi,1}=0.33,\quad u_{\psi,2} = 0.5,\quad \gamma =-1.0,\quad  u_{\phi\psi }  = 1.0,\quad  \omega_{CDW,\tilde{{{K}}}}^2 = 0.8 
\label{eq:parameter}
\ea 
\textcolor{black}{This set of parameters is found via the following procedure. We first let $u_{\phi}=u_{\phi\psi} = -\gamma =1.0$. We then pick a relatively large $u_{\psi,2} = 0.5 u_{\phi}$ to suppress the double-$\bar{K}$ phase. We note that $u_{\psi,2}$ appears in the free energy (\cref{eq:saddle_point_fe}) in the form of $u_{\psi,2} \rho_1^2 \rho_2^2$, then a single-$\bar{K}$ phase will be favored by a relatively large and positive $u_{\psi,2}$ by making $\rho_1^2\rho_2^2=0$. We next pick different choices of $u_{\psi,1}, \omega^2_{CDW,\bar{K}}$ with $u_{\psi,1} \in [0,1], \omega^2_{CDW,\bar{K}} \in [0,1]$ and identify one set of the parameters that could satisfy \cref{eq:pd_assumption}.
In principle, the choice of parameters has some arbitrariness here. But by satisfying \cref{eq:pd_assumption}, we are able to mimic the experimental setup that the high-temperature phase is disorder. 
}
In addition, in this parameter region, the single-$\tilde{{{K}}}$ solution has lower energy than the double-$\tilde{{{K}}}$ solution. 
We first discuss the phase diagram at the saddle point level by tuning $\omega_{soft,H}^2$. This leads to two phases (\cref{fig:op_fluct}, right panel)
\begin{itemize}
    \item Disorder phase at $\omega_{soft,H}^2 >  0$
    \item $H$-order phase at $\omega_{soft,H}^2< 0$
\end{itemize}
with a second-order phase transition happens at $\omega_{soft,H}^2=0$ as shown in \cref{fig:op_fluct} (left panel).

We now investigate the effect of Gaussian fluctuations and the phase diagram with Gaussian fluctuations. The contribution to the free energy from Gaussian fluctuation (from \cref{eq:one_loop_free_energy}) takes the form of 
\ba 
F_1 = \frac{1}{2\beta} \sum_{\mathbf{q},\Omega_n}\log\bigg\{\bigg[ \Omega_n^2 +\omega^2_{soft,H+\mathbf{q}}+6u_\phi \rho_0^2 + u_{\psi \phi} (\rho_1^2+\rho_2^2)\bigg] +
\log\bigg[ \Omega_n^2 +\omega^2_{soft,H+\mathbf{q}}+2u_\phi \rho_0^2 + u_{\psi \phi} (\rho_1^2+\rho_2^2)\bigg] \bigg\}
\ea 
Therefore, we find the contribution to the free energy from Gaussian fluctuation of three solutions are
\ba 
\text{Disorder}:&F_{dis,1} = \frac{1}{\beta} \sum_{\mathbf{q},\Omega_n}\bigg\{\log\bigg[ \Omega_n^2+\omega_{soft,H}^2+ c^2|\mathbf{q}|^2 \bigg]  \bigg\}\nonumber\\ 
H\text{-order}:&F_{H,1} = \frac{1}{2\beta} \sum_{\mathbf{q},\Omega_n}\bigg\{\log\bigg[ \Omega_n^2-2\omega_{soft,H}^2+ c^2|\mathbf{q}|^2 \bigg] +
\log\bigg[ \Omega_n^2 +c^2|\mathbf{q}|^2 \bigg] \bigg\}\nonumber\\ 
\tilde{{{K}}}\text{-order}:&F_{\tilde{{{K}}}, 1} = \frac{1}{\beta} \sum_{\mathbf{q},\Omega_n}\bigg\{\log\bigg[ \Omega_n^2+\omega_{soft,H}^2+M^2 + c^2|\mathbf{q}|^2 \bigg] \bigg\}
\label{eq:f1_gaus}
\ea 
where 
\ba 
M^2 = u_{\psi \phi}(\rho_{1,opt}^2+\rho_{2,opt}^2) 
\ea 
and $F_{dis,1},F_{H,1},F_{\tilde{{{K}}},1}$ denote the contribution to the free energy from the Gaussian fluctuations of disorder phase, $H$-order phase and $\tilde{{{K}}}$-order phase respectively.

Here, we comment on the stability of the phases under Gaussian fluctuations. From \cref{eq:gasu_thy}, the dispersion of fluctuation fields $y,z$ are 
\ba 
E_{1,\mathbf{q}}^2 =  \omega_{soft,H}^2 + c^2 |\mathbf{q}|^2 +6u_\phi \rho_0^2 + u_{\psi \phi} (\rho_1^2+\rho_2^2)  ,\quad E_{2,\mathbf{q}}^2 =\omega_{soft,H}^2+ c^2|\mathbf{q}|^2 +2u_\phi \rho_0^2 + u_{\psi \phi} (\rho_1^2+\rho_2^2) 
\ea  
Only when both $E_{1,\mathbf{q}}^2 \ge 0, E_{2,\mathbf{q}}^2\ge 0$, the phases can be stable under Gaussian fluctuation. Otherwise, $y,z$ fields could condense and destroy the phases that are obtained from the saddle-point equation. \textcolor{black}{For three set types of saddle-point solutions (disorder, $H$-order, $\tilde{{{K}}}$-order), the following stability conditions can be obtained by requiring $E_{1,\mathbf{q}}^2 \ge 0, E_{2,\mathbf{q}}^2\ge 0$}
\begin{itemize}
    \item Disorder: 
    \ba 
     \omega_{soft,H}^2 \ge 0  \label{eq:stab_dis_ord}
    \ea 
    \item $H$ order:
   \ba 
     \omega_{soft,H}^2\le 0 \label{eq:stab_h_ord}
   \ea 
    \item $\tilde{{{K}}}$ order:
    \ba 
    M^2 + \omega_{soft,H}^2 \ge 0  \label{eq:stab_kb_ord}
    \ea 
\end{itemize}
\textcolor{black}{We also point out that these conditions indicate whether the saddle-point solution is stable under Gaussian fluctuations. The stable saddle-point solution does not necessarily characterize the phase of the system. To find the phase of the system, one needs to compare the free energies of the stable saddle-point solutions and find the one with the lowest free energy.}

Combining both the saddle-point contribution and Gaussian-fluctuation contribution, the final free energy of three phases are (\cref{eq:f0dis}, 
\cref{eq:f0H}, \cref{eq:f0kb}, \cref{eq:f1_gaus})
\ba 
&F_{dis} = F_{dis,0} +F_{dis,1}=\frac{1}{\beta} \sum_{\mathbf{q},\Omega_n}\log\bigg\{\bigg[ \Omega_n^2+\omega_{soft,H}^2+ c^2|\mathbf{q}|^2 \bigg]  \bigg\}\nonumber\\ 
&F_{H} =F_{H,0}+F_{H,1} =-N\frac{\omega_{soft,H}^4}{4u_\phi} +\frac{1}{2\beta} \sum_{\mathbf{q},\Omega_n}\log\bigg\{\bigg[ \Omega_n^2-2\omega_{soft,H}^2 + c^2|\mathbf{q}|^2 \bigg] +
\log\bigg[ \Omega_n^2 +c^2|\mathbf{q}|^2 \bigg] \bigg\}\nonumber\\ 
& F_{\tilde{{{K}}}}=F_{\tilde{{{K}}},0} +F_{\tilde{{{K}}},1} =F_{\tilde{{{K}}},0}+\frac{1}{\beta} \sum_{\mathbf{q},\Omega_n}\log\bigg\{\bigg[ \Omega_n^2+\omega_{soft,H}^2+M^2 + c^2|\mathbf{q}|^2 \bigg] \bigg\}
\ea 
where $F_{dis}, F_H, F_{\tilde{{{K}}}}$ denote the free energy (with both saddle-point contribution and Gaussian-fluctuation contribution included) of disorder phase, $H$-order phase and $\tilde{{{K}}}$-order phase respectively. Also $F_{\tilde{{{K}}},0}$ is treated as a constant, since it does not depend on our tuning parameters $\omega_{soft,H}^2$.

We now analyze the phase diagram.
We start from $\omega_{soft,H}^2 \ge 0$. In this case, $H$-order solution does not exist (\cref{eq:stab_h_ord}), we compare the free energy of $\tilde{{{K}}}$ order and disorder.
 \ba 
 \Delta F_{\tilde{{{K}}}, dis} = F_{\tilde{{{K}}}} - F_{dis} = F_{\tilde{{{K}}},0} +\frac{1}{\beta} \sum_{\mathbf{q},\Omega_n}\log\bigg[ \frac{ \Omega_n^2+m^2+M^2 + c^2|\mathbf{q}|^2 }{
\Omega_n^2+m^2 + c^2|\mathbf{q}|^2
 }
 \bigg] 
 \ea 
 We note that $F_{\tilde{{{K}}},0} >0$ from our assumption (\cref{eq:pd_assumption}), and 
 \ba 
 \frac{ \Omega_n^2+\omega_{soft,H}^2+M^2 + c^2|\mathbf{q}|^2 }{
\Omega_n^2+\omega_{soft,H}^2 + c^2|\mathbf{q}|^2
 } \ge 1 \Rightarrow \log 
 \bigg[ \frac{ \Omega_n^2+\omega_{soft,H}^2+M^2 + c^2|\mathbf{q}|^2 }{
\Omega_n^2+\omega_{soft,H}^2 + c^2|\mathbf{q}|^2
 }  \bigg] \ge 0 
 \ea 
 Thus 
 \ba 
 \Delta F_{\tilde{{{K}}}, dis} > 0 
 \label{eq: m2_ge_0}
 \ea 
 In other words, the system stays in the disorder phase at $\omega_{soft,H}^2> 0$.

We next consider $\omega_{soft,H}^2=0$, at $\omega_{soft,H}^2=0$, the order parameter of $H$-order phase $\rho_0=\sqrt{-\omega_{soft,H}^2/(2u_\phi)}=0$, and $H$-order solution is equivalent to disorder solution. We compare the energy between $H$-order solution with $\tilde{{{K}}}$ solution
\ba 
\Delta F_{\tilde{{{K}}}, H}\bigg|_{\omega_{soft,H}^2=0} = F_{\tilde{{{K}}}} \bigg|_{\omega_{soft,H}^2=0}- F_{H} \bigg|_{\omega_{soft,H}^2=0}= F_{\tilde{{{K}}},0} +\frac{1}{\beta} \sum_{\mathbf{q},\Omega_n}\log\bigg[ \frac{ \Omega_n^2+M^2 + c^2|\mathbf{q}|^2 }{
\Omega_n^2+ c^2|\mathbf{q}|^2}\bigg] 
\ea 
Since $F_{\tilde{{{K}}},0} >0 $ and 
\ba \log\bigg[ \frac{ \Omega_n^2+M^2 + c^2|\mathbf{q}|^2 }{
\Omega_n^2+ c^2|\mathbf{q}|^2}\bigg] > 0 
\ea 
we have 
\ba 
\Delta F_{\tilde{{{K}}}, H}\bigg|_{\omega_{soft,H}^2=0}  > 0
\label{eq:del_F_m2_0}
\ea 
and the system remains in the disorder phase at $\omega_{soft,H}^2=0$. 

 We next consider the case of $-M^2 < \omega_{soft,H}^2 < 0 $. 
\textcolor{black}{According to \cref{eq:stab_kb_ord}, $\tilde{{{K}}}$-order solution is stable under Gaussian fluctuations in this parameter region (because $\omega_{soft,H}^2 >-M^2$. $H$-order is also stable under Gaussian fluctuation (\cref{eq:stab_h_ord}).}
 \textcolor{black}{However, from \cref{eq:stab_dis_ord}, the disorder phase is no-longer stable under Gaussian fluctuations.}
Therefore, we only need to compare the energy between $\tilde{{{K}}}$ order and $H$ order \textcolor{black}{which are two stable solutions under Gaussian fluctuations}. 
 \ba 
 \Delta F_{\tilde{{{K}}}, H} = F_{\tilde{{{K}}}} -F_{H} = F_{\tilde{{{K}}},0} -F_{H,0} + \frac{1}{2\beta}\sum_{\mathbf{q},\Omega_n} \log
 \bigg(
 \frac{ 
 (\Omega_n^2 + c^2|\mathbf{q}|^2 +\omega_{soft,H}^2 +M^2 )^2 
 }{(\Omega_n^2 + c^2 |\mathbf{q}|^2 -2\omega_{soft,H}^2 ) (\Omega_n^2+c^2 |\mathbf{q}|^2)}
 \bigg) 
 \label{eq:df_order}
 \ea 
To understand the behavior of $\Delta F_{\tilde{{{K}}},H}$ 
, we define the following function 
\ba
G(-\omega_{soft,H}^2) = \Delta F_{\tilde{{{K}}}, H} = F_{\tilde{{{K}}},0} -F_{H,0} + \frac{1}{2\beta}\sum_{\mathbf{q},\Omega_n} \log
 \bigg(
 \frac{ 
 (\Omega_n^2 + c^2|\mathbf{q}|^2 +\omega_{soft,H}^2 +M^2 )^2 
 }{(\Omega_n^2 + c^2 |\mathbf{q}|^2 -2\omega_{soft,H}^2 ) (\Omega_n^2+c^2 |\mathbf{q}|^2)}
 \bigg) 
 \label{eq:def_G}
\ea 
We analyze the behavior of $G(-\omega_{soft,H}^2)$ in the region of $-M^2 \le -\omega_{soft,H}^2 \le 0$. 
When $G(-\omega_{soft,H}^2) < 0$, the system stays in $\tilde{{{K}}}$ order phase, and when $G(\omega_{soft,H}^2) >0 $, the system stays in the $H$ order phase. In addition, from \cref{eq:del_F_m2_0}, $G(0)>0$, which means, in the small region near $\omega_{soft,H}^2=0$ (and also with $\omega_{soft,H}^2<0$) the system stays in the $H$ order phase. 

We now demonstrate that as we reduce $\omega_{soft,H}^2$ from zero, $G(-\omega_{soft,H}^2)$ will change sign and a phase transition from the $H$ order phase to the $\tilde{{{K}}}$ order phase happens. 
To prove this, we will show $\lim_{\delta \rightarrow 0^+} G(-M^2+\delta) < 0$. 
\ba 
\lim_{\delta \rightarrow 0^+}G(-M^2+\delta)=  F_{\tilde{{{K}}},0}-F_{H,0} +\lim_{\delta \rightarrow 0^+}\frac{1}{2\beta}\sum_{\mathbf{q},\Omega_n}\log 
\bigg( 
\frac{(\Omega_n^2 + c^2 |\mathbf{q}|^2 + \delta ) }{(\Omega_n^2 + c^2 |\mathbf{q}|^2 + 2M^2)}
\label{eq:del_f}
\bigg) 
\ea 
Since we are interested in the finite-temperature transition with experimentally observed transition temperature $T\approx 8 $meV, we find $\Omega_{n=0} = 0$meV and $\Omega_{n>1}\ge \Omega_{n=1}=2\pi T  \approx 49$meV. Since $49$meV is much larger than the bandwidth of the lowest-energy phonon (around $5$meV). We thus only keep the $\Omega_{n=0}$ contribution (In the limit of $\Omega_n \rightarrow \infty$, the $\log( \frac{\Omega_n^2 + c^2 |\mathbf{q}|^2 + 2M^2}{\Omega_n^2 + c^2 |\mathbf{q}|^2 }) \rightarrow 0 $). We thus drop the contribution from $\Omega_{n\ne 0}$ and have
\ba 
 \lim_{\delta \rightarrow 0^+}G(-M^2+\delta) = &F_{\tilde{{{K}}},0} -F_{H,0}+\lim_{\delta \rightarrow 0^+}\frac{1}{2\beta}\sum_{\mathbf{q},\Omega_n}\log 
\bigg( 
\frac{(\Omega_n^2 + c^2 |\mathbf{q}|^2 + \delta) }{(\Omega_n^2 + c^2 |\mathbf{q}|^2 + 2M^2)}
\bigg) \nonumber\\ 
\approx& F_{\tilde{{{K}}},0} -F_{H,0}+\lim_{\delta \rightarrow 0^+}\frac{1}{2\beta}\sum_{\mathbf{q}}\log 
\bigg(  
\frac{( c^2 |\mathbf{q}|^2 + \delta) }{( c^2 |\mathbf{q}|^2 + 2M^2)}
\bigg) 
\nonumber\\ 
\approx& F_{\tilde{{{K}}},0} -F_{H,0}+\lim_{\delta \rightarrow 0^+}\frac{N}{2\beta A}\int_{|\mathbf{q}|<\Lambda} \log 
\bigg( 
\frac{( c^2 |\mathbf{q}|^2 + \delta) }{( c^2 |\mathbf{q}|^2 + 2M^2)}
\bigg) d^3 \mathbf{q} \nonumber\\ 
=& F_{\tilde{{{K}}},0} -F_{H,0}+\lim_{\delta \rightarrow 0^+}\frac{2\pi N}{\beta A}\int_0^{\Lambda} \log 
\bigg( 
\frac{( c^2 q^2 + \delta) }{( c^2 q^2  + 2M^2)}
\bigg) q^2 d q
\label{eq:f1_diff}
\ea 
where we have replaced the summation in the momentum space with the integral and $A$ is the area of the first Brillouin zone, $\Lambda$ is the momentum cutoff. We now calculate the integral. We first define the function 
\ba 
f(\delta/M^2) = \frac{2\pi N}{\beta A}\int_0^{\Lambda} \log 
\bigg( 
\frac{( c^2 q^2 + \delta) }{( c^2 q^2  + 2M^2)}
\bigg) q^2 d q
\label{eq:f1_diff_2}
\ea 
We evaluate \textcolor{black}{ $f(\delta/M^2)$ }
in the limit of small $\delta/M^2$. $\log(x)$ is an increasing function of $x$, so $f(\delta)$ is an increasing function of $\delta$. This allows us to set an upper limit of $\lim_{\delta \rightarrow 0 } f(\delta)$ 
\ba 
\lim_{\delta \rightarrow
 0} f(\delta/M^2)  \le  f(D^2/M^2)
 \label{eq:bound}
\ea 
where 
\ba 
D= c\Lambda 
\label{eq:phonon_band}
\ea 
We note that $D$ characterizes the ``bandwidth'' of the phonon dispersion near $H$ ($D$ is not the exact bandwidth of the lowest phonon band over the whole Brillouin zone because of the momentum cutoff, but it characterizes the dispersion of the lowest phonon band near $H$ point). We next calculate
\ba 
f(D^2/M^2) =&\frac{2\pi N}{\beta A}\int_0^{\Lambda} \log 
\bigg( 
\frac{( c^2 q^2 + c^2\Lambda^2 ) }{( c^2 q^2  + 2M^2)}
\bigg) q^2 d q \nonumber\\ 
=&\frac{2\pi N}{\beta A}\int_0^{\Lambda} 
\bigg\{ 
\log(D^2/(2M^2) ) +\sum_{n=1}^\infty \frac{(-1)^{n}}{n}\bigg[\bigg(\frac{cq}{\sqrt{2}M}\bigg)^{2n}  - \bigg(\frac{ cq}{D} \bigg)^{2n}\bigg] \bigg\}
q^2 d q \nonumber\\ 
=& \frac{2\pi N \Lambda^3 }{3\beta A}\log(\frac{D^2}{2M^2}) 
+\frac{2\pi N\Lambda^3}{3\beta A}\bigg[
2 -\frac{4M^2}{D^2} +\frac{4\sqrt{2}M^3}{D^3}\tan^{-1}(\frac{D}{\sqrt{2}M}) - \frac{\pi}{2} -\log(1+\frac{D^2}{2M^2})
\bigg] 
\label{eq:eval_f1_0}
\ea 
It is crucial to notice that the lowest-energy phonon band is extremely flat near the $H$ point (\cref{fig:flat_phonon_H}), which indicates a relatively small $D$. Then we can expand \cref{eq:eval_f1_0} in powers of $D^2/M^2$
\ba 
f(D^2/M^2) \approx \frac{2\pi N\Lambda^3}{\beta A}\log(D^2/(2M^2) )+O(D^{0})
\label{eq:f_small_D_behavior}
\ea 
where we observe a log divergence in the limit of small bandwidth $D$ (to the negative infinity). 
\textcolor{black}{Here, we also mention that we use $f(D^2/M)$ to set an upper bound of $f(\delta/M^2)$ for the purpose of convenience. Since $\delta, D^2$ are all small parameters, it will be more convenient to only keep one of them when we discuss the behaviors of $f(\delta/M^2)$ at small $\delta$ and small $D$. Since $\delta$ plays the role of tuning parameters that can be arbitrarily small, we can assume $\delta <D^2$ and consider $f(\delta/M^2) <f(D^2/M^2)$. Then we only need to keep track of the single small parameter $D$.}

From \cref{eq:f_small_D_behavior}, we find, near $\omega_{soft,H}^2 = -M^2+0^+$, the free energy difference between $\tilde{{{K}}}$ order state and $H$ order state follows (from \cref{eq:del_f}, \cref{eq:f1_diff_2}, \cref{eq:bound}, \cref{eq:eval_f1_0}) 
\ba 
\lim_{\delta \rightarrow 0^+}G(-M^2+\delta) = &\lim_{\delta \rightarrow 0^+}\bigg( F_{\tilde{{{K}}},0} -F_{H,0}  +f(\delta /M^2)\bigg)\nonumber\\ 
\le & F_{\tilde{{{K}}},0} -F_{H,0}\bigg|_{\omega_{soft,H}^2=-M^2}  +f(D^2/M^2) \nonumber\\ 
\approx & F_{\tilde{{{K}}},0} -F_{H,0}\bigg|_{\omega_{soft,H}^2=-M^2} +2\pi N\Lambda^3/(\beta A)\log( D^2/(2M^2)) +O(D^0)
\ea 
Since $F_{\tilde{{{K}}},0}-F_{H,0}$ at $\omega_{soft,H}^2=-M^2$ are constant numbers that are independent of $D$, we conclude that, in the limit of the flat phonon bands $(D\rightarrow 0)$, $G(-M^2+0^+)  < 0 $ due to the log divergence of $f(D^2/M^2)$. This concludes our proof that, whenever the phonon band is flat enough with a small $D$, there will be a parameter region within $-M^2 <\omega_{soft,H}^2<0$ such that $\tilde{{{K}}}$ phase has the lowest free energy ($G(\omega_{soft,H}^2)<0$).

We now provide an estimation of the transition point between $\tilde{{{K}}}$ ordered phase and $H$ ordered phase. We consider a small and negative $\omega_{soft,H}^2$ and perform an expansion of the following free energy difference
\ba 
\Delta  F_{\tilde{{{K}}},H}= F_{\tilde{{{K}}},0} -F_{H,0} +\frac{1}{2\beta}\sum_{|\mathbf{q}|<\Lambda}\log\bigg( 
 \frac{ 
 (c^2|\mathbf{q}|^2 +\omega_{soft,H}^2 +M^2 )^2 
 }{(c^2 |\mathbf{q}|^2 -2\omega_{soft,H}^2 ) (\Omega_n^2+c^2 |\mathbf{q}|^2)}
 \bigg)  
 \label{eq:Delta_FKH}
\ea 
\textcolor{black}{where we only keep $\Omega_n=0$ contribution.}
To calculate $\Delta F_{\tilde{{{K}}},H}$, we first introduce the following functions 
\ba 
h(-\omega_{soft,H}^2) = \frac{1}{2\beta}\sum_{|\mathbf{q}|<\Lambda}\log\bigg( 
 \frac{ 
 ( c^2|\mathbf{q}|^2 +\omega_{soft,H}^2 +M^2 )^2 
 }{(c^2 |\mathbf{q}|^2 -2\omega_{soft,H}^2) (c^2 |\mathbf{q}|^2)}
 \bigg) 
\ea 
We investigate behavior of $h(x)$ at small $x$ by performing an expansion in powers of $x$
\ba 
h(x) =& h(0) + \frac{1}{2\beta} \sum_{|\mathbf{q}|<\Lambda} \frac{-2x}{c^2|\mathbf{q}|^2 +M^2} - \frac{1}{2\beta} \sum_{|\mathbf{q}|<\Lambda}\frac{1}{c^2|\mathbf{q}|^2 }(2x) \nonumber\\ 
=&h(0) + \frac{N}{2\beta A}\int_{|\mathbf{q}|<\Lambda} 
\bigg( \frac{-2x}{c^2 |\mathbf{q}|^2 +M^2} -\frac{2x}{c^2|\mathbf{q}|^2} \bigg) d^3\mathbf{q}\nonumber \\ 
=& h(0) +\frac{-8x \pi N}{\beta A} \int_0^{\Lambda}
\bigg( 
\frac{1}{c^2 q^2 + M^2} +\frac{1}{c^2q^2}
\bigg) q^2 dq  \nonumber\\ 
=& h(0) -
\frac{ 8 \pi N \Lambda^3 }{\beta AD^2 } \bigg(2 - \frac{M}{D}\tan^{-1}(\frac{D}{M}) \bigg)x
\ea 
Then the free energy difference is 
\ba 
\Delta F_{\tilde{{{K}}},H} \approx (F_{\tilde{{{K}}},0}-F_{H,0} +h(0) ) -
\frac{ 8 \pi N \Lambda^3 }{\beta AD^2 } \bigg(2 - \frac{M}{D}\tan^{-1}(\frac{D}{M}) \bigg)(-\omega_{soft,H}^2)
\label{eq:dela_f}
\ea 
We let $\Delta F_0 = (F_{\tilde{{{K}}},0}-F_{H,0} +h(0) ) $, which denotes the energy difference between $H$ order solution and $\tilde{{{K}}}$ order solution at $\omega_{soft,H}^2=0$, then 
\ba
\Delta F_{\tilde{{{K}}},H}  = \Delta F_0 -
\frac{ 8 \pi N \Lambda^3 }{\beta AD^2 } \bigg(2 - \frac{M}{D}\tan^{-1}(\frac{D}{M}) \bigg)(-\omega_{soft,H}^2)
\ea 
We also estimate $\Delta F_0$ in power of $\omega_{soft,H}^2$
\ba 
\Delta F_0 = F_{\tilde{{{K}}},0} +h(0) + O(\omega_{soft,H}^4)
\label{eq:delta_f0}
\ea 
where $h(0)$ is
\ba 
h(0) =&  \frac{1}{2\beta}\sum_{|\mathbf{q}|<\Lambda}\log\bigg( 
 \frac{ 
 ( c^2|\mathbf{q}|^2 +M^2 )^2 
 }{(c^2 |\mathbf{q}|^2  ) (c^2 |\mathbf{q}|^2)}
 \bigg) \approx \frac{4\pi N}{\beta A} \int_0^{\Lambda} \log(  \frac{ 
 ( c^2|\mathbf{q}|^2 +M^2 )
 }{(c^2 |\mathbf{q}|^2  )})q^2 dq \nonumber\\ 
 =&\frac{4\pi N\Lambda^3}{3\beta A D^3}\bigg( 
 2DM^2 -2M^3 \tan^{-1}(D/M) -D^3\log(D^2/(D^2+M^2))
 \bigg) 
 \ea 
 In the flat band limit with small $D/M$, we find 
 \ba 
 h(0) \approx \frac{4\pi N\Lambda^3}{9\beta A}(2- 6\log(D/M))
 \label{eq:g0_behavior}
 \ea

The transition point between $H$-order phase and $\tilde{{{K}}}$-order phase locates at (\cref{eq:delta_f0}, \cref{eq:g0_behavior} and \cref{eq:dela_f}) 
\ba 
\Delta F_{\tilde{{{K}}},H}= 0 \Rightarrow &-\omega_{soft,H}^2 \approx  D^2 \frac{\beta A \Delta F_0}{8\pi N\Lambda^3 }\frac{1}{2 - \frac{M}{D}\tan^{-1}(\frac{D}{M})} \nonumber\\ 
\Rightarrow& -\omega_{soft,H}^2 \approx 
 D^2 \frac{\beta A }{8\pi N\Lambda^3 }\frac{1}{2 - \frac{M}{D}\tan^{-1}(\frac{D}{M})}
\bigg( 
F_{\tilde{{{K}}},0} +h(0) 
\bigg) 
\ea 
It is worth mentioning that, in the flat band limit with a small $D/M$, the transition point is 
\ba 
-\omega_{soft,H}^2 \approx  D^2 \bigg( C -\frac{1}{3}\log(D/M)
\bigg) 
\ea 
where the constant $C$
\ba 
C =\frac{1}{9}  +\frac{\beta A}{8\pi\Lambda^3N} F_{\tilde{{{K}}},0}
\ea 
We note that the transition points $ D^2 \bigg( C -\frac{1}{3}\log(D/M)\bigg) \rightarrow 0 $ as $D^2\rightarrow 0 $. 
Clearly, if the phonon bands near $H$ are flat enough, a transition to the $\tilde{{{K}}}$ ordered phase will happen almost immediately after the collapsing of the lowest-energy phonon mode at $H$($\omega_{soft,H}^2=0$). 

\textcolor{black}{ 
To confirm our analytical result, we also perform numerical calculations by treating $\omega_{soft,H}^2$ as our tuning parameter and using the parameters given in \cref{eq:parameter}. The phase of the system is determined as follows, for $\omega_{soft,H}^2>0$, the system is in the disordered phase as we proved around \cref{eq: m2_ge_0}. For $\omega_{soft,H}^2\le 0$, we numerically calculate $\Delta F_{\tilde{{{K}}},H}$ (\cref{eq:Delta_FKH}). If $\Delta F_{\tilde{{{K}}},H}<0$ the system locates at $\tilde{{{K}}}$-order phase. If $\Delta F_{\tilde{{{K}}},H}>0$, the system locates at $H$-order phase. In \cref{fig:op_fluct} (b), (c), we perform calculations at two different phonon velocities $c$. We can observe, for small enough $c$ (equivalently, small enough $D=c\Lambda$), the transition happens almost immediately after the phonon collapsing point $\omega_{soft,H}^2=0$. For comparison, in \cref{fig:op_fluct}, we also show the result from saddle-point approximation without Gaussian fluctuations where we do not observe any transition to the $\tilde{{{K}}}$-order phase. 
}

In conclusion, we demonstrate that, due to the flatness of the phonon bands and the cubic term of the $\tilde{{{K}}}$ order parameter, a first-order transition to the $\tilde{{{K}}}$ ordered phase will almost immediately happen after the collapsing of the phonon modes at $H$. Several things are essential to understand this transition
\begin{itemize}
    \item There is a symmetry-allowed cubic term for $\tilde{{{K}}}$ order which leads to a first-order transition.
    \item The dispersion of the lowest-energy phonon is flat near $H$ point which leads to strong order-parameter fluctuations and stabilizes a $\tilde{{{K}}}$ order.  
\end{itemize}

We also comment on the validity of our approach based on saddle-point solutions with Gaussian fluctuations. Clearly, near the transition point, the flatness of phonon modes will invalidate the bare saddle-point solution in which the fluctuations are totally ignored. However, if the fluctuations are so strong (for example, exactly at the flat band limit with $c=0$), high-order contributions/fluctuations beyond Gaussian fluctuations could also be important. Therefore, an exact and better approach should be the numerical simulation of the effective theory given in \cref{eq:Lall} which is beyond our current consideration. However, as a proof of principle, our current approach has already demonstrated the Gaussian fluctuation of the $H$ order parameter alone could already destroy the $H$ order phase and stabilize the $\tilde{{{K}}}$ phase, which provides a possible explanation of the puzzling experimental observation. 

Finally, we comment that both $\tilde{{{K}}}$ and $H$ order parameters are formed by the phonon from the lowest-energy phonon band (\cref{fig:flat_phonon_H}), with close energy in the phonon dispersion. However, the $H$ point gives leading order instability which explains why $H$-phonon first collapses. Because the lowest-energy phonon band is very flat near $H$ point, all phonon with momentum near $H$ (including $\tilde{{{K}}}$) becomes soft and is likely to generate CDW instabilities. However, the flatness will also produce strong fluctuations, which makes it hard to stabilize the condensation of the $H$ phonon which produces leading order instability. But, it is important to notice that, for all the momenta near $H$ point (within the region of the flat band), $\tilde{{{K}}}$ is the only momentum that allows a cubic term in the effective field theory (another candidate is $K$ point which has higher energy than $\tilde{{{K}}}$ in the phonon dispersion). The existence of cubic terms makes the transition between the order and disorder phase to be first-order, which will suppress the strong fluctuations of the flat phonon band and stabilize the $\tilde{{{K}}}$ order. In summary, $\tilde{{{K}}}$ is a natural choice of the CDW phase for the following two reasons
\begin{itemize}
    \item Phonon at $\tilde{{{K}}}$ is part of the flat-band phonon near $H$ point with close energy (to $H$-phonon) in the phonon dispersion. 
    \item For the phonon at $\tilde{{{K}}}$, a cubic term is allowed in the effective field theory, which drives the transition to be first-order-like and suppresses the strong fluctuations from the flat phonon band. 
\end{itemize}

\section{CDW order parameters}\label{app:sec:CDW_ops}

In this \siSection{}, we analyze both the electronic and phonon order parameters (OPs) induced by the CDW phase transition. 
For the electronic OPs, we first introduce a new method to extract mean-field translational symmetry-breaking OPs from the tight-binding (TB) model of the CDW phase, and then add these OPs to the non-CDW TB model. We find the OPs given by the onsite term of the mirror-even orbitals $m_e$ of triangular Sn $p_z$ orbitals and the hopping term between $m_e$ and V $d_{z^2},d_{yz},d_{x^2-y^2}$ have the largest contributions to the band splittings close to the Fermi level, and by adding them to the non-CDW TB we can quantitatively reproduce the band structure of the CDW phase. 
For the phonon OPs, we extract the displacements of atoms in the CDW phase from the experiment structure and show that they are consistent with the calculated condensated lowest-energy phonon mode at $\bar{K}$ point.

\subsection{The electronic order parameter of the CDW phase}\label{app:Sec:CDW_ops:electronic}

\subsubsection{Algorithm}
We now discuss the electronic order parameter of the CDW phase by adopting a weak-coupling mean-field approach and matching it to the DFT computed structure. At the mean-field level, we assume the following term is induced by the CDW order 
\ba 
{ \hat{H} }_{CDW} = \sum_{\mathbf{Q},\mathbf{k},\alpha \gamma,\sigma} c_{\mathbf{k}+\mathbf{Q},\alpha\sigma}^\dag \Phi_{\alpha\gamma,\mathbf{k},\mathbf{Q}} c_{\mathbf{k},\gamma \sigma}
\ea 
The CDW phase has translational symmetry corresponding to the primitive vector $\bm{P}_1,\bm{P}_2,\bm{P}_3$. 
The $c_{\mathbf{k},\alpha \sigma}$ operators defined in the non-CDW unit cell with basis $\{\bm{a}_i\}_{i=1,2,3}$ transform under  translational symmetry as
\ba 
T_\mathbf{R} c_{\mathbf{k},\alpha \sigma} T_\mathbf{R}^\dag = c_{\mathbf{k},\alpha \sigma} e^{i\mathbf{k}\cdot \mathbf{R}},
\quad \mathbf{R} \in \mathbb{Z}\bm{a}_1+\mathbb{Z}\bm{a}_2+\mathbb{Z}\bm{a}_3
\ea 
Here we aim to define the CDW order parameter that breaks the original translational symmetry (corresponding to $\{\bm{a}_i\}_{i=1,2,3}$ but preserves the translational symmetry with respect to the $\{\bm{P}_i\}_{i=1,2,3}$. 
We require that 
\ba 
T_\mathbf{R} { \hat{H} }_{CDW}T_\mathbf{R}^\dag = { \hat{H} }_{CDW},\quad \mathbf{R} \in \mathbb{Z}\bm{P}_1+\mathbb{Z}\bm{P}_2+\mathbb{Z}\bm{P}_3
\ea 
which gives
\ba 
&T_\mathbf{R} { \hat{H} }_{CDW}T_\mathbf{R}^\dag = \sum_{\mathbf{k},\mathbf{Q},\alpha\gamma,\sigma}c_{\mathbf{k}+\mathbf{Q},\alpha\sigma}^\dag \Phi_{\alpha\gamma,\mathbf{k},\mathbf{Q}}e^{-i\mathbf{Q} \cdot \mathbf{R}} c_{\mathbf{k},\gamma \sigma} = { \hat{H} }_{CDW}
,\quad \mathbf{R} \in \mathbb{Z}\bm{P}_1+\mathbb{Z}\bm{P}_2+\mathbb{Z}\bm{P}_3
\nonumber \\ 
\Rightarrow& \Phi_{\alpha\gamma,\mathbf{k},\mathbf{Q}} = 
\Phi_{\alpha\gamma,\mathbf{k},\mathbf{Q}}e^{-i\mathbf{Q} \cdot \mathbf{R}},\quad 
\mathbf{R} \in \mathbb{Z}\bm{P}_1+\mathbb{Z}\bm{P}_2+\mathbb{Z}\bm{P}_3
\ea 
Consequently, we have 
\ba 
\Phi_{\alpha\gamma, \mathbf{k},\mathbf{Q}} =0 \quad \text{ if } \mathbf{Q} \notin \mathbb{Z}\tilde{\bm{b}}_1 +\mathbb{Z}\tilde{\bm{b}}_2 +\mathbb{Z}\tilde{\bm{b}}_3
\ea 
where $\tilde{\bm{b}}_i$ are the reciprocal lattice vectors with 
\ba 
\tilde{\bm{b}}_i\cdot \bm{P}_j =2\pi  \delta_{i,j} 
\ea 
In the non-CDW phase, $\Phi_{\alpha\gamma,\mathbf{k},\mathbf{Q}}$ can also be nonzero when $\mathbf{Q} \in \mathbb{Z}{\bm{b}}_1 +\mathbb{Z}{\bm{b}}_2 +\mathbb{Z}{\bm{b}}_3 $. $\Phi_{\alpha\gamma,\mathbf{k},\mathbf{Q}}$ are mean-field order parameters that correspond to onsite or hopping terms and couple electron operators at the same momentum $\mathbf{k}$.Therefore, the CDW order parameter can be characterized by 
\ba 
\Phi_{\alpha\gamma,\mathbf{k},\mathbf{Q}} \quad \text{with}\quad \mathbf{Q} \in \mathbb{Z}\tilde{\bm{b}}_1 +\mathbb{Z}\tilde{\bm{b}}_2 +\mathbb{Z}\tilde{\bm{b}}_3 \quad \text{ and }\quad \mathbf{Q} \notin \mathbb{Z}{\bm{b}}_1 +\mathbb{Z}{\bm{b}}_2 +\mathbb{Z}{\bm{b}}_3 
\ea 
We comment that, here, we only consider the translational symmetry. In practice, there could be additional order parameter that preserves translational symmetry but break other symmetries, such as rotational symmetry, time-reversal symmetry, etc. 
The translational order parameters are the simplest but turn out to be adequate to reproduce the CDW band structure as shown in the following sections.

In order to identify the corresponding term in the tight-binding model derived in the CDW phase, we perform a transformation between the operators in the CDW phase and in the non-CDW phase. The primitive cell basis in the CDW and non-CDW phases are related by the $M$ matrix defined \cref{eq:basis_transf_a}, with $\det(M)=3$ by ignoring negligible changes in lattice constants. 
In the following, we first consider a simplifed case where the CDW unit cell is tripled along the $a_3$ direction of the non-CDW unit cell to demonstrate the procedure of obtaining order parameters. Then we consider the general case of CDW unit cell defined in \cref{eq:basis_transf_a}.

\paragraph{\textbf{Simplified case}}
We first consider the simplifed case, by assuming that the CDW primitive cell is a simple tripled unit cell along the $a_3$ direction, and thus the lattice constants satisfy $a'=a, c'=3c$ and the atoms in the unit cell have the no displacements compared with the non-CDW phase. 

In the non-CDW phase, we take the original electron operators 
\ba 
\text{non-CDW:}\quad c_{\mathbf{R},\alpha\sigma}, \quad \mathbf{R} \in L_{\text{non-CDW}}
\ea 
where $L_{\text{non-CDW}}$ denotes the lattice formed by the unit cell of the non-CDW phase.
For the CDW phase, we divide the atoms in the unit cell into three layers by $\bm{a}_3$ and take the following electron operator
\ba 
\text{CDW}:\quad \tilde{c}_{\tilde{\mathbf{R}},i,\alpha \sigma},\quad \tilde{\mathbf{R}} \in L_{\text{CDW}}
\ea 
where $L_{\text{CDW}} $ denotes the lattice formed by the unit cell of the CDW phase, and the additional index $i=1,2,3$ denotes the three copies of original orbitals $\alpha$ in the tripled unit cells. In addition, $\tilde{c}_{\tilde{\mathbf{R}},i,\alpha \sigma}$
denote the electron operators at $\tilde{\mathbf{R}} +\mathbf{R}_{i,\alpha} +\mathbf{r}_\alpha$ where $\mathbf{R}_{i,\alpha}$ denote the shift of $i$-th copy of $\alpha$-th orbital, with 
$\mathbf{R}_{i=1,\alpha}=\bm{0}, \mathbf{R}_{i=2,\alpha} = \bm{a}_3, \mathbf{R}_{i=3,\alpha} = 2\bm{a}_3$ for all $\alpha$, where $\bm{a}_3=\frac{1}{3}(\bm{P}_1+\bm{P}_2+\bm{P}_3)$.
In the simplified case, we have 
\ba 
\tilde{c}_{\tilde{\mathbf{R}},i,\alpha \sigma} = c_{\tilde{\mathbf{R}}+\mathbf{R}_{i,\alpha},\alpha \sigma}
\label{eq:c_to_ctilde}
\ea 

We next introduce the corresponding Fourier transformation
\ba 
&c_{\mathbf{k},\alpha \sigma} = \frac{1}{\sqrt{N}}\sum_{\mathbf{R} \in L_{non-CDW}} e^{-i\mathbf{k}\cdot \mathbf{R}} c_{\mathbf{R},\alpha\sigma} \nonumber \\ 
&\tilde{c}_{\mathbf{k},i,\alpha \sigma} = \frac{\sqrt{3}}{\sqrt{N}}\sum_{\tilde{\mathbf{R}} \in L_{CDW}} e^{-i\mathbf{k}\cdot \tilde{\mathbf{R}}}\tilde{c}_{\tilde{\mathbf{R}},i,\alpha \sigma}
\label{eq:fourier_non_cdw_cdw}
\ea 
This Fourier transform is usually called the lattice gauge, which does not involve the sublattice shifting.Since we have triple the size of the unit cell in the CDW phase, there will be an additional $\sqrt{3}$ factor in the definition of $\tilde{c}_{\mathbf{k},i,\alpha\sigma}$. We also comment that, here, in order to simplify the notation, we do not include the sublattice shifting within the unit cell in the definition of Fourier transformation 
The periodicity conditions now are 
\ba 
&c_{\mathbf{k}+\mathbf{G},\alpha\sigma} = c_{\mathbf{k},\alpha\sigma},\quad \mathbf{G} \in \mathbb{Z}\bm{b}_1 + \mathbb{Z}\bm{b}_2 
+ \mathbb{Z}\bm{b}_3 \nonumber \\ 
&\tilde{c}_{\mathbf{k}+\tilde{\mathbf{G}},i,\alpha\sigma} = \tilde{c}_{\mathbf{k},i,\alpha\sigma},\quad \tilde{\mathbf{G}} \in \mathbb{Z}\tilde{\bm{b}}_1 + \mathbb{Z}\tilde{\bm{b}}_2 
+ \mathbb{Z}\tilde{\bm{b}}_3 .
\ea

Combining \cref{eq:c_to_ctilde} and \cref{eq:fourier_non_cdw_cdw}, we find 
\ba 
c_{\mathbf{k},\alpha \sigma} =& \frac{1}{\sqrt{N}}\sum_{\tilde{\mathbf{R} }\in L_{CDW}}\sum_i  e^{-i\mathbf{k}\cdot (\tilde{\mathbf{R}} +\mathbf{R}_{i,\alpha} )}
c_{\tilde{\mathbf{R}}+\mathbf{R}_{i,\alpha},\alpha \sigma} \nonumber \\ 
=& \frac{1}{\sqrt{N}}\sum_{\tilde{\mathbf{R} }\in L_{CDW}}\sum_i  e^{-i\mathbf{k}\cdot (\tilde{\mathbf{R}} +\mathbf{R}_{i,\alpha} )}
\tilde{c}_{\tilde{\mathbf{R}},i,\alpha \sigma} \nonumber \\ 
=& \frac{1}{\sqrt{3}}\sum_{i}\tilde{c}_{\mathbf{k},i,\alpha\sigma} e^{-i\mathbf{k}\cdot \mathbf{R}_{i,\alpha} }
\ea 
Without loss of generality, we can pick such a unit cell of CDW phase that  
Then we let $\mathbf{Q} = \tilde{\bm{b}}_3 $ and find 
\ba 
&c_{\mathbf{k},\alpha\sigma} = \frac{1}{\sqrt{3}} \sum_{j}\tilde{c}_{\mathbf{k},j,\alpha\sigma} e^{-i\mathbf{k}\cdot \mathbf{R}_{j,\alpha} } 
\nonumber \\ 
&c_{\mathbf{k}+\mathbf{Q},\alpha\sigma} = \frac{1}{\sqrt{3}} \sum_{j}\tilde{c}_{\mathbf{k},j,\alpha\sigma} e^{-i\mathbf{k}\cdot \mathbf{R}_{j,\alpha} } e^{-i\frac{2\pi (j-1)}{3} }
\nonumber \\ 
&c_{\mathbf{k}+2\mathbf{Q},\alpha\sigma} = \frac{1}{\sqrt{3}} \sum_{j}\tilde{c}_{\mathbf{k},i,\alpha\sigma} e^{-i\mathbf{k}\cdot \mathbf{R}_{j,\alpha} } e^{-i\frac{4\pi(j-1)}{3}}
\ea 
We can also represent $\tilde{c}$ with $c$
\ba 
&\tilde{c}_{\mathbf{k},1,\alpha\sigma} =\frac{ e^{i\mathbf{k}\cdot \mathbf{R}_{1,\alpha} }}{\sqrt{3}}
\bigg( c_{\mathbf{k},\alpha\sigma} + c_{\mathbf{k}+\mathbf{Q},\alpha\sigma} + c_{\mathbf{k}+2\mathbf{Q},\alpha\sigma} \bigg) \nonumber \\ 
&\tilde{c}_{\mathbf{k},2,\alpha\sigma} =\frac{ e^{i\mathbf{k}\cdot \mathbf{R}_{2,\alpha} }}{\sqrt{3}}
\bigg( c_{\mathbf{k},\alpha\sigma} + e^{i2\pi/3}c_{\mathbf{k}+\mathbf{Q},\alpha\sigma} + e^{-i2\pi/3}c_{\mathbf{k}+2\mathbf{Q},\alpha\sigma} \bigg) \nonumber\\ 
&\tilde{c}_{\mathbf{k},3,\alpha\sigma} =\frac{ e^{i\mathbf{k}\cdot \mathbf{R}_{3,\alpha} }}{\sqrt{3}}
\bigg( c_{\mathbf{k},\alpha\sigma} + e^{-i2\pi/3}c_{\mathbf{k}+\mathbf{Q},\alpha\sigma} + e^{i2\pi/3} c_{\mathbf{k}+2\mathbf{Q},\alpha\sigma} \bigg),
\ea 
or equivalently, 
\ba
\begin{bmatrix}
	\tilde{c}_{\mathbf{k},1,\alpha\sigma}\\ 
	\tilde{c}_{\mathbf{k},2,\alpha\sigma}\\ 
	\tilde{c}_{\mathbf{k},3,\alpha\sigma} 
\end{bmatrix}=S_{\mathbf{k}}
\begin{bmatrix}
	c_{\mathbf{k},\alpha\sigma} \\ 
	c_{\mathbf{k}+\mathbf{Q},\alpha\sigma} \\ 
	c_{\mathbf{k}+2\mathbf{Q},\alpha\sigma} 
\end{bmatrix},\quad
S_{\mathbf{k}}=\frac{1}{\sqrt{3}}
\begin{bmatrix}
	1 & 1 & 1\\ 
	e^{i\mathbf{k}\cdot \mathbf{R}_{2,\alpha}} & 
	e^{i\mathbf{k}\cdot \mathbf{R}_{2,\alpha}} e^{i\frac{2\pi}{3}} & 
	e^{i\mathbf{k}\cdot \mathbf{R}_{2,\alpha}} e^{-i\frac{2\pi}{3}} \\ 
	e^{i\mathbf{k}\cdot \mathbf{R}_{3,\alpha}} & 
	e^{i\mathbf{k}\cdot \mathbf{R}_{3,\alpha}} e^{-i\frac{2\pi}{3}} & 
	e^{i\mathbf{k}\cdot \mathbf{R}_{3,\alpha}} e^{i\frac{2\pi}{3}} \\ 
\end{bmatrix}
\label{Eq:Sk_CDW2nonCDW}
\ea

We then transform the tight-binding model defined in the $\tilde{c}$ basis of CDW phase obtained by performing ab-initio on the CDW state to a new tight-binding model defined in the $c_{\mathbf{k},\alpha\sigma},c_{\mathbf{k}+\mathbf{Q},\alpha\sigma},c_{\mathbf{k}+2\mathbf{Q},\alpha\sigma}$ basis. The term that corresponds to $c_{\mathbf{k},\alpha\sigma}^\dag c_{\mathbf{k}+\mathbf{Q},\alpha'\sigma'},c_{\mathbf{k},\alpha\sigma}^\dag c_{\mathbf{k}+2\mathbf{Q},\alpha'\sigma'},c_{\mathbf{k}+\mathbf{Q},\alpha\sigma}^\dag c_{\mathbf{k}+2\mathbf{Q},\alpha'\sigma'}$ and their Hermitian conjugate will describe the nature of CDW phase.

Here, we also derive the transformation $S_{\mathbf{k}}$ under another Fourier transformation convention with sublattice shifts, which is usually called the atomic gauge:
\ba 
&c_{\mathbf{k},\alpha \sigma} = \frac{1}{\sqrt{N}}\sum_{\mathbf{R} \in L_{non-CDW}} e^{-i\mathbf{k}\cdot (\mathbf{R}+\mathbf{r}_\alpha)} c_{\mathbf{R},\alpha\sigma} \nonumber \\ 
&\tilde{c}_{\mathbf{k},i,\alpha \sigma} = \frac{\sqrt{3}}{\sqrt{N}}\sum_{\tilde{\mathbf{R}} \in L_{CDW}} e^{-i\mathbf{k}\cdot (\tilde{\mathbf{R}}+\mathbf{R}_{i,\alpha}+\mathbf{r}_\alpha)} \tilde{c}_{\tilde{\mathbf{R}},i,\alpha \sigma}
\ea 
where $\mathbf{r}_\alpha$ is the sublattice shift within the non-CDW unit cell. 
Then 
\ba 
c_{\mathbf{k},\alpha \sigma} =& \frac{1}{\sqrt{N}}\sum_{\tilde{\mathbf{R} }\in L_{CDW}}\sum_i  e^{-i\mathbf{k}\cdot (\tilde{\mathbf{R}} +\mathbf{R}_{i,\alpha}+\mathbf{r}_\alpha )}
c_{\tilde{\mathbf{R}}+\mathbf{R}_{i,\alpha},\alpha \sigma} \nonumber \\ 
=& \frac{1}{\sqrt{N}}\sum_{\tilde{\mathbf{R} }\in L_{CDW}}\sum_i  e^{-i\mathbf{k}\cdot (\tilde{\mathbf{R}} +\mathbf{R}_{i,\alpha}+\mathbf{r}_\alpha )}
\tilde{c}_{\tilde{\mathbf{R}},i,\alpha \sigma} \nonumber \\ 
=& \frac{1}{\sqrt{3}}\sum_{i}\tilde{c}_{\mathbf{k},i,\alpha\sigma}
\ea 
By taking $\mathbf{Q} = \tilde{\bm{b}}_3 $, we have 
\ba 
&c_{\mathbf{k},\alpha\sigma} = \frac{1}{\sqrt{3}} \sum_{j}\tilde{c}_{\mathbf{k},j,\alpha\sigma}
\nonumber \\ 
&c_{\mathbf{k}+\mathbf{Q},\alpha\sigma} = \frac{1}{\sqrt{3}} \sum_{j}\tilde{c}_{\mathbf{k},j,\alpha\sigma} e^{-i\mathbf{Q}\cdot \mathbf{r}_{\alpha}} e^{-i\frac{2\pi (j-1)}{3} }
\nonumber \\ 
&c_{\mathbf{k}+2\mathbf{Q},\alpha\sigma} = \frac{1}{\sqrt{3}} \sum_{j}\tilde{c}_{\mathbf{k},i,\alpha\sigma} e^{-2i\mathbf{Q}\cdot \mathbf{r}_{\alpha}} e^{-i\frac{4\pi(j-1)}{3}}
\ea 
Finally, we express $\tilde{c}_{\mathbf{k},i,\alpha\sigma}$ using $c_{\mathbf{k}+n\mathbf{Q},\alpha\sigma}$
\ba
\begin{bmatrix}
	\tilde{c}_{\mathbf{k},1,\alpha\sigma}\\ 
	\tilde{c}_{\mathbf{k},2,\alpha\sigma}\\ 
	\tilde{c}_{\mathbf{k},3,\alpha\sigma} 
\end{bmatrix}=S_{\mathbf{k}}
\begin{bmatrix}
	c_{\mathbf{k},\alpha\sigma} \\ 
	c_{\mathbf{k}+\mathbf{Q},\alpha\sigma} \\ 
	c_{\mathbf{k}+2\mathbf{Q},\alpha\sigma} 
\end{bmatrix},\quad
S_{\mathbf{k}}=\frac{1}{\sqrt{3}}
\begin{bmatrix}
    1 & e^{i\mathbf{Q}\cdot \mathbf{r}_{\alpha}} & e^{2i\mathbf{Q}\cdot \mathbf{r}_{\alpha}}\\ 
    1 & 
    e^{i\mathbf{Q}\cdot \mathbf{r}_{\alpha}} e^{i\frac{2\pi}{3}} & 
    e^{2i\mathbf{Q}\cdot \mathbf{r}_{\alpha}} e^{-i\frac{2\pi}{3}} \\ 
    1 & 
    e^{i\mathbf{Q}\cdot \mathbf{r}_{\alpha}} e^{-i\frac{2\pi}{3}} & 
    e^{2i\mathbf{Q}\cdot \mathbf{r}_{\alpha}} e^{i\frac{2\pi}{3}} \\ 
\end{bmatrix}
\ea

\paragraph{\textbf{General case}}
In ScV$_6$Sn$_6$, the atom positions have displacements less than $0.2 \AA$ in the CDW unit cell compared with the non-CDW unit cell, while the lattice constants $a'\approx \sqrt{3}a, c'\approx 3c$ have errors less than $0.3\AA$. We ignore these negligible errors and map the unit cell basis and atom positions in CDW and non-CDW phases.
The CDW primitive cell basis $\bm{P}_i$ defined in \cref{eq:basis_transf_a} are related to the non-CDW basis $\bm{a}_i$ through the following $M$ matrix:
\ba
\begin{bmatrix}
	\bm{P}_1\\ 
	\bm{P}_2\\ 
	\bm{P}_3 
\end{bmatrix}=M
\begin{bmatrix}
	\bm{a}_1\\ 
	\bm{a}_2\\ 
	\bm{a}_3 
\end{bmatrix}, \quad
M=
\begin{bmatrix}
	0 & 1 & 1\\ 
	-1 & -1 & 1 \\ 
	1 & 0 & 1 
\end{bmatrix}
\ea

We use the Smith normal form method in order to obtain a new set of regularized CDW and non-CDW basis that satisfy the tripled relation. 
The Smith normal form decomposition of $M$ is:
\ba
M=L^{-1}\Lambda R^{-1},
\ea
where
\ba
L^{-1}=
\begin{bmatrix}
	0 & 1 & 0\\ 
	-1 & -1 &1 \\ 
	1 & 0 & 0 
\end{bmatrix},
\Lambda=
\begin{bmatrix}
	1 & 0 & 0\\ 
	0 & 1 & 0 \\ 
	0 & 0 & 3
\end{bmatrix},
R^{-1}=
\begin{bmatrix}
	1 & 0 & 1\\ 
	0 & 1 & 1 \\ 
	0 & 0 & 1
\end{bmatrix}.
\ea
Thus
\ba
L\begin{bmatrix}
	\bm{P}_1\\ 
	\bm{P}_2\\ 
	\bm{P}_3 
\end{bmatrix}=
\Lambda R^{-1}
\begin{bmatrix}
	\bm{a}_1\\ 
	\bm{a}_2\\ 
	\bm{a}_3 
\end{bmatrix}
\ea
The diagonal elements of $\Lambda$ give information about the supercell, i.e., the CDW primitive cell can be seen as a $\mathbb{Z}_1\times\mathbb{Z}_1\times\mathbb{Z}_3$ supercell.
We define a new set of basis $\bm{P}_{i=1,2,3}^\prime$ for CDW phase and $\bm{a}_{i=1,2,3}^\prime$ non-CDW phase:
\ba
\begin{bmatrix}
	\bm{P}_1^\prime\\ 
	\bm{P}_2^\prime\\ 
	\bm{P}_3^\prime 
\end{bmatrix}=
L\begin{bmatrix}
	\bm{P}_1\\ 
	\bm{P}_2\\ 
	\bm{P}_3 
\end{bmatrix}
=
\begin{bmatrix}
	0 & 0 & 1\\ 
	1 & 0 & 0 \\ 
	1 & 1 & 1
\end{bmatrix}
\begin{bmatrix}
	\bm{P}_1\\ 
	\bm{P}_2\\ 
	\bm{P}_3 
\end{bmatrix},\quad
\begin{bmatrix}
	\bm{a}_1^\prime\\ 
	\bm{a}_2^\prime\\ 
	\bm{a}_3^\prime 
\end{bmatrix}=R^{-1}
\begin{bmatrix}
	\bm{a}_1\\ 
	\bm{a}_2\\ 
	\bm{a}_3 
\end{bmatrix}
=
\begin{bmatrix}
	0 & 0 & 1\\ 
	1 & 0 & 0 \\ 
	\frac{1}{3} & \frac{1}{3} & \frac{1}{3}
\end{bmatrix}
\begin{bmatrix}
	\bm{P}_1\\ 
	\bm{P}_2\\ 
	\bm{P}_3 
\end{bmatrix}.
\ea
Under this new basis, the CDW unit cell can be seen as a simple tripled supercell along $\bm{a}_3^\prime$, i.e.,
\ba
\begin{bmatrix}
	\bm{P}_1^\prime\\ 
	\bm{P}_2^\prime\\ 
	\bm{P}_3^\prime 
\end{bmatrix}=\Lambda
\begin{bmatrix}
	\bm{a}_1^\prime\\ 
	\bm{a}_2^\prime\\ 
	\bm{a}_3^\prime
\end{bmatrix}
\label{Eq:new_cdw_basis}
\ea
Denote the BZ basis of $\bm{a}_i$, $\bm{a}_i^\prime$, $\bm{P}_i$, $\bm{P}_i^\prime$ (row vectors) as $\bm{b}_i$, $\bm{b}_i^\prime$, $\tilde{\bm{b}}_i$, $\tilde{\bm{b}}_i^\prime$ (column vectors), respectively. They satisfy the following transformations:
\ba
\begin{bmatrix}
	\tilde{\bm{b}}_1^\prime & \tilde{\bm{b}}_2^\prime & \tilde{\bm{b}}_3^\prime
\end{bmatrix}
=
\begin{bmatrix}
	\tilde{\bm{b}}_1 & \tilde{\bm{b}}_2 & \tilde{\bm{b}}_3
\end{bmatrix} L^{-1},\quad
\begin{bmatrix}
	\tilde{\bm{b}}_1 & \tilde{\bm{b}}_2 & \tilde{\bm{b}}_3
\end{bmatrix}
=
\begin{bmatrix}
	\bm{b}_1 & \bm{b}_2 & \bm{b}_3 
\end{bmatrix} M^{-1} \Rightarrow
\begin{bmatrix}
	\tilde{\bm{b}}_1^\prime & \tilde{\bm{b}}_2^\prime & \tilde{\bm{b}}_3^\prime
\end{bmatrix}
=\begin{bmatrix}
	\bm{b}_1 & \bm{b}_2 & \bm{b}_3 
\end{bmatrix}M^{-1} L^{-1}
\ea

Under the new basis $\mathbf{P}_i^\prime$ and $\bm{a}_i^\prime$, the transformation $S_{\mathbf{k}}$ from CDW to non-CDW basis defined in \cref{Eq:Sk_CDW2nonCDW} can be applied, with $\mathbf{R}_{i=1,\alpha}=\bm{0}, \mathbf{R}_{i=2,\alpha} = \bm{a}_3^\prime=\bm{a}_3, \mathbf{R}_{i=3,\alpha} = 2\bm{a}_3^\prime$ for all $\alpha$, and 
\ba
\bm{Q}=\bm{b}_3^\prime=-\frac{1}{3}\bm{b}_1-\frac{1}{3}\bm{b}_2 + \frac{1}{3}\bm{b}_3
\ea
which is equivalent to the CDW $\bm{Q}=\frac{1}{3}(\bm{b}_1+\bm{b}_2+\bm{b}_3)$ through $C_{2z}$, which is a symmetry of SG 191 of the non-CDW structure.

We mention some technical details when extracting order parameters from TB Hamiltonians in CDW and non-CDW phases:
\begin{itemize}
	\item The TB basis, i.e., Wannier orbitals, in the new CDW primitive cell $\bm{P}_i^\prime$ need to be sorted properly s.t. orbitals in three layers have the same order as the non-CDW cell, related by $\bm{R}_{i,\alpha}$. This leads to a rearrangement of the TB basis. 
	\item Wannier orbitals in the CDW basis $\bm{P}_i^\prime$ need to add proper lattice shifts $\tilde{\mathbf{R}}_{0,\alpha}$ in $\bm{P}_i^\prime$ s.t. orbitals in three layers are related exactly by $\bm{R}_{i,\alpha}$, which is an assumption made when deriving $S_{\mathbf{k}}$. 
	This leads to a gauge transformation of phase $e^{i\mathbf{k}\cdot \tilde{\mathbf{R}}_{0,\alpha}}$ for each orbital. 
	\item Wannier orbitals in the CDW and non-CDW phases may have different spread and gauges, and a unitary basis transformation is necessary to unify the basis in order to compare the TB Hamiltonian. We leave more details in the next paragraph. 
	\item The Wannier orbitals in the non-CDW cell need to add proper lattice shifts $\mathbf{R}_{0,\alpha}$ in $\bm{a}_i$ in order to align with the Wannier orbitals in the first layer of CDW cell $\bm{P}_i^\prime$. This also leads to a gauge transformation of phase $e^{i\mathbf{k}\cdot \mathbf{R}_{0,\alpha}}$ for each non-CDW orbital. 
\end{itemize}

To simplify these basis transformations, one can also rearrange the unit cell basis as in \cref{Eq:new_cdw_basis} and atomic positions according to $\mathbf{R}_{i,\alpha}$, and redo DFT computations. In this case, the $S_{\mathbf{k}}$ transformation in the simplified case be applied directly.

\paragraph{\textbf{Transformation of Wannier functions in CDW and non-CDW phases}}

In practice, the TB models for CDW and non-CDW phases are obtained separately using \textit{Wannier90} to fit \textit{ab-initio} band structures. Thus the Wannier bases in the two phases are not exactly the same, with possibly different spread and gauge, and need to be unified in order to compare the TB.

First, assume the \textit{ab-initio} Bloch eigen states in non-CDW and CDW phases have the plane expansion (assume $C_{n\mathbf{k}}^{\mathbf{G}}$ have already been normalized):
\ba
\text{non-CDW:}\quad 
\psi_{n\mathbf{k}}^{0}(\mathbf{r})&=\sum_{\mathbf{G}} e^{i(\mathbf{k}+\mathbf{G})\cdot \mathbf{r}} C_{n\mathbf{k}}^{\mathbf{G}} \nonumber\\
\text{CDW:}\quad 
\tilde{\psi}^{0}_{n\tilde{\mathbf{k}}}(\mathbf{r})&=\sum_{\tilde{\mathbf{G}}} e^{i(\tilde{\mathbf{k}}+\tilde{\mathbf{G}})\cdot \mathbf{r}} \tilde{C}_{n\tilde{\mathbf{k}}}^{\tilde{\mathbf{G}}}
\ea
\textit{Wannier90} outputs a unitary transformation $U^{\mathbf{k}}$ to mix the Bloch states at each $\mathbf{k}$ and results in Bloch functions that are smooth in $\mathbf{k}$:
\ba
\text{non-CDW:}\quad 
\psi_{n\mathbf{k}}(\mathbf{r})&=\sum_m U_{nm}^{\mathbf{k}} \psi_{m\mathbf{k}}^{0}(\mathbf{r}) \nonumber\\
\text{CDW:}\quad 
\tilde{\psi}_{n\tilde{\mathbf{k}}}(\mathbf{r})&=
\sum_m U_{nm}^{\mathbf{k}} \tilde{\psi}^{0}_{m\tilde{\mathbf{k}}}(\mathbf{r})
\ea
This $U^{\mathbf{k}}$ is composed of two transformations, i.e., 
\ba
U^{\mathbf{k}} = U^{\mathbf{k}}_{ml} U^{\mathbf{k}}_{dis}
\ea
where $U^{\mathbf{k}}_{dis}$ of dimension $N_w\times N_b$ is the unitary transformation for disentanglement, and $U^{\mathbf{k}}_{ml}$ of dimension $N_b\times N_b$ is for obtaining maximally localized Wannier functions, with $N_w$ being the number of Wannier orbitals in the unit cell and $N_b$ the number of bands used for constructing maximally localized Wannier functions (MLWFs).

The MLWFs are defined as the Fourier transform:
\ba
\text{non-CDW:}\quad 
W_{n,\mathbf{R}}(\mathbf{r})&= \frac{V_{\text{nC}}}{(2\pi)^3} \int_{\text{non-CDW BZ}} d\mathbf{k}\  \psi_{n\mathbf{k}}(\mathbf{r}) e^{-i\mathbf{k}\cdot \mathbf{R}} \nonumber \\
\text{CDW:}\quad 
\tilde{W}_{n,\tilde{\mathbf{R}}}(\mathbf{r})&= \frac{V_{\text{C}}}{(2\pi)^3} \int_{\text{CDW BZ}} d\tilde{\mathbf{k}}\  \tilde{\psi}_{n\tilde{\mathbf{k}}}(\mathbf{r}) e^{-i\tilde{\mathbf{k}} \cdot \tilde{\mathbf{R}}} 
\ea
In practice, the integral over BZ is replaced by a summation over a discretized mesh in the BZ:
\ba
\text{non-CDW:}\quad 
W_{n,\mathbf{R}}(\mathbf{r})&= \frac{1}{\sqrt{N_{\text{nC}}}} \sum_{\mathbf{k}}  \psi_{n\mathbf{k}}(\mathbf{r}) e^{-i\mathbf{k}\cdot \mathbf{R}} \nonumber\\
\text{CDW:}\quad 
\tilde{W}_{n,\tilde{\mathbf{R}}}(\mathbf{r})&= \frac{1}{\sqrt{N_{\text{C}}}} \sum_{\tilde{\mathbf{k}}}  \tilde{\psi}_{n\tilde{\mathbf{k}}}(\mathbf{r}) e^{-i\tilde{\mathbf{k}} \cdot \tilde{\mathbf{R}}}
\ea
where $N_{\text{nC}}$ ($N_{\text{C}}$) is the number of kpoints in the non-CDW (CDW) BZ.

We aim to compute the overlap matrix between Wannier functions in CDW and non-CDW phases:
\ba
S_{mn}(\tilde{\mathbf{R}},\mathbf{R}) &= \langle \tilde{W}_{m,\tilde{\mathbf{R}}}|W_{n,\mathbf{R}} \rangle \nonumber\\
&= \frac{1}{\sqrt{N_{\text{C}}N_{\text{nC}}}} \sum_{\tilde{\mathbf{k}},\mathbf{k}} 
e^{i\tilde{\mathbf{k}} \cdot \tilde{\mathbf{R}}-i\mathbf{k}\cdot \mathbf{R}} \langle \tilde{\psi}_{m\tilde{\mathbf{k}}} |\psi_{n\mathbf{k}} \rangle
\ea
where the overlap between Bloch functions are nonzero only when $\tilde{\mathbf{k}}=\mathbf{k}-\tilde{\mathbf{G}}_0$, with $\tilde{\mathbf{G}}_0$ being a reciprocal lattice vector in the CDW BZ:
\ba
\langle \tilde{\psi}_{m\tilde{\mathbf{k}}} |\psi_{n\mathbf{k}} \rangle 
&= \sum_{\mathbf{G},\tilde{\mathbf{G}}} \tilde{C}_{m\tilde{\mathbf{k}}}^{\tilde{\mathbf{G}},*} C_{n\mathbf{k}}^{\mathbf{G}} \delta_{\tilde{\mathbf{G}},\mathbf{G}+\tilde{\mathbf{G}}_0} \nonumber\\
&=  \sum_{\mathbf{G}} \tilde{C}_{m\tilde{\mathbf{k}}}^{\mathbf{G}+\tilde{\mathbf{G}}_0,*} C_{n\mathbf{k}}^{\mathbf{G}} 
\ea

For ScV$_6$Sn$_6$, the CDW BZ basis $\tilde{\bm{b}}_i$ and non-CDW BZ basis $\bm{b}_i$ are related by
\ba
\begin{bmatrix}
	\tilde{\bm{b}}_1 & \tilde{\bm{b}}_2 & \tilde{\bm{b}}_3
\end{bmatrix}M
=
\begin{bmatrix}
	\bm{b}_1 & \bm{b}_2 & \bm{b}_3 
\end{bmatrix}
\ea
Thus a reciprocal lattice vector $\mathbf{G}$ in the non-CDW BZ corresponds to $\tilde{\mathbf{G}}=M\mathbf{G}$ in the CDW BZ. 
We then outline the steps to obtain the overlap matrix between Wannier functions in the case when basis of CDW and non-CDW satisfied \cref{Eq:new_cdw_basis}:
\begin{enumerate}
	\item First, perform \textit{ab-initio} computations for CDW and non-CDW phases, by setting a $N\times N\times N$ $\Gamma$-centered uniform kmesh in the CDW BZ, and a $N\times N\times 3N$ $\Gamma$-centered uniform kmesh in the non-CDW BZ. In this way, each kpoint $\tilde{\mathbf{k}}$ in the CDW BZ has exactly three corresponding $\mathbf{k}_i$ in the non-CDW BZ, i.e., $\tilde{\mathbf{k}}=\mathbf{k}_i-\tilde{\mathbf{G}}_i$ $(i=1,2,3)$. 
\item Compute MLWFs using \textit{Wannier90}, and extract the $U^{\mathbf{k}}$ matrix to expand the MLWFs into plane wave basis.
	\item Compute the overlap between Wannier functions in CDW and non-CDW phases. As the CDW primitive cell has 3-times orbitals compared with the non-CDW cell, we set $\tilde{\mathbf{R}}=0$ and choose appropriate non-CDW $\mathbf{R}$ s.t. $\tilde{W}_{m,0}$ and $W_{n,\mathbf{R}}$ have the closest Wannier center. In this way, the overlap matrix has dimension $3N_w\times 3N_w$, where $N_w$ is the number of Wannier orbitals in the non-CDW primitive cell.
\end{enumerate}

The overlap matrix $S_{mn}$ may not be strictly unitary, and one can perform the singular value decomposition to extract a unitary matrix from it. Using the unitary overlap matrix, the Wannier orbitals in the CDW phase can be transformed into the same Wannier orbitals as in the non-CDW phase. 

The method above can be used when the \textit{ab-initio} wavefunctions are fully expanded in the plane wave basis. However, in practice, many DFT software uses pseudo-potential\cite{HAM79, VAN90, KRE99} and only the pseudo-wavefunctions are expanded using plane waves while the all-electron wavefunctions have singular parts that are not expanded, which makes the overlap between Wannier functions difficult to compute. However, we find that when the initial guess of Wannier functions in CDW and non-CDW phases are taken as the same and proper local coordinate systems are taken on Kagome V atoms, the resultant Wannier TBs are very close which means the overlap matrix between Wannier functions is an almost identity matrix. Thus we skip the computation of the overlap matrix of Wannier functions and assume it as an identity matrix, which turns out to be a good approximation.

\subsubsection{Analytic expressions for the order parameters}
In this subsection, we derive the analytic expressions of the dominant order parameters and compare their values. 
We first consider the order parameters given by the onsite terms of mirror-even orbital $m_e$ of triangular Sn. Assume the onsite term of $m_e$ in the CDW phase is 
\begin{equation}
H^{0}(\mathbf{k})=
\begin{bmatrix}
\mu_{m_e^1} & & \\ 
& \mu_{m_e^2} & \\ 
& & \mu_{m_e^3} \\ 
\end{bmatrix}
\end{equation}
where $\mu_{e^i}$ denote the onsite energies of three $m_e$ orbitals in the CDW primitive cell. We then transform $H^{0}(\mathbf{k})$ using $S_{\mathbf{k}}$ defined in \cref{Eq:Sk_CDW2nonCDW} (in lattice gauge):
\begin{equation} 
\begin{aligned}
H^{1}(\mathbf{k})&=S_{\mathbf{k}} H^{0}(\mathbf{k}) S_{\mathbf{k}}^{-1}=
\frac{1}{3}(\mu_{m_e^1}+\mu_{m_e^2}+\mu_{m_e^3})\bm{I}_3 + 
\begin{bmatrix}
0 & O_1 & O_1^* \\ 
& 0 & O_1 \\ 
c.c & & 0 \\ 
\end{bmatrix}\\
O_1 &= \frac{1}{3}(\mu_{m_e^1}+\mu_{m_e^2}e^{i\frac{2\pi}{3}}+\mu_{m_e^3}e^{-i\frac{2\pi}{3}}) \\
\end{aligned}
\label{Eq:order_param_onsite}
\end{equation}
$O_1$ are evaluated using \textit{ab-initio} onsite energy values in \cref{tab:mirror_even_onsite}, with results given in \cref{tab:order_param_values}.  
Note that when $\mu_{m_e^1}=\mu_{m_e^2}=\mu_{m_e^3}$, the off-diagonal terms are all zero and return to the non-CDW case with equal onsite energy.

\begin{table}[htbp]
\begin{tabular}{c|c|ccc}
\hline\hline
Phase             & non-CDW & \multicolumn{3}{c}{CDW}  \\\hline
$\mu_{p_z^1}$     & 0.792   & 0.948  & 0.587  & 1.053  \\
$\mu_{p_z^2}$     & 0.792   & 1.053  & 0.587  & 0.948  \\
$t_{p_z^1,p_z^2}$ & 1.846   & 2.196  & 1.284  & 2.196  \\
$t_{p_z^2,p_z^1}$ & 1.846   & 2.196  & 1.284  & 2.196  \\\hline
$\mu_{m_e}$       & -1.054  & -1.196 & -0.697 & -1.196 \\
\hline\hline
\end{tabular}
\caption{Onsite energy $\mu_{p_z^i}$ and NN hoppings $t_{p_z^i,p_z^j}$ of Sn$^T$ $p_z$ orbitals in the non-CDW and CDW phase. The last three columns correspond to the Sn$^T$ $p_z$ orbitals in the three layers of the CDW primitive cell. In the last row, we list the onsite energy of the mirror even orbitals $m_e$ defined in \cref{eq:ele_even_odd}, which has the expression $\mu_{m_e}=\frac{1}{2}(\mu_{p_z^1}+\mu_{p_z^2})-t_{p_z^1,p_z^2}$. The onsite energies of three $m_e$ orbitals in the CDW unit cell are different due to different displacements of Sn$^T$ atoms as listed in \cref{Tab:atom_pos_compare_CDW} and \cref{Tab:atom_pos_compare_CDW_allatom}, while the average onsite energy $\bar{\mu}_{m_e}=-1.030$ in the CDW phase is very close to the non-CDW phase. }
\label{tab:mirror_even_onsite}
\end{table}

\begin{table}[htbp]
\begin{tabular}{c|ccc|ccc|ccc}
\hline\hline
Hopping & $d_{z^2}^1$ & $d_{yz}^1$ & $d_{x^2-y^2}^1$ & $d_{z^2}^2$ & $d_{yz}^2$ & $d_{x^2-y^2}^2$ & 
$d_{z^2}^e$ & $d_{yz}^e$ & $d_{x^2-y^2}^e $\\ \hline
$m_e^1$ & 0.250       & 0.346      & 0.173           & 0.283       & -0.363     & 0.163  
& 0.345  & 0.501 & 0.238     \\ \hline
$m_e^2$ & 0.163       & 0.380      & 0.117           & 0.163       & -0.380     & 0.117  
& 0.230 & 0.538 & 0.167 \\ \hline
$m_e^3$ & 0.238       & 0.363      & 0.163           & 0.250       & -0.346     & 0.173     
& 0.345 & 0.501 & 0.238 \\ \hline
CDW average & 0.217 & 0.363 & 0.151 & 0.217 & -0.363 & 0.151  & 0.307 & 0.513 & 0.214
\\\hline
non-CDW & 0.212 & 0.356 & 0.150 & 0.212 & -0.356 & 0.150 
& 0.300 & 0.503 & 0.212 \\
\hline\hline
\end{tabular}
\caption{Hopping values between mirror even $m_e$ orbitals of Sn$^T$ $p_z$ and $d$ orbitals of V. We use $m_e^{i=1,2,3}$ to denote $m_e$ in three layers, $d_{z^2}^{i},d_{yz}^{i},d_{x^2-y^2}^{i}$ ($i=1,2$) the $d$ orbitals of V in two Kagome sub-layers in each layer, and $d^{e}_i$ the mirror even orbital formed by $d_i$ orbitals in two sub-layers, defined in \cref{Eq:mirror_even_Vd}. The hopping values are averaged over six $C_6$-related V atoms in each Kagome layer. Note that $C_{2x}$ symmetry relates orbitals in two Kagome sub-layers and further constrains the hopping, i.e., $t_{m_e^1,d^1}=t_{m_e^3,d^2}$ (an extra minus sign exists for $d_{yz}$). 
The translational symmetry-breaking order parameters can be computed for each column using \cref{Eq:order_param_Vhop}. 
In the last two rows, we also list the averaged hopping values in the CDW phase and the hopping in the non-CDW phase, which are very close.
}
\label{tab:hop_V_me}
\end{table}

\begin{table}[htbp]
\scriptsize
\begin{tabular}{c|c|ccc|ccc|ccc}
\hline\hline
OP      & $\mu_{m_e}$       & $t_{m_e,d_{z^2}^1}$   & $t_{m_e,d_{yz}^1}$    & $t_{m_e,d_{x^2-y^2}^1}$  & $t_{m_e,d_{z^2}^2}$   & $t_{m_e,d_{yz}^2}$    & $t_{m_e,d_{x^2-y^2}^2}$ 
& $t_{m_e,d_{z^2}^e}$   & $t_{m_e,d_{yz}^e}$    & $t_{m_e,d_{x^2-y^2}^e}$ 
\\ \hline
$O_1$   & $-0.083 - 0.144i$ & $0.017-0.022i$ & $-0.009 + 0.005i$ & $0.011-0.013i$ & $0.010 - 0.025i$ & $-0.010i$ & $0.006 - 0.016i$   
& $0.019 -0.033i$ & $-0.006 + 0.011i$ & $0.012-0.021i$
\\ \hline
$|O_1|$ & $0.166$  & $0.027$ & $0.010$ & $0.017$ & $0.027$ & $0.010$ & $0.017$  
& $0.038$ & $0.012$ & $0.024$ \\ 
\hline\hline
\end{tabular}
\caption{Summary of translational symmetry-breaking order parameters (OPs). $\mu_{m_e}$ denotes the onsite energy of the mirror even orbital, and $t_{m_e,d_i}$ denotes the hopping from $m_e$ to $d_i$ orbitals. $d^{i=1,2}_i$ denotes the $d_i$ orbital on the two Kagome sub-layers, and $d^{e}_i$ the mirror even orbital formed by $d_i$ orbitals in two sub-layers, defined in \cref{Eq:mirror_even_Vd}. The absolute values $|O_1|$ for mirror-odd $d$ bases $t_{m_e,d_{z^2}^o}$, $t_{m_e,d_{yz}^o}$, $t_{m_e,d_{x^2-y^2}^o}$ are 0.005, 0.007, 0.004, respectively, which are much smaller than the values of the mirror-even bases. 
}
\label{tab:order_param_values}
\end{table}

We then consider the order parameters given by the NN hopping between $m_e$ and a $d$ orbital of V. Assume the Hamiltonian in the CDW phase is 
\begin{equation}
H^{0}(\mathbf{k})=
\begin{bmatrix}
\mu_{m_e^1} & S_{m_ed}^1(\mathbf{k}) & & & & \\ 
S_{m_ed}^{1,\dagger}(\mathbf{k}) & \mu_{d^1} & & & & \\ 
 & &\mu_{m_e^2} & S_{m_ed}^2(\mathbf{k}) & & \\ 
 & & S_{m_ed}^{2,\dagger}(\mathbf{k}) & \mu_{d^2} &\\ 
 & & & & \mu_{m_e^3} & S_{m_ed}^{3}(\mathbf{k})\\ 
 & & & &  S_{m_ed}^{3,\dagger}(\mathbf{k}) & \mu_{d^3}\\ 
\end{bmatrix}
\end{equation}
where the $m_e$ and $d$ orbitals are sorted into three layers (labeled by $1,2,3$) in the CDW primitive cell, and we limit the hopping to the nearest neighbor which makes orbitals in three layers decoupled. 
Then
\begin{equation}
\begin{aligned}
H^{1}(\mathbf{k})=S_{\mathbf{k}} H^{0}(\mathbf{k}) S_{\mathbf{k}}^{-1} &=
\begin{bmatrix}
\bar{\mu}_{m_e} & \bar{S}_{m_ed}(\mathbf{k}) & & & & \\ 
\bar{S}_{m_ed}^{\dagger}(\mathbf{k}) & \bar{\mu}_{d} & & & & \\ 
 & &\bar{\mu}_{m_e} & \bar{S}_{m_ed}(\mathbf{k}) & & \\ 
 & & \bar{S}_{m_ed}^{\dagger}(\mathbf{k}) & \bar{\mu}_{d} &\\ 
 & & & & \bar{\mu}_{m_e} & \bar{S}_{m_ed}(\mathbf{k})\\ 
 & & & &  \bar{S}_{m_ed}^{\dagger}(\mathbf{k}) & \bar{\mu}_{d} \\ 
\end{bmatrix}\\
&+
\begin{bmatrix}
0& 0& O_1(\mu_{m_e}) & O_1(S_{m_ed}(\mathbf{k})) & 
O_1^*(\mu_{m_e}) & O_1^*(S_{m_ed}(\mathbf{k}))\\ 
0& 0& O_1(S_{m_ed}^{\dagger}(\mathbf{k})) & O_1(\mu_{d}) & 
O_1^*(S_{m_ed}^{\dagger}(\mathbf{k})) & O_1^*(\mu_{d}) \\ 
 & & 0& 0& O_1(\mu_{m_e}) & O_1(S_{m_ed}(\mathbf{k}))\\ 
 & & 0& 0& O_1(S_{m_ed}^{\dagger}(\mathbf{k})) & O_1(\mu_{d})\\ 
 & & & & 0&0\\ 
 h.c. & & & & 0&0\\ 
\end{bmatrix}\\
\end{aligned}
\label{Eq:order_param_Vhop}
\end{equation}
where $\bar{\mu}=\frac{1}{3}(\mu_1+\mu_2+\mu_3)$ is the averaged onsite energy (similar for $\bar{S}_{m_ed}(\mathbf{k})$), and 
$O_1(\mu)=\frac{1}{3}(\mu_{1}+\mu_{2}e^{i\frac{2\pi}{3}}+\mu_{3}e^{-i\frac{2\pi}{3}})$ (similar for $O_1(S_{m_ed}(\mathbf{k}))$) are defined similarly as in \cref{Eq:order_param_onsite}. 
In \cref{Eq:order_param_Vhop}, the three diagonal blocks are the averaged onsite and hoppings that should be very close to the non-CDW values, and off-diagonal blocks are order parameters induced by translational symmetry breaking, i.e., the inequivalent onsite energy and hopping values in three layers of the CDW primitive cell. The order parameter for a larger set of orbitals and longer-range hoppings can be derived similarly. 

\begin{figure}[h]
    \centering
    \includegraphics[width=0.8\textwidth]{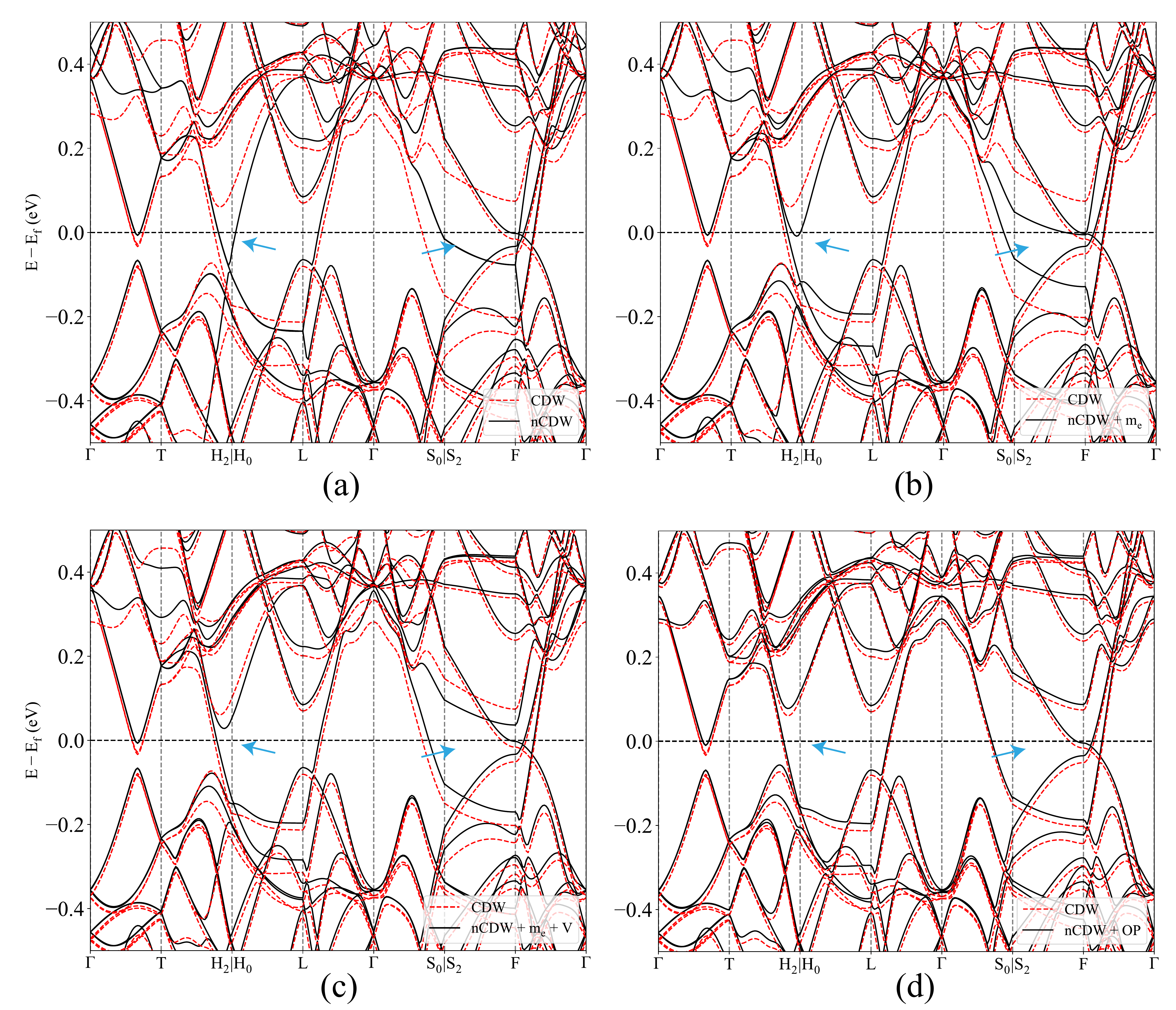}
    \caption{(a) Comparison of CDW bands (red) and folded non-CDW bands (black) by $\mathbf{Q}=\frac{1}{3}(\bm{b}_1+\bm{b}_2+\bm{b}_3)$. Two blue arrows mark the main differences where the CDW bands have gap openings.
    (b) Comparison of CDW bands (red) and non-CDW bands (black) with order parameters of the onsite energies of the mirror even orbitals $m_e$ of triangular Sn$^T$ $p_z$ orbitals. The added order parameters split the bands marked by the blue arrows. 
    (c) Comparison of CDW bands (red) and non-CDW bands (black) with order parameters of the onsite energies of the mirror-even orbitals $m_e$, and the NN hoppings between $m_e$ and mirror-even $d_{z^2}, d_{yz}, d_{x^2-y^2}$ orbitals of V defined in \cref{Eq:mirror_even_Vd}. The NN hoppings with V further enlarge the band splitting and can quantitatively reproduce the band structure near $E_f$ and thus reproduce the Fermi surface in the CDW phase.
    (d) Comparison of CDW bands (red) and non-CDW bands (black) with all translational symmetry-breaking order parameters (OPs). The bands show good agreement.
    }
    \label{fig:order_param}
\end{figure}

We then compute \textit{ab-initio} values of these hopping order parameters. 
For each Sn$^T$ atom, there are six V atoms circling around it, with the same distance in the non-CDW phase enforced by the $C_6$ symmetry. In the CDW phase, however, the six V atoms split into two groups related by the $C_3$ symmetry, with a distance difference less than $0.005\AA$ as $C_6$ is only weakly broken. Thus we treat the six V atoms circling Sn$^T$ as equivalent. In \cref{tab:hop_V_me}, we list the hopping value from $m_e^{i}$ ($i=1,2,3$ denote $m_e$ in three layers) to $d_{z^2}^{i},d_{yz}^{i},d_{x^2-y^2}^{i}$ ($i=1,2$ denote $d$ orbitals in two Kagome sub-layers in each layer) orbitals of V. The hopping values are averaged over six $C_6$-related V atoms in each Kagome layer (note for $d_{yz}$, the $C_{2z}$-related hoppings have opposite signs, and we take their absolute value to average). 
The translational symmetry-breaking order parameters are listed in \cref{tab:order_param_values}. The onsite order parameters of V and hopping order parameters of $d_{xz}, d_{xy}$ have negligible values and are omitted. 
It can be seen that the order parameters of onsite terms of $m_e$ are about 10 times larger than that of the hopping terms between $m_e$ and $d$ orbitals of V. However, as the number of V atoms is 6 times the number of $m_e$ in the unit cell, the band splittings in \cref{fig:order_param} caused by these two sets of order parameters are comparable.

One could also introduce a mirror-even basis for the $d$-electrons at V
\ba 
&d_{\mathbf{R},i, z^2, e, \sigma} = \frac{1}{\sqrt{2}} (d_{\mathbf{R}, i, z^2,1,\sigma} + d_{\mathbf{R},i,z^2, 2,\sigma}),\quad  
d_{\mathbf{R},i, yz,e,\sigma} = \frac{1}{\sqrt{2}} (d_{\mathbf{R}, i, yz, 1,\sigma} - d_{\mathbf{R},i, yz, 2,\sigma}) \nonumber,\\ 
&
d_{\mathbf{R}, i, x^2-y^2, e ,\sigma} = \frac{1}{\sqrt{2}} (d_{\mathbf{R},i, x^2-y^2, 1,\sigma} + d_{\mathbf{R}, i, x^2-y^2, 2,\sigma}),
\label{Eq:mirror_even_Vd}
\ea 
where $i=1,2,3$ denote Kagome sublattice, $1,2$ denote two Kagome sub-layers. 
Their hoppings with $m_e$ are also listed in \cref{tab:hop_V_me}, and order parameters are listed in \cref{tab:order_param_values}. These mirror-even $d$ bases have more significant order parameters compared with the mirror-odd bases, as shown in \cref{tab:order_param_values}.

In summary, the mean-field Hamiltonian of CDW phase $H_{CDW}$ can be approximately described by 
\ba 
H_{CDW} \approx  H_{nCDW} + H_{onsite} +H_{Bond}
\ea 
where $H_{nCDW}$ is the single-particle Hamiltonian of the non-CDW phase, $H_{onsite}$ and $H_{Bond}$ characterize the mean-field contribution of CDW order parameters, where $H_{onsite}$ describes the on-site term of mirror-even orbital of triangular Sn atom and $H_{Bond}$ describes the bond term between mirror-even orbital of triangular Sn atom and mirror-even $d$ orbital of V atom. 
The onsite term can be written as 
\ba 
H_{onsite} = \sum_{\mathbf{R},\sigma} \Phi_{onsite, \mathbf{R}} c_{\mathbf{R},e,\sigma}^\dag c_{\mathbf{R},e,\sigma} 
\ea 
where $\mathbf{R}$ denotes the position vector of the non-CDW unit cell. $\Phi_{onsite}$ generates a CDW modulation and takes the form of  
\ba 
\Phi_{onsite,\mathbf{R}} = \rho_{o} \cos(\mathbf{Q}\cdot\mathbf{R} +\phi)
\label{Eq:me_onsite_modulation}
\ea 
where the value of $\rho_{o},\phi$ are given in \cref{tab:hop_modulation_value}, and $\mathbf{Q}=(\frac{1}{3},\frac{1}{3},\frac{1}{3})$ is the CDW wavevector. Note that the $\mathbf{R}$-independent constant onsite term is given by $H_{nCDW}$. 
As for the bond term, we have 
\ba 
H_{Bond}=\sum_{(\mathbf{R},\mathbf{R}',i)\in NN,\alpha,\sigma}\rho_{B,\alpha} B_{\mathbf{R},\alpha} c_{\mathbf{R},e,\sigma}^\dag d_{\mathbf{R},i,\alpha,e,\sigma}
\ea 
where $\alpha=z^2,yz,x^2-y^2$ denote orbital, $i=1,2,3$ denote Kagome sublattice, and $N.N.$ labels the set of nearest-neighbor bonds between mirror-even $p_z$ orbital of triangular Sn and mirror-even $d$ orbital of V. $B_{\mathbf{R},e}$ describe the mean-field order parameters which have a modulation with wavevector $\mathbf{Q}=(\frac{1}{3},\frac{1}{3},\frac{1}{3})$
\ba 
B_{\mathbf{R},\alpha} = \rho_{B,\alpha} \cos(\mathbf{Q}\cdot\mathbf{R} + \phi)
\label{Eq:me_hop_modulation}
\ea 
where the values of $\rho_{B,\alpha}, \phi$ are given in \cref{tab:hop_modulation_value}.

\begin{table}[h]
\begin{tabular}{c|c|ccc|c}
	\hline\hline
	Parameter      & $\rho_o$  & $\rho_{B,d_{z^2}}$   & $\rho_{B,d_{yz}}$    & $\rho_{B, x^2-y^2}$ & $\phi$
		\\ \hline
	Value & 0.333 &  -0.077 & 0.025 & -0.047 & $-2\pi/3$ \\ 
	   \hline\hline
    \end{tabular}
\caption{Parameters values of the CDW-modulated onsite energy of mirror-even orbital $m_e$ of triangular Sn and the bond term between $m_e$ and mirror-even $d$ orbitals of V, defined in \cref{Eq:me_onsite_modulation} and \cref{Eq:me_hop_modulation}.}
\label{tab:hop_modulation_value}
\end{table}

\subsubsection{Results}
In this part, we analyze the band structure of the CDW phase and find the additional terms induced by the CDW phase, which can be understood as the mean-field order parameters. We compute the order parameters using the algorithm introduced in the previous subsection.

In \cref{fig:order_param}(a), we first compare the band structures of CDW and non-CDW phases (directly folded by $\mathbf{Q}=\frac{1}{3}(\bm{b}_1+\bm{b}_2+\bm{b}_3)$), which are both computed using the experimental structures\cite{ARA22}. Two blue arrows mark the gap opening in the CDW phase. The $T-H_2$ path in the CDW BZ is equivalent to the $(0,0,1/2)-(0.27, 0, 1/2)$ path (in basis $\bm{b}_{i=1,2,3}$) in the non-CDW BZ, i.e., $A-L$ direction, while the $\Gamma-S_0|S_2-F$ path is equivalent to $(1/3,1/3,1/3)-(1/3,0,1/3)-(1/6,1/6,1/3)$ path (also in basis $\bm{b}_{i=1,2,3}$) in the non-CDW BZ. Thus the non-CDW bands marked by two blue arrows contribute to the large Fermi surface in \cref{fig:fermi_surface}(a) and are gapped in the CDW phase. This can also be verified from the unfolded CDW bands in \cref{fig:ScV6Sn6_struct_band}(d) where the CWD bands are gapped on the $k_3=\pi$ plane in the non-CDW BZ.

We then extract the order parameters and add them to the non-CDW TB. 
We first consider the order parameters given by the onsite term of the mirror-even orbitals $m_e$ formed by the $p_z$ of Sn$^T$, defined in \cref{eq:ele_even_odd}.
In \cref{fig:order_param}(b), we show the non-CDW band structure by adding order parameters given by onsite energy of $m_e$, where gap openings appear at bands marked by blue arrows. However, the band splittings are not large enough as the split bands still cross $E_f$ and disagree with the CDW bands.

We then further consider the order parameters given by the NN hoppings between $m_e$ and mirror-even $d_{z^2}, d_{yz}, d_{x^2-y^2}$ orbitals of V, with band structure shown in \cref{fig:order_param}(c) ($d_{xz}$ and $d_{xy}$ of V have negligible effects and are not considered). The band splittings at two blue arrows are enlarged. Although the resultant bands cannot match the CDW bands exactly, they quantitatively reproduce the CDW bands near $E_f$ and thus can reproduce the Fermi surface in the CDW phase.

At last, we add all translational symmetry-breaking order parameters as shown in \cref{fig:order_param}(d). The bands show good agreement.

\subsection{The phonon order parameter of the CDW phase}\label{app:sec:CDW_ops:phonon}
In this \siSection{}, we analyze the experimentally-obtained equilibrium displacements of the atoms in the CDW phase of ScV$_6$Sn$_6${} and show that they are consistent with the condensation of its lowest-energy phonon at the $\bar{\mathrm{K}}$ point. 

\subsubsection{Atomic displacements in the CDW phase}\label{app:sec:CDW_ops:phonon:atomic_displacement}

We first note that the unit cell of the CDW phase is only \emph{approximately} (but not \emph{exactly}) commensurate with the unit cell of the normal phase. To be specific, the primitive lattice vectors $\vec{P}_i$ (with $1 \leq i \leq 3$) describing the unit cell of the CDW phase defined in \cref{eq:basis_transf_a} can be written as the following linear combination of lattice vectors in the normal phase
\begin{equation}
	\begin{bmatrix}
		\vec{P}_1\\ 
		\vec{P}_2\\ 
		\vec{P}_3 
	\end{bmatrix}=\begin{bmatrix}
		 0 & 0.994002 & 0.995313 \\
 -0.994002 & -0.994002 & 0.995313 \\
 0.994002 & 0 & 0.995313 	\end{bmatrix}
	\begin{bmatrix}
		\vec{a}_1\\ 
		\vec{a}_2\\ 
		\vec{a}_3 
	\end{bmatrix}\approx\begin{bmatrix}
		0 & 1 & 1 \\
		-1 & -1 & 1 \\
		1 & 0 & 1
	\end{bmatrix}
	\begin{bmatrix}
	\vec{a}_1\\ 
	\vec{a}_2\\ 
	\vec{a}_3 
	\end{bmatrix}.
	\label{app:eqn:cdw_lat_vecs_exact_and_approx}
\end{equation}
For the normal phase we use the relaxed crystal structure from \cref{Tab:atom_pos_compare}, whereas for the CDW phase, we employ the crystal structure measured experimentally in this work from \cref{Tab:atom_pos_compare_CDW_allatom,Tab:atom_pos_compare_CDW}. Within the CDW phase, the equilibrium positions of each of the atoms (henceforth denoted by $\vec{r}'_{i}$, with $1 \leq i \leq 3 \times 13=39$) can be written as linear combinations of the CDW lattice vectors as follows
\begin{equation}
	\label{app:eqn:atoms_CDW_lin_combo}
	\vec{r}'_{i} = \sum_{j=1}^{3} A_{ij} \vec{P}_j \quad \text{for $1 \leq i \leq 39$},
\end{equation}
where the real matrix $A_{ij}$ can be read from \cref{Tab:atom_pos_compare_CDW_allatom}. In what follows, we will neglect the small deviations (of approximately $0.5\%$) from \emph{perfect} commensurability between the unit cells of the CDW and normal phases of ScV$_6$Sn$_6${} and take the lattice vectors of the former to be exact integer linear combinations of those of the latter, as implied by the approximation of \cref{app:eqn:cdw_lat_vecs_exact_and_approx}. At the same time, we will take the position of the atoms in the CDW phase to be given by the same linear combinations from \cref{app:eqn:atoms_CDW_lin_combo}, but within the perfectly commensurate CDW unit cell. 

Assuming a perfectly commensurate CDW structure, we can map every atom from the CDW unit cell to the corresponding atom in the normal phase unit cell. This is because the atoms are only slightly displaced in the CDW structure relative to the normal phase. Assume that $\mathcal{I}_{i}$ (with $1 \leq \mathcal{I}_i \leq 13$) is the index within the normal phase unit cell of the $i$-th atom from the CDW unit cell (where $1 \leq i \leq 39$). The mapping $\mathcal{I}_{i}$ can be found as follows: 
\begin{enumerate}
	\item Take the $i$-th atom from the CDW unit cell whose displacement from the CDW unit cell origin is given by $\vec{r}_i'$.
	\item Since the CDW and normal phase unit cells have the same origin, we then find the atom $j$ within the normal phase unit cell that is located at \emph{approximately} the same absolute position (modulo a \emph{normal} lattice vector translation). From a mathematical standpoint, this equivalent to finding $j$ such that 
	\begin{equation}
		\label{app:eqn:min_for_CDW_mapping}
		\min_{\vec{R}} \abs{\vec{r}_i'-\vec{R}-\vec{r}_j},
	\end{equation}
	is minimized. In \cref{app:eqn:min_for_CDW_mapping}, $\vec{R}$ are lattice vectors within the normal phase. We then conclude that $\mathcal{I}_i = j$. For later use, we also denote the lattice vector for which the minimum of \cref{app:eqn:min_for_CDW_mapping} is attained by $\vec{R}_i$. 
	\item We repeat the procedure described above until all the atoms in the CDW unit cell have been mapped to those in the normal phase unit cell. It is worth noting that, relative to the normal (high-temperature) phase, the atoms are only slightly displaced in the CDW one, \textit{i.e.}{}, 
	\begin{equation}
		\abs{\vec{r}'_i - \vec{R}_i - \vec{r}_{\mathcal{I}_i}} \lesssim \SI{0.16}{\angstrom}, \quad \text{for any $1\leq i \leq 39$}.
	\end{equation}
\end{enumerate} 
Since the CDW unit cell is three times larger than the normal phase unit cell, the mapping $\mathcal{I}_{i}$ for $1 \leq i \leq 39$ will map exactly three distinct types of atoms in the CDW unit cell to the same type of atoms in the normal unit cell.

We now construct the displacement field corresponding to the CDW structure. We let $\Delta_{j \nu} \left( \vec{R} \right)$ denote the displacement of the $j$-th atom from unit cell $\vec{R}$ (indexed within the normal phase unit cell) along the Cartesian direction $\nu$. Using the mapping established above, the $i$-th atom in the CDW unit cell (located at position $\vec{r}'_i$ in the CDW phase) is the $\mathcal{I}_i$-th atom in the normal phase. As such, we find
\begin{equation}
	\label{app:eqn:disp_in_CDW_phase}
	\Delta_{\mathcal{I}_i \nu} \left( \vec{R}_i + \sum_{m=1}^{3} n_m \vec{P}_m \right) = \left[ \vec{r}'_i - \vec{R}_i - \vec{r}_{\mathcal{I}_i} \right]_{\nu}, \quad \text{for any $1 \leq i \leq 39$, $n_1,n_2,n_3 \in \mathbb{Z}$, and $\nu \in \lbrace x,y,z\rbrace$}.
\end{equation}
We also define the Fourier-transformed displacement field according to \cref{eq:mod_exp} which is given by 
\begin{equation}
	\label{app:eqn:disp_in_CDW_phase_ft_definition}
	\Delta_{j\nu} \left( \vec{k} \right) = \frac{1}{\sqrt{N}} \sum_{\vec{R}} \Delta_{j \nu} \left( \vec{R} \right) e^{-i \vec{k} \cdot \left( \vec{R} + \vec{r}_j \right) },
\end{equation}
which, upon substituting \cref{app:eqn:disp_in_CDW_phase}, becomes
\begin{align}
	\Delta_{j \nu} \left( \vec{k} \right) &= \frac{1}{\sqrt{N}} \sum_{n_1,n_2,n_3} \sum_{i=1}^{39} \Delta_{j \nu} \left( \vec{R}_i + \sum_{m=1}^{3} n_m \vec{P}_m \right) e^{-i \vec{k} \cdot \left( \vec{R}_i + \vec{r}_j + \sum_{m=1}^{3} n_m \vec{P}_m \right)} \delta_{\mathcal{I}_i,j} \nonumber \\
	&= \left( \frac{1}{\sqrt{3}} \sum_{i=1}^{39} \left[ \vec{r}'_i - \vec{R}_i - \vec{r}_{\mathcal{I}_i} \right]_{\nu}  e^{-i \vec{k} \cdot \left( \vec{R}_i + \vec{r}_j \right)} \delta_{\mathcal{I}_i,j} \right) \sum_{m_1,m_2,m_3} \delta_{\vec{k} \cdot \vec{P}_1, 2 \pi m_1} \delta_{\vec{k} \cdot \vec{P}_2, 2 \pi m_2} \delta_{\vec{k} \cdot \vec{P}_3, 2 \pi m_3}. \label{app:eqn:disp_in_CDW_phase_ft}
\end{align}
The only non-equivalent $\vec{k}$ vectors for which the left hand side of \cref{app:eqn:disp_in_CDW_phase_ft} is nonzero are $\vec{k}' = \pm 2\pi \left( \frac{1}{3}, \frac{1}{3}, -\frac{1}{3} \right)$. Finally, for later use in \cref{app:sec:CDW_ops:phonon:phonon_overlan}, we also define the displacement field in the rescaled coordinates introduced in \cref{eq:def_tilde_up}
\begin{equation}
	\label{app:eqn:disp_in_CDW_phase_ft_rescaled}
	\tilde{\Delta}_{i \mu} \left( \vec{k} \right) = \sqrt{M_i} \Delta_{i \mu} \left( \vec{k} \right).
\end{equation}

\subsubsection{Relation with the phonon spectrum}\label{app:sec:CDW_ops:phonon:phonon_overlan}

\begin{figure}[t]
	\centering
	\includegraphics[width=0.5\textwidth]{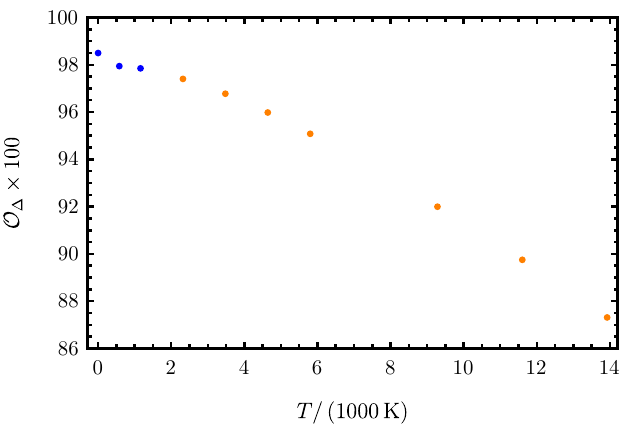}
	\caption{Overlap between the atomic displacement in the CDW phase and the lowest-energy phonon at $\bar{\mathrm{K}}$. We use the experimentally-measured crystal structure of the CDW phase shown in \cref{Tab:atom_pos_compare_CDW_allatom,Tab:atom_pos_compare_CDW} to extract the atomic displacements in the CDW phase. The wave function of the lowest-energy phonon at $\bar{\mathrm{K}}$ is determined from \textit{ab-initio} methods as explained in \cref{app:sec:phonon_spectrum:ab_initio} for various temperatures. The overlap $\mathcal{O}_{\Delta}$ is calculated according \cref{app:eqn:cdw_overlap}. The blue (orange) points correspond to temperatures below (above) the CDW transition.}
	\label{fig:cdw_overlap}
\end{figure}

Since the Fourier transformation of the CDW displacement field is only nonzero at two of the four $\bar{\mathrm{K}}$ points, we conclude that the CDW can be thought of as arising from the condensation of some phonon (or linear combination of phonons modes) with momenta $\vec{k}'= \pm 2\pi \left( \frac{1}{3}, \frac{1}{3}, -\frac{1}{3} \right)$\footnote{Note that condensation at the other two of the four $\bar{\mathrm{K}}$ points, having momenta $\vec{k}''= \pm 2\pi \left( \frac{1}{3}, \frac{1}{3}, \frac{1}{3} \right)$, will lead to an equivalent CDW phase, related to the one measured in the experiment by a $m_z$ transformation.}. In what follows, we will show that the phonon whose condensation leads to the CDW phase is primarily the lowest energy phonon at $\bar{\mathrm{K}}$.

Using the same notation as in \cref{app:sec:ele_phon_model:hamiltonian}, we consider the coherent phonon state corresponding to the $n'$-th phonon mode at momentum $\vec{k}'$,  
\begin{equation}
	\ket{\beta,\vec{k}',n'} = \exp \left(\beta b^{\dagger}_{\vec{k}',n'} - \beta^{*} b_{\vec{k}',n'} \right) \ket{0} = e^{-\frac{\abs{\beta}^2}{2}}\sum_{m=0}^{\infty} \frac{\beta^m}{\sqrt{m!}} \left( b^{\dagger}_{\vec{k}',n'} \right)^m \ket{0},
\end{equation}
where $\beta$ is some complex number. The state $\ket{\beta,\vec{k}',n'}$ is a linear superposition of states with different numbers of $b^{\dagger}_{\vec{k}',n'}$ phonon excitations and is an eigenstate of $b_{\vec{k}',n'}$ of eigenvalue $\beta$
\begin{equation}
	b_{\vec{k}',n'} \ket{\beta,\vec{k}',n'} = \beta \ket{\beta,\vec{k}',n'}.
\end{equation}
Using \cref{app:eqn:ft_convention_phonon,eq:u_to_b}, we can evaluate the expectation value the of atomic displacement operators
\begin{equation}
	u_{i\mu} \left( \vec{R} \right) = \sum_{\vec{R}} \sum_{n} \sqrt{\frac{1}{2 \omega_{n} (\vec{q}) M_i}} U_{i\mu,n} (\vec{q}) \left( b_{\vec{q},n} + b^{\dagger}_{-\vec{q},n}  \right) e^{i \vec{q} \cdot \left( \vec{R} + \vec{r}_i \right)},
\end{equation}
on the coherent state $\ket{\beta,\vec{k}',n'}$ to find
\begin{equation}
	\label{app:eqn:coherent_stat_spatial_variation}
	\bra{\beta,\vec{k}',n'} u_{i\mu} \left( \vec{R} \right) \ket{\beta,\vec{k}',n'} = 2 \Re \left[ \sqrt{\frac{1}{2 \omega_{n'} (\vec{k}') M_i}} \beta U_{i\mu,n'} (\vec{k}') e^{i \vec{k}' \cdot \left( \vec{R} + \vec{r}_i \right)} \right],
\end{equation}
which corresponds to a macroscopic oscillation of the atoms according to the phonon with wave function $U_{i\mu,n'} (\vec{k}')$. In comparison, the atomic displacement in the CDW phase can be written from \cref{app:eqn:disp_in_CDW_phase_ft_definition,app:eqn:disp_in_CDW_phase_ft,app:eqn:disp_in_CDW_phase_ft_rescaled} as 
\begin{equation}
	\Delta_{i\mu} \left( \vec{R} \right) = 2 \Re \left[ \frac{1}{\sqrt{M_i}} \tilde{\Delta}_{i\mu} (\vec{k}') e^{i \vec{k}' \cdot \left( \vec{R} + \vec{r}_i \right)} \right],
\end{equation}
which, having identical spatial periodicity to \cref{app:eqn:coherent_stat_spatial_variation}, shows that, indeed, the atomic displacement in the CDW phase arise from the condensation of a phonon (or linear combination of phonons) at the $\vec{k}'$ point. To further show that out of the phonon modes at $\vec{k}'$, $\Delta_{i\mu} \left( \vec{R} \right)$ corresponds to the condensation of the lowest-energy one, we plot the overlap 
\begin{equation}
	\label{app:eqn:cdw_overlap}
	\mathcal{O}_{\Delta} = \frac{\abs{\sum_{i,\mu} U^{*}_{i\mu,n'} (\vec{k}') \tilde{\Delta}_{i\mu} (\vec{k}')}^2}{\sum_{i,\mu} \tilde{\Delta}^{*}_{i\mu,n'} (\vec{k}') \tilde{\Delta}_{i\mu} (\vec{k}')},
\end{equation}
in \cref{fig:cdw_overlap} as a function of temperature. In \cref{app:eqn:cdw_overlap}, $U_{i\mu,n'} (\vec{k}')$ is the eigenstates of the lowest energy phonon at momenum $\vec{k}'$, as obtained through \cref{app:eqn:eigs_of_phon_dyn_mat} from the \textit{ab-initio} dynamical matrix of the system. We find that the overlap is large ($\mathcal{O}_{\Delta} \gtrsim 0.86$) for the entire temperature range we consider, attaining a maximum at zero temperature where $\mathcal{O}_{\Delta} \approx  0.98$. This confirms that the CDW phase indeed arises from the condensation of the lowest-energy phonon mode at the $\bar{\mathrm{K}}$ point.

\section{Effective simple models of the low-energy phonon spectra of ScV$_6$Sn$_6${} and YV$_6$Sn$_6${}}\label{app:sec:three_band_model}

The quadratic phonon Hamiltonian obtained directly from \textit{ab-initio} methods features 39 bands (\textit{i.e.}{}, three bands for each of the 13 atoms within the unit cell) for both ScV$_6$Sn$_6${} and YV$_6$Sn$_6${}. To gain an analytical understanding of the low-energy phonon spectrum, in this \siSection{}, we derive an effective three-band model describing the soft phonon of ScV$_6$Sn$_6${}, as well as three- and four-band generalizations describing the low-energy phonon spectrum of YV$_6$Sn$_6${}. We start by observing that the low-energy part of the spectrum (which includes the soft phonon) is mainly contributed by the motion of the Sn$^T$ atoms along the $z$-direction (away from the $\Gamma$ point) and by the longitudinal $z$-directed acoustic mode (near the $\Gamma$ point). We construct a Wannier orbital that (together with the $z$-directed motion of the Sn$^T$ atoms) correctly describes the acoustic mode near the $\Gamma$ point. By projecting the spectrum onto this Wannier orbital and the $z$-directed displacement of the two Sn$^T$ atoms, we derive a three-band effective phonon Hamiltonian. The hopping parameters of the resulting Hamiltonian are fitted to the \textit{ab-initio} spectrum at various temperatures such that the soft-phonon is accurately reproduced within the simplified model in both energetics and wave function. Finally, for YV$_6$Sn$_6${}, we derive two types of effective models: a three-band model (similar to the effective Hamiltonian describing the soft phonon mode of ScV$_6$Sn$_6${}), and a four-band model which additionally takes into account the $z$-directed displacement of the Y atoms. For all the effective models derived in this \siSection{}, we provide both a ``full'' model which very closely reproduces the low-energy phonon spectrum (but features numerous hopping parameters), as well as ``simplified'' model which reproduces the spectrum only qualitatively (but otherwise features fewer hopping parameters).

\subsection{Wannier phonon orbitals}\label{app:sec:three_band_model:orbitals}

\subsubsection{Symmetry of phonon bands}\label{app:sec:three_band_model:orbitals:symmetry}

As discussed around \cref{app:eqn:repres_mat_sym}, from a symmetry standpoint, it is useful to think about the displacements of the $i$-th atom along the three Cartesian directions as being $p_x$, $p_y$, and $p_z$ orbitals (located at the Wyckoff position of the corresponding atom), on which the dynamical matrix introduced in \cref{app:eqn:dyn_matrix} acts as a tight-binding Hamiltonian~\cite{XU22}. Consequently, if $U_{i\mu,n} \left(\vec{k} \right)$ is an eigenvector of $D\left( \vec{k} \right)$ corresponding to the frequency $\omega_{n}\left( \vec{k} \right)$, then $\sum_{i,\mu} \mathcal{D}_{j\nu,i\mu}^{g} U^{(*)}_{i\mu,n} \left( \vec{k} \right)$ will be an eigenvector of $D\left( g\vec{k} \right)$ corresponding to the same frequency $\omega_{n}\left( \vec{k} \right)$. We can therefore introduce the (unitary) sewing matrix 
\begin{equation}
	\label{app:eqn:sewing_mat_def}
	\left[ \mathcal{B}^{g}\left( \vec{k} \right) \right]_{nm} = \sum_{i,\nu,j,\mu} U^{*}_{j \nu,n} \left( g\vec{k} \right) \mathcal{D}_{j\nu,i\mu}^{g} U^{(*)}_{i\mu,m} \left( \vec{k} \right),
\end{equation}
such that 
\begin{equation}
	\label{app:eqn:sewing_mat_action}
	\sum_{i,\nu} \mathcal{D}_{j\nu,i\mu}^{g} U^{(*)}_{i\mu,m} \left( \vec{k} \right) = \sum_{n} \left[ \mathcal{B}^{g}\left( \vec{k} \right) \right]_{nm} U_{j \nu,n} \left( g\vec{k} \right), \quad \text{with} \left[ \mathcal{B}^{g}\left( \vec{k} \right) \right]_{nm} = 0 \quad \text{if} \quad \omega_{n}\left( \vec{k} \right) \neq \omega_{m}\left( \vec{k} \right).
\end{equation}

We now restrict to the symmetry operations $g$ that belong to the little group of some momentum point $\vec{k}$ (\textit{i.e.}{}, for which $g \vec{k} = \vec{k}$). The corresponding sewing matrix is block diagonal (with the blocks corresponding to the eigensubspaces of the dynamical matrix $D\left( \vec{k} \right)$ with distinct eigenvalues) and forms a (generally reducible) representation of the little group of $\vec{k}$~\cite{zotero-4159,BRA17,ELC17,VER17}{}. As such, all phonon bands can be labeled at all momentum points by the irreducible representation (irreps) of the corresponding little group.

\subsubsection{Wannier basis}\label{app:sec:three_band_model:orbitals:wannier_basis}

\begin{figure}[t]
	\centering
	\includegraphics[width=\textwidth]{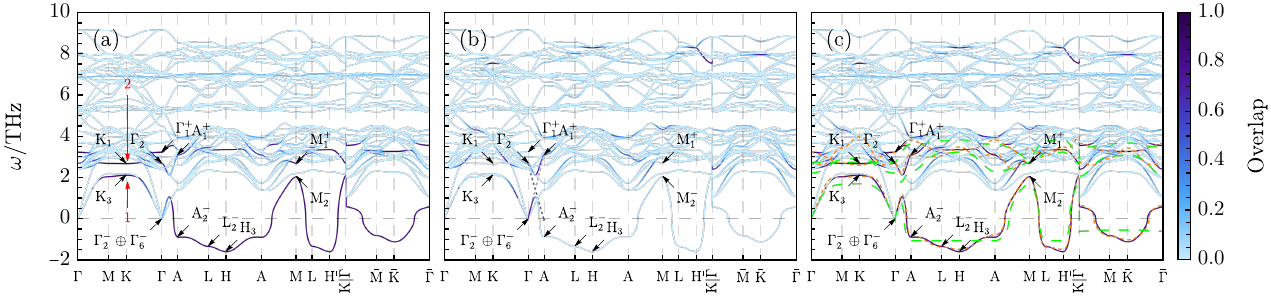}
	\caption{Effective three-band model for the soft phonon mode of ScV$_6$Sn$_6${}. In (a)-(c), the phonon bands of ScV$_6$Sn$_6${} obtained from \textit{ab-initio} calculations (see \cref{app:sec:phonon_spectrum:ab_initio}) are shown in shades of blue. The irreps at high-symmetry momenta are indicated by black arrows for some of the low-energy bands. The colormap corresponds to the overlap between the phonon modes obtained from \textit{ab-initio} calculations and the $z$-directed displacement of the Sn$^T$ atoms in (a), the effective Wannier orbital of the $z$-directed acoustic mode from \cref{app:eqn:wan_3} in (b), and the Hillbert subspace spanned by \emph{both} the $z$-directed displacement of the Sn$^T$ atoms \emph{and} the effective Wannier orbital of the $z$-directed acoustic mode from \cref{app:eqn:wan_3} in (c). In (a), the two red arrows indicate the almost-flat acoustic (1) and optical (2) phonon modes that are primarily contributed by the $z$-directed motion of the Sc$^T$ atoms. The dashed lines between the $\Gamma$ and $\mathrm{A}$ points in (b) depict schematically the avoided crossing between the optical phonon mainly contributed by the $z$-directed motion of the Sc$^T$ atoms (dark gray) and the acoustic $z$-directed phonon (light gray). Finally, the dashed orange (green) lines in (c) depict the band structure of the full (simplified) effective three-band model derived in \cref{app:sec:three_band_model:Hamiltonian,app:sec:three_band_model:results}.}
	\label{fig:zero_temperature_phonon_model}
\end{figure}

Our main goal in this \siSection{} is to construct an effective, yet accurate phonon Hamiltonian that \emph{includes} the soft phonon mode of ScV$_6$Sn$_6${}, but otherwise features \emph{far fewer bands} (ideally, the minimum necessary) than the phonon Hamiltonian derived directly from \textit{ab-initio} calculation. To do so, we start by identifying a small Hilbert space which includes the soft phonon mode. Additionally, we will require that the resulting restricted Hilbert space features a localized real space representation, or, in other words, that it admits an exponentially localized (Wannier) basis which obeys the symmetries of the space group. In this way, we will be able to find short-range effective force constant matrix elements between these Wannier phonon states. Using symmetry considerations, we will then construct an effective Hamiltonian within this small Hilbert space and require that the energy \emph{and} the wave function of the soft phonon is accurately reproduced by this effective Hamiltonian.   

As shown in by the zero-temperature phonon spectrum from \cref{fig:zero_temperature_phonon_model}(a), the soft phonon mode is primarily contributed by the $z$-component vibrations of the Sn$^T$ atoms throughout the entire Brillouin zone, except for a small region around the $\Gamma$ point. In the $k_z = 0$ plane and away from the $\Gamma$ point, the $z$-component vibrations of the Sn$^T$ atoms form two very flat bands corresponding to an optical ($\omega_{n} \left( \vec{k} \right) \approx \SI{2.75}{\tera\hertz}$) and an acoustic ($\omega_{n} \left( \vec{k} \right) \approx \SI{2}{\tera\hertz}$) phonon mode, which are indicated by the red arrows in \cref{fig:zero_temperature_phonon_model}(a). Along the $\Gamma-\mathrm{A}$ line, the optical phonon softens and undergoes an avoided crossing with one of the acoustic mode. We note that there are three acoustic modes (corresponding to the three spatial dimensions), which have zero frequency at the $\Gamma$ point and which transform as the $\Gamma_2^{-} \oplus \Gamma_6^{-}$ representation (as will be shown below). Remembering that the acoustic phonon modes correspond to the in-phase motion of all the atoms within the unit cell, we can write the corresponding (un-normalized) phonon eigenvectors as 
\begin{equation}
	\label{app:eqn:all_acoustic_modes_at_gamma}
	U_{i\mu,1} \left( \vec{0} \right) \propto \delta_{\mu,1} \sqrt{M_i}, \quad
	U_{i\mu,2} \left( \vec{0} \right) \propto \delta_{\mu,2} \sqrt{M_i}, \quad 
	U_{i\mu,3} \left( \vec{0} \right) \propto \delta_{\mu,3} \sqrt{M_i},
\end{equation}  
which correspond to the $x$-, $y$-, and $z$-directed acoustic phonons. Using \cref{app:eqn:sewing_mat_def}, one can directly compute the sewing matrices corresponding to the symmetries of the little group of $\vec{k}=\vec{0}$ within the zero-frequency eigenspace spanned by the phonon modes from \cref{app:eqn:all_acoustic_modes_at_gamma}. We find that the $z$-direct acoustic phonon transforms as the $\Gamma_{2}^{-}$ irrep, whereas the in-plane acoustic phonon mode (\textit{i.e.}{}, the $x$- and $y$-directed acoustic phonons) form the two-dimensional $\Gamma_{6}^{-}$ irrep~\cite{zotero-4159,BRA17,ELC17,VER17}{}. Along the $\Gamma-\mathrm{A}$ line, the irreps corresponding to acoustic modes subduce according to~\cite{zotero-4159,BRA17,ELC17,VER17}{}
\begin{equation}
	\Gamma_{2}^{-} \to \Delta_{1} \quad \text{and} \quad
	\Gamma_{6}^{-} \to \Delta_{6},
\end{equation} 
meaning that only the $z$-directed acoustic mode can hybridize with the $z$-directed displacement of the Sn$^{T}$ atoms, which induce two one-dimensional $\Delta_{1}$ irreps, along the $\Gamma-\mathrm{A}$ line\footnote{The $C_{3z}$ and $\mathcal{T}$ symmetries are enough to suppress the coupling between the in-plane acoustic modes and the $z$-directed displacement of the Sn$^T$ atoms.}~\cite{zotero-4159,BRA17,ELC17,VER17}{}. As such, to construct an accurate effective Hamiltonian for the soft phonon mode, the restricted Hilbert space we consider must include \emph{both} the $z$-component vibrations of the Sn$^T$ atoms \emph{and} the $z$-directed acoustic phonon (but not the two in-plane acoustic phonons). 

\begin{figure}[t]
	\centering
	\includegraphics[width=0.33\textwidth]{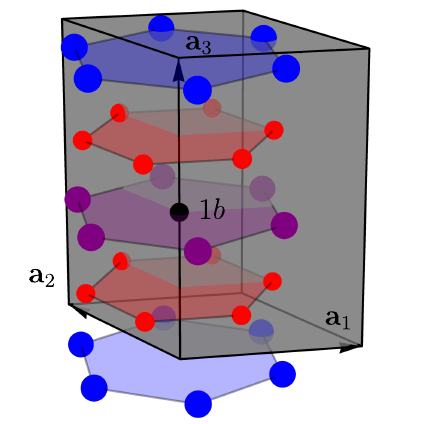}
	\caption{Compact localized state used as a precursor to the third Wannier function from \cref{app:eqn:wan_3}. The ScV$_6$Sn$_6${} unit cell in the pristine phase is shown by the gray parallelogram, with the $1b$ Wyckoff position indicated by the black dot. Considering the atoms nearest to the $1b$ Wyckoff position shown, we take symmetric linear combinations of the $z$-directed displacement of the V atoms (Red), the Sn$^H$ atoms located in the plane containing the unit cell origin (blue), and the Sn$^H$ atoms located in the plane containing the $1b$ Wyckoff position (purple). The resulting compact localized state, given by \cref{app:eqn:trial_for_w3}, transforms according to an $A_{2u}$ irrep located at the $1b$ Wyckoff position.}
	\label{fig:wannierAcoustic}
\end{figure}

We let $w_{i\mu,n} \left( \vec{k} \right)$ denote the momentum-space Wannier functions indexed by $1 \leq n \leq 3$ and spanning the Hillbert space on which we will derive our effective three-band model\footnote{We note that $w_{i\mu,n} \left( \vec{k} \right)$ are \emph{not} the trial Wannier functions that are usually supplied to Wannier90 to obtain tight-binding models for a subset of \textit{ab-initio} bands (as was done for example in \cref{app:sec:electronic_band_structure:comp_details}), but rather the \emph{actual} Wannier functions on which we will define our effective Hamiltonian.}. We also define $\boldsymbol{\rho}_n$ to be the displacement of the Wannier center of $w_{i\mu,n} \left( \vec{k} \right)$ from the unit cell origin. As mentioned above, the first two Wannier functions are given by the $z$-directed displacements of the Sn$^T$ atoms 
\begin{align}
	w_{i\mu,1} \left( \vec{k} \right) &= \delta_{i,8} \delta_{\mu,3}, \label{app:eqn:wan_1}\\
	w_{i\mu,2} \left( \vec{k} \right) &= \delta_{i,9} \delta_{\mu,3}, \label{app:eqn:wan_2}
\end{align}
whose Wannier centers are given by the positions of the Sn$^T$ atoms, \textit{i.e.}{} $\boldsymbol{\rho}_1 = \vec{r}_8$ and $\boldsymbol{\rho}_2 = \vec{r}_9$, as given in \cref{app:eqn:atom_position_vecs}. Unlike the $z$-component of the Sn$^T$ displacements, the acoustic $z$-directed phonon mode does not correspond to an atomic limit. Moreover, the mere notion of acoustic phonon is also only rigorously defined in the vicinity of the $\Gamma$ point, where it corresponds to the simultaneous displacement of all atoms within the unit cell phonon and whose (un-normalized) wave function is given approximately by 
\begin{equation}
	\label{app:eqn:acoustic_near_gamma}
	U_{i\mu,3} \left( \vec{k} \right) \underset{\sim}{\propto} \delta_{\mu,3} \sqrt{M_i}, \quad \text{for} \quad \vec{k} \approx \vec{0},
\end{equation}
where the relation ``$\underset{\sim}{\propto}$'' corresponds to two vectors that are approximately equal up to normalization prefactors. The acoustic phonon mode will be supported only partially by the two Wannier functions from \cref{app:eqn:wan_1,app:eqn:wan_2}, the rest of which will be supported on all the other atoms. Indeed, ''projecting out`` the first two Wannier functions from the $z$-directed acoustic phonon near the $\Gamma$ point leaves us with a contribution
\begin{equation}
	\label{app:eqn:remaining_acoustic}
	\sum_{i,\mu} \left[ \delta_{i,j} \delta_{\mu,\nu} - \sum_{n=1}^{2} w_{j \nu,n} \left( \vec{k} \right) w^{*}_{i \mu,n} \left( \vec{k} \right) \right] U_{i\mu,3} \left( \vec{k} \right) \underset{\sim}{\propto} \delta_{\nu,3} \sqrt{M_j} \left( 1 - \delta_{j,8} - \delta_{j,9} \right), \quad \text{for} \quad \vec{k} \approx \vec{0}.
\end{equation}
which must be supported by the third Wannier function $w_{i\mu,3} \left( \vec{k} \right)$, if the three Wannier functions are to correctly capture the low-energy phonon spectrum. To complete the Wannier basis, we therefore want to find the Wannier function $w_{i\mu,3} \left( \vec{k} \right)$ that gives the correct $\vec{k} \approx 0$ limit,
\begin{equation}
	\label{app:eqn:requirement_w3_strong}
	w_{j\nu,3} \left( \vec{k} \right) \underset{\sim}{\propto} \delta_{\nu,3} \sqrt{M_j} \left( 1 - \delta_{j,8} - \delta_{j,9} \right), \quad \text{for} \quad \vec{k} \approx \vec{0},
\end{equation}
such that when the other Wannier states [$w_{i\mu,n} \left( \vec{k} \right)$ with $n=1,2$] are added, the basis can reproduce the in-phase $z$-directed motion of all atoms and, hence, the acoustic mode. Away from the $\Gamma$ point, the low-energy phonon is well-supported by the Wannier states from \cref{app:eqn:wan_1,app:eqn:wan_2}, as shown in \cref{fig:zero_temperature_phonon_model}(a), so we will not be concerned with the \emph{exact} form of the $w_{j\nu,3} \left( \vec{k} \right)$ function and only require that it corresponds to a symmetric, exponentially-localized Wannier orbital (as will be explained below). Finally, we note that the atomic mass of Sc is much smaller than the atomic masses of V and Sn, so we are justified in making a further approximation and completely ignoring the Sc contribution to the acoustic phonon by requiring that
\begin{equation}
	\label{app:eqn:requirement_w3_weak}
	w_{j\nu,3} \left( \vec{k} \right) \underset{\sim}{\propto} \delta_{\nu,3} \sqrt{M_j} \left( 1  - \delta_{j,1} - \delta_{j,8} - \delta_{j,9} \right), \quad \text{for} \quad \vec{k} \approx \vec{0},
\end{equation}
where $j=1$ corresponds to the Sc atom.

To extend \cref{app:eqn:requirement_w3_weak} throughout the entire Brillouin zone (and thus construct the third Wannier orbital completely), we will invoke the following symmetry arguments. As discussed above and shown in \cref{fig:zero_temperature_phonon_model}(a), the $w_{i\nu,1} \left( \vec{k} \right)$ and $w_{i\nu,2} \left( \vec{k} \right)$ orbitals fully support the low-energy phonon throughout the entire Brillouin zone, except in the vicinity of the $\Gamma$ point. Moreover, the soft phonon in the $k_z=\pi$ plane, can be understood as stemming from the optical phonon contributed by the $z$-directed displacement of the Sn$^T$ atoms, shown by the red arrow in \cref{fig:zero_temperature_phonon_model}(a), which drops in energy along the $\Gamma - \mathrm{A}$ line and becomes imaginary in frequency in the $k_z=\pi$ plane. The avoided crossing between the $z$-directed acoustic mode and the optical phonon going soft is superimposed schematically in \cref{fig:zero_temperature_phonon_model}(b). For the avoided crossing, the branch starting at the $\Gamma_{1}^{+}$ irrep in the optical phonon and ending at the $\mathrm{A}_{2}^{-}$ irrep in the soft phonon is already well supported by the first two Wannier orbitals from \cref{app:eqn:wan_1,app:eqn:wan_2}. On the other hand, the acoustic branch starting at the $\Gamma_{2}^{-}$ irrep at zero energy and ending at the $\mathrm{A}_{1}^{+}$ irrep does not have a good overlap with the first two Wannier functions, because the acoustic modes are comprised of the in-phase motion of \emph{all} atoms in the unit cell (not just that of Sn$^T$). \Cref{app:eqn:requirement_w3_weak} requires that $w_{i\nu,3} \left( \vec{k} \right)$ provides the missing part of the acoustic phonon at the $\Gamma$ point. By visually inspecting the avoided crossing in \cref{fig:zero_temperature_phonon_model}(b), we also find that the $z$-directed acoustic mode should transform according to the $\mathrm{A}_{1}^{+}$ irrep at the $\mathrm{A}$ point. Remembering that $w_{i\nu,3} \left( \vec{k} \right)$ is an atomic limit (by virtue of being a Wannier function), we find that the former should obey the following constraints:
\begin{itemize}
	\item It should provide the missing part of the acoustic phonon in the vicinity of the $\Gamma$ point, as implied by \cref{app:eqn:requirement_w3_weak}.
	\item It should induce the $\Gamma_{2}^{-}$ irrep at the $\Gamma$ point (so that it can overlap with the $z$-directed acoustic phonon).
	\item It should induce the $\mathrm{A}_{1}^{+}$ irrep at the $\mathrm{A}$ point (so that it correctly captures the avoided crossing along the $\Gamma-\mathrm{A}$ line). 
\end{itemize}

There is only one atomic limit that can induce a single $\Gamma_{2}^{-}$ irrep and a single $\mathrm{A}_{1}^{+}$ irrep~\cite{zotero-4159,BRA17,ELC17,VER17}{}: $p_z$ orbital ($A_{2u}$ irrep) located at the $1b$ Wyckoff position. With the Wannier center aand symmetry properties of $w_{i\mu, 3}\left( \vec{k} \right)$ identified, we now use \cref{app:eqn:requirement_w3_weak} to fix its form: we want $w_{i\mu, 3}\left( \vec{0} \right)$ to be comprised of the in-phase $z$-directed motion of the Sn$^H$ and V atoms. To build such a Wannier function, we start from a real space picture and leverage the idea of compact localized states used in the context of crystalline flat bands~\cite{BER08}. Specifically, we take symmetric combinations of the $z$-directed displacements of the Sn$^H$ and V atoms located closest to the $1b$ Wyckoff position, as indicated in \cref{fig:wannierAcoustic},
\begin{align}
	u^{\text{CLS}} \left( \vec{R} \right) &\propto \delta_{\mu,3} \bigg[ c_1 \sum_{j=0}^{1} \left( u_{(2+3j)\mu} \left( \vec{R} \right) + u_{(3+3j)\mu} \left( \vec{R} \right) + u_{(4+3j)\mu} \left( \vec{R} \right) \right. \nonumber \\
	& \quad \left. + u_{(2+3j)\mu} \left( \vec{R} - \vec{a}_1 \right) + u_{(3+3j)\mu} \left( \vec{R} -  \vec{a}_2 \right) + u_{(4+3j)\mu} \left( \vec{R} - \vec{a}_1 - \vec{a}_2 \right) \right) \nonumber \\
	& + c_2 \left( 
	u_{13 \mu} \left( \vec{R} \right) + u_{12 \mu} \left( \vec{R} \right) + 
	u_{13 \mu} \left( \vec{R} - \vec{a}_1 \right) \right. \nonumber \\
	& \quad + \left. u_{12 \mu} \left( \vec{R}  - \vec{a}_1 -  \vec{a}_2 \right) + 
	u_{13 \mu} \left( \vec{R} - \vec{a}_1 - \vec{a}_2 \right) + u_{12 \mu} \left( \vec{R}  -  \vec{a}_2 \right)  \right) \nonumber \\
	& + c_3 \sum_{j=0}^{1}\left( 
	u_{11 \mu} \left( \vec{R} + j \vec{a}_3 \right) + u_{10 \mu} \left( \vec{R} + j \vec{a}_3 \right) + 
	u_{11 \mu} \left( \vec{R} - \vec{a}_1  + j \vec{a}_3 \right) \right. \nonumber \\
	& \quad +  \left. u_{10 \mu} \left( \vec{R}  - \vec{a}_1 -  \vec{a}_2  + j \vec{a}_3 \right) + 
	u_{11 \mu} \left( \vec{R} - \vec{a}_1 - \vec{a}_2  + j \vec{a}_3 \right) + u_{10 \mu} \left( \vec{R}  -  \vec{a}_2  + j \vec{a}_3 \right)  \right) \bigg], \label{app:eqn:trial_for_w3} 
\end{align}
where $c_1$, $c_2$, and $c_3$ are real coefficients which will be fixed below. The compact localized state $u^{\text{CLS}} \left( \vec{R} \right)$ transforms as a $A_{2u}$ irrep located at the $1b$ Wyckoff position. Note that $u^{\text{CLS}} \left( \vec{R} \right)$ itself is \emph{not} the third Wannier function, as the states $u^{\text{CLS}} \left( \vec{R} \right)$ corresponding to different unit cells $\vec{R}$ are only linearly independent\footnote{Linear independence of $u^{\text{CLS}} \left( \vec{R} \right)$ is harder to see in real space. It will become apparent once we Fourier transform to obtain the third Wannier function in \cref{app:eqn:wan_3}.}, but not orthonormal. Nevertheless, by Fourier transforming and normalizing the resulting wave function, one can use $u^{\text{CLS}} \left( \vec{R} \right)$ to build the analytical expression for $w_{i\nu,3} \left( \vec{k} \right)$.

We start by fixing the $c_1$, $c_2$, and $c_3$ coefficients: we require that the zero momentum Fourier transformation of $u^{\text{CLS}} \left( \vec{R} \right)$ 
\begin{equation}
	\sum_{\vec{R}} u^{\text{CLS}} \left( \vec{R} \right) = \delta_{\mu,3} \sum_{\vec{R}} \left[2 c_1 \sum_{j=2}^{7} u_{j\mu} \left( \vec{R} \right) + 3c_2 \left( u_{12\mu} \left( \vec{R} \right) + u_{13\mu} \left( \vec{R} \right) \right) + 6c_3 \left( u_{10\mu} \left( \vec{R} \right) + u_{11\mu} \left( \vec{R} \right) \right) \right]
\end{equation}
consists of in-phase $z$-directed displacements of the V and Sn$^H$ atoms (corresponding to the $z$-directed acoustic phonon), which results in the following solution
\begin{equation}
	c_1 = \frac{1}{2}, \quad
	c_2 = \frac{1}{3}, \quad
	c_3 = \frac{1}{6}.
\end{equation}
Now, we can obtain the third Wannier orbital by Fourier transforming $u^{\text{CLS}} \left( \vec{R} \right)$ from \cref{app:eqn:trial_for_w3} and normalizing the corresponding Bloch function at every momentum
\begin{align}
	w_{j\nu,3} \left( \vec{k} \right) &= \frac{\delta_{\nu,3} \sqrt{M_j}}{\mathcal{N}\left( \vec{k} \right)} e^{-i\vec{k} \cdot \left( \vec{r}_j - \boldsymbol{\rho}_3 \right)} \bigg\lbrace \frac{1}{2} \sum_{l=0}^{1} \left[ \delta_{j,2+3l} \left( 1 + e^{i \vec{k} \cdot \vec{a}_1} \right) + \delta_{j,3+3l} \left( 1 + e^{i \vec{k} \cdot \vec{a}_2} \right) + \delta_{j,4+3l} \left( 1 + e^{i\vec{k} \cdot \left(\vec{a}_1 + \vec{a}_2 \right)} \right)  \right] \nonumber \\
	& + \frac{1}{3} \left[ 
	\delta_{j,13} \left( 1 + e^{i \vec{k} \cdot \vec{a}_1} + e^{i \vec{k} \cdot \left(\vec{a}_1 + \vec{a}_2 \right)} \right) +
	\delta_{j,12} \left( 1 + e^{i \vec{k} \cdot \vec{a}_2} + e^{i \vec{k} \cdot \left(\vec{a}_1 + \vec{a}_2 \right)} \right) \right] \nonumber \\
	& + \frac{1}{6} \sum_{l=0}^{1} e^{-i \vec{k} \cdot j \vec{a}_3}\left[ 
	\delta_{j,11} \left( 1 + e^{i \vec{k} \cdot \vec{a}_1} + e^{i \vec{k} \cdot \left(\vec{a}_1 + \vec{a}_2 \right)} \right) +
	\delta_{j,10} \left( 1 + e^{i \vec{k} \cdot \vec{a}_2} + e^{i \vec{k} \cdot \left(\vec{a}_1 + \vec{a}_2 \right)} \right) \right] \bigg\rbrace, \label{app:eqn:wan_3}
\end{align}
where $\boldsymbol{\rho}_3 = \frac{1}{2} \vec{a}_3$ denotes the Wannier center of the third Wannier orbital. The normalization constant $\mathcal{N} \left( \vec{k} \right)$ from \cref{app:eqn:wan_3} is fixed by requiring that $w_{j\nu,3} \left( \vec{k} \right)$ is properly normalized, \textit{i.e.}{} 
\begin{equation}
	\sum_{j}\sum_{\mu} w_{j\mu,3} \left( \vec{k} \right) w^{*}_{j\mu,3} \left( \vec{k} \right) = 1, \quad \text{for any} \quad \vec{k}.
\end{equation}
Note that $\mathcal{N} \left( \vec{k} \right)$ is non-vanishing throughout the entire Brillouin zone (which implies that the compact localized states from \cref{app:eqn:trial_for_w3} corresponding to different unit cells were indeed linearly independent). Moreover, the orbital corresponding to $w_{j\nu,3} \left( \vec{k} \right)$ is exponentially localized by virtue of $w_{j\nu,3} \left( \vec{k} \right)$ being an analytical function of $\vec{k}$ of the correct periodicity. As seen in \cref{fig:zero_temperature_phonon_model}(b), the $w_{j\nu,3} \left( \vec{k} \right)$ Wannier function fully supports the $z$-directed acoustic phonon mode near the $\Gamma$ point. All together, the three Wannier functions fully support the low-energy phonon, as shown in \cref{fig:zero_temperature_phonon_model}(c).

\subsection{Wannier phonon Hamiltonian}\label{app:sec:three_band_model:Hamiltonian}
\begin{figure}[!t]
	\centering
	\includegraphics[width=0.6\textwidth]{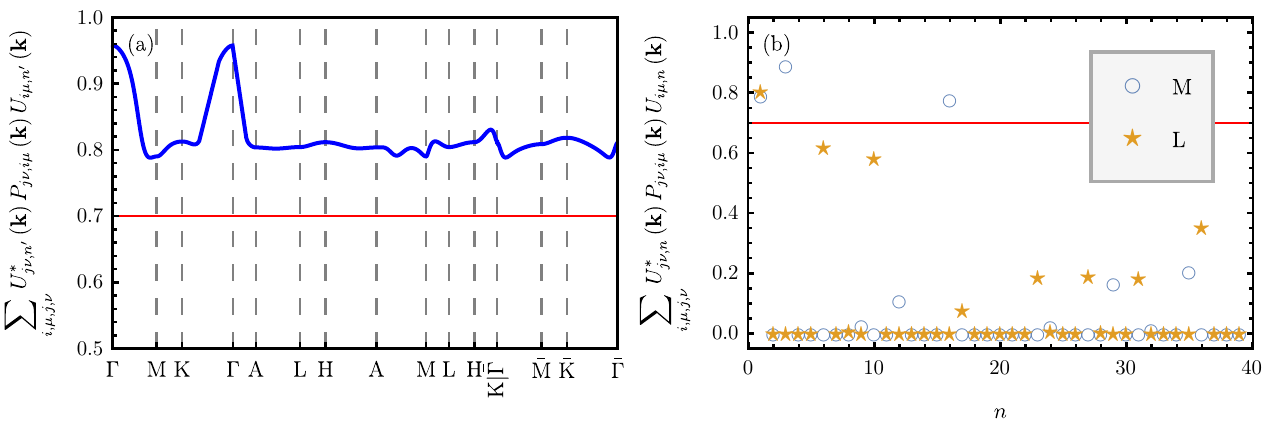}
	\caption{Overlap between the various phonon eigenstates of ScV$_6$Sn$_6${} and the Wannier basis from \cref{app:sec:three_band_model:orbitals:wannier_basis}. In (a) we show the weight of the lowest-energy phonon band of ScV$_6$Sn$_6${} on the three-orbital Wannier basis. The overlap between the Wannier states and all the 31 phonon eigenstates at the $\mathrm{M}$ and $\mathrm{L}$ points is shown in (b). In both panels, the phonon eigenstates are obtained from the \textit{ab-initio} dynamical matrix at zero temperature. The red guide line indicates the threshold $\epsilon_{\text{overlap}} = 0.7$ for including the states into the fitting procedure, as explained around \cref{app:eqn:fit_thresh}.}
	\label{fig:understanding}
\end{figure}
\begin{figure}[!t]
	\centering
	\includegraphics[width=\textwidth]{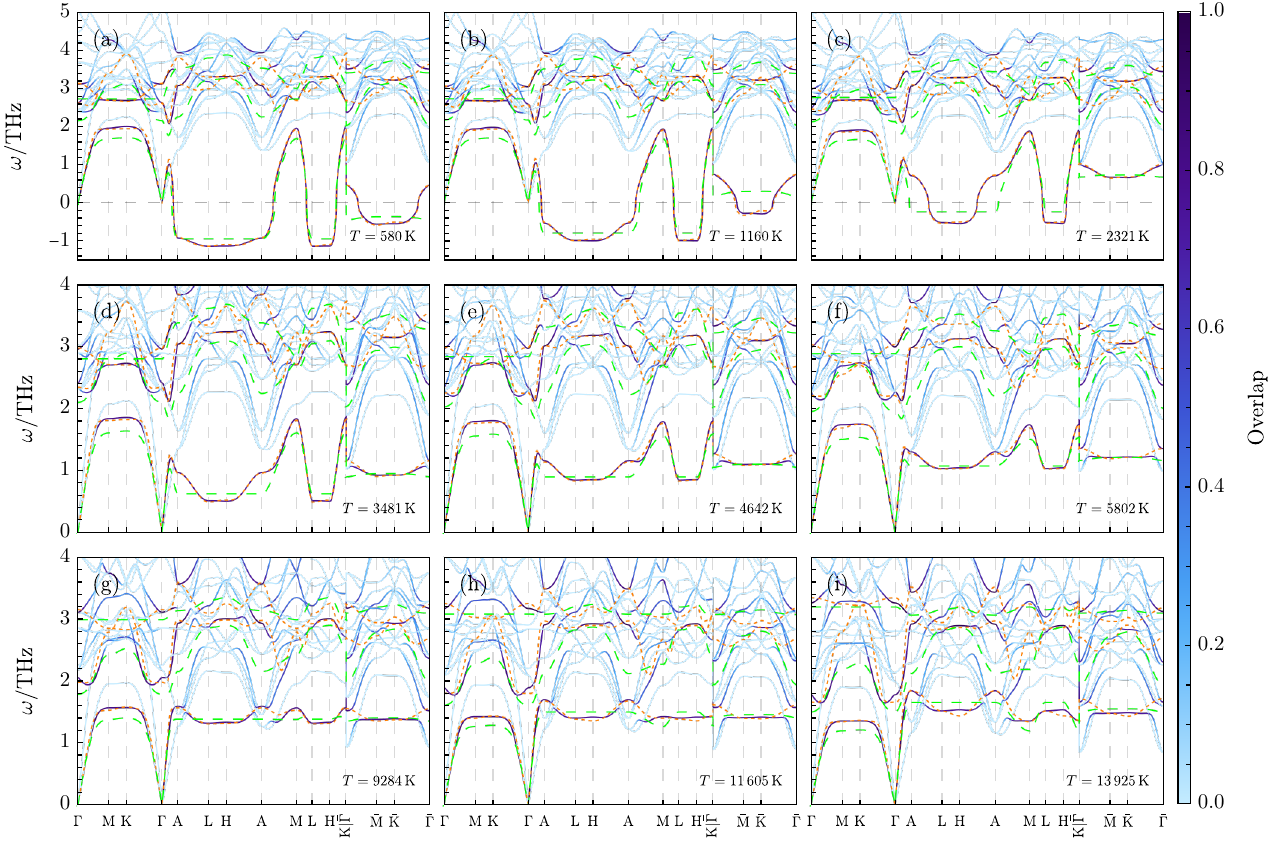}
	\caption{Effective three-band model of the low-energy phonon spectrum of ScV$_6$Sn$_6${} at finite temperature. The phonon bands obtained from \textit{ab-initio} calculations are shown in blue and are colored according to their weight on the three-orbital Wannier basis derived in \cref{app:sec:three_band_model:orbitals}. The temperature used in the simulations is shown for each panel. The dispersion of the full (simplified) effective three-band model is shown by the orange (green) dashed lines.}
	\label{fig:fin_temperature_phonon_model_Sc}
\end{figure}
\begin{figure}[!t]
	\centering
	\includegraphics[width=\textwidth]{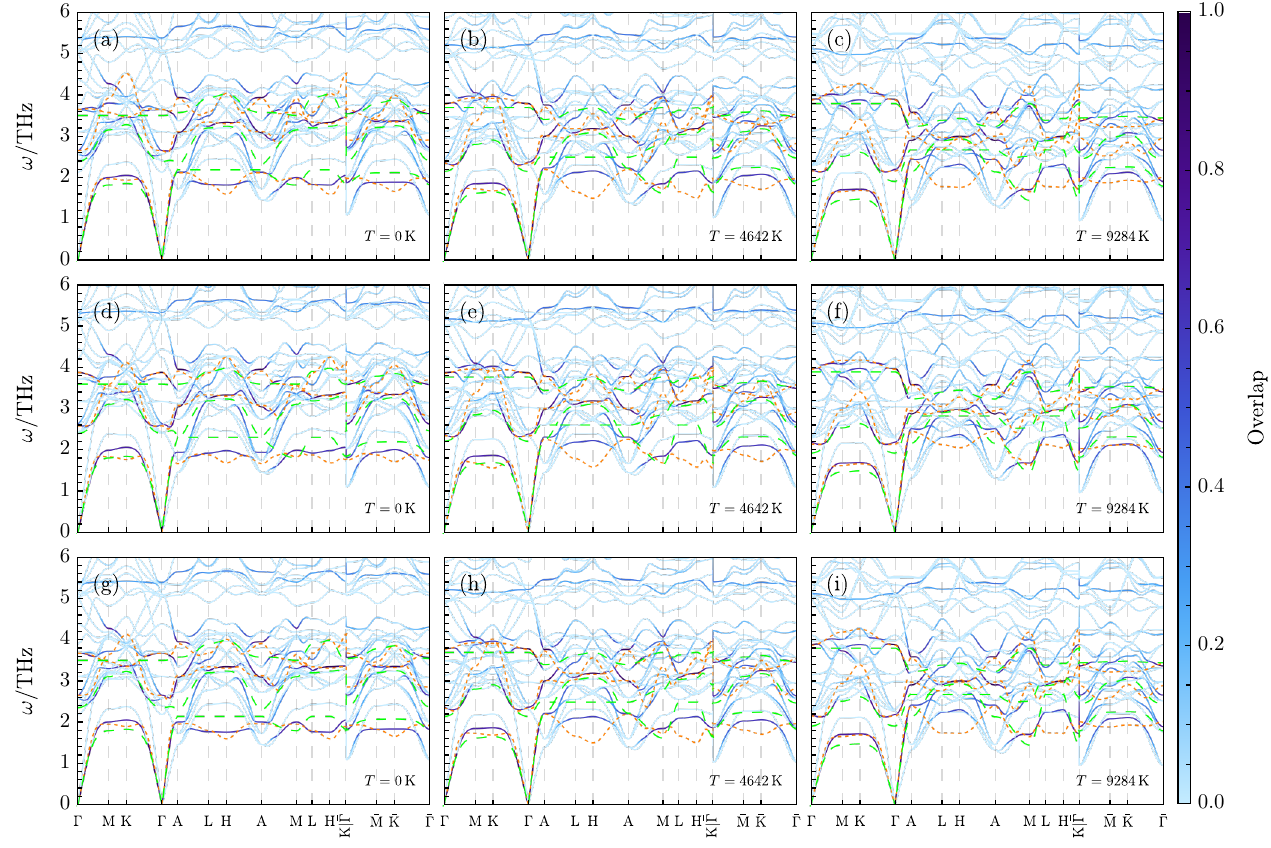}
	\caption{Effective three-band model of the low-energy phonon spectrum of YV$_6$Sn$_6${} at finite temperature. The phonon bands obtained from \textit{ab-initio} calculations are shown in blue and are colored according to their weight on the three-orbital Wannier basis derived in \cref{app:sec:three_band_model:orbitals}. The temperature used in the simulations is shown for each panel. The dispersion of the full (simplified) effective three-band model is shown by the orange (green) dashed lines. For the crystal structure of YV$_6$Sn$_6${}, we use the relaxed structure from Ref.~\cite{POK21} in (a)-(c), the experimentally-measured structure from Ref.~\cite{ROM11} in (d)-(f), and the relaxed structure from Ref.~\cite{ROM11} in (g)-(i). }
	\label{fig:fin_temperature_phonon_model_Y_3}
\end{figure}
\begin{figure}[!t]
	\centering
	\includegraphics[width=\textwidth]{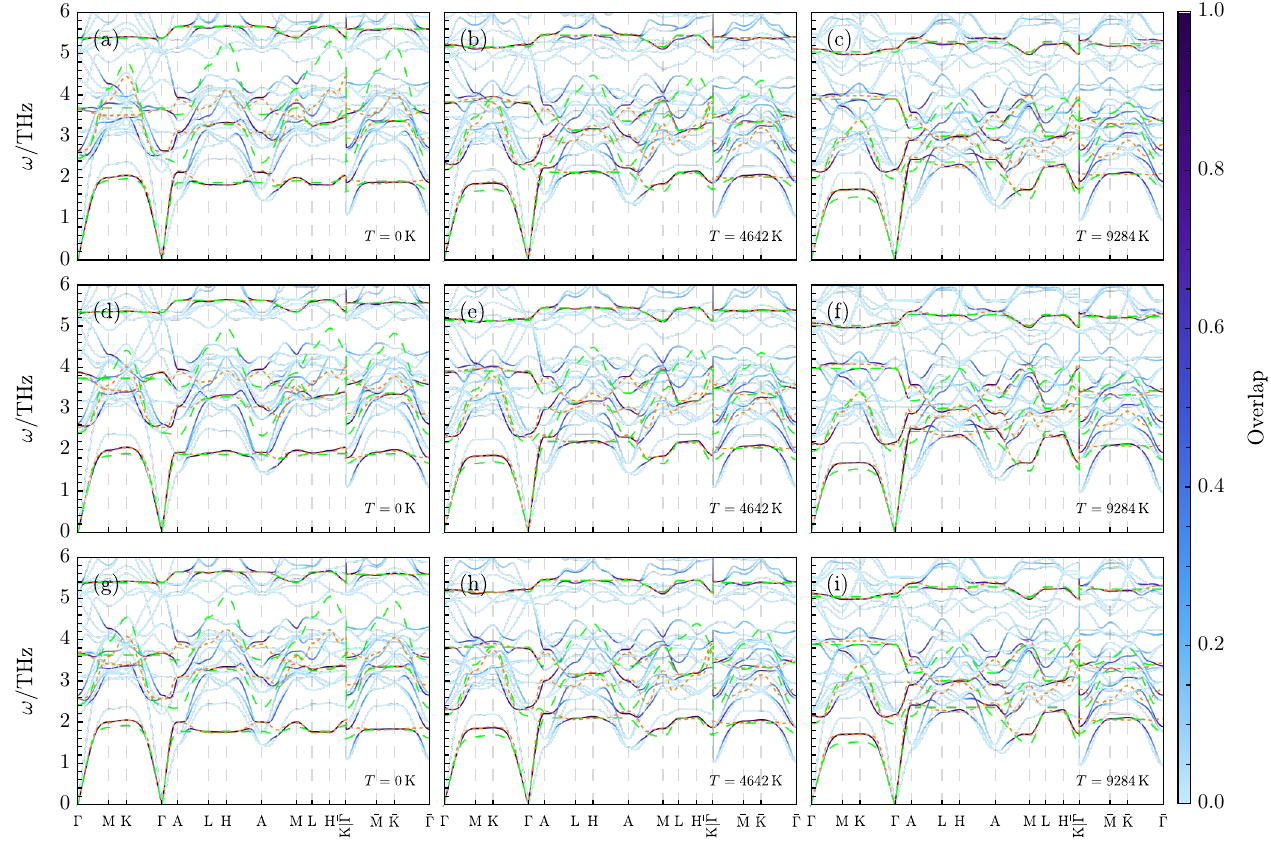}
	\caption{Effective four-band model of the low-energy phonon spectrum of YV$_6$Sn$_6${} at finite temperature. The phonon bands obtained from \textit{ab-initio} calculations are shown in blue and are colored according to their weight on the four-orbital Wannier basis derived in \cref{app:sec:three_band_model:results:Y}. The temperature used in the simulations is shown for each panel. The dispersion of the full (simplified) effective four-band model is shown by the orange (green) dashed lines. For the crystal structure of YV$_6$Sn$_6${}, we use the relaxed structure from Ref.~\cite{POK21} in (a)-(c), the experimentally-measured structure from Ref.~\cite{ROM11} in (d)-(f), and the relaxed structure from Ref.~\cite{ROM11} in (g)-(i). }
	\label{fig:fin_temperature_phonon_model_Y_4}
\end{figure}

With the Wannier basis at hand, we now proceed to find the effective Hamiltonian describing low-energy spectrum of ScV$_6$Sn$_6${} within the Wannier basis defined in \cref{app:sec:three_band_model:orbitals:wannier_basis}. We start by deriving and parameterizing the most general symmetry-preserving tight-binding Hamiltonian within the phonon Wannier basis. The hopping parameters of the effective Hamiltonian are then fitted, such that the latter correctly reproduces the low-energy phonon in both energy and wave function, as obtained from \textit{ab-initio} calculations.

\subsubsection{Symmetry constraints and parameterization}\label{app:sec:three_band_model:Hamiltonian:parameterization}

We let $h_{nm} \left( \vec{k} \right)$ (with $1 \leq n,m \leq 3$) denote the effective phonon Hamiltonian defined within the Wannier basis introduced in \cref{app:sec:three_band_model:orbitals:wannier_basis}, which describes the low-energy phonon spectrum of ScV$_6$Sn$_6${}\footnote{Although we are working with phonons, we will find it more intuitive to use electronic terminology such as ``tight-binding Hamiltonian'' or ``hopping amplitude''.}.  Since the Wannier basis derived in \cref{app:sec:three_band_model:orbitals:wannier_basis} is exponentially localized, the effective phonon Hamiltonian can be written as a tight-binding Hamiltonian with a few nonzero hopping amplitudes $t_{nm} \left( \vec{R} \right)$,
\begin{equation}
	\label{app:eqn:wan_hamiltonian}
	h_{nm} \left( \vec{k} \right) = \sum_{\vec{R}} t_{nm} \left( \vec{R} \right) e^{-i \vec{k} \cdot \left( \vec{R} + \boldsymbol{\rho}_n - \boldsymbol{\rho}_m \right)}.
\end{equation}
In \cref{app:eqn:wan_hamiltonian}, $t_{nm} \left( \vec{R} \right)$ denotes the hopping amplitude between the $m$-th Wannier function located at $\boldsymbol{\rho}_m$ and the $n$-th Wannier function located at $\vec{R} + \boldsymbol{\rho}_n$. 

The hopping amplitudes are variational parameters which will be fixed according to the method described in \cref{app:sec:three_band_model:Hamiltonian:fitting}. At the same time, the hopping amplitudes are also constrained by the crystalline symmetries of the system. Under a crystalline symmetry $g$, the Wannier functions introduced in \cref{app:sec:three_band_model:orbitals:wannier_basis} transform according to 
\begin{equation}
	\label{app:eqn:sym_wan_trafo}
	\sum_{i,\nu} \mathcal{D}_{j\nu,i\mu}^{g} w^{(*)}_{i\mu,m} \left( \vec{k} \right) = \sum_{n=1}^{3} \left[ \mathcal{B}_{w}^{g} \right]_{nm} w_{j \nu,n} \left( g\vec{k} \right), \quad \text{for} \quad  1 \leq m \leq 3,
\end{equation}
where the unitary matrix $\mathcal{B}_{w}^{g}$ is momentum independent by virtue of $w_{i \mu,m} \left( \vec{k} \right)$ being atomic limits and the symmetry group of ScV$_6$Sn$_6${} being symmorphic. The representation matrices $\mathcal{B}_{w}^{g}$ have been tabulated by Ref.~\cite{zotero-4159} for all atomic limits of all Shubnikov Space Groups. As the effective phonon Hamiltonian should be symmetric under the symmetries of ScV$_6$Sn$_6${}, the corresponding matrix elements and hopping amplitudes must obey
\begin{align}
	\label{app:eqn:constraint_symmetry_dynamical}
	\sum_{n,m} \left[ \mathcal{B}_{w}^{g} \right]_{n'n} h_{nm}^{(*)}\left( \vec{k} \right) \left[ \mathcal{B}_{w}^{g} \right]^{*}_{m'm} &= h_{n'm'} \left( g \vec{k} \right), \\
	\sum_{n,m} \left[ \mathcal{B}_{w}^{g} \right]_{n'n} t^{(*)}_{nm} \left( g^{-1} \left( \vec{R} + \boldsymbol{\rho}_{n'} - \boldsymbol{\rho}_{m'} \right) - \left( \boldsymbol{\rho}_{n} - \boldsymbol{\rho}_{m} \right) \right) \left[ \mathcal{B}_{w}^{g} \right]^{*}_{m'm} &= t_{n'm'} \left( \vec{R} \right),
\end{align}
In particular, as a result of $\mathcal{T}$ symmetry, all hopping amplitudes are real. 

\subsubsection{Fitting the hopping amplitudes}\label{app:sec:three_band_model:Hamiltonian:fitting}

The key property required from $h_{nm} \left( \vec{k} \right)$ is that it should correctly reproduce the soft phonon mode in both energetics and wave function. We will now explain exactly what this entails: assume that the soft phonon wave function and frequency at a given momentum $\vec{k}$ are given, respectively, by $U_{i \mu,n'} \left( \vec{k} \right)$ and $\omega_{n'} \left( \vec{k} \right)$, both of which are obtained through \cref{app:eqn:eigs_of_phon_dyn_mat} from the \textit{ab-initio} dynamical matrix of the system. We will then \emph{require} that the effective Hamiltonian $h_{nm} \left( \vec{k} \right)$ has a corresponding eigenvector $W_{n,a'} \left( \vec{k} \right)$ (where $1 \leq a' \leq 3$) with an eigenvalue $\omega^2_{n'} \left( \vec{k} \right)$, \textit{i.e.}{}
\begin{equation}
    \label{app:eqn:match_wan_energy_exact}
    \sum_{m=1}^{3} h_{nm} \left( \vec{k} \right) W_{m,a'} \left( \vec{k} \right) = 
    \omega^2_{n'} \left( \vec{k} \right) W_{n,a'} \left( \vec{k} \right),
\end{equation}
such that 
\begin{equation}
    \label{app:eqn:match_wan_wavf_exact}
    U_{j\nu,n'} \left( \vec{k} \right) = \sum_{n=1}^{3} w_{j\nu,n} \left( \vec{k} \right) W_{n,a'} \left( \vec{k} \right) .
\end{equation}
In simpler terms, \cref{app:eqn:match_wan_energy_exact,app:eqn:match_wan_wavf_exact} state that at each $\vec{k}$, the effective Hamiltonian has one eigenvector that matches the energy and wave function of the lowest-energy phonon mode. Since the effective Hamiltonian is defined in the Wannier basis (as opposed to the atomic displacement basis of the \textit{ab-initio} dynamical matrix), the eigenvector $W_{n,a'} \left( \vec{k} \right)$ is related to the $U_{j\nu,n'} \left( \vec{k} \right)$ eigenvector obtained from \textit{ab-initio} methods through the projection operation from \cref{app:eqn:match_wan_wavf_exact}.

By enforcing \cref{app:eqn:match_wan_energy_exact,app:eqn:match_wan_wavf_exact} at all momenta, we ensure that the soft phonon mode will be correctly captured within our effective Hamiltonian. In practice, however, the lowest-energy phonon mode will not be \emph{exactly} fully supported by the Wannier basis derived in \cref{app:sec:three_band_model:orbitals:wannier_basis}. Take, for instance, the $z$-directed acoustic phonon mode around the $\Gamma$ point: being an acoustic mode, this phonon comprises of the $z$-directed displacements of \emph{all} the atoms within the unit cell, including the Sc atom. However, the latter is not included in our Wannier basis, meaning that \cref{app:eqn:match_wan_wavf_exact} cannot be satisfied exactly at the $\Gamma$ point for the acoustic mode. To get around this problem, we relax the condition in \cref{app:eqn:match_wan_wavf_exact} such that it only holds for the projection of $ U_{j\nu,n'} \left( \vec{k} \right)$ into the Wannier basis at that particular momentum. Defining $P_{i\mu,j\nu} \left(\vec{k}\right)$ to be the projector onto the Wannier basis at $\vec{k}$
\begin{equation}
    P_{i\mu,j\nu} \left(\vec{k}\right) = \sum_{n=0}^{3} w_{i\mu,n} \left( \vec{k} \right) w^{*}_{j\mu,n} \left( \vec{k} \right),
\end{equation}
\cref{app:eqn:match_wan_wavf_exact} is modified to
\begin{equation}
    \label{app:eqn:match_wan_wavf_approx}
    \sum_{i,\mu}  P_{i\mu,j\nu} \left(\vec{k}\right) U_{i\mu,n'} \left( \vec{k} \right) \propto \sum_{n=1}^{3} W_{n,a'} \left( \vec{k} \right) w_{j\nu,n} \left( \vec{k} \right).
\end{equation}
Moreover, in order to reproduce more features of the low-energy phonon spectrum of ScV$_6$Sn$_6${} (in addition to the lowest-energy phonon) -- such as, for example, the bands describing the avoided crossing in \cref{fig:zero_temperature_phonon_model}(b) -- we will require that \cref{app:eqn:match_wan_energy_exact,app:eqn:match_wan_wavf_approx} are satisfied not only for the soft phonon mode, but for \emph{any} state $U_{i\mu,n'} \left( \vec{k} \right)$ having sufficient overlap with the Wannier basis. Mathematically, we take the states with sufficient overlap with the Wannier basis to be the states $U_{i\mu,n'} \left( \vec{k} \right)$ for which
\begin{equation}
	\label{app:eqn:fit_thresh}
    \sum_{i,\mu,j,\nu} U^*_{j\nu,n'} \left( \vec{k} \right) P_{j\nu,i\mu} \left( \vec{k} \right) U_{i\mu,n'} \left( \vec{k} \right) \geq \epsilon_{\text{overlap}},
\end{equation}
where the overlap threshold was set to $\epsilon_{\text{overlap}} = 0.7$. In \cref{fig:understanding}(a), we show the overlap between the lowest-energy phonon band and the Wannier basis. We find that the latter is indeed very well supported by the three Wannier orbitals, with an overlap larger than $\epsilon_{\text{overlap}} = 0.7$ throughout the entire Brillouin zone. The number of higher energy states having sufficient overlap with the Wannier basis does vary throughout the Brillouin zone. In \cref{fig:understanding}(b), we show that at the $\mathrm{M}$ point, there are a total of three states in the \textit{ab-initio} spectrum having an overlap larger than $\epsilon_{\text{overlap}} = 0.7$ with the Wannier basis, whereas at the $\mathrm{L}$ point, there is only one such state. At a given $\vec{k}$-point, the states with insufficient overlap with the Wannier basis cannot be approximated by the effective Hamiltonian, and are thus discarded from the optimization.  

As such, our method for fitting the hopping parameters from \cref{app:eqn:wan_hamiltonian} can be summarized as follows:
\begin{enumerate}
	\item Discretize the Brillouin zone and find all the phonon eigenstates $U_{i\mu,n'} \left( \vec{k} \right)$ (obtained from the \textit{ab-initio} dynamical matrix) with sufficient weight on the Wannier basis, as implied by \cref{app:eqn:match_wan_wavf_approx}. Denote the set of these states to be 
	\begin{equation}
		\mathcal{U} = \set{ U_{i\mu,n'} \left( \vec{k} \right) \mid  \sum_{i,\mu,j,\nu} U^*_{j\nu,n'} \left( \vec{k} \right) P_{j\nu,i\mu} \left( \vec{k} \right) U_{i\mu,n'} \left( \vec{k} \right) \geq \epsilon_{\text{overlap}} }
	\end{equation}
	\item Project every such state on the Wannier basis and require that the projection is eigenvector of the effective Hamiltonian. More specifically, for every $U_{i\mu,n'} \left( \vec{k} \right) \in \mathcal{U}$, compute  
	\begin{equation}
		\label{app:eqn:projection_on_wannier_basis}
		W_{n,a'} \left( \vec{k} \right) = \sum_{j,\nu} w^{*}_{j\nu,n} \left( \vec{k} \right) U_{j\nu,n'} \left( \vec{k} \right),
	\end{equation}
	and require that \cref{app:eqn:match_wan_energy_exact} is satisfied for $W_{n,a'} \left( \vec{k} \right)$ (\textit{i.e.}{}, $W_{n,a'} \left( \vec{k} \right)$ is an eigenvector of $h_{nm} \left( \vec{k} \right)$ with eigenvalue $\omega^2_{n'} \left( \vec{k} \right)$). 
	\item The aforementioned constraints on the effective phonon Hamiltonian form an overdetermined linear system of equation, because the number of free parameters -- the hopping amplitudes $t_{nm} \left( \vec{R} \right)$ -- is much smaller than the number of independent equations. As such, we solve for the hopping amplitudes using least squares optimization. First, define the set of all projections in the Wannier basis to be 
	\begin{equation}
		\mathcal{W} = \set{ W_{n,a'} \left( \vec{k} \right) \mid  	W_{n,a'} \left( \vec{k} \right) = \sum_{j,\nu} w^{*}_{j\nu,n} \left( \vec{k} \right) U_{j\nu,n'} \left( \vec{k} \right) \text{ with } U_{j\nu,n'} \left( \vec{k} \right) \in \mathcal{U}}.
	\end{equation}
	The hopping amplitudes are obtained through the following minimization procedure
	\begin{equation}
		\label{app:eqn:fitting_phonon_minimization}
		\min_{t_{nm} \left( \vec{R} \right)} \left( \sum_{W_{m,a'} \left( \vec{k} \right) \in \mathcal{W}} \sum_{n=1}^{3} \abs{ \sum_{m=1}^{3} \left( h_{nm} \left( \vec{k} \right) - \omega^2_{n'} \left( \vec{k} \right) \delta_{nm} \right) W_{m,a'} \left( \vec{k} \right)}^2 \right),
	\end{equation}
	which ensures that the effective Hamiltonian reproduces, as much as possible, the low-energy phonon spectrum of ScV$_6$Sn$_6${} in both energetics and wave function. In other words, for every $W_{m,a'} \left( \vec{k} \right) \in \mathcal{W}$, we define an error $\sum_{m=1}^{3} \left( h_{nm} \left( \vec{k} \right) - \omega^2_{n'} \left( \vec{k} \right) \delta_{nm} \right) W_{m,a'}$ which quantifies how far $W_{m,a'}$ is from an eigenvector of $h_{nm} \left( \vec{k} \right)$ with eigenvalue $\omega^2_{n'}$. By minimizing the sum of the squares of these reminders, we ensure that all vectors $W_{m,a'} \left( \vec{k} \right) \in \mathcal{W}$ are as close as possible to being eigenvectors of the effective Hamiltonian\footnote{We note that the number of eigenvectors we want the effective Hamiltonian to reproduce is not necessarily three at every $\vec{k}$ point. In \cref{fig:understanding}(b), we show that, while at the $\mathrm{M}$ point, there are three states $U_{i\mu,n'} \left( \vec{k} \right)$ having sufficient overlap with the Wannier basis, at the $\mathrm{L}$ point, there is only one such state (\textit{i.e.}{}, the lowest-energy one). As such, one cannot construct the effective Hamiltonian at every $\vec{k}$ point from its eigenstates and fit the hopping amplitudes to it. Instead we have to employ \cref{app:eqn:fitting_phonon_minimization} which does not need all three eigenvectors of the effective Hamiltonian at all $\vec{k}$ points.}.
\end{enumerate}

\subsubsection{Imposing the sound speed constraints exactly}\label{app:sec:three_band_model:Hamiltonian:sound_speed}

In the phonon spectrum of ScV$_6$Sn$_6${} at the $\Gamma$ point (as well as in any stable crystalline material), the acoustic $z$-directed phonon has a frequency of \emph{exactly} zero. In the procedure described in \cref{app:sec:three_band_model:Hamiltonian:parameterization}, this constraints is imposed only \emph{approximately} by the minimization in \cref{app:eqn:fitting_phonon_minimization}, which sometimes results in imaginary phonon frequencies at the $\Gamma$ point within the effective phonon Hamiltonian. To get around this problem, we will impose a series of constraints on the hopping amplitudes such that the low-energy spectrum around the $\Gamma$ point is \emph{exactly} reproduced by $h_{nm} \left( \vec{k} \right)$. 

We start by considering a $\vec{k} \cdot \vec{p}$ expansion of the dynamical matrix around the $\Gamma$ point for the $z$-directed acoustic mode. The wave function of the latter is given by \cref{app:eqn:all_acoustic_modes_at_gamma}. Within the $\vec{k} \cdot \vec{p}$ expansion approximation, the frequency of the acoustic mode is given by 
\begin{equation}
	\omega_3^{2} \left( \vec{k} \right) = \sum_{i,\mu,j,\nu} U^{*}_{i\mu,3} \left( \vec{0} \right) D_{i \mu, j\nu} \left( \vec{k} \right) U_{j\nu,3} \left( \vec{0} \right).
\end{equation}
Under any symmetry transformation $g$, \cref{app:eqn:constraint_symmetry_dynamical,app:eqn:sewing_mat_action} require that 
\begin{equation}
	\omega_3^{2} \left( \vec{k} \right) = \omega_3^{2} \left( g \vec{k} \right),
\end{equation}
which, combined with the fact that $\omega_3 \left( \vec{0} \right) =0$, result in the following expansion for the frequency of the $z$-directed phonon mode
\begin{equation}
	\label{app:eqn:expansion_acc_frequency}
	\omega_3^{2} \left( \vec{k} \right) = c_1 \left( k_x^2 + k_y^2 \right) + c_2 k_z^2 + \mathcal{O} \left( \abs{\vec{k}}^3 \right),
\end{equation} 
where the real coefficients $c_1$ and $c_2$ depend on the dynamical matrix obtained from \textit{ab-initio} methods according to
\begin{align}
	c_1 &= \frac{1}{2} \sum_{i,\mu,j,\nu} U^{*}_{i\mu,3} \left( \vec{0} \right) \eval{\frac{\partial^2 D_{i \mu, j\nu} \left( \vec{k} \right)}{\partial k_x^2}}_{\vec{k}=0}  U_{j\nu,3} \left( \vec{0} \right), \\
	c_2 &= \frac{1}{2} \sum_{i,\mu,j,\nu} U^{*}_{i\mu,3} \left( \vec{0} \right) \eval{\frac{\partial^2 D_{i \mu, j\nu} \left( \vec{k} \right)}{\partial k_z^2}}_{\vec{k}=0}  U_{j\nu,3} \left( \vec{0} \right).
\end{align} 

One can also perform a similar $\vec{k} \cdot \vec{p}$ expansion on the effective Hamiltonian matrix. Letting 
\begin{equation}
	\label{app:eqn:projection_on_wannier_basis_acoustic}
	W_{n,1} \left( \vec{0} \right) = \sum_{j,\nu} w^{*}_{j\nu,n} \left( \vec{0} \right) U_{j\nu,3} \left( \vec{0} \right),
\end{equation}
be the projection of the $z$-directed acoustic phonon at $\Gamma$ on the Wannier basis, the frequency of the $z$-directed acoustic mode within the effective Hamiltonian -- denoted by $\omega_{\text{acc, eff}} \left( \vec{k} \right)$ -- obeys
\begin{equation}
	\omega^2_{\text{acc, eff}} \left( \vec{k} \right) = \sum_{n,m} W^{*}_{n,1} \left( \vec{0} \right) h_{nm} \left( \vec{k} \right) W_{m,1} \left( \vec{0} \right).
\end{equation}
As a consequence of the $P6/mmm$ symmetry of the effective Hamiltonian imposed by \cref{app:eqn:constraint_symmetry_dynamical}, 
\begin{equation}
	\omega^2_{\text{acc, eff}} \left( \vec{k} \right) = \omega^2_{\text{acc, eff}} \left( g \vec{k} \right),
\end{equation} 
for any crystalline symmetry $g$, which results in the following expansion near the $\Gamma$ point 
\begin{equation}
	\label{app:eqn:expansion_effective_acc_frequency}
	\omega^2_{\text{acc, eff}} \left( \vec{k} \right) = c'_0 + c'_1 \left( k_x^2 + k_y^2 \right) + c'_2 k_z^2 + \mathcal{O} \left( \abs{\vec{k}}^3 \right).
\end{equation} 
In \cref{app:eqn:expansion_effective_acc_frequency}, the real coefficients $c'_0$, $c'_1$, and $c'_2$ are linear functions of the hopping amplitudes $t_{nm} \left( \vec{R} \right)$
\begin{alignat}{3}
	& c'_0 &&= \sum_{n,m} W^{*}_{n,1} \left( \vec{0} \right) h_{nm} \left( \vec{0} \right) W_{m,1} \left( \vec{0} \right) &&=  \sum_{n,m,\vec{R}} W^{*}_{n,1} \left( \vec{0} \right) t_{nm} \left( \vec{R} \right) W_{m,1} \left( \vec{0} \right) , \\
	& c'_1 &&= \frac{1}{2} \sum_{n,m} W^{*}_{n,1} \left( \vec{0} \right) \eval{\frac{\partial^2 h_{nm} \left( \vec{k} \right)}{\partial k_x^2}}_{\vec{k}=\vec{0}} W_{m,1} \left( \vec{0} \right) &&= -\frac{1}{2}  \sum_{n,m,\vec{R}} W^{*}_{n,1} \left( \vec{0} \right) \left[ \vec{R} + \boldsymbol{\rho}_n - \boldsymbol{\rho}_m \right]^2_x t_{nm} \left( \vec{R} \right) W_{m,1} \left( \vec{0} \right), \\
	& c'_2 &&= \frac{1}{2} \sum_{n,m} W^{*}_{n,1} \left( \vec{0} \right) \eval{\frac{\partial^2 h_{nm} \left( \vec{k} \right)}{\partial k_z^2}}_{\vec{k}=\vec{0}} W_{m,1} \left( \vec{0} \right) &&= -\frac{1}{2}  \sum_{n,m,\vec{R}} W^{*}_{n,1} \left( \vec{0} \right) \left[ \vec{R} + \boldsymbol{\rho}_n - \boldsymbol{\rho}_m \right]^2_z t_{nm} \left( \vec{R} \right) W_{m,1} \left( \vec{0} \right).
\end{alignat} 

By matching \cref{app:eqn:expansion_acc_frequency,app:eqn:expansion_effective_acc_frequency}, we obtain
\begin{equation}
	\label{app:eqn:ss_constraints}
	c'_0 = 0, \quad
	c'_1 = c_1, \quad
	c'_2 = c_2, 
\end{equation}
which impose three linear constraints on the real hopping amplitudes $t_{nm} \left( \vec{R} \right)$. We note that \cref{app:eqn:ss_constraints} is equivalent to requiring that the effective phonon model reproduces the speeds of sound of ScV$_6$Sn$_6${} for the $z$-directed acoustic phonon. In practice, we first impose the constraints of \cref{app:eqn:ss_constraints} on the hopping amplitudes. The resulting (underdetermined) system of linear equations is then solved exactly, allowing us to write three of the hopping parameters in terms of the remaining ones. Only then, we fix remaining free parameters using the fitting procedure outlined in \cref{app:sec:three_band_model:Hamiltonian:fitting}. 

\subsection{Numerical results}\label{app:sec:three_band_model:results}

Having derived the effective phonon Hamiltonian within the Wannier basis, we now apply the technique outlined in \cref{app:sec:three_band_model:Hamiltonian} to numerically obtain the hopping parameters from the \textit{ab-initio} phonon spectra of both ScV$_6$Sn$_6${} and YV$_6$Sn$_6${}.

\subsubsection{ScV$_6$Sn$_6${}}\label{app:sec:three_band_model:results:sc}
For ScV$_6$Sn$_6${}, we construct two different types of models within the three-orbital Wannier basis: a ``full'' three-band effective model with a larger number of hopping parameters (28 parameters), as well as a ``simplified'' three-band model which features only seven nonzero hopping parameters. The simple model is obtained from the full one, by eliminating the least significant hopping parameters and refitting the rest. The effective Hamiltonian of the simple model is given by 
\begin{equation}
	\label{app:eqn:simple_3_band}
	\resizebox{0.85\hsize}{!}{$h \left( \vec{k} \right) = \begin{bmatrix}
		 t_1 & \left(t_6+t_2 e^{-i k_z}\right) e^{-2 i (z_{\mathrm{Sn}^T}-1) k_z} & t_3 e^{-i \left(z_{\mathrm{Sn}^T}-\frac{1}{2}\right) k_z} \\
 \left(t_6+t_2 e^{i k_z}\right) e^{2 i (z_{\mathrm{Sn}^T}-1) k_z} & t_1 & t_3 e^{\frac{1}{2} i (2 z_{\mathrm{Sn}^T}-1) k_z} \\
 t_3 e^{\frac{1}{2} i (2 z_{\mathrm{Sn}^T}-1) k_z} & t_3 e^{-i \left(z_{\mathrm{Sn}^T}-\frac{1}{2}\right) k_z} & 2 t_5 \left(2 \cos \left(\frac{k_x}{2}\right) \cos \left(\frac{\sqrt{3} k_y}{2}\right)+\cos \left(k_x\right)\right)+2 t_7 \cos \left(k_z\right)+t_4 \\ 	\end{bmatrix}
	$},
\end{equation} 
where $z_{\mathrm {Sn}^T}$ is the displacement of the first Sn$^T$ atom along the $z$-direction and we have denoted $t_1=t_{1 1}\big( (0, 0, 0 ) \big)$, $t_2=t_{1 2}\big( (0, 0, 0 ) \big)$, $t_3=t_{1 3}\big( (0, 0, 0 ) \big)$, $t_4=t_{3 3}\big( (0, 0, 0 ) \big)$, $t_5=t_{3 3}\big( (0, 1, 0 ) \big)$, $t_6=t_{2 1}\big( (0, 0, 1 ) \big)$, and $t_7=t_{3 3}\big( (0, 0, 1 ) \big)$ \unskip{}, for brevity. The simple three-band model from \cref{app:eqn:simple_3_band} features only $z$-directed hoppings for the first two Wannier orbitals (which are essentially the $z$-directed displacements of the two Sn$^T$ atoms), as well as both in-plane and out-of-plane hoppings for the third Wannier orbital. The values of the hopping parameters for both the full and the simple model are listed in \cref{tab:HoppingEffPhonon}. Since the soft phonon mode is mostly supported by the $z$-directed displacement of Sn$^T$, the lack of direct in-plane hopping for the first two Wannier orbitals explains the relative flatness of the soft phonon mode on the $k_z=\pi$ plane, as show in \cref{fig:zero_temperature_phonon_model}(c) at zero temperature. The model also correctly reproduces the avoided crossing between the acoustic and optical phonon on the $\Gamma-\mathrm{A}$ line. With a larger number of hopping parameters, the agreement can be rendered quantitative. We also use the same method to construct effective low-energy phonon models for ScV$_6$Sn$_6${} at finite temperature. As shown in \cref{fig:fin_temperature_phonon_model_Sc}, the full (simplified) three-band model quantitatively (qualitatively) matches the low-energy ScV$_6$Sn$_6${} phonon spectrum across the entire temperature range.

\subsubsection{YV$_6$Sn$_6${}}\label{app:sec:three_band_model:results:Y}

For completeness, we also fit the three-band Hamiltonian obtained in \cref{app:sec:three_band_model:Hamiltonian} to the low-energy spectrum of YV$_6$Sn$_6${} at both zero and finite temperatures. The values of the corresponding hopping parameters are presented in \cref{tab:HoppingEffPhonon}, while the spectra of the low-energy model is shown in \cref{fig:fin_temperature_phonon_model_Y_3}, for both the full and simplified models. We find that the three-band model still matches the \textit{ab-initio} spectrum, although the agreement is not as good as in the case of ScV$_6$Sn$_6${}. This is due to the larger atomic mass of Y compared to Sc, which therefore contributes significantly to the low-energy phonon spectrum. 

To account for the effects of the displacement of the Y atom, we also construct a \emph{four-band} effective Wannier model, in which we add a fourth Wannier orbital corresponding to the $z$-directed displacement of the $Y$ atom,
\begin{align}
	w_{i\mu,4} \left( \vec{k} \right) &= \delta_{i,1} \delta_{\mu,3}, \label{app:eqn:wan_4}
\end{align}
to the three-band model from \cref{app:sec:three_band_model:Hamiltonian}. Similarly to the three-band model, we also construct full and simplified versions of the effective four-band Hamiltonian featuring, respectively, nine and 42 nonzero hopping amplitudes. The Hamiltonian on the simplified model is given by
\begin{equation}
	\label{app:eqn:simple_4_band}
	\resizebox{0.85\hsize}{!}{$h \left( \vec{k} \right) = \begin{bmatrix}
			 t_1 & \left(t_{10}+t_2 e^{-i k_z}\right) e^{-2 i (z_{\mathrm{Sn}^T}-1) k_z} & t_3 e^{-i \left(z_{\mathrm{Sn}^T}-\frac{1}{2}\right) k_z} & \left(t_4+t_5 e^{i k_z}\right) e^{-i z_{\mathrm{Sn}^T} k_z} \\
 \left(t_{10}+t_2 e^{i k_z}\right) e^{2 i (z_{\mathrm{Sn}^T}-1) k_z} & t_1 & t_3 e^{\frac{1}{2} i (2 z_{\mathrm{Sn}^T}-1) k_z} & \left(t_4+t_5 e^{-i k_z}\right) e^{i z_{\mathrm{Sn}^T} k_z} \\
 t_3 e^{\frac{1}{2} i (2 z_{\mathrm{Sn}^T}-1) k_z} & t_3 e^{-i \left(z_{\mathrm{Sn}^T}-\frac{1}{2}\right) k_z} & 2 \left(2 \cos \left(\frac{k_x}{2}\right) \cos \left(\frac{\sqrt{3} k_y}{2}\right) \left(t_{12} \left(2 \cos \left(k_x\right)-1\right)+t_9\right)+t_9 \cos \left(k_x\right)+t_{12} \cos \left(\sqrt{3} k_y\right)+t_{11} \cos \left(k_z\right)\right)+t_6 & 2 t_7 \cos \left(\frac{k_z}{2}\right) \\
 \left(t_4+t_5 e^{-i k_z}\right) e^{i z_{\mathrm{Sn}^T} k_z} & \left(t_4+t_5 e^{i k_z}\right) e^{-i z_{\mathrm{Sn}^T} k_z} & 2 t_7 \cos \left(\frac{k_z}{2}\right) & t_8 \\ 		\end{bmatrix}
		$},
\end{equation} 
where, again, we have denoted $t_1=t_{1 1}\big( (0, 0, 0 ) \big)$, $t_2=t_{1 2}\big( (0, 0, 0 ) \big)$, $t_3=t_{1 3}\big( (0, 0, 0 ) \big)$, $t_4=t_{1 4}\big( (0, 0, 0 ) \big)$, $t_5=t_{2 4}\big( (0, 0, 0 ) \big)$, $t_6=t_{3 3}\big( (0, 0, 0 ) \big)$, $t_7=t_{3 4}\big( (0, 0, 0 ) \big)$, $t_8=t_{4 4}\big( (0, 0, 0 ) \big)$, $t_9=t_{3 3}\big( (0, 1, 0 ) \big)$, $t_{10}=t_{2 1}\big( (0, 0, 1 ) \big)$, $t_{11}=t_{3 3}\big( (0, 0, 1 ) \big)$, and $t_{12}=t_{3 3}\big( (1, 2, 0 ) \big)$ \unskip{}, for brevity. Similarly to the three-band model, we find that only the effective acoustic Wannier orbitals features in plane hopping within the simplified model. The values of the hopping amplitudes are summarized in \cref{tab:HoppingEffPhonon} for both zero and finite temperature. From \cref{fig:fin_temperature_phonon_model_Y_4}, we see that the four-band model more closely reproduces the band structure of YV$_6$Sn$_6${}, compared to the three-band one.

\FloatBarrier

\clearpage
\newcolumntype{C}[1]{>{\let\temp=\\\raggedleft\let\\=\temp}p{\dimexpr(0.6\linewidth) / #1 \relax}}
\setlength\tabcolsep{0 pt}
\setlength{\LTpost}{0pt}
\setlength{\LTpre}{0pt}
\bigskip
\begin{longtable}[!h]{| p{\dimexpr(0.2\linewidth)} |*{10}{C{10}|}}
	\hline
	
	\multicolumn{11}{|c|}{\textbf{Full three-band model of ScV$_6$Sn$_6${}}} \\
	\hline
	\multirow{2}{\dimexpr(\linewidth)}{Hopping Parameter ($\si{\tera\hertz^2}$)} & 
	\multicolumn{10}{c|}{$T / \si{\kelvin}$} \\ \cline{2-11}
	& $0$ & $580$ & $1160$ & $2321$ & $3481$ & $4642$ & $5802$ & $9284$ & $11605$ & $13925$
	\\ \hline

	$t_{3 3}\big( (0, 0, 0 ) \big)$ & $ 8.87$ & $ 8.63$ & $ 8.50$ & $ 8.05$ & $ 8.81$ & $ 8.61$ & $ 7.30$ & $ 6.94$ & $ 6.58$ & $ 6.63$\\ 
\hline
$t_{1 1}\big( (0, 0, 0 ) \big)$ & $ 5.74$ & $ 5.64$ & $ 5.67$ & $ 5.73$ & $ 5.71$ & $ 5.69$ & $ 5.64$ & $ 5.55$ & $ 5.63$ & $ 5.93$\\ 
\hline
$t_{2 1}\big( (0, 0, 1 ) \big)$ & $-4.15$ & $-4.03$ & $-3.94$ & $-3.79$ & $-3.66$ & $-3.53$ & $-3.42$ & $-3.29$ & $-3.46$ & $-3.65$\\ 
\hline
$t_{1 2}\big( (0, 0, 0 ) \big)$ & $ 2.46$ & $ 2.26$ & $ 2.18$ & $ 1.88$ & $ 1.60$ & $ 1.31$ & $ 1.01$ & $ 0.32$ & $-0.14$ & $-0.59$\\ 
\hline
$t_{3 3}\big( (0, 0, 1 ) \big)\times 10^{1}$ & $-2.47$ & $-1.80$ & $-2.63$ & $-1.64$ & $-8.31$ & $-8.39$ & $-15.09$ & $-9.48$ & $-10.12$ & $-8.21$\\ 
\hline
$t_{3 3}\big( (0, 1, 0 ) \big)\times 10^{1}$ & $-7.80$ & $-7.16$ & $-6.85$ & $-6.41$ & $-7.56$ & $-7.59$ & $-3.55$ & $-4.60$ & $-4.18$ & $-4.86$\\ 
\hline
$t_{1 3}\big( (0, 0, 0 ) \big)\times 10^{1}$ & $-7.98$ & $-7.20$ & $-7.09$ & $-6.63$ & $-6.87$ & $-6.72$ & $-6.05$ & $-4.59$ & $-3.83$ & $-2.97$\\ 
\hline
$t_{3 3}\big( (0, 0, 2 ) \big)\times 10^{1}$ & $ 3.98$ & $ 3.99$ & $ 4.34$ & $ 4.17$ & $ 4.49$ & $ 4.18$ & $ 4.09$ & $ 4.83$ & $ 4.82$ & $ 5.04$\\ 
\hline
$t_{1 2}\big( (0, 1, 1 ) \big)\times 10^{1}$ & $-3.98$ & $-4.06$ & $-4.07$ & $-4.22$ & $-3.20$ & $-2.99$ & $-1.66$ & $-2.82$ & $-2.70$ & $-3.24$\\ 
\hline
$t_{3 3}\big( (1, 2, 0 ) \big)\times 10^{1}$ & $ 3.29$ & $ 2.85$ & $ 2.85$ & $ 2.98$ & $ 2.69$ & $ 2.59$ & $ 0.27$ & $ 1.74$ & $ 1.64$ & $ 2.42$\\ 
\hline
$t_{2 3}\big( (0, 0, 1 ) \big)\times 10^{1}$ & $-2.35$ & $-2.27$ & $-2.12$ & $-2.11$ & $-1.67$ & $-1.62$ & $-1.10$ & $-1.43$ & $-1.28$ & $-1.50$\\ 
\hline
$t_{1 3}\big( (0, 1, 0 ) \big)\times 10^{1}$ & $-1.92$ & $-1.84$ & $-1.88$ & $-1.79$ & $-1.73$ & $-1.57$ & $-1.71$ & $-1.22$ & $-0.99$ & $-0.85$\\ 
\hline
$t_{1 1}\big( (0, 1, 0 ) \big)\times 10^{1}$ & $ 2.11$ & $ 1.61$ & $ 1.59$ & $ 1.42$ & $ 1.27$ & $ 1.18$ & $ 1.19$ & $ 1.23$ & $ 1.18$ & $ 1.02$\\ 
\hline
$t_{1 1}\big( (0, 0, 1 ) \big)\times 10^{1}$ & $ 3.42$ & $ 2.56$ & $ 1.68$ & $ 0.67$ & $ 0.10$ & $-0.09$ & $-0.12$ & $ 0.02$ & $ 0.91$ & $ 1.81$\\ 
\hline
$t_{1 3}\big( (0, 0, 1 ) \big)\times 10^{2}$ & $-12.49$ & $-8.13$ & $-4.89$ & $-2.79$ & $ 6.25$ & $ 6.67$ & $ 10.46$ & $ 12.45$ & $ 11.78$ & $ 14.17$\\ 
\hline
$t_{1 3}\big( (0, 0, 2 ) \big)\times 10^{2}$ & $-3.95$ & $-4.47$ & $-6.82$ & $-6.02$ & $-8.21$ & $-8.75$ & $-10.25$ & $-11.80$ & $-11.73$ & $-7.83$\\ 
\hline
$t_{2 1}\big( (0, 1, 1 ) \big)\times 10^{2}$ & $-7.03$ & $-7.23$ & $-6.11$ & $-4.76$ & $-4.34$ & $-4.31$ & $-4.68$ & $-4.74$ & $-3.63$ & $-1.65$\\ 
\hline
$t_{2 1}\big( (0, 0, 2 ) \big)\times 10^{2}$ & $-7.52$ & $-5.41$ & $-4.33$ & $-2.31$ & $-1.35$ & $-0.91$ & $-0.54$ & $ 0.84$ & $-6.06$ & $-11.65$\\ 
\hline
$t_{2 3}\big( (0, 0, 2 ) \big)\times 10^{2}$ & $ 4.30$ & $ 4.38$ & $ 2.55$ & $ 2.34$ & $-0.01$ & $ 2.01$ & $ 2.00$ & $-2.06$ & $ 0.62$ & $-2.58$\\ 
\hline
$t_{1 2}\big( (0, 1, 0 ) \big)\times 10^{2}$ & $-7.85$ & $-0.99$ & $-0.88$ & $-0.42$ & $ 0.42$ & $-0.15$ & $-0.71$ & $-2.34$ & $-2.28$ & $-1.79$\\ 
\hline
$t_{1 2}\big( (0, 0, 1 ) \big)\times 10^{2}$ & $-1.06$ & $-1.02$ & $-0.96$ & $ 0.32$ & $ 1.02$ & $ 1.74$ & $-0.12$ & $-3.95$ & $-3.24$ & $-4.15$\\ 
\hline
$t_{3 3}\big( (0, 1, 1 ) \big)\times 10^{2}$ & $ 1.31$ & $ 2.17$ & $ 2.27$ & $ 1.87$ & $ 2.34$ & $ 1.55$ & $ 0.14$ & $ 0.75$ & $ 0.49$ & $ 2.10$\\ 
\hline
$t_{1 1}\big( (0, 0, 2 ) \big)\times 10^{2}$ & $-1.86$ & $-1.80$ & $-1.27$ & $-1.57$ & $-2.48$ & $-2.99$ & $-1.72$ & $ 0.76$ & $-0.05$ & $ 0.38$\\ 
\hline
$t_{1 1}\big( (0, 1, 1 ) \big)\times 10^{2}$ & $ 2.76$ & $ 3.39$ & $ 2.14$ & $ 1.04$ & $ 0.37$ & $ 0.48$ & $ 0.78$ & $ 0.49$ & $-0.43$ & $-2.11$\\ 
\hline
$t_{1 1}\big( (1, 2, 0 ) \big)\times 10^{2}$ & $ 0.26$ & $ 1.92$ & $ 1.94$ & $ 1.61$ & $ 1.10$ & $ 0.85$ & $ 1.57$ & $ 1.68$ & $ 0.96$ & $-0.04$\\ 
\hline
$t_{1 2}\big( (0, 0, 2 ) \big)\times 10^{2}$ & $-0.79$ & $-2.29$ & $-1.69$ & $-1.45$ & $-1.58$ & $-1.05$ & $-0.41$ & $-0.37$ & $ 1.31$ & $ 0.90$\\ 
\hline
$t_{1 3}\big( (1, 2, 0 ) \big)\times 10^{2}$ & $ 0.75$ & $-0.98$ & $-0.68$ & $-0.61$ & $ 0.41$ & $ 0.92$ & $ 3.23$ & $ 1.23$ & $ 1.17$ & $ 0.08$\\ 
\hline
$t_{2 3}\big( (0, 1, 1 ) \big)\times 10^{3}$ & $ 3.20$ & $-1.17$ & $ 8.49$ & $ 17.02$ & $ 16.51$ & $ 17.57$ & $ 8.63$ & $ 9.34$ & $ 4.26$ & $-7.00$\\ 
\hline
$t_{1 3}\big( (0, 1, 1 ) \big)\times 10^{3}$ & $ 12.71$ & $ 8.72$ & $ 7.93$ & $ 9.30$ & $ 0.73$ & $-2.38$ & $-3.94$ & $-8.57$ & $-4.47$ & $ 1.09$\\ 
\hline
$t_{1 2}\big( (1, 2, 0 ) \big)\times 10^{3}$ & $ 23.18$ & $ 1.16$ & $ 2.44$ & $ 0.82$ & $-2.11$ & $-3.15$ & $ 6.41$ & $ 3.78$ & $ 5.01$ & $ 10.40$ \\ 
	\hline
	
	\multicolumn{11}{|c|}{\textbf{Simplified three-band model of ScV$_6$Sn$_6${}}} \\
	\hline
	\multirow{2}{\dimexpr(\linewidth)}{Hopping Parameter ($\si{\tera\hertz^2}$)} & 
	\multicolumn{10}{c|}{$T / \si{\kelvin}$} \\ \cline{2-11}
	& $0$ & $580$ & $1160$ & $2321$ & $3481$ & $4642$ & $5802$ & $9284$ & $11605$ & $13925$
	\\ \hline

	$t_{3 3}\big( (0, 0, 0 ) \big)$ & $ 8.15$ & $ 8.87$ & $ 8.60$ & $ 8.15$ & $ 7.78$ & $ 7.45$ & $ 7.10$ & $ 6.32$ & $ 5.83$ & $ 5.60$\\ 
\hline
$t_{1 1}\big( (0, 0, 0 ) \big)$ & $ 5.71$ & $ 5.58$ & $ 5.61$ & $ 5.64$ & $ 5.63$ & $ 5.62$ & $ 5.64$ & $ 5.68$ & $ 5.74$ & $ 5.97$\\ 
\hline
$t_{2 1}\big( (0, 0, 1 ) \big)$ & $-4.33$ & $-4.20$ & $-4.08$ & $-3.89$ & $-3.73$ & $-3.63$ & $-3.59$ & $-3.50$ & $-3.61$ & $-3.72$\\ 
\hline
$t_{1 2}\big( (0, 0, 0 ) \big)$ & $ 2.45$ & $ 2.24$ & $ 2.14$ & $ 1.79$ & $ 1.50$ & $ 1.18$ & $ 0.89$ & $ 0.26$ & $-0.13$ & $-0.51$\\ 
\hline
$t_{1 3}\big( (0, 0, 0 ) \big)$ & $-1.49$ & $-1.41$ & $-1.43$ & $-1.38$ & $-1.32$ & $-1.23$ & $-1.14$ & $-0.95$ & $-0.78$ & $-0.68$\\ 
\hline
$t_{3 3}\big( (0, 0, 1 ) \big)$ & $-1.20$ & $-1.16$ & $-1.17$ & $-1.17$ & $-1.15$ & $-1.14$ & $-1.12$ & $-1.07$ & $-1.04$ & $-1.05$\\ 
\hline
$t_{3 3}\big( (0, 1, 0 ) \big)\times 10^{1}$ & $-7.65$ & $-9.07$ & $-8.58$ & $-7.91$ & $-7.39$ & $-7.01$ & $-6.63$ & $-5.73$ & $-5.24$ & $-4.97$ \\ 
	\hline
\end{longtable}\addtocounter{table}{-1}\begin{longtable}[!h]{| p{\dimexpr(0.2\linewidth)} |*{9}{C{9}|}}
	
	\multicolumn{10}{|c|}{\textbf{Full three-band model of YV$_6$Sn$_6${}}} \\
	\hline
	\multirow{3}{\dimexpr(\linewidth)}{Hopping Parameter ($\si{\tera\hertz^2}$)} & 
	\multicolumn{3}{c|}{Relaxed~\cite{POK21}} &
	\multicolumn{3}{c|}{Experimental~\cite{ROM11}} &
	\multicolumn{3}{c|}{Relaxed~\cite{ROM11}} \\ \cline{2-10}
	& 
	\multicolumn{3}{c|}{$T / \si{\kelvin}$} &
	\multicolumn{3}{c|}{$T / \si{\kelvin}$} &
	\multicolumn{3}{c|}{$T / \si{\kelvin}$} \\ \cline{2-10}
	& $0$ & $4642$ & $9284$
	& $0$ & $4642$ & $9284$
	& $0$ & $4642$ & $9284$
	\\ \hline
	$t_{3 3}\big( (0, 0, 0 ) \big)$ & $ 11.20$ & $ 8.37$ & $ 6.67$ & $ 11.14$ & $ 9.26$ & $ 7.42$ & $ 10.65$ & $ 8.35$ & $ 6.77$\\ 
\hline
$t_{1 1}\big( (0, 0, 0 ) \big)$ & $ 8.30$ & $ 8.17$ & $ 8.34$ & $ 8.07$ & $ 8.32$ & $ 8.52$ & $ 8.01$ & $ 8.10$ & $ 8.25$\\ 
\hline
$t_{2 1}\big( (0, 0, 1 ) \big)$ & $-3.96$ & $-4.41$ & $-4.38$ & $-4.22$ & $-4.51$ & $-4.11$ & $-3.99$ & $-4.36$ & $-4.37$\\ 
\hline
$t_{1 2}\big( (0, 0, 0 ) \big)$ & $ 0.08$ & $-0.92$ & $-1.90$ & $-0.01$ & $-1.05$ & $-2.13$ & $ 0.25$ & $-0.92$ & $-1.85$\\ 
\hline
$t_{1 3}\big( (0, 0, 0 ) \big)\times 10^{1}$ & $-12.66$ & $-7.18$ & $-5.80$ & $-11.13$ & $-8.03$ & $-6.62$ & $-10.94$ & $-7.06$ & $-5.74$\\ 
\hline
$t_{3 3}\big( (0, 1, 0 ) \big)\times 10^{1}$ & $-12.11$ & $-8.38$ & $-4.18$ & $-11.38$ & $-9.37$ & $-5.55$ & $-10.29$ & $-8.39$ & $-4.62$\\ 
\hline
$t_{3 3}\big( (0, 0, 1 ) \big)\times 10^{1}$ & $-3.50$ & $-3.85$ & $-4.16$ & $-11.28$ & $-7.66$ & $-7.22$ & $-7.94$ & $-3.90$ & $-4.79$\\ 
\hline
$t_{1 1}\big( (0, 0, 1 ) \big)\times 10^{1}$ & $ 3.23$ & $ 6.56$ & $ 8.73$ & $ 3.08$ & $ 5.81$ & $ 7.24$ & $ 2.45$ & $ 6.36$ & $ 8.76$\\ 
\hline
$t_{2 3}\big( (0, 0, 1 ) \big)\times 10^{1}$ & $-7.55$ & $-5.24$ & $-3.37$ & $-4.10$ & $-3.48$ & $-3.06$ & $-4.70$ & $-5.49$ & $-3.21$\\ 
\hline
$t_{3 3}\big( (0, 0, 2 ) \big)\times 10^{1}$ & $ 4.76$ & $ 4.42$ & $ 4.45$ & $ 4.52$ & $ 4.42$ & $ 4.02$ & $ 4.79$ & $ 4.38$ & $ 4.23$\\ 
\hline
$t_{1 2}\big( (0, 1, 1 ) \big)\times 10^{1}$ & $-4.05$ & $-3.71$ & $-3.75$ & $-2.71$ & $-3.22$ & $-3.03$ & $-3.27$ & $-3.70$ & $-3.45$\\ 
\hline
$t_{3 3}\big( (1, 2, 0 ) \big)\times 10^{1}$ & $ 3.88$ & $ 3.83$ & $ 2.30$ & $ 3.09$ & $ 3.60$ & $ 2.11$ & $ 2.87$ & $ 3.86$ & $ 2.24$\\ 
\hline
$t_{1 2}\big( (0, 0, 1 ) \big)\times 10^{1}$ & $-0.43$ & $-2.54$ & $-4.38$ & $-1.02$ & $-2.24$ & $-4.60$ & $-0.21$ & $-2.59$ & $-4.48$\\ 
\hline
$t_{2 1}\big( (0, 0, 2 ) \big)\times 10^{1}$ & $-1.21$ & $-2.07$ & $-2.43$ & $-2.14$ & $-2.54$ & $-2.69$ & $-1.54$ & $-1.91$ & $-2.25$\\ 
\hline
$t_{1 3}\big( (0, 0, 2 ) \big)\times 10^{1}$ & $-1.47$ & $-1.40$ & $-1.82$ & $-1.18$ & $-1.84$ & $-1.21$ & $-1.28$ & $-1.29$ & $-1.45$\\ 
\hline
$t_{1 1}\big( (0, 1, 0 ) \big)\times 10^{1}$ & $ 1.57$ & $ 0.96$ & $ 0.49$ & $ 1.76$ & $ 1.38$ & $ 0.61$ & $ 1.79$ & $ 0.93$ & $ 0.41$\\ 
\hline
$t_{1 3}\big( (0, 1, 0 ) \big)\times 10^{1}$ & $-0.86$ & $-0.62$ & $-1.20$ & $-1.25$ & $-0.97$ & $-1.23$ & $-1.42$ & $-0.61$ & $-1.19$\\ 
\hline
$t_{2 3}\big( (0, 0, 2 ) \big)\times 10^{2}$ & $ 5.01$ & $ 7.94$ & $ 16.81$ & $ 4.56$ & $ 11.92$ & $ 9.13$ & $ 3.78$ & $ 6.23$ & $ 11.52$\\ 
\hline
$t_{1 3}\big( (0, 0, 1 ) \big)\times 10^{2}$ & $ 6.06$ & $-3.86$ & $ 9.15$ & $ 11.60$ & $-0.86$ & $ 14.83$ & $ 4.67$ & $-0.91$ & $ 9.20$\\ 
\hline
$t_{1 1}\big( (0, 1, 1 ) \big)\times 10^{2}$ & $-1.35$ & $-5.61$ & $-9.52$ & $ 5.57$ & $-4.11$ & $-6.04$ & $ 1.37$ & $-5.72$ & $-9.28$\\ 
\hline
$t_{2 1}\big( (0, 1, 1 ) \big)\times 10^{2}$ & $ 1.00$ & $ 4.82$ & $ 7.51$ & $-7.39$ & $ 5.57$ & $ 0.51$ & $-1.63$ & $ 4.57$ & $ 7.60$\\ 
\hline
$t_{3 3}\big( (0, 1, 1 ) \big)\times 10^{2}$ & $ 8.83$ & $ 6.77$ & $-0.02$ & $ 5.74$ & $ 4.03$ & $ 2.00$ & $ 5.06$ & $ 7.55$ & $ 0.60$\\ 
\hline
$t_{1 3}\big( (0, 1, 1 ) \big)\times 10^{2}$ & $-0.51$ & $ 4.36$ & $ 7.38$ & $-1.14$ & $ 2.92$ & $ 7.00$ & $-0.91$ & $ 4.61$ & $ 6.96$\\ 
\hline
$t_{1 3}\big( (1, 2, 0 ) \big)\times 10^{2}$ & $-3.04$ & $-8.55$ & $-0.39$ & $-1.46$ & $-5.31$ & $ 0.53$ & $ 0.28$ & $-8.85$ & $-0.30$\\ 
\hline
$t_{1 2}\big( (0, 1, 0 ) \big)\times 10^{2}$ & $-2.38$ & $ 1.26$ & $ 6.34$ & $-1.30$ & $ 1.92$ & $ 5.89$ & $-1.48$ & $ 1.21$ & $ 6.76$\\ 
\hline
$t_{1 1}\big( (1, 2, 0 ) \big)\times 10^{2}$ & $-4.97$ & $-0.46$ & $ 2.39$ & $-5.28$ & $-3.81$ & $ 0.01$ & $-4.71$ & $-0.47$ & $ 2.47$\\ 
\hline
$t_{1 2}\big( (1, 2, 0 ) \big)\times 10^{2}$ & $ 3.07$ & $ 2.16$ & $-0.96$ & $ 3.34$ & $ 1.25$ & $ 1.60$ & $ 1.90$ & $ 2.56$ & $-1.01$\\ 
\hline
$t_{2 3}\big( (0, 1, 1 ) \big)\times 10^{2}$ & $ 1.75$ & $ 2.54$ & $ 2.90$ & $-0.41$ & $ 3.90$ & $ 1.10$ & $ 0.50$ & $ 2.06$ & $ 2.17$\\ 
\hline
$t_{1 2}\big( (0, 0, 2 ) \big)\times 10^{2}$ & $-1.98$ & $-2.12$ & $-1.79$ & $-1.75$ & $ 1.03$ & $-1.62$ & $-0.41$ & $-2.26$ & $ 0.26$\\ 
\hline
$t_{1 1}\big( (0, 0, 2 ) \big)\times 10^{3}$ & $-1.15$ & $-14.51$ & $ 3.38$ & $-9.51$ & $-10.92$ & $-1.55$ & $-4.45$ & $-3.37$ & $-12.91$
	\\ \hline

	\multicolumn{10}{|c|}{\textbf{Simplified three-band model of YV$_6$Sn$_6${}}} \\
	\hline
	\multirow{3}{\dimexpr(\linewidth)}{Hopping Parameter ($\si{\tera\hertz^2}$)} & 
	\multicolumn{3}{c|}{Relaxed~\cite{POK21}} &
	\multicolumn{3}{c|}{Experimental~\cite{ROM11}} &
	\multicolumn{3}{c|}{Relaxed~\cite{ROM11}} \\ \cline{2-10}
	& 
	\multicolumn{3}{c|}{$T / \si{\kelvin}$} &
	\multicolumn{3}{c|}{$T / \si{\kelvin}$} &
	\multicolumn{3}{c|}{$T / \si{\kelvin}$} \\ \cline{2-10}
	& $0$ & $4642$ & $9284$
	& $0$ & $4642$ & $9284$
	& $0$ & $4642$ & $9284$
	\\ \hline
	$t_{1 1}\big( (0, 0, 0 ) \big)$ & $ 8.39$ & $ 8.67$ & $ 8.64$ & $ 8.69$ & $ 8.97$ & $ 9.00$ & $ 8.28$ & $ 8.59$ & $ 8.60$\\ 
\hline
$t_{3 3}\big( (0, 0, 0 ) \big)$ & $ 9.56$ & $ 7.80$ & $ 6.57$ & $ 9.15$ & $ 7.97$ & $ 6.66$ & $ 9.26$ & $ 7.76$ & $ 6.54$\\ 
\hline
$t_{2 1}\big( (0, 0, 1 ) \big)$ & $-3.77$ & $-3.75$ & $-3.64$ & $-3.82$ & $-3.70$ & $-3.59$ & $-3.83$ & $-3.70$ & $-3.60$\\ 
\hline
$t_{1 3}\big( (0, 0, 0 ) \big)$ & $-1.73$ & $-1.39$ & $-1.10$ & $-1.73$ & $-1.46$ & $-1.11$ & $-1.67$ & $-1.38$ & $-1.10$\\ 
\hline
$t_{1 2}\big( (0, 0, 0 ) \big)$ & $-0.20$ & $-1.35$ & $-2.19$ & $-0.44$ & $-1.52$ & $-2.57$ & $-0.17$ & $-1.34$ & $-2.18$\\ 
\hline
$t_{3 3}\big( (0, 0, 1 ) \big)$ & $-1.25$ & $-1.14$ & $-1.05$ & $-1.22$ & $-1.16$ & $-1.07$ & $-1.23$ & $-1.13$ & $-1.04$\\ 
\hline
$t_{3 3}\big( (0, 1, 0 ) \big)\times 10^{1}$ & $-9.51$ & $-7.40$ & $-6.03$ & $-8.94$ & $-7.52$ & $-6.10$ & $-9.17$ & $-7.38$ & $-6.01$
	\\ \hline

	\multicolumn{10}{|c|}{\textbf{Full four-band model of YV$_6$Sn$_6${}}} \\
	\hline
	\multirow{3}{\dimexpr(\linewidth)}{Hopping Parameter ($\si{\tera\hertz^2}$)} & 
	\multicolumn{3}{c|}{Relaxed~\cite{POK21}} &
	\multicolumn{3}{c|}{Experimental~\cite{ROM11}} &
	\multicolumn{3}{c|}{Relaxed~\cite{ROM11}} \\ \cline{2-10}
	& 
	\multicolumn{3}{c|}{$T / \si{\kelvin}$} &
	\multicolumn{3}{c|}{$T / \si{\kelvin}$} &
	\multicolumn{3}{c|}{$T / \si{\kelvin}$} \\ \cline{2-10}
	& $0$ & $4642$ & $9284$
	& $0$ & $4642$ & $9284$
	& $0$ & $4642$ & $9284$
	\\ \hline
	$t_{4 4}\big( (0, 0, 0 ) \big)$ & $ 21.75$ & $ 19.91$ & $ 19.25$ & $ 21.11$ & $ 19.42$ & $ 18.81$ & $ 21.57$ & $ 19.75$ & $ 19.12$\\ 
\hline
$t_{1 1}\big( (0, 0, 0 ) \big)$ & $ 12.50$ & $ 12.58$ & $ 12.16$ & $ 12.48$ & $ 12.84$ & $ 12.46$ & $ 12.27$ & $ 12.41$ & $ 12.06$\\ 
\hline
$t_{3 3}\big( (0, 0, 0 ) \big)$ & $ 11.69$ & $ 8.11$ & $ 6.71$ & $ 10.58$ & $ 8.38$ & $ 6.90$ & $ 11.42$ & $ 7.86$ & $ 6.63$\\ 
\hline
$t_{2 4}\big( (0, 0, 0 ) \big)$ & $-8.80$ & $-8.04$ & $-7.37$ & $-8.72$ & $-7.98$ & $-7.33$ & $-8.79$ & $-8.00$ & $-7.34$\\ 
\hline
$t_{1 2}\big( (0, 0, 0 ) \big)$ & $-0.96$ & $-2.16$ & $-2.85$ & $-1.09$ & $-2.40$ & $-3.18$ & $-0.71$ & $-2.04$ & $-2.77$\\ 
\hline
$t_{3 3}\big( (0, 1, 0 ) \big)\times 10^{1}$ & $-12.32$ & $-6.63$ & $-4.64$ & $-9.51$ & $-6.73$ & $-4.60$ & $-11.28$ & $-6.09$ & $-4.53$\\ 
\hline
$t_{1 3}\big( (0, 0, 0 ) \big)\times 10^{1}$ & $-10.16$ & $-6.57$ & $-4.65$ & $-10.30$ & $-7.25$ & $-5.78$ & $-9.53$ & $-6.46$ & $-4.66$\\ 
\hline
$t_{1 4}\big( (0, 0, 0 ) \big)\times 10^{1}$ & $ 9.07$ & $ 6.48$ & $ 3.37$ & $ 8.84$ & $ 6.29$ & $ 3.36$ & $ 8.92$ & $ 6.40$ & $ 3.28$\\ 
\hline
$t_{3 3}\big( (0, 0, 1 ) \big)\times 10^{1}$ & $-4.32$ & $-2.02$ & $-4.11$ & $-7.40$ & $-3.71$ & $-5.05$ & $-10.90$ & $-2.19$ & $-3.99$\\ 
\hline
$t_{3 3}\big( (0, 0, 2 ) \big)\times 10^{1}$ & $ 4.32$ & $ 4.05$ & $ 4.02$ & $ 3.97$ & $ 4.23$ & $ 4.02$ & $ 4.08$ & $ 4.03$ & $ 4.04$\\ 
\hline
$t_{3 3}\big( (0, 1, 1 ) \big)\times 10^{1}$ & $-3.64$ & $-3.88$ & $-3.31$ & $-3.10$ & $-3.73$ & $-3.24$ & $-2.46$ & $-3.86$ & $-3.34$\\ 
\hline
$t_{3 4}\big( (0, 0, 0 ) \big)\times 10^{1}$ & $-5.75$ & $-2.47$ & $-1.94$ & $-4.07$ & $-2.15$ & $-1.13$ & $-5.21$ & $-2.18$ & $-1.76$\\ 
\hline
$t_{3 3}\big( (1, 2, 0 ) \big)\times 10^{1}$ & $ 3.59$ & $ 2.90$ & $ 2.44$ & $ 2.47$ & $ 2.73$ & $ 2.18$ & $ 2.77$ & $ 2.77$ & $ 2.44$\\ 
\hline
$t_{2 1}\big( (0, 0, 1 ) \big)\times 10^{1}$ & $ 3.83$ & $-1.28$ & $-3.33$ & $ 4.16$ & $-0.32$ & $-3.47$ & $ 3.60$ & $-0.86$ & $-3.40$\\ 
\hline
$t_{2 3}\big( (0, 0, 1 ) \big)\times 10^{1}$ & $-1.79$ & $-2.73$ & $-3.26$ & $-0.98$ & $-2.31$ & $-2.98$ & $-1.03$ & $-2.53$ & $-3.28$\\ 
\hline
$t_{1 3}\big( (0, 1, 0 ) \big)\times 10^{1}$ & $-1.86$ & $-1.74$ & $-1.38$ & $-2.04$ & $-1.84$ & $-1.44$ & $-1.92$ & $-1.75$ & $-1.36$\\ 
\hline
$t_{1 1}\big( (0, 0, 1 ) \big)\times 10^{1}$ & $-1.67$ & $ 0.96$ & $ 1.72$ & $-1.64$ & $ 0.72$ & $ 1.83$ & $-1.69$ & $ 0.72$ & $ 1.68$\\ 
\hline
$t_{1 1}\big( (0, 1, 0 ) \big)\times 10^{1}$ & $ 1.47$ & $ 0.74$ & $ 1.16$ & $ 1.65$ & $ 0.82$ & $ 1.05$ & $ 1.63$ & $ 0.81$ & $ 1.12$\\ 
\hline
$t_{3 4}\big( (0, 0, 1 ) \big)\times 10^{2}$ & $ 15.73$ & $ 4.78$ & $ 1.43$ & $ 14.46$ & $ 4.64$ & $-1.99$ & $ 19.44$ & $ 3.96$ & $ 0.81$\\ 
\hline
$t_{1 3}\big( (0, 0, 1 ) \big)\times 10^{2}$ & $-5.56$ & $ 3.72$ & $ 11.52$ & $-0.67$ & $ 7.32$ & $ 12.74$ & $-2.64$ & $ 2.29$ & $ 11.37$\\ 
\hline
$t_{2 1}\big( (0, 1, 1 ) \big)\times 10^{2}$ & $ 1.73$ & $ 6.66$ & $ 8.07$ & $-8.71$ & $ 4.43$ & $ 6.96$ & $-2.63$ & $ 5.90$ & $ 7.77$\\ 
\hline
$t_{1 3}\big( (0, 0, 2 ) \big)\times 10^{2}$ & $-9.96$ & $-3.94$ & $-1.81$ & $-7.59$ & $-5.28$ & $-4.42$ & $-9.78$ & $-4.21$ & $-2.65$\\ 
\hline
$t_{2 1}\big( (0, 0, 2 ) \big)\times 10^{2}$ & $ 6.67$ & $-8.29$ & $-5.37$ & $ 4.21$ & $-5.34$ & $-5.47$ & $ 2.52$ & $-5.74$ & $-4.71$\\ 
\hline
$t_{4 4}\big( (0, 1, 0 ) \big)\times 10^{2}$ & $-5.35$ & $-1.37$ & $ 6.23$ & $-7.74$ & $-2.68$ & $ 4.16$ & $-5.50$ & $-1.44$ & $ 5.78$\\ 
\hline
$t_{4 4}\big( (0, 0, 1 ) \big)\times 10^{2}$ & $ 11.34$ & $ 1.68$ & $ 0.49$ & $ 8.77$ & $ 3.01$ & $ 0.92$ & $ 10.48$ & $ 2.68$ & $ 0.79$\\ 
\hline
$t_{1 2}\big( (0, 1, 0 ) \big)\times 10^{2}$ & $-0.15$ & $ 6.82$ & $ 2.30$ & $-4.67$ & $ 6.53$ & $ 3.56$ & $-2.12$ & $ 6.15$ & $ 2.48$\\ 
\hline
$t_{1 2}\big( (0, 0, 1 ) \big)\times 10^{2}$ & $ 5.02$ & $-2.37$ & $-3.14$ & $ 3.32$ & $-5.83$ & $-5.11$ & $ 6.08$ & $-1.57$ & $-2.18$\\ 
\hline
$t_{2 3}\big( (0, 0, 2 ) \big)\times 10^{2}$ & $-4.66$ & $-2.68$ & $-5.85$ & $-4.74$ & $-4.62$ & $-1.80$ & $-3.24$ & $-1.68$ & $-4.65$\\ 
\hline
$t_{2 4}\big( (0, 0, 1 ) \big)\times 10^{2}$ & $-4.41$ & $-3.41$ & $ 1.16$ & $-3.76$ & $-3.82$ & $ 0.35$ & $-4.94$ & $-5.03$ & $ 1.10$\\ 
\hline
$t_{2 4}\big( (0, 1, 0 ) \big)\times 10^{2}$ & $ 1.59$ & $-3.09$ & $-5.78$ & $-0.49$ & $-1.69$ & $-3.47$ & $ 1.01$ & $-3.00$ & $-5.37$\\ 
\hline
$t_{1 1}\big( (0, 1, 1 ) \big)\times 10^{2}$ & $-1.50$ & $-3.06$ & $-3.72$ & $ 3.72$ & $-3.00$ & $-3.74$ & $ 0.61$ & $-2.62$ & $-3.46$\\ 
\hline
$t_{3 4}\big( (0, 1, 0 ) \big)\times 10^{2}$ & $-1.91$ & $-1.45$ & $-3.29$ & $-2.84$ & $-1.17$ & $-3.50$ & $-2.32$ & $-1.53$ & $-3.31$\\ 
\hline
$t_{1 4}\big( (0, 0, 1 ) \big)\times 10^{2}$ & $-0.61$ & $-2.26$ & $-3.74$ & $-1.40$ & $-1.67$ & $-2.68$ & $-2.29$ & $-1.76$ & $-3.21$\\ 
\hline
$t_{1 4}\big( (0, 1, 0 ) \big)\times 10^{2}$ & $-1.45$ & $-3.50$ & $-1.25$ & $ 0.83$ & $-3.45$ & $-2.17$ & $-1.31$ & $-3.45$ & $-1.51$\\ 
\hline
$t_{1 3}\big( (1, 2, 0 ) \big)\times 10^{2}$ & $ 3.72$ & $ 1.45$ & $-0.14$ & $ 4.74$ & $ 2.09$ & $ 1.13$ & $ 3.30$ & $ 1.09$ & $-0.18$\\ 
\hline
$t_{1 1}\big( (1, 2, 0 ) \big)\times 10^{2}$ & $ 2.45$ & $ 2.78$ & $ 0.44$ & $ 2.98$ & $ 2.19$ & $-0.31$ & $ 1.69$ & $ 2.93$ & $ 0.24$\\ 
\hline
$t_{4 4}\big( (0, 0, 2 ) \big)\times 10^{2}$ & $-1.09$ & $ 2.17$ & $ 2.15$ & $-1.26$ & $ 0.17$ & $ 2.08$ & $-2.36$ & $ 1.28$ & $ 1.87$\\ 
\hline
$t_{2 3}\big( (0, 1, 1 ) \big)\times 10^{2}$ & $-2.09$ & $ 0.75$ & $ 2.10$ & $-0.44$ & $ 0.91$ & $ 2.16$ & $-1.79$ & $ 0.47$ & $ 1.91$\\ 
\hline
$t_{1 2}\big( (0, 0, 2 ) \big)\times 10^{2}$ & $-1.59$ & $ 0.70$ & $-1.46$ & $-2.42$ & $ 0.24$ & $-1.59$ & $-1.53$ & $ 0.27$ & $-1.52$\\ 
\hline
$t_{1 1}\big( (0, 0, 2 ) \big)\times 10^{2}$ & $-1.43$ & $ 1.84$ & $ 0.58$ & $-0.70$ & $ 2.75$ & $ 0.63$ & $-1.40$ & $ 1.68$ & $ 0.10$\\ 
\hline
$t_{1 3}\big( (0, 1, 1 ) \big)\times 10^{2}$ & $ 3.42$ & $ 0.52$ & $ 0.01$ & $ 1.58$ & $ 0.79$ & $ 0.66$ & $ 2.11$ & $ 1.05$ & $ 0.25$\\ 
\hline
$t_{4 4}\big( (0, 1, 1 ) \big)\times 10^{2}$ & $-2.03$ & $-0.59$ & $-0.84$ & $-1.34$ & $-0.79$ & $-0.60$ & $-2.38$ & $-0.79$ & $-0.76$\\ 
\hline
$t_{1 2}\big( (0, 1, 1 ) \big)\times 10^{3}$ & $-7.17$ & $ 8.98$ & $ 8.02$ & $-13.88$ & $ 10.78$ & $ 6.03$ & $-14.88$ & $ 5.04$ & $ 6.45$\\ 
\hline
$t_{4 4}\big( (1, 2, 0 ) \big)\times 10^{3}$ & $ 15.87$ & $-1.22$ & $-7.07$ & $ 17.59$ & $ 4.99$ & $-2.09$ & $ 15.45$ & $ 1.79$ & $-6.84$\\ 
\hline
$t_{1 2}\big( (1, 2, 0 ) \big)\times 10^{3}$ & $-2.84$ & $-3.64$ & $ 8.07$ & $ 2.48$ & $-6.58$ & $ 7.67$ & $ 3.61$ & $-5.90$ & $ 7.58$
	\\ \hline

	\multicolumn{10}{|c|}{\textbf{Simplified four-band model of YV$_6$Sn$_6${}}} \\
	\hline
	\multirow{3}{\dimexpr(\linewidth)}{Hopping Parameter ($\si{\tera\hertz^2}$)} & 
	\multicolumn{3}{c|}{Relaxed~\cite{POK21}} &
	\multicolumn{3}{c|}{Experimental~\cite{ROM11}} &
	\multicolumn{3}{c|}{Relaxed~\cite{ROM11}} \\ \cline{2-10}
	& 
	\multicolumn{3}{c|}{$T / \si{\kelvin}$} &
	\multicolumn{3}{c|}{$T / \si{\kelvin}$} &
	\multicolumn{3}{c|}{$T / \si{\kelvin}$} \\ \cline{2-10}
	& $0$ & $4642$ & $9284$
	& $0$ & $4642$ & $9284$
	& $0$ & $4642$ & $9284$
	\\ \hline
	$t_{4 4}\big( (0, 0, 0 ) \big)$ & $ 21.78$ & $ 19.88$ & $ 19.20$ & $ 21.10$ & $ 19.39$ & $ 18.75$ & $ 21.59$ & $ 19.72$ & $ 19.06$\\ 
\hline
$t_{1 1}\big( (0, 0, 0 ) \big)$ & $ 12.64$ & $ 12.55$ & $ 12.17$ & $ 12.72$ & $ 12.83$ & $ 12.45$ & $ 12.32$ & $ 12.40$ & $ 12.05$\\ 
\hline
$t_{3 3}\big( (0, 0, 0 ) \big)$ & $ 15.49$ & $ 11.25$ & $ 8.77$ & $ 13.67$ & $ 11.26$ & $ 8.91$ & $ 14.18$ & $ 10.85$ & $ 8.69$\\ 
\hline
$t_{2 4}\big( (0, 0, 0 ) \big)$ & $-8.83$ & $-8.12$ & $-7.48$ & $-8.75$ & $-8.05$ & $-7.40$ & $-8.84$ & $-8.09$ & $-7.43$\\ 
\hline
$t_{1 2}\big( (0, 0, 0 ) \big)$ & $-1.18$ & $-2.11$ & $-2.77$ & $-1.33$ & $-2.37$ & $-3.05$ & $-0.96$ & $-2.04$ & $-2.68$\\ 
\hline
$t_{3 3}\big( (0, 1, 0 ) \big)$ & $-2.32$ & $-1.53$ & $-1.10$ & $-1.92$ & $-1.51$ & $-1.12$ & $-2.04$ & $-1.44$ & $-1.09$\\ 
\hline
$t_{1 3}\big( (0, 0, 0 ) \big)$ & $-1.83$ & $-1.46$ & $-1.11$ & $-1.80$ & $-1.50$ & $-1.13$ & $-1.78$ & $-1.44$ & $-1.10$\\ 
\hline
$t_{3 3}\big( (0, 0, 1 ) \big)$ & $-1.25$ & $-1.12$ & $-1.00$ & $-1.23$ & $-1.16$ & $-1.03$ & $-1.23$ & $-1.12$ & $-0.99$\\ 
\hline
$t_{1 4}\big( (0, 0, 0 ) \big)\times 10^{1}$ & $ 8.85$ & $ 5.83$ & $ 2.79$ & $ 8.49$ & $ 5.59$ & $ 2.63$ & $ 8.53$ & $ 5.57$ & $ 2.57$\\ 
\hline
$t_{3 4}\big( (0, 0, 0 ) \big)\times 10^{1}$ & $-5.77$ & $-4.15$ & $-4.33$ & $-4.78$ & $-3.50$ & $-3.83$ & $-5.30$ & $-3.92$ & $-4.20$\\ 
\hline
$t_{3 3}\big( (1, 2, 0 ) \big)\times 10^{1}$ & $ 4.56$ & $ 2.62$ & $ 1.66$ & $ 3.43$ & $ 2.52$ & $ 1.71$ & $ 3.73$ & $ 2.33$ & $ 1.63$\\ 
\hline
$t_{2 1}\big( (0, 0, 1 ) \big)\times 10^{1}$ & $ 1.16$ & $-1.66$ & $-3.31$ & $ 0.79$ & $-1.22$ & $-3.21$ & $ 1.07$ & $-1.60$ & $-3.52$
	\\ \hline

	\caption{\label{tab:HoppingEffPhonon} Hopping amplitudes for the effective three- and four-band models of the low-energy phonon spectrum of ScV$_6$Sn$_6${} and YV$_6$Sn$_6${}. We only list the symmetry-non-redundant hopping amplitudes (\textit{i.e.}{}, if multiple hopping amplitudes are equal due crystalline symmetries, as implied by \cref{app:eqn:constraint_symmetry_dynamical}, then only one of them is listed). All the nonzero hopping amplitudes can be restored through \cref{app:eqn:constraint_symmetry_dynamical}.}
\end{longtable}

\section{Brief theoretical introduction to ARPES}\label{app:sec:arpes}

In this \siSection{}, we provide a brief theoretical overview of angle-resolved photoemission spectroscopy (ARPES) and summarize how the experimentally-measured photoemission signal can be related to the band structure characteristics of a compound, placing a special emphasis on ScV$_6$Sn$_6${}. We refer the reader to Ref.~\cite{DAM03,MOS17} for a more in-depth treatment. We start from the experimental setup used in the ARPES experiment and consider the process in which the absorption of a photon causes one of the electrons from the sample to scatter and reach the detector. By summing over all possible processes that result in a scattered electron of a given momentum and energy, we show that the ARPES photocurrent is given by the contraction between the sample's spectral function and some one-body matrix elements (which will be derived below). We then discuss the one-body matrix elements and show how the presence of discrete symmetries can render them exactly zero, leading to the so called ARPES selection rules. Finally, we discuss the implications that these selection rules have on the ARPES signal of ScV$_6$Sn$_6${} by means of an example.

\subsection{The ARPES photocurrent}\label{app:sec:arpes:photocurrent}

\subsubsection{The electron-photon Hamiltonian}\label{app:sec:arpes:photocurrent:ele_phot}

\begin{figure}[t]
	\centering
	\includegraphics[width=0.3\textwidth]{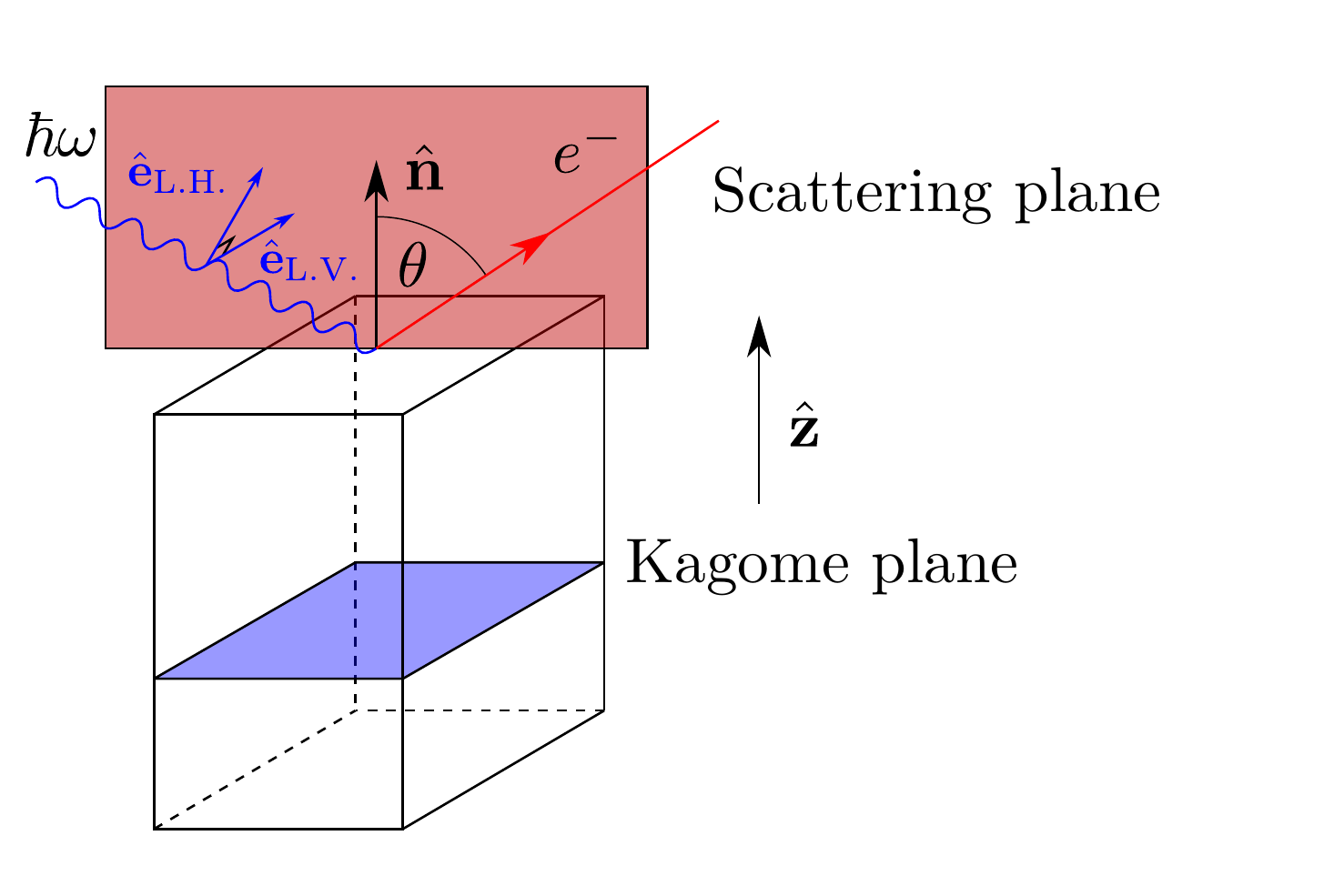}
	\caption{Setup used in ARPES experiments of ScV$_6$Sn$_6${}. The ScV$_6$Sn$_6${} sample is cleaved parallel to the V kagome plane (\textit{i.e.}{}, perpendicularly to the $\hat{\vec{z}}$ direction). Incoming photons of energy $\hbar \omega$ (shown by the blue wavy line) cause electrons to scatter at an angle $\theta$ to the sample normal $\hat{\vec{n}} \parallel \hat{\vec{z}}$ (shown by the red line). The plane containing the incoming photon, scattered electron, and normal $\hat{\vec{n}}$ is known as the scattering plane (and is perpendicular to the surface of the crystal). Relative to the scattering plane, the light can either be in-plane linearly polarized [known as linearly horizontal (L.H.) polarized light, whose polarization unit vector is given by $\hat{\vec{e}}_{\mathrm{L.H.}}$], or linearly polarized perpendicular to the scattering plane [known as linearly vertical (L.V.) polarized light, whose polarization vector is given by $\hat{\vec{e}}_{\mathrm{L.V.}}$]. }
	\label{fig:experiment_arpes}
\end{figure}

A schematic depiction of the experimental setup used in ARPES experiments is shown in \cref{fig:experiment_arpes}: an incoming photon of energy $\hbar \omega$ is absorbed by the ScV$_6$Sn$_6${} crystal causing an electron to scatter off at an angle $\theta$ to the crystal surface normal $\hat{\vec{n}}$. The absorption of the photon by the sample is a process for which the electromagnetic field is macroscopic and, as a result, we can treat the electromagnetic field classically (as opposed to quantum mechanically)~\cite{SHA13}. The Hamiltonian describing our system is given simply by 
\begin{equation}
	\hat{H} \left( t \right) = \hat{H}_0 + \hat{H}_{\mathrm{int}} \left(t \right),
\end{equation}
where $\hat{H}_0$ is the Hamiltonian of the ScV$_6$Sn$_6${} crystal and $\hat{H}_{\mathrm{int}} \left(t \right)$ is the time-dependent perturbation governing the interaction between the electrons and the (classical) electromagnetic field. The electron-photon interaction Hamiltonian is given by
\begin{equation}
	\label{app:eqn:phot_ele_coupling}
	\hat{H}_{\mathrm{int}} \left(t \right) = \sum_{i=1}^{N} \frac{e}{2m_e} \left( \vec{p}_i \cdot \vec{A} \left( \vec{r}_i,t \right) + \vec{A} \left( \vec{r}_i,t \right) \cdot \vec{p}_i \right) +  \frac{e}{2 m_e} \vec{A}^2 \left( \vec{r}_i,t \right),
\end{equation}
where $e$ and $m_e$ are the electron's mass and charge, respectively, $\vec{p}_i$ and $\vec{r}_i$ are the momentum and position operators of the $i$-th electron, and $\vec{A} \left( \vec{r},t \right)$ is the vector potential of the electromagnetic field. In \cref{app:eqn:phot_ele_coupling} and in what follows, we have also employed the Coulomb gauge (in which $\nabla \cdot \vec{A} = 0$, and the electromagnetic scalar potential vanishes). Using the fact that in the Coulomb gauge
\begin{equation}
	\vec{p} \cdot \vec{A} \left( \vec{r}, t \right) \psi(\vec{r},t) = -i\hbar \nabla \cdot \left[ \vec{A} \left( \vec{r}, t \right) \psi(\vec{r},t) \right] = -i\hbar \psi(\vec{r},t) \nabla \cdot  \vec{A} \left( \vec{r}, t \right) - i \hbar \vec{A} \left( \vec{r}, t \right) \cdot \nabla  \psi(\vec{r},t) = \vec{A} \left( \vec{r}, t \right) \cdot \vec{p} \psi(\vec{r},t),
\end{equation}
for any scalar wave function $\psi(\vec{r},t)$ and also neglecting the second order electromagnetic contribution~\cite{DAM03}, we find that the electron-photon interaction Hamiltonian is given by
\begin{equation}
	\label{app:eqn:phot_ele_coupling_simple}
	\hat{H}_{\mathrm{int}} \left(t \right) \approx \sum_{i=1}^{N} \frac{e}{m_e} \vec{A} \left( \vec{r}_i,t \right) \cdot \vec{p}_i.
\end{equation}

\subsubsection{Photon-induced electron scattering rate}\label{app:sec:arpes:photocurrent:phot_ind_scat}
We take the vector potential of the incident light to be given by
\begin{equation}
	\label{app:eqn:light_vector_pot}
	\vec{A} \left( \vec{r}, t \right) = \vec{A}_0 \cos \left( \vec{k}_{\text{ph}} \cdot \vec{r} - \omega t \right),
\end{equation}
which corresponds to light with wave vector $\vec{k}_{\text{ph}}$, whose polarization is along the direction of $\vec{A}_0$ (and $\vec{k}_{\text{ph}} \cdot \vec{A}_0 = 0$). The corresponding photons cause an optical excitation from $\ket{\Psi^N_i}$ -- the $N$-electron initial state of the system (in which all $N$ electrons are bound inside the crystal) -- to $\ket{\Psi^N_f}$ -- the $N$-electron final state (in which $N-1$ electrons remain bound inside the sample, and one electron is scattered outside the crystal). By Fermi's Golden Rule, the transition probability per unit time between the initial and final states is given by~\cite{SHA13,DAM03,PAV14} 
\begin{equation}
	\label{app:eqn:fermi_golden}
	w_{fi} = \frac{2 \pi}{\hbar} \abs{\bra{\Psi^N_f} \hat{H}_{\mathrm{int}} \ket{\Psi^N_i}}^2 \delta \left(E^N_f - E^N_i - \hbar \omega \right),
\end{equation}
where $E^N_i$ and $E^N_f$ are the energies of the initial and final states, respectively, and
\begin{equation}
	\label{app:eqn:phot_ele_coupling_time_indep}
	\hat{H}_{\mathrm{int}} = \sum_{i=1}^{N} \frac{e}{2m_e} e^{i \vec{k}_{\text{ph}} \cdot \vec{r}_i } \vec{A}_0 \cdot \vec{p}_i
\end{equation}
is the \emph{time-independent} perturbation operator. In deriving \cref{app:eqn:fermi_golden} from Fermi's Golden Rule, we have used the fact that $E^N_f>E^N_i$ (the system with the scattered electron has a higher energy than the system in which the electron is bound). To move forward, we now assume the sudden approximation~\cite{PAV14}, which states that the $N$-electron final state $\ket{\Psi^N_f}$ is the direct product between a free electron state (the scattered electron of momentum $\hbar \vec{k}_e$ and spin $s$) and an excited many-body $(N-1)$-electron state within the crystal ($\ket{\Phi^{N-1}}$), \textit{i.e.}{}, 
\begin{equation}
	\label{app:eqn:final_state}
	\ket{\Psi^N_f} = \hat{c}^\dagger_{\vec{k}_e,s} \ket{\Phi^{N-1}}.
\end{equation}
In \cref{app:eqn:final_state}, the $\hat{c}^\dagger_{\vec{k}_e,s}$ operator creates an electron of spin $s=\uparrow,\downarrow$ and energy $E_e$, which has momentum $\hbar \vec{k}_e$ at \emph{infinite distance from the sample}. Additionally, we \emph{assume} that $\hat{c}_{\vec{k}_e,s} \ket{\Psi^N_i} = 0$ (there are no scattered electrons within the initial state). With these simplifications, \cref{app:eqn:fermi_golden} becomes
\begin{align}
	w_{fi} &= \frac{2 \pi}{\hbar} \abs{\bra{\Phi^{N-1}} \hat{c}_{\vec{k}_e,s} \hat{H}_{\text{int}}\ket{\Psi^N_i}}^2 \delta \left(E^N_f - E^N_i - \hbar \omega \right) \nonumber \\
	&= \frac{2 \pi}{\hbar} \abs{\bra{\Phi^{N-1}} \hat{H}_{\text{int}} \hat{c}_{\vec{k}_e,s} + \left[ \hat{c}_{\vec{k}_e,s}, \hat{H}_{\text{int}} \right] \ket{\Psi^N_i}}^2 \delta \left(E^N_f - E^N_i - \hbar \omega \right) \nonumber \\
	&= \frac{2 \pi}{\hbar} \abs{\bra{\Phi^{N-1}} \left[ \hat{c}_{\vec{k}_e,s}, \hat{H}_{\text{int}} \right] \ket{\Psi^N_i}}^2 \delta \left(E_e + E^{N-1}_{\Phi} - E^N_i - \hbar \omega \right), \label{app:eqn:fermi_golden_interm_1}
\end{align}
where $E_f^N = E_e + E^{N-1}_{\Phi}$, since, by the sudden approximation, the scattered electron and the $N-1$ electrons within the crystal do not interact with one another.  

We now assume that the crystalline system features an exponentially-localized Wannier orbital basis. We take $\hat{c}^\dagger_{\vec{R},\alpha}$ to be the creation operator corresponding to the $\alpha$-th (spinful) Wannier orbital within unit cell $\vec{R}$, whose Wannier center's displacement from the unit cell origin is given by $\boldsymbol{\rho}_{\alpha}$. We note that the time-independent perturbation Hamiltonian $\hat{H}_{\text{int}}$ is a one-body term. As a result, its second-quantized expression is a quadratic fermion operator written generically as
\begin{equation}
	\label{app:eqn:H_phonon_electron_interaction}
	\hat{H}_{\text{int}} = \sum_{m,n} \Delta_{mn} \hat{c}^\dagger_{m} \hat{c}_{n},
\end{equation}
where $\hat{c}^\dagger_{n}$ are fermion creation operators corresponding to some complete basis in which the matrix elements of $\hat{H}_{\text{int}}$ are given by $\Delta_{mn} = \bra{0} \hat{c}_{m} \hat{H}_{\text{int}} \hat{c}^\dagger_{n} \ket{0}$. Letting $\ket{\psi_{\alpha} \left( \vec{R} \right)} \equiv \hat{c}^\dagger_{\vec{R},\alpha} \ket{0}$ and $\ket{\psi_{\vec{k}_e,s}} \equiv \hat{c}^\dagger_{\vec{k}_e,s} \ket{0}$, \cref{app:eqn:H_phonon_electron_interaction} implies that the commutator in \cref{app:eqn:fermi_golden_interm_1} can be written as 
\begin{equation}
	\label{app:eqn:commutator}
	\left[ \hat{c}_{\vec{k}_e,s}, \hat{H}_{\text{int}} \right] = \sum_{\vec{R},\alpha} \left[\Delta\left(\vec{A}_0 \right)\right]_{\vec{k}_e,s;\vec{R},\alpha} \hat{c}_{\vec{R},\alpha} + (\dots) ,
\end{equation}
where we have defined
\begin{equation}
	\left[\Delta\left(\vec{A}_0 \right)\right]_{\vec{k}_e,s;\vec{R},\alpha} = 
	\bra{\psi_{\vec{k}_e,s}}
	\hat{H}_{\text{int}}
	\ket{\psi_{\alpha} \left( \vec{R} \right)} = \mel**{\psi_{\vec{k}_e,s}}{
		\frac{e}{2 m_e}e^{i \vec{k}_{\text{ph}} \cdot \vec{r} } \vec{A}_0 \cdot \vec{p}
	}{\psi_{\alpha} \left( \vec{R} \right)}. \label{app:eqn:sel_rule_mat_elem_unapprox}
\end{equation}
The right-hand side of \cref{app:eqn:commutator} contains other terms which are proportional to annihilation operators corresponding to scattered electron states with various momenta. These additional terms have been denoted generically by $(\dots)$, and are unimportant for our discussion, as they annihilate the initial state $\ket{\Psi^N_i}$ (which does not contain any scattered electron states). Finally, in \cref{app:eqn:sel_rule_mat_elem_unapprox} and in what follows, we will leave the dependence of the matrix elements on the photon wave vector $\vec{k}_{\text{ph}}$ implicit to keep notation brief. 

Because the $\ket{\psi_{\alpha} \left( \vec{R} \right)}$ orbital is exponentially localized around its Wannier center (which is located at $\vec{R} + \boldsymbol{\rho}_{\alpha}$) and the localization length is much smaller the the incoming photon wavelength, we can make the dipole approximation~\cite{SHA13} and replace the operator $\vec{r}$, by the Wannier center of the orbital $\ket{\psi_{\alpha} \left( \vec{R} \right)}$. Under the dipole approximation, \cref{app:eqn:sel_rule_mat_elem_unapprox} becomes\footnote{it is worth mentioning that $\ket{\psi_{\vec{k}_e,s}}$ denotes a scattered electron state behaving as a plane wave only at infinity. In other words, because $\ket{\psi_{\alpha} \left( \vec{R} \right)}$ is localized within the sample, where $\ket{\psi_{\vec{k}_e,s}}$ is not a plane wave,
	\begin{equation}
		\mel**{\psi_{\vec{k}_e,s}}{\vec{A}_0 \cdot \vec{p}}{\psi_{\alpha} \left( \vec{R} \right)} \neq \hbar \vec{A}_0 \cdot \vec{k}_e \bra{\psi_{\vec{k}_e,s}}\ket{\psi_{\alpha} \left( \vec{R} \right)}.
	\end{equation}}
\begin{equation}
	\left[\Delta\left(\vec{A}_0 \right)\right]_{\vec{k}_e,s;\vec{R},\alpha} = \frac{e}{2 m_e}e^{i \vec{k}_{\text{ph}} \cdot \left( \vec{R} + \boldsymbol{\rho}_{\alpha} \right) } \mel**{\psi_{\vec{k}_e,s}}{
		\vec{A}_0 \cdot \vec{p}
	}{\psi_{\alpha} \left( \vec{R} \right)}. \label{app:eqn:matrix_elem_wannier}
\end{equation}
Combining \cref{app:eqn:fermi_golden_interm_1,app:eqn:matrix_elem_wannier}, we find that the transition probability per unit time becomes
\begin{equation}
	w_{fi} = \sum_{\substack{\vec{R}_1,\alpha_1 \\ \vec{R}_2, \alpha_2}}\bra{\Phi^{N-1}} \hat{c}_{\vec{R}_1,\alpha_1} \ket{\Psi^N_i} \bra{\Psi^N_i} \hat{c}^\dagger_{\vec{R}_2,\alpha_2} \ket{\Phi^{N-1}} \Delta_{\vec{k}_e,s;\vec{R}_1,\alpha_1} \Delta^*_{\vec{k}_e,s;\vec{R}_2,\alpha_2}  \delta \left(E_e + E^{N-1}_{\Phi} - E^N_i - \hbar \omega \right), \label{app:eqn:fermi_golden_interm_2}
\end{equation}
where here and in what follows we will leave the dependence of the matrix elements on $\vec{A}_0$ implicit for the sake of brevity.

\subsubsection{The relation between the ARPES photocurrent and the spectral function}\label{app:sec:arpes:photocurrent:sp_func}

An ARPES experiment measures the photocurrent stemming from \emph{all} electrons with a given momentum $\vec{k}_e$. This photocurrent can be obtained by first summing the transition probability from \cref{app:eqn:fermi_golden_interm_2} over all the possible initial states, weighted by the corresponding Boltzmann factor. Additionally, we also need to sum over all the possible final $N$-electron states that have one scattered electron with momentum $\vec{k}_e$. In \cref{app:eqn:fermi_golden_interm_2}, this latter summation is equivalent to summing over all $(N-1)$-electron states $\ket{\Phi^{N-1}}$. As such, the total photocurrent is given by
\begin{align}
	\label{app:eqn:photocurrent}
	I \left( \vec{k}_e, E_e, \omega \right) &= \sum_{\substack{\vec{R}_1,\alpha_1 \\ \vec{R}_2, \alpha_2}} \sum_{\Psi^{N}_i,\Phi^{N-1},s} \bra{\Phi^{N-1}} \hat{c}_{\vec{R}_1,\alpha_1} \ket{\Psi^N_i} \bra{\Psi^N_i} \hat{c}^\dagger_{\vec{R}_2,\alpha_2} \ket{\Phi^{N-1}} \nonumber \\
	&\times \frac{e^{-\beta E_i^{N}}}{Z} \Delta_{\vec{k}_e,s;\vec{R}_1,\alpha_1} \Delta^*_{\vec{k}_e,s;\vec{R}_2,\alpha_2}  \delta \left(E_e + E^{N-1}_{\Phi} - E^N_i - \hbar \omega \right),
\end{align}  
where $\beta = 1/ \left(k_B T \right)$ and $Z$ is the system's partition function. We now note that whereas the translation symmetry of the crystal is broken along the sample normal, the system still features two-dimensional discrete translation symmetry parallel to the sample surface. Denoting by $\vec{v}^{\parallel}$ and $v^{\perp}$ the parallel and perpendicular components of a given vector $\vec{v}$, we can Fourier transform the Wannier orbitals along the sample plane and define
\begin{equation}
	\label{app:eqn:hyb_bas}
	\hat{c}^\dagger_{\vec{k}^{\parallel},R^{\perp},\alpha} = \frac{1}{\sqrt{N^{\parallel}}} \sum_{\vec{R}^{\parallel}} \hat{c}^\dagger_{\vec{R},\alpha} e^{i \vec{k}^{\parallel} \cdot \left(  \vec{R}^{\parallel} + \boldsymbol{\rho}^{\parallel}_{\alpha} \right)},
\end{equation}
where $N^{\parallel}$ denotes the number of unit cells along the sample surface. Using the hybrid basis from \cref{app:eqn:hyb_bas}, we find that the photocurrent from \cref{app:eqn:photocurrent} can be written as
\begin{align}
	I \left( \vec{k}_e, E_e, \omega \right) &= \sum_{\vec{k}^{\parallel}_1,\vec{k}^{\parallel}_2}\sum_{\substack{\vec{R}_1,\alpha_1 \\ \vec{R}_2, \alpha_2}} \sum_{\Psi^{N}_i,\Phi^{N-1},s} \bra{\Phi^{N-1}} \hat{c}_{\vec{k}_1^{\parallel},R_1^{\perp},\alpha_1} \ket{\Psi^N_i} \bra{\Psi^N_i} \hat{c}^\dagger_{\vec{k}_2^{\parallel},R_2^{\perp},\alpha_2} \ket{\Phi^{N-1}} \nonumber \\	
	&\times  
	e^{i \vec{k}_1^{\parallel} \cdot \left(  \vec{R}_1^{\parallel} + \boldsymbol{\rho}^{\parallel}_{\alpha_1} \right)}
	e^{-i \vec{k}_2^{\parallel} \cdot \left(  \vec{R}_2^{\parallel} + \boldsymbol{\rho}^{\parallel}_{\alpha_2} \right)} 
	\frac{e^{-\beta E_i^{N}}}{Z} \Delta_{\vec{k}_e,s;\vec{R}_1,\alpha_1} \Delta^*_{\vec{k}_e,s;\vec{R}_2,\alpha_2}  \delta \left(E_e + E^{N-1}_{\Phi} - E^N_i - \hbar \omega \right). 	\label{app:eqn:photocurrent_hyb_1}
\end{align}  
As the system still features crystalline translation symmetry along the sample surface, the many-body states $\ket{\Psi^N_i}$ and $\ket{\Phi^{N-1}}$ can be chosen to be states of definite momentum, which implies that the only non-vanishing terms in \cref{app:eqn:photocurrent_hyb_1} are the ones for which $\vec{k}^{\parallel}_1 = \vec{k}^{\parallel}_2$ 
\begin{align}
	I \left( \vec{k}_e, E_e, \omega \right) &= \sum_{\vec{k}^{\parallel}}\sum_{\substack{\vec{R}_1,\alpha_1 \\ \vec{R}_2, \alpha_2}} \sum_{\Psi^{N}_i,\Phi^{N-1},s} \bra{\Phi^{N-1}} \hat{c}_{\vec{k}^{\parallel},R_1^{\perp},\alpha_1} \ket{\Psi^N_i} \bra{\Psi^N_i} \hat{c}^\dagger_{\vec{k}^{\parallel},R_2^{\perp},\alpha_2} \ket{\Phi^{N-1}} \nonumber \\	
	&\times  
	\frac{1}{N_{\parallel}}
	e^{i \vec{k}^{\parallel} \cdot \left(  \vec{R}_1^{\parallel} + \boldsymbol{\rho}^{\parallel}_{\alpha_1} \right)}
	e^{-i \vec{k}^{\parallel} \cdot \left(  \vec{R}_2^{\parallel} + \boldsymbol{\rho}^{\parallel}_{\alpha_2} \right)} 
	\frac{e^{-\beta E_i^{N}}}{Z} \Delta_{\vec{k}_e,s;\vec{R}_1,\alpha_1} \Delta^*_{\vec{k}_e,s;\vec{R}_2,\alpha_2}  \delta \left(E_e + E^{N-1}_{\Phi} - E^N_i - \hbar \omega \right). 	\label{app:eqn:photocurrent_hyb_2}
\end{align}  

Remembering that the spectral function of the system is given by~\cite{MAH00}
\begin{align}
	\mathcal{A}_{R_1^{\perp},\alpha_1; R_2^{\perp}, \alpha_2} \left( \vec{k}^{\parallel}, \hbar \Omega \right) =& \frac{2 \pi}{\hbar} \sum_{\Psi^{N}_i,\Phi^{N-1}} \bra{\Phi^{N-1}} \hat{c}_{\vec{k}^{\parallel},R_1^{\perp},\alpha_1} \ket{\Psi^N_i} \bra{\Psi^N_i} \hat{c}^\dagger_{\vec{k}^{\parallel},R_2^{\perp},\alpha_2} \ket{\Phi^{N-1}} \nonumber \\
	&\times \frac{e^{-\beta E_i^{N}}+e^{-\beta E^{N-1}_{\Phi}}}{Z}\delta \left(E^{N-1}_{\Phi} - E^N_i - \hbar \Omega \right),
	\label{app:eqn:def_sp_func_hyb_bas}
\end{align}
and defining the Fourier-transformed matrix elements
\begin{equation}
	\label{app:eqn:ft_mat_elem}
	\left[\Delta \left(\vec{A}_0 \right)\right]_{\vec{k}_e,s;\vec{k}^{\parallel},R^{\perp},\alpha} = \frac{1}{\sqrt{N_{\parallel}}} \sum_{\vec{R}^{\parallel}} \left[\Delta \left(\vec{A}_0 \right)\right]_{\vec{k}_e,s;\vec{R},\alpha} e^{i \vec{k}^{\parallel} \cdot \left(  \vec{R}^{\parallel} + \boldsymbol{\rho}^{\parallel}_{\alpha} \right)},
\end{equation}
we find that the photocurrent reads as
\begin{align}
	\label{app:eqn:photocurrent_sp_func_1}
	I \left( \vec{k}_e, E_e, \omega \right) = \sum_{\vec{k}^{\parallel},s} \sum_{\substack{R^{\perp}_1,\alpha_1 \\ R^{\perp}_2, \alpha_2}}  \mathcal{A}_{R_1^{\perp},\alpha_1; R_2^{\perp}, \alpha_2} \left( \vec{k}^{\parallel}, \hbar \omega - E_e \right) \Delta_{\vec{k}_e,s;\vec{k}^{\parallel},R_1^{\perp},\alpha_1}
	\Delta^*_{\vec{k}_e,s;\vec{k}^{\parallel},R_2^{\perp},\alpha_2} f \left( \hbar \omega - E_e \right),
\end{align} 
where $f (E)$ is the Fermi distribution function. To simplify the expression in \cref{app:eqn:photocurrent_sp_func_1}, we turn our attention to the Fourier-transformed matrix elements, which, upon combining \cref{app:eqn:ft_mat_elem,app:eqn:matrix_elem_wannier}, become 
\begin{align}
	\Delta_{\vec{k}_e,s;\vec{k}^{\parallel},R^{\perp},\alpha} &= \frac{1}{\sqrt{N_{\parallel}}} \frac{e}{2 m_e} \sum_{\vec{R}^{\parallel}} e^{i \vec{k}_{\text{ph}} \cdot \left( \vec{R} + \boldsymbol{\rho}_{\alpha} \right) }  
	e^{i \vec{k}^{\parallel} \cdot \left(  \vec{R}^{\parallel} + \boldsymbol{\rho}^{\parallel}_{\alpha} \right)}
	\mel**{\psi_{\vec{k}_e,s}}{\vec{A}_0 \cdot \vec{p}}{\psi_{\alpha} \left( \vec{R}^{\parallel},R^{\perp} \right)} \nonumber \\
	&= \frac{1}{\sqrt{N_{\parallel}}} \frac{e}{2 m_e} \sum_{\vec{R}^{\parallel}} e^{i \vec{k}_{\text{ph}} \cdot \left( \vec{R} + \boldsymbol{\rho}_{\alpha} \right) }  
	e^{i \vec{k}^{\parallel} \cdot \left(  \vec{R}^{\parallel} + \boldsymbol{\rho}^{\parallel}_{\alpha} \right)}
	e^{-i \vec{k}_e^{\parallel} \cdot \vec{R}^{\parallel} }
	\mel**{\psi_{\vec{k}_e,s}}{\vec{A}_0 \cdot \vec{p}}{\psi_{\alpha} \left( \vec{0},R^{\perp} \right)} \nonumber \\
	&= \sqrt{N_{\parallel}} \frac{e}{2 m_e} \sum_{\vec{G}^{\parallel}} \delta_{\vec{k}^{\parallel}_e + \vec{G}^{\parallel},\vec{k}^{\parallel} + \vec{k}^{\parallel}_{\text{ph}} } 
	e^{i k^{\perp}_{\text{ph}} \left( R^{\perp} + \rho^{\perp}_{\alpha} \right) }  
	e^{i \left( \vec{k}_e^{\parallel} + \vec{G}^{\parallel} \right) \cdot \boldsymbol{\rho}^{\parallel}_{\alpha}}
	\mel**{\psi_{\vec{k}_e,s}}{\vec{A}_0 \cdot \vec{p}}{\psi_{\alpha} \left( \vec{0},R^{\perp} \right)} \nonumber \\
	&= \frac{1}{\sqrt{N_{\parallel}}} \frac{e}{2 m_e} \sum_{\vec{G}^{\parallel}} \delta_{\vec{k}^{\parallel}_e + \vec{G}^{\parallel},\vec{k}^{\parallel} + \vec{k}^{\parallel}_{\text{ph}} } 
	e^{i k^{\perp}_{\text{ph}} \left( R^{\perp} + \rho^{\perp}_{\alpha} \right) }  
	e^{i \left( \vec{k}_e^{\parallel} + \vec{G}^{\parallel} \right) \cdot \boldsymbol{\rho}^{\parallel}_{\alpha}}
	\sum_{\vec{k}^{\prime \parallel}}
	\mel**{\psi_{\vec{k}_e,s}}{\vec{A}_0 \cdot \vec{p}}{\psi_{\alpha} \left( \vec{k}^{\prime \parallel},R^{\perp} \right)} e^{-i \vec{k}^{\prime \parallel} \cdot \boldsymbol{\rho}^{\parallel}_{\alpha} } \nonumber \\
	&= \frac{1}{\sqrt{N_{\parallel}}} \frac{e}{2 m_e} \sum_{\vec{G}^{\parallel}} \delta_{\vec{k}^{\parallel}_e + \vec{G}^{\parallel},\vec{k}^{\parallel} + \vec{k}^{\parallel}_{\text{ph}} } 
	e^{i k^{\perp}_{\text{ph}} \left( R^{\perp} + \rho^{\perp}_{\alpha} \right) }  
	\mel**{\psi_{\vec{k}_e,s}}{\vec{A}_0 \cdot \vec{p}}{\psi_{\alpha} \left( \vec{k}_e^{\parallel} + \vec{G}^{\parallel}, R^{\perp} \right)} \label{app:eqn:ft_mat_elem_1},
\end{align}
where 
\begin{equation}
	\label{app:eqn:ket_momentum}
	\ket{\psi_{\alpha} \left( \vec{k}^{\parallel},R^{\perp} \right)} = \hat{c}^\dagger_{\vec{k}^{\parallel},R^{\perp},\alpha} \ket{0},
\end{equation}
and $\vec{G}^{\parallel}$ are reciprocal lattice vectors parallel to the sample's surface. The reader is reminded that the dependence of the matrix elements in \cref{app:eqn:ft_mat_elem_1} on the vector potential amplitude $\vec{A}_0$ and the wave vector $\vec{k}_{\text{ph}}$ has been left implicit to keep the notation brief, as mentioned around \cref{app:eqn:sel_rule_mat_elem_unapprox,app:eqn:fermi_golden_interm_2}. In going from the first to the second line of \cref{app:eqn:ft_mat_elem_1}, we have employed the properties of the electron states under lattice translations parallel to the sample surface. More specifically, letting $\mathcal{T}_{\vec{R}^{\parallel}}$ denote the operator which translates by a lattice vector $\vec{R}^{\parallel}$, we find that the final electron obeys\footnote{The system features discrete translation symmetry parallel to the sample surface. As a result, \emph{all} electronic energy eigenstates can be labeled by their eigenvalues under translation. Letting $\ket{\vec{r},s}$ denote a position state (corresponding to an electron localized of spin $s$ localized at $\vec{r}$), we find that for $\vec{r}$ sufficiently far away from the sample $\bra{\vec{r},s}\ket{\psi_{\vec{k}_e,s}} \propto e^{i \vec{k}_e \cdot \vec{r}}$ (\textit{i.e.}{}, the state $\ket{\psi_{\vec{k}_e,s}}$ corresponds to a plane wave of momentum $\hbar \vec{k}_e$ at large enough distances from the sample). As a result, $\mel**{\vec{r},s}{\mathcal{T}_{\vec{R}^{\parallel}}}{\psi_{\vec{k}_e,s}} = \bra{\vec{r} - \vec{R}^{\parallel},s} \ket{\psi_{\vec{k}_e,s} } \propto e^{i \vec{k}_e \cdot \left( \vec{r} - \vec{R}^{\parallel}\right)}$, from which one can determine that the eigenvalue of $ \ket{\psi_{\vec{k}_e,s}}$ under translation is given by \cref{app:sec:eigenvalue_translation_scat_state}. }
\begin{equation}
	\label{app:sec:eigenvalue_translation_scat_state}
	\mathcal{T}_{\vec{R}^{\parallel}} \ket{\psi_{\vec{k}_e,s}} = e^{-i \vec{k}_e^{\parallel} \cdot \vec{R}^{\parallel} } \ket{\psi_{\vec{k}_e,s}},
\end{equation}
which implies that
\begin{align}
	&\mel**{\psi_{\vec{k}_e,s}}{\vec{A}_0 \cdot \vec{p}}{\psi_{\alpha} \left( \vec{R}^{\parallel},R^{\perp} \right)} 
	= \mel**{\psi_{\vec{k}_e,s}}{\vec{A}_0 \cdot \vec{p} \mathcal{T}_{\vec{R}^{\parallel}}}{\psi_{\alpha} \left( \vec{0},R^{\perp} \right)} \nonumber\\
	= &\mel**{\psi_{\vec{k}_e,s}}{\mathcal{T}_{\vec{R}^{\parallel}} \vec{A}_0 \cdot \vec{p}}{\psi_{\alpha} \left( \vec{0},R^{\perp} \right)}
	= e^{-i \vec{k}_e^{\parallel} \cdot \vec{R}^{\parallel} } \mel**{\psi_{\vec{k}_e,s}}{\vec{A}_0 \cdot \vec{p}}{\psi_{\alpha} \left( \vec{0},R^{\perp} \right)}
\end{align}

In what follows, we note that the wave vector of the incoming photon ($\vec{k}_{\text{ph}}$) is much smaller than that of the electron, and can therefore be ignored with respect to the latter\footnote{The typical photons used in ARPES experiments have $\abs{\vec{k}_{\text{ph}}} \ll \abs{\vec{k}_e}$, which allows us to ignore $\vec{k}_{\text{ph}}$ in the $\delta$-function from \cref{app:eqn:ft_mat_elem_1}~\cite{DAM03}. On the other hand, the photon might penetrate deep into the sample surface meaning that $k_{\text{ph}}^{\perp} R^{\perp}$ is not necessarily small and, therefore, will not be ignored.}~\cite{DAM03}. As a result, \cref{app:eqn:ft_mat_elem} can be approximated to 
\begin{equation}
	\left[\Delta \left( \vec{A}_0 \right) \right]_{\vec{k}_e,s;\vec{k}^{\parallel},R^{\perp},\alpha} = \frac{1}{\sqrt{N_{\parallel}}} \frac{e}{2 m_e} \sum_{\vec{G}^{\parallel}} \delta_{\vec{k}^{\parallel}_e + \vec{G}^{\parallel},\vec{k}^{\parallel}} 
	e^{i k^{\perp}_{\text{ph}} \left( R^{\perp} + \rho^{\perp}_{\alpha} \right) }  
	\mel**{\psi_{\vec{k}_e,s}}{\vec{A}_0 \cdot \vec{p}}{\psi_{\alpha} \left( \vec{k}_e^{\parallel} + \vec{G}^{\parallel}, R^{\perp} \right)} \label{app:eqn:ft_mat_elem_2}.	
\end{equation}
Using embedding relation for the hybrid basis from \cref{app:eqn:hyb_bas}, 
\begin{equation}
	\label{app:eqn:emb_hyb_bas}
	\hat{c}^\dagger_{\vec{k}^{\parallel} + \vec{G}^{\parallel},R^{\perp},\alpha} = \hat{c}^\dagger_{\vec{k}^{\parallel}, R^{\perp},\alpha} e^{i \vec{G}^{\parallel} \cdot \boldsymbol{\rho}^{\parallel}_{\alpha} },
\end{equation}
we can rewrite \cref{app:eqn:ft_mat_elem_2} as 
\begin{equation}
	\left[\Delta \left( \vec{A}_0 \right) \right]_{\vec{k}_e,s;\vec{k}^{\parallel},R^{\perp},\alpha} = \sum_{\vec{G}^{\parallel}} \delta_{\vec{k}^{\parallel}_e + \vec{G}^{\parallel},\vec{k}^{\parallel}} 
	\left[\Delta \left( \vec{A}_0 \right) \right]_{\vec{k}_e,s;\vec{k}_e^{\parallel},R^{\perp},\alpha} 
	e^{i \vec{G}^{\parallel} \cdot \boldsymbol{\rho}^{\parallel}_{\alpha}}.
	\label{app:eqn:ft_mat_elem_2_interm_1}
\end{equation}
Substituting \cref{app:eqn:ft_mat_elem_2_interm_1} into the expression for the photocurrent from \cref{app:eqn:photocurrent_sp_func_1} leads to 
\begin{align}
	I \left( \vec{k}_e, E_e, \omega \right) &= \sum_{\substack{\vec{k}^{\parallel},s \\ \vec{G}^{\parallel}}} \sum_{\substack{R^{\perp}_1,\alpha_1 \\ R^{\perp}_2, \alpha_2}} \delta_{\vec{k}^{\parallel}_e + \vec{G}^{\parallel},\vec{k}^{\parallel}} \mathcal{A}_{R_1^{\perp},\alpha_1; R_2^{\perp}, \alpha_2} \left( \vec{k}^{\parallel}, \hbar \omega - E_e \right) f \left( \hbar \omega - E_e \right) \nonumber \\
	&\times \Delta_{\vec{k}_e,s;\vec{k}_e^{\parallel},R_1^{\perp},\alpha_1}
	\Delta^*_{\vec{k}_e,s;\vec{k}_e^{\parallel},R_2^{\perp},\alpha_2} e^{i \vec{G}^{\parallel} \cdot \left( \boldsymbol{\rho}^{\parallel}_{\alpha_1} - \boldsymbol{\rho}^{\parallel}_{\alpha_2} \right)} \nonumber \\
	&= \sum_{s} \sum_{\substack{R^{\perp}_1,\alpha_1 \\ R^{\perp}_2, \alpha_2}}  \mathcal{A}_{R_1^{\perp},\alpha_1; R_2^{\perp}, \alpha_2} \left( \vec{k}^{\parallel}_e + \vec{G}^{\prime\parallel}, \hbar \omega - E_e \right) \Delta_{\vec{k}_e,s;\vec{k}_e^{\parallel},R_1^{\perp},\alpha_1}
	\Delta^*_{\vec{k}_e,s;\vec{k}_e^{\parallel},R_2^{\perp},\alpha_2} e^{i \vec{G}^{\prime\parallel} \cdot \left( \boldsymbol{\rho}^{\parallel}_{\alpha_1} - \boldsymbol{\rho}^{\parallel}_{\alpha_2} \right)} f \left( \hbar \omega - E_e \right) \nonumber \\
	&= \sum_{s} \sum_{\substack{R^{\perp}_1,\alpha_1 \\ R^{\perp}_2, \alpha_2}}  \mathcal{A}_{R_1^{\perp},\alpha_1; R_2^{\perp}, \alpha_2} \left( \vec{k}_e^{\parallel}, \hbar \omega - E_e \right) \Delta_{\vec{k}_e,s;\vec{k}_e^{\parallel},R_1^{\perp},\alpha_1}
	\Delta^*_{\vec{k}_e,s;\vec{k}_e^{\parallel},R_2^{\perp},\alpha_2} f \left( \hbar \omega - E_e \right). \label{app:eqn:photocurrent_sp_func_2}
\end{align} 
In going from the first two rows of \cref{app:eqn:photocurrent_sp_func_2} to the third one, we have used the fact that the summation over $\vec{k}^{\parallel}$ runs over the first Brillouin zone, implying that there exist a unique reciprocal lattice vector $\vec{G}^{\prime\parallel}$, such that $\vec{k}_e^{\parallel} + \vec{G}^{\prime\parallel}$ is in the first Brillouin zone. Finally, to obtain the last line of \cref{app:eqn:photocurrent_sp_func_2}, we have employed the periodicity properties of the spectral function
\begin{equation}
	\mathcal{A}_{R_1^{\perp},\alpha_1; R_2^{\perp}, \alpha_2} \left( \vec{k}^{\parallel} + \vec{G}^{\parallel}, \hbar \omega - E_e \right) = 
	\mathcal{A}_{R_1^{\perp},\alpha_1; R_2^{\perp}, \alpha_2} \left( \vec{k}^{\parallel}, \hbar \omega - E_e \right) e^{i \vec{G}^{\parallel} \cdot \left( \boldsymbol{\rho}^{\parallel}_{\alpha_2} - \boldsymbol{\rho}^{\parallel}_{\alpha_1} \right)},
\end{equation} 
which follow straightforwardly from its definition in \cref{app:eqn:def_sp_func_hyb_bas} and the embedding relation \cref{app:eqn:emb_hyb_bas}.

\Cref{app:eqn:photocurrent_sp_func_2} is the main result of this \siSection{} and shows that the ARPES photocurrent is proportional to the spectral function of the crystal, up to some one-body matrix elements 
\begin{equation}
	\label{app:eqn:photocurrent_sp_func_3}
	\left[ \Delta \left( \vec{A}_0 \right) \right]_{\vec{k}_e,s;\vec{k}_e^{\parallel},R^{\perp},\alpha} = \frac{1}{\sqrt{N_{\parallel}}} \frac{e}{2 m_e} e^{i k^{\perp}_{\text{ph}} \left( R^{\perp} + \rho^{\perp}_{\alpha} \right) }  
	\mel**{\psi_{\vec{k}_e,s}}{\vec{A}_0 \cdot \vec{p}}{\psi_{\alpha} \left( \vec{k}_e^{\parallel}, R^{\perp} \right)}.
\end{equation}
For a given $\vec{k}_e$, the spectral function $\mathcal{A}_{R_1^{\perp},\alpha_1; R_2^{\perp}, \alpha_2} \left( \vec{k}_e^{\parallel}, E \right) $ is sharply peaked at those energies that correspond to bulk and surface states of the crystal with parallel momentum equal to $\vec{k}_e^{\parallel}$. As a result, ARPES experiments can be used to directly reconstruct the band structure of a material, as we will explain below.  

\subsubsection{The ARPES photocurrent in the band basis}\label{app:sec:arpes:photocurrent:band_basis}

To gain some intuition on \cref{app:eqn:photocurrent_sp_func_2}, we \emph{assume} the existence of a ``band basis''\footnote{This is not necessarily the case if the system we study is not a Fermi liquid.}. In particular, we assume that for every $\vec{k}^{\parallel}$, we can recombine the operators $\hat{c}^\dagger_{\vec{k}^{\parallel},R^{\perp},\alpha}$ into charge-$(\pm 1)$ excitation operators 
\begin{equation}
	\hat{c}^\dagger_{\vec{k}^{\parallel},n} = \sum_{R^{\perp},\alpha} U_{n;R^{\perp},\alpha} \left( \vec{k}^{\parallel} \right) \hat{c}^\dagger_{\vec{k}^{\parallel},R^{\perp},\alpha},
\end{equation}
where $n$ is the charge-$(\pm 1)$ excitation band index and $U_{n;R^{\perp},\alpha} \left( \vec{k}^{\parallel} \right)$ is a unitary matrix. In the non-interacting case, $\hat{c}^\dagger_{\vec{k}^{\parallel},n}$ are simply the fermion operators corresponding to the energy band basis (with $n$ labeling the band). In the interacting case, $\hat{c}^\dagger_{\vec{k}^{\parallel},n}$ are the fermions that diagonalize the \textit{e.g.}{} Hartree-Fock Hamiltonian. 

The advantage of working within the band basis will become apparent in \cref{app:eqn:photocurrent_band_approx}, but for now we note that in terms of these new operators, the spectral function of the system is given by 
\begin{align}
	\mathcal{A}_{nm} \left( \vec{k}^{\parallel}, \hbar \Omega \right) =& \frac{2 \pi}{\hbar} \sum_{\Psi^{N}_i,\Phi^{N-1}} \bra{\Phi^{N-1}} \hat{c}_{\vec{k}^{\parallel},n} \ket{\Psi^N_i} \bra{\Psi^N_i} \hat{c}^\dagger_{\vec{k}^{\parallel},m} \ket{\Phi^{N-1}} \frac{e^{-\beta E_i^{N}}+e^{-\beta E^{N-1}_{\Phi}}}{Z}\delta \left(E^{N-1}_{\Phi} - E^N_i - \hbar \Omega \right), \nonumber \\
	=& \sum_{\substack{R^{\perp}_1,\alpha_1 \\ R^{\perp}_2, \alpha_2}} U^*_{n;R_1^{\perp},\alpha_1} \left( \vec{k}^{\parallel} \right) U_{m;R_2^{\perp},\alpha_2} \left( \vec{k}^{\parallel} \right)  \mathcal{A}_{R_1^{\perp},\alpha_1; R_2^{\perp}, \alpha_2} \left( \vec{k}^{\parallel}, \hbar \Omega \right).
\end{align}
We can also define the matrix elements $\Delta_{\vec{k}_e,s;\vec{k}_e^{\parallel},R^{\perp},\alpha}$ in the band basis
\begin{align}
	\left[ \Delta \left( \vec{A}_0 \right) \right]_{\vec{k}_e,s;\vec{k}_e^{\parallel},n} &= \sum_{R^{\perp},\alpha} U_{n;R^{\perp},\alpha} \left( \vec{k}_e^{\parallel} \right) \left[ \Delta \left( \vec{A}_0 \right) \right]_{\vec{k}_e,s;\vec{k}_e^{\parallel},R^{\perp},\alpha} \nonumber \\
	& = \frac{1}{\sqrt{N_{\parallel}}} \frac{e}{2 m_e}   
	\mel**{\psi_{\vec{k}_e,s}}{
		\vec{A}_0 \cdot \vec{p}
		\hat{\mathcal{O}}
	}{\psi_{n} \left( \vec{k}_e^{\parallel}\right)}, 
	\label{app:eqn:photocurrent_sp_func_band}
\end{align}
where, similarly to \cref{app:eqn:ket_momentum}, we have introduced 
\begin{equation}
	\label{app:eqn:ket_band_basis_momentum}
	\ket{\psi_{n} \left( \vec{k}^{\parallel},R^{\perp} \right)} = \hat{c}^\dagger_{\vec{k}^{\parallel},n} \ket{0},
\end{equation}
as well as the operator
\begin{equation}
	\label{app:eqn:cal_o_operator}
	\hat{\mathcal{O}} = \sum_{R^{\perp},\alpha} e^{i k^{\perp}_{\text{ph}} \left( R^{\perp} + \rho^{\perp}_{\alpha} \right)}
	\ket{\psi_{\alpha} \left( \vec{k}_e^{\parallel}, R^{\perp} \right)}
	\bra{\psi_{\alpha} \left( \vec{k}_e^{\parallel}, R^{\perp} \right)}.
\end{equation}
In deriving \cref{app:eqn:photocurrent_sp_func_band}, we have employed the fact that 
\begin{equation}
	U_{n;R^{\perp},\alpha} = \bra{\psi_{\alpha} \left( \vec{k}_e^{\parallel}, R^{\perp} \right)}\ket{\psi_{n} \left( \vec{k}_e^{\parallel}\right)}.
\end{equation}
Finally, we note that in the band basis, the expression of the photocurrent is analogous to \cref{app:eqn:photocurrent_sp_func_2} 
\begin{equation}
	\label{app:eqn:photocurrent_band}
	I \left( \vec{k}_e, E_e, \omega \right) = \sum_{s} \sum_{n,m}  \mathcal{A}_{nm} \left( \vec{k}_e^{\parallel}, \hbar \omega - E_e \right) \Delta_{\vec{k}_e,s;\vec{k}_e^{\parallel},n}
	\Delta^*_{\vec{k}_e,s;\vec{k}_e^{\parallel},m} f \left( \hbar \omega - E_e \right).
\end{equation}

We take the operator $\hat{c}_{\vec{k}_e^{\parallel},n'}$ to create a charge-(-1) excitation in the system with energy $\epsilon_{n'}\left(\vec{k}_e^{\parallel} \right)$. To be completely general, we also consider the case in which this charge-(-1) excitation is degenerate such that \emph{all} the operators $\hat{c}_{\vec{k}_e^{\parallel},m}$ for $m$ in some set $\aleph$ (that contains $n'$) create distinct charge-(-1) excitations with identical energies $\epsilon_{m} \left( \vec{k}_e^{\parallel} \right) = \epsilon_{n'}\left(\vec{k}_e^{\parallel} \right)$. Since the nondegenerate case is equivalent to the degenerate one with a single-element set $\aleph$, we will work in the degenerate case in what follows. Near the $\abs{\epsilon_{n'}\left(\vec{k}_e^{\parallel} \right)}$ excitation energy, the spectral function has a simple form in the band basis
\begin{equation}
	\mathcal{A}_{nm} \left( \vec{k}_e^{\parallel}, \hbar \Omega \right) \approx \delta_{nm} A \left( \hbar \Omega + \epsilon_{n'}\left(\vec{k}_e^{\parallel} \right)  \right), \quad \text{for} \quad n,m \in \aleph,
\end{equation}
where $A \left( E \right)$ is some real function sharply peaked at $E=0$ (and is, in fact, proportional to a Dirac $\delta$-function in the non-interacting limit). In other words, the spectral function is sharply peaked near the charge-$(-1)$ excitation energies and is diagonal within the corresponding degenerate subspace. The photocurrent contribution stemming from the charge-$(-1)$ excitations created by $\hat{c}_{\vec{k}_e^{\parallel},m}$ for $m \in \aleph$ is given by 
\begin{equation}
	\label{app:eqn:photocurrent_band_approx}
	I \left( \vec{k}_e, E_e, \omega \right) \bigg\lvert_{E_e \approx \hbar \omega + \epsilon_{n'}\left(\vec{k}_e^{\parallel} \right)} \approx  A \left( \hbar \omega + \epsilon_{n'}\left(\vec{k}_e^{\parallel} \right) - E_e \right) f \left( \hbar \omega - E_e \right) \sum_{\substack{s=\uparrow,\downarrow \\ n \in \aleph}} \abs{\left[\Delta \left( \vec{A}_0 \right) \right]_{\vec{k}_e,s;\vec{k}_e^{\parallel},n}}^2.
\end{equation}
Therefore, for a given scattered electron momentum $\vec{k}_e$, the photocurrent intensity peaks at those energies $E_e \approx  \hbar \omega + \epsilon_{n'}\left(\vec{k}_e^{\parallel} \right)$ which correspond to charge-$(-1)$ excitations of energy $\epsilon_{n'}\left(\vec{k}_e^{\parallel} \right)$. The matrix elements $\left[\Delta \left( \vec{A}_0 \right) \right]_{\vec{k}_e,s;\vec{k}_e^{\parallel},n}$ will also influence the intensity of the measured photocurrent and, in some cases, will lead to a \emph{complete} suppression of the ARPES signal as a result of so-called ``ARPES selection rules''. As we will show in the following \cref{app:sec:arpes:selection_rules}, the latter arise from the crystalline symmetries of the sample, which can render some of the matrix elements $\left[\Delta \left( \vec{A}_0 \right) \right]_{\vec{k}_e,s;\vec{k}_e^{\parallel},n}$ exactly zero.

\subsection{ARPES selection rules}\label{app:sec:arpes:selection_rules}

As shown in \cref{app:eqn:photocurrent_band_approx} the photocurrent contribution from some (possibly degenerate) bulk or surface charge-$(-1)$ excitations created by $\hat{c}_{\vec{k}^{\parallel}_e,n}$ (with $n\in \aleph$) depends on the matrix elements $\left[\Delta \left( \vec{A}_0 \right) \right]_{\vec{k}_e,s;\vec{k}_e^{\parallel},n}$. In this \siSection{}, we discuss the constraints imposed by the crystalline symmetries of the system on these matrix elements.

We assume that the scattering geometry is such that the sample-scattered electron system is invariant under a unitary symmetry $g$. By assumption, $g$ is a symmetry of the \emph{whole} system (and not just of the infinite crystal) and, therefore, $g$ must keep invariant \emph{both} the sample normal and the scattered electron wave vector, \textit{i.e.}{} $g \vec{k}_e = \vec{k}_e$ and $g \hat{\vec{n}} = \hat{\vec{n}}$. Moreover, because $g$ is a symmetry of the crystalline sample, one must have that the action of $g$ in the charge-(-1) excitation eigensubspace is closed, \textit{i.e.}{}
\begin{equation}
	g \ket{\psi_{n} \left( \vec{k}^{\parallel}_e \right)} = \sum_{m \in \aleph} \left[B \left(g, \vec{k}^{\parallel}_e \right)\right]_{mn} \ket{\psi_{m} \left( \vec{k}^{\parallel}_e \right)}, \quad \text{for} \quad m \in \aleph,
\end{equation}
with the corresponding unitary sewing matrix being given by 
\begin{equation}
	\left[B \left(g, \vec{k}^{\parallel}_e \right)\right]_{mn} = \bra{\psi_{m} \left( \vec{k}^{\parallel}_e \right)} g \ket{\psi_{n} \left( \vec{k}^{\parallel}_e \right)}, \quad \text{for} \quad n,m \in \aleph.
\end{equation}
Similarly, the action of $g$ on the scattered electron states is given by 
\begin{equation}
	g\ket{\psi_{\vec{k}_e,s}} = \sum_{s'} \left[R\left(g\right)\right]_{s',s} \ket{\psi_{\vec{k}_e,s'}},
\end{equation}
where $R\left(g\right)$ is the $\mathrm{SU}(2)$ spin rotation matrix corresponding to $g$.

As a result of the symmetry $g$, the matrix elements $\left[\Delta \left( \vec{A}_0 \right) \right]_{\vec{k}_e,s;\vec{k}_e^{\parallel},n}$ obey the following constraints
\begin{align}
	\left[\Delta \left( \vec{A}_0 \right) \right]_{\vec{k}_e,s;\vec{k}_e^{\parallel},n} &= \frac{1}{\sqrt{N_{\parallel}}} \frac{e}{2 m_e}   
	\mel**{\psi_{\vec{k}_e,s}}{g^{-1}
		\vec{A}_0 \cdot \left( g \vec{p} g^{-1}\right) 
		\left(g \hat{\mathcal{O}} g^{-1}\right) g
	}{\psi_{n} \left( \vec{k}_e^{\parallel}\right)} \nonumber \\
	& = \frac{1}{\sqrt{N_{\parallel}}} \frac{e}{2 m_e}   
	\sum_{s',m} \left[R\left(g\right)\right]^{*}_{s',s} \left[B \left(g, \vec{k}^{\parallel}_e \right)\right]_{mn} \mel**{\psi_{\vec{k}_e,s'}}{
		\left(g\vec{A}_0 \right) \cdot \vec{p} 
		\hat{\mathcal{O}} 
	}{\psi_{m} \left( \vec{k}_e^{\parallel}\right)} \nonumber \\
	& = \sum_{s',m} \left[R\left(g\right)\right]^{*}_{s',s} \left[B \left(g, \vec{k}^{\parallel}_e \right)\right]_{mn} 
	\left[\Delta \left( g\vec{A}_0 \right) \right]_{\vec{k}_e,s';\vec{k}_e^{\parallel},m}, \label{app:eqn:constraint_delta_1}
\end{align}
where we have used the fact that the $\mathcal{O}$ operator defined in \cref{app:eqn:cal_o_operator} is invariant under the symmetry $g$
\begin{align}
	g\hat{\mathcal{O}}g^{-1} &= \sum_{R^{\perp},\alpha} e^{i k^{\perp}_{\text{ph}} \left( R^{\perp} + \rho^{\perp}_{\alpha} \right)}
	g\ket{\psi_{\alpha} \left( \vec{k}_e^{\parallel}, R^{\perp} \right)}
	\bra{\psi_{\alpha} \left( \vec{k}_e^{\parallel}, R^{\perp} \right)}g^{-1} \nonumber \\
	&= \sum_{R^{\perp},\alpha,\beta,\beta'} e^{i k^{\perp}_{\text{ph}} \left( R^{\perp} + \rho^{\perp}_{\alpha} \right)}
	\left[D(g)\right]_{\beta \alpha}\ket{\psi_{\beta} \left( \vec{k}_e^{\parallel}, R^{\perp} \right)}
	\bra{\psi_{\beta'} \left( \vec{k}_e^{\parallel}, R^{\perp} \right)} \left[D(g)\right]^{*}_{\beta' \alpha} \nonumber \\
	&= \sum_{R^{\perp},\beta} e^{i k^{\perp}_{\text{ph}} \left( R^{\perp} + \rho^{\perp}_{\beta} \right)}
	\ket{\psi_{\beta} \left( \vec{k}_e^{\parallel}, R^{\perp} \right)}
	\bra{\psi_{\beta} \left( \vec{k}_e^{\parallel}, R^{\perp} \right)}. \label{app:eqn:invariance_of_cal_o}
\end{align}
In deriving \cref{app:eqn:invariance_of_cal_o}, $D(g)$ denotes the unitary representation matrix of the $g$ symmetry on the Wannier basis defined near \cref{app:eqn:commutator}, which obeys
\begin{equation}
	\left[D(g)\right]_{\alpha \beta} = 0 \quad \text{if} \quad \rho^{\perp}_{\alpha} \neq \rho^{\perp}_{\beta},
\end{equation}
\textit{i.e.}{}, since $g$ preserves the normal vector $\hat{\vec{n}}$ then it cannot exchange orbitals that have different displacements along $\hat{\vec{n}}$. 

\Cref{app:eqn:constraint_delta_1} can be rewritten in a simpler form by remembering that $\left[\Delta \left( \vec{A}_0 \right) \right]_{\vec{k}_e,s;\vec{k}_e^{\parallel},n}$ is a linear function of $\vec{A}_0$ and, moreover, holds for \emph{any} choice of $\vec{A}_0$. Letting $\hat{\vec{e}}_i$ ($1 \leq i \leq 3$) be the three Cartesian unit vectors and defining the $\mathrm{SO}(3)$ representation matrix of the symmetry $g$
\begin{equation}
	\left[\mathcal{R}(g)\right]_{ji} = \hat{\vec{e}}_j \cdot g \hat{\vec{e}}_i,
\end{equation}
we find that, by letting $\vec{A}_0 \propto \vec{e}_i$, \cref{app:eqn:constraint_delta_1} can be recast as
\begin{equation}
	\left[\Delta \left( \hat{\vec{e}}_i \right) \right]_{\vec{k}_e,s;\vec{k}_e^{\parallel},n} = \sum_{s',m,j} \left[\mathcal{R}(g)\right]_{ji} \left[R\left(g\right)\right]^{*}_{s',s} \left[B \left(g, \vec{k}^{\parallel}_e \right)\right]_{mn} 
	\left[\Delta \left( \hat{\vec{e}}_j \right) \right]_{\vec{k}_e,s';\vec{k}_e^{\parallel},m}.  \label{app:eqn:constraint_delta_2}
\end{equation}
\Cref{app:eqn:constraint_delta_2} is very powerful as it can be used to directly infer the ARPES selection rules. Consider a certain scattering geometry with a given $\vec{k}_e$ that probes the states $\ket{\psi_{n} \left( \vec{k}_e^{\parallel}\right)}$ from the sample. The symmetries $g$ of the sample-scattered electron system will form a group $\mathcal{G}$. Under the symmetries of $\mathcal{G}$, the rank-three tensor $\left[\Delta \left( \hat{\vec{e}}_i \right) \right]_{\vec{k}_e,s;\vec{k}_e^{\parallel},n}$ (viewed as a tensor in the $s$, $n$, and $i$ indices) will transform as the direct product of three representations of the group $\mathcal{G}$, namely $R(g)$, $B\left(g,\vec{k}_e^{\parallel} \right)$, and $\mathcal{R}(g)$. If the representation $R^*(g) \otimes B\left(g,\vec{k}_e^{\parallel} \right) \otimes \mathcal{R}(g)$ does \emph{not} contain the trivial (identity) irrep of the group $\mathcal{G}$, then the matrix elements $\left[\Delta \left( \hat{\vec{e}}_i \right) \right]_{\vec{k}_e,s;\vec{k}_e^{\parallel},n}$ will be zero by symmetry and the corresponding ARPES signal will not be detected. This criterion is the most general form of the ARPES selection rules.

\subsection{ARPES selection rules in ScV$_6$Sn$_6${}}\label{app:sec:arpes:selection_rules_inSVS}

As discussed in \cref{app:sec:crystal_struct_nCDW}, ScV$_6$Sn$_6${} has \emph{exact} $P6/mmm$ symmetry in the pristine phase. For the purpose of deriving the ARPES selection rules in this compound, we will assume that it still features an \emph{approximate} $P6/mmm$ symmetry in the CDW phase. Moreover, symmetry-enforced selection rules leading to the complete suppression of the ARPES signal can only arise whenever the sample-scattered electron system has a non-trivial symmetry group. For the scattering geometry shown in \cref{fig:experiment_arpes} (\textit{i.e.}, with the sample crystal being cleaved parallel to the V kagome plane), we distinguish two different cases\footnote{Note that because the parallel component of the photon momentum $\vec{k}^{\parallel}_{\text{ph}}$ was neglected in \cref{app:eqn:ft_mat_elem_2}, the incidence angle of the light does not play any role in determining the symmetries of the sample-scattered electron system.}:
\begin{enumerate}
	\item $\theta = 0$: For normal photoemission, the sample-scattered electron system features $C_{6v}$ symmetry. Normal photoemission corresponds to $\vec{k}_e^{\parallel} = \vec{0}$ and therefore probes the dispersion at the $\Gamma'$ point.
	\item $\theta \neq 0$ and $\vec{k}_e^{\parallel}$ along the $\Gamma' - \mathrm{M}'$ or $\Gamma' - \mathrm{K}'$ directions: In this case the sample-scattered electron is symmetric with respect to mirror reflections in the scattering plane. 
\end{enumerate}
Throughout this \siSection{}, we will denote by $\Gamma'$, $\mathrm{K}'$, and $\mathrm{M}'$ the high symmetry points of the two-dimensional Brillouin zone of the sample.

Because SOC is not significant in ScV$_6$Sn$_6${}, we will also assume the system features $\mathrm{SU}(2)$ symmetry, which is equivalent to using spinless representations for $R(g)$ and $B\left(g, \vec{k}^{\parallel}_e \right)$ in \cref{app:eqn:constraint_delta_2}. More precisely, in the spinless case the photocurrent intensity becomes
\begin{equation}
	\label{app:eqn:photocurrent_band_spinless}
	I \left( \vec{k}_e, E_e, \omega \right) = \sum_{n,m}  \mathcal{A}_{nm} \left( \vec{k}_e^{\parallel}, \hbar \omega - E_e \right) \left[ \Delta \left( \vec{A}_0 \right) \right]_{\vec{k}_e;\vec{k}_e^{\parallel},n}
	\left[ \Delta \left( \vec{A}_0 \right) \right]^*_{\vec{k}_e;\vec{k}_e^{\parallel},m} f \left( \hbar \omega - E_e \right),
\end{equation}
where the spinless matrix elements are given by
\begin{align}
	\left[ \Delta \left( \vec{A}_0 \right) \right]_{\vec{k}_e,s;\vec{k}_e^{\parallel},n} = \frac{1}{\sqrt{N_{\parallel}}} \frac{e}{2 m_e}   
	\mel**{\psi_{\vec{k}_e}}{
		\vec{A}_0 \cdot \vec{p}
		\hat{\mathcal{O}}
	}{\psi_{n} \left( \vec{k}_e^{\parallel}\right)}, 
	\label{app:eqn:photocurrent_sp_func_band_spinless}
\end{align}
with $\ket{\psi_{\vec{k}_e}}$ and $\ket{\psi_{n} \left( \vec{k}_e^{\parallel}\right)}$ denoting spinless electron states. In particular, the spinless matrix elements will obey the following symmetry constraint
\begin{equation}
	\left[\Delta \left( \hat{\vec{e}}_i \right) \right]_{\vec{k}_e;\vec{k}_e^{\parallel},n} = \sum_{m,j} \left[\mathcal{R}(g)\right]_{ji} \left[B \left(g, \vec{k}^{\parallel}_e \right)\right]_{mn} 
	\left[\Delta \left( \hat{\vec{e}}_j \right) \right]_{\vec{k}_e;\vec{k}_e^{\parallel},m},  \label{app:eqn:constraint_delta_spinless}
\end{equation}
meaning that for the selection rules we only need to consider whether the $\mathcal{R}(g) \otimes B\left(g, \vec{k}^{\parallel}_e \right)$ representation contains the trivial irrep. This former criterion can be applied straightforwardly using a character table, such as the one shown in \cref{tab:char_table_c6v} for the $C_{6v}$ group: the character of the $\mathcal{R}(g) \otimes B\left(g, \vec{k}^{\parallel}_e \right)$ representation is given by 
\begin{equation}
	\chi_{\mathcal{R}(g) \otimes B\left(g, \vec{k}^{\parallel}_e \right)} = \chi_{\mathcal{R}(g)} \chi_{B\left(g, \vec{k}^{\parallel}_e \right)},
\end{equation} 
where $\chi_{\mathcal{R}(g)}$ and $\chi_{B\left(g, \vec{k}^{\parallel}_e \right)}$ are, respectively, the characters of the $\mathcal{R}(g)$ and $B\left(g, \vec{k}^{\parallel}_e \right)$ representations. Remembering that the character of the trivial irrep is one for every group element, we find that $\mathcal{R}(g) \otimes B\left(g, \vec{k}^{\parallel}_e \right)$ will contain the trivial irrep only if 
\begin{equation}
	\label{app:eqn:simple_selection_crit}
	\sum_{g} \chi_{\mathcal{R}(g) \otimes B\left(g, \vec{k}^{\parallel}_e \right)} = \sum_{g} \chi_{\mathcal{R}(g)} \chi_{B\left(g, \vec{k}^{\parallel}_e \right)} \neq 0.
\end{equation}
We will now consider the two scattering geometries that feature selection rules.

\subsubsection{$\theta = 0$ (normal) photoemission}

\begin{table}[t]
	\begin{tabular}{|c|c|c|c|c|c|c|c|c|}
		\hline
		Irrep & $\mathbbm{1}$ & $2C_{6z}$ & $2C_{3z}$ & $C_{2z}$ & $3m_{v}$ & $3m_{h}$ & Linear Functions & Quadratic Functions \\
		\hline
		$A_1$ & $+1$ & $+1$ & $+1$ & $+1$ & $+1$ & $+1$ & $z$ & $x^2+y^2$, $z^2$ \\
		\hline
		$A_2$ & $+1$ & $+1$ & $+1$ & $+1$ & $-1$ & $-1$ & - & - \\
		\hline
		$B_1$ & $+1$ & $-1$ & $+1$ & $-1$ & $+1$ & $-1$ & - & - \\
		\hline
		$B_2$ & $+1$ & $-1$ & $+1$ & $-1$ & $-1$ & $+1$ & - & - \\
		\hline
		$E_1$ & $+2$ & $+1$ & $-1$ & $-2$ & $0$ & $0$ & $(x,y)$ & $(xz,yz)$ \\
		\hline
		$E_2$ & $+2$ & $-1$ & $-1$ & $+2$ & $0$ & $0$ & - & $(x^2-y^2,xy)$ \\
		\hline
	\end{tabular}
	\caption{Character table of the $C_{6v}$ group. Where applicable, we also list the linear and quadratic functions that transform according to the different irreps.\label{tab:char_table_c6v}}
\end{table}

\begin{figure}[t]
	\centering
	\includegraphics[width=0.95\textwidth]{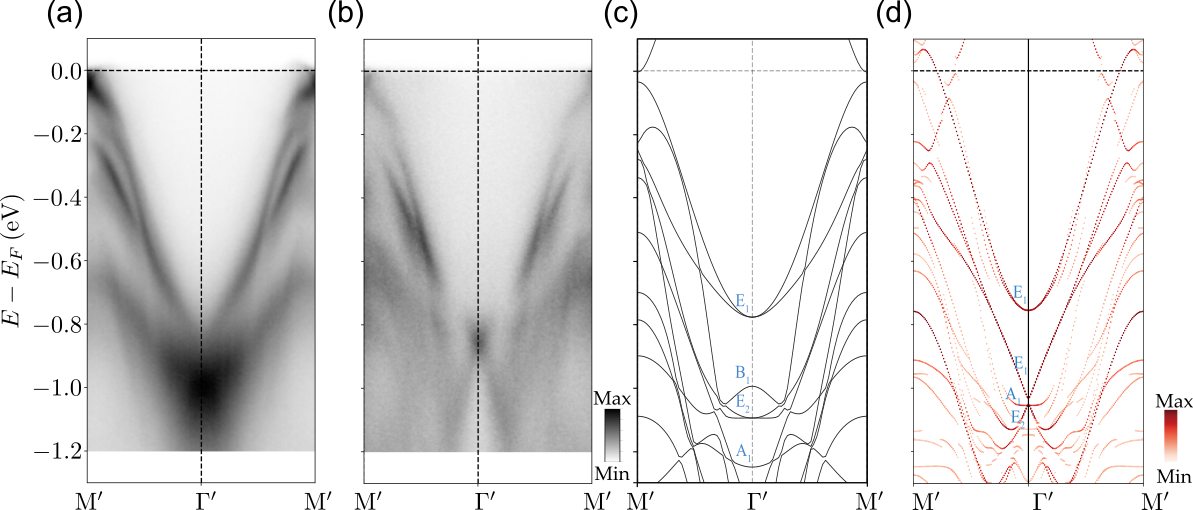}
	\caption{ARPES selection rules at normal incidence. The experimental ARPES spectra measured along the high-symmetry line $\mathrm{M}'-\Gamma'-\mathrm{M}'$ in the $k_z/(2\pi) \approx 0.1$ at $T= \SI{7}{\kelvin}$ are shown in (a) and (b) for L.H. and L.V. light, respectively. The \textit{ab-initio} bulk band structure of ScV$_6$Sn$_6${} at $k_z/(2\pi)=0$ is plotted in (c), while (d) illustrates the \textit{ab-initio} dispersion of the surface states. The irreps at the $\Gamma'$ point (whose little group is $C_{6v}$) are indicated in blue. Figure adapted from Ref.~\cite{KOR23}.}
	\label{fig:arpes}
\end{figure}

For electrons scattering perpendicular to the crystal surface, the system has $C_{6v}$ symmetry. The spinless matrix elements split into two independent sets of components which transform among themselves under the symetries of the group
\begin{align}
	\left[\Delta \left( \hat{\vec{z}} \right) \right]_{\vec{k}_e;\vec{k}_e^{\parallel},n} &= \sum_{m} \left[B \left(g, \vec{k}^{\parallel}_e \right)\right]_{mn} 
	\left[\Delta \left( \hat{\vec{z}} \right) \right]_{\vec{k}_e;\vec{k}_e^{\parallel},m}, \\
	\left[\Delta \left( \hat{\vec{e}}^{\parallel}_i \right) \right]_{\vec{k}_e;\vec{k}_e^{\parallel},n} &= \sum_{m,j} \left[\mathcal{R}^{\parallel}(g)\right]_{ji} \left[B \left(g, \vec{k}^{\parallel}_e \right)\right]_{mn} 
	\left[\Delta \left( \hat{\vec{e}}^{\parallel}_j \right) \right]_{\vec{k}_e;\vec{k}_e^{\parallel},m}, 
\end{align}
where $\hat{\vec{e}}^{\parallel}_i \in \left\lbrace \hat{\vec{x}}, \hat{\vec{y}} \right\rbrace$ and $\mathcal{R}^{\parallel}(g)$ is the $E_1$ representation of $C_{6v}$. From \cref{tab:char_table_c6v} and using \cref{app:eqn:simple_selection_crit}, we conclude that:
\begin{itemize}
	\item $\left[\Delta \left( \hat{\vec{z}} \right) \right]_{\vec{k}_e;\vec{k}_e^{\parallel},n}$ can only be nonzero if $B \left(g, \vec{k}^{\parallel}_e \right)$ is the $A_1$ (trivial) irrep.
	\item $\left[\Delta \left( \hat{\vec{e}}^{\parallel}_i \right) \right]_{\vec{k}_e;\vec{k}_e^{\parallel},n}$ can only be nonzero if $B \left(g, \vec{k}^{\parallel}_e \right)$ is the $E_1$ irrep.
\end{itemize}
As discussed around \cref{app:eqn:light_vector_pot}, the polarization vector of the incoming photon is parallel to $\vec{A}_0$. As depicted in \cref{fig:experiment_arpes}, the polarization vector of linearly vertical (L.V.) polarized light $\hat{\vec{e}}_{\mathrm{L.V.}}$ has only components perpendicular to the sample normal $\hat{\vec{n}}$, whereas the polarization vector of linearly horizontal (L.H.) polarized light $\hat{\vec{e}}_{\mathrm{L.H.}}$ components both perpendicular and parallel to the sample normal\footnote{The reader is reminded that at normal photoemission ($\theta=0$), the photon momentum is \emph{not} necessarily perpendicular to the sample surface, meaning that the polarization vector of L.H. light can have components both parallel and perpendicular to the sample normal.}. The ARPES signal under L.V. light will only have contribution from the $\left[\Delta \left( \hat{\vec{e}}^{\parallel}_i \right) \right]_{\vec{k}_e;\vec{k}_e^{\parallel},n}$ matrix elements, whereas the the one obtained under L.H. light will have contributions from both the $\left[\Delta \left( \hat{\vec{e}}^{\parallel}_i \right) \right]_{\vec{k}_e;\vec{k}_e^{\parallel},n}$ and the $\left[\Delta \left( \hat{\vec{z}} \right) \right]_{\vec{k}_e;\vec{k}_e^{\parallel},n}$ matrix elements. As a result, we find that:
\begin{itemize}
	\item Under L.V. light at $\theta = 0$, the electronic states will give rise to an ARPES signal only if they transform as the $E_1$ irrep.
	\item Under L.H. light at $\theta = 0$, the electronic states will give rise to an ARPES signal only if they transform as one of the $A_1$ or $E_1$ irreps.
\end{itemize}

An example of the selection rules at normal photoemission is shown in \cref{fig:arpes}. A more detailed comparison between the ARPES spectra and the band structures obtained from \textit{ab-initio} methods is given in Ref.~\cite{KOR23}. We consider the ARPES spectra along the high-symmetry line $\mathrm{M}' - \Gamma' -\mathrm{M}'$ in the $k_z/(2\pi) \approx 0.1$ plane~\cite{KOR23}. Inspecting the \textit{ab-initio} bulk and surface band structures from \cref{fig:arpes}(c) and (d) reveals bulk and surface states transforming according to the $E_1$ irrep at $E-E_F \approx -\SI{0.7}{\electronvolt}$, which are visible in the ARPES spectra measured under both L.H. and L.V. light from \cref{fig:arpes}(a) and (b), respectively. As shown in \cref{fig:arpes}(d), ScV$_6$Sn$_6${} features two surface states at $E-E_F \approx -\SI{1.0}{\electronvolt}$, transforming as the $E_1$ and $A_1$ irreps. These surface states lead to a strong signal under L.H. light in \cref{fig:arpes}(a). In \cref{fig:arpes}(b), the surface state transforming according to the $A_1$ irrep becomes invisible under L.V. light (according to the selection rules stated above), leading to a significantly diminished ARPES signal at $E-E_F \approx -\SI{1.0}{\electronvolt}$. Both the bulk and the surface band structures from \cref{fig:arpes} (c) and (d) also feature states transforming according to irreps other than $A_1$ and $E_1$, but these are not visible in the ARPES spectrum, as a result of the selection rules.

\subsubsection{$\theta \neq 0$ emission with $\vec{k}_e^{\parallel}$ along the $\Gamma' - \mathrm{M}'$ or $\Gamma' - \mathrm{K}'$ directions}
Whenever $\vec{k}_e^{\parallel}$ is along the $\Gamma' - \mathrm{M}'$ or $\Gamma' - \mathrm{K}'$ directions of the crystal, the system features mirror symmetry with respect to reflections perpendicular to the scattering plane $m_{\mathrm{scat}}$. The spinless matrix elements corresponding to the two possible photon polarization directions transform as 
\begin{align}
	\left[\Delta \left( \hat{\vec{e}}_{\mathrm{L.V.}} \right) \right]_{\vec{k}_e;\vec{k}_e^{\parallel},n} &= - \left[B \left(m_{\mathrm{scat}}, \vec{k}^{\parallel}_e \right)\right]_{nn} 
	\left[\Delta \left( \hat{\vec{e}}_{\mathrm{L.V.}} \right) \right]_{\vec{k}_e;\vec{k}_e^{\parallel},n}, \\
	\left[\Delta \left( \hat{\vec{e}}_{\mathrm{L.H.}} \right) \right]_{\vec{k}_e;\vec{k}_e^{\parallel},n} &= \left[B \left(m_{\mathrm{scat}}, \vec{k}^{\parallel}_e \right)\right]_{nn} 
	\left[\Delta \left( \hat{\vec{e}}_{\mathrm{L.H.}} \right) \right]_{\vec{k}_e;\vec{k}_e^{\parallel},n}, 
\end{align}
where we have used the fact that with just mirror symmetry, the irreps are one-dimensional. We then conclude that: 
\begin{itemize}
	\item Under L.V. light and $\vec{k}_e^{\parallel}$ along the $\Gamma' - \mathrm{M}'$ or $\Gamma' - \mathrm{K}'$ directions, the electronic states will give rise to ARPES signal only if they are antisymmetric with respect to mirror reflections perpendicular to the scattering plane (\cref{fig:arpes} (b)).
	\item Under L.H. light and $\vec{k}_e^{\parallel}$ along the $\Gamma' - \mathrm{M}'$ or $\Gamma' - \mathrm{K}'$ directions, the electronic states will give rise to ARPES signal only if they are symmetric with respect to mirror reflections perpendicular to the scattering plane (\cref{fig:arpes} (a)).
\end{itemize}


\begin{thebibliography}{118}%
\makeatletter
\providecommand \@ifxundefined [1]{%
 \@ifx{#1\undefined}
}%
\providecommand \@ifnum [1]{%
 \ifnum #1\expandafter \@firstoftwo
 \else \expandafter \@secondoftwo
 \fi
}%
\providecommand \@ifx [1]{%
 \ifx #1\expandafter \@firstoftwo
 \else \expandafter \@secondoftwo
 \fi
}%
\providecommand \natexlab [1]{#1}%
\providecommand \enquote  [1]{``#1''}%
\providecommand \bibnamefont  [1]{#1}%
\providecommand \bibfnamefont [1]{#1}%
\providecommand \citenamefont [1]{#1}%
\providecommand \href@noop [0]{\@secondoftwo}%
\providecommand \href [0]{\begingroup \@sanitize@url \@href}%
\providecommand \@href[1]{\@@startlink{#1}\@@href}%
\providecommand \@@href[1]{\endgroup#1\@@endlink}%
\providecommand \@sanitize@url [0]{\catcode `\\12\catcode `\$12\catcode
  `\&12\catcode `\#12\catcode `\^12\catcode `\_12\catcode `\%12\relax}%
\providecommand \@@startlink[1]{}%
\providecommand \@@endlink[0]{}%
\providecommand \url  [0]{\begingroup\@sanitize@url \@url }%
\providecommand \@url [1]{\endgroup\@href {#1}{\urlprefix }}%
\providecommand \urlprefix  [0]{URL }%
\providecommand \Eprint [0]{\href }%
\providecommand \doibase [0]{https://doi.org/}%
\providecommand \selectlanguage [0]{\@gobble}%
\providecommand \bibinfo  [0]{\@secondoftwo}%
\providecommand \bibfield  [0]{\@secondoftwo}%
\providecommand \translation [1]{[#1]}%
\providecommand \BibitemOpen [0]{}%
\providecommand \bibitemStop [0]{}%
\providecommand \bibitemNoStop [0]{.\EOS\space}%
\providecommand \EOS [0]{\spacefactor3000\relax}%
\providecommand \BibitemShut  [1]{\csname bibitem#1\endcsname}%
\let\auto@bib@innerbib\@empty
\bibitem [{\citenamefont {Jiang}\ \emph {et~al.}(2023)\citenamefont {Jiang},
  \citenamefont {Hu}, \citenamefont {Calugaru}, \citenamefont {Xu},\ and\
  \citenamefont {Bernevig}}]{JIA23}%
  \BibitemOpen
  \bibfield  {author} {\bibinfo {author} {\bibfnamefont {Y.}~\bibnamefont
  {Jiang}}, \bibinfo {author} {\bibfnamefont {H.}~\bibnamefont {Hu}}, \bibinfo
  {author} {\bibfnamefont {D.}~\bibnamefont {Calugaru}}, \bibinfo {author}
  {\bibfnamefont {Y.}~\bibnamefont {Xu}},\ and\ \bibinfo {author}
  {\bibfnamefont {B.~A.}\ \bibnamefont {Bernevig}},\ }\href@noop {} {\bibfield
  {journal} {\bibinfo  {journal} {To be published}\ } (\bibinfo {year}
  {2023})}\BibitemShut {NoStop}%
\bibitem [{\citenamefont {Yu}\ \emph {et~al.}(2023)\citenamefont {Yu},
  \citenamefont {Ciccarino}, \citenamefont {Bianco}, \citenamefont {Errea},
  \citenamefont {Narang},\ and\ \citenamefont {Bernevig}}]{YU23}%
  \BibitemOpen
  \bibfield  {author} {\bibinfo {author} {\bibfnamefont {J.}~\bibnamefont
  {Yu}}, \bibinfo {author} {\bibfnamefont {C.~J.}\ \bibnamefont {Ciccarino}},
  \bibinfo {author} {\bibfnamefont {R.}~\bibnamefont {Bianco}}, \bibinfo
  {author} {\bibfnamefont {I.}~\bibnamefont {Errea}}, \bibinfo {author}
  {\bibfnamefont {P.}~\bibnamefont {Narang}},\ and\ \bibinfo {author}
  {\bibfnamefont {B.~A.}\ \bibnamefont {Bernevig}},\ }\href
  {https://doi.org/10.48550/arXiv.2305.02340} {\bibinfo {title} {Nontrivial
  {{Quantum Geometry}} and the {{Strength}} of {{Electron-Phonon Coupling}}}}
  (\bibinfo {year} {2023}),\ \Eprint {https://arxiv.org/abs/2305.02340}
  {arxiv:2305.02340 [cond-mat]} \BibitemShut {NoStop}%
\bibitem [{\citenamefont {Korshunov}\ \emph {et~al.}(2023)\citenamefont
  {Korshunov}, \citenamefont {Hu}, \citenamefont {Subires}, \citenamefont
  {Jiang}, \citenamefont {C{\u a}lug{\u a}ru}, \citenamefont {Feng},
  \citenamefont {Rajapitamahuni}, \citenamefont {Yi}, \citenamefont
  {Roychowdhury}, \citenamefont {Vergniory}, \citenamefont {Strempfer},
  \citenamefont {Shekhar}, \citenamefont {Vescovo}, \citenamefont {Chernyshov},
  \citenamefont {Said}, \citenamefont {Bosak}, \citenamefont {Felser},
  \citenamefont {Bernevig},\ and\ \citenamefont {{Blanco-Canosa}}}]{KOR23}%
  \BibitemOpen
  \bibfield  {author} {\bibinfo {author} {\bibfnamefont {A.}~\bibnamefont
  {Korshunov}}, \bibinfo {author} {\bibfnamefont {H.}~\bibnamefont {Hu}},
  \bibinfo {author} {\bibfnamefont {D.}~\bibnamefont {Subires}}, \bibinfo
  {author} {\bibfnamefont {Y.}~\bibnamefont {Jiang}}, \bibinfo {author}
  {\bibfnamefont {D.}~\bibnamefont {C{\u a}lug{\u a}ru}}, \bibinfo {author}
  {\bibfnamefont {X.}~\bibnamefont {Feng}}, \bibinfo {author} {\bibfnamefont
  {A.}~\bibnamefont {Rajapitamahuni}}, \bibinfo {author} {\bibfnamefont
  {C.}~\bibnamefont {Yi}}, \bibinfo {author} {\bibfnamefont {S.}~\bibnamefont
  {Roychowdhury}}, \bibinfo {author} {\bibfnamefont {M.~G.}\ \bibnamefont
  {Vergniory}}, \bibinfo {author} {\bibfnamefont {J.}~\bibnamefont
  {Strempfer}}, \bibinfo {author} {\bibfnamefont {C.}~\bibnamefont {Shekhar}},
  \bibinfo {author} {\bibfnamefont {E.}~\bibnamefont {Vescovo}}, \bibinfo
  {author} {\bibfnamefont {D.}~\bibnamefont {Chernyshov}}, \bibinfo {author}
  {\bibfnamefont {A.~H.}\ \bibnamefont {Said}}, \bibinfo {author}
  {\bibfnamefont {A.}~\bibnamefont {Bosak}}, \bibinfo {author} {\bibfnamefont
  {C.}~\bibnamefont {Felser}}, \bibinfo {author} {\bibfnamefont {B.~A.}\
  \bibnamefont {Bernevig}},\ and\ \bibinfo {author} {\bibfnamefont
  {S.}~\bibnamefont {{Blanco-Canosa}}},\ }\href
  {https://doi.org/10.48550/arXiv.2304.09173} {\bibinfo {title} {Softening of a
  flat phonon mode in the kagome {{ScV}}$_6${{Sn}}$_6$}} (\bibinfo {year}
  {2023}),\ \Eprint {https://arxiv.org/abs/2304.09173} {arxiv:2304.09173
  [cond-mat]} \BibitemShut {NoStop}%
\bibitem [{\citenamefont {Ortiz}\ \emph {et~al.}(2019)\citenamefont {Ortiz},
  \citenamefont {Gomes}, \citenamefont {Morey}, \citenamefont {Winiarski},
  \citenamefont {Bordelon}, \citenamefont {Mangum}, \citenamefont {Oswald},
  \citenamefont {{Rodriguez-Rivera}}, \citenamefont {Neilson}, \citenamefont
  {Wilson}, \citenamefont {Ertekin}, \citenamefont {McQueen},\ and\
  \citenamefont {Toberer}}]{ORT19}%
  \BibitemOpen
  \bibfield  {author} {\bibinfo {author} {\bibfnamefont {B.~R.}\ \bibnamefont
  {Ortiz}}, \bibinfo {author} {\bibfnamefont {L.~C.}\ \bibnamefont {Gomes}},
  \bibinfo {author} {\bibfnamefont {J.~R.}\ \bibnamefont {Morey}}, \bibinfo
  {author} {\bibfnamefont {M.}~\bibnamefont {Winiarski}}, \bibinfo {author}
  {\bibfnamefont {M.}~\bibnamefont {Bordelon}}, \bibinfo {author}
  {\bibfnamefont {J.~S.}\ \bibnamefont {Mangum}}, \bibinfo {author}
  {\bibfnamefont {I.~W.~H.}\ \bibnamefont {Oswald}}, \bibinfo {author}
  {\bibfnamefont {J.~A.}\ \bibnamefont {{Rodriguez-Rivera}}}, \bibinfo {author}
  {\bibfnamefont {J.~R.}\ \bibnamefont {Neilson}}, \bibinfo {author}
  {\bibfnamefont {S.~D.}\ \bibnamefont {Wilson}}, \bibinfo {author}
  {\bibfnamefont {E.}~\bibnamefont {Ertekin}}, \bibinfo {author} {\bibfnamefont
  {T.~M.}\ \bibnamefont {McQueen}},\ and\ \bibinfo {author} {\bibfnamefont
  {E.~S.}\ \bibnamefont {Toberer}},\ }\href
  {https://doi.org/10.1103/PhysRevMaterials.3.094407} {\bibfield  {journal}
  {\bibinfo  {journal} {Phys. Rev. Mater.}\ }\textbf {\bibinfo {volume} {3}},\
  \bibinfo {pages} {094407} (\bibinfo {year} {2019})}\BibitemShut {NoStop}%
\bibitem [{\citenamefont {Cho}\ \emph {et~al.}(2021)\citenamefont {Cho},
  \citenamefont {Ma}, \citenamefont {Xia}, \citenamefont {Yang}, \citenamefont
  {Liu}, \citenamefont {Huang}, \citenamefont {Jiang}, \citenamefont {Lu},
  \citenamefont {Liu}, \citenamefont {Liu}, \citenamefont {Li}, \citenamefont
  {Wang}, \citenamefont {Liu}, \citenamefont {Jia}, \citenamefont {Guo},
  \citenamefont {Liu},\ and\ \citenamefont {Shen}}]{CHO21b}%
  \BibitemOpen
  \bibfield  {author} {\bibinfo {author} {\bibfnamefont {S.}~\bibnamefont
  {Cho}}, \bibinfo {author} {\bibfnamefont {H.}~\bibnamefont {Ma}}, \bibinfo
  {author} {\bibfnamefont {W.}~\bibnamefont {Xia}}, \bibinfo {author}
  {\bibfnamefont {Y.}~\bibnamefont {Yang}}, \bibinfo {author} {\bibfnamefont
  {Z.}~\bibnamefont {Liu}}, \bibinfo {author} {\bibfnamefont {Z.}~\bibnamefont
  {Huang}}, \bibinfo {author} {\bibfnamefont {Z.}~\bibnamefont {Jiang}},
  \bibinfo {author} {\bibfnamefont {X.}~\bibnamefont {Lu}}, \bibinfo {author}
  {\bibfnamefont {J.}~\bibnamefont {Liu}}, \bibinfo {author} {\bibfnamefont
  {Z.}~\bibnamefont {Liu}}, \bibinfo {author} {\bibfnamefont {J.}~\bibnamefont
  {Li}}, \bibinfo {author} {\bibfnamefont {J.}~\bibnamefont {Wang}}, \bibinfo
  {author} {\bibfnamefont {Y.}~\bibnamefont {Liu}}, \bibinfo {author}
  {\bibfnamefont {J.}~\bibnamefont {Jia}}, \bibinfo {author} {\bibfnamefont
  {Y.}~\bibnamefont {Guo}}, \bibinfo {author} {\bibfnamefont {J.}~\bibnamefont
  {Liu}},\ and\ \bibinfo {author} {\bibfnamefont {D.}~\bibnamefont {Shen}},\
  }\href {https://doi.org/10.1103/PhysRevLett.127.236401} {\bibfield  {journal}
  {\bibinfo  {journal} {Phys. Rev. Lett.}\ }\textbf {\bibinfo {volume} {127}},\
  \bibinfo {pages} {236401} (\bibinfo {year} {2021})}\BibitemShut {NoStop}%
\bibitem [{\citenamefont {Denner}\ \emph {et~al.}(2021)\citenamefont {Denner},
  \citenamefont {Thomale},\ and\ \citenamefont {Neupert}}]{DEN21}%
  \BibitemOpen
  \bibfield  {author} {\bibinfo {author} {\bibfnamefont {M.~M.}\ \bibnamefont
  {Denner}}, \bibinfo {author} {\bibfnamefont {R.}~\bibnamefont {Thomale}},\
  and\ \bibinfo {author} {\bibfnamefont {T.}~\bibnamefont {Neupert}},\ }\href
  {https://doi.org/10.1103/PhysRevLett.127.217601} {\bibfield  {journal}
  {\bibinfo  {journal} {Phys. Rev. Lett.}\ }\textbf {\bibinfo {volume} {127}},\
  \bibinfo {pages} {217601} (\bibinfo {year} {2021})}\BibitemShut {NoStop}%
\bibitem [{\citenamefont {Heritage}\ \emph {et~al.}(2020)\citenamefont
  {Heritage}, \citenamefont {Bryant}, \citenamefont {Fenner}, \citenamefont
  {Wills}, \citenamefont {Aeppli},\ and\ \citenamefont {Soh}}]{HER20a}%
  \BibitemOpen
  \bibfield  {author} {\bibinfo {author} {\bibfnamefont {K.}~\bibnamefont
  {Heritage}}, \bibinfo {author} {\bibfnamefont {B.}~\bibnamefont {Bryant}},
  \bibinfo {author} {\bibfnamefont {L.~A.}\ \bibnamefont {Fenner}}, \bibinfo
  {author} {\bibfnamefont {A.~S.}\ \bibnamefont {Wills}}, \bibinfo {author}
  {\bibfnamefont {G.}~\bibnamefont {Aeppli}},\ and\ \bibinfo {author}
  {\bibfnamefont {Y.-A.}\ \bibnamefont {Soh}},\ }\href
  {https://doi.org/10.1002/adfm.201909163} {\bibfield  {journal} {\bibinfo
  {journal} {Adv. Funct. Mater.}\ }\textbf {\bibinfo {volume} {30}},\ \bibinfo
  {pages} {1909163} (\bibinfo {year} {2020})}\BibitemShut {NoStop}%
\bibitem [{\citenamefont {Ishikawa}\ \emph {et~al.}(2021)\citenamefont
  {Ishikawa}, \citenamefont {Yajima}, \citenamefont {Kawamura}, \citenamefont
  {Mitamura},\ and\ \citenamefont {Kindo}}]{ISH21}%
  \BibitemOpen
  \bibfield  {author} {\bibinfo {author} {\bibfnamefont {H.}~\bibnamefont
  {Ishikawa}}, \bibinfo {author} {\bibfnamefont {T.}~\bibnamefont {Yajima}},
  \bibinfo {author} {\bibfnamefont {M.}~\bibnamefont {Kawamura}}, \bibinfo
  {author} {\bibfnamefont {H.}~\bibnamefont {Mitamura}},\ and\ \bibinfo
  {author} {\bibfnamefont {K.}~\bibnamefont {Kindo}},\ }\href
  {https://doi.org/10.7566/JPSJ.90.124704} {\bibfield  {journal} {\bibinfo
  {journal} {J. Phys. Soc. Jpn.}\ }\textbf {\bibinfo {volume} {90}},\ \bibinfo
  {pages} {124704} (\bibinfo {year} {2021})}\BibitemShut {NoStop}%
\bibitem [{\citenamefont {Ortiz}\ \emph {et~al.}(2020)\citenamefont {Ortiz},
  \citenamefont {Teicher}, \citenamefont {Hu}, \citenamefont {Zuo},
  \citenamefont {Sarte}, \citenamefont {Schueller}, \citenamefont {Abeykoon},
  \citenamefont {Krogstad}, \citenamefont {Rosenkranz}, \citenamefont {Osborn},
  \citenamefont {Seshadri}, \citenamefont {Balents}, \citenamefont {He},\ and\
  \citenamefont {Wilson}}]{ORT20}%
  \BibitemOpen
  \bibfield  {author} {\bibinfo {author} {\bibfnamefont {B.~R.}\ \bibnamefont
  {Ortiz}}, \bibinfo {author} {\bibfnamefont {S.~M.~L.}\ \bibnamefont
  {Teicher}}, \bibinfo {author} {\bibfnamefont {Y.}~\bibnamefont {Hu}},
  \bibinfo {author} {\bibfnamefont {J.~L.}\ \bibnamefont {Zuo}}, \bibinfo
  {author} {\bibfnamefont {P.~M.}\ \bibnamefont {Sarte}}, \bibinfo {author}
  {\bibfnamefont {E.~C.}\ \bibnamefont {Schueller}}, \bibinfo {author}
  {\bibfnamefont {A.~M.~M.}\ \bibnamefont {Abeykoon}}, \bibinfo {author}
  {\bibfnamefont {M.~J.}\ \bibnamefont {Krogstad}}, \bibinfo {author}
  {\bibfnamefont {S.}~\bibnamefont {Rosenkranz}}, \bibinfo {author}
  {\bibfnamefont {R.}~\bibnamefont {Osborn}}, \bibinfo {author} {\bibfnamefont
  {R.}~\bibnamefont {Seshadri}}, \bibinfo {author} {\bibfnamefont
  {L.}~\bibnamefont {Balents}}, \bibinfo {author} {\bibfnamefont
  {J.}~\bibnamefont {He}},\ and\ \bibinfo {author} {\bibfnamefont {S.~D.}\
  \bibnamefont {Wilson}},\ }\href
  {https://doi.org/10.1103/PhysRevLett.125.247002} {\bibfield  {journal}
  {\bibinfo  {journal} {Phys. Rev. Lett.}\ }\textbf {\bibinfo {volume} {125}},\
  \bibinfo {pages} {247002} (\bibinfo {year} {2020})}\BibitemShut {NoStop}%
\bibitem [{\citenamefont {Kang}\ \emph {et~al.}(2021)\citenamefont {Kang},
  \citenamefont {Fang}, \citenamefont {Kim}, \citenamefont {Ortiz},
  \citenamefont {Ryu}, \citenamefont {Kim}, \citenamefont {Yoo}, \citenamefont
  {Sangiovanni}, \citenamefont {Di~Sante}, \citenamefont {Park}, \citenamefont
  {Jozwiak}, \citenamefont {Bostwick}, \citenamefont {Rotenberg}, \citenamefont
  {Kaxiras}, \citenamefont {Wilson}, \citenamefont {Park},\ and\ \citenamefont
  {Comin}}]{KAN21a}%
  \BibitemOpen
  \bibfield  {author} {\bibinfo {author} {\bibfnamefont {M.}~\bibnamefont
  {Kang}}, \bibinfo {author} {\bibfnamefont {S.}~\bibnamefont {Fang}}, \bibinfo
  {author} {\bibfnamefont {J.-K.}\ \bibnamefont {Kim}}, \bibinfo {author}
  {\bibfnamefont {B.~R.}\ \bibnamefont {Ortiz}}, \bibinfo {author}
  {\bibfnamefont {S.~H.}\ \bibnamefont {Ryu}}, \bibinfo {author} {\bibfnamefont
  {J.}~\bibnamefont {Kim}}, \bibinfo {author} {\bibfnamefont {J.}~\bibnamefont
  {Yoo}}, \bibinfo {author} {\bibfnamefont {G.}~\bibnamefont {Sangiovanni}},
  \bibinfo {author} {\bibfnamefont {D.}~\bibnamefont {Di~Sante}}, \bibinfo
  {author} {\bibfnamefont {B.-G.}\ \bibnamefont {Park}}, \bibinfo {author}
  {\bibfnamefont {C.}~\bibnamefont {Jozwiak}}, \bibinfo {author} {\bibfnamefont
  {A.}~\bibnamefont {Bostwick}}, \bibinfo {author} {\bibfnamefont
  {E.}~\bibnamefont {Rotenberg}}, \bibinfo {author} {\bibfnamefont
  {E.}~\bibnamefont {Kaxiras}}, \bibinfo {author} {\bibfnamefont {S.~D.}\
  \bibnamefont {Wilson}}, \bibinfo {author} {\bibfnamefont {J.-H.}\
  \bibnamefont {Park}},\ and\ \bibinfo {author} {\bibfnamefont
  {R.}~\bibnamefont {Comin}},\ }\href
  {https://doi.org/10.48550/arXiv.2105.01689} {\bibinfo {title} {Twofold van
  {{Hove}} singularity and origin of charge order in topological kagome
  superconductor {{CsV}}$_3${{Sb}}$_5$}} (\bibinfo {year} {2021}),\ \Eprint
  {https://arxiv.org/abs/2105.01689} {arxiv:2105.01689 [cond-mat]} \BibitemShut
  {NoStop}%
\bibitem [{\citenamefont {Li}\ \emph {et~al.}(2021{\natexlab{a}})\citenamefont
  {Li}, \citenamefont {Wang}, \citenamefont {Wang}, \citenamefont {Yuan},
  \citenamefont {Song}, \citenamefont {Lou}, \citenamefont {Liu}, \citenamefont
  {Huang}, \citenamefont {Liu}, \citenamefont {Lei}, \citenamefont {Yin},\ and\
  \citenamefont {Wang}}]{LI21b}%
  \BibitemOpen
  \bibfield  {author} {\bibinfo {author} {\bibfnamefont {M.}~\bibnamefont
  {Li}}, \bibinfo {author} {\bibfnamefont {Q.}~\bibnamefont {Wang}}, \bibinfo
  {author} {\bibfnamefont {G.}~\bibnamefont {Wang}}, \bibinfo {author}
  {\bibfnamefont {Z.}~\bibnamefont {Yuan}}, \bibinfo {author} {\bibfnamefont
  {W.}~\bibnamefont {Song}}, \bibinfo {author} {\bibfnamefont {R.}~\bibnamefont
  {Lou}}, \bibinfo {author} {\bibfnamefont {Z.}~\bibnamefont {Liu}}, \bibinfo
  {author} {\bibfnamefont {Y.}~\bibnamefont {Huang}}, \bibinfo {author}
  {\bibfnamefont {Z.}~\bibnamefont {Liu}}, \bibinfo {author} {\bibfnamefont
  {H.}~\bibnamefont {Lei}}, \bibinfo {author} {\bibfnamefont {Z.}~\bibnamefont
  {Yin}},\ and\ \bibinfo {author} {\bibfnamefont {S.}~\bibnamefont {Wang}},\
  }\href {https://doi.org/10.1038/s41467-021-23536-8} {\bibfield  {journal}
  {\bibinfo  {journal} {Nat. Commun.}\ }\textbf {\bibinfo {volume} {12}},\
  \bibinfo {pages} {3129} (\bibinfo {year} {2021}{\natexlab{a}})}\BibitemShut
  {NoStop}%
\bibitem [{\citenamefont {Li}\ \emph {et~al.}(2021{\natexlab{b}})\citenamefont
  {Li}, \citenamefont {{Reig-i-Plessis}}, \citenamefont {Liu}, \citenamefont
  {Wu}, \citenamefont {Wang}, \citenamefont {Hallas}, \citenamefont {Stone},
  \citenamefont {Broholm},\ and\ \citenamefont {Aronson}}]{LI21c}%
  \BibitemOpen
  \bibfield  {author} {\bibinfo {author} {\bibfnamefont {X.~Y.}\ \bibnamefont
  {Li}}, \bibinfo {author} {\bibfnamefont {D.}~\bibnamefont
  {{Reig-i-Plessis}}}, \bibinfo {author} {\bibfnamefont {P.-F.}\ \bibnamefont
  {Liu}}, \bibinfo {author} {\bibfnamefont {S.}~\bibnamefont {Wu}}, \bibinfo
  {author} {\bibfnamefont {B.-T.}\ \bibnamefont {Wang}}, \bibinfo {author}
  {\bibfnamefont {A.~M.}\ \bibnamefont {Hallas}}, \bibinfo {author}
  {\bibfnamefont {M.~B.}\ \bibnamefont {Stone}}, \bibinfo {author}
  {\bibfnamefont {C.}~\bibnamefont {Broholm}},\ and\ \bibinfo {author}
  {\bibfnamefont {M.~C.}\ \bibnamefont {Aronson}},\ }\href
  {https://doi.org/10.1103/PhysRevB.104.134305} {\bibfield  {journal} {\bibinfo
   {journal} {Phys. Rev. B}\ }\textbf {\bibinfo {volume} {104}},\ \bibinfo
  {pages} {134305} (\bibinfo {year} {2021}{\natexlab{b}})}\BibitemShut
  {NoStop}%
\bibitem [{\citenamefont {Lin}\ and\ \citenamefont
  {Nandkishore}(2021)}]{LIN21}%
  \BibitemOpen
  \bibfield  {author} {\bibinfo {author} {\bibfnamefont {Y.-P.}\ \bibnamefont
  {Lin}}\ and\ \bibinfo {author} {\bibfnamefont {R.~M.}\ \bibnamefont
  {Nandkishore}},\ }\href {https://doi.org/10.1103/PhysRevB.104.045122}
  {\bibfield  {journal} {\bibinfo  {journal} {Phys. Rev. B}\ }\textbf {\bibinfo
  {volume} {104}},\ \bibinfo {pages} {045122} (\bibinfo {year}
  {2021})}\BibitemShut {NoStop}%
\bibitem [{\citenamefont {Liu}\ \emph {et~al.}(2023)\citenamefont {Liu},
  \citenamefont {Lyu}, \citenamefont {Liu}, \citenamefont {Zhang},
  \citenamefont {Yang}, \citenamefont {Du}, \citenamefont {Wang}, \citenamefont
  {Wei},\ and\ \citenamefont {Liu}}]{LIU23}%
  \BibitemOpen
  \bibfield  {author} {\bibinfo {author} {\bibfnamefont {Y.}~\bibnamefont
  {Liu}}, \bibinfo {author} {\bibfnamefont {M.}~\bibnamefont {Lyu}}, \bibinfo
  {author} {\bibfnamefont {J.}~\bibnamefont {Liu}}, \bibinfo {author}
  {\bibfnamefont {S.}~\bibnamefont {Zhang}}, \bibinfo {author} {\bibfnamefont
  {J.}~\bibnamefont {Yang}}, \bibinfo {author} {\bibfnamefont {Z.}~\bibnamefont
  {Du}}, \bibinfo {author} {\bibfnamefont {B.}~\bibnamefont {Wang}}, \bibinfo
  {author} {\bibfnamefont {H.}~\bibnamefont {Wei}},\ and\ \bibinfo {author}
  {\bibfnamefont {E.}~\bibnamefont {Liu}},\ }\href
  {https://doi.org/10.1088/0256-307X/40/4/047102} {\bibfield  {journal}
  {\bibinfo  {journal} {Chinese Phys. Lett.}\ }\textbf {\bibinfo {volume}
  {40}},\ \bibinfo {pages} {047102} (\bibinfo {year} {2023})}\BibitemShut
  {NoStop}%
\bibitem [{\citenamefont {Neupert}\ \emph {et~al.}(2022)\citenamefont
  {Neupert}, \citenamefont {Denner}, \citenamefont {Yin}, \citenamefont
  {Thomale},\ and\ \citenamefont {Hasan}}]{NEU22}%
  \BibitemOpen
  \bibfield  {author} {\bibinfo {author} {\bibfnamefont {T.}~\bibnamefont
  {Neupert}}, \bibinfo {author} {\bibfnamefont {M.~M.}\ \bibnamefont {Denner}},
  \bibinfo {author} {\bibfnamefont {J.-X.}\ \bibnamefont {Yin}}, \bibinfo
  {author} {\bibfnamefont {R.}~\bibnamefont {Thomale}},\ and\ \bibinfo {author}
  {\bibfnamefont {M.~Z.}\ \bibnamefont {Hasan}},\ }\href
  {https://doi.org/10.1038/s41567-021-01404-y} {\bibfield  {journal} {\bibinfo
  {journal} {Nat. Phys.}\ }\textbf {\bibinfo {volume} {18}},\ \bibinfo {pages}
  {137} (\bibinfo {year} {2022})}\BibitemShut {NoStop}%
\bibitem [{\citenamefont {Ortiz}\ \emph
  {et~al.}(2021{\natexlab{a}})\citenamefont {Ortiz}, \citenamefont {Teicher},
  \citenamefont {Kautzsch}, \citenamefont {Sarte}, \citenamefont {Ratcliff},
  \citenamefont {Harter}, \citenamefont {Ruff}, \citenamefont {Seshadri},\ and\
  \citenamefont {Wilson}}]{ORT21}%
  \BibitemOpen
  \bibfield  {author} {\bibinfo {author} {\bibfnamefont {B.~R.}\ \bibnamefont
  {Ortiz}}, \bibinfo {author} {\bibfnamefont {S.~M.~L.}\ \bibnamefont
  {Teicher}}, \bibinfo {author} {\bibfnamefont {L.}~\bibnamefont {Kautzsch}},
  \bibinfo {author} {\bibfnamefont {P.~M.}\ \bibnamefont {Sarte}}, \bibinfo
  {author} {\bibfnamefont {N.}~\bibnamefont {Ratcliff}}, \bibinfo {author}
  {\bibfnamefont {J.}~\bibnamefont {Harter}}, \bibinfo {author} {\bibfnamefont
  {J.~P.~C.}\ \bibnamefont {Ruff}}, \bibinfo {author} {\bibfnamefont
  {R.}~\bibnamefont {Seshadri}},\ and\ \bibinfo {author} {\bibfnamefont
  {S.~D.}\ \bibnamefont {Wilson}},\ }\href
  {https://doi.org/10.1103/PhysRevX.11.041030} {\bibfield  {journal} {\bibinfo
  {journal} {Phys. Rev. X}\ }\textbf {\bibinfo {volume} {11}},\ \bibinfo
  {pages} {041030} (\bibinfo {year} {2021}{\natexlab{a}})}\BibitemShut
  {NoStop}%
\bibitem [{\citenamefont {Pal}\ \emph {et~al.}(2022)\citenamefont {Pal},
  \citenamefont {Hazra}, \citenamefont {G{\"o}bel}, \citenamefont {Jeon},
  \citenamefont {Pandeya}, \citenamefont {Chakraborty}, \citenamefont {Busch},
  \citenamefont {Srivastava}, \citenamefont {Deniz}, \citenamefont {Taylor},
  \citenamefont {Meyerheim}, \citenamefont {Mertig}, \citenamefont {Yang},\
  and\ \citenamefont {Parkin}}]{PAL22}%
  \BibitemOpen
  \bibfield  {author} {\bibinfo {author} {\bibfnamefont {B.}~\bibnamefont
  {Pal}}, \bibinfo {author} {\bibfnamefont {B.~K.}\ \bibnamefont {Hazra}},
  \bibinfo {author} {\bibfnamefont {B.}~\bibnamefont {G{\"o}bel}}, \bibinfo
  {author} {\bibfnamefont {J.-C.}\ \bibnamefont {Jeon}}, \bibinfo {author}
  {\bibfnamefont {A.~K.}\ \bibnamefont {Pandeya}}, \bibinfo {author}
  {\bibfnamefont {A.}~\bibnamefont {Chakraborty}}, \bibinfo {author}
  {\bibfnamefont {O.}~\bibnamefont {Busch}}, \bibinfo {author} {\bibfnamefont
  {A.~K.}\ \bibnamefont {Srivastava}}, \bibinfo {author} {\bibfnamefont
  {H.}~\bibnamefont {Deniz}}, \bibinfo {author} {\bibfnamefont {J.~M.}\
  \bibnamefont {Taylor}}, \bibinfo {author} {\bibfnamefont {H.}~\bibnamefont
  {Meyerheim}}, \bibinfo {author} {\bibfnamefont {I.}~\bibnamefont {Mertig}},
  \bibinfo {author} {\bibfnamefont {S.-H.}\ \bibnamefont {Yang}},\ and\
  \bibinfo {author} {\bibfnamefont {S.~S.~P.}\ \bibnamefont {Parkin}},\ }\href
  {https://doi.org/10.1126/sciadv.abo5930} {\bibfield  {journal} {\bibinfo
  {journal} {Sci. Adv.}\ }\textbf {\bibinfo {volume} {8}},\ \bibinfo {pages}
  {eabo5930} (\bibinfo {year} {2022})}\BibitemShut {NoStop}%
\bibitem [{\citenamefont {Zhao}\ \emph {et~al.}(2021)\citenamefont {Zhao},
  \citenamefont {Li}, \citenamefont {Ortiz}, \citenamefont {Teicher},
  \citenamefont {Park}, \citenamefont {Ye}, \citenamefont {Wang}, \citenamefont
  {Balents}, \citenamefont {Wilson},\ and\ \citenamefont {Zeljkovic}}]{ZHA21b}%
  \BibitemOpen
  \bibfield  {author} {\bibinfo {author} {\bibfnamefont {H.}~\bibnamefont
  {Zhao}}, \bibinfo {author} {\bibfnamefont {H.}~\bibnamefont {Li}}, \bibinfo
  {author} {\bibfnamefont {B.~R.}\ \bibnamefont {Ortiz}}, \bibinfo {author}
  {\bibfnamefont {S.~M.~L.}\ \bibnamefont {Teicher}}, \bibinfo {author}
  {\bibfnamefont {T.}~\bibnamefont {Park}}, \bibinfo {author} {\bibfnamefont
  {M.}~\bibnamefont {Ye}}, \bibinfo {author} {\bibfnamefont {Z.}~\bibnamefont
  {Wang}}, \bibinfo {author} {\bibfnamefont {L.}~\bibnamefont {Balents}},
  \bibinfo {author} {\bibfnamefont {S.~D.}\ \bibnamefont {Wilson}},\ and\
  \bibinfo {author} {\bibfnamefont {I.}~\bibnamefont {Zeljkovic}},\ }\href
  {https://doi.org/10.1038/s41586-021-03946-w} {\bibfield  {journal} {\bibinfo
  {journal} {Nature}\ }\textbf {\bibinfo {volume} {599}},\ \bibinfo {pages}
  {216} (\bibinfo {year} {2021})}\BibitemShut {NoStop}%
\bibitem [{\citenamefont {Zhang}\ \emph
  {et~al.}(2022{\natexlab{a}})\citenamefont {Zhang}, \citenamefont {Ni},
  \citenamefont {Datta}, \citenamefont {Wang}, \citenamefont {Yao},\ and\
  \citenamefont {Cao}}]{ZHA22b}%
  \BibitemOpen
  \bibfield  {author} {\bibinfo {author} {\bibfnamefont {Y.-F.}\ \bibnamefont
  {Zhang}}, \bibinfo {author} {\bibfnamefont {X.-S.}\ \bibnamefont {Ni}},
  \bibinfo {author} {\bibfnamefont {T.}~\bibnamefont {Datta}}, \bibinfo
  {author} {\bibfnamefont {M.}~\bibnamefont {Wang}}, \bibinfo {author}
  {\bibfnamefont {D.-X.}\ \bibnamefont {Yao}},\ and\ \bibinfo {author}
  {\bibfnamefont {K.}~\bibnamefont {Cao}},\ }\href
  {https://doi.org/10.1103/PhysRevB.106.184422} {\bibfield  {journal} {\bibinfo
   {journal} {Phys. Rev. B}\ }\textbf {\bibinfo {volume} {106}},\ \bibinfo
  {pages} {184422} (\bibinfo {year} {2022}{\natexlab{a}})}\BibitemShut
  {NoStop}%
\bibitem [{\citenamefont {Chen}\ \emph {et~al.}(2022)\citenamefont {Chen},
  \citenamefont {Chen}, \citenamefont {Schnelle}, \citenamefont {Felser},\ and\
  \citenamefont {Gaulin}}]{CHE22a}%
  \BibitemOpen
  \bibfield  {author} {\bibinfo {author} {\bibfnamefont {Q.}~\bibnamefont
  {Chen}}, \bibinfo {author} {\bibfnamefont {D.}~\bibnamefont {Chen}}, \bibinfo
  {author} {\bibfnamefont {W.}~\bibnamefont {Schnelle}}, \bibinfo {author}
  {\bibfnamefont {C.}~\bibnamefont {Felser}},\ and\ \bibinfo {author}
  {\bibfnamefont {B.~D.}\ \bibnamefont {Gaulin}},\ }\href
  {https://doi.org/10.1103/PhysRevLett.129.056401} {\bibfield  {journal}
  {\bibinfo  {journal} {Phys. Rev. Lett.}\ }\textbf {\bibinfo {volume} {129}},\
  \bibinfo {pages} {056401} (\bibinfo {year} {2022})}\BibitemShut {NoStop}%
\bibitem [{\citenamefont {Diego}\ \emph {et~al.}(2021)\citenamefont {Diego},
  \citenamefont {Said}, \citenamefont {Mahatha}, \citenamefont {Bianco},
  \citenamefont {Monacelli}, \citenamefont {Calandra}, \citenamefont {Mauri},
  \citenamefont {Rossnagel}, \citenamefont {Errea},\ and\ \citenamefont
  {{Blanco-Canosa}}}]{DIE21}%
  \BibitemOpen
  \bibfield  {author} {\bibinfo {author} {\bibfnamefont {J.}~\bibnamefont
  {Diego}}, \bibinfo {author} {\bibfnamefont {A.~H.}\ \bibnamefont {Said}},
  \bibinfo {author} {\bibfnamefont {S.~K.}\ \bibnamefont {Mahatha}}, \bibinfo
  {author} {\bibfnamefont {R.}~\bibnamefont {Bianco}}, \bibinfo {author}
  {\bibfnamefont {L.}~\bibnamefont {Monacelli}}, \bibinfo {author}
  {\bibfnamefont {M.}~\bibnamefont {Calandra}}, \bibinfo {author}
  {\bibfnamefont {F.}~\bibnamefont {Mauri}}, \bibinfo {author} {\bibfnamefont
  {K.}~\bibnamefont {Rossnagel}}, \bibinfo {author} {\bibfnamefont
  {I.}~\bibnamefont {Errea}},\ and\ \bibinfo {author} {\bibfnamefont
  {S.}~\bibnamefont {{Blanco-Canosa}}},\ }\href
  {https://doi.org/10.1038/s41467-020-20829-2} {\bibfield  {journal} {\bibinfo
  {journal} {Nat. Commun.}\ }\textbf {\bibinfo {volume} {12}},\ \bibinfo
  {pages} {598} (\bibinfo {year} {2021})}\BibitemShut {NoStop}%
\bibitem [{\citenamefont {Ferrari}\ \emph {et~al.}(2022)\citenamefont
  {Ferrari}, \citenamefont {Becca},\ and\ \citenamefont {Valent{\'i}}}]{FER22}%
  \BibitemOpen
  \bibfield  {author} {\bibinfo {author} {\bibfnamefont {F.}~\bibnamefont
  {Ferrari}}, \bibinfo {author} {\bibfnamefont {F.}~\bibnamefont {Becca}},\
  and\ \bibinfo {author} {\bibfnamefont {R.}~\bibnamefont {Valent{\'i}}},\
  }\href {https://doi.org/10.1103/PhysRevB.106.L081107} {\bibfield  {journal}
  {\bibinfo  {journal} {Phys. Rev. B}\ }\textbf {\bibinfo {volume} {106}},\
  \bibinfo {pages} {L081107} (\bibinfo {year} {2022})}\BibitemShut {NoStop}%
\bibitem [{\citenamefont {Jiang}\ \emph {et~al.}(2021)\citenamefont {Jiang},
  \citenamefont {Yin}, \citenamefont {Denner}, \citenamefont {Shumiya},
  \citenamefont {Ortiz}, \citenamefont {Xu}, \citenamefont {Guguchia},
  \citenamefont {He}, \citenamefont {Hossain}, \citenamefont {Liu},
  \citenamefont {Ruff}, \citenamefont {Kautzsch}, \citenamefont {Zhang},
  \citenamefont {Chang}, \citenamefont {Belopolski}, \citenamefont {Zhang},
  \citenamefont {Cochran}, \citenamefont {Multer}, \citenamefont {Litskevich},
  \citenamefont {Cheng}, \citenamefont {Yang}, \citenamefont {Wang},
  \citenamefont {Thomale}, \citenamefont {Neupert}, \citenamefont {Wilson},\
  and\ \citenamefont {Hasan}}]{JIA21}%
  \BibitemOpen
  \bibfield  {author} {\bibinfo {author} {\bibfnamefont {Y.-X.}\ \bibnamefont
  {Jiang}}, \bibinfo {author} {\bibfnamefont {J.-X.}\ \bibnamefont {Yin}},
  \bibinfo {author} {\bibfnamefont {M.~M.}\ \bibnamefont {Denner}}, \bibinfo
  {author} {\bibfnamefont {N.}~\bibnamefont {Shumiya}}, \bibinfo {author}
  {\bibfnamefont {B.~R.}\ \bibnamefont {Ortiz}}, \bibinfo {author}
  {\bibfnamefont {G.}~\bibnamefont {Xu}}, \bibinfo {author} {\bibfnamefont
  {Z.}~\bibnamefont {Guguchia}}, \bibinfo {author} {\bibfnamefont
  {J.}~\bibnamefont {He}}, \bibinfo {author} {\bibfnamefont {M.~S.}\
  \bibnamefont {Hossain}}, \bibinfo {author} {\bibfnamefont {X.}~\bibnamefont
  {Liu}}, \bibinfo {author} {\bibfnamefont {J.}~\bibnamefont {Ruff}}, \bibinfo
  {author} {\bibfnamefont {L.}~\bibnamefont {Kautzsch}}, \bibinfo {author}
  {\bibfnamefont {S.~S.}\ \bibnamefont {Zhang}}, \bibinfo {author}
  {\bibfnamefont {G.}~\bibnamefont {Chang}}, \bibinfo {author} {\bibfnamefont
  {I.}~\bibnamefont {Belopolski}}, \bibinfo {author} {\bibfnamefont
  {Q.}~\bibnamefont {Zhang}}, \bibinfo {author} {\bibfnamefont {T.~A.}\
  \bibnamefont {Cochran}}, \bibinfo {author} {\bibfnamefont {D.}~\bibnamefont
  {Multer}}, \bibinfo {author} {\bibfnamefont {M.}~\bibnamefont {Litskevich}},
  \bibinfo {author} {\bibfnamefont {Z.-J.}\ \bibnamefont {Cheng}}, \bibinfo
  {author} {\bibfnamefont {X.~P.}\ \bibnamefont {Yang}}, \bibinfo {author}
  {\bibfnamefont {Z.}~\bibnamefont {Wang}}, \bibinfo {author} {\bibfnamefont
  {R.}~\bibnamefont {Thomale}}, \bibinfo {author} {\bibfnamefont
  {T.}~\bibnamefont {Neupert}}, \bibinfo {author} {\bibfnamefont {S.~D.}\
  \bibnamefont {Wilson}},\ and\ \bibinfo {author} {\bibfnamefont {M.~Z.}\
  \bibnamefont {Hasan}},\ }\href {https://doi.org/10.1038/s41563-021-01034-y}
  {\bibfield  {journal} {\bibinfo  {journal} {Nat. Mater.}\ }\textbf {\bibinfo
  {volume} {20}},\ \bibinfo {pages} {1353} (\bibinfo {year}
  {2021})}\BibitemShut {NoStop}%
\bibitem [{\citenamefont {Kenney}\ \emph {et~al.}(2021)\citenamefont {Kenney},
  \citenamefont {Ortiz}, \citenamefont {Wang}, \citenamefont {Wilson},\ and\
  \citenamefont {Graf}}]{KEN21a}%
  \BibitemOpen
  \bibfield  {author} {\bibinfo {author} {\bibfnamefont {E.~M.}\ \bibnamefont
  {Kenney}}, \bibinfo {author} {\bibfnamefont {B.~R.}\ \bibnamefont {Ortiz}},
  \bibinfo {author} {\bibfnamefont {C.}~\bibnamefont {Wang}}, \bibinfo {author}
  {\bibfnamefont {S.~D.}\ \bibnamefont {Wilson}},\ and\ \bibinfo {author}
  {\bibfnamefont {M.~J.}\ \bibnamefont {Graf}},\ }\href
  {https://doi.org/10.1088/1361-648X/abe8f9} {\bibfield  {journal} {\bibinfo
  {journal} {J. Phys.: Condens. Matter}\ }\textbf {\bibinfo {volume} {33}},\
  \bibinfo {pages} {235801} (\bibinfo {year} {2021})}\BibitemShut {NoStop}%
\bibitem [{\citenamefont {Li}\ \emph {et~al.}(2021{\natexlab{c}})\citenamefont
  {Li}, \citenamefont {Zhang}, \citenamefont {Yilmaz}, \citenamefont {Pai},
  \citenamefont {Marvinney}, \citenamefont {Said}, \citenamefont {Yin},
  \citenamefont {Gong}, \citenamefont {Tu}, \citenamefont {Vescovo},
  \citenamefont {Nelson}, \citenamefont {Moore}, \citenamefont {Murakami},
  \citenamefont {Lei}, \citenamefont {Lee}, \citenamefont {Lawrie},\ and\
  \citenamefont {Miao}}]{LI21a}%
  \BibitemOpen
  \bibfield  {author} {\bibinfo {author} {\bibfnamefont {H.}~\bibnamefont
  {Li}}, \bibinfo {author} {\bibfnamefont {T.~T.}\ \bibnamefont {Zhang}},
  \bibinfo {author} {\bibfnamefont {T.}~\bibnamefont {Yilmaz}}, \bibinfo
  {author} {\bibfnamefont {Y.~Y.}\ \bibnamefont {Pai}}, \bibinfo {author}
  {\bibfnamefont {C.~E.}\ \bibnamefont {Marvinney}}, \bibinfo {author}
  {\bibfnamefont {A.}~\bibnamefont {Said}}, \bibinfo {author} {\bibfnamefont
  {Q.~W.}\ \bibnamefont {Yin}}, \bibinfo {author} {\bibfnamefont {C.~S.}\
  \bibnamefont {Gong}}, \bibinfo {author} {\bibfnamefont {Z.~J.}\ \bibnamefont
  {Tu}}, \bibinfo {author} {\bibfnamefont {E.}~\bibnamefont {Vescovo}},
  \bibinfo {author} {\bibfnamefont {C.~S.}\ \bibnamefont {Nelson}}, \bibinfo
  {author} {\bibfnamefont {R.~G.}\ \bibnamefont {Moore}}, \bibinfo {author}
  {\bibfnamefont {S.}~\bibnamefont {Murakami}}, \bibinfo {author}
  {\bibfnamefont {H.~C.}\ \bibnamefont {Lei}}, \bibinfo {author} {\bibfnamefont
  {H.~N.}\ \bibnamefont {Lee}}, \bibinfo {author} {\bibfnamefont {B.~J.}\
  \bibnamefont {Lawrie}},\ and\ \bibinfo {author} {\bibfnamefont
  {H.}~\bibnamefont {Miao}},\ }\href
  {https://doi.org/10.1103/PhysRevX.11.031050} {\bibfield  {journal} {\bibinfo
  {journal} {Phys. Rev. X}\ }\textbf {\bibinfo {volume} {11}},\ \bibinfo
  {pages} {031050} (\bibinfo {year} {2021}{\natexlab{c}})}\BibitemShut
  {NoStop}%
\bibitem [{\citenamefont {Li}\ \emph {et~al.}(2023)\citenamefont {Li},
  \citenamefont {Liu}, \citenamefont {Kim},\ and\ \citenamefont {Kee}}]{LI23}%
  \BibitemOpen
  \bibfield  {author} {\bibinfo {author} {\bibfnamefont {H.}~\bibnamefont
  {Li}}, \bibinfo {author} {\bibfnamefont {X.}~\bibnamefont {Liu}}, \bibinfo
  {author} {\bibfnamefont {Y.~B.}\ \bibnamefont {Kim}},\ and\ \bibinfo {author}
  {\bibfnamefont {H.-Y.}\ \bibnamefont {Kee}},\ }\href
  {https://doi.org/10.48550/arXiv.2302.10178} {\bibinfo {title} {Origin of
  $\pi$-shifted three-dimensional charge density waves in kagome metal
  {{AV}}$_3${{Sb}}$_5$}} (\bibinfo {year} {2023}),\ \Eprint
  {https://arxiv.org/abs/2302.10178} {arxiv:2302.10178 [cond-mat]} \BibitemShut
  {NoStop}%
\bibitem [{\citenamefont {Liang}\ \emph {et~al.}(2021)\citenamefont {Liang},
  \citenamefont {Hou}, \citenamefont {Zhang}, \citenamefont {Ma}, \citenamefont
  {Wu}, \citenamefont {Zhang}, \citenamefont {Yu}, \citenamefont {Ying},
  \citenamefont {Jiang}, \citenamefont {Shan}, \citenamefont {Wang},\ and\
  \citenamefont {Chen}}]{LIA21a}%
  \BibitemOpen
  \bibfield  {author} {\bibinfo {author} {\bibfnamefont {Z.}~\bibnamefont
  {Liang}}, \bibinfo {author} {\bibfnamefont {X.}~\bibnamefont {Hou}}, \bibinfo
  {author} {\bibfnamefont {F.}~\bibnamefont {Zhang}}, \bibinfo {author}
  {\bibfnamefont {W.}~\bibnamefont {Ma}}, \bibinfo {author} {\bibfnamefont
  {P.}~\bibnamefont {Wu}}, \bibinfo {author} {\bibfnamefont {Z.}~\bibnamefont
  {Zhang}}, \bibinfo {author} {\bibfnamefont {F.}~\bibnamefont {Yu}}, \bibinfo
  {author} {\bibfnamefont {J.-J.}\ \bibnamefont {Ying}}, \bibinfo {author}
  {\bibfnamefont {K.}~\bibnamefont {Jiang}}, \bibinfo {author} {\bibfnamefont
  {L.}~\bibnamefont {Shan}}, \bibinfo {author} {\bibfnamefont {Z.}~\bibnamefont
  {Wang}},\ and\ \bibinfo {author} {\bibfnamefont {X.-H.}\ \bibnamefont
  {Chen}},\ }\href {https://doi.org/10.1103/PhysRevX.11.031026} {\bibfield
  {journal} {\bibinfo  {journal} {Phys. Rev. X}\ }\textbf {\bibinfo {volume}
  {11}},\ \bibinfo {pages} {031026} (\bibinfo {year} {2021})}\BibitemShut
  {NoStop}%
\bibitem [{\citenamefont {Liu}\ \emph {et~al.}(2021{\natexlab{a}})\citenamefont
  {Liu}, \citenamefont {Zhao}, \citenamefont {Yin}, \citenamefont {Gong},
  \citenamefont {Tu}, \citenamefont {Li}, \citenamefont {Song}, \citenamefont
  {Liu}, \citenamefont {Shen}, \citenamefont {Huang}, \citenamefont {Liu},
  \citenamefont {Lei},\ and\ \citenamefont {Wang}}]{LIU21b}%
  \BibitemOpen
  \bibfield  {author} {\bibinfo {author} {\bibfnamefont {Z.}~\bibnamefont
  {Liu}}, \bibinfo {author} {\bibfnamefont {N.}~\bibnamefont {Zhao}}, \bibinfo
  {author} {\bibfnamefont {Q.}~\bibnamefont {Yin}}, \bibinfo {author}
  {\bibfnamefont {C.}~\bibnamefont {Gong}}, \bibinfo {author} {\bibfnamefont
  {Z.}~\bibnamefont {Tu}}, \bibinfo {author} {\bibfnamefont {M.}~\bibnamefont
  {Li}}, \bibinfo {author} {\bibfnamefont {W.}~\bibnamefont {Song}}, \bibinfo
  {author} {\bibfnamefont {Z.}~\bibnamefont {Liu}}, \bibinfo {author}
  {\bibfnamefont {D.}~\bibnamefont {Shen}}, \bibinfo {author} {\bibfnamefont
  {Y.}~\bibnamefont {Huang}}, \bibinfo {author} {\bibfnamefont
  {K.}~\bibnamefont {Liu}}, \bibinfo {author} {\bibfnamefont {H.}~\bibnamefont
  {Lei}},\ and\ \bibinfo {author} {\bibfnamefont {S.}~\bibnamefont {Wang}},\
  }\href {https://doi.org/10.1103/PhysRevX.11.041010} {\bibfield  {journal}
  {\bibinfo  {journal} {Phys. Rev. X}\ }\textbf {\bibinfo {volume} {11}},\
  \bibinfo {pages} {041010} (\bibinfo {year} {2021}{\natexlab{a}})}\BibitemShut
  {NoStop}%
\bibitem [{\citenamefont {Luo}\ \emph {et~al.}(2022)\citenamefont {Luo},
  \citenamefont {Gao}, \citenamefont {Liu}, \citenamefont {Gu}, \citenamefont
  {Wu}, \citenamefont {Yi}, \citenamefont {Jia}, \citenamefont {Wu},
  \citenamefont {Luo}, \citenamefont {Xu}, \citenamefont {Zhao}, \citenamefont
  {Wang}, \citenamefont {Mao}, \citenamefont {Liu}, \citenamefont {Zhu},
  \citenamefont {Shi}, \citenamefont {Jiang}, \citenamefont {Hu}, \citenamefont
  {Xu},\ and\ \citenamefont {Zhou}}]{LUO22}%
  \BibitemOpen
  \bibfield  {author} {\bibinfo {author} {\bibfnamefont {H.}~\bibnamefont
  {Luo}}, \bibinfo {author} {\bibfnamefont {Q.}~\bibnamefont {Gao}}, \bibinfo
  {author} {\bibfnamefont {H.}~\bibnamefont {Liu}}, \bibinfo {author}
  {\bibfnamefont {Y.}~\bibnamefont {Gu}}, \bibinfo {author} {\bibfnamefont
  {D.}~\bibnamefont {Wu}}, \bibinfo {author} {\bibfnamefont {C.}~\bibnamefont
  {Yi}}, \bibinfo {author} {\bibfnamefont {J.}~\bibnamefont {Jia}}, \bibinfo
  {author} {\bibfnamefont {S.}~\bibnamefont {Wu}}, \bibinfo {author}
  {\bibfnamefont {X.}~\bibnamefont {Luo}}, \bibinfo {author} {\bibfnamefont
  {Y.}~\bibnamefont {Xu}}, \bibinfo {author} {\bibfnamefont {L.}~\bibnamefont
  {Zhao}}, \bibinfo {author} {\bibfnamefont {Q.}~\bibnamefont {Wang}}, \bibinfo
  {author} {\bibfnamefont {H.}~\bibnamefont {Mao}}, \bibinfo {author}
  {\bibfnamefont {G.}~\bibnamefont {Liu}}, \bibinfo {author} {\bibfnamefont
  {Z.}~\bibnamefont {Zhu}}, \bibinfo {author} {\bibfnamefont {Y.}~\bibnamefont
  {Shi}}, \bibinfo {author} {\bibfnamefont {K.}~\bibnamefont {Jiang}}, \bibinfo
  {author} {\bibfnamefont {J.}~\bibnamefont {Hu}}, \bibinfo {author}
  {\bibfnamefont {Z.}~\bibnamefont {Xu}},\ and\ \bibinfo {author}
  {\bibfnamefont {X.~J.}\ \bibnamefont {Zhou}},\ }\href
  {https://doi.org/10.1038/s41467-021-27946-6} {\bibfield  {journal} {\bibinfo
  {journal} {Nat. Commun.}\ }\textbf {\bibinfo {volume} {13}},\ \bibinfo
  {pages} {273} (\bibinfo {year} {2022})}\BibitemShut {NoStop}%
\bibitem [{\citenamefont {Mielke}\ \emph {et~al.}(2022)\citenamefont {Mielke},
  \citenamefont {Das}, \citenamefont {Yin}, \citenamefont {Liu}, \citenamefont
  {Gupta}, \citenamefont {Jiang}, \citenamefont {Medarde}, \citenamefont {Wu},
  \citenamefont {Lei}, \citenamefont {Chang}, \citenamefont {Dai},
  \citenamefont {Si}, \citenamefont {Miao}, \citenamefont {Thomale},
  \citenamefont {Neupert}, \citenamefont {Shi}, \citenamefont {Khasanov},
  \citenamefont {Hasan}, \citenamefont {Luetkens},\ and\ \citenamefont
  {Guguchia}}]{MIE22}%
  \BibitemOpen
  \bibfield  {author} {\bibinfo {author} {\bibfnamefont {C.}~\bibnamefont
  {Mielke}}, \bibinfo {author} {\bibfnamefont {D.}~\bibnamefont {Das}},
  \bibinfo {author} {\bibfnamefont {J.-X.}\ \bibnamefont {Yin}}, \bibinfo
  {author} {\bibfnamefont {H.}~\bibnamefont {Liu}}, \bibinfo {author}
  {\bibfnamefont {R.}~\bibnamefont {Gupta}}, \bibinfo {author} {\bibfnamefont
  {Y.-X.}\ \bibnamefont {Jiang}}, \bibinfo {author} {\bibfnamefont
  {M.}~\bibnamefont {Medarde}}, \bibinfo {author} {\bibfnamefont
  {X.}~\bibnamefont {Wu}}, \bibinfo {author} {\bibfnamefont {H.~C.}\
  \bibnamefont {Lei}}, \bibinfo {author} {\bibfnamefont {J.}~\bibnamefont
  {Chang}}, \bibinfo {author} {\bibfnamefont {P.}~\bibnamefont {Dai}}, \bibinfo
  {author} {\bibfnamefont {Q.}~\bibnamefont {Si}}, \bibinfo {author}
  {\bibfnamefont {H.}~\bibnamefont {Miao}}, \bibinfo {author} {\bibfnamefont
  {R.}~\bibnamefont {Thomale}}, \bibinfo {author} {\bibfnamefont
  {T.}~\bibnamefont {Neupert}}, \bibinfo {author} {\bibfnamefont
  {Y.}~\bibnamefont {Shi}}, \bibinfo {author} {\bibfnamefont {R.}~\bibnamefont
  {Khasanov}}, \bibinfo {author} {\bibfnamefont {M.~Z.}\ \bibnamefont {Hasan}},
  \bibinfo {author} {\bibfnamefont {H.}~\bibnamefont {Luetkens}},\ and\
  \bibinfo {author} {\bibfnamefont {Z.}~\bibnamefont {Guguchia}},\ }\href
  {https://doi.org/10.1038/s41586-021-04327-z} {\bibfield  {journal} {\bibinfo
  {journal} {Nature}\ }\textbf {\bibinfo {volume} {602}},\ \bibinfo {pages}
  {245} (\bibinfo {year} {2022})}\BibitemShut {NoStop}%
\bibitem [{\citenamefont {Ratcliff}\ \emph {et~al.}(2021)\citenamefont
  {Ratcliff}, \citenamefont {Hallett}, \citenamefont {Ortiz}, \citenamefont
  {Wilson},\ and\ \citenamefont {Harter}}]{RAT21}%
  \BibitemOpen
  \bibfield  {author} {\bibinfo {author} {\bibfnamefont {N.}~\bibnamefont
  {Ratcliff}}, \bibinfo {author} {\bibfnamefont {L.}~\bibnamefont {Hallett}},
  \bibinfo {author} {\bibfnamefont {B.~R.}\ \bibnamefont {Ortiz}}, \bibinfo
  {author} {\bibfnamefont {S.~D.}\ \bibnamefont {Wilson}},\ and\ \bibinfo
  {author} {\bibfnamefont {J.~W.}\ \bibnamefont {Harter}},\ }\href
  {https://doi.org/10.1103/PhysRevMaterials.5.L111801} {\bibfield  {journal}
  {\bibinfo  {journal} {Phys. Rev. Mater.}\ }\textbf {\bibinfo {volume} {5}},\
  \bibinfo {pages} {L111801} (\bibinfo {year} {2021})}\BibitemShut {NoStop}%
\bibitem [{\citenamefont {Shumiya}\ \emph {et~al.}(2021)\citenamefont
  {Shumiya}, \citenamefont {Hossain}, \citenamefont {Yin}, \citenamefont
  {Jiang}, \citenamefont {Ortiz}, \citenamefont {Liu}, \citenamefont {Shi},
  \citenamefont {Yin}, \citenamefont {Lei}, \citenamefont {Zhang},
  \citenamefont {Chang}, \citenamefont {Zhang}, \citenamefont {Cochran},
  \citenamefont {Multer}, \citenamefont {Litskevich}, \citenamefont {Cheng},
  \citenamefont {Yang}, \citenamefont {Guguchia}, \citenamefont {Wilson},\ and\
  \citenamefont {Hasan}}]{SHU21}%
  \BibitemOpen
  \bibfield  {author} {\bibinfo {author} {\bibfnamefont {N.}~\bibnamefont
  {Shumiya}}, \bibinfo {author} {\bibfnamefont {M.~S.}\ \bibnamefont
  {Hossain}}, \bibinfo {author} {\bibfnamefont {J.-X.}\ \bibnamefont {Yin}},
  \bibinfo {author} {\bibfnamefont {Y.-X.}\ \bibnamefont {Jiang}}, \bibinfo
  {author} {\bibfnamefont {B.~R.}\ \bibnamefont {Ortiz}}, \bibinfo {author}
  {\bibfnamefont {H.}~\bibnamefont {Liu}}, \bibinfo {author} {\bibfnamefont
  {Y.}~\bibnamefont {Shi}}, \bibinfo {author} {\bibfnamefont {Q.}~\bibnamefont
  {Yin}}, \bibinfo {author} {\bibfnamefont {H.}~\bibnamefont {Lei}}, \bibinfo
  {author} {\bibfnamefont {S.~S.}\ \bibnamefont {Zhang}}, \bibinfo {author}
  {\bibfnamefont {G.}~\bibnamefont {Chang}}, \bibinfo {author} {\bibfnamefont
  {Q.}~\bibnamefont {Zhang}}, \bibinfo {author} {\bibfnamefont {T.~A.}\
  \bibnamefont {Cochran}}, \bibinfo {author} {\bibfnamefont {D.}~\bibnamefont
  {Multer}}, \bibinfo {author} {\bibfnamefont {M.}~\bibnamefont {Litskevich}},
  \bibinfo {author} {\bibfnamefont {Z.-J.}\ \bibnamefont {Cheng}}, \bibinfo
  {author} {\bibfnamefont {X.~P.}\ \bibnamefont {Yang}}, \bibinfo {author}
  {\bibfnamefont {Z.}~\bibnamefont {Guguchia}}, \bibinfo {author}
  {\bibfnamefont {S.~D.}\ \bibnamefont {Wilson}},\ and\ \bibinfo {author}
  {\bibfnamefont {M.~Z.}\ \bibnamefont {Hasan}},\ }\href
  {https://doi.org/10.1103/PhysRevB.104.035131} {\bibfield  {journal} {\bibinfo
   {journal} {Phys. Rev. B}\ }\textbf {\bibinfo {volume} {104}},\ \bibinfo
  {pages} {035131} (\bibinfo {year} {2021})}\BibitemShut {NoStop}%
\bibitem [{\citenamefont {Song}\ \emph
  {et~al.}(2021{\natexlab{a}})\citenamefont {Song}, \citenamefont {Zheng},
  \citenamefont {Yu}, \citenamefont {Li}, \citenamefont {Nie}, \citenamefont
  {Shan}, \citenamefont {Zhao}, \citenamefont {Li}, \citenamefont {Kang},
  \citenamefont {Wu}, \citenamefont {Zhou}, \citenamefont {Sun}, \citenamefont
  {Liu}, \citenamefont {Luo}, \citenamefont {Wang}, \citenamefont {Ying},
  \citenamefont {Wan}, \citenamefont {Wu},\ and\ \citenamefont
  {Chen}}]{SON21a}%
  \BibitemOpen
  \bibfield  {author} {\bibinfo {author} {\bibfnamefont {D.~W.}\ \bibnamefont
  {Song}}, \bibinfo {author} {\bibfnamefont {L.~X.}\ \bibnamefont {Zheng}},
  \bibinfo {author} {\bibfnamefont {F.~H.}\ \bibnamefont {Yu}}, \bibinfo
  {author} {\bibfnamefont {J.}~\bibnamefont {Li}}, \bibinfo {author}
  {\bibfnamefont {L.~P.}\ \bibnamefont {Nie}}, \bibinfo {author} {\bibfnamefont
  {M.}~\bibnamefont {Shan}}, \bibinfo {author} {\bibfnamefont {D.}~\bibnamefont
  {Zhao}}, \bibinfo {author} {\bibfnamefont {S.~J.}\ \bibnamefont {Li}},
  \bibinfo {author} {\bibfnamefont {B.~L.}\ \bibnamefont {Kang}}, \bibinfo
  {author} {\bibfnamefont {Z.~M.}\ \bibnamefont {Wu}}, \bibinfo {author}
  {\bibfnamefont {Y.~B.}\ \bibnamefont {Zhou}}, \bibinfo {author}
  {\bibfnamefont {K.~L.}\ \bibnamefont {Sun}}, \bibinfo {author} {\bibfnamefont
  {K.}~\bibnamefont {Liu}}, \bibinfo {author} {\bibfnamefont {X.~G.}\
  \bibnamefont {Luo}}, \bibinfo {author} {\bibfnamefont {Z.~Y.}\ \bibnamefont
  {Wang}}, \bibinfo {author} {\bibfnamefont {J.~J.}\ \bibnamefont {Ying}},
  \bibinfo {author} {\bibfnamefont {X.~G.}\ \bibnamefont {Wan}}, \bibinfo
  {author} {\bibfnamefont {T.}~\bibnamefont {Wu}},\ and\ \bibinfo {author}
  {\bibfnamefont {X.~H.}\ \bibnamefont {Chen}},\ }\href
  {https://doi.org/10.48550/arXiv.2104.09173} {\bibinfo {title} {Orbital
  ordering and fluctuations in a kagome superconductor {{CsV}}$_3${{Sb}}$_5$}}
  (\bibinfo {year} {2021}{\natexlab{a}}),\ \Eprint
  {https://arxiv.org/abs/2104.09173} {arxiv:2104.09173 [cond-mat]} \BibitemShut
  {NoStop}%
\bibitem [{\citenamefont {Setty}\ \emph {et~al.}(2021)\citenamefont {Setty},
  \citenamefont {Hu}, \citenamefont {Chen},\ and\ \citenamefont {Si}}]{SET21}%
  \BibitemOpen
  \bibfield  {author} {\bibinfo {author} {\bibfnamefont {C.}~\bibnamefont
  {Setty}}, \bibinfo {author} {\bibfnamefont {H.}~\bibnamefont {Hu}}, \bibinfo
  {author} {\bibfnamefont {L.}~\bibnamefont {Chen}},\ and\ \bibinfo {author}
  {\bibfnamefont {Q.}~\bibnamefont {Si}},\ }\href
  {https://doi.org/10.48550/arXiv.2105.15204} {\bibinfo {title} {Electron
  correlations and {{T-breaking}} density wave order in a $\mathbb{Z}_2$ kagome
  metal}} (\bibinfo {year} {2021}),\ \Eprint {https://arxiv.org/abs/2105.15204}
  {arxiv:2105.15204 [cond-mat]} \BibitemShut {NoStop}%
\bibitem [{\citenamefont {Tan}\ \emph {et~al.}(2021)\citenamefont {Tan},
  \citenamefont {Liu}, \citenamefont {Wang},\ and\ \citenamefont
  {Yan}}]{TAN21}%
  \BibitemOpen
  \bibfield  {author} {\bibinfo {author} {\bibfnamefont {H.}~\bibnamefont
  {Tan}}, \bibinfo {author} {\bibfnamefont {Y.}~\bibnamefont {Liu}}, \bibinfo
  {author} {\bibfnamefont {Z.}~\bibnamefont {Wang}},\ and\ \bibinfo {author}
  {\bibfnamefont {B.}~\bibnamefont {Yan}},\ }\href
  {https://doi.org/10.1103/PhysRevLett.127.046401} {\bibfield  {journal}
  {\bibinfo  {journal} {Phys. Rev. Lett.}\ }\textbf {\bibinfo {volume} {127}},\
  \bibinfo {pages} {046401} (\bibinfo {year} {2021})}\BibitemShut {NoStop}%
\bibitem [{\citenamefont {Tsirlin}\ \emph {et~al.}(2022)\citenamefont
  {Tsirlin}, \citenamefont {Fertey}, \citenamefont {Ortiz}, \citenamefont
  {Klis}, \citenamefont {Merkl}, \citenamefont {Dressel}, \citenamefont
  {Wilson},\ and\ \citenamefont {Uykur}}]{TSI22}%
  \BibitemOpen
  \bibfield  {author} {\bibinfo {author} {\bibfnamefont {A.}~\bibnamefont
  {Tsirlin}}, \bibinfo {author} {\bibfnamefont {P.}~\bibnamefont {Fertey}},
  \bibinfo {author} {\bibfnamefont {B.~R.}\ \bibnamefont {Ortiz}}, \bibinfo
  {author} {\bibfnamefont {B.}~\bibnamefont {Klis}}, \bibinfo {author}
  {\bibfnamefont {V.}~\bibnamefont {Merkl}}, \bibinfo {author} {\bibfnamefont
  {M.}~\bibnamefont {Dressel}}, \bibinfo {author} {\bibfnamefont
  {S.}~\bibnamefont {Wilson}},\ and\ \bibinfo {author} {\bibfnamefont
  {E.}~\bibnamefont {Uykur}},\ }\href
  {https://doi.org/10.21468/SciPostPhys.12.2.049} {\bibfield  {journal}
  {\bibinfo  {journal} {SciPost Phys.}\ }\textbf {\bibinfo {volume} {12}},\
  \bibinfo {pages} {049} (\bibinfo {year} {2022})}\BibitemShut {NoStop}%
\bibitem [{\citenamefont {Tsvelik}\ and\ \citenamefont {Sarkar}(2023)}]{TSV23}%
  \BibitemOpen
  \bibfield  {author} {\bibinfo {author} {\bibfnamefont {A.~M.}\ \bibnamefont
  {Tsvelik}}\ and\ \bibinfo {author} {\bibfnamefont {S.}~\bibnamefont
  {Sarkar}},\ }\href {https://doi.org/10.48550/arXiv.2304.01122} {\bibinfo
  {title} {Charge-density wave fluctuation driven composite order in the
  layered {{Kagome Metals}}}} (\bibinfo {year} {2023}),\ \Eprint
  {https://arxiv.org/abs/2304.01122} {arxiv:2304.01122 [cond-mat]} \BibitemShut
  {NoStop}%
\bibitem [{\citenamefont {Uykur}\ \emph {et~al.}(2021)\citenamefont {Uykur},
  \citenamefont {Ortiz}, \citenamefont {Iakutkina}, \citenamefont {Wenzel},
  \citenamefont {Wilson}, \citenamefont {Dressel},\ and\ \citenamefont
  {Tsirlin}}]{UYK21}%
  \BibitemOpen
  \bibfield  {author} {\bibinfo {author} {\bibfnamefont {E.}~\bibnamefont
  {Uykur}}, \bibinfo {author} {\bibfnamefont {B.~R.}\ \bibnamefont {Ortiz}},
  \bibinfo {author} {\bibfnamefont {O.}~\bibnamefont {Iakutkina}}, \bibinfo
  {author} {\bibfnamefont {M.}~\bibnamefont {Wenzel}}, \bibinfo {author}
  {\bibfnamefont {S.~D.}\ \bibnamefont {Wilson}}, \bibinfo {author}
  {\bibfnamefont {M.}~\bibnamefont {Dressel}},\ and\ \bibinfo {author}
  {\bibfnamefont {A.~A.}\ \bibnamefont {Tsirlin}},\ }\href
  {https://doi.org/10.1103/PhysRevB.104.045130} {\bibfield  {journal} {\bibinfo
   {journal} {Phys. Rev. B}\ }\textbf {\bibinfo {volume} {104}},\ \bibinfo
  {pages} {045130} (\bibinfo {year} {2021})}\BibitemShut {NoStop}%
\bibitem [{\citenamefont {Uykur}\ \emph {et~al.}(2022)\citenamefont {Uykur},
  \citenamefont {Ortiz}, \citenamefont {Wilson}, \citenamefont {Dressel},\ and\
  \citenamefont {Tsirlin}}]{UYK22}%
  \BibitemOpen
  \bibfield  {author} {\bibinfo {author} {\bibfnamefont {E.}~\bibnamefont
  {Uykur}}, \bibinfo {author} {\bibfnamefont {B.~R.}\ \bibnamefont {Ortiz}},
  \bibinfo {author} {\bibfnamefont {S.~D.}\ \bibnamefont {Wilson}}, \bibinfo
  {author} {\bibfnamefont {M.}~\bibnamefont {Dressel}},\ and\ \bibinfo {author}
  {\bibfnamefont {A.~A.}\ \bibnamefont {Tsirlin}},\ }\href
  {https://doi.org/10.1038/s41535-021-00420-8} {\bibfield  {journal} {\bibinfo
  {journal} {npj Quantum Mater.}\ }\textbf {\bibinfo {volume} {7}},\ \bibinfo
  {pages} {1} (\bibinfo {year} {2022})}\BibitemShut {NoStop}%
\bibitem [{\citenamefont {Wang}\ \emph
  {et~al.}(2021{\natexlab{a}})\citenamefont {Wang}, \citenamefont {Ma},
  \citenamefont {Zhang}, \citenamefont {Yang}, \citenamefont {Zhao},
  \citenamefont {Ou}, \citenamefont {Zhu}, \citenamefont {Ni}, \citenamefont
  {Lu}, \citenamefont {Chen}, \citenamefont {Jiang}, \citenamefont {Yu},
  \citenamefont {Zhang}, \citenamefont {Dong}, \citenamefont {Hu},
  \citenamefont {Gao},\ and\ \citenamefont {Zhao}}]{WAN21b}%
  \BibitemOpen
  \bibfield  {author} {\bibinfo {author} {\bibfnamefont {Z.}~\bibnamefont
  {Wang}}, \bibinfo {author} {\bibfnamefont {S.}~\bibnamefont {Ma}}, \bibinfo
  {author} {\bibfnamefont {Y.}~\bibnamefont {Zhang}}, \bibinfo {author}
  {\bibfnamefont {H.}~\bibnamefont {Yang}}, \bibinfo {author} {\bibfnamefont
  {Z.}~\bibnamefont {Zhao}}, \bibinfo {author} {\bibfnamefont {Y.}~\bibnamefont
  {Ou}}, \bibinfo {author} {\bibfnamefont {Y.}~\bibnamefont {Zhu}}, \bibinfo
  {author} {\bibfnamefont {S.}~\bibnamefont {Ni}}, \bibinfo {author}
  {\bibfnamefont {Z.}~\bibnamefont {Lu}}, \bibinfo {author} {\bibfnamefont
  {H.}~\bibnamefont {Chen}}, \bibinfo {author} {\bibfnamefont {K.}~\bibnamefont
  {Jiang}}, \bibinfo {author} {\bibfnamefont {L.}~\bibnamefont {Yu}}, \bibinfo
  {author} {\bibfnamefont {Y.}~\bibnamefont {Zhang}}, \bibinfo {author}
  {\bibfnamefont {X.}~\bibnamefont {Dong}}, \bibinfo {author} {\bibfnamefont
  {J.}~\bibnamefont {Hu}}, \bibinfo {author} {\bibfnamefont {H.-J.}\
  \bibnamefont {Gao}},\ and\ \bibinfo {author} {\bibfnamefont {Z.}~\bibnamefont
  {Zhao}},\ }\href {https://doi.org/10.48550/arXiv.2104.05556} {\bibinfo
  {title} {Distinctive momentum dependent charge-density-wave gap observed in
  {{CsV}}$_3${{Sb}}$_5$ superconductor with topological {{Kagome}} lattice}}
  (\bibinfo {year} {2021}{\natexlab{a}}),\ \Eprint
  {https://arxiv.org/abs/2104.05556} {arxiv:2104.05556 [cond-mat]} \BibitemShut
  {NoStop}%
\bibitem [{\citenamefont {Wang}\ \emph
  {et~al.}(2021{\natexlab{b}})\citenamefont {Wang}, \citenamefont {Jiang},
  \citenamefont {Yin}, \citenamefont {Li}, \citenamefont {Wang}, \citenamefont
  {Huang}, \citenamefont {Shao}, \citenamefont {Liu}, \citenamefont {Zhu},
  \citenamefont {Shumiya}, \citenamefont {Hossain}, \citenamefont {Liu},
  \citenamefont {Shi}, \citenamefont {Duan}, \citenamefont {Li}, \citenamefont
  {Chang}, \citenamefont {Dai}, \citenamefont {Ye}, \citenamefont {Xu},
  \citenamefont {Wang}, \citenamefont {Zheng}, \citenamefont {Jia},
  \citenamefont {Hasan},\ and\ \citenamefont {Yao}}]{WAN21c}%
  \BibitemOpen
  \bibfield  {author} {\bibinfo {author} {\bibfnamefont {Z.}~\bibnamefont
  {Wang}}, \bibinfo {author} {\bibfnamefont {Y.-X.}\ \bibnamefont {Jiang}},
  \bibinfo {author} {\bibfnamefont {J.-X.}\ \bibnamefont {Yin}}, \bibinfo
  {author} {\bibfnamefont {Y.}~\bibnamefont {Li}}, \bibinfo {author}
  {\bibfnamefont {G.-Y.}\ \bibnamefont {Wang}}, \bibinfo {author}
  {\bibfnamefont {H.-L.}\ \bibnamefont {Huang}}, \bibinfo {author}
  {\bibfnamefont {S.}~\bibnamefont {Shao}}, \bibinfo {author} {\bibfnamefont
  {J.}~\bibnamefont {Liu}}, \bibinfo {author} {\bibfnamefont {P.}~\bibnamefont
  {Zhu}}, \bibinfo {author} {\bibfnamefont {N.}~\bibnamefont {Shumiya}},
  \bibinfo {author} {\bibfnamefont {M.~S.}\ \bibnamefont {Hossain}}, \bibinfo
  {author} {\bibfnamefont {H.}~\bibnamefont {Liu}}, \bibinfo {author}
  {\bibfnamefont {Y.}~\bibnamefont {Shi}}, \bibinfo {author} {\bibfnamefont
  {J.}~\bibnamefont {Duan}}, \bibinfo {author} {\bibfnamefont {X.}~\bibnamefont
  {Li}}, \bibinfo {author} {\bibfnamefont {G.}~\bibnamefont {Chang}}, \bibinfo
  {author} {\bibfnamefont {P.}~\bibnamefont {Dai}}, \bibinfo {author}
  {\bibfnamefont {Z.}~\bibnamefont {Ye}}, \bibinfo {author} {\bibfnamefont
  {G.}~\bibnamefont {Xu}}, \bibinfo {author} {\bibfnamefont {Y.}~\bibnamefont
  {Wang}}, \bibinfo {author} {\bibfnamefont {H.}~\bibnamefont {Zheng}},
  \bibinfo {author} {\bibfnamefont {J.}~\bibnamefont {Jia}}, \bibinfo {author}
  {\bibfnamefont {M.~Z.}\ \bibnamefont {Hasan}},\ and\ \bibinfo {author}
  {\bibfnamefont {Y.}~\bibnamefont {Yao}},\ }\href
  {https://doi.org/10.1103/PhysRevB.104.075148} {\bibfield  {journal} {\bibinfo
   {journal} {Phys. Rev. B}\ }\textbf {\bibinfo {volume} {104}},\ \bibinfo
  {pages} {075148} (\bibinfo {year} {2021}{\natexlab{b}})}\BibitemShut
  {NoStop}%
\bibitem [{\citenamefont {Wang}\ \emph
  {et~al.}(2021{\natexlab{c}})\citenamefont {Wang}, \citenamefont {Wu},
  \citenamefont {Yin}, \citenamefont {Gong}, \citenamefont {Tu}, \citenamefont
  {Lin}, \citenamefont {Liu}, \citenamefont {Shi}, \citenamefont {Zhang},
  \citenamefont {Wu}, \citenamefont {Lei}, \citenamefont {Dong},\ and\
  \citenamefont {Wang}}]{WAN21d}%
  \BibitemOpen
  \bibfield  {author} {\bibinfo {author} {\bibfnamefont {Z.~X.}\ \bibnamefont
  {Wang}}, \bibinfo {author} {\bibfnamefont {Q.}~\bibnamefont {Wu}}, \bibinfo
  {author} {\bibfnamefont {Q.~W.}\ \bibnamefont {Yin}}, \bibinfo {author}
  {\bibfnamefont {C.~S.}\ \bibnamefont {Gong}}, \bibinfo {author}
  {\bibfnamefont {Z.~J.}\ \bibnamefont {Tu}}, \bibinfo {author} {\bibfnamefont
  {T.}~\bibnamefont {Lin}}, \bibinfo {author} {\bibfnamefont {Q.~M.}\
  \bibnamefont {Liu}}, \bibinfo {author} {\bibfnamefont {L.~Y.}\ \bibnamefont
  {Shi}}, \bibinfo {author} {\bibfnamefont {S.~J.}\ \bibnamefont {Zhang}},
  \bibinfo {author} {\bibfnamefont {D.}~\bibnamefont {Wu}}, \bibinfo {author}
  {\bibfnamefont {H.~C.}\ \bibnamefont {Lei}}, \bibinfo {author} {\bibfnamefont
  {T.}~\bibnamefont {Dong}},\ and\ \bibinfo {author} {\bibfnamefont {N.~L.}\
  \bibnamefont {Wang}},\ }\href {https://doi.org/10.1103/PhysRevB.104.165110}
  {\bibfield  {journal} {\bibinfo  {journal} {Phys. Rev. B}\ }\textbf {\bibinfo
  {volume} {104}},\ \bibinfo {pages} {165110} (\bibinfo {year}
  {2021}{\natexlab{c}})}\BibitemShut {NoStop}%
\bibitem [{\citenamefont {Wang}\ \emph
  {et~al.}(2021{\natexlab{d}})\citenamefont {Wang}, \citenamefont {Kong},
  \citenamefont {Shi}, \citenamefont {Pei}, \citenamefont {Wen}, \citenamefont
  {Gao}, \citenamefont {Zhao}, \citenamefont {Yin}, \citenamefont {Wu},
  \citenamefont {Li}, \citenamefont {Lei}, \citenamefont {Li}, \citenamefont
  {Chen}, \citenamefont {Yan},\ and\ \citenamefont {Qi}}]{WAN21g}%
  \BibitemOpen
  \bibfield  {author} {\bibinfo {author} {\bibfnamefont {Q.}~\bibnamefont
  {Wang}}, \bibinfo {author} {\bibfnamefont {P.}~\bibnamefont {Kong}}, \bibinfo
  {author} {\bibfnamefont {W.}~\bibnamefont {Shi}}, \bibinfo {author}
  {\bibfnamefont {C.}~\bibnamefont {Pei}}, \bibinfo {author} {\bibfnamefont
  {C.}~\bibnamefont {Wen}}, \bibinfo {author} {\bibfnamefont {L.}~\bibnamefont
  {Gao}}, \bibinfo {author} {\bibfnamefont {Y.}~\bibnamefont {Zhao}}, \bibinfo
  {author} {\bibfnamefont {Q.}~\bibnamefont {Yin}}, \bibinfo {author}
  {\bibfnamefont {Y.}~\bibnamefont {Wu}}, \bibinfo {author} {\bibfnamefont
  {G.}~\bibnamefont {Li}}, \bibinfo {author} {\bibfnamefont {H.}~\bibnamefont
  {Lei}}, \bibinfo {author} {\bibfnamefont {J.}~\bibnamefont {Li}}, \bibinfo
  {author} {\bibfnamefont {Y.}~\bibnamefont {Chen}}, \bibinfo {author}
  {\bibfnamefont {S.}~\bibnamefont {Yan}},\ and\ \bibinfo {author}
  {\bibfnamefont {Y.}~\bibnamefont {Qi}},\ }\href
  {https://doi.org/10.1002/adma.202102813} {\bibfield  {journal} {\bibinfo
  {journal} {Advanced Materials}\ }\textbf {\bibinfo {volume} {33}},\ \bibinfo
  {pages} {2102813} (\bibinfo {year} {2021}{\natexlab{d}})}\BibitemShut
  {NoStop}%
\bibitem [{\citenamefont {Yu}\ \emph {et~al.}(2021{\natexlab{a}})\citenamefont
  {Yu}, \citenamefont {Wu}, \citenamefont {Wang}, \citenamefont {Lei},
  \citenamefont {Zhuo}, \citenamefont {Ying},\ and\ \citenamefont
  {Chen}}]{YU21}%
  \BibitemOpen
  \bibfield  {author} {\bibinfo {author} {\bibfnamefont {F.~H.}\ \bibnamefont
  {Yu}}, \bibinfo {author} {\bibfnamefont {T.}~\bibnamefont {Wu}}, \bibinfo
  {author} {\bibfnamefont {Z.~Y.}\ \bibnamefont {Wang}}, \bibinfo {author}
  {\bibfnamefont {B.}~\bibnamefont {Lei}}, \bibinfo {author} {\bibfnamefont
  {W.~Z.}\ \bibnamefont {Zhuo}}, \bibinfo {author} {\bibfnamefont {J.~J.}\
  \bibnamefont {Ying}},\ and\ \bibinfo {author} {\bibfnamefont {X.~H.}\
  \bibnamefont {Chen}},\ }\href {https://doi.org/10.1103/PhysRevB.104.L041103}
  {\bibfield  {journal} {\bibinfo  {journal} {Phys. Rev. B}\ }\textbf {\bibinfo
  {volume} {104}},\ \bibinfo {pages} {L041103} (\bibinfo {year}
  {2021}{\natexlab{a}})}\BibitemShut {NoStop}%
\bibitem [{\citenamefont {Zhu}\ \emph {et~al.}(2022)\citenamefont {Zhu},
  \citenamefont {Yang}, \citenamefont {Xia}, \citenamefont {Yin}, \citenamefont
  {Wang}, \citenamefont {Zhao}, \citenamefont {Dai}, \citenamefont {Tu},
  \citenamefont {Song}, \citenamefont {Tao}, \citenamefont {Tu}, \citenamefont
  {Gong}, \citenamefont {Lei}, \citenamefont {Guo},\ and\ \citenamefont
  {Li}}]{ZHU22}%
  \BibitemOpen
  \bibfield  {author} {\bibinfo {author} {\bibfnamefont {C.~C.}\ \bibnamefont
  {Zhu}}, \bibinfo {author} {\bibfnamefont {X.~F.}\ \bibnamefont {Yang}},
  \bibinfo {author} {\bibfnamefont {W.}~\bibnamefont {Xia}}, \bibinfo {author}
  {\bibfnamefont {Q.~W.}\ \bibnamefont {Yin}}, \bibinfo {author} {\bibfnamefont
  {L.~S.}\ \bibnamefont {Wang}}, \bibinfo {author} {\bibfnamefont {C.~C.}\
  \bibnamefont {Zhao}}, \bibinfo {author} {\bibfnamefont {D.~Z.}\ \bibnamefont
  {Dai}}, \bibinfo {author} {\bibfnamefont {C.~P.}\ \bibnamefont {Tu}},
  \bibinfo {author} {\bibfnamefont {B.~Q.}\ \bibnamefont {Song}}, \bibinfo
  {author} {\bibfnamefont {Z.~C.}\ \bibnamefont {Tao}}, \bibinfo {author}
  {\bibfnamefont {Z.~J.}\ \bibnamefont {Tu}}, \bibinfo {author} {\bibfnamefont
  {C.~S.}\ \bibnamefont {Gong}}, \bibinfo {author} {\bibfnamefont {H.~C.}\
  \bibnamefont {Lei}}, \bibinfo {author} {\bibfnamefont {Y.~F.}\ \bibnamefont
  {Guo}},\ and\ \bibinfo {author} {\bibfnamefont {S.~Y.}\ \bibnamefont {Li}},\
  }\href {https://doi.org/10.1103/PhysRevB.105.094507} {\bibfield  {journal}
  {\bibinfo  {journal} {Phys. Rev. B}\ }\textbf {\bibinfo {volume} {105}},\
  \bibinfo {pages} {094507} (\bibinfo {year} {2022})}\BibitemShut {NoStop}%
\bibitem [{\citenamefont {Chen}\ \emph
  {et~al.}(2021{\natexlab{a}})\citenamefont {Chen}, \citenamefont {Yang},
  \citenamefont {Hu}, \citenamefont {Zhao}, \citenamefont {Yuan}, \citenamefont
  {Xing}, \citenamefont {Qian}, \citenamefont {Huang}, \citenamefont {Li},
  \citenamefont {Ye}, \citenamefont {Ma}, \citenamefont {Ni}, \citenamefont
  {Zhang}, \citenamefont {Yin}, \citenamefont {Gong}, \citenamefont {Tu},
  \citenamefont {Lei}, \citenamefont {Tan}, \citenamefont {Zhou}, \citenamefont
  {Shen}, \citenamefont {Dong}, \citenamefont {Yan}, \citenamefont {Wang},\
  and\ \citenamefont {Gao}}]{CHE21a}%
  \BibitemOpen
  \bibfield  {author} {\bibinfo {author} {\bibfnamefont {H.}~\bibnamefont
  {Chen}}, \bibinfo {author} {\bibfnamefont {H.}~\bibnamefont {Yang}}, \bibinfo
  {author} {\bibfnamefont {B.}~\bibnamefont {Hu}}, \bibinfo {author}
  {\bibfnamefont {Z.}~\bibnamefont {Zhao}}, \bibinfo {author} {\bibfnamefont
  {J.}~\bibnamefont {Yuan}}, \bibinfo {author} {\bibfnamefont {Y.}~\bibnamefont
  {Xing}}, \bibinfo {author} {\bibfnamefont {G.}~\bibnamefont {Qian}}, \bibinfo
  {author} {\bibfnamefont {Z.}~\bibnamefont {Huang}}, \bibinfo {author}
  {\bibfnamefont {G.}~\bibnamefont {Li}}, \bibinfo {author} {\bibfnamefont
  {Y.}~\bibnamefont {Ye}}, \bibinfo {author} {\bibfnamefont {S.}~\bibnamefont
  {Ma}}, \bibinfo {author} {\bibfnamefont {S.}~\bibnamefont {Ni}}, \bibinfo
  {author} {\bibfnamefont {H.}~\bibnamefont {Zhang}}, \bibinfo {author}
  {\bibfnamefont {Q.}~\bibnamefont {Yin}}, \bibinfo {author} {\bibfnamefont
  {C.}~\bibnamefont {Gong}}, \bibinfo {author} {\bibfnamefont {Z.}~\bibnamefont
  {Tu}}, \bibinfo {author} {\bibfnamefont {H.}~\bibnamefont {Lei}}, \bibinfo
  {author} {\bibfnamefont {H.}~\bibnamefont {Tan}}, \bibinfo {author}
  {\bibfnamefont {S.}~\bibnamefont {Zhou}}, \bibinfo {author} {\bibfnamefont
  {C.}~\bibnamefont {Shen}}, \bibinfo {author} {\bibfnamefont {X.}~\bibnamefont
  {Dong}}, \bibinfo {author} {\bibfnamefont {B.}~\bibnamefont {Yan}}, \bibinfo
  {author} {\bibfnamefont {Z.}~\bibnamefont {Wang}},\ and\ \bibinfo {author}
  {\bibfnamefont {H.-J.}\ \bibnamefont {Gao}},\ }\href
  {https://doi.org/10.1038/s41586-021-03983-5} {\bibfield  {journal} {\bibinfo
  {journal} {Nature}\ }\textbf {\bibinfo {volume} {599}},\ \bibinfo {pages}
  {222} (\bibinfo {year} {2021}{\natexlab{a}})}\BibitemShut {NoStop}%
\bibitem [{\citenamefont {Chen}\ \emph
  {et~al.}(2021{\natexlab{b}})\citenamefont {Chen}, \citenamefont {Wang},
  \citenamefont {Yin}, \citenamefont {Gu}, \citenamefont {Jiang}, \citenamefont
  {Tu}, \citenamefont {Gong}, \citenamefont {Uwatoko}, \citenamefont {Sun},
  \citenamefont {Lei}, \citenamefont {Hu},\ and\ \citenamefont
  {Cheng}}]{CHE21b}%
  \BibitemOpen
  \bibfield  {author} {\bibinfo {author} {\bibfnamefont {K.~Y.}\ \bibnamefont
  {Chen}}, \bibinfo {author} {\bibfnamefont {N.~N.}\ \bibnamefont {Wang}},
  \bibinfo {author} {\bibfnamefont {Q.~W.}\ \bibnamefont {Yin}}, \bibinfo
  {author} {\bibfnamefont {Y.~H.}\ \bibnamefont {Gu}}, \bibinfo {author}
  {\bibfnamefont {K.}~\bibnamefont {Jiang}}, \bibinfo {author} {\bibfnamefont
  {Z.~J.}\ \bibnamefont {Tu}}, \bibinfo {author} {\bibfnamefont {C.~S.}\
  \bibnamefont {Gong}}, \bibinfo {author} {\bibfnamefont {Y.}~\bibnamefont
  {Uwatoko}}, \bibinfo {author} {\bibfnamefont {J.~P.}\ \bibnamefont {Sun}},
  \bibinfo {author} {\bibfnamefont {H.~C.}\ \bibnamefont {Lei}}, \bibinfo
  {author} {\bibfnamefont {J.~P.}\ \bibnamefont {Hu}},\ and\ \bibinfo {author}
  {\bibfnamefont {J.-G.}\ \bibnamefont {Cheng}},\ }\href
  {https://doi.org/10.1103/PhysRevLett.126.247001} {\bibfield  {journal}
  {\bibinfo  {journal} {Phys. Rev. Lett.}\ }\textbf {\bibinfo {volume} {126}},\
  \bibinfo {pages} {247001} (\bibinfo {year} {2021}{\natexlab{b}})}\BibitemShut
  {NoStop}%
\bibitem [{\citenamefont {Du}\ \emph {et~al.}(2021)\citenamefont {Du},
  \citenamefont {Luo}, \citenamefont {Ortiz}, \citenamefont {Chen},
  \citenamefont {Duan}, \citenamefont {Zhang}, \citenamefont {Lu},
  \citenamefont {Wilson}, \citenamefont {Song},\ and\ \citenamefont
  {Yuan}}]{DU21}%
  \BibitemOpen
  \bibfield  {author} {\bibinfo {author} {\bibfnamefont {F.}~\bibnamefont
  {Du}}, \bibinfo {author} {\bibfnamefont {S.}~\bibnamefont {Luo}}, \bibinfo
  {author} {\bibfnamefont {B.~R.}\ \bibnamefont {Ortiz}}, \bibinfo {author}
  {\bibfnamefont {Y.}~\bibnamefont {Chen}}, \bibinfo {author} {\bibfnamefont
  {W.}~\bibnamefont {Duan}}, \bibinfo {author} {\bibfnamefont {D.}~\bibnamefont
  {Zhang}}, \bibinfo {author} {\bibfnamefont {X.}~\bibnamefont {Lu}}, \bibinfo
  {author} {\bibfnamefont {S.~D.}\ \bibnamefont {Wilson}}, \bibinfo {author}
  {\bibfnamefont {Y.}~\bibnamefont {Song}},\ and\ \bibinfo {author}
  {\bibfnamefont {H.}~\bibnamefont {Yuan}},\ }\href
  {https://doi.org/10.1103/PhysRevB.103.L220504} {\bibfield  {journal}
  {\bibinfo  {journal} {Phys. Rev. B}\ }\textbf {\bibinfo {volume} {103}},\
  \bibinfo {pages} {L220504} (\bibinfo {year} {2021})}\BibitemShut {NoStop}%
\bibitem [{\citenamefont {Duan}\ \emph {et~al.}(2021)\citenamefont {Duan},
  \citenamefont {Nie}, \citenamefont {Luo}, \citenamefont {Yu}, \citenamefont
  {Ortiz}, \citenamefont {Yin}, \citenamefont {Su}, \citenamefont {Du},
  \citenamefont {Wang}, \citenamefont {Chen}, \citenamefont {Lu}, \citenamefont
  {Ying}, \citenamefont {Wilson}, \citenamefont {Chen}, \citenamefont {Song},\
  and\ \citenamefont {Yuan}}]{DUA21}%
  \BibitemOpen
  \bibfield  {author} {\bibinfo {author} {\bibfnamefont {W.}~\bibnamefont
  {Duan}}, \bibinfo {author} {\bibfnamefont {Z.}~\bibnamefont {Nie}}, \bibinfo
  {author} {\bibfnamefont {S.}~\bibnamefont {Luo}}, \bibinfo {author}
  {\bibfnamefont {F.}~\bibnamefont {Yu}}, \bibinfo {author} {\bibfnamefont
  {B.~R.}\ \bibnamefont {Ortiz}}, \bibinfo {author} {\bibfnamefont
  {L.}~\bibnamefont {Yin}}, \bibinfo {author} {\bibfnamefont {H.}~\bibnamefont
  {Su}}, \bibinfo {author} {\bibfnamefont {F.}~\bibnamefont {Du}}, \bibinfo
  {author} {\bibfnamefont {A.}~\bibnamefont {Wang}}, \bibinfo {author}
  {\bibfnamefont {Y.}~\bibnamefont {Chen}}, \bibinfo {author} {\bibfnamefont
  {X.}~\bibnamefont {Lu}}, \bibinfo {author} {\bibfnamefont {J.}~\bibnamefont
  {Ying}}, \bibinfo {author} {\bibfnamefont {S.~D.}\ \bibnamefont {Wilson}},
  \bibinfo {author} {\bibfnamefont {X.}~\bibnamefont {Chen}}, \bibinfo {author}
  {\bibfnamefont {Y.}~\bibnamefont {Song}},\ and\ \bibinfo {author}
  {\bibfnamefont {H.}~\bibnamefont {Yuan}},\ }\href
  {https://doi.org/10.1007/s11433-021-1747-7} {\bibfield  {journal} {\bibinfo
  {journal} {Sci. China Phys. Mech. Astron.}\ }\textbf {\bibinfo {volume}
  {64}},\ \bibinfo {pages} {107462} (\bibinfo {year} {2021})}\BibitemShut
  {NoStop}%
\bibitem [{\citenamefont {Feng}\ \emph {et~al.}(2021)\citenamefont {Feng},
  \citenamefont {Jiang}, \citenamefont {Wang},\ and\ \citenamefont
  {Hu}}]{FEN21}%
  \BibitemOpen
  \bibfield  {author} {\bibinfo {author} {\bibfnamefont {X.}~\bibnamefont
  {Feng}}, \bibinfo {author} {\bibfnamefont {K.}~\bibnamefont {Jiang}},
  \bibinfo {author} {\bibfnamefont {Z.}~\bibnamefont {Wang}},\ and\ \bibinfo
  {author} {\bibfnamefont {J.}~\bibnamefont {Hu}},\ }\href
  {https://doi.org/10.1016/j.scib.2021.04.043} {\bibfield  {journal} {\bibinfo
  {journal} {Science Bulletin}\ }\textbf {\bibinfo {volume} {66}},\ \bibinfo
  {pages} {1384} (\bibinfo {year} {2021})}\BibitemShut {NoStop}%
\bibitem [{\citenamefont {Kang}\ \emph
  {et~al.}(2023{\natexlab{a}})\citenamefont {Kang}, \citenamefont {Fang},
  \citenamefont {Yoo}, \citenamefont {Ortiz}, \citenamefont {Oey},
  \citenamefont {Choi}, \citenamefont {Ryu}, \citenamefont {Kim}, \citenamefont
  {Jozwiak}, \citenamefont {Bostwick}, \citenamefont {Rotenberg}, \citenamefont
  {Kaxiras}, \citenamefont {Checkelsky}, \citenamefont {Wilson}, \citenamefont
  {Park},\ and\ \citenamefont {Comin}}]{KAN23a}%
  \BibitemOpen
  \bibfield  {author} {\bibinfo {author} {\bibfnamefont {M.}~\bibnamefont
  {Kang}}, \bibinfo {author} {\bibfnamefont {S.}~\bibnamefont {Fang}}, \bibinfo
  {author} {\bibfnamefont {J.}~\bibnamefont {Yoo}}, \bibinfo {author}
  {\bibfnamefont {B.~R.}\ \bibnamefont {Ortiz}}, \bibinfo {author}
  {\bibfnamefont {Y.~M.}\ \bibnamefont {Oey}}, \bibinfo {author} {\bibfnamefont
  {J.}~\bibnamefont {Choi}}, \bibinfo {author} {\bibfnamefont {S.~H.}\
  \bibnamefont {Ryu}}, \bibinfo {author} {\bibfnamefont {J.}~\bibnamefont
  {Kim}}, \bibinfo {author} {\bibfnamefont {C.}~\bibnamefont {Jozwiak}},
  \bibinfo {author} {\bibfnamefont {A.}~\bibnamefont {Bostwick}}, \bibinfo
  {author} {\bibfnamefont {E.}~\bibnamefont {Rotenberg}}, \bibinfo {author}
  {\bibfnamefont {E.}~\bibnamefont {Kaxiras}}, \bibinfo {author} {\bibfnamefont
  {J.~G.}\ \bibnamefont {Checkelsky}}, \bibinfo {author} {\bibfnamefont
  {S.~D.}\ \bibnamefont {Wilson}}, \bibinfo {author} {\bibfnamefont {J.-H.}\
  \bibnamefont {Park}},\ and\ \bibinfo {author} {\bibfnamefont
  {R.}~\bibnamefont {Comin}},\ }\href
  {https://doi.org/10.1038/s41563-022-01375-2} {\bibfield  {journal} {\bibinfo
  {journal} {Nat. Mater.}\ }\textbf {\bibinfo {volume} {22}},\ \bibinfo {pages}
  {186} (\bibinfo {year} {2023}{\natexlab{a}})}\BibitemShut {NoStop}%
\bibitem [{\citenamefont {Li}\ \emph {et~al.}(2022)\citenamefont {Li},
  \citenamefont {Zhao}, \citenamefont {Ortiz}, \citenamefont {Park},
  \citenamefont {Ye}, \citenamefont {Balents}, \citenamefont {Wang},
  \citenamefont {Wilson},\ and\ \citenamefont {Zeljkovic}}]{LI22a}%
  \BibitemOpen
  \bibfield  {author} {\bibinfo {author} {\bibfnamefont {H.}~\bibnamefont
  {Li}}, \bibinfo {author} {\bibfnamefont {H.}~\bibnamefont {Zhao}}, \bibinfo
  {author} {\bibfnamefont {B.~R.}\ \bibnamefont {Ortiz}}, \bibinfo {author}
  {\bibfnamefont {T.}~\bibnamefont {Park}}, \bibinfo {author} {\bibfnamefont
  {M.}~\bibnamefont {Ye}}, \bibinfo {author} {\bibfnamefont {L.}~\bibnamefont
  {Balents}}, \bibinfo {author} {\bibfnamefont {Z.}~\bibnamefont {Wang}},
  \bibinfo {author} {\bibfnamefont {S.~D.}\ \bibnamefont {Wilson}},\ and\
  \bibinfo {author} {\bibfnamefont {I.}~\bibnamefont {Zeljkovic}},\ }\href
  {https://doi.org/10.1038/s41567-021-01479-7} {\bibfield  {journal} {\bibinfo
  {journal} {Nat. Phys.}\ }\textbf {\bibinfo {volume} {18}},\ \bibinfo {pages}
  {265} (\bibinfo {year} {2022})}\BibitemShut {NoStop}%
\bibitem [{\citenamefont {Liu}\ \emph {et~al.}(2021{\natexlab{b}})\citenamefont
  {Liu}, \citenamefont {Wang}, \citenamefont {Cai}, \citenamefont {Hao},
  \citenamefont {Ma}, \citenamefont {Wang}, \citenamefont {Liu}, \citenamefont
  {Chen}, \citenamefont {Zhou}, \citenamefont {Wang}, \citenamefont {Wang},
  \citenamefont {He}, \citenamefont {Liu}, \citenamefont {Cui}, \citenamefont
  {Wang}, \citenamefont {Huang}, \citenamefont {Chen},\ and\ \citenamefont
  {Mei}}]{LIU21d}%
  \BibitemOpen
  \bibfield  {author} {\bibinfo {author} {\bibfnamefont {Y.}~\bibnamefont
  {Liu}}, \bibinfo {author} {\bibfnamefont {Y.}~\bibnamefont {Wang}}, \bibinfo
  {author} {\bibfnamefont {Y.}~\bibnamefont {Cai}}, \bibinfo {author}
  {\bibfnamefont {Z.}~\bibnamefont {Hao}}, \bibinfo {author} {\bibfnamefont
  {X.-M.}\ \bibnamefont {Ma}}, \bibinfo {author} {\bibfnamefont
  {L.}~\bibnamefont {Wang}}, \bibinfo {author} {\bibfnamefont {C.}~\bibnamefont
  {Liu}}, \bibinfo {author} {\bibfnamefont {J.}~\bibnamefont {Chen}}, \bibinfo
  {author} {\bibfnamefont {L.}~\bibnamefont {Zhou}}, \bibinfo {author}
  {\bibfnamefont {J.}~\bibnamefont {Wang}}, \bibinfo {author} {\bibfnamefont
  {S.}~\bibnamefont {Wang}}, \bibinfo {author} {\bibfnamefont {H.}~\bibnamefont
  {He}}, \bibinfo {author} {\bibfnamefont {Y.}~\bibnamefont {Liu}}, \bibinfo
  {author} {\bibfnamefont {S.}~\bibnamefont {Cui}}, \bibinfo {author}
  {\bibfnamefont {J.}~\bibnamefont {Wang}}, \bibinfo {author} {\bibfnamefont
  {B.}~\bibnamefont {Huang}}, \bibinfo {author} {\bibfnamefont
  {C.}~\bibnamefont {Chen}},\ and\ \bibinfo {author} {\bibfnamefont {J.-W.}\
  \bibnamefont {Mei}},\ }\href {https://doi.org/10.48550/arXiv.2110.12651}
  {\bibinfo {title} {Doping evolution of superconductivity, charge order and
  band topology in hole-doped topological kagome superconductors
  {{Cs}}({{V}}$_{1-x}${{Ti}}$_x$)$_3${{Sb}}$_5$}} (\bibinfo {year}
  {2021}{\natexlab{b}}),\ \Eprint {https://arxiv.org/abs/2110.12651}
  {arxiv:2110.12651 [cond-mat]} \BibitemShut {NoStop}%
\bibitem [{\citenamefont {Mu}\ \emph {et~al.}(2021)\citenamefont {Mu},
  \citenamefont {Yin}, \citenamefont {Tu}, \citenamefont {Gong}, \citenamefont
  {Lei}, \citenamefont {Li},\ and\ \citenamefont {Luo}}]{MU21}%
  \BibitemOpen
  \bibfield  {author} {\bibinfo {author} {\bibfnamefont {C.}~\bibnamefont
  {Mu}}, \bibinfo {author} {\bibfnamefont {Q.}~\bibnamefont {Yin}}, \bibinfo
  {author} {\bibfnamefont {Z.}~\bibnamefont {Tu}}, \bibinfo {author}
  {\bibfnamefont {C.}~\bibnamefont {Gong}}, \bibinfo {author} {\bibfnamefont
  {H.}~\bibnamefont {Lei}}, \bibinfo {author} {\bibfnamefont {Z.}~\bibnamefont
  {Li}},\ and\ \bibinfo {author} {\bibfnamefont {J.}~\bibnamefont {Luo}},\
  }\href {https://doi.org/10.1088/0256-307X/38/7/077402} {\bibfield  {journal}
  {\bibinfo  {journal} {Chinese Phys. Lett.}\ }\textbf {\bibinfo {volume}
  {38}},\ \bibinfo {pages} {077402} (\bibinfo {year} {2021})}\BibitemShut
  {NoStop}%
\bibitem [{\citenamefont {Nakayama}\ \emph {et~al.}(2021)\citenamefont
  {Nakayama}, \citenamefont {Li}, \citenamefont {Kato}, \citenamefont {Liu},
  \citenamefont {Wang}, \citenamefont {Takahashi}, \citenamefont {Yao},\ and\
  \citenamefont {Sato}}]{NAK21}%
  \BibitemOpen
  \bibfield  {author} {\bibinfo {author} {\bibfnamefont {K.}~\bibnamefont
  {Nakayama}}, \bibinfo {author} {\bibfnamefont {Y.}~\bibnamefont {Li}},
  \bibinfo {author} {\bibfnamefont {T.}~\bibnamefont {Kato}}, \bibinfo {author}
  {\bibfnamefont {M.}~\bibnamefont {Liu}}, \bibinfo {author} {\bibfnamefont
  {Z.}~\bibnamefont {Wang}}, \bibinfo {author} {\bibfnamefont {T.}~\bibnamefont
  {Takahashi}}, \bibinfo {author} {\bibfnamefont {Y.}~\bibnamefont {Yao}},\
  and\ \bibinfo {author} {\bibfnamefont {T.}~\bibnamefont {Sato}},\ }\href
  {https://doi.org/10.1103/PhysRevB.104.L161112} {\bibfield  {journal}
  {\bibinfo  {journal} {Phys. Rev. B}\ }\textbf {\bibinfo {volume} {104}},\
  \bibinfo {pages} {L161112} (\bibinfo {year} {2021})}\BibitemShut {NoStop}%
\bibitem [{\citenamefont {Ni}\ \emph {et~al.}(2021)\citenamefont {Ni},
  \citenamefont {Ma}, \citenamefont {Zhang}, \citenamefont {Yuan},
  \citenamefont {Yang}, \citenamefont {Lu}, \citenamefont {Wang}, \citenamefont
  {Sun}, \citenamefont {Zhao}, \citenamefont {Li}, \citenamefont {Liu},
  \citenamefont {Zhang}, \citenamefont {Chen}, \citenamefont {Jin},
  \citenamefont {Cheng}, \citenamefont {Yu}, \citenamefont {Zhou},
  \citenamefont {Dong}, \citenamefont {Hu}, \citenamefont {Gao},\ and\
  \citenamefont {Zhao}}]{NI21}%
  \BibitemOpen
  \bibfield  {author} {\bibinfo {author} {\bibfnamefont {S.}~\bibnamefont
  {Ni}}, \bibinfo {author} {\bibfnamefont {S.}~\bibnamefont {Ma}}, \bibinfo
  {author} {\bibfnamefont {Y.}~\bibnamefont {Zhang}}, \bibinfo {author}
  {\bibfnamefont {J.}~\bibnamefont {Yuan}}, \bibinfo {author} {\bibfnamefont
  {H.}~\bibnamefont {Yang}}, \bibinfo {author} {\bibfnamefont {Z.}~\bibnamefont
  {Lu}}, \bibinfo {author} {\bibfnamefont {N.}~\bibnamefont {Wang}}, \bibinfo
  {author} {\bibfnamefont {J.}~\bibnamefont {Sun}}, \bibinfo {author}
  {\bibfnamefont {Z.}~\bibnamefont {Zhao}}, \bibinfo {author} {\bibfnamefont
  {D.}~\bibnamefont {Li}}, \bibinfo {author} {\bibfnamefont {S.}~\bibnamefont
  {Liu}}, \bibinfo {author} {\bibfnamefont {H.}~\bibnamefont {Zhang}}, \bibinfo
  {author} {\bibfnamefont {H.}~\bibnamefont {Chen}}, \bibinfo {author}
  {\bibfnamefont {K.}~\bibnamefont {Jin}}, \bibinfo {author} {\bibfnamefont
  {J.}~\bibnamefont {Cheng}}, \bibinfo {author} {\bibfnamefont
  {L.}~\bibnamefont {Yu}}, \bibinfo {author} {\bibfnamefont {F.}~\bibnamefont
  {Zhou}}, \bibinfo {author} {\bibfnamefont {X.}~\bibnamefont {Dong}}, \bibinfo
  {author} {\bibfnamefont {J.}~\bibnamefont {Hu}}, \bibinfo {author}
  {\bibfnamefont {H.-J.}\ \bibnamefont {Gao}},\ and\ \bibinfo {author}
  {\bibfnamefont {Z.}~\bibnamefont {Zhao}},\ }\href
  {https://doi.org/10.1088/0256-307X/38/5/057403} {\bibfield  {journal}
  {\bibinfo  {journal} {Chinese Phys. Lett.}\ }\textbf {\bibinfo {volume}
  {38}},\ \bibinfo {pages} {057403} (\bibinfo {year} {2021})}\BibitemShut
  {NoStop}%
\bibitem [{\citenamefont {Ortiz}\ \emph
  {et~al.}(2021{\natexlab{b}})\citenamefont {Ortiz}, \citenamefont {Sarte},
  \citenamefont {Kenney}, \citenamefont {Graf}, \citenamefont {Teicher},
  \citenamefont {Seshadri},\ and\ \citenamefont {Wilson}}]{ORT21a}%
  \BibitemOpen
  \bibfield  {author} {\bibinfo {author} {\bibfnamefont {B.~R.}\ \bibnamefont
  {Ortiz}}, \bibinfo {author} {\bibfnamefont {P.~M.}\ \bibnamefont {Sarte}},
  \bibinfo {author} {\bibfnamefont {E.~M.}\ \bibnamefont {Kenney}}, \bibinfo
  {author} {\bibfnamefont {M.~J.}\ \bibnamefont {Graf}}, \bibinfo {author}
  {\bibfnamefont {S.~M.~L.}\ \bibnamefont {Teicher}}, \bibinfo {author}
  {\bibfnamefont {R.}~\bibnamefont {Seshadri}},\ and\ \bibinfo {author}
  {\bibfnamefont {S.~D.}\ \bibnamefont {Wilson}},\ }\href
  {https://doi.org/10.1103/PhysRevMaterials.5.034801} {\bibfield  {journal}
  {\bibinfo  {journal} {Phys. Rev. Mater.}\ }\textbf {\bibinfo {volume} {5}},\
  \bibinfo {pages} {034801} (\bibinfo {year} {2021}{\natexlab{b}})}\BibitemShut
  {NoStop}%
\bibitem [{\citenamefont {Shrestha}\ \emph {et~al.}(2022)\citenamefont
  {Shrestha}, \citenamefont {Chapai}, \citenamefont {Pokharel}, \citenamefont
  {Miertschin}, \citenamefont {Nguyen}, \citenamefont {Zhou}, \citenamefont
  {Chung}, \citenamefont {Kanatzidis}, \citenamefont {Mitchell}, \citenamefont
  {Welp}, \citenamefont {Popovi{\'c}}, \citenamefont {Graf}, \citenamefont
  {Lorenz},\ and\ \citenamefont {Kwok}}]{SHR22}%
  \BibitemOpen
  \bibfield  {author} {\bibinfo {author} {\bibfnamefont {K.}~\bibnamefont
  {Shrestha}}, \bibinfo {author} {\bibfnamefont {R.}~\bibnamefont {Chapai}},
  \bibinfo {author} {\bibfnamefont {B.~K.}\ \bibnamefont {Pokharel}}, \bibinfo
  {author} {\bibfnamefont {D.}~\bibnamefont {Miertschin}}, \bibinfo {author}
  {\bibfnamefont {T.}~\bibnamefont {Nguyen}}, \bibinfo {author} {\bibfnamefont
  {X.}~\bibnamefont {Zhou}}, \bibinfo {author} {\bibfnamefont {D.~Y.}\
  \bibnamefont {Chung}}, \bibinfo {author} {\bibfnamefont {M.~G.}\ \bibnamefont
  {Kanatzidis}}, \bibinfo {author} {\bibfnamefont {J.~F.}\ \bibnamefont
  {Mitchell}}, \bibinfo {author} {\bibfnamefont {U.}~\bibnamefont {Welp}},
  \bibinfo {author} {\bibfnamefont {D.}~\bibnamefont {Popovi{\'c}}}, \bibinfo
  {author} {\bibfnamefont {D.~E.}\ \bibnamefont {Graf}}, \bibinfo {author}
  {\bibfnamefont {B.}~\bibnamefont {Lorenz}},\ and\ \bibinfo {author}
  {\bibfnamefont {W.~K.}\ \bibnamefont {Kwok}},\ }\href
  {https://doi.org/10.1103/PhysRevB.105.024508} {\bibfield  {journal} {\bibinfo
   {journal} {Phys. Rev. B}\ }\textbf {\bibinfo {volume} {105}},\ \bibinfo
  {pages} {024508} (\bibinfo {year} {2022})}\BibitemShut {NoStop}%
\bibitem [{\citenamefont {Song}\ \emph
  {et~al.}(2021{\natexlab{b}})\citenamefont {Song}, \citenamefont {Kong},
  \citenamefont {Xia}, \citenamefont {Yin}, \citenamefont {Tu}, \citenamefont
  {Zhao}, \citenamefont {Dai}, \citenamefont {Meng}, \citenamefont {Tao},
  \citenamefont {Tu}, \citenamefont {Gong}, \citenamefont {Lei}, \citenamefont
  {Guo}, \citenamefont {Yang},\ and\ \citenamefont {Li}}]{SON21b}%
  \BibitemOpen
  \bibfield  {author} {\bibinfo {author} {\bibfnamefont {B.~Q.}\ \bibnamefont
  {Song}}, \bibinfo {author} {\bibfnamefont {X.~M.}\ \bibnamefont {Kong}},
  \bibinfo {author} {\bibfnamefont {W.}~\bibnamefont {Xia}}, \bibinfo {author}
  {\bibfnamefont {Q.~W.}\ \bibnamefont {Yin}}, \bibinfo {author} {\bibfnamefont
  {C.~P.}\ \bibnamefont {Tu}}, \bibinfo {author} {\bibfnamefont {C.~C.}\
  \bibnamefont {Zhao}}, \bibinfo {author} {\bibfnamefont {D.~Z.}\ \bibnamefont
  {Dai}}, \bibinfo {author} {\bibfnamefont {K.}~\bibnamefont {Meng}}, \bibinfo
  {author} {\bibfnamefont {Z.~C.}\ \bibnamefont {Tao}}, \bibinfo {author}
  {\bibfnamefont {Z.~J.}\ \bibnamefont {Tu}}, \bibinfo {author} {\bibfnamefont
  {C.~S.}\ \bibnamefont {Gong}}, \bibinfo {author} {\bibfnamefont {H.~C.}\
  \bibnamefont {Lei}}, \bibinfo {author} {\bibfnamefont {Y.~F.}\ \bibnamefont
  {Guo}}, \bibinfo {author} {\bibfnamefont {X.~F.}\ \bibnamefont {Yang}},\ and\
  \bibinfo {author} {\bibfnamefont {S.~Y.}\ \bibnamefont {Li}},\ }\href
  {https://doi.org/10.48550/arXiv.2105.09248} {\bibinfo {title} {Competing
  superconductivity and charge-density wave in {{Kagome}} metal
  {{CsV}}$_3${{Sb}}$_5$: Evidence from their evolutions with sample thickness}}
  (\bibinfo {year} {2021}{\natexlab{b}}),\ \Eprint
  {https://arxiv.org/abs/2105.09248} {arxiv:2105.09248 [cond-mat]} \BibitemShut
  {NoStop}%
\bibitem [{\citenamefont {Wang}\ \emph
  {et~al.}(2021{\natexlab{e}})\citenamefont {Wang}, \citenamefont {Chen},
  \citenamefont {Yin}, \citenamefont {Ma}, \citenamefont {Pan}, \citenamefont
  {Yang}, \citenamefont {Ji}, \citenamefont {Wu}, \citenamefont {Shan},
  \citenamefont {Xu}, \citenamefont {Tu}, \citenamefont {Gong}, \citenamefont
  {Liu}, \citenamefont {Li}, \citenamefont {Uwatoko}, \citenamefont {Dong},
  \citenamefont {Lei}, \citenamefont {Sun},\ and\ \citenamefont
  {Cheng}}]{WAN21e}%
  \BibitemOpen
  \bibfield  {author} {\bibinfo {author} {\bibfnamefont {N.~N.}\ \bibnamefont
  {Wang}}, \bibinfo {author} {\bibfnamefont {K.~Y.}\ \bibnamefont {Chen}},
  \bibinfo {author} {\bibfnamefont {Q.~W.}\ \bibnamefont {Yin}}, \bibinfo
  {author} {\bibfnamefont {Y.~N.~N.}\ \bibnamefont {Ma}}, \bibinfo {author}
  {\bibfnamefont {B.~Y.}\ \bibnamefont {Pan}}, \bibinfo {author} {\bibfnamefont
  {X.}~\bibnamefont {Yang}}, \bibinfo {author} {\bibfnamefont {X.~Y.}\
  \bibnamefont {Ji}}, \bibinfo {author} {\bibfnamefont {S.~L.}\ \bibnamefont
  {Wu}}, \bibinfo {author} {\bibfnamefont {P.~F.}\ \bibnamefont {Shan}},
  \bibinfo {author} {\bibfnamefont {S.~X.}\ \bibnamefont {Xu}}, \bibinfo
  {author} {\bibfnamefont {Z.~J.}\ \bibnamefont {Tu}}, \bibinfo {author}
  {\bibfnamefont {C.~S.}\ \bibnamefont {Gong}}, \bibinfo {author}
  {\bibfnamefont {G.~T.}\ \bibnamefont {Liu}}, \bibinfo {author} {\bibfnamefont
  {G.}~\bibnamefont {Li}}, \bibinfo {author} {\bibfnamefont {Y.}~\bibnamefont
  {Uwatoko}}, \bibinfo {author} {\bibfnamefont {X.~L.}\ \bibnamefont {Dong}},
  \bibinfo {author} {\bibfnamefont {H.~C.}\ \bibnamefont {Lei}}, \bibinfo
  {author} {\bibfnamefont {J.~P.}\ \bibnamefont {Sun}},\ and\ \bibinfo {author}
  {\bibfnamefont {J.-G.}\ \bibnamefont {Cheng}},\ }\href
  {https://doi.org/10.1103/PhysRevResearch.3.043018} {\bibfield  {journal}
  {\bibinfo  {journal} {Phys. Rev. Res.}\ }\textbf {\bibinfo {volume} {3}},\
  \bibinfo {pages} {043018} (\bibinfo {year} {2021}{\natexlab{e}})}\BibitemShut
  {NoStop}%
\bibitem [{\citenamefont {Wang}\ \emph
  {et~al.}(2021{\natexlab{f}})\citenamefont {Wang}, \citenamefont {Yu},
  \citenamefont {Zhang}, \citenamefont {Liu}, \citenamefont {Li}, \citenamefont
  {Peng}, \citenamefont {Di}, \citenamefont {Jiang},\ and\ \citenamefont
  {Mu}}]{WAN21f}%
  \BibitemOpen
  \bibfield  {author} {\bibinfo {author} {\bibfnamefont {T.}~\bibnamefont
  {Wang}}, \bibinfo {author} {\bibfnamefont {A.}~\bibnamefont {Yu}}, \bibinfo
  {author} {\bibfnamefont {H.}~\bibnamefont {Zhang}}, \bibinfo {author}
  {\bibfnamefont {Y.}~\bibnamefont {Liu}}, \bibinfo {author} {\bibfnamefont
  {W.}~\bibnamefont {Li}}, \bibinfo {author} {\bibfnamefont {W.}~\bibnamefont
  {Peng}}, \bibinfo {author} {\bibfnamefont {Z.}~\bibnamefont {Di}}, \bibinfo
  {author} {\bibfnamefont {D.}~\bibnamefont {Jiang}},\ and\ \bibinfo {author}
  {\bibfnamefont {G.}~\bibnamefont {Mu}},\ }\href
  {https://doi.org/10.48550/arXiv.2105.07732} {\bibinfo {title} {Enhancement of
  the superconductivity and quantum metallic state in the thin film of
  superconducting {{Kagome}} metal {{KV}}$_3${{Sb}}$_5$}} (\bibinfo {year}
  {2021}{\natexlab{f}}),\ \Eprint {https://arxiv.org/abs/2105.07732}
  {arxiv:2105.07732 [cond-mat]} \BibitemShut {NoStop}%
\bibitem [{\citenamefont {Wang}\ \emph {et~al.}(2023)\citenamefont {Wang},
  \citenamefont {Yang}, \citenamefont {Sivakumar}, \citenamefont {Ortiz},
  \citenamefont {Teicher}, \citenamefont {Wu}, \citenamefont {Srivastava},
  \citenamefont {Garg}, \citenamefont {Liu}, \citenamefont {Parkin},
  \citenamefont {Toberer}, \citenamefont {McQueen}, \citenamefont {Wilson},\
  and\ \citenamefont {Ali}}]{WAN23}%
  \BibitemOpen
  \bibfield  {author} {\bibinfo {author} {\bibfnamefont {Y.}~\bibnamefont
  {Wang}}, \bibinfo {author} {\bibfnamefont {S.}~\bibnamefont {Yang}}, \bibinfo
  {author} {\bibfnamefont {P.~K.}\ \bibnamefont {Sivakumar}}, \bibinfo {author}
  {\bibfnamefont {B.~R.}\ \bibnamefont {Ortiz}}, \bibinfo {author}
  {\bibfnamefont {S.~M.~L.}\ \bibnamefont {Teicher}}, \bibinfo {author}
  {\bibfnamefont {H.}~\bibnamefont {Wu}}, \bibinfo {author} {\bibfnamefont
  {A.~K.}\ \bibnamefont {Srivastava}}, \bibinfo {author} {\bibfnamefont
  {C.}~\bibnamefont {Garg}}, \bibinfo {author} {\bibfnamefont {D.}~\bibnamefont
  {Liu}}, \bibinfo {author} {\bibfnamefont {S.~S.~P.}\ \bibnamefont {Parkin}},
  \bibinfo {author} {\bibfnamefont {E.~S.}\ \bibnamefont {Toberer}}, \bibinfo
  {author} {\bibfnamefont {T.}~\bibnamefont {McQueen}}, \bibinfo {author}
  {\bibfnamefont {S.~D.}\ \bibnamefont {Wilson}},\ and\ \bibinfo {author}
  {\bibfnamefont {M.~N.}\ \bibnamefont {Ali}},\ }\href
  {https://doi.org/10.48550/arXiv.2012.05898} {\bibinfo {title} {Anisotropic
  proximity-induced superconductivity and edge supercurrent in {{Kagome}}
  metal, {{K1-xV3Sb5}}}} (\bibinfo {year} {2023}),\ \Eprint
  {https://arxiv.org/abs/2012.05898} {arxiv:2012.05898 [cond-mat]} \BibitemShut
  {NoStop}%
\bibitem [{\citenamefont {Wu}\ \emph {et~al.}(2021)\citenamefont {Wu},
  \citenamefont {Schwemmer}, \citenamefont {M{\"u}ller}, \citenamefont
  {Consiglio}, \citenamefont {Sangiovanni}, \citenamefont {Di~Sante},
  \citenamefont {Iqbal}, \citenamefont {Hanke}, \citenamefont {Schnyder},
  \citenamefont {Denner}, \citenamefont {Fischer}, \citenamefont {Neupert},\
  and\ \citenamefont {Thomale}}]{WU21b}%
  \BibitemOpen
  \bibfield  {author} {\bibinfo {author} {\bibfnamefont {X.}~\bibnamefont
  {Wu}}, \bibinfo {author} {\bibfnamefont {T.}~\bibnamefont {Schwemmer}},
  \bibinfo {author} {\bibfnamefont {T.}~\bibnamefont {M{\"u}ller}}, \bibinfo
  {author} {\bibfnamefont {A.}~\bibnamefont {Consiglio}}, \bibinfo {author}
  {\bibfnamefont {G.}~\bibnamefont {Sangiovanni}}, \bibinfo {author}
  {\bibfnamefont {D.}~\bibnamefont {Di~Sante}}, \bibinfo {author}
  {\bibfnamefont {Y.}~\bibnamefont {Iqbal}}, \bibinfo {author} {\bibfnamefont
  {W.}~\bibnamefont {Hanke}}, \bibinfo {author} {\bibfnamefont {A.~P.}\
  \bibnamefont {Schnyder}}, \bibinfo {author} {\bibfnamefont {M.~M.}\
  \bibnamefont {Denner}}, \bibinfo {author} {\bibfnamefont {M.~H.}\
  \bibnamefont {Fischer}}, \bibinfo {author} {\bibfnamefont {T.}~\bibnamefont
  {Neupert}},\ and\ \bibinfo {author} {\bibfnamefont {R.}~\bibnamefont
  {Thomale}},\ }\href {https://doi.org/10.1103/PhysRevLett.127.177001}
  {\bibfield  {journal} {\bibinfo  {journal} {Phys. Rev. Lett.}\ }\textbf
  {\bibinfo {volume} {127}},\ \bibinfo {pages} {177001} (\bibinfo {year}
  {2021})}\BibitemShut {NoStop}%
\bibitem [{\citenamefont {Xiang}\ \emph {et~al.}(2021)\citenamefont {Xiang},
  \citenamefont {Li}, \citenamefont {Li}, \citenamefont {Xie}, \citenamefont
  {Yang}, \citenamefont {Wang}, \citenamefont {Yao},\ and\ \citenamefont
  {Wen}}]{XIA21}%
  \BibitemOpen
  \bibfield  {author} {\bibinfo {author} {\bibfnamefont {Y.}~\bibnamefont
  {Xiang}}, \bibinfo {author} {\bibfnamefont {Q.}~\bibnamefont {Li}}, \bibinfo
  {author} {\bibfnamefont {Y.}~\bibnamefont {Li}}, \bibinfo {author}
  {\bibfnamefont {W.}~\bibnamefont {Xie}}, \bibinfo {author} {\bibfnamefont
  {H.}~\bibnamefont {Yang}}, \bibinfo {author} {\bibfnamefont {Z.}~\bibnamefont
  {Wang}}, \bibinfo {author} {\bibfnamefont {Y.}~\bibnamefont {Yao}},\ and\
  \bibinfo {author} {\bibfnamefont {H.-H.}\ \bibnamefont {Wen}},\ }\href
  {https://doi.org/10.1038/s41467-021-27084-z} {\bibfield  {journal} {\bibinfo
  {journal} {Nat. Commun.}\ }\textbf {\bibinfo {volume} {12}},\ \bibinfo
  {pages} {6727} (\bibinfo {year} {2021})}\BibitemShut {NoStop}%
\bibitem [{\citenamefont {Xu}\ \emph {et~al.}(2021)\citenamefont {Xu},
  \citenamefont {Yan}, \citenamefont {Yin}, \citenamefont {Xia}, \citenamefont
  {Fang}, \citenamefont {Chen}, \citenamefont {Li}, \citenamefont {Yang},
  \citenamefont {Guo},\ and\ \citenamefont {Feng}}]{XU21b}%
  \BibitemOpen
  \bibfield  {author} {\bibinfo {author} {\bibfnamefont {H.-S.}\ \bibnamefont
  {Xu}}, \bibinfo {author} {\bibfnamefont {Y.-J.}\ \bibnamefont {Yan}},
  \bibinfo {author} {\bibfnamefont {R.}~\bibnamefont {Yin}}, \bibinfo {author}
  {\bibfnamefont {W.}~\bibnamefont {Xia}}, \bibinfo {author} {\bibfnamefont
  {S.}~\bibnamefont {Fang}}, \bibinfo {author} {\bibfnamefont {Z.}~\bibnamefont
  {Chen}}, \bibinfo {author} {\bibfnamefont {Y.}~\bibnamefont {Li}}, \bibinfo
  {author} {\bibfnamefont {W.}~\bibnamefont {Yang}}, \bibinfo {author}
  {\bibfnamefont {Y.}~\bibnamefont {Guo}},\ and\ \bibinfo {author}
  {\bibfnamefont {D.-L.}\ \bibnamefont {Feng}},\ }\href
  {https://doi.org/10.1103/PhysRevLett.127.187004} {\bibfield  {journal}
  {\bibinfo  {journal} {Phys. Rev. Lett.}\ }\textbf {\bibinfo {volume} {127}},\
  \bibinfo {pages} {187004} (\bibinfo {year} {2021})}\BibitemShut {NoStop}%
\bibitem [{\citenamefont {Yin}\ \emph {et~al.}(2021{\natexlab{a}})\citenamefont
  {Yin}, \citenamefont {Tu}, \citenamefont {Gong}, \citenamefont {Fu},
  \citenamefont {Yan},\ and\ \citenamefont {Lei}}]{YIN21}%
  \BibitemOpen
  \bibfield  {author} {\bibinfo {author} {\bibfnamefont {Q.}~\bibnamefont
  {Yin}}, \bibinfo {author} {\bibfnamefont {Z.}~\bibnamefont {Tu}}, \bibinfo
  {author} {\bibfnamefont {C.}~\bibnamefont {Gong}}, \bibinfo {author}
  {\bibfnamefont {Y.}~\bibnamefont {Fu}}, \bibinfo {author} {\bibfnamefont
  {S.}~\bibnamefont {Yan}},\ and\ \bibinfo {author} {\bibfnamefont {a.~H.}\
  \bibnamefont {Lei}},\ }\href {https://doi.org/10.1088/0256-307X/38/3/037403}
  {\bibfield  {journal} {\bibinfo  {journal} {Chin. Phys. Lett.}\ }\textbf
  {\bibinfo {volume} {38}},\ \bibinfo {pages} {037403} (\bibinfo {year}
  {2021}{\natexlab{a}})}\BibitemShut {NoStop}%
\bibitem [{\citenamefont {Yin}\ \emph {et~al.}(2021{\natexlab{b}})\citenamefont
  {Yin}, \citenamefont {Zhang}, \citenamefont {Chen}, \citenamefont {Ye},
  \citenamefont {Yu}, \citenamefont {Ortiz}, \citenamefont {Luo}, \citenamefont
  {Duan}, \citenamefont {Su}, \citenamefont {Ying}, \citenamefont {Wilson},
  \citenamefont {Chen}, \citenamefont {Yuan}, \citenamefont {Song},\ and\
  \citenamefont {Lu}}]{YIN21a}%
  \BibitemOpen
  \bibfield  {author} {\bibinfo {author} {\bibfnamefont {L.}~\bibnamefont
  {Yin}}, \bibinfo {author} {\bibfnamefont {D.}~\bibnamefont {Zhang}}, \bibinfo
  {author} {\bibfnamefont {C.}~\bibnamefont {Chen}}, \bibinfo {author}
  {\bibfnamefont {G.}~\bibnamefont {Ye}}, \bibinfo {author} {\bibfnamefont
  {F.}~\bibnamefont {Yu}}, \bibinfo {author} {\bibfnamefont {B.~R.}\
  \bibnamefont {Ortiz}}, \bibinfo {author} {\bibfnamefont {S.}~\bibnamefont
  {Luo}}, \bibinfo {author} {\bibfnamefont {W.}~\bibnamefont {Duan}}, \bibinfo
  {author} {\bibfnamefont {H.}~\bibnamefont {Su}}, \bibinfo {author}
  {\bibfnamefont {J.}~\bibnamefont {Ying}}, \bibinfo {author} {\bibfnamefont
  {S.~D.}\ \bibnamefont {Wilson}}, \bibinfo {author} {\bibfnamefont
  {X.}~\bibnamefont {Chen}}, \bibinfo {author} {\bibfnamefont {H.}~\bibnamefont
  {Yuan}}, \bibinfo {author} {\bibfnamefont {Y.}~\bibnamefont {Song}},\ and\
  \bibinfo {author} {\bibfnamefont {X.}~\bibnamefont {Lu}},\ }\href
  {https://doi.org/10.1103/PhysRevB.104.174507} {\bibfield  {journal} {\bibinfo
   {journal} {Phys. Rev. B}\ }\textbf {\bibinfo {volume} {104}},\ \bibinfo
  {pages} {174507} (\bibinfo {year} {2021}{\natexlab{b}})}\BibitemShut
  {NoStop}%
\bibitem [{\citenamefont {Yu}\ \emph {et~al.}(2021{\natexlab{b}})\citenamefont
  {Yu}, \citenamefont {Ma}, \citenamefont {Zhuo}, \citenamefont {Liu},
  \citenamefont {Wen}, \citenamefont {Lei}, \citenamefont {Ying},\ and\
  \citenamefont {Chen}}]{YU21a}%
  \BibitemOpen
  \bibfield  {author} {\bibinfo {author} {\bibfnamefont {F.~H.}\ \bibnamefont
  {Yu}}, \bibinfo {author} {\bibfnamefont {D.~H.}\ \bibnamefont {Ma}}, \bibinfo
  {author} {\bibfnamefont {W.~Z.}\ \bibnamefont {Zhuo}}, \bibinfo {author}
  {\bibfnamefont {S.~Q.}\ \bibnamefont {Liu}}, \bibinfo {author} {\bibfnamefont
  {X.~K.}\ \bibnamefont {Wen}}, \bibinfo {author} {\bibfnamefont
  {B.}~\bibnamefont {Lei}}, \bibinfo {author} {\bibfnamefont {J.~J.}\
  \bibnamefont {Ying}},\ and\ \bibinfo {author} {\bibfnamefont {X.~H.}\
  \bibnamefont {Chen}},\ }\href {https://doi.org/10.1038/s41467-021-23928-w}
  {\bibfield  {journal} {\bibinfo  {journal} {Nat. Commun.}\ }\textbf {\bibinfo
  {volume} {12}},\ \bibinfo {pages} {3645} (\bibinfo {year}
  {2021}{\natexlab{b}})}\BibitemShut {NoStop}%
\bibitem [{\citenamefont {Zhang}\ \emph
  {et~al.}(2022{\natexlab{b}})\citenamefont {Zhang}, \citenamefont {Hou},
  \citenamefont {Xia}, \citenamefont {Xu}, \citenamefont {Yang}, \citenamefont
  {Wang}, \citenamefont {Liu}, \citenamefont {Shen}, \citenamefont {Zhang},
  \citenamefont {Dong}, \citenamefont {Uwatoko}, \citenamefont {Sun},
  \citenamefont {Wang}, \citenamefont {Guo},\ and\ \citenamefont
  {Cheng}}]{ZHA22a}%
  \BibitemOpen
  \bibfield  {author} {\bibinfo {author} {\bibfnamefont {X.}~\bibnamefont
  {Zhang}}, \bibinfo {author} {\bibfnamefont {J.}~\bibnamefont {Hou}}, \bibinfo
  {author} {\bibfnamefont {W.}~\bibnamefont {Xia}}, \bibinfo {author}
  {\bibfnamefont {Z.}~\bibnamefont {Xu}}, \bibinfo {author} {\bibfnamefont
  {P.}~\bibnamefont {Yang}}, \bibinfo {author} {\bibfnamefont {A.}~\bibnamefont
  {Wang}}, \bibinfo {author} {\bibfnamefont {Z.}~\bibnamefont {Liu}}, \bibinfo
  {author} {\bibfnamefont {J.}~\bibnamefont {Shen}}, \bibinfo {author}
  {\bibfnamefont {H.}~\bibnamefont {Zhang}}, \bibinfo {author} {\bibfnamefont
  {X.}~\bibnamefont {Dong}}, \bibinfo {author} {\bibfnamefont {Y.}~\bibnamefont
  {Uwatoko}}, \bibinfo {author} {\bibfnamefont {J.}~\bibnamefont {Sun}},
  \bibinfo {author} {\bibfnamefont {B.}~\bibnamefont {Wang}}, \bibinfo {author}
  {\bibfnamefont {Y.}~\bibnamefont {Guo}},\ and\ \bibinfo {author}
  {\bibfnamefont {J.}~\bibnamefont {Cheng}},\ }\href
  {https://doi.org/10.3390/ma15207372} {\bibfield  {journal} {\bibinfo
  {journal} {Materials}\ }\textbf {\bibinfo {volume} {15}},\ \bibinfo {pages}
  {7372} (\bibinfo {year} {2022}{\natexlab{b}})}\BibitemShut {NoStop}%
\bibitem [{\citenamefont {Pokharel}\ \emph {et~al.}(2021)\citenamefont
  {Pokharel}, \citenamefont {Teicher}, \citenamefont {Ortiz}, \citenamefont
  {Sarte}, \citenamefont {Wu}, \citenamefont {Peng}, \citenamefont {He},
  \citenamefont {Seshadri},\ and\ \citenamefont {Wilson}}]{POK21}%
  \BibitemOpen
  \bibfield  {author} {\bibinfo {author} {\bibfnamefont {G.}~\bibnamefont
  {Pokharel}}, \bibinfo {author} {\bibfnamefont {S.~M.~L.}\ \bibnamefont
  {Teicher}}, \bibinfo {author} {\bibfnamefont {B.~R.}\ \bibnamefont {Ortiz}},
  \bibinfo {author} {\bibfnamefont {P.~M.}\ \bibnamefont {Sarte}}, \bibinfo
  {author} {\bibfnamefont {G.}~\bibnamefont {Wu}}, \bibinfo {author}
  {\bibfnamefont {S.}~\bibnamefont {Peng}}, \bibinfo {author} {\bibfnamefont
  {J.}~\bibnamefont {He}}, \bibinfo {author} {\bibfnamefont {R.}~\bibnamefont
  {Seshadri}},\ and\ \bibinfo {author} {\bibfnamefont {S.~D.}\ \bibnamefont
  {Wilson}},\ }\href {https://doi.org/10.1103/PhysRevB.104.235139} {\bibfield
  {journal} {\bibinfo  {journal} {Phys. Rev. B}\ }\textbf {\bibinfo {volume}
  {104}},\ \bibinfo {pages} {235139} (\bibinfo {year} {2021})}\BibitemShut
  {NoStop}%
\bibitem [{\citenamefont {Romaka}\ \emph {et~al.}(2011)\citenamefont {Romaka},
  \citenamefont {Stadnyk}, \citenamefont {Romaka}, \citenamefont {Demchenko},
  \citenamefont {Stadnyshyn},\ and\ \citenamefont {Konyk}}]{ROM11}%
  \BibitemOpen
  \bibfield  {author} {\bibinfo {author} {\bibfnamefont {L.}~\bibnamefont
  {Romaka}}, \bibinfo {author} {\bibfnamefont {{\relax Yu}.}~\bibnamefont
  {Stadnyk}}, \bibinfo {author} {\bibfnamefont {V.~V.}\ \bibnamefont {Romaka}},
  \bibinfo {author} {\bibfnamefont {P.}~\bibnamefont {Demchenko}}, \bibinfo
  {author} {\bibfnamefont {M.}~\bibnamefont {Stadnyshyn}},\ and\ \bibinfo
  {author} {\bibfnamefont {M.}~\bibnamefont {Konyk}},\ }\href
  {https://doi.org/10.1016/j.jallcom.2011.06.095} {\bibfield  {journal}
  {\bibinfo  {journal} {J. Alloys Compd.}\ }\textbf {\bibinfo {volume} {509}},\
  \bibinfo {pages} {8862} (\bibinfo {year} {2011})}\BibitemShut {NoStop}%
\bibitem [{\citenamefont {Yin}\ \emph {et~al.}(2020)\citenamefont {Yin},
  \citenamefont {Ma}, \citenamefont {Cochran}, \citenamefont {Xu},
  \citenamefont {Zhang}, \citenamefont {Tien}, \citenamefont {Shumiya},
  \citenamefont {Cheng}, \citenamefont {Jiang}, \citenamefont {Lian},
  \citenamefont {Song}, \citenamefont {Chang}, \citenamefont {Belopolski},
  \citenamefont {Multer}, \citenamefont {Litskevich}, \citenamefont {Cheng},
  \citenamefont {Yang}, \citenamefont {Swidler}, \citenamefont {Zhou},
  \citenamefont {Lin}, \citenamefont {Neupert}, \citenamefont {Wang},
  \citenamefont {Yao}, \citenamefont {Chang}, \citenamefont {Jia},\ and\
  \citenamefont {Zahid~Hasan}}]{YIN20a}%
  \BibitemOpen
  \bibfield  {author} {\bibinfo {author} {\bibfnamefont {J.-X.}\ \bibnamefont
  {Yin}}, \bibinfo {author} {\bibfnamefont {W.}~\bibnamefont {Ma}}, \bibinfo
  {author} {\bibfnamefont {T.~A.}\ \bibnamefont {Cochran}}, \bibinfo {author}
  {\bibfnamefont {X.}~\bibnamefont {Xu}}, \bibinfo {author} {\bibfnamefont
  {S.~S.}\ \bibnamefont {Zhang}}, \bibinfo {author} {\bibfnamefont {H.-J.}\
  \bibnamefont {Tien}}, \bibinfo {author} {\bibfnamefont {N.}~\bibnamefont
  {Shumiya}}, \bibinfo {author} {\bibfnamefont {G.}~\bibnamefont {Cheng}},
  \bibinfo {author} {\bibfnamefont {K.}~\bibnamefont {Jiang}}, \bibinfo
  {author} {\bibfnamefont {B.}~\bibnamefont {Lian}}, \bibinfo {author}
  {\bibfnamefont {Z.}~\bibnamefont {Song}}, \bibinfo {author} {\bibfnamefont
  {G.}~\bibnamefont {Chang}}, \bibinfo {author} {\bibfnamefont
  {I.}~\bibnamefont {Belopolski}}, \bibinfo {author} {\bibfnamefont
  {D.}~\bibnamefont {Multer}}, \bibinfo {author} {\bibfnamefont
  {M.}~\bibnamefont {Litskevich}}, \bibinfo {author} {\bibfnamefont {Z.-J.}\
  \bibnamefont {Cheng}}, \bibinfo {author} {\bibfnamefont {X.~P.}\ \bibnamefont
  {Yang}}, \bibinfo {author} {\bibfnamefont {B.}~\bibnamefont {Swidler}},
  \bibinfo {author} {\bibfnamefont {H.}~\bibnamefont {Zhou}}, \bibinfo {author}
  {\bibfnamefont {H.}~\bibnamefont {Lin}}, \bibinfo {author} {\bibfnamefont
  {T.}~\bibnamefont {Neupert}}, \bibinfo {author} {\bibfnamefont
  {Z.}~\bibnamefont {Wang}}, \bibinfo {author} {\bibfnamefont {N.}~\bibnamefont
  {Yao}}, \bibinfo {author} {\bibfnamefont {T.-R.}\ \bibnamefont {Chang}},
  \bibinfo {author} {\bibfnamefont {S.}~\bibnamefont {Jia}},\ and\ \bibinfo
  {author} {\bibfnamefont {M.}~\bibnamefont {Zahid~Hasan}},\ }\href
  {https://doi.org/10.1038/s41586-020-2482-7} {\bibfield  {journal} {\bibinfo
  {journal} {Nature}\ }\textbf {\bibinfo {volume} {583}},\ \bibinfo {pages}
  {533} (\bibinfo {year} {2020})}\BibitemShut {NoStop}%
\bibitem [{\citenamefont {Chen}\ \emph
  {et~al.}(2021{\natexlab{c}})\citenamefont {Chen}, \citenamefont {Le},
  \citenamefont {Fu}, \citenamefont {Lin}, \citenamefont {Schnelle},
  \citenamefont {Sun},\ and\ \citenamefont {Felser}}]{CHE21c}%
  \BibitemOpen
  \bibfield  {author} {\bibinfo {author} {\bibfnamefont {D.}~\bibnamefont
  {Chen}}, \bibinfo {author} {\bibfnamefont {C.}~\bibnamefont {Le}}, \bibinfo
  {author} {\bibfnamefont {C.}~\bibnamefont {Fu}}, \bibinfo {author}
  {\bibfnamefont {H.}~\bibnamefont {Lin}}, \bibinfo {author} {\bibfnamefont
  {W.}~\bibnamefont {Schnelle}}, \bibinfo {author} {\bibfnamefont
  {Y.}~\bibnamefont {Sun}},\ and\ \bibinfo {author} {\bibfnamefont
  {C.}~\bibnamefont {Felser}},\ }\href
  {https://doi.org/10.1103/PhysRevB.103.144410} {\bibfield  {journal} {\bibinfo
   {journal} {Phys. Rev. B}\ }\textbf {\bibinfo {volume} {103}},\ \bibinfo
  {pages} {144410} (\bibinfo {year} {2021}{\natexlab{c}})}\BibitemShut
  {NoStop}%
\bibitem [{\citenamefont {Arachchige}\ \emph {et~al.}(2022)\citenamefont
  {Arachchige}, \citenamefont {Meier}, \citenamefont {Marshall}, \citenamefont
  {Matsuoka}, \citenamefont {Xue}, \citenamefont {McGuire}, \citenamefont
  {Hermann}, \citenamefont {Cao},\ and\ \citenamefont {Mandrus}}]{ARA22}%
  \BibitemOpen
  \bibfield  {author} {\bibinfo {author} {\bibfnamefont {H.~W.~S.}\
  \bibnamefont {Arachchige}}, \bibinfo {author} {\bibfnamefont {W.~R.}\
  \bibnamefont {Meier}}, \bibinfo {author} {\bibfnamefont {M.}~\bibnamefont
  {Marshall}}, \bibinfo {author} {\bibfnamefont {T.}~\bibnamefont {Matsuoka}},
  \bibinfo {author} {\bibfnamefont {R.}~\bibnamefont {Xue}}, \bibinfo {author}
  {\bibfnamefont {M.~A.}\ \bibnamefont {McGuire}}, \bibinfo {author}
  {\bibfnamefont {R.~P.}\ \bibnamefont {Hermann}}, \bibinfo {author}
  {\bibfnamefont {H.}~\bibnamefont {Cao}},\ and\ \bibinfo {author}
  {\bibfnamefont {D.}~\bibnamefont {Mandrus}},\ }\href
  {https://doi.org/10.1103/PhysRevLett.129.216402} {\bibfield  {journal}
  {\bibinfo  {journal} {Phys. Rev. Lett.}\ }\textbf {\bibinfo {volume} {129}},\
  \bibinfo {pages} {216402} (\bibinfo {year} {2022})}\BibitemShut {NoStop}%
\bibitem [{\citenamefont {Tan}\ and\ \citenamefont {Yan}(2023)}]{TAN23}%
  \BibitemOpen
  \bibfield  {author} {\bibinfo {author} {\bibfnamefont {H.}~\bibnamefont
  {Tan}}\ and\ \bibinfo {author} {\bibfnamefont {B.}~\bibnamefont {Yan}},\
  }\href {https://doi.org/10.48550/arXiv.2302.07922} {\bibinfo {title}
  {Abundant lattice instability in kagome metal {{ScV}}$_6${{Sn}}$_6$}}
  (\bibinfo {year} {2023}),\ \Eprint {https://arxiv.org/abs/2302.07922}
  {arxiv:2302.07922 [cond-mat]} \BibitemShut {NoStop}%
\bibitem [{\citenamefont {Hu}\ \emph {et~al.}(2023{\natexlab{a}})\citenamefont
  {Hu}, \citenamefont {Pi}, \citenamefont {Xu}, \citenamefont {Yue},
  \citenamefont {Wu}, \citenamefont {Liu}, \citenamefont {Zhang}, \citenamefont
  {Li}, \citenamefont {Zhou}, \citenamefont {Yuan}, \citenamefont {Wu},
  \citenamefont {Dong}, \citenamefont {Weng},\ and\ \citenamefont
  {Wang}}]{HU23d}%
  \BibitemOpen
  \bibfield  {author} {\bibinfo {author} {\bibfnamefont {T.}~\bibnamefont
  {Hu}}, \bibinfo {author} {\bibfnamefont {H.}~\bibnamefont {Pi}}, \bibinfo
  {author} {\bibfnamefont {S.}~\bibnamefont {Xu}}, \bibinfo {author}
  {\bibfnamefont {L.}~\bibnamefont {Yue}}, \bibinfo {author} {\bibfnamefont
  {Q.}~\bibnamefont {Wu}}, \bibinfo {author} {\bibfnamefont {Q.}~\bibnamefont
  {Liu}}, \bibinfo {author} {\bibfnamefont {S.}~\bibnamefont {Zhang}}, \bibinfo
  {author} {\bibfnamefont {R.}~\bibnamefont {Li}}, \bibinfo {author}
  {\bibfnamefont {X.}~\bibnamefont {Zhou}}, \bibinfo {author} {\bibfnamefont
  {J.}~\bibnamefont {Yuan}}, \bibinfo {author} {\bibfnamefont {D.}~\bibnamefont
  {Wu}}, \bibinfo {author} {\bibfnamefont {T.}~\bibnamefont {Dong}}, \bibinfo
  {author} {\bibfnamefont {H.}~\bibnamefont {Weng}},\ and\ \bibinfo {author}
  {\bibfnamefont {N.}~\bibnamefont {Wang}},\ }\href
  {https://doi.org/10.1103/PhysRevB.107.165119} {\bibfield  {journal} {\bibinfo
   {journal} {Phys. Rev. B}\ }\textbf {\bibinfo {volume} {107}},\ \bibinfo
  {pages} {165119} (\bibinfo {year} {2023}{\natexlab{a}})}\BibitemShut
  {NoStop}%
\bibitem [{\citenamefont {Lee}\ \emph {et~al.}(2023)\citenamefont {Lee},
  \citenamefont {Won}, \citenamefont {Kim}, \citenamefont {Yoo}, \citenamefont
  {Park}, \citenamefont {Denlinger}, \citenamefont {Jozwiak}, \citenamefont
  {Bostwick}, \citenamefont {Rotenberg}, \citenamefont {Comin}, \citenamefont
  {Kang},\ and\ \citenamefont {Park}}]{LEE23}%
  \BibitemOpen
  \bibfield  {author} {\bibinfo {author} {\bibfnamefont {S.}~\bibnamefont
  {Lee}}, \bibinfo {author} {\bibfnamefont {C.}~\bibnamefont {Won}}, \bibinfo
  {author} {\bibfnamefont {J.}~\bibnamefont {Kim}}, \bibinfo {author}
  {\bibfnamefont {J.}~\bibnamefont {Yoo}}, \bibinfo {author} {\bibfnamefont
  {S.}~\bibnamefont {Park}}, \bibinfo {author} {\bibfnamefont {J.}~\bibnamefont
  {Denlinger}}, \bibinfo {author} {\bibfnamefont {C.}~\bibnamefont {Jozwiak}},
  \bibinfo {author} {\bibfnamefont {A.}~\bibnamefont {Bostwick}}, \bibinfo
  {author} {\bibfnamefont {E.}~\bibnamefont {Rotenberg}}, \bibinfo {author}
  {\bibfnamefont {R.}~\bibnamefont {Comin}}, \bibinfo {author} {\bibfnamefont
  {M.}~\bibnamefont {Kang}},\ and\ \bibinfo {author} {\bibfnamefont {J.-H.}\
  \bibnamefont {Park}},\ }\href {https://doi.org/10.48550/arXiv.2304.11820}
  {\bibinfo {title} {Nature of charge density wave in kagome metal
  {{ScV}}$_6${{Sn}}$_6$}} (\bibinfo {year} {2023}),\ \Eprint
  {https://arxiv.org/abs/2304.11820} {arxiv:2304.11820 [cond-mat]} \BibitemShut
  {NoStop}%
\bibitem [{\citenamefont {Tuniz}\ \emph {et~al.}(2023)\citenamefont {Tuniz},
  \citenamefont {Consiglio}, \citenamefont {Puntel}, \citenamefont {Bigi},
  \citenamefont {Enzner}, \citenamefont {Pokharel}, \citenamefont {Orgiani},
  \citenamefont {Bronsch}, \citenamefont {Parmigiani}, \citenamefont
  {Polewczyk}, \citenamefont {King}, \citenamefont {Wells}, \citenamefont
  {Zeljkovic}, \citenamefont {Carrara}, \citenamefont {Rossi}, \citenamefont
  {Fujii}, \citenamefont {Vobornik}, \citenamefont {Wilson}, \citenamefont
  {Thomale}, \citenamefont {Wehling}, \citenamefont {Sangiovanni},
  \citenamefont {Panaccione}, \citenamefont {Cilento}, \citenamefont
  {Di~Sante},\ and\ \citenamefont {Mazzola}}]{TUN23}%
  \BibitemOpen
  \bibfield  {author} {\bibinfo {author} {\bibfnamefont {M.}~\bibnamefont
  {Tuniz}}, \bibinfo {author} {\bibfnamefont {A.}~\bibnamefont {Consiglio}},
  \bibinfo {author} {\bibfnamefont {D.}~\bibnamefont {Puntel}}, \bibinfo
  {author} {\bibfnamefont {C.}~\bibnamefont {Bigi}}, \bibinfo {author}
  {\bibfnamefont {S.}~\bibnamefont {Enzner}}, \bibinfo {author} {\bibfnamefont
  {G.}~\bibnamefont {Pokharel}}, \bibinfo {author} {\bibfnamefont
  {P.}~\bibnamefont {Orgiani}}, \bibinfo {author} {\bibfnamefont
  {W.}~\bibnamefont {Bronsch}}, \bibinfo {author} {\bibfnamefont
  {F.}~\bibnamefont {Parmigiani}}, \bibinfo {author} {\bibfnamefont
  {V.}~\bibnamefont {Polewczyk}}, \bibinfo {author} {\bibfnamefont {P.~D.~C.}\
  \bibnamefont {King}}, \bibinfo {author} {\bibfnamefont {J.~W.}\ \bibnamefont
  {Wells}}, \bibinfo {author} {\bibfnamefont {I.}~\bibnamefont {Zeljkovic}},
  \bibinfo {author} {\bibfnamefont {P.}~\bibnamefont {Carrara}}, \bibinfo
  {author} {\bibfnamefont {G.}~\bibnamefont {Rossi}}, \bibinfo {author}
  {\bibfnamefont {J.}~\bibnamefont {Fujii}}, \bibinfo {author} {\bibfnamefont
  {I.}~\bibnamefont {Vobornik}}, \bibinfo {author} {\bibfnamefont {S.~D.}\
  \bibnamefont {Wilson}}, \bibinfo {author} {\bibfnamefont {R.}~\bibnamefont
  {Thomale}}, \bibinfo {author} {\bibfnamefont {T.}~\bibnamefont {Wehling}},
  \bibinfo {author} {\bibfnamefont {G.}~\bibnamefont {Sangiovanni}}, \bibinfo
  {author} {\bibfnamefont {G.}~\bibnamefont {Panaccione}}, \bibinfo {author}
  {\bibfnamefont {F.}~\bibnamefont {Cilento}}, \bibinfo {author} {\bibfnamefont
  {D.}~\bibnamefont {Di~Sante}},\ and\ \bibinfo {author} {\bibfnamefont
  {F.}~\bibnamefont {Mazzola}},\ }\href
  {https://doi.org/10.48550/arXiv.2302.10699} {\bibinfo {title} {Dynamics and
  {{Resilience}} of the {{Charge Density Wave}} in a bilayer kagome metal}}
  (\bibinfo {year} {2023}),\ \Eprint {https://arxiv.org/abs/2302.10699}
  {arxiv:2302.10699 [cond-mat]} \BibitemShut {NoStop}%
\bibitem [{\citenamefont {Cao}\ \emph {et~al.}(2023)\citenamefont {Cao},
  \citenamefont {Xu}, \citenamefont {Fukui}, \citenamefont {Manjo},
  \citenamefont {Shi}, \citenamefont {Liu}, \citenamefont {Cao},\ and\
  \citenamefont {Song}}]{CAO23}%
  \BibitemOpen
  \bibfield  {author} {\bibinfo {author} {\bibfnamefont {S.}~\bibnamefont
  {Cao}}, \bibinfo {author} {\bibfnamefont {C.}~\bibnamefont {Xu}}, \bibinfo
  {author} {\bibfnamefont {H.}~\bibnamefont {Fukui}}, \bibinfo {author}
  {\bibfnamefont {T.}~\bibnamefont {Manjo}}, \bibinfo {author} {\bibfnamefont
  {M.}~\bibnamefont {Shi}}, \bibinfo {author} {\bibfnamefont {Y.}~\bibnamefont
  {Liu}}, \bibinfo {author} {\bibfnamefont {C.}~\bibnamefont {Cao}},\ and\
  \bibinfo {author} {\bibfnamefont {Y.}~\bibnamefont {Song}},\ }\href
  {https://doi.org/10.48550/arXiv.2304.08197} {\bibinfo {title} {Competing
  charge-density wave instabilities in the kagome metal {{ScV}}$_6${{Sn}}$_6$}}
  (\bibinfo {year} {2023}),\ \Eprint {https://arxiv.org/abs/2304.08197}
  {arxiv:2304.08197 [cond-mat]} \BibitemShut {NoStop}%
\bibitem [{\citenamefont {Hu}\ \emph {et~al.}(2023{\natexlab{b}})\citenamefont
  {Hu}, \citenamefont {Ma}, \citenamefont {Li}, \citenamefont {Gawryluk},
  \citenamefont {Hu}, \citenamefont {Teyssier}, \citenamefont {Multian},
  \citenamefont {Yin}, \citenamefont {Jiang}, \citenamefont {Xu}, \citenamefont
  {Shin}, \citenamefont {Plokhikh}, \citenamefont {Han}, \citenamefont {Plumb},
  \citenamefont {Liu}, \citenamefont {Yin}, \citenamefont {Guguchia},
  \citenamefont {Zhao}, \citenamefont {Schnyder}, \citenamefont {Wu},
  \citenamefont {Pomjakushina}, \citenamefont {Hasan}, \citenamefont {Wang},\
  and\ \citenamefont {Shi}}]{HU23e}%
  \BibitemOpen
  \bibfield  {author} {\bibinfo {author} {\bibfnamefont {Y.}~\bibnamefont
  {Hu}}, \bibinfo {author} {\bibfnamefont {J.}~\bibnamefont {Ma}}, \bibinfo
  {author} {\bibfnamefont {Y.}~\bibnamefont {Li}}, \bibinfo {author}
  {\bibfnamefont {D.~J.}\ \bibnamefont {Gawryluk}}, \bibinfo {author}
  {\bibfnamefont {T.}~\bibnamefont {Hu}}, \bibinfo {author} {\bibfnamefont
  {J.}~\bibnamefont {Teyssier}}, \bibinfo {author} {\bibfnamefont
  {V.}~\bibnamefont {Multian}}, \bibinfo {author} {\bibfnamefont
  {Z.}~\bibnamefont {Yin}}, \bibinfo {author} {\bibfnamefont {Y.}~\bibnamefont
  {Jiang}}, \bibinfo {author} {\bibfnamefont {S.}~\bibnamefont {Xu}}, \bibinfo
  {author} {\bibfnamefont {S.}~\bibnamefont {Shin}}, \bibinfo {author}
  {\bibfnamefont {I.}~\bibnamefont {Plokhikh}}, \bibinfo {author}
  {\bibfnamefont {X.}~\bibnamefont {Han}}, \bibinfo {author} {\bibfnamefont
  {N.~C.}\ \bibnamefont {Plumb}}, \bibinfo {author} {\bibfnamefont
  {Y.}~\bibnamefont {Liu}}, \bibinfo {author} {\bibfnamefont {J.}~\bibnamefont
  {Yin}}, \bibinfo {author} {\bibfnamefont {Z.}~\bibnamefont {Guguchia}},
  \bibinfo {author} {\bibfnamefont {Y.}~\bibnamefont {Zhao}}, \bibinfo {author}
  {\bibfnamefont {A.~P.}\ \bibnamefont {Schnyder}}, \bibinfo {author}
  {\bibfnamefont {X.}~\bibnamefont {Wu}}, \bibinfo {author} {\bibfnamefont
  {E.}~\bibnamefont {Pomjakushina}}, \bibinfo {author} {\bibfnamefont {M.~Z.}\
  \bibnamefont {Hasan}}, \bibinfo {author} {\bibfnamefont {N.}~\bibnamefont
  {Wang}},\ and\ \bibinfo {author} {\bibfnamefont {M.}~\bibnamefont {Shi}},\
  }\href {https://doi.org/10.48550/arXiv.2304.06431} {\bibinfo {title} {Phonon
  promoted charge density wave in topological kagome metal
  {{ScV}}$_6${{Sn}}$_6$}} (\bibinfo {year} {2023}{\natexlab{b}}),\ \Eprint
  {https://arxiv.org/abs/2304.06431} {arxiv:2304.06431 [cond-mat]} \BibitemShut
  {NoStop}%
\bibitem [{\citenamefont {Kang}\ \emph
  {et~al.}(2023{\natexlab{b}})\citenamefont {Kang}, \citenamefont {Li},
  \citenamefont {Meier}, \citenamefont {Villanova}, \citenamefont {Hus},
  \citenamefont {Jeon}, \citenamefont {Arachchige}, \citenamefont {Lu},
  \citenamefont {Gai}, \citenamefont {Denlinger}, \citenamefont {Moore},
  \citenamefont {Yoon},\ and\ \citenamefont {Mandrus}}]{KAN23}%
  \BibitemOpen
  \bibfield  {author} {\bibinfo {author} {\bibfnamefont {S.-H.}\ \bibnamefont
  {Kang}}, \bibinfo {author} {\bibfnamefont {H.}~\bibnamefont {Li}}, \bibinfo
  {author} {\bibfnamefont {W.~R.}\ \bibnamefont {Meier}}, \bibinfo {author}
  {\bibfnamefont {J.~W.}\ \bibnamefont {Villanova}}, \bibinfo {author}
  {\bibfnamefont {S.}~\bibnamefont {Hus}}, \bibinfo {author} {\bibfnamefont
  {H.}~\bibnamefont {Jeon}}, \bibinfo {author} {\bibfnamefont {H.~W.~S.}\
  \bibnamefont {Arachchige}}, \bibinfo {author} {\bibfnamefont
  {Q.}~\bibnamefont {Lu}}, \bibinfo {author} {\bibfnamefont {Z.}~\bibnamefont
  {Gai}}, \bibinfo {author} {\bibfnamefont {J.}~\bibnamefont {Denlinger}},
  \bibinfo {author} {\bibfnamefont {R.}~\bibnamefont {Moore}}, \bibinfo
  {author} {\bibfnamefont {M.}~\bibnamefont {Yoon}},\ and\ \bibinfo {author}
  {\bibfnamefont {D.}~\bibnamefont {Mandrus}},\ }\href
  {https://doi.org/10.48550/arXiv.2302.14041} {\bibinfo {title} {Emergence of a
  new band and the {{Lifshitz}} transition in kagome metal
  {{ScV}}$_6${{Sn}}$_6$ with charge density wave}} (\bibinfo {year}
  {2023}{\natexlab{b}}),\ \Eprint {https://arxiv.org/abs/2302.14041}
  {arxiv:2302.14041 [cond-mat]} \BibitemShut {NoStop}%
\bibitem [{\citenamefont {Gu}\ \emph {et~al.}(2023)\citenamefont {Gu},
  \citenamefont {Ritz}, \citenamefont {Meier}, \citenamefont {Blockmon},
  \citenamefont {Smith}, \citenamefont {Madhogaria}, \citenamefont {Mozaffari},
  \citenamefont {Mandrus}, \citenamefont {Birol},\ and\ \citenamefont
  {Musfeldt}}]{GU23}%
  \BibitemOpen
  \bibfield  {author} {\bibinfo {author} {\bibfnamefont {Y.}~\bibnamefont
  {Gu}}, \bibinfo {author} {\bibfnamefont {E.}~\bibnamefont {Ritz}}, \bibinfo
  {author} {\bibfnamefont {W.~R.}\ \bibnamefont {Meier}}, \bibinfo {author}
  {\bibfnamefont {A.}~\bibnamefont {Blockmon}}, \bibinfo {author}
  {\bibfnamefont {K.}~\bibnamefont {Smith}}, \bibinfo {author} {\bibfnamefont
  {R.~P.}\ \bibnamefont {Madhogaria}}, \bibinfo {author} {\bibfnamefont
  {S.}~\bibnamefont {Mozaffari}}, \bibinfo {author} {\bibfnamefont
  {D.}~\bibnamefont {Mandrus}}, \bibinfo {author} {\bibfnamefont
  {T.}~\bibnamefont {Birol}},\ and\ \bibinfo {author} {\bibfnamefont {J.~L.}\
  \bibnamefont {Musfeldt}},\ }\href {https://doi.org/10.48550/arXiv.2305.01086}
  {\bibinfo {title} {Origin and stability of the charge density wave in
  {{ScV}}$_6${{Sn}}$_6$}} (\bibinfo {year} {2023}),\ \Eprint
  {https://arxiv.org/abs/2305.01086} {arxiv:2305.01086 [cond-mat]} \BibitemShut
  {NoStop}%
\bibitem [{\citenamefont {Guguchia}\ \emph {et~al.}(2023)\citenamefont
  {Guguchia}, \citenamefont {Gawryluk}, \citenamefont {Shin}, \citenamefont
  {Hao}, \citenamefont {Mielke~III}, \citenamefont {Das}, \citenamefont
  {Plokhikh}, \citenamefont {Liborio}, \citenamefont {Shenton}, \citenamefont
  {Hu}, \citenamefont {Sazgari}, \citenamefont {Medarde}, \citenamefont {Deng},
  \citenamefont {Cai}, \citenamefont {Chen}, \citenamefont {Jiang},
  \citenamefont {Amato}, \citenamefont {Shi}, \citenamefont {Hasan},
  \citenamefont {Yin}, \citenamefont {Khasanov}, \citenamefont {Pomjakushina},\
  and\ \citenamefont {Luetkens}}]{GUG23}%
  \BibitemOpen
  \bibfield  {author} {\bibinfo {author} {\bibfnamefont {Z.}~\bibnamefont
  {Guguchia}}, \bibinfo {author} {\bibfnamefont {D.~J.}\ \bibnamefont
  {Gawryluk}}, \bibinfo {author} {\bibfnamefont {S.}~\bibnamefont {Shin}},
  \bibinfo {author} {\bibfnamefont {Z.}~\bibnamefont {Hao}}, \bibinfo {author}
  {\bibfnamefont {C.}~\bibnamefont {Mielke~III}}, \bibinfo {author}
  {\bibfnamefont {D.}~\bibnamefont {Das}}, \bibinfo {author} {\bibfnamefont
  {I.}~\bibnamefont {Plokhikh}}, \bibinfo {author} {\bibfnamefont
  {L.}~\bibnamefont {Liborio}}, \bibinfo {author} {\bibfnamefont
  {K.}~\bibnamefont {Shenton}}, \bibinfo {author} {\bibfnamefont
  {Y.}~\bibnamefont {Hu}}, \bibinfo {author} {\bibfnamefont {V.}~\bibnamefont
  {Sazgari}}, \bibinfo {author} {\bibfnamefont {M.}~\bibnamefont {Medarde}},
  \bibinfo {author} {\bibfnamefont {H.}~\bibnamefont {Deng}}, \bibinfo {author}
  {\bibfnamefont {Y.}~\bibnamefont {Cai}}, \bibinfo {author} {\bibfnamefont
  {C.}~\bibnamefont {Chen}}, \bibinfo {author} {\bibfnamefont {Y.}~\bibnamefont
  {Jiang}}, \bibinfo {author} {\bibfnamefont {A.}~\bibnamefont {Amato}},
  \bibinfo {author} {\bibfnamefont {M.}~\bibnamefont {Shi}}, \bibinfo {author}
  {\bibfnamefont {M.~Z.}\ \bibnamefont {Hasan}}, \bibinfo {author}
  {\bibfnamefont {J.-X.}\ \bibnamefont {Yin}}, \bibinfo {author} {\bibfnamefont
  {R.}~\bibnamefont {Khasanov}}, \bibinfo {author} {\bibfnamefont
  {E.}~\bibnamefont {Pomjakushina}},\ and\ \bibinfo {author} {\bibfnamefont
  {H.}~\bibnamefont {Luetkens}},\ }\href
  {https://doi.org/10.48550/arXiv.2304.06436} {\bibinfo {title} {Hidden
  magnetism uncovered in charge ordered bilayer kagome material
  {{ScV}}$_6${{Sn}}$_6$}} (\bibinfo {year} {2023}),\ \Eprint
  {https://arxiv.org/abs/2304.06436} {arxiv:2304.06436 [cond-mat]} \BibitemShut
  {NoStop}%
\bibitem [{\citenamefont {Cheng}\ \emph {et~al.}(2023)\citenamefont {Cheng},
  \citenamefont {Ren}, \citenamefont {Li}, \citenamefont {Oh}, \citenamefont
  {Tan}, \citenamefont {Pokharel}, \citenamefont {DeStefano}, \citenamefont
  {Rosenberg}, \citenamefont {Guo}, \citenamefont {Zhang}, \citenamefont {Yue},
  \citenamefont {Lee}, \citenamefont {Gorovikov}, \citenamefont {Zonno},
  \citenamefont {Hashimoto}, \citenamefont {Lu}, \citenamefont {Ke},
  \citenamefont {Mazzola}, \citenamefont {Kono}, \citenamefont {Birgeneau},
  \citenamefont {Chu}, \citenamefont {Wilson}, \citenamefont {Wang},
  \citenamefont {Yan}, \citenamefont {Yi},\ and\ \citenamefont
  {Zeljkovic}}]{CHE23}%
  \BibitemOpen
  \bibfield  {author} {\bibinfo {author} {\bibfnamefont {S.}~\bibnamefont
  {Cheng}}, \bibinfo {author} {\bibfnamefont {Z.}~\bibnamefont {Ren}}, \bibinfo
  {author} {\bibfnamefont {H.}~\bibnamefont {Li}}, \bibinfo {author}
  {\bibfnamefont {J.}~\bibnamefont {Oh}}, \bibinfo {author} {\bibfnamefont
  {H.}~\bibnamefont {Tan}}, \bibinfo {author} {\bibfnamefont {G.}~\bibnamefont
  {Pokharel}}, \bibinfo {author} {\bibfnamefont {J.~M.}\ \bibnamefont
  {DeStefano}}, \bibinfo {author} {\bibfnamefont {E.}~\bibnamefont
  {Rosenberg}}, \bibinfo {author} {\bibfnamefont {Y.}~\bibnamefont {Guo}},
  \bibinfo {author} {\bibfnamefont {Y.}~\bibnamefont {Zhang}}, \bibinfo
  {author} {\bibfnamefont {Z.}~\bibnamefont {Yue}}, \bibinfo {author}
  {\bibfnamefont {Y.}~\bibnamefont {Lee}}, \bibinfo {author} {\bibfnamefont
  {S.}~\bibnamefont {Gorovikov}}, \bibinfo {author} {\bibfnamefont
  {M.}~\bibnamefont {Zonno}}, \bibinfo {author} {\bibfnamefont
  {M.}~\bibnamefont {Hashimoto}}, \bibinfo {author} {\bibfnamefont
  {D.}~\bibnamefont {Lu}}, \bibinfo {author} {\bibfnamefont {L.}~\bibnamefont
  {Ke}}, \bibinfo {author} {\bibfnamefont {F.}~\bibnamefont {Mazzola}},
  \bibinfo {author} {\bibfnamefont {J.}~\bibnamefont {Kono}}, \bibinfo {author}
  {\bibfnamefont {R.~J.}\ \bibnamefont {Birgeneau}}, \bibinfo {author}
  {\bibfnamefont {J.-H.}\ \bibnamefont {Chu}}, \bibinfo {author} {\bibfnamefont
  {S.~D.}\ \bibnamefont {Wilson}}, \bibinfo {author} {\bibfnamefont
  {Z.}~\bibnamefont {Wang}}, \bibinfo {author} {\bibfnamefont {B.}~\bibnamefont
  {Yan}}, \bibinfo {author} {\bibfnamefont {M.}~\bibnamefont {Yi}},\ and\
  \bibinfo {author} {\bibfnamefont {I.}~\bibnamefont {Zeljkovic}},\ }\href
  {https://doi.org/10.48550/arXiv.2302.12227} {\bibinfo {title} {Nanoscale
  visualization and spectral fingerprints of the charge order in {{ScV6Sn6}}
  distinct from other kagome metals}} (\bibinfo {year} {2023}),\ \Eprint
  {https://arxiv.org/abs/2302.12227} {arxiv:2302.12227 [cond-mat]} \BibitemShut
  {NoStop}%
\bibitem [{\citenamefont {Mozaffari}\ \emph {et~al.}(2023)\citenamefont
  {Mozaffari}, \citenamefont {Meier}, \citenamefont {Madhogaria}, \citenamefont
  {Kang}, \citenamefont {Villanova}, \citenamefont {Arachchige}, \citenamefont
  {Zheng}, \citenamefont {Zhu}, \citenamefont {Chen}, \citenamefont {Jenkins},
  \citenamefont {Zhang}, \citenamefont {Chan}, \citenamefont {Li},
  \citenamefont {Yoon}, \citenamefont {Zhang},\ and\ \citenamefont
  {Mandrus}}]{MOZ23}%
  \BibitemOpen
  \bibfield  {author} {\bibinfo {author} {\bibfnamefont {S.}~\bibnamefont
  {Mozaffari}}, \bibinfo {author} {\bibfnamefont {W.~R.}\ \bibnamefont
  {Meier}}, \bibinfo {author} {\bibfnamefont {R.~P.}\ \bibnamefont
  {Madhogaria}}, \bibinfo {author} {\bibfnamefont {S.-H.}\ \bibnamefont
  {Kang}}, \bibinfo {author} {\bibfnamefont {J.~W.}\ \bibnamefont {Villanova}},
  \bibinfo {author} {\bibfnamefont {H.~W.~S.}\ \bibnamefont {Arachchige}},
  \bibinfo {author} {\bibfnamefont {G.}~\bibnamefont {Zheng}}, \bibinfo
  {author} {\bibfnamefont {Y.}~\bibnamefont {Zhu}}, \bibinfo {author}
  {\bibfnamefont {K.-W.}\ \bibnamefont {Chen}}, \bibinfo {author}
  {\bibfnamefont {K.}~\bibnamefont {Jenkins}}, \bibinfo {author} {\bibfnamefont
  {D.}~\bibnamefont {Zhang}}, \bibinfo {author} {\bibfnamefont
  {A.}~\bibnamefont {Chan}}, \bibinfo {author} {\bibfnamefont {L.}~\bibnamefont
  {Li}}, \bibinfo {author} {\bibfnamefont {M.}~\bibnamefont {Yoon}}, \bibinfo
  {author} {\bibfnamefont {Y.}~\bibnamefont {Zhang}},\ and\ \bibinfo {author}
  {\bibfnamefont {D.~G.}\ \bibnamefont {Mandrus}},\ }\href
  {https://doi.org/10.48550/arXiv.2305.02393} {\bibinfo {title} {Universal
  sublinear resistivity in vanadium kagome materials hosting charge density
  waves}} (\bibinfo {year} {2023}),\ \Eprint {https://arxiv.org/abs/2305.02393}
  {arxiv:2305.02393 [cond-mat]} \BibitemShut {NoStop}%
\bibitem [{\citenamefont {Yi}\ \emph {et~al.}(2023)\citenamefont {Yi},
  \citenamefont {Feng}, \citenamefont {Yanda}, \citenamefont {Roychowdhury},
  \citenamefont {Felser},\ and\ \citenamefont {Shekhar}}]{YI23}%
  \BibitemOpen
  \bibfield  {author} {\bibinfo {author} {\bibfnamefont {C.}~\bibnamefont
  {Yi}}, \bibinfo {author} {\bibfnamefont {X.}~\bibnamefont {Feng}}, \bibinfo
  {author} {\bibfnamefont {P.}~\bibnamefont {Yanda}}, \bibinfo {author}
  {\bibfnamefont {S.}~\bibnamefont {Roychowdhury}}, \bibinfo {author}
  {\bibfnamefont {C.}~\bibnamefont {Felser}},\ and\ \bibinfo {author}
  {\bibfnamefont {C.}~\bibnamefont {Shekhar}},\ }\href
  {https://doi.org/10.48550/arXiv.2305.04683} {\bibinfo {title} {Charge density
  wave induced anomalous {{Hall}} effect in kagome {{ScV}}$_6${{Sn}}$_6$}}
  (\bibinfo {year} {2023}),\ \Eprint {https://arxiv.org/abs/2305.04683}
  {arxiv:2305.04683 [cond-mat]} \BibitemShut {NoStop}%
\bibitem [{\citenamefont {Su}\ \emph {et~al.}(1979)\citenamefont {Su},
  \citenamefont {Schrieffer},\ and\ \citenamefont {Heeger}}]{SU79}%
  \BibitemOpen
  \bibfield  {author} {\bibinfo {author} {\bibfnamefont {W.~P.}\ \bibnamefont
  {Su}}, \bibinfo {author} {\bibfnamefont {J.~R.}\ \bibnamefont {Schrieffer}},\
  and\ \bibinfo {author} {\bibfnamefont {A.~J.}\ \bibnamefont {Heeger}},\
  }\href {https://doi.org/10.1103/PhysRevLett.42.1698} {\bibfield  {journal}
  {\bibinfo  {journal} {Phys. Rev. Lett.}\ }\textbf {\bibinfo {volume} {42}},\
  \bibinfo {pages} {1698} (\bibinfo {year} {1979})}\BibitemShut {NoStop}%
\bibitem [{zot()}]{zotero-4159}%
  \BibitemOpen
  \href@noop {} {\bibinfo {title} {Bilbao {{Crystallographic Server}}}},\
  \bibinfo {howpublished} {https://www.cryst.ehu.es/}\BibitemShut {NoStop}%
\bibitem [{\citenamefont {Bradlyn}\ \emph {et~al.}(2017)\citenamefont
  {Bradlyn}, \citenamefont {Elcoro}, \citenamefont {Cano}, \citenamefont
  {Vergniory}, \citenamefont {Wang}, \citenamefont {Felser}, \citenamefont
  {Aroyo},\ and\ \citenamefont {Bernevig}}]{BRA17}%
  \BibitemOpen
  \bibfield  {author} {\bibinfo {author} {\bibfnamefont {B.}~\bibnamefont
  {Bradlyn}}, \bibinfo {author} {\bibfnamefont {L.}~\bibnamefont {Elcoro}},
  \bibinfo {author} {\bibfnamefont {J.}~\bibnamefont {Cano}}, \bibinfo {author}
  {\bibfnamefont {M.~G.}\ \bibnamefont {Vergniory}}, \bibinfo {author}
  {\bibfnamefont {Z.}~\bibnamefont {Wang}}, \bibinfo {author} {\bibfnamefont
  {C.}~\bibnamefont {Felser}}, \bibinfo {author} {\bibfnamefont {M.~I.}\
  \bibnamefont {Aroyo}},\ and\ \bibinfo {author} {\bibfnamefont {B.~A.}\
  \bibnamefont {Bernevig}},\ }\href {https://doi.org/10.1038/nature23268}
  {\bibfield  {journal} {\bibinfo  {journal} {Nature}\ }\textbf {\bibinfo
  {volume} {547}},\ \bibinfo {pages} {298} (\bibinfo {year}
  {2017})}\BibitemShut {NoStop}%
\bibitem [{\citenamefont {Elcoro}\ \emph {et~al.}(2017)\citenamefont {Elcoro},
  \citenamefont {Bradlyn}, \citenamefont {Wang}, \citenamefont {Vergniory},
  \citenamefont {Cano}, \citenamefont {Felser}, \citenamefont {Bernevig},
  \citenamefont {Orobengoa}, \citenamefont {de~la Flor},\ and\ \citenamefont
  {Aroyo}}]{ELC17}%
  \BibitemOpen
  \bibfield  {author} {\bibinfo {author} {\bibfnamefont {L.}~\bibnamefont
  {Elcoro}}, \bibinfo {author} {\bibfnamefont {B.}~\bibnamefont {Bradlyn}},
  \bibinfo {author} {\bibfnamefont {Z.}~\bibnamefont {Wang}}, \bibinfo {author}
  {\bibfnamefont {M.~G.}\ \bibnamefont {Vergniory}}, \bibinfo {author}
  {\bibfnamefont {J.}~\bibnamefont {Cano}}, \bibinfo {author} {\bibfnamefont
  {C.}~\bibnamefont {Felser}}, \bibinfo {author} {\bibfnamefont {B.~A.}\
  \bibnamefont {Bernevig}}, \bibinfo {author} {\bibfnamefont {D.}~\bibnamefont
  {Orobengoa}}, \bibinfo {author} {\bibfnamefont {G.}~\bibnamefont {de~la
  Flor}},\ and\ \bibinfo {author} {\bibfnamefont {M.~I.}\ \bibnamefont
  {Aroyo}},\ }\href {https://doi.org/10.1107/S1600576717011712} {\bibfield
  {journal} {\bibinfo  {journal} {J. Appl. Cryst.}\ }\textbf {\bibinfo {volume}
  {50}},\ \bibinfo {pages} {1457} (\bibinfo {year} {2017})}\BibitemShut
  {NoStop}%
\bibitem [{\citenamefont {Vergniory}\ \emph {et~al.}(2017)\citenamefont
  {Vergniory}, \citenamefont {Elcoro}, \citenamefont {Wang}, \citenamefont
  {Cano}, \citenamefont {Felser}, \citenamefont {Aroyo}, \citenamefont
  {Bernevig},\ and\ \citenamefont {Bradlyn}}]{VER17}%
  \BibitemOpen
  \bibfield  {author} {\bibinfo {author} {\bibfnamefont {M.~G.}\ \bibnamefont
  {Vergniory}}, \bibinfo {author} {\bibfnamefont {L.}~\bibnamefont {Elcoro}},
  \bibinfo {author} {\bibfnamefont {Z.}~\bibnamefont {Wang}}, \bibinfo {author}
  {\bibfnamefont {J.}~\bibnamefont {Cano}}, \bibinfo {author} {\bibfnamefont
  {C.}~\bibnamefont {Felser}}, \bibinfo {author} {\bibfnamefont {M.~I.}\
  \bibnamefont {Aroyo}}, \bibinfo {author} {\bibfnamefont {B.~A.}\ \bibnamefont
  {Bernevig}},\ and\ \bibinfo {author} {\bibfnamefont {B.}~\bibnamefont
  {Bradlyn}},\ }\href {https://doi.org/10.1103/PhysRevE.96.023310} {\bibfield
  {journal} {\bibinfo  {journal} {Phys. Rev. E}\ }\textbf {\bibinfo {volume}
  {96}},\ \bibinfo {pages} {023310} (\bibinfo {year} {2017})}\BibitemShut
  {NoStop}%
\bibitem [{\citenamefont {Auerbach}(1994)}]{AUE94}%
  \BibitemOpen
  \bibfield  {author} {\bibinfo {author} {\bibfnamefont {A.}~\bibnamefont
  {Auerbach}},\ }\href {https://doi.org/10.1007/978-1-4612-0869-3} {\emph
  {\bibinfo {title} {Interacting {{Electrons}} and {{Quantum Magnetism}}}}},\
  edited by\ \bibinfo {editor} {\bibfnamefont {J.~L.}\ \bibnamefont {Birman}},
  \bibinfo {editor} {\bibfnamefont {J.~W.}\ \bibnamefont {Lynn}}, \bibinfo
  {editor} {\bibfnamefont {M.~P.}\ \bibnamefont {Silverman}}, \bibinfo {editor}
  {\bibfnamefont {H.~E.}\ \bibnamefont {Stanley}},\ and\ \bibinfo {editor}
  {\bibfnamefont {M.}~\bibnamefont {Voloshin}},\ Graduate {{Texts}} in
  {{Contemporary Physics}}\ (\bibinfo  {publisher} {{Springer New York}},\
  \bibinfo {address} {{New York, NY}},\ \bibinfo {year} {1994})\BibitemShut
  {NoStop}%
\bibitem [{\citenamefont {Kresse}\ and\ \citenamefont {Hafner}(1994)}]{KRE94}%
  \BibitemOpen
  \bibfield  {author} {\bibinfo {author} {\bibfnamefont {G.}~\bibnamefont
  {Kresse}}\ and\ \bibinfo {author} {\bibfnamefont {J.}~\bibnamefont
  {Hafner}},\ }\href {https://doi.org/10.1103/PhysRevB.49.14251} {\bibfield
  {journal} {\bibinfo  {journal} {Phys. Rev. B}\ }\textbf {\bibinfo {volume}
  {49}},\ \bibinfo {pages} {14251} (\bibinfo {year} {1994})}\BibitemShut
  {NoStop}%
\bibitem [{\citenamefont {Kresse}\ and\ \citenamefont
  {Furthm{\"u}ller}(1996{\natexlab{a}})}]{KRE96c}%
  \BibitemOpen
  \bibfield  {author} {\bibinfo {author} {\bibfnamefont {G.}~\bibnamefont
  {Kresse}}\ and\ \bibinfo {author} {\bibfnamefont {J.}~\bibnamefont
  {Furthm{\"u}ller}},\ }\href {https://doi.org/10.1103/PhysRevB.54.11169}
  {\bibfield  {journal} {\bibinfo  {journal} {Phys. Rev. B}\ }\textbf {\bibinfo
  {volume} {54}},\ \bibinfo {pages} {11169} (\bibinfo {year}
  {1996}{\natexlab{a}})}\BibitemShut {NoStop}%
\bibitem [{\citenamefont {Kresse}\ and\ \citenamefont
  {Furthm{\"u}ller}(1996{\natexlab{b}})}]{KRE96b}%
  \BibitemOpen
  \bibfield  {author} {\bibinfo {author} {\bibfnamefont {G.}~\bibnamefont
  {Kresse}}\ and\ \bibinfo {author} {\bibfnamefont {J.}~\bibnamefont
  {Furthm{\"u}ller}},\ }\href {https://doi.org/10.1016/0927-0256(96)00008-0}
  {\bibfield  {journal} {\bibinfo  {journal} {Computational Materials Science}\
  }\textbf {\bibinfo {volume} {6}},\ \bibinfo {pages} {15} (\bibinfo {year}
  {1996}{\natexlab{b}})}\BibitemShut {NoStop}%
\bibitem [{\citenamefont {Giannozzi}\ \emph {et~al.}(2009)\citenamefont
  {Giannozzi}, \citenamefont {Baroni}, \citenamefont {Bonini}, \citenamefont
  {Calandra}, \citenamefont {Car}, \citenamefont {Cavazzoni}, \citenamefont
  {Ceresoli}, \citenamefont {Chiarotti}, \citenamefont {Cococcioni},
  \citenamefont {Dabo}, \citenamefont {Corso}, \citenamefont {de~Gironcoli},
  \citenamefont {Fabris}, \citenamefont {Fratesi}, \citenamefont {Gebauer},
  \citenamefont {Gerstmann}, \citenamefont {Gougoussis}, \citenamefont
  {Kokalj}, \citenamefont {Lazzeri}, \citenamefont {{Martin-Samos}},
  \citenamefont {Marzari}, \citenamefont {Mauri}, \citenamefont {Mazzarello},
  \citenamefont {Paolini}, \citenamefont {Pasquarello}, \citenamefont
  {Paulatto}, \citenamefont {Sbraccia}, \citenamefont {Scandolo}, \citenamefont
  {Sclauzero}, \citenamefont {Seitsonen}, \citenamefont {Smogunov},
  \citenamefont {Umari},\ and\ \citenamefont {Wentzcovitch}}]{GIA09}%
  \BibitemOpen
  \bibfield  {author} {\bibinfo {author} {\bibfnamefont {P.}~\bibnamefont
  {Giannozzi}}, \bibinfo {author} {\bibfnamefont {S.}~\bibnamefont {Baroni}},
  \bibinfo {author} {\bibfnamefont {N.}~\bibnamefont {Bonini}}, \bibinfo
  {author} {\bibfnamefont {M.}~\bibnamefont {Calandra}}, \bibinfo {author}
  {\bibfnamefont {R.}~\bibnamefont {Car}}, \bibinfo {author} {\bibfnamefont
  {C.}~\bibnamefont {Cavazzoni}}, \bibinfo {author} {\bibfnamefont
  {D.}~\bibnamefont {Ceresoli}}, \bibinfo {author} {\bibfnamefont {G.~L.}\
  \bibnamefont {Chiarotti}}, \bibinfo {author} {\bibfnamefont {M.}~\bibnamefont
  {Cococcioni}}, \bibinfo {author} {\bibfnamefont {I.}~\bibnamefont {Dabo}},
  \bibinfo {author} {\bibfnamefont {A.~D.}\ \bibnamefont {Corso}}, \bibinfo
  {author} {\bibfnamefont {S.}~\bibnamefont {de~Gironcoli}}, \bibinfo {author}
  {\bibfnamefont {S.}~\bibnamefont {Fabris}}, \bibinfo {author} {\bibfnamefont
  {G.}~\bibnamefont {Fratesi}}, \bibinfo {author} {\bibfnamefont
  {R.}~\bibnamefont {Gebauer}}, \bibinfo {author} {\bibfnamefont
  {U.}~\bibnamefont {Gerstmann}}, \bibinfo {author} {\bibfnamefont
  {C.}~\bibnamefont {Gougoussis}}, \bibinfo {author} {\bibfnamefont
  {A.}~\bibnamefont {Kokalj}}, \bibinfo {author} {\bibfnamefont
  {M.}~\bibnamefont {Lazzeri}}, \bibinfo {author} {\bibfnamefont
  {L.}~\bibnamefont {{Martin-Samos}}}, \bibinfo {author} {\bibfnamefont
  {N.}~\bibnamefont {Marzari}}, \bibinfo {author} {\bibfnamefont
  {F.}~\bibnamefont {Mauri}}, \bibinfo {author} {\bibfnamefont
  {R.}~\bibnamefont {Mazzarello}}, \bibinfo {author} {\bibfnamefont
  {S.}~\bibnamefont {Paolini}}, \bibinfo {author} {\bibfnamefont
  {A.}~\bibnamefont {Pasquarello}}, \bibinfo {author} {\bibfnamefont
  {L.}~\bibnamefont {Paulatto}}, \bibinfo {author} {\bibfnamefont
  {C.}~\bibnamefont {Sbraccia}}, \bibinfo {author} {\bibfnamefont
  {S.}~\bibnamefont {Scandolo}}, \bibinfo {author} {\bibfnamefont
  {G.}~\bibnamefont {Sclauzero}}, \bibinfo {author} {\bibfnamefont {A.~P.}\
  \bibnamefont {Seitsonen}}, \bibinfo {author} {\bibfnamefont {A.}~\bibnamefont
  {Smogunov}}, \bibinfo {author} {\bibfnamefont {P.}~\bibnamefont {Umari}},\
  and\ \bibinfo {author} {\bibfnamefont {R.~M.}\ \bibnamefont {Wentzcovitch}},\
  }\href {https://doi.org/10.1088/0953-8984/21/39/395502} {\bibfield  {journal}
  {\bibinfo  {journal} {J. Phys.: Condens. Matter}\ }\textbf {\bibinfo {volume}
  {21}},\ \bibinfo {pages} {395502} (\bibinfo {year} {2009})}\BibitemShut
  {NoStop}%
\bibitem [{\citenamefont {Perdew}\ \emph {et~al.}(1996)\citenamefont {Perdew},
  \citenamefont {Burke},\ and\ \citenamefont {Ernzerhof}}]{PER96a}%
  \BibitemOpen
  \bibfield  {author} {\bibinfo {author} {\bibfnamefont {J.~P.}\ \bibnamefont
  {Perdew}}, \bibinfo {author} {\bibfnamefont {K.}~\bibnamefont {Burke}},\ and\
  \bibinfo {author} {\bibfnamefont {M.}~\bibnamefont {Ernzerhof}},\ }\href
  {https://doi.org/10.1103/PhysRevLett.77.3865} {\bibfield  {journal} {\bibinfo
   {journal} {Phys. Rev. Lett.}\ }\textbf {\bibinfo {volume} {77}},\ \bibinfo
  {pages} {3865} (\bibinfo {year} {1996})}\BibitemShut {NoStop}%
\bibitem [{\citenamefont {Togo}\ and\ \citenamefont {Tanaka}(2015)}]{TOG15}%
  \BibitemOpen
  \bibfield  {author} {\bibinfo {author} {\bibfnamefont {A.}~\bibnamefont
  {Togo}}\ and\ \bibinfo {author} {\bibfnamefont {I.}~\bibnamefont {Tanaka}},\
  }\href {https://doi.org/10.1016/j.scriptamat.2015.07.021} {\bibfield
  {journal} {\bibinfo  {journal} {Scr. Mater.}\ }\textbf {\bibinfo {volume}
  {108}},\ \bibinfo {pages} {1} (\bibinfo {year} {2015})}\BibitemShut {NoStop}%
\bibitem [{\citenamefont {Prandini}\ \emph {et~al.}(2018)\citenamefont
  {Prandini}, \citenamefont {Marrazzo}, \citenamefont {Castelli}, \citenamefont
  {Mounet},\ and\ \citenamefont {Marzari}}]{PRA18}%
  \BibitemOpen
  \bibfield  {author} {\bibinfo {author} {\bibfnamefont {G.}~\bibnamefont
  {Prandini}}, \bibinfo {author} {\bibfnamefont {A.}~\bibnamefont {Marrazzo}},
  \bibinfo {author} {\bibfnamefont {I.~E.}\ \bibnamefont {Castelli}}, \bibinfo
  {author} {\bibfnamefont {N.}~\bibnamefont {Mounet}},\ and\ \bibinfo {author}
  {\bibfnamefont {N.}~\bibnamefont {Marzari}},\ }\href
  {https://doi.org/10.1038/s41524-018-0127-2} {\bibfield  {journal} {\bibinfo
  {journal} {npj Comput Mater}\ }\textbf {\bibinfo {volume} {4}},\ \bibinfo
  {pages} {1} (\bibinfo {year} {2018})}\BibitemShut {NoStop}%
\bibitem [{\citenamefont {Kresse}\ and\ \citenamefont
  {Hafner}(1993{\natexlab{a}})}]{KRE93}%
  \BibitemOpen
  \bibfield  {author} {\bibinfo {author} {\bibfnamefont {G.}~\bibnamefont
  {Kresse}}\ and\ \bibinfo {author} {\bibfnamefont {J.}~\bibnamefont
  {Hafner}},\ }\href {https://doi.org/10.1103/PhysRevB.48.13115} {\bibfield
  {journal} {\bibinfo  {journal} {Phys. Rev. B}\ }\textbf {\bibinfo {volume}
  {48}},\ \bibinfo {pages} {13115} (\bibinfo {year}
  {1993}{\natexlab{a}})}\BibitemShut {NoStop}%
\bibitem [{\citenamefont {Kresse}\ and\ \citenamefont
  {Hafner}(1993{\natexlab{b}})}]{KRE93a}%
  \BibitemOpen
  \bibfield  {author} {\bibinfo {author} {\bibfnamefont {G.}~\bibnamefont
  {Kresse}}\ and\ \bibinfo {author} {\bibfnamefont {J.}~\bibnamefont
  {Hafner}},\ }\href {https://doi.org/10.1103/PhysRevB.47.558} {\bibfield
  {journal} {\bibinfo  {journal} {Phys. Rev. B}\ }\textbf {\bibinfo {volume}
  {47}},\ \bibinfo {pages} {558} (\bibinfo {year}
  {1993}{\natexlab{b}})}\BibitemShut {NoStop}%
\bibitem [{\citenamefont {Marzari}\ and\ \citenamefont
  {Vanderbilt}(1997)}]{MAR97c}%
  \BibitemOpen
  \bibfield  {author} {\bibinfo {author} {\bibfnamefont {N.}~\bibnamefont
  {Marzari}}\ and\ \bibinfo {author} {\bibfnamefont {D.}~\bibnamefont
  {Vanderbilt}},\ }\href {https://doi.org/10.1103/PhysRevB.56.12847} {\bibfield
   {journal} {\bibinfo  {journal} {Phys. Rev. B}\ }\textbf {\bibinfo {volume}
  {56}},\ \bibinfo {pages} {12847} (\bibinfo {year} {1997})}\BibitemShut
  {NoStop}%
\bibitem [{\citenamefont {Souza}\ \emph {et~al.}(2001)\citenamefont {Souza},
  \citenamefont {Marzari},\ and\ \citenamefont {Vanderbilt}}]{SOU01b}%
  \BibitemOpen
  \bibfield  {author} {\bibinfo {author} {\bibfnamefont {I.}~\bibnamefont
  {Souza}}, \bibinfo {author} {\bibfnamefont {N.}~\bibnamefont {Marzari}},\
  and\ \bibinfo {author} {\bibfnamefont {D.}~\bibnamefont {Vanderbilt}},\
  }\href {https://doi.org/10.1103/PhysRevB.65.035109} {\bibfield  {journal}
  {\bibinfo  {journal} {Phys. Rev. B}\ }\textbf {\bibinfo {volume} {65}},\
  \bibinfo {pages} {035109} (\bibinfo {year} {2001})}\BibitemShut {NoStop}%
\bibitem [{\citenamefont {Marzari}\ \emph {et~al.}(2012)\citenamefont
  {Marzari}, \citenamefont {Mostofi}, \citenamefont {Yates}, \citenamefont
  {Souza},\ and\ \citenamefont {Vanderbilt}}]{MAR12a}%
  \BibitemOpen
  \bibfield  {author} {\bibinfo {author} {\bibfnamefont {N.}~\bibnamefont
  {Marzari}}, \bibinfo {author} {\bibfnamefont {A.~A.}\ \bibnamefont
  {Mostofi}}, \bibinfo {author} {\bibfnamefont {J.~R.}\ \bibnamefont {Yates}},
  \bibinfo {author} {\bibfnamefont {I.}~\bibnamefont {Souza}},\ and\ \bibinfo
  {author} {\bibfnamefont {D.}~\bibnamefont {Vanderbilt}},\ }\href
  {https://doi.org/10.1103/RevModPhys.84.1419} {\bibfield  {journal} {\bibinfo
  {journal} {Rev. Mod. Phys.}\ }\textbf {\bibinfo {volume} {84}},\ \bibinfo
  {pages} {1419} (\bibinfo {year} {2012})}\BibitemShut {NoStop}%
\bibitem [{\citenamefont {Pizzi}\ \emph {et~al.}(2020)\citenamefont {Pizzi},
  \citenamefont {Vitale}, \citenamefont {Arita}, \citenamefont {Bl{\"u}gel},
  \citenamefont {Freimuth}, \citenamefont {G{\'e}ranton}, \citenamefont
  {Gibertini}, \citenamefont {Gresch}, \citenamefont {Johnson}, \citenamefont
  {Koretsune}, \citenamefont {{Iba{\~n}ez-Azpiroz}}, \citenamefont {Lee},
  \citenamefont {Lihm}, \citenamefont {Marchand}, \citenamefont {Marrazzo},
  \citenamefont {Mokrousov}, \citenamefont {Mustafa}, \citenamefont {Nohara},
  \citenamefont {Nomura}, \citenamefont {Paulatto}, \citenamefont {Ponc{\'e}},
  \citenamefont {Ponweiser}, \citenamefont {Qiao}, \citenamefont {Th{\"o}le},
  \citenamefont {Tsirkin}, \citenamefont {Wierzbowska}, \citenamefont
  {Marzari}, \citenamefont {Vanderbilt}, \citenamefont {Souza}, \citenamefont
  {Mostofi},\ and\ \citenamefont {Yates}}]{PIZ20a}%
  \BibitemOpen
  \bibfield  {author} {\bibinfo {author} {\bibfnamefont {G.}~\bibnamefont
  {Pizzi}}, \bibinfo {author} {\bibfnamefont {V.}~\bibnamefont {Vitale}},
  \bibinfo {author} {\bibfnamefont {R.}~\bibnamefont {Arita}}, \bibinfo
  {author} {\bibfnamefont {S.}~\bibnamefont {Bl{\"u}gel}}, \bibinfo {author}
  {\bibfnamefont {F.}~\bibnamefont {Freimuth}}, \bibinfo {author}
  {\bibfnamefont {G.}~\bibnamefont {G{\'e}ranton}}, \bibinfo {author}
  {\bibfnamefont {M.}~\bibnamefont {Gibertini}}, \bibinfo {author}
  {\bibfnamefont {D.}~\bibnamefont {Gresch}}, \bibinfo {author} {\bibfnamefont
  {C.}~\bibnamefont {Johnson}}, \bibinfo {author} {\bibfnamefont
  {T.}~\bibnamefont {Koretsune}}, \bibinfo {author} {\bibfnamefont
  {J.}~\bibnamefont {{Iba{\~n}ez-Azpiroz}}}, \bibinfo {author} {\bibfnamefont
  {H.}~\bibnamefont {Lee}}, \bibinfo {author} {\bibfnamefont {J.-M.}\
  \bibnamefont {Lihm}}, \bibinfo {author} {\bibfnamefont {D.}~\bibnamefont
  {Marchand}}, \bibinfo {author} {\bibfnamefont {A.}~\bibnamefont {Marrazzo}},
  \bibinfo {author} {\bibfnamefont {Y.}~\bibnamefont {Mokrousov}}, \bibinfo
  {author} {\bibfnamefont {J.~I.}\ \bibnamefont {Mustafa}}, \bibinfo {author}
  {\bibfnamefont {Y.}~\bibnamefont {Nohara}}, \bibinfo {author} {\bibfnamefont
  {Y.}~\bibnamefont {Nomura}}, \bibinfo {author} {\bibfnamefont
  {L.}~\bibnamefont {Paulatto}}, \bibinfo {author} {\bibfnamefont
  {S.}~\bibnamefont {Ponc{\'e}}}, \bibinfo {author} {\bibfnamefont
  {T.}~\bibnamefont {Ponweiser}}, \bibinfo {author} {\bibfnamefont
  {J.}~\bibnamefont {Qiao}}, \bibinfo {author} {\bibfnamefont {F.}~\bibnamefont
  {Th{\"o}le}}, \bibinfo {author} {\bibfnamefont {S.~S.}\ \bibnamefont
  {Tsirkin}}, \bibinfo {author} {\bibfnamefont {M.}~\bibnamefont
  {Wierzbowska}}, \bibinfo {author} {\bibfnamefont {N.}~\bibnamefont
  {Marzari}}, \bibinfo {author} {\bibfnamefont {D.}~\bibnamefont {Vanderbilt}},
  \bibinfo {author} {\bibfnamefont {I.}~\bibnamefont {Souza}}, \bibinfo
  {author} {\bibfnamefont {A.~A.}\ \bibnamefont {Mostofi}},\ and\ \bibinfo
  {author} {\bibfnamefont {J.~R.}\ \bibnamefont {Yates}},\ }\href
  {https://doi.org/10.1088/1361-648X/ab51ff} {\bibfield  {journal} {\bibinfo
  {journal} {J. Phys.: Condens. Matter}\ }\textbf {\bibinfo {volume} {32}},\
  \bibinfo {pages} {165902} (\bibinfo {year} {2020})}\BibitemShut {NoStop}%
\bibitem [{\citenamefont {Wu}\ \emph {et~al.}(2018)\citenamefont {Wu},
  \citenamefont {Zhang}, \citenamefont {Song}, \citenamefont {Troyer},\ and\
  \citenamefont {Soluyanov}}]{WU18b}%
  \BibitemOpen
  \bibfield  {author} {\bibinfo {author} {\bibfnamefont {Q.}~\bibnamefont
  {Wu}}, \bibinfo {author} {\bibfnamefont {S.}~\bibnamefont {Zhang}}, \bibinfo
  {author} {\bibfnamefont {H.-F.}\ \bibnamefont {Song}}, \bibinfo {author}
  {\bibfnamefont {M.}~\bibnamefont {Troyer}},\ and\ \bibinfo {author}
  {\bibfnamefont {A.~A.}\ \bibnamefont {Soluyanov}},\ }\href
  {https://doi.org/10.1016/j.cpc.2017.09.033} {\bibfield  {journal} {\bibinfo
  {journal} {Comput. Phys. Commun.}\ }\textbf {\bibinfo {volume} {224}},\
  \bibinfo {pages} {405} (\bibinfo {year} {2018})}\BibitemShut {NoStop}%
\bibitem [{\citenamefont {Ganose}\ \emph {et~al.}(2021)\citenamefont {Ganose},
  \citenamefont {Searle}, \citenamefont {Jain},\ and\ \citenamefont
  {Griffin}}]{GAN21}%
  \BibitemOpen
  \bibfield  {author} {\bibinfo {author} {\bibfnamefont {A.~M.}\ \bibnamefont
  {Ganose}}, \bibinfo {author} {\bibfnamefont {A.}~\bibnamefont {Searle}},
  \bibinfo {author} {\bibfnamefont {A.}~\bibnamefont {Jain}},\ and\ \bibinfo
  {author} {\bibfnamefont {S.~M.}\ \bibnamefont {Griffin}},\ }\href
  {https://doi.org/10.21105/joss.03089} {\bibfield  {journal} {\bibinfo
  {journal} {J. Open Source Softw.}\ }\textbf {\bibinfo {volume} {6}},\
  \bibinfo {pages} {3089} (\bibinfo {year} {2021})}\BibitemShut {NoStop}%
\bibitem [{\citenamefont {Hu}\ \emph {et~al.}(2023{\natexlab{c}})\citenamefont
  {Hu}, \citenamefont {Bernevig},\ and\ \citenamefont {Tsvelik}}]{HU23a}%
  \BibitemOpen
  \bibfield  {author} {\bibinfo {author} {\bibfnamefont {H.}~\bibnamefont
  {Hu}}, \bibinfo {author} {\bibfnamefont {B.~A.}\ \bibnamefont {Bernevig}},\
  and\ \bibinfo {author} {\bibfnamefont {A.~M.}\ \bibnamefont {Tsvelik}},\
  }\href {https://doi.org/10.48550/arXiv.2301.04669} {\bibinfo {title} {Kondo
  {{Lattice Model}} of {{Magic-Angle Twisted-Bilayer Graphene}}: {{Hund}}'s
  {{Rule}}, {{Local-Moment Fluctuations}}, and {{Low-Energy Effective
  Theory}}}} (\bibinfo {year} {2023}{\natexlab{c}}),\ \Eprint
  {https://arxiv.org/abs/2301.04669} {arxiv:2301.04669 [cond-mat]} \BibitemShut
  {NoStop}%
\bibitem [{\citenamefont {Hamann}\ \emph {et~al.}(1979)\citenamefont {Hamann},
  \citenamefont {Schl{\"u}ter},\ and\ \citenamefont {Chiang}}]{HAM79}%
  \BibitemOpen
  \bibfield  {author} {\bibinfo {author} {\bibfnamefont {D.~R.}\ \bibnamefont
  {Hamann}}, \bibinfo {author} {\bibfnamefont {M.}~\bibnamefont
  {Schl{\"u}ter}},\ and\ \bibinfo {author} {\bibfnamefont {C.}~\bibnamefont
  {Chiang}},\ }\href {https://doi.org/10.1103/PhysRevLett.43.1494} {\bibfield
  {journal} {\bibinfo  {journal} {Phys. Rev. Lett.}\ }\textbf {\bibinfo
  {volume} {43}},\ \bibinfo {pages} {1494} (\bibinfo {year}
  {1979})}\BibitemShut {NoStop}%
\bibitem [{\citenamefont {Vanderbilt}(1990)}]{VAN90}%
  \BibitemOpen
  \bibfield  {author} {\bibinfo {author} {\bibfnamefont {D.}~\bibnamefont
  {Vanderbilt}},\ }\href {https://doi.org/10.1103/PhysRevB.41.7892} {\bibfield
  {journal} {\bibinfo  {journal} {Phys. Rev. B}\ }\textbf {\bibinfo {volume}
  {41}},\ \bibinfo {pages} {7892} (\bibinfo {year} {1990})}\BibitemShut
  {NoStop}%
\bibitem [{\citenamefont {Kresse}\ and\ \citenamefont {Joubert}(1999)}]{KRE99}%
  \BibitemOpen
  \bibfield  {author} {\bibinfo {author} {\bibfnamefont {G.}~\bibnamefont
  {Kresse}}\ and\ \bibinfo {author} {\bibfnamefont {D.}~\bibnamefont
  {Joubert}},\ }\href {https://doi.org/10.1103/PhysRevB.59.1758} {\bibfield
  {journal} {\bibinfo  {journal} {Phys. Rev. B}\ }\textbf {\bibinfo {volume}
  {59}},\ \bibinfo {pages} {1758} (\bibinfo {year} {1999})}\BibitemShut
  {NoStop}%
\bibitem [{\citenamefont {Xu}\ \emph {et~al.}(2022)\citenamefont {Xu},
  \citenamefont {Vergniory}, \citenamefont {Ma}, \citenamefont {Ma{\~n}es},
  \citenamefont {Song}, \citenamefont {Bernevig}, \citenamefont {Regnault},\
  and\ \citenamefont {Elcoro}}]{XU22}%
  \BibitemOpen
  \bibfield  {author} {\bibinfo {author} {\bibfnamefont {Y.}~\bibnamefont
  {Xu}}, \bibinfo {author} {\bibfnamefont {M.~G.}\ \bibnamefont {Vergniory}},
  \bibinfo {author} {\bibfnamefont {D.-S.}\ \bibnamefont {Ma}}, \bibinfo
  {author} {\bibfnamefont {J.~L.}\ \bibnamefont {Ma{\~n}es}}, \bibinfo {author}
  {\bibfnamefont {Z.-D.}\ \bibnamefont {Song}}, \bibinfo {author}
  {\bibfnamefont {B.~A.}\ \bibnamefont {Bernevig}}, \bibinfo {author}
  {\bibfnamefont {N.}~\bibnamefont {Regnault}},\ and\ \bibinfo {author}
  {\bibfnamefont {L.}~\bibnamefont {Elcoro}},\ }\href
  {https://doi.org/10.48550/arXiv.2211.11776} {\bibinfo {title} {Catalogue of
  topological phonon materials}} (\bibinfo {year} {2022}),\ \Eprint
  {https://arxiv.org/abs/2211.11776} {arxiv:2211.11776 [cond-mat]} \BibitemShut
  {NoStop}%
\bibitem [{\citenamefont {Bergman}\ \emph {et~al.}(2008)\citenamefont
  {Bergman}, \citenamefont {Wu},\ and\ \citenamefont {Balents}}]{BER08}%
  \BibitemOpen
  \bibfield  {author} {\bibinfo {author} {\bibfnamefont {D.~L.}\ \bibnamefont
  {Bergman}}, \bibinfo {author} {\bibfnamefont {C.}~\bibnamefont {Wu}},\ and\
  \bibinfo {author} {\bibfnamefont {L.}~\bibnamefont {Balents}},\ }\href
  {https://doi.org/10.1103/PhysRevB.78.125104} {\bibfield  {journal} {\bibinfo
  {journal} {Phys. Rev. B}\ }\textbf {\bibinfo {volume} {78}},\ \bibinfo
  {pages} {125104} (\bibinfo {year} {2008})}\BibitemShut {NoStop}%
\bibitem [{\citenamefont {Damascelli}\ \emph {et~al.}(2003)\citenamefont
  {Damascelli}, \citenamefont {Hussain},\ and\ \citenamefont {Shen}}]{DAM03}%
  \BibitemOpen
  \bibfield  {author} {\bibinfo {author} {\bibfnamefont {A.}~\bibnamefont
  {Damascelli}}, \bibinfo {author} {\bibfnamefont {Z.}~\bibnamefont
  {Hussain}},\ and\ \bibinfo {author} {\bibfnamefont {Z.-X.}\ \bibnamefont
  {Shen}},\ }\href {https://doi.org/10.1103/RevModPhys.75.473} {\bibfield
  {journal} {\bibinfo  {journal} {Rev. Mod. Phys.}\ }\textbf {\bibinfo {volume}
  {75}},\ \bibinfo {pages} {473} (\bibinfo {year} {2003})}\BibitemShut
  {NoStop}%
\bibitem [{\citenamefont {Moser}(2017)}]{MOS17}%
  \BibitemOpen
  \bibfield  {author} {\bibinfo {author} {\bibfnamefont {S.}~\bibnamefont
  {Moser}},\ }\href {https://doi.org/10.1016/j.elspec.2016.11.007} {\bibfield
  {journal} {\bibinfo  {journal} {J. Electron Spectros. Relat. Phenomena}\
  }\textbf {\bibinfo {volume} {214}},\ \bibinfo {pages} {29} (\bibinfo {year}
  {2017})}\BibitemShut {NoStop}%
\bibitem [{\citenamefont {Shankar}(2013)}]{SHA13}%
  \BibitemOpen
  \bibfield  {author} {\bibinfo {author} {\bibfnamefont {R.}~\bibnamefont
  {Shankar}},\ }\href@noop {} {\emph {\bibinfo {title} {Principles of {{Quantum
  Mechanics}}}}},\ \bibinfo {edition} {softcover reprint of the original 1st
  ed. 1980 edition}\ ed.\ (\bibinfo  {publisher} {{Springer}},\ \bibinfo {year}
  {2013})\BibitemShut {NoStop}%
\bibitem [{\citenamefont {Pavarini}\ \emph {et~al.}(2014)\citenamefont
  {Pavarini}, \citenamefont {Koch}, \citenamefont {Vollhardt},\ and\
  \citenamefont {Lichtenstein}}]{PAV14}%
  \BibitemOpen
  \bibfield  {author} {\bibinfo {author} {\bibfnamefont {E.}~\bibnamefont
  {Pavarini}}, \bibinfo {author} {\bibfnamefont {E.}~\bibnamefont {Koch}},
  \bibinfo {author} {\bibfnamefont {D.}~\bibnamefont {Vollhardt}},\ and\
  \bibinfo {author} {\bibfnamefont {A.}~\bibnamefont {Lichtenstein}},\
  }\href@noop {} {\emph {\bibinfo {title} {{{DMFT}} at 25: {{Infinite
  Dimensions}}: {{Lecture Notes}} of the {{Autumn School}} on {{Correlated
  Electrons}} 2014}}}\ (\bibinfo  {publisher} {{Forschungszentrum J\"ulich}},\
  \bibinfo {year} {2014})\BibitemShut {NoStop}%
\bibitem [{\citenamefont {Mahan}(2000)}]{MAH00}%
  \BibitemOpen
  \bibfield  {author} {\bibinfo {author} {\bibfnamefont {G.~D.}\ \bibnamefont
  {Mahan}},\ }\href@noop {} {\emph {\bibinfo {title} {Many Particle
  Physics}}},\ \bibinfo {edition} {3rd}\ ed.,\ Physics of Solids and Liquids\
  (\bibinfo  {publisher} {{Springer Science + Business Media, LLC}},\ \bibinfo
  {address} {{New York}},\ \bibinfo {year} {2000})\BibitemShut {NoStop}%
\end{thebibliography}
\end{document}